\documentclass[12pt,letterpaper]{article}
\pdfoutput=1
\usepackage[dvipdfmx]{graphicx} 
\usepackage{bmpsize}

\usepackage{lmodern}
\usepackage[english]{babel}
\usepackage{authblk}
\usepackage{longtable}
\usepackage{soul}
\usepackage{cite}
\usepackage{url}
\usepackage{amsmath}
\usepackage{amssymb}
\usepackage{hyperref}
\usepackage[T1]{fontenc}
\usepackage[utf8]{inputenc}
\usepackage{todonotes}
\usepackage{adjustbox}
\usepackage{graphicx}
\usepackage{subfigure}
\usepackage{multirow}
\usepackage{booktabs}
\usepackage[tableposition=top,small]{caption}
\usepackage{lineno}
\hypersetup{
    colorlinks=true,
    linkcolor=cyan,
    filecolor=magenta,      
    urlcolor=blue,
}
\expandafter\def\expandafter\UrlBreaks\expandafter{\UrlBreaks% save the current one
  \do\a\do\b\do\c\do\d\do\e\do\f\do\g\do\h\do\i\do\j%
  \do\k\do\l\do\m\do\n\do\o\do\p\do\q\do\r\do\s\do\t%
  \do\u\do\v\do\w\do\x\do\y\do\z\do\A\do\B\do\C\do\D%
  \do\E\do\F\do\G\do\H\do\I\do\J\do\K\do\L\do\M\do\N%
  \do\O\do\P\do\Q\do\R\do\S\do\T\do\U\do\V\do\W\do\X%
  \do\Y\do\Z\do\*\do\-\do\~\do\'\do\"\do\-}%

\PassOptionsToPackage{unicode}{hyperref}
\PassOptionsToPackage{naturalnames}{hyperref}

\newcommand{\eq}{Eq.}

\newcommand{\Refs}{Refs.}
\newcommand{\Sec}{Section}

\newcommand{\Tab}{Tab.}

\newcommand{\equ}[1]{\eq~(\ref{equ:#1})}

\newcommand{\figu}[1]{Fig.~\ref{fig:#1}}

\newcommand{\parenbar}[1]{\ensuremath{\overset{\scriptscriptstyle{(-)}}{#1}}}
\newcommand{\gsim}      {\mbox{\raisebox{-0.4ex}{$\;\stackrel{>}{\scriptstyle \sim}\;$}}}
\newcommand{\lsim}      {\mbox{\raisebox{-0.4ex}{$\;\stackrel{<}{\scriptstyle \sim}\;$}}}

\makeatletter         
\def\@maketitle{
\begin{center}
{\Huge \bfseries \sffamily \@title }\\[4ex] 
\includegraphics[width = 40mm,height=40mm]{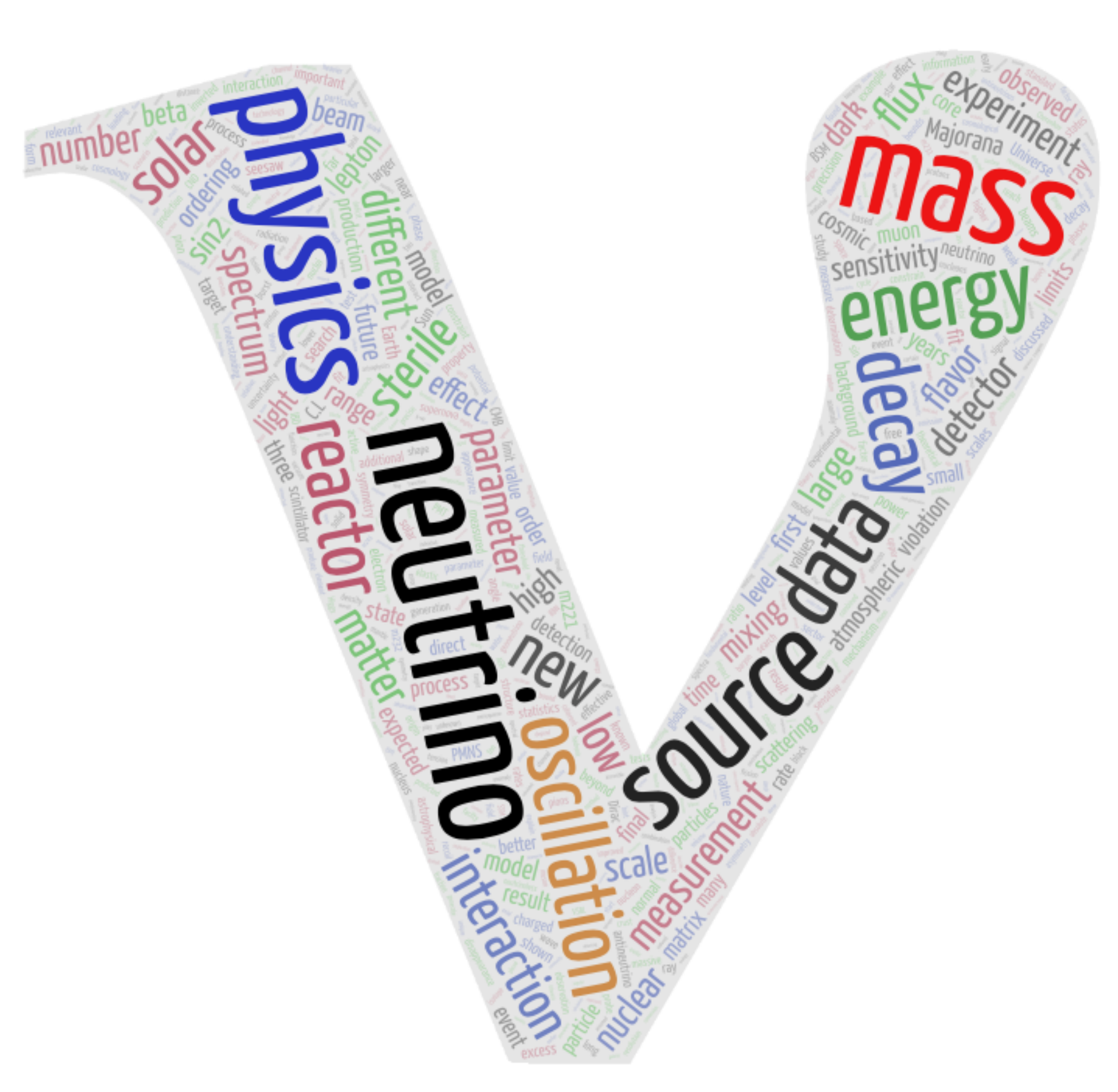}\\[2ex]
{\Large  \@author}\\[4ex] 
\end{center}}
\makeatother

\title{{\bf \Large Status and Perspectives of Neutrino Physics}}

\author[1]{\bf  M.\ Sajjad Athar}
\author[2]{\bf Steven W.\ Barwick}
\author[3,4]{\bf Thomas~Brunner}
\author[5]{\bf Jun Cao}
\author[6]{\bf Mikhail~Danilov}
\author[7]{\bf Kunio~Inoue}
\author[8]{\bf Takaaki~Kajita}
\author[9,10]{\bf Marek~Kowalski}
\author[11]{\bf Manfred~Lindner}
\author[12]{\bf Kenneth~R.~Long}
\author[13]{\bf Nathalie~Palanque-Delabrouille}
\author[11]{\bf Werner~Rodejohann}
\author[14]{\bf Heidi~Schellman}
\author[15]{\bf Kate~Scholberg}
\author[16]{\bf Seon-Hee~Seo}
\author[4,17,18]{\bf Nigel~J.T.~Smith}
\author[10]{\bf Walter~Winter}
\author[19]{\bf Geralyn~P.~Zeller}
\author[20]{{\bf Renata~Zukanovich~Funchal}}
\affil[1]{\it Department of Physics, Aligarh Muslim University, Aligarh-202002, India}
\affil[2]{\it Dept.\ of Physics and Astronomy, University of California, Irvine, CA 92697, USA}
\affil[3]{\it Physics Department, McGill University, Montréal, Québec H3A 2T8, Canada}
\affil[4]{\it TRIUMF, Vancouver, British Columbia V6T 2A3, Canada}
\affil[5]{\it Institute of High Energy Physics, Chinese Academy of Sciences, Beijing 100049, China}
\affil[6]{\it Lebedev Physical Institute of the Russian Academy of Sciences, 53 Leninskiy Prospekt, Moscow, 119991, Russia}
\affil[7]{\it Research Center for Neutrino Science, Tohoku University, Sendai 980-8578, Japan}
\affil[8]{\it Research Center for Cosmic Neutrinos, Institute for Cosmic Ray Research (ICRR), The University of Tokyo, 5-1-5, Kashiwanoha, Kashiwa, Chiba, 277-8582, Japan}
\affil[9]{\it Institut f\"ur Physik, Humboldt-Universit\"at zu Berlin, D-12489 Berlin, Germany}
\affil[10]{\it DESY, D-15738 Zeuthen, Germany}
\affil[11]{\it Max-Planck-Institut f\"ur Kernphysik, Saupfercheckweg 1, 69117 Heidelberg, Germany}
\affil[12]{\it Imperial College London, Department of Physics, London, United Kingdom}
\affil[13]{\it IRFU, CEA, Universite Paris-Saclay, F-91191 Gif-sur-Yvette}
\affil[14]{\it Department of Physics, Oregon State University, Corvallis OR, 97331, USA}
\affil[15]{\it Duke University, Durham, NC 27708, USA}
\affil[16]{\it Center for Underground Physics, Institute for Basic Science (IBS), 55, Expo-ro, Yuseong-gu, Daejeon, 34126, Korea}
\affil[17]{\it Department of Physics and Astronomy, Laurentian University, Sudbury, Ontario, P3E 2C6, Canada}
\affil[18]{\it SNOLAB, Lively, Ontario, P3Y 1M3, Canada}
\affil[19]{\it Fermi National Accelerator Laboratory, Batavia, IL 60510, USA}
\affil[20]{\it Instituto de Fisica, Universidade de Sao Paulo, C.P.\ 66.318, 05315-970 Sao Paulo, Brazil}

\newpage

\date{\today}

\newpage

\begin{document}

%\linenumbers
\maketitle

\begin{abstract}
\noindent
This review demonstrates the unique role of the neutrino by discussing in detail the physics of and with neutrinos. We deal with  neutrino sources, neutrino oscillations, absolute masses, interactions, the possible existence of sterile neutrinos, and theoretical implications. In addition, synergies of neutrino physics with other research fields are found, and requirements to continue successful neutrino physics in the future, in terms of technological developments and adequate infrastructures, are stressed. 
\end{abstract}

\newpage
\tableofcontents
\clearpage

\section*{Glossary}
\begin{longtable}{lp{12.0cm}}
AGN & Active Galactic Nucleus \\  
BBN & Big Bang Nucleosynthesis \\
BSM & Beyond the Standard Model \\
CKM  & Cabibbo-Kobayashi-Maskawa matrix (quark mixing matrix)\\ 
CMB & Cosmic Microwave Background \\
CE$\nu$NS   & Coherent Elastic Neutrino-Nucleus Scattering\\
CC & Charged Current \\
C.L. & Confidence Level \\
CNO & Carbon-Nitrogen-Oxygen\\
CP & Charge Parity \\
DM & Dark Matter\\
DSNB   & Diffuse Supernova Neutrino Background\\
 eV & electron Volt\\
GRB & Gamma Ray Burst\\
IBD  & Inverse Beta Decay\\
IO & Inverted Mass Ordering\\
 LMA & Large Mixing Angle \\
LBL & long baseline\\
LHC & Large Hadron Collider \\
LS & Liquid Scintillator\\
LFV & Lepton Flavor Violation \\
MO & Neutrino Mass Ordering \\
MSW  & Mikheyev-Smirnov-Wolfenstein effect\\
NC & Neutral Current \\
NSI & Non-Standard Interactions \\
NO &  Normal Mass Ordering\\
PMNS & Pontecorvo-Maki-Nakagawa-Saki matrix (lepton mixing matrix)\\
PMT & Photo Multiplier Tube \\ 
QE & Quasi-Elastic \\
RAA & Reactor Antineutrino Anomaly \\
R\&D & Research and Developement \\
SBL & very long baseline\\
SM & Standard Model \\
 SN & Supernova\\
t  & metric ton (tonne)\\ 
  TPC & Time Projection Chamber\\
   VSBL & Very Short Baseline\\
 W & Watt \\

$P(\nu_\alpha \to \nu_\beta)$ & neutrino oscillation probability \\
$U_{\alpha i}$ & PMNS matrix element, $\alpha = e, \mu, \tau$, $i=1,2,3$\\
   $\theta_{12}$ & mixing angle mainly for solar and LBL reactor neutrinos\\
$\theta_{13}$ & mixing angle mainly for SBL reactor and LBL accelerator neutrinos\\
$\theta_{23}$ & mixing angle mainly for atmospheric and LBL accelerator neutrinos\\
$\Delta m^2_{21}$ & mass-squared difference mainly for solar and LBL reactor neutrinos\\
$\Delta m^2_{31/2}$ & mass-squared difference mainly for accelerator, atmospheric and SBL reactor neutrinos\\
$\delta_{\rm CP} $ &  (Dirac) CP Phase in neutrino oscillations \\
$\alpha$, $\beta$ & (Majorana) CP phases in neutrinoless double beta decay\\
$ m_{1,2,3}$ &  active neutrino mass eigenstates \\
$\nu_e$ & electron neutrino, $SU(2)$-partner of electron\\
$\nu_\mu$ & muon neutrino, $SU(2)$-partner of muon\\ 
$\nu_\tau$ & tauon neutrino, $SU(2)$-partner of tauon\\  
$\nu_s$ & hypothetical sterile neutrino \\
$m_4$ & mass of hypothetical sterile neutrino\\
$\Delta m^2_{41}$   & mass-squared difference involving sterile neutrino\\
$N_{\rm eff} $ & effective number of neutrino families in cosmology\\

  $0\nu\beta\beta$ & neutrinoless double beta decay $(A,Z) \to (A,Z+2) + 2 e^-$\\
   & \\
   & \\
   & \\
   & \\
   & \\
   & \\
   & \\
   & \\
   & \\
   & \\
   & \\
   & \\
   & \\
   & \\
   & \\
   & \\
   & \\
   & \\
   & \\
   & \\
   & \\
\end{longtable}

\clearpage

\clearpage

{\bf Preamble:} 
This review of the field of neutrino physics emerged from a report written by a panel on the request of IUPAP (International Union of Pure and Applied Physics). The mandate, the panel members and the report can be found on the web page of the panel at \url{https://www.iupapneutrinopanel.org}.
The report is available at 
\url{https://www.iupapneutrinopanel.org/wp-content/uploads/2021/10/IUPAP_Neutrino_Panel_Report.pdf}
and at 
\url{https://iupap.org/who-we-are/internal-organization/commissions/c11-particles-and-fields/c11-reports/}
on the web pages of IUPAP.

For completeness, the Executive Summary of the IUPAP report can be found at the end of this document. 

%\end{document}

\section{Introduction}
\label{sec:motivation}

Neutrinos are very special particles which have led again and again to surprising and important discoveries, a number of which were recognized with Noble prizes. Neutrinos were theoretically invented in 1930 by Pauli to preserve energy-momentum conservation and their first experimental detection in 1956 by a team lead by Reines and Cowan at the Savannah River reactor was another landmark. Later it was found that three versions (flavors) exist, which was again a major discovery. Next, solar neutrinos showed oscillations on their way to Earth, which is a quantum mechanical effect, something usually only relevant on atomic scales. Neutrinos were found to have very tiny masses, which is so far the only solid evidence for particle physics beyond the Standard Model and  has important consequences for structures in the Universe. There are numerous other topics where it is already known that neutrinos play an important role, but there are also very good reasons and maybe even indications that more surprising results may show up in the future. 

Starting from what we know so far, namely three massive neutrinos which mix, one can organize neutrino research topics into two main directions: First, all known neutrino sources, artificial or natural, can be used to learn about the properties of neutrinos and their interactions. This leads to numerous unique and very important insights into the Standard Model of Particle Physics (SM) and into completely new physics Beyond the Standard Model (BSM). Second, neutrinos allow unique and important insights into the sources of neutrinos. Neutrinos from the Sun allow one, for example, to better understand in detail how stars work and evolve. The explosion of supernovae is another topic to which neutrinos can provide important contributions. Both of these main directions have various inter-dependencies and further connections to other fields. Neutrino physics unites thus a remarkably wide set of scientific communities. Besides astroparticle physics, particle physics, astronomy and cosmology, neutrino physics also has strong connections to nuclear physics, geology and even material science. 

These two main directions might also be called the physics {\em of} and the physics {\em with} neutrinos. Regarding the physics of neutrinos, it is useful to summarize the parameters that govern neutrino physics.  As all SM fermions, neutrinos come in three generations, that is, there are three flavor states $\nu_e$, $\nu_\mu$ and $\nu_\tau$, which live together with their charged lepton counterparts $e^-$, $\mu^-$ and $\tau^-$ in weak interaction doublets.  The neutrinos have well-defined quantum numbers under the SM gauge symmetries, which fix their interactions with the $W$ and $Z$ bosons of the electroweak interactions.  
Diagonalising the mass matrices of leptons and neutrinos yields the three known charged lepton masses. In addition to those, three neutrino masses $m_{1,2,3}$ are present, corresponding to the mass states $\nu_{1,2,3}$. Another consequence of diagonalisation is the existence of the PMNS matrix denoted here by $U$, which is the analogue of the CKM matrix in the quark sector; $U$ implies, for instance, that the electron-neutrino is a linear combination of the three mass states, $\nu_e = U_{ei} \nu_i$. 
For vanishing neutrino masses the PMNS matrix would be the identity matrix, because one can  identify their interaction eigenstates with the corresponding mass eigenstates up to phase redefinition. 
The PMNS matrix contains three mixing angles, $\theta_{12}$, $\theta_{13}$ and $\theta_{23}$, plus a phase $\delta_{\rm CP}$ responsible for CP violation. In case neutrinos are their own antiparticles, i.e.\ if they are Majorana fermions, two additional phases exist (denoted for instance by $\alpha$ and $\beta$), which only appear in lepton-number violating processes, and in particular  do not influence neutrino oscillations. 

These standard parameters are summarized in Tab.\ \ref{tab:sum1}, together with the main methods and neutrino sources to determine them. One subtlety exists here, namely it is not clear whether the mass state that is 
mostly composed of the first-generation electron neutrino state is the heaviest or the lightest one. 
This is the question of the mass ordering, which can be normal or inverted. In the established notation of the field the normal mass ordering corresponds to $m_3 > m_2 > m_1$, or $\Delta m^2_{31} > 0$, while the inverted mass ordering corresponds to $m_2 > m_1 > m_3$, or $\Delta m^2_{31} < 0$. Here the notation normal and inverted compares the situation to the quark sector, in which the mass state which is mostly composed of the first-generation up-quark is the lightest one. 
 
Apart from this standard paradigm of three massive (Majorana) neutrinos mixing with each other, more neutrino states may exist, which must be sterile, i.e.\ not participating in SM interactions except for via mixing with the  active states. Additional parameters such as magnetic moments may exist, or neutrinos may participate in new interactions beyond the SM. Furthermore, the mechanism that generates neutrino mass may come with new particles, energy states and parameters, whose main methods of determination needs to be discussed for each model individually.

\begin{table}[t]
  \centering
  \begin{tabular}{l|c|c|c}
  \hline
  \hline
 Parameter   & Main method(s)  &  Source(s) & Status  \\
  \hline \hline
$\theta_{12}$ & Oscillations & solar, reactor & known \\ 
$\theta_{23}$ & Oscillations & atmospheric, accelerator &  known\\
$\theta_{13}$ & Oscillations  & reactor, accelerator &  known \\ \hline 
$\delta_{\rm CP} $ & Oscillations & accelerator &  hints \\
$\alpha$, $\beta$ & Rare processes & double beta decay &  unknown \\ \hline
$\Delta m_{21}^2$ & Oscillations & reactor, solar & known \\
$|\Delta m_{31}^2|$ & Oscillations & reactor, accelerator, atmospheric & known \\
Ordering (sgn $\Delta m_{31}^2$) & Oscillations & reactor, accelerator, atmospheric & hints \\
$m_{1,2,3}$ & Kinematics & $\beta$ decay, cosmology & limits \\
   \hline \hline
   
  \end{tabular}
  \caption{Standard neutrino parameters, the main method(s) to determine them, the most important source(s) for the determination and the current status. Except the phases $\alpha$ and $\beta$ (for the case of Majorana neutrinos), all unknown parameters are expected to be determined within the next 10 years.}
  \label{tab:sum1}
\end{table}

Thus, the main questions of the physics of neutrinos relate to particle physics and address topics which can roughly be grouped as follows: 

\begin{itemize}
\item
 {\bf What are the properties of the neutrinos?\!} This includes ``expected'' properties, such as neutrino masses and mixings, the pattern and scale of the neutrino masses, the origin and nature of the neutrino mass terms,  as well as BSM properties, such as possible magnetic moments. For example, the flavor structure of the SM leptons seems to be very distinct from that of quarks, which indicates fundamental differences which cannot be captured by the SM. A key question is if there is leptonic CP violation, as this may be an indicator for neutrinos playing a role in generating the observed baryon asymmetry of the Universe (baryo/leptogenesis). Another and possibly related question is  whether neutrinos are Dirac or Majorana fermions, i.e., if lepton number is a  conserved or violated symmetry of nature. 
\item
 {\bf How do neutrinos interact?\!} On the SM side, the impact of nuclear physics is the main challenge here. Often the uncertainties of cross sections of neutrinos with the target material influence the precise determination of neutrino parameters. 
 In turn, neutrino interactions can help to refine nuclear models. There are also possible new  BSM interactions of neutrinos which are frequently described by effective four-fermion interactions (so-called ``non-standard interactions''), which need to be tested and which may have phenomenological impact on the extraction of neutrino properties. Neutrino interactions can be also tested at extremely high (PeV to EeV) energies, where BSM effects may most naturally contribute, using astrophysical neutrinos.
\item
 {\bf How many neutrino generations are there?\!} Sterile neutrinos (neutrinos which are not participating in weak interactions), may exist at different energy scales with implications in cosmology (eV scales), as warm dark matter candidates (keV scales) or even in baryogenesis (GeV scales and beyond). Since there have been experimental indications for neutrinos at eV mass-scale in short-baseline experiments, and the existence of sterile neutrinos has profound implications for our understanding of particle physics, sterile neutrinos need to be further tested. It is also an interesting theoretical question if neutrinos can solve the remaining puzzles in particle physics, such as the dark matter problem.
 \end{itemize}

The use of neutrinos with neutrinos, i.e., as probe of sources, can roughly be grouped as being sensitive to various extreme properties.  

\begin{itemize}
\item {\bf Extreme distances:\!}
The role of neutrinos as messengers is probably most evident in astrophysics; examples are the detection of neutrinos from supernova 1987A and of solar neutrinos, including the very recent confirmation of the existence of the carbon-nitrogen-oxygen %(CNO) 
fusion cycle in the Sun. Neutrinos can, however, see the Universe beyond our local environment, and are, in fact, the only known high-energy messengers which can directly penetrate through the whole Universe. In contrast, gamma-rays interact with the cosmic background radiation and charged cosmic rays are deflected by extragalactic magnetic fields. They are an indicator for the origin of the cosmic rays because they are produced in the interaction of cosmic rays with matter and radiation. While so far most of the electromagnetic signatures detected in astrophysics have been described by accelerated electrons and their radiation processes, the origin of cosmic rays remains a mystery. 
%-- which in some cases has been indirectly inferred by electromagnetic spectral slopes of features which cannot be described otherwise. 
Recent indications for neutrinos from jets in Active Galactic Nuclei and from the astrophysical phenomena accompanying the tidal disruption of massive stars approaching a black hole can be therefore interpreted as the first direct evidence for the origin of cosmic rays at PeV energies. Since ultra-high-energy cosmic rays %(UHECRs) 
interact with the cosmic background radiation, secondary neutrinos originating from such processes (frequently called ``cosmogenic neutrinos'') are an indicator for the composition of cosmic rays as well. 
\item{\bf Extreme environments:\!}
Apart from their role as messengers to study their sources, neutrinos and their properties govern the physics of astrophysical objects and even the whole Universe in a wide range of processes: they control the explosions of core-collapse supernovae, drive the winds from neutron star merger accretion disks, determine the ratio between protons and neutrons in astrophysical outflows which generate heavy elements, re-distribute energy in the formation of large-scale structure in the early Universe and are a key player in the primordial plasma and the formation of light elements. Because of the  importance of neutrinos in such extreme environments, it is also natural to expect that astrophysical environments constrain potential BSM properties of  neutrinos; prominent examples are cosmological constraints on neutrino mass and on the effective number of neutrinos constraining models with sterile neutrinos. 
\item{\bf Extreme past:\!}
The detection of primordial neutrinos (neutrinos that have decoupled from the primordial plasma in the early Universe), sometimes also referred to as ``big bang neutrinos'' or cosmic neutrino background), is therefore often perceived as the Holy Grail of neutrino (detection) physics. Detecting the effects of massive neutrinos in cosmological data sets is also probing physics at early times in the cosmological evolution. 
%Why would one do that at some point in the future? While a possible answer is: ``to better understand the Universe'', another one may be: ``because we can!''.
 \end{itemize}

The above attempt to classify the vast set of topics of neutrino physics unavoidably leaves some interesting topics out. For instance, by detecting neutrinos produced in radioactive decay chains of heavy elements found inside our planet, one can study the Earth's interior. 
%Neutrinos are, for example, also used to study the interior of the Earth. Radioactive decays of heavy elements can be detected by the neutrinos they produce, such as 
The isotopes $^{238}$U and $^{232}$Th are especially interesting because they produce neutrinos beyond the inverse beta decay threshold. This information can be used to learn about the magnitude and distribution of the Earth's radioactivity -- and may even be used for an independent determination of the Earth's age. On the other hand, neutrinos interact with Earth matter by coherent forward scattering affecting oscillations (MeV to GeV energies) and by increasing cross sections (beyond TeV energies), which can be used to study the interior of the Earth in terms of composition, density and structure. Another example is the use of neutrinos for nuclear non-proliferation, as the burning material of nuclear reactors can be tested via measurements of the neutrinos they emit. 

This rough overview shows that neutrino physics is a very wide field which connects very different scientific communities with vastly different scientific techniques and methods. This includes a huge range of energies spanning over 30 orders of magnitude, distance scales ranging from thousands of Megaparsec down to $10^{-20}$ meters or even below, experiments with high event rates and  a small number of events in huge experiments. Theoretical physics is here very important since it helps to combine results from completely different experiments, including non-neutrino experiments,  into one coherent overall physics picture. The combination leads to very important tests of the SM and to very powerful searches for new BSM physics which often cannot be done by the individual experiments. Theory is also important in guiding experiments by calculating the expected signals of BSM scenarios and to point out detection strategies within one experiment or by the combination of different experiments. 

Neutrino physics evolves in an exciting and promising way, but with a wide spectrum of technologies, growing detector sizes and time scales. In order to fully exploit the unique potential of neutrinos,  coordination is needed. 
This review sketches the status quo of neutrino physics, points out  future directions and recommendations on the need to balance different types and sizes of experiments and to make best use of resources by looking for synergies in R\&D efforts and in large-scale experiments.
%led IUPAP to set up a neutrino panel with the mandate {\sl\bf to promote international cooperation in the development of an experimental program to study the properties of neutrinos and to promote international collaboration in the development of future neutrino experiments to establish the properties of neutrinos}.
%This white paper is the report of this panel sketching the status quo, pointing to future directions and recommendations on the need to balance different types and sizes of experiments and to make best use of resources by looking for synergies in R\&D efforts and in large-scale experiments. 

The structure of this document follows therefore an experiment-driven approach: We first discuss the sources of the neutrinos, along with their physics aspects, then we discuss neutrino oscillations and absolute neutrino masses, which are the main particle physics-oriented targets. We then come back to SM physics and discuss neutrino interactions including their nuclear physics aspects. The possible existence of light sterile neutrinos and their potential consequences is also discussed. The document includes also a discussion on new technologies and cross-over topics to other fields. In the end of the document we outline the physics implications of present and future results and  their connection to various beyond-the-SM theories.

\clearpage

\section{Physics of Neutrino Sources} 
{Contributing additional authors: Stephen T.\ Dye (Hawaii Pacific U.), Livia Ludhova (RWTH Aachen and Forschungszentrum Jülich), Irene Tamborra (NBI), Christopher G.\ Tully (Princeton U.)}
\label{chap:sources}
%\documentclass{article}
%\usepackage[utf8]{inputenc}
%\usepackage{adjustbox}
%\usepackage{graphicx}

%\newcommand{\shs}[1]{\textcolor{blue}{#1}}

%\title{Neutrino Panel WG4: \\
%Physics of Neutrino Sources}
%\author{M. Kowalski, K. Scholberg, S. H. Seo}
%\date{\today}

%\begin{document}

%\maketitle
\label{WG4}

\subsection{Introduction}
\label{wg4_intro}

Neutrinos are the $2^{\rm nd}$ most abundant particles after photons in the visible Universe, yet very hard to work with because they only interact very weakly. Reactors are in addition a very intense man-made source of electron antineutrinos ($\sim2\times10^{20} \, \bar{\nu}_e$ per Giga-Watt thermal power, GW$_{\rm th}$). The Earth can also be considered as a ``natural'' reactor emitting electron antineutrinos from beta decay of radioactive elements present in the mantle and core of the Earth. 
Electron neutrinos are produced in the core of the Sun through nuclear fusion processes and reach us in enormous numbers ($\sim 7 \times 10^{10} \, \nu_{e}$/$\rm cm^2$/sec). Detecting these neutrinos has allowed us to study the core of the Sun in real time together with photons produced in the past. 
Neutrinos play important roles in supernova (SN) bursts and can give an early warning to optical observations of SNe because the weakness of their interaction with matter allows them to escape the stellar envelope faster than photons.
Even earlier alerts preceding both SN burst neutrinos and gravitational waves, may be possible for nearby SNe by detecting neutrinos from the Si-burning stage. 
Neutrinos from the diffuse supernova background (DSNB) have not yet been observed, but once detected with sufficient statistics could shed light on the cosmic evolution and the star formation rate in the Universe.  
Neutrinos produced in the early Universe can be also detected but are very challenging due to their very low energy. 
%($< $1~eV). \\

Neutrinos are also produced in the upper atmosphere when cosmic rays interact with the atoms forming the Earth's atmosphere. 
The first observation of neutrino oscillation was achieved in 1998 by Super-Kamiokande using atmospheric neutrinos. 
Atmospheric neutrinos up to $\sim$ 10~TeV act as background to  neutrinos from astrophysical origin such as AGNs (Active Galactic Nuclei) and GRBs (Gamma Ray Burst). Such astrophysical neutrinos are produced in violent environments and provide a unique source of information on the acceleration mechanisms and origin of cosmic rays.  So-called cosmogenic neutrinos are produced in the interaction of cosmic rays with the cosmic microwave background.  
Accelerator neutrinos are produced in similar processes as atmospheric neutrinos, and are nowadays the workhorse for the determination of unknown neutrino parameters. These artificial neutrino beams are targeted over up to 1000 km to huge detectors located underground. 

\begin{figure}[ht]
\begin{center}
\includegraphics[width=1.0\textwidth]{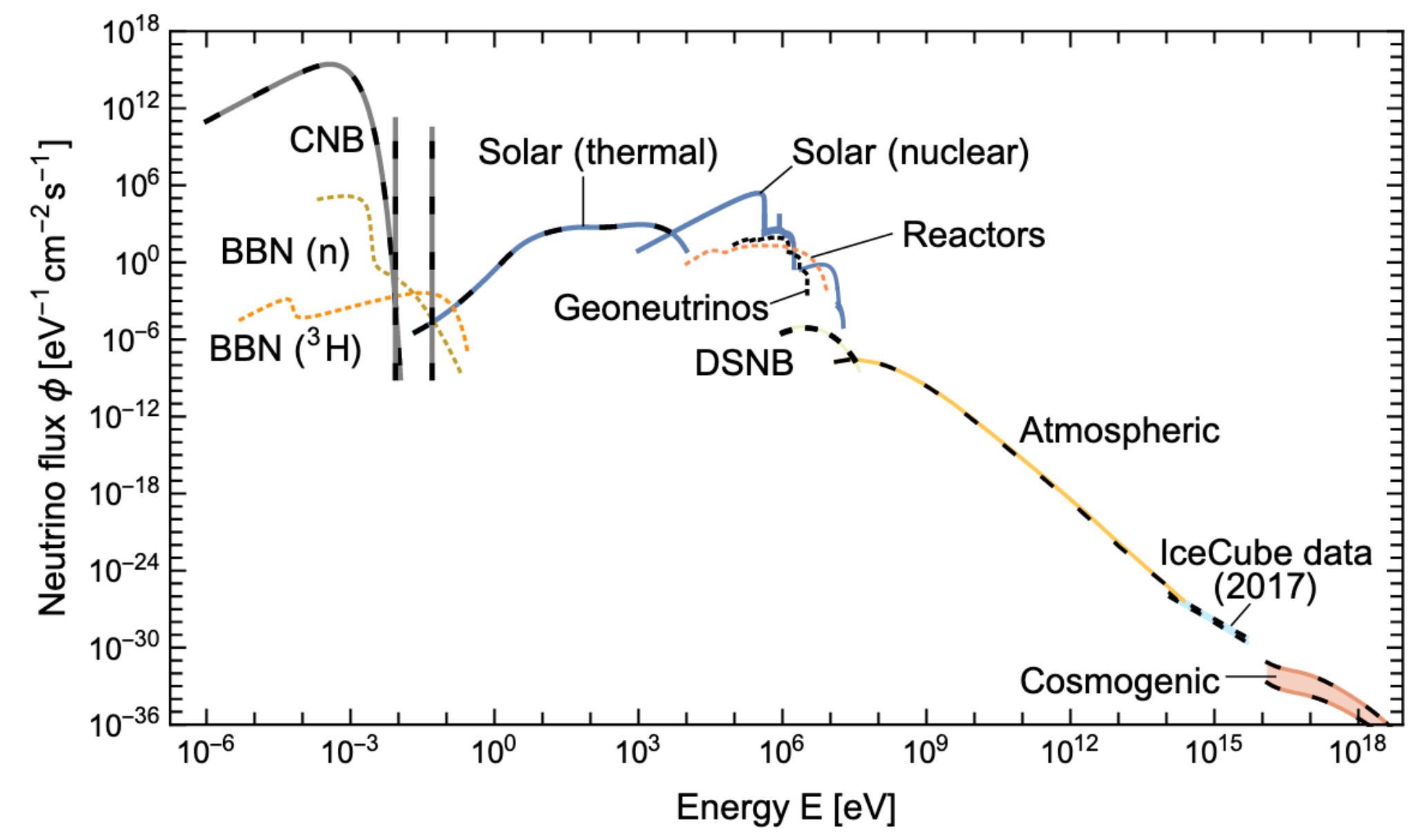}
\end{center}
\caption{Neutrino sources and corresponding energies and fluxes on Earth, taken from Ref.~\cite{Vitagliano:2019yzm}. The abbreviations are Big Bang Nucleosynthesis (BBN), Cosmic Neutrino Background (CNB) and DSNB (Diffuse Supernova Neutrino Background). Nuclear solar neutrinos are produced by $pp$ and CNO cycles, thermal solar neutrinos are produced from processes like bremsstrahlung or plasmon decay. 
See later sections for more on the various neutrino sources.}
\label{f:nuFlux2}
\end{figure}

Figure~\ref{f:nuFlux2} shows the sources of neutrinos and their fluxes vs.\  energies, demonstrating the vast amount of sources that can be probed by neutrino physics. 
In this chapter, the current status and future prospects on the physics of neutrino sources are discussed along with the relevant experiments. 

\iffalse
\begin{figure}[h]
\begin{center}
\includegraphics[width=0.7\textwidth]{WG4/WG4_figs/nu_flux_sources.pdf}
\end{center}
\caption{Energies and fluxes of neutrinos from various sources~\cite{Vitagliano:2019yzm}.}
\label{f:nuFlux}
\end{figure} 
\fi

\subsection{Reactor Neutrinos }
\label{sec:reactorflux}
%================================================
\subsubsection{Introduction}

Reactor antineutrinos ($\bar{\nu}_e$)\footnote{In this section, reactor antineutrinos will be called reactor neutrinos for brevity.} were used to  discover neutrinos in 1956. However, the reactor neutrino flux itself is still not fully understood, due to the well known anomalies observed in both its absolute flux and spectral shape.

In commercial light-water reactors, where low-enriched ($3-5$\%) $^{235}$U is used as fuel, there are four main isotopes, $^{235}$U, $^{239}$Pu, $^{238}$U, and $^{241}$Pu, which contribute to the production of more than 99\% $\bar{\nu}_e$ from beta decays in the decay chain of these isotopes' fission products. 
On average, each fission releases about 200~MeV energy~\cite{Kopeikin:2004cn,Ma:2012bm} and produces six $\bar{\nu}_e$ with energy up to about 10~MeV.

The reactor neutrino flux is calculated or simulated by using the following equation: 
\begin{eqnarray}
\Phi(E_\nu) = \frac{P_{\rm th}}{\sum_{i}^{\textrm{isotopes}}f_i\times E_i}\sum_{i}^{\textrm{isotopes}} f_i\times\phi_i(E_\nu),
\end{eqnarray}
where $P_{\rm th}$, $f_i$, $E_i$, and $\phi_i(E_{\nu})$ represent, respectively, 
reactor thermal power, 
fission fraction of each isotope determined by reactor core simulation, 
energy released per fission, 
and neutrino spectrum of each fission isotope~\cite{Huber:2011wv,Mueller:2011nm}. 
The fission fraction, $f_i$, changes as a function of time 
while the reactor thermal power $P_{\rm th}$ is usually kept constant unless they are turned off
for fuel exchange or maintenance.

\begin{figure}[t]
\begin{center}
\includegraphics[width=0.55\textwidth]{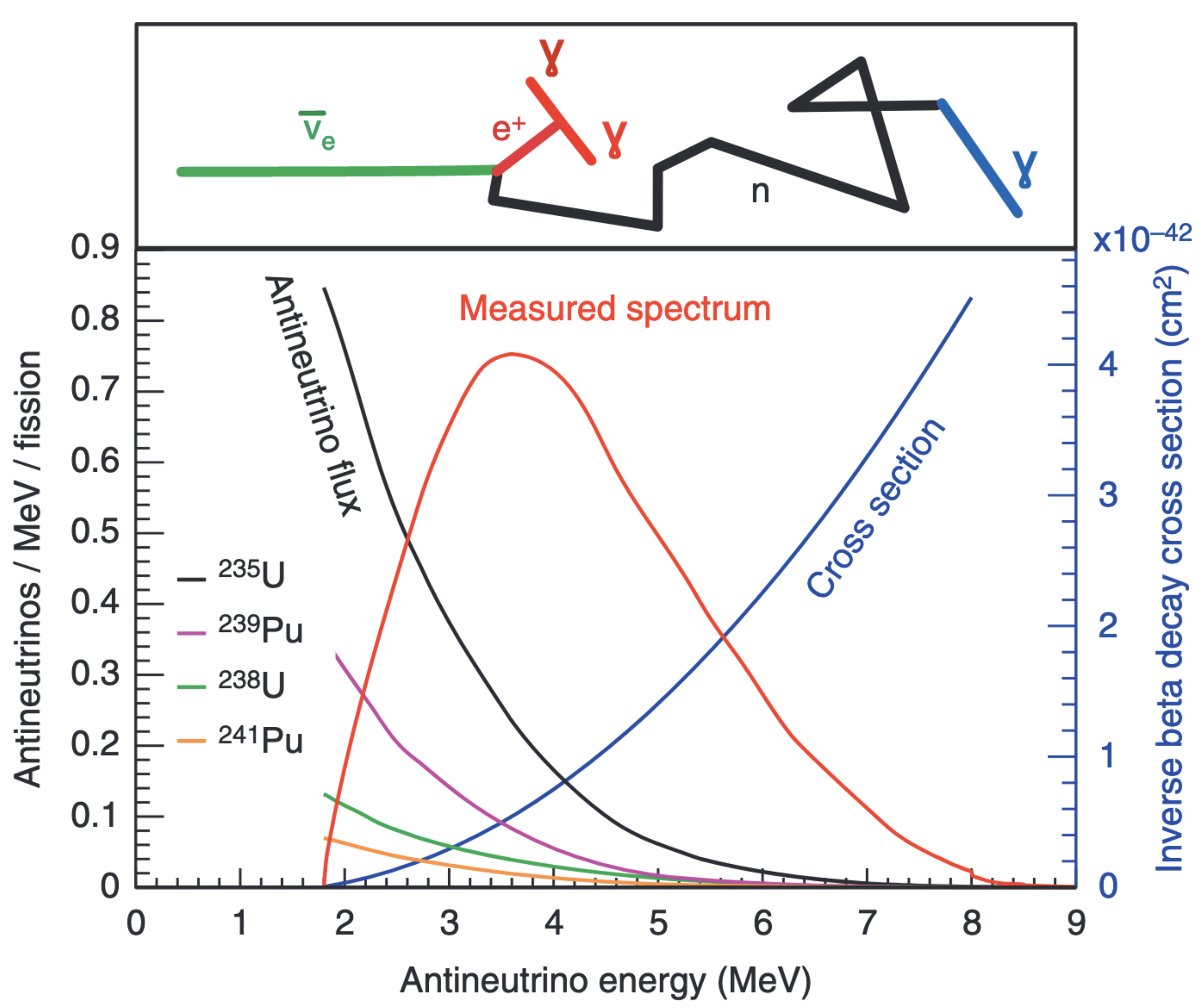}
\end{center}
\caption{Reactor neutrino fluxes for the four main isotopes (black, violet, green and orange), IBD cross section (blue) and corresponding measurable neutrino spectrum (red) from Ref.~\cite{Vogel:2015wua}.}
\label{f:flux_xs}
\end{figure}

Figure~\ref{f:flux_xs} illustrates reactor neutrino fluxes, the relevant cross section of inverse beta decay (IBD), and the corresponding reactor neutrino spectrum. 
Reactor neutrinos are usually detected through an IBD process, $\bar{\nu}_e + p \rightarrow e^+ + n$, 
and only  neutrinos with energy larger than 1.8~MeV can participate in the IBD process. 
Once the IBD process occurs the positron carries away most of the original neutrino energy
and the neutron scatters around until it is thermalized and then captured by a nucleus. Loading liquid scintillators with gadolinium (Gd) or other metals such as Li or Cd, significantly improves the neutron capture efficiency. 
The positron annihilates immediately producing a prompt signal, and the neutron captured by a H (Gd) nucleus produces a delayed signal of 2.2~MeV ($\sim$ 8~MeV).  
Depending on whether H or Gd capture the neutron, the average time of the delayed signal is different, $\sim$ 200~$\mu$sec or $\sim$ 28~$\mu$sec (for 0.1\% Gd by weight), respectively.

\iffalse
After the discovery of neutrinos in 1956, in the 70's, there were more than 20 reactor neutrino experiments at very short baseline ($< 100$~m) searching for neutrino oscillation. In the 80's, a team by Reines observed an evidence ($2-3\sigma$) of neutrino oscillation which was not confirmed by the other  experiments at that time. 
In the mid 90's, Chooz experiment as well as Palo Verde also did not find the oscillation. In 2002, KamLAND~\cite{Eguchi:2002dm} observed a reactor neutrino disappearance at an average baseline of 180~km, and combination of the KamLAND disappearance with the flavor conversion observed in solar neutrino experiments rule out all hypotheses but the large mixing angle neutrino oscillation solution. Two cycles of oscillatory behavior of reactor neutrino survival probability over $L/E$ by KamLAND has been observed later with higher statistics and resulted in the precise measurement of $\Delta m^2_{21}$.
\fi
The historical development of reactor neutrino experiments is found in Ref.~\cite{Bemporad:2001qy}. 
Modern reactor neutrino experiments, Double Chooz, Daya Bay and RENO, in the mid 2000 started building two or more identical detectors at near and far sites to reduce systematic uncertainties, which was required to measure the at the time unknown smallest neutrino mixing angle $\theta_{13}$. In 2012 the first discovery of non-zero $\theta_{13}$ was made by Daya Bay~\cite{An:2012eh} and RENO~\cite{Ahn:2012nd}, independently, with earlier indications from T2K~\cite{Catanesi:2013fxa}, MINOS~\cite{Adamson:2011qu} and Double Chooz~\cite{Abe:2011fz}.

%=====================================================
\subsubsection{Current Status and open Questions}

Neutrino oscillations have been very well understood by measuring so far the neutrino mixing parameters  $\theta_{12}, \theta_{23}, \theta_{13}$, $\Delta m^2_{21}$ and $|\Delta m^2_{31}|$  
(see Section \ref{sec:osc} for more details). Especially the  measurement of a not too small $\theta_{13}$ using reactor neutrinos in 2012 has opened the possibility to measure CP violation in the  lepton sector using the next generation of neutrino detectors currently being constructed. 
Detailed measurements of flux and shape of the emitted neutrino spectrum from reactors showed a discrepancy from the expected spectra. These discrepancies will be called here ``reactor $\bar{\nu}_e$ flux anomaly'' and ``shape anomaly'' (or ``5~MeV excess''). In the following subsections these two well known anomalies are discussed. 
Understanding these two anomalies would lead to a better understanding of reactor neutrinos. 

%----------------------------------------------------------
\paragraph{Absolute Reactor Flux Anomaly}

\label{sec:FluxAnomaly}

\begin{figure}[t]
\begin{center}
\includegraphics[width=0.65\textwidth]{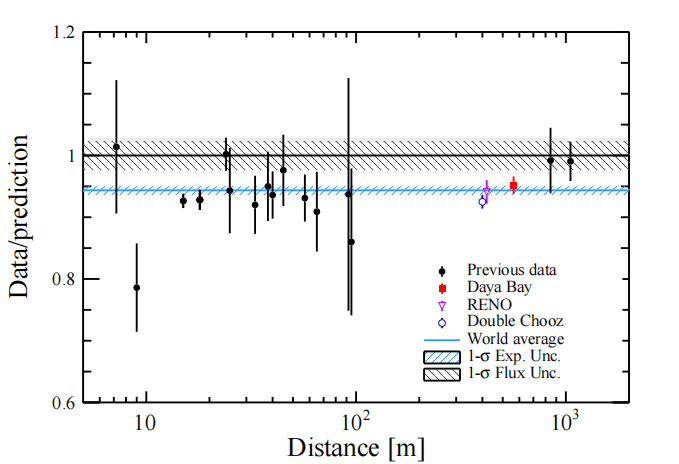}
\end{center}
\caption{
The ratio of measured reactor antineutrino yield to the Huber-Mueller theoretical prediction as a function of the distance from  reactor to detector.  The blue shaded region represents the global average and its 1$\sigma$ uncertainty. }
\label{f:abs_flux}
\end{figure}

Until 2011, there had been a 3\% deficit of the measured reactor neutrino flux compared to the  predicted one in  very short baseline (VSBL), i.e.\ $<$ 100 m, reactor neutrino experiments. 
In 2011 Mueller et al.\ \cite{Mueller:2011nm} re-evaluated the prediction of reactor neutrino spectrum for the four main isotopes and found that the deficit has further increased to about 6\%. 
Huber's independent re-evaluation also confirmed the result of \cite{Mueller:2011nm} for $^{235}$U, $^{239}$Pu and $^{241}$Pu isotopes, for averaged antineutrino energy spectra. Short-baseline, $\mathcal{O}$(1~km), reactor neutrino experiments  Daya Bay, Double Chooz and RENO have measured the absolute reactor neutrino flux using their near detectors and observed a flux deficit of 0.952 $\pm$ 0.014 (exp)~\cite{Adey:2018qct}, $0.925\pm 0.002$ (stat) $\pm$ 0.010 (exp) \cite{DoubleChooz:2019qbj}
and 0.940 $\pm$ 0.001 (stat) $\pm$ 0.020 (syst) \cite{RENO:2020Nu}, respectively, in comparison with the Huber-Mueller model. All measurements share the same additional model uncertainty of $\pm 0.023$. 
This 6\% deficit is called the reactor antineutrino anomaly (RAA), and one of the immediately  suggested explanations was that it is caused by neutrino oscillations from active to eV-scale sterile neutrinos. The various hints and aspects of such sterile neutrinos are separately discussed in Sec.\ \ref{sec:Neff}. A summary of measurements can be found in Fig.\ \ref{f:abs_flux}. 
Unlike past VSBL experiments, most of modern VSBL experiments use relative spectral shape distortion to search for eV-scale sterile neutrinos rather than absolute flux measurements.

\begin{figure}[t]
\begin{center}
\includegraphics[width=0.47\textwidth]{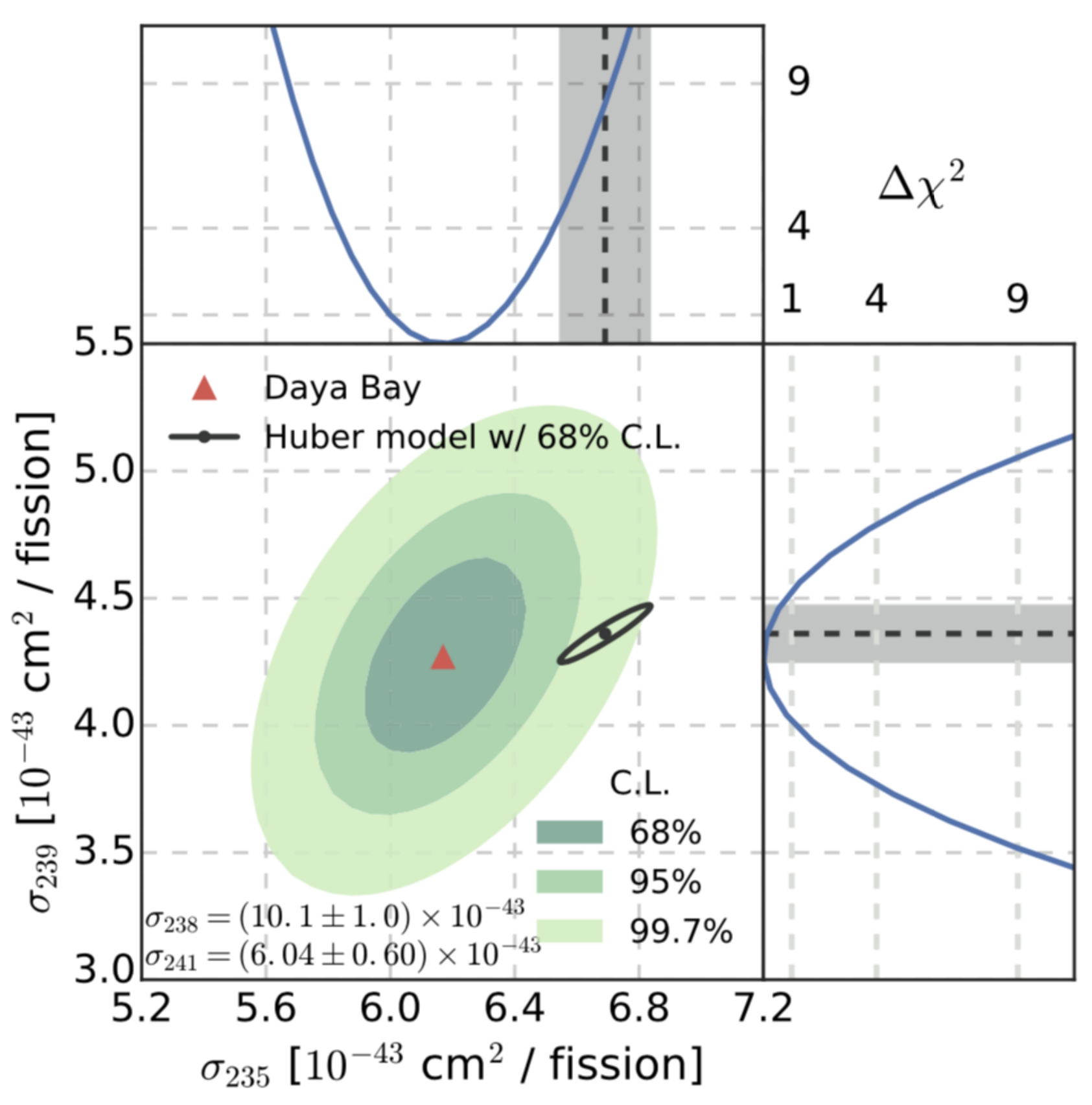}
\includegraphics[width=0.47\textwidth]{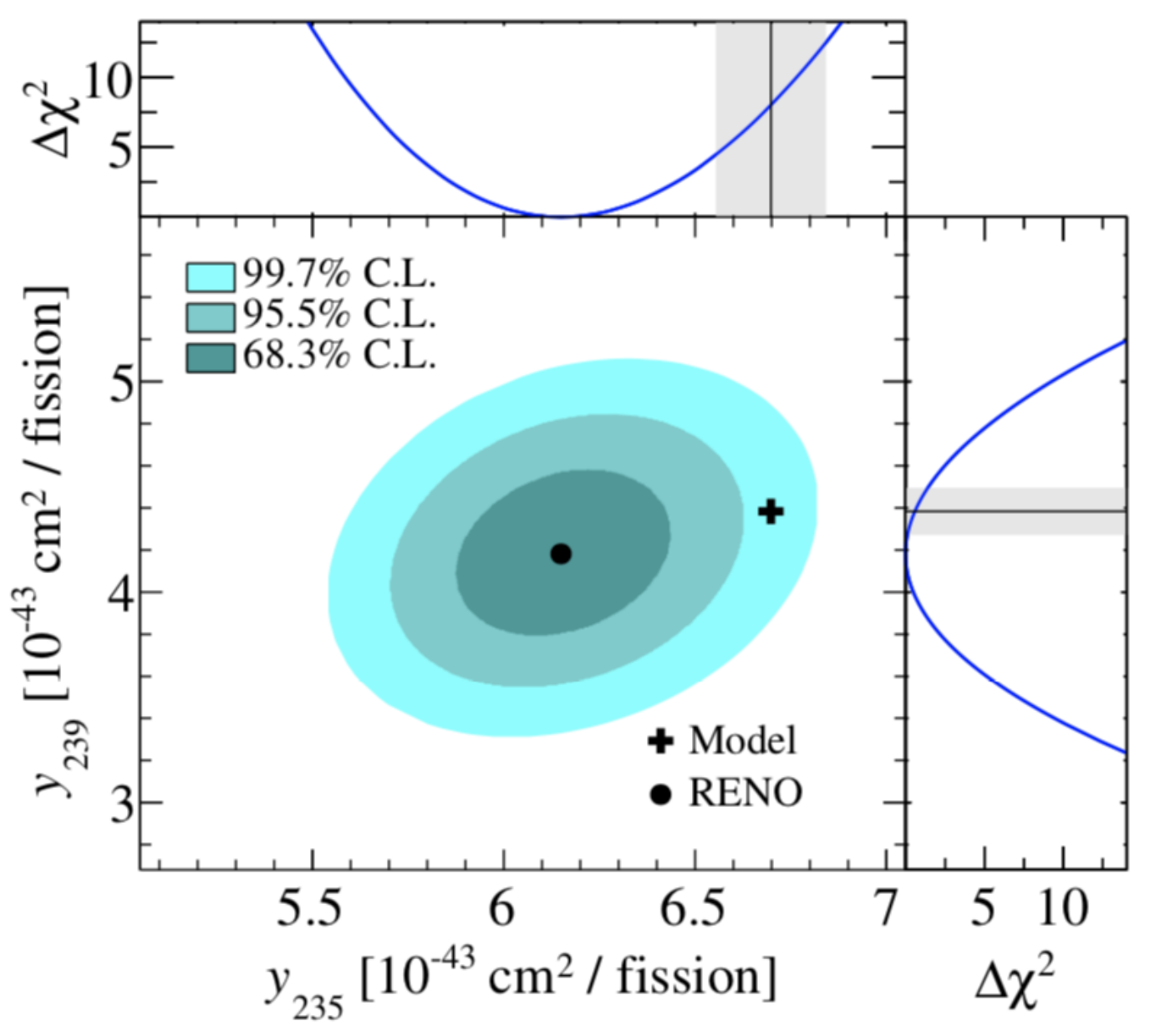}
\end{center}
\caption{IBD yield per fission from $^{235}$U vs.\ $^{239}$Pu by Daya Bay (left) and RENO (right) from Refs.\ \cite{An:2017osx} and \cite{RENO:2018pwo}, respectively.}
\label{f:U235_yield}
\end{figure}

Towards solving the flux anomaly, Daya Bay and RENO  independently observed that the predicted IBD yield per fission from $^{235}$U is higher than the measured ones at 3$\sigma$ levels (see Fig.~\ref{f:U235_yield}). More precise measurements on the IBD yield per fission from $^{235}$U and $^{239}$Pu would be necessary to fully clarify the situation, even though the two independent experiments showed very similar results. 
In parallel, two new reactor neutrino flux calculations~\cite{Hayen:2018uyg,Estienne:2019ujo}  were performed. However, the disagreement with measurement has  not been resolved, as one model predicts more~\cite{Hayen:2018uyg} and the other less~\cite{Estienne:2019ujo} flux than that of H-M model. 
Recently a new measurement of the ratio of beta-spectra from $^{235}$U and $^{239}$Pu was presented \cite{Kopeikin:2021ugh}, which is (1.054 $\pm$ 0.0002) times smaller than the ILL result used for predictions of the reactor antineutrino flux. This reduces the predicted antineutrino flux very close to the experimental results \cite{Kopeikin:2021ugh}, and would mean that the basis of the RAA is in question. 
%----------------------------------------------------------
\paragraph{Shape Anomaly}

\begin{figure}[t!]
\begin{center}
\includegraphics[width=0.65\textwidth]{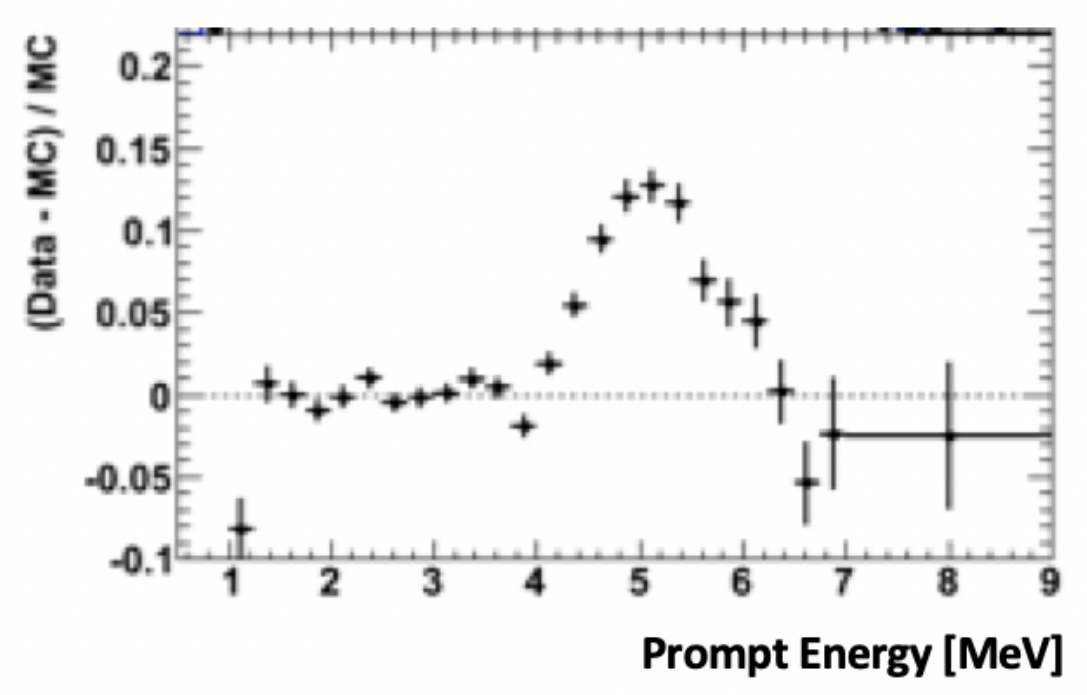}
\end{center}
\begin{center}
\includegraphics[width=0.65\textwidth]{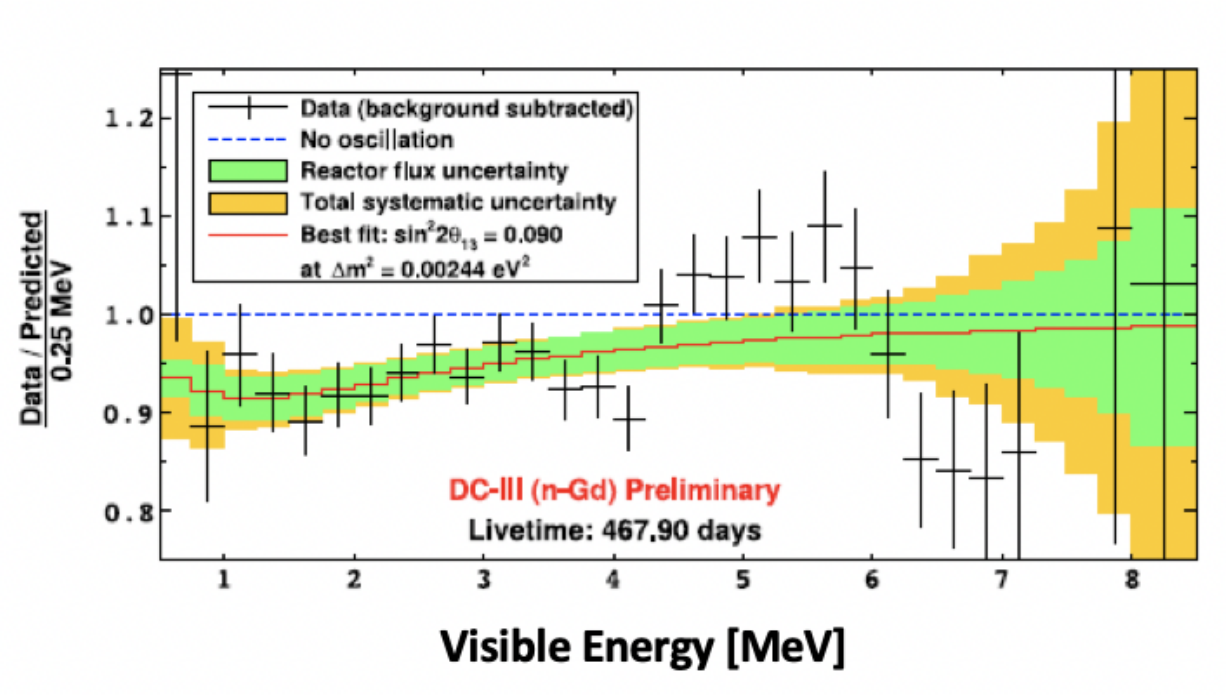}
\end{center}
\begin{center}
\includegraphics[width=0.65\textwidth]{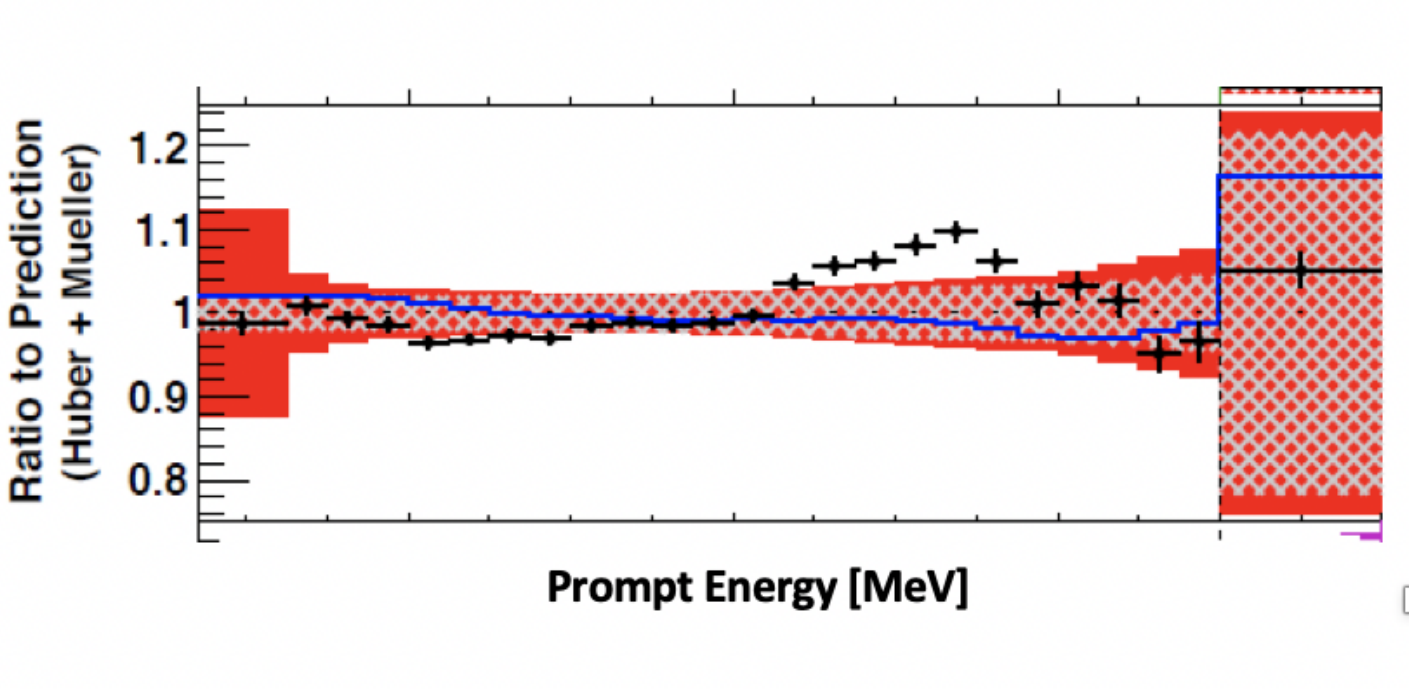}
\end{center}
\caption{The 5~MeV excess measurements in 2014 by RENO \cite{Seo:2014xei}, Double Chooz \cite{Abe:2014bwa} and Daya Bay \cite{An:2015nua} (from top to bottom).}
\label{f:5MeV_excess}
\end{figure}

The shape anomaly, or so-called the ``5~MeV excess'', was first reported by in 2014 \cite{Seo:2014xei,Abe:2014bwa,AC}. 
Later in 2014 Daya Bay \cite{An:2015nua} also confirmed the 5~MeV excess compared to Huber-Mueller model (see Fig.~\ref{f:5MeV_excess}). The origin of the  excess, however, has not been fully identified yet. 
In 2014 it was also shown that the excess is correlated to the reactor thermal power \cite{Seo:2014xei,Abe:2014bwa,AC}, implying that this is very likely caused by reactor neutrinos unpredicted by the model. 
Recently, RENO and Daya Bay showed 3.2$\sigma$ \cite{RENO:2020Nu} and 4$\sigma$ \cite{Adey:2019ywk} evidences of the correlation between the 5 MeV excess and the $^{235}$U flux. 

Most of the VSBL reactor neutrino experiments (see Section \ref{sec:sterile_rea}) use research reactors, in which $^{235}$U is highly enriched. They are expected to nail down the correlation between the excess and $^{235}$U. Among the VSBL experiments, NEOS using a commercial reactor has clearly observed the  excess for the first time in 2017, thanks to  high statistics and good energy resolution. Recently, PROSPECT also showed the 5~MeV excess with increased data (total of 96 calendar days of reactor-ON data) and disfavored it being from only (no) $^{235}$U at 2.4 (2.2)$\sigma$ CL \cite{Andriamirado:2020erz}. 
STEREO has released the first measurement \cite{AlmazanMolina:2020jlh} of the antineutrino energy spectrum from $^{235}$U at the ILL reactor and found an excess of 12.1 $\pm 3.4$\% ($3.5\sigma$) at 5.3~MeV neutrino energy, which is a little bit lower in energy (0.5 MeV) than other experiments.

There have been many efforts to understand the ``5~MeV excess'' in the nuclear theory community by re-evaluating reactor neutrino flux and energy spectrum in two different methods: summation (or ab-initio) and conversion methods. 
A summation method is based on nuclear databases and sums up beta decay spectra from all possible fission products in the databases to obtain neutrino spectra \cite{Hayes:2015yka,Estienne:2019ujo}. 
A conversion method is based on the measured beta spectra from $^{235}$U, $^{239}$Pu, and $^{241}$Pu fission at ILL in Grenoble, France in the 1980's and converts the beta spectra to neutrino spectra \cite{Huber:2011wv,Mueller:2011nm}. 
The conversion method is known to be more precise but there is only a single measurement of the beta spectra at ILL. The Huber-Mueller model is also based on the conversion method for $^{235}$U, $^{239}$Pu, and $^{241}$Pu isotopes. \\
Along with the other theoretical and experimental studies being performed by various groups, the International Atomic Energy Agency (IAEA) is also interested in  understanding  both the 5~MeV excess and the deficit of the absolute neutrino flux, and within 5 to 10 year-time scale better understanding of these anomalies is to be expected. 

%====================================================
\subsubsection{\label{sec:reac_fut}Future Outlook}

Double Chooz and Daya Bay have shut down in 2018 and December 2020, respectively, and RENO is taking data through 2021. More data and upcoming improved analyses of existing data  will be useful to understand the two anomalies discussed in the previous section. The JUNO collaboration plans to install a 3~ton Gadolinium liquid scintillator 
(GdLS) detector with $4\pi$ photo-coverage of Silicon Photomultipliers
(SiPMs) operating at $-50^{\rm{o}}$ C at a very short baseline (30~m) from a reactor (4.6~GW$_{\rm th}$) in the Taishan Nuclear Power Plant. This project is called JUNO-TAO (Taishan Antineutrino Observatory) or TAO, and its main goal is to measure the reactor neutrino spectrum very precisely with very good energy resolution, to understand the fundamental physics of the very complex beta decay processes in reactors, clarify the origin of the 5~MeV excess, as well as sterile neutrino search~\cite{junotaocdr}. The energy resolution goal is $< 2$\% at 1~MeV, and its operation is expected to start in 2022.

Neutrino detectors can be also used for a non-proliferation purpose by detecting neutrinos from remote (un-)known reactors. Unlike other traditional nuclear monitoring methods, neutrino detectors provide no interference with reactors, and therefore monitoring can be done anytime, 24-hour year round. More details on the nuclear monitoring is described in Sec.\ \ref{wg5_tech}.

\label{sec:source_rea}

\subsection{Accelerator Neutrinos }
\label{wg4_accel}

\subsubsection{Conventional Beams}
\begin{figure}[t!]
\begin{center}
  \includegraphics[width=\textwidth] {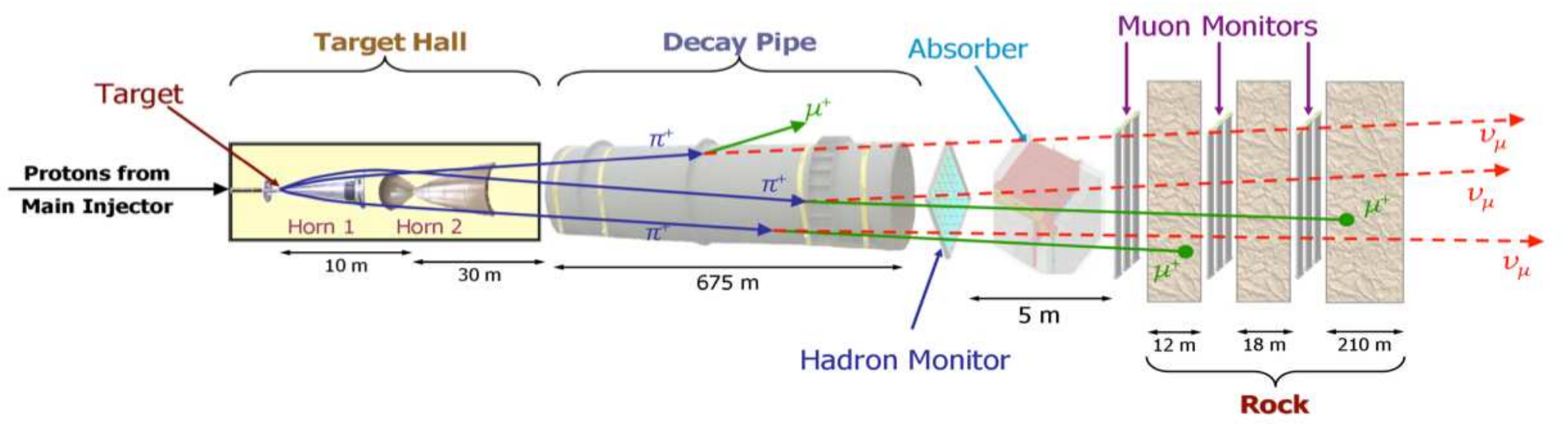}
\end{center}
\caption{The NuMI beamline at Fermilab as described in \protect{\cite{Adamson:2014vgd}.}
}\label{fig:WG4_NuMI}
\end{figure}

\begin{figure}[t!]
\begin{center}
 \includegraphics[height=.35\textwidth]{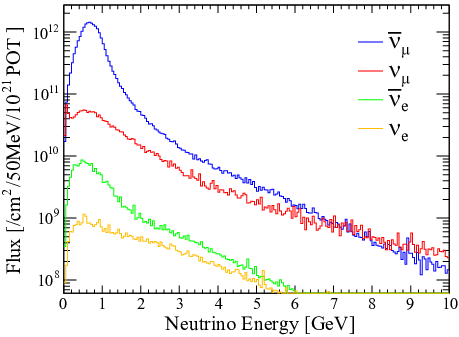}
    \includegraphics[height=.38\textwidth]{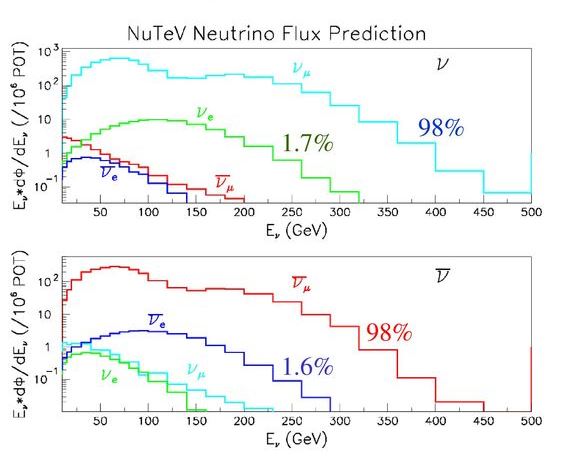}
\end{center}
\caption{Composition of the low energy T2K \protect{\cite{Abe:2020iop}} (left) and high energy  (800  GeV protons) TeV II neutrino beam \protect{\cite{Naples:2003fe}} at Fermilab (right, the relative contribution to the flux is indicated).  }\label{fig:WG4_TeVII}
\end{figure}

 Accelerator-based  neutrino sources had their start in the early 1960's \cite{Schwartz:1960hg} and led to the discovery that the electron and muon neutrino are distinct particles. Most modern neutrino sources  use the same basic concepts as those original experiments\footnote{For a historical review please see reference \cite{Dore:2018ldz}. }. A beam of protons is aimed at a target and produces  charged pions and kaons that are focused to create a collimated sign-selected beam which then enters a decay region where they decay to  neutrinos and the appropriate lepton species. 
The remaining hadrons  hit an absorber, while the muons are ranged out in rock or other material leaving a neutrino beam.  Positive pion decays are dominated by 2-body decays to muon neutrinos ($\nu_\mu$) and antimuons ($\mu^+$), while negative pions produce antineutrinos ($\bar{\nu}_\mu$) and muons ($\mu^-$).  Kaon  and muon decays produce a small contamination of electron neutrinos.  Figure \ref{fig:WG4_NuMI} shows the NuMI \cite{Adamson:2014vgd} beamline at Fermilab.  Figure \ref{fig:WG4_TeVII} illustrates the composition of the pion-dominated  low-energy T2K \cite{Abe:2020iop} antineutrino and   the high-energy TeV II \cite{Naples:2003fe} neutrino beams created, respectively,  by 30 and 800 GeV protons at J-PARC and Fermilab.  In the TeV II beam a lower energy peak around 70 GeV due to pion decay is present.  Kaons in the secondary beam produce a second peak in the muon neutrino distribution around 200 GeV as well as high energy electron neutrinos.  

\begin{figure}[t!]
\begin{center}
  \includegraphics[height=.4\textwidth] {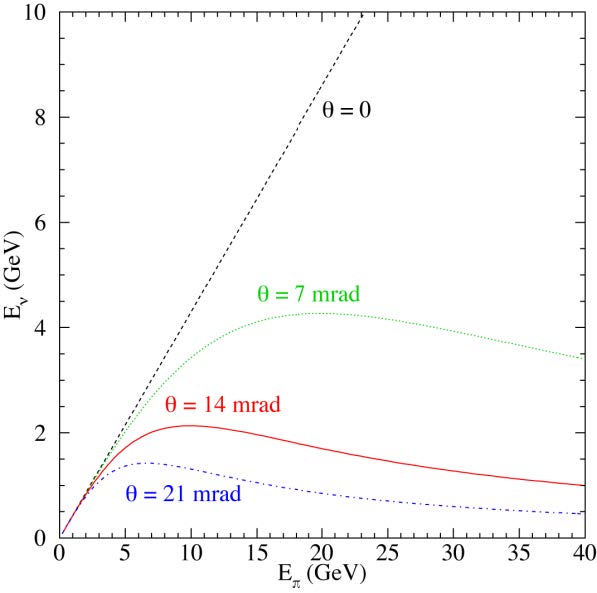}
   \includegraphics[height=.4\textwidth] {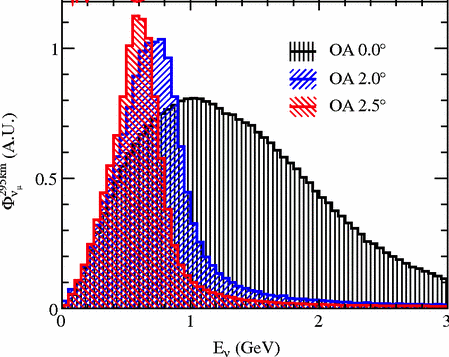}
\end{center}
\caption{Dependence of neutrino energy on pion energy and angle relative to the secondary beam axis for the NuMI beamline \cite{Muether:2013gxa} (left). Energy spectra for the off-axis T2K \cite{PhysRevD.87.012001} beamline (right). The on-axis spectrum is shown, along with the optimized off-axis distribution.}
\label{fig:WG4_Axis}
\end{figure}

Most accelerator-based neutrino beams follow this same basic design and produce neutrino beams with energies between 0.5 GeV (J-PARC) and 500 GeV (TeV II).  A maximum integrated flux is achieved by using the broad energy spectrum on-axis beam, while a narrower neutrino energy spectrum can be achieved by using an off-axis configuration that takes advantage of  the  Jacobian peak (see Fig.\ \ref{fig:WG4_Axis}) in neutrino energy from decays transverse to the direction of motion.  

An important variant is neutrino beams generated by stopping pions and kaons, often referred to as Decay At Rest (DAR), which produce a well-defined neutrino energy spectrum.  This technique with low energy pion beams has been used to generate 30 MeV muon neutrinos by the LSND  experiment at
Los Alamos (LANSCE),  KARMEN at ISIS (RAL) and  at the SNS at Oakridge.   Kaon DAR production of neutrinos was recently demonstrated \cite{Aguilar-Arevalo:2018ylq} by parasitic use of the NuMI beam dump and the MiniBooNE detector located on the surface above the dump. 

\subsubsection{Novel Neutrino Beams}
\begin{figure}[t!]
\begin{center}
  \includegraphics[height=.6\textwidth] {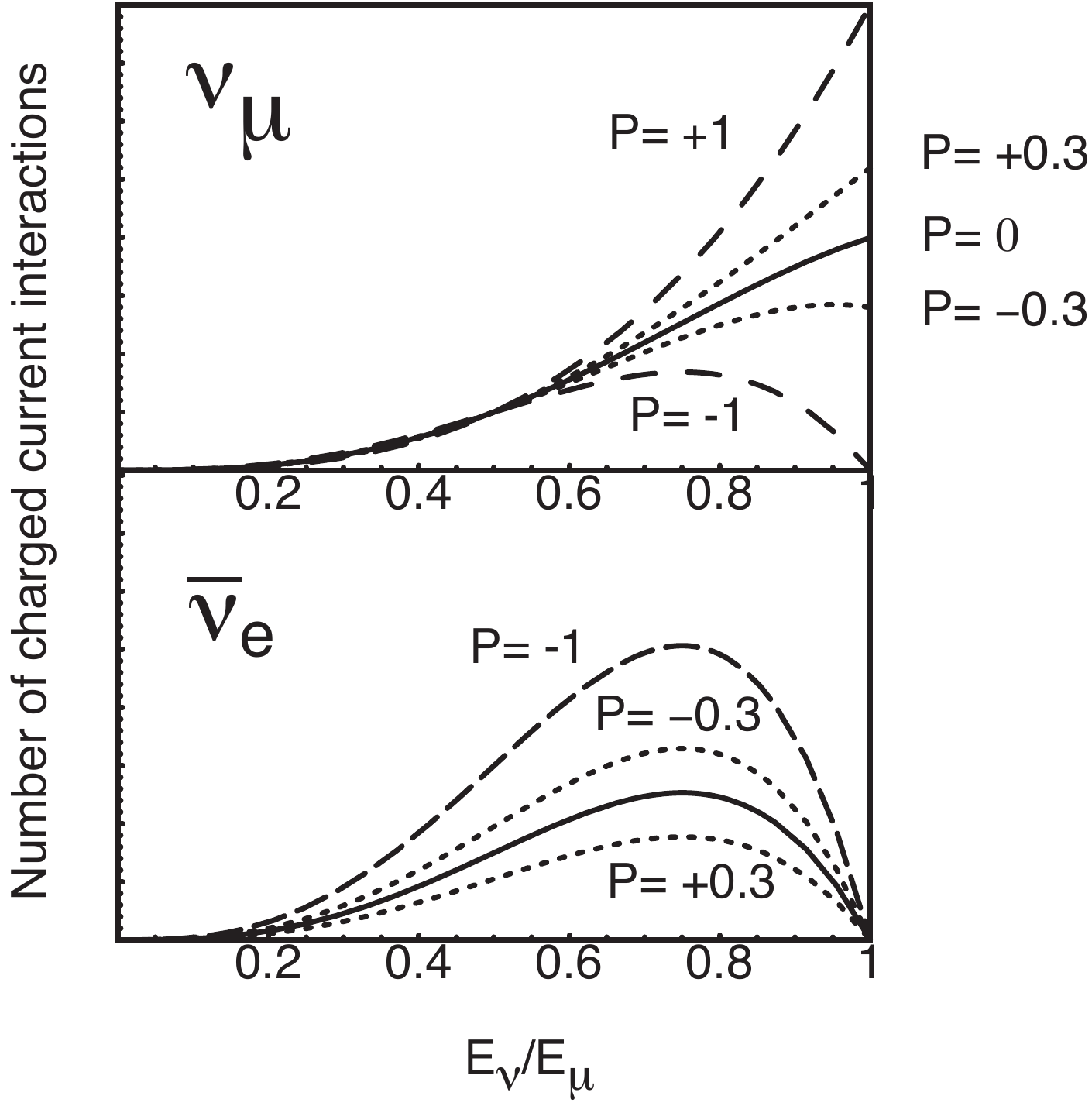}
  \end{center}
\caption{Neutrino spectra from a  20 GeV muon storage ring. For a given muon polarization, $P$, the muon and electron neutrino spectra are fixed by the beam energy. Taken from \cite{Albright:2000xi}. }
\label{fig:WG4_NF}
\end{figure}

The conventional beams described above rely on the decay of pions and kaons, but other beta decay processes can produce neutrino beams as well. 
Muon storage rings, in principle, can produce intense and very well defined neutrino beams from the 3-body decay $\mu^- \rightarrow \bar{\nu}_e  e^- \nu_\mu$.  Decays of a monochromatic muon beam generated in the straight section of a muon storage ring result in very well defined neutrino spectra, see Fig.\  \ref{fig:WG4_NF}.  Using this principle, high flux ``neutrino factories''  have been proposed \cite{Albright:2000xi,DeRujula:1998umv,Choubey:2011zzq} but await the development of cooled muon beams to be practical.  A first step is the $\nu$STORM \cite{Adey:2013pio,Ahdida:2020whw} experiment recently proposed at CERN \cite{Ahdida:2020whw}. %{\bf WR: here 1-2 more sentences would be good!}

Radioactive ion storage rings have been proposed as a source of electron neutrino ``beta-beams'' \cite{Huber:2005jk, Benedikt:2011za}, stopped ions have been proposed for an ``IsoDAR'' beam \cite{Bungau:2012ys,Conrad:2013sqa}. 
%{\bf WR: here a few more sentences would be good!}
 See also the end of Sec.\ \ref{sec:acc_fut} and Sec.\ \ref{wg5_tech}  for a discussion.

\label{sec:source_acc}

\subsection{Solar Neutrinos }
\label{subsec:WG4solar}

\begin{figure}[t!]
    \centering
    \includegraphics[width=0.9\textwidth]{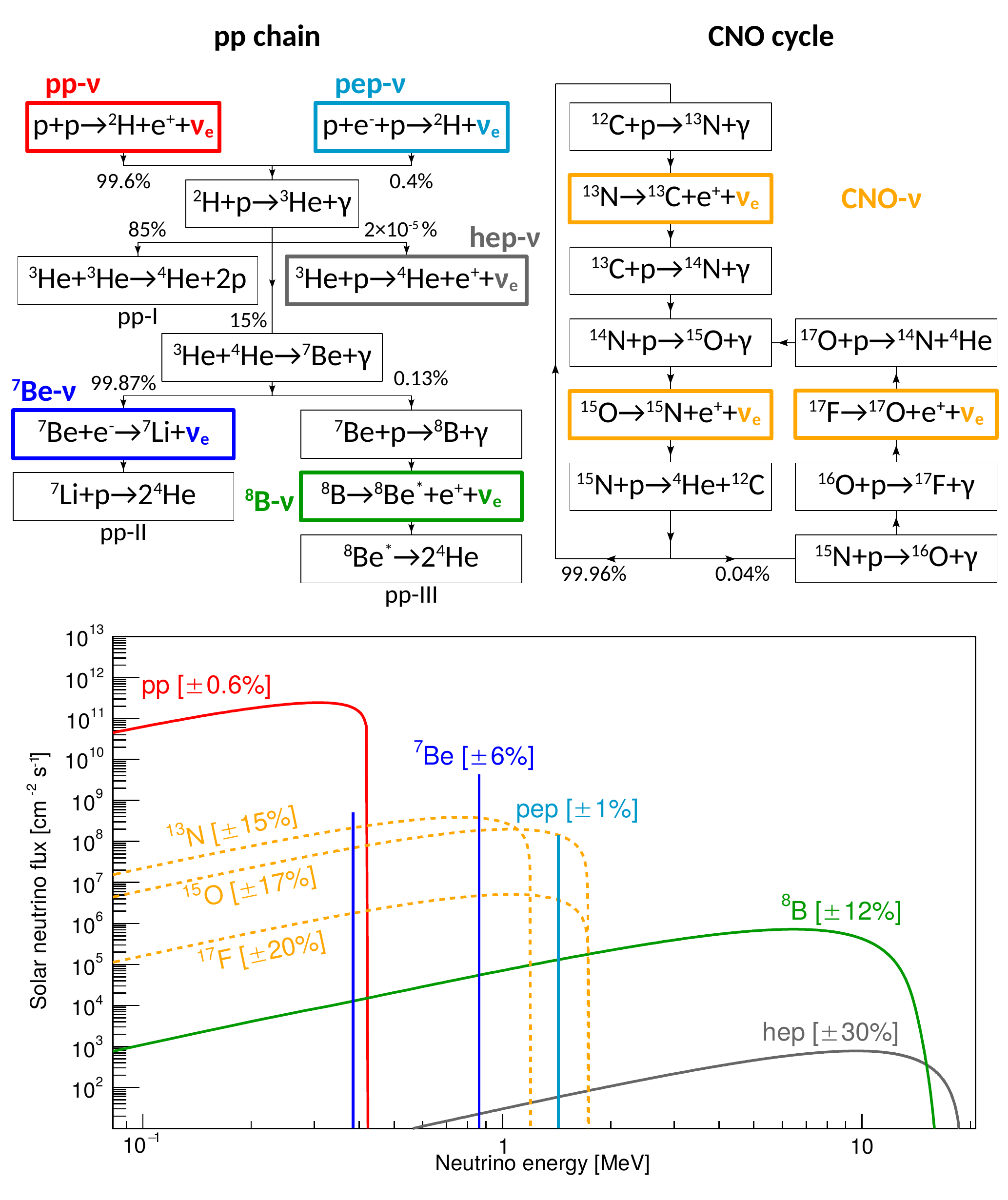}
    \caption{Top: Nuclear fusion sequences occurring in the core of the Sun: schematic view of the $pp$-chain and the CNO cycle. Bottom: Solar neutrino energy spectrum with fluxes from~\cite{Vinyoles_2017}. The flux (vertical scale) is given in cm$^{-2}$ s$^{-1}$ MeV$^{-1}$ for continuous sources and in cm$^{-2}$ s$^{-1}$ for mono-energetic ones. Taken  from~\cite{Agostini:2018uly}.}
    \label{fig:SolarSpectrum}
\end{figure}

Solar neutrinos are emitted during the fusion of protons to helium nuclei taking place in the solar core. This fusion, $$4p + 2 e^- \to  ^4{\rm He} + 2 \nu_e + 26.73 \, {\rm MeV},$$ is the energy source of our star. The dominant fusion process is the $pp$-chain, while a small, order-1\% fraction of solar energy is produced in the so-called CNO cycle\footnote{Less important are neutrinos in the keV energy range produced from thermal processes such as   bremsstrahlung, plasmon decay or pair emission, see Ref.\   \cite{Vitagliano:2017odj}}. In the latter process, expected to be the dominant energy source for stars at least 1.3 times heavier than the Sun, the fusion is catalyzed by the presence of heavier elements, namely carbon, oxygen, and nitrogen. Figure~\ref{fig:SolarSpectrum} shows in its top part the schemes of the $pp$-chain and of the CNO cycle, while its lower part shows the energy spectrum of solar neutrinos. The flux of solar neutrinos is dominated by  $pp$ neutrinos (order of $10^{10}$ s$^{-1}$ cm$^{-2}$) having a continuous energy spectrum with a 420 keV endpoint. In the $pp$-chain, also mono-energetic $^7$Be (0.384 MeV and 0.862 MeV) and $pep$ (1.44 MeV) neutrinos are produced, as well as $^8$B neutrinos characterized by lower flux (order of $10^{6}$ s$^{-1}$ cm$^{-2}$) and a continuous energy spectrum extending up to about 15 MeV. 
Measurements of solar neutrinos interaction rates, with energy-dependent deficits with respect to solar model calculations, were the first hint of neutrino  oscillations. For the history of solar neutrinos, see Ref.\ \cite{Bahcall:2002ng}. The determination of the relevant solar neutrino parameters is discussed in Sec.\ \ref{WG1:Sect:Solars}.

\begin{table}[t!]
  \centering
  \begin{tabular}{l|c|c|c|c}
  \hline
  \hline
 Solar $\nu$  & B16-GS98 (HZ) & B16-AGSS09met (LZ)& Measurement & Exp \\
  \hline \hline
$pp$-cycle &  & & & \\  \hline \hline
$pp$ & $5.98 (1.0 \pm 0.006)$ & $6.03 (1.0 \pm 0.005)$ & $6.1 \pm 0.5 ^{+0.3}_{-0.5}$~\cite{Agostini:2018uly}  & $10^{10}$ \\ [4pt]
$^7$Be & $4.93 (1.0 \pm 0.06)$ & $4.50 (1.0 \pm 0.06) $ & $4.99 \pm 0.11 ^{+0.06}_{-0.08}$~\cite{Agostini:2018uly} & $10^{9}$ \\[4pt]
& & & $5.82 \pm 0.98$~\cite{Gando:2014wjd} & $10^{9}$\\[4pt]
$pep$ & $1.44 (1.0 \pm 0.01)$ & $1.46 (1.0 \pm 0.009)$ & $1.27\pm 0.19 ^{+0.08}_{-0.12}$~\cite{Agostini:2018uly} & $10^{8}$ \\[4pt]
$^8$B & $5.46 (1.0 \pm 0.12) $ & $4.50 (1.0 \pm 0.12)$ & 5.4 $\pm$ 0.02 $\pm$ 0.1~\cite{SuperK:nu2020}& $10^{6}$\\[4pt]
& & & $5.25 \pm 0.16 ^{+0.11}_{-0.13}$~\cite{Aharmim:2011vm}& $10^{6}$\\[4pt]
   &  & & $ 5.68^{+0.39}_{-0.41}{}^{+0.03}_{-0.03}$~\cite{Agostini:2018uly} & $10^{6}$  \\[4pt]
    &  & & $5.95 ^{+0.75}_{-0.71}{}^{+0.28}_{-0.30}$~\cite{SNO+B8} & $10^{6}$ \\[4pt]
$hep$ & $7.98 (1.0 \pm 0.30)$ & $8.25 (1.0 \pm 0.12)$ & $< 23 $ (90\% C.L.)~\cite{Aharmim_2006} & $10^{3}$\\[4pt]
& & & $< 150$ (90\% C.L.)~\cite{PhysRevD.73.112001} & $10^{3}$  \\[4pt]
& & & $< 180$ (90\% C.L.)~\cite{B8BX} & $10^{3}$\\[4pt]
\hline \hline 
CNO& $4.88 (1.0 \pm 0.11)$ & $3.51 (1.0 \pm 0.11)$ & $ 7.0^{+3.0}_{-2.0}$~\cite{BXCNO} & $10^{8}$\\[4pt]
   \hline \hline
  \end{tabular}
  \caption{Solar neutrino fluxes predicted by the Standard Solar Models B16-GS98 (High Metallicity) and  B16-AGSS09met (Low Metallicity)~\cite{Vinyoles_2017} and as measured by various experiments in units of cm$^{-2}$ s$^{-1}$ (with the exponential factor given in the last column). For the measured fluxes, the first error is statistical and the second error systematical.}
  \label{tab:SolarFluxes}
\end{table}

Solar neutrinos were first detected by radiochemical detection methods~\cite{RevModPhys.75.985, ABDURASHITOV1994234,ANSELMANN1992376} revealing information about integral fluxes above a certain threshold. The technologies using water Cherenkov and liquid-scintillator detectors provide an opportunity to perform real-time precision spectroscopy of solar neutrinos. Table~\ref{tab:SolarFluxes} summarizes solar neutrino fluxes as predicted by the Standard Solar Model (SSM)~\cite{Vinyoles_2017} compared to the existing measurements.

\begin{figure}[t]
    \centering
    \includegraphics[width=0.9\textwidth]{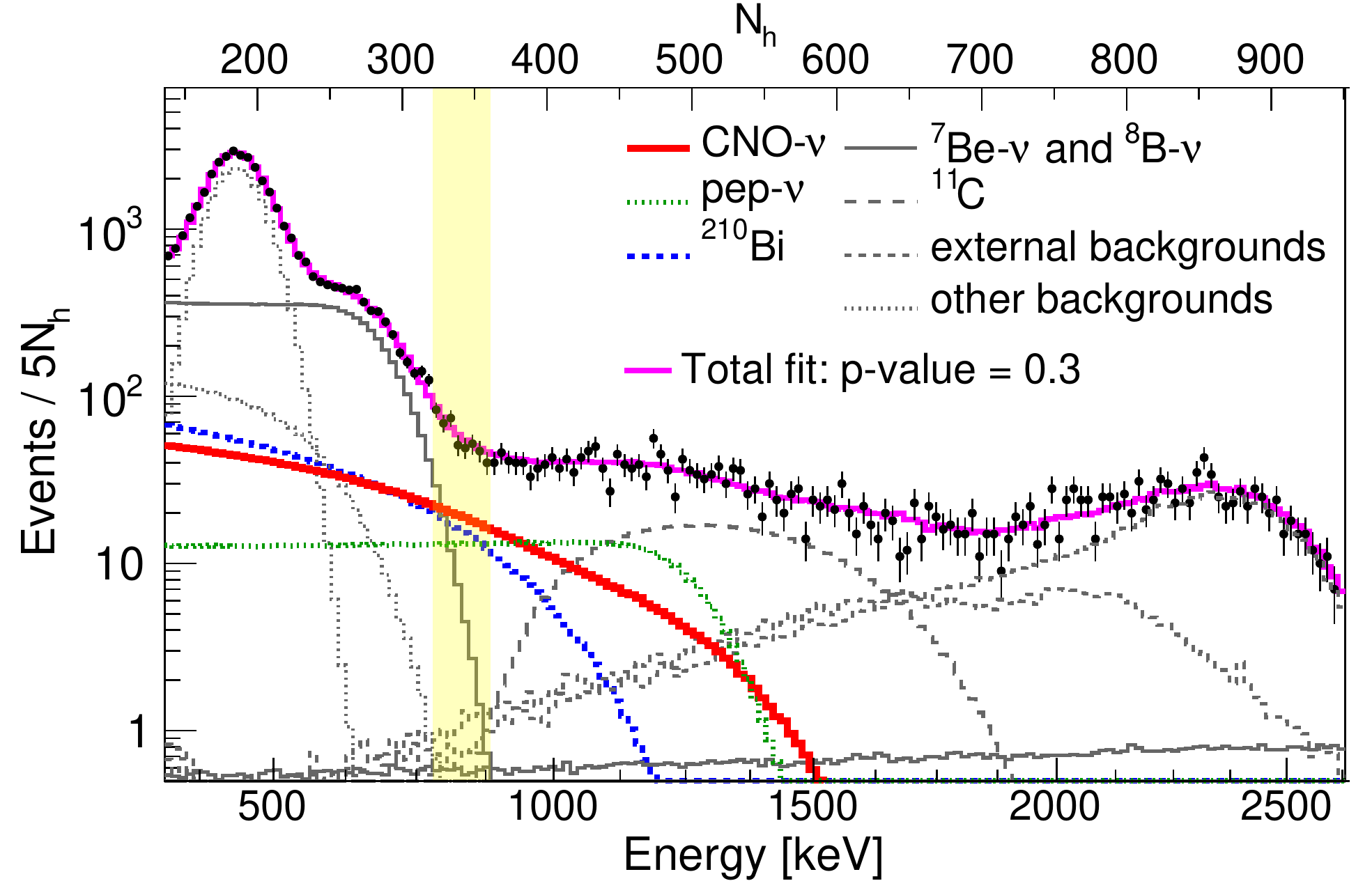}
    \caption{Spectral fit of the Borexino Phase-III data that lead to the observation of solar neutrinos from the CNO cycle shown in red. Taken from~\cite{BXCNO}.}
    \label{fig:Borex-CNO}
\end{figure}

The most precise measurements of the $pp$, $^7$Be, and $pep$ neutrinos are provided by Borexino (LNGS, Italy)~\cite{Agostini:2018uly} characterized by unprecedented levels of radiopurity of its liquid scintillator. High precision measurements of $^8$B neutrinos come from water Cherenkov detectors SNO (Sudbury, Canada)~\cite{Aharmim:2011vm} and Super-Kamiokande (Kamioka, Japan)~\cite{Abe:2016nxk,SuperK:nu2020}.
For the $hep$ neutrinos with a very low expected flux only upper limits exist: the most stringent from SNO~\cite{Aharmim_2006} is still about a factor of three higher than the expected SSM flux. Detection of neutrinos from the CNO fusion cycle is complicated due to the degeneracy of its spectral shape with those of $^{210}$Bi contaminating the liquid scintillator and of $pep$ solar neutrinos. Additionally, the cosmogenic $^{11}$C background from muon spallation further complicates the CNO solar neutrino detection. Borexino has recently observed for the first time CNO solar neutrinos with 5$\sigma$ C.L.~\cite{BXCNO}, as shown in Fig.~\ref{fig:Borex-CNO}. This was achieved through the upper-limit constraint on the $^{210}$Bi contamination of the liquid scintillator, made possible thanks to an exceptional thermal stabilisation of the detector achieved over five years.

Measurements of solar neutrinos are a source of information about the neutrino properties. They also provide a direct insight about the core of the Sun. The implications of solar neutrino measurements towards our understanding of neutrino oscillations are discussed in Sec.~\ref{WG1:Sect:Solars}. These include determination of the $\theta_{12}$ mixing angle, $\Delta m ^2_{21}$ mass splitting, study of the matter effects in the Sun (energy dependence of the survival probability) and during their journey while traversing
through the Earth (day-night effect). Concerning  solar physics, in the first place, neutrinos are the only direct evidence that indeed nuclear fusion is powering our closest star. In addition, by comparing the 
luminosity of photons and neutrinos emitted from the Sun, one may study the thermal equilibrium of the Sun on a time scale of $10^5$ years. This is the time it takes photons to escape from the solar core, while neutrinos can freely traverse the dense solar matter. 
Special attention is  needed to understand the so-called ``metallicity problem'' in solar physics, where by metallicity we mean the solar abundances of elements heavier than hydrogen -- an important input to the Standard Solar Models. The fact is that newer, more precise spectroscopy measurements of the solar photosphere yielded lower abundances of these elements, which spoiled the earlier agreement between the helioseismology measurements and the SSM predictions of the radial dependency of the speed of  sound waves across the Sun. This argument is discussed in detail in~\cite{Vinyoles_2017}, where also the SSM predictions for solar neutrino fluxes are given separately for  low (LZ) and high (HZ) metallicity (see also Tab.\ \ref{tab:SolarFluxes}). The metallicity influences the opacity of the Sun and consequently also the temperature in the core and the fusion rates. There is a sizeable difference of 9\% and 19\% between the HZ and LZ SSM predictions of $^7$Be and $^8$B fluxes, respectively. This fact is at the base of a slight preference toward HZ SSM reported by Borexino~\cite{Agostini:2018uly} (based on its measurement of both $^7$Be and $^8$B neutrinos, see Tab.\ \ref{tab:SolarFluxes}), that weakens if global fits of all solar neutrino data are considered. The largest difference between the fluxes predicted by the LZ and HZ SSM results for the CNO cycle, which is directly catalyzed by heavy elements and amounts to about 32\%. Thus, the Borexino observation of  CNO solar neutrinos~\cite{BXCNO} paves the way towards the solution of this problem. Borexino might increase the precision of its CNO measurement, which is also among the scientific goals of  SNO+~\cite{Andringa:2015tza}, currently filling its detector with liquid scintillator at SNOLAB in Sudbury, Canada. The future Jinping solar neutrino experiment~\cite{Beacom_2017} aims to perform precision spectroscopy of  CNO neutrinos, taking advantage of its shielding against cosmic muons in the world's deepest laboratory located at Jinping in China. Detection of low flux $hep$ neutrinos and collection of a high statistics low energy $^8$B neutrinos are the goals of  next generation large volume detectors. JUNO~\cite{An:2015jdp}, the 20 kton liquid scintillator detector under construction in south China, might be able to measure $^8$B neutrinos down to 2.5\,MeV~\cite{JUNO_B8}. The future water-Cherenkov Hyper-Kamiokande detector plans 187\,kton fiducial volume for detection of $^8$B neutrino and $hep$ neutrinos~\cite{Yano:2020aap}. Next generation  experiments with novel techniques also aim at measuring solar neutrinos. THEIA~\cite{askins2019theia} considers a few-tens-of-kton-scale detector filled with water-based liquid
scintillator, combining the advantages of water (directional Cherenkov light) and liquid scintillator (high light yield) detectors. The  two-phase liquid argon time projection chambers  under development for direct Dark Matter WIMP
searches (DarkSide-20k~\cite{DS20k} and its long time-scale successor Argo) are also considered for solar neutrino spectroscopy~\cite{Franco_2016}.  The DARWIN~\cite{Aalbers:2016jon} project aims at the realisation of a future astroparticle observatory in Europe. While its main goal is the direct detection of dark matter in a sensitive time projection chamber using a multi-ton target of liquid xenon, it would also be capable of a precision spectroscopy of low-energy neutrinos, especially $pp$ neutrinos~\cite{Aalbers:2020gsn} (see Sec.~\ref{WG1:SubSec:Solar:Projected}).

\label{sec:source_sol}

\subsection{Supernova Neutrinos}

Core-collapse supernovae originate from the death of massive stars and are among the most powerful sources of neutrinos: about $10^{58}$ neutrinos are emitted during the burst. 
This brilliant burst of neutrinos  has been observed just once, in 1987A.  This core collapse of a
  blue supergiant in the Large Magellanic Cloud, about 55~kpc away
  from us resulted in a supernova and a $\sim$ 10-second long burst of
  few-tens-of-MeV neutrinos observed in water Cherenkov and
  scintillator detectors~\cite{Bionta:1987qt,Hirata:1987hu,Alekseev:1987ej}.  The number of neutrino
  interactions seen was small, and the recorded neutrino events were primarily $\bar{\nu}_e$
  seen via inverse beta decay; nevertheless the observation was sufficient to confirm our
  understanding of the general mechanism of core collapse.  
  
  Neutrinos play a fundamental role in supernovae, transporting energy and lepton number. According to our current understanding, the supernova explosion occurs through the delayed neutrino-driven explosion mechanism~\cite{Bethe:1984ux}, i.e.\  neutrinos revive the stalled shockwave, triggering the explosion. Hydrodynamical simulations of supernovae have recently achieved the three-dimensional frontier~\cite{Mezzacappa:2020oyq}. However, despite the level of sophistication reached by neutrino transport, neutrino flavor mixing is not included in hydrodynamical simulations. In addition, 
  magneto-hydrodynamical effects are yet to be explored. 

 Stellar core collapses are expected to happen a few times a century
 in the Milky Way, and perhaps twice as often within a
 MPc radius, including Andromeda and the Local Group.
The next observed burst of neutrinos from a core collapse will bring a
wealth of information about the astrophysics of core collapse.
Because neutrinos carry information from deep inside the supernova
thanks to the weakness of their interactions, we will learn about the explosion processes.  The neutrino fluxes will track
accretion-related phenomena and asymmetries, as well as the sloshing of
material, the so-called ``standing accretion shock instability''~\cite{Lin:2019wwm}.
Neutrinos will also tell us about non-explosions -- an unknown fraction
of core collapses fail to result in spectacular fireworks.
Neutrinos will be emitted in similar numbers whether or not there is an eventual fizzle, with a sharp
neutrino flux cutoff signaling the formation of a black hole~\cite{Li:2020ujl}.

\begin{figure}[t]
\begin{center}
\includegraphics[width=1.0\textwidth]{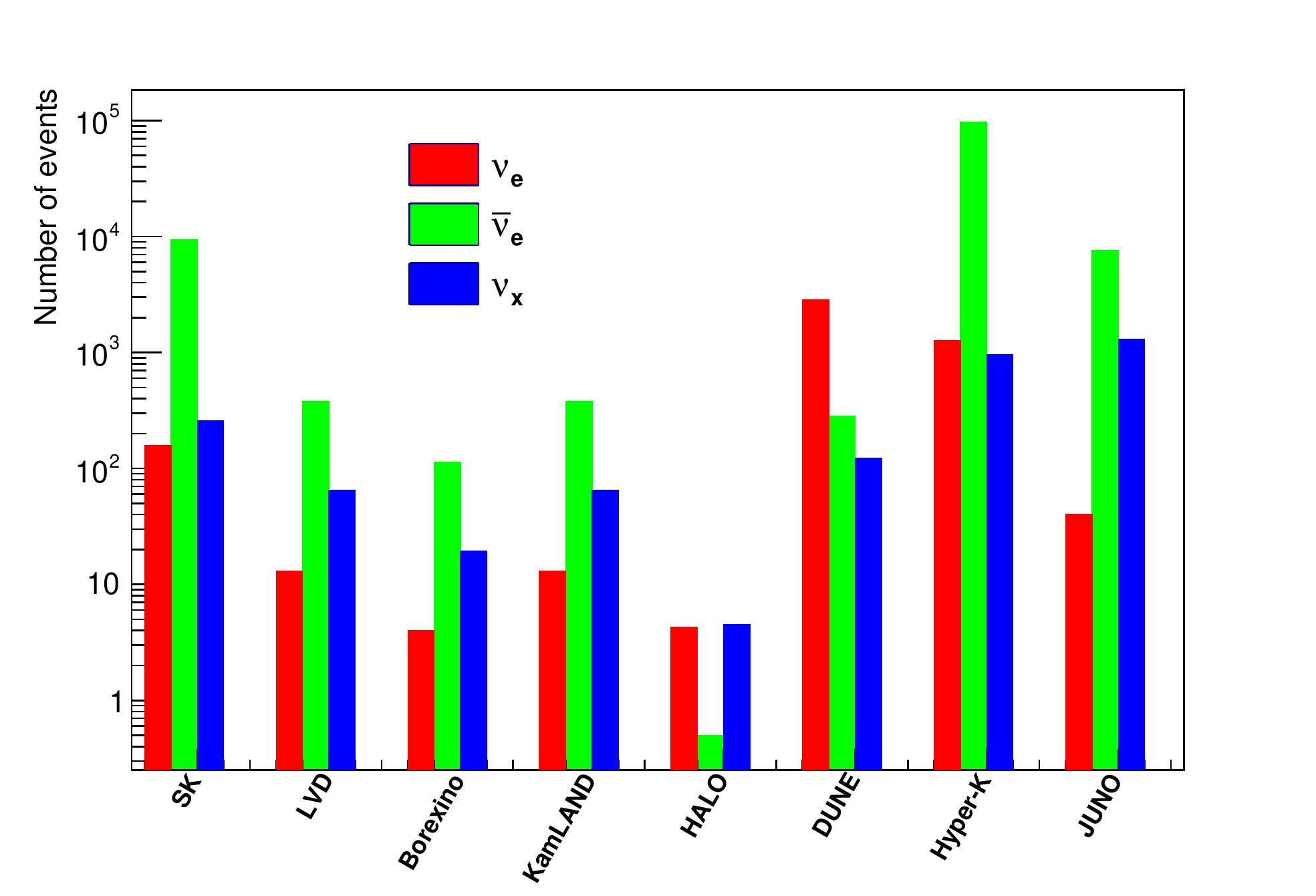}
\end{center}
\caption{Approximate predicted event counts for each flavor for several current and future detectors for an observed core-collapse burst at 10~kpc.}
\label{fig:snb}
\end{figure}

Particle physics is in play also during the core collapse -- neutrinos
oscillate, which will modulate the flavor content emitted from the
supernova.  Furthermore, the neutrino density may be so high that
neutrino-neutrino interactions come into play,
resulting in complex, exotic non-linear effects on the flavor content. 
Our understanding of neutrino mixing in the source is  
preliminary~\cite{Mirizzi:2015eza, Duan:2010bg, Tamborra:2020cul, Chakraborty:2016yeg} and subject of an active field of research.  Neutrino-neutrino interactions are responsible for making the flavor evolution non-linear and, contrary to common intuition, neutrinos of different energies collectively oscillate to another flavor. In addition, the flavor evolution is crucially affected by the neutrino angular divergence. A recent development in the field concerns the possible occurrence of flavor conversions triggered by pairwise scattering of neutrinos among themselves~\cite{Tamborra:2020cul,Chakraborty:2016yeg}. 
If this should be the case, then it would be necessary to include neutrino flavor conversions in hydrodynamical simulations. 
%These
%are the so-called ``collective oscillations,'' which can result in
%non-thermal, flavor-dependent spectral signatures as a function of
%time.  
 The observed fluxes may depend strongly on the neutrino mass ordering~\cite{Scholberg:2017czd}, though currently the believe is that answering this question independently from terrestrial oscillation experiments is unlikely. The effects are also important for nucleosynthesis in the neutrino-driven supernova wind.

Beyond-the-SM particle physics can imprint itself on the signal.  The existence of new particles would result in some fraction of energy emission into new degrees of freedom,  which would modify the energy emission scale of the neutrino burst; the observed burst of 1987A has resulted in a number of limits on exotic physics~\cite{Schramm:1990pf}, and a future high-statistics observation will enable yet more stringent limits or possibly point the way to new physics.
The observables that will give us a window on both core-collapse and
particle physics associated with the event are the neutrino flavors
and energy spectra as a function of time~\cite{Mirizzi:2015eza}.  

While gravitational-energy-powered core collapses are known signals a
few times a century, 
other types of astrophysical events, such as Type I (thermonuclear) supernovae will emit neutrinos as well,
although will need to be relatively near to be in range for neutrino detection~\cite{Wright:2016gar,Wright:2016xma}.

Many detectors worldwide, most with an array of other physics goals, are
sensitive to a core-collapse burst within at least a few tens of kpc
range~\cite{Scholberg:2012id}.  Figure~\ref{fig:snb} shows a summary of approximate event counts expected in current and future large detectors. Current detectors are primarily sensitive to the $\bar{\nu}_e$
component of the burst, via inverse beta decay (IBD) on free protons
as the dominant interaction channel.  These include liquid scintillator detectors,
such as KamLAND~\cite{Eguchi:2002dm}, LVD~\cite{Agafonova:2014leu}, Borexino~\cite{Monzani:2006jg}, and Daya Bay~\cite{Wei:2015qga}.  Water Cherenkov
detectors, like Super-Kamiokande, also have dominant sensitivity to $\bar{\nu}_e$ via IBD on free protons~\cite{Ikeda:2007sa,Abe:2016waf}.
``Long-string'' water Cherenkov detectors such as IceCube also have
primary sensitivity to $\bar{\nu}_e$, but detect an integrated
Cherenkov glow as a coincident single-photoelectron count rate
increase, rather than detecting interactions as fully reconstructed
events~\cite{Abbasi:2011ss,Aartsen:2014njl}.  HALO observes supernova neutrino interactions via ejected
final-state neutrons from charged-current $\bar{\nu}_e$ interactions
on lead~\cite{Duba:2008zz}.  All of these detectors have subdominant neutrino interaction
channels as well.  Neutrino-electron elastic scattering is notable 
as a highly anisotropic interaction; detectors able to exploit the anisotropy (e.g., Cherenkov ring-imaging detectors) can use this interaction to point back to a supernova.
The future large-scale detectors planned for the next decade will enhance  statistics and have richer flavor sensitivity.  JUNO~\cite{An:2015jdp}, at 20-kton scale, will increase the scintillator signal by a factor of $\sim 20$.  The 374-kton Hyper-K detector~\cite{Abe:2018uyc} will provide vast statistics. Upgrades to IceCube, as well as KM3NeT~\cite{Adrian-Martinez:2016fdl}, will also improve the time profile information. Notably, DUNE~\cite{Abi:2020lpk} in its 40 kton of LArTPCs (Liquid Argon Time Projection Chambers), will provide unique sensitivity to the $\nu_e$ flavor, thanks to a relatively high  charged-current cross section on argon nuclei.   Smaller LArTPC detectors such as MicroBooNE~\cite{Abratenko:2020hfy} will have sensitivity as well.
Furthermore, there are opportunities for more NC sensitivity, to the entire flavor profile, via elastic scattering (ES) interactions: scintillator detectors are sensitive to ES on protons, whereas a new generation of DM detectors (for example the 40-ton DARWIN~\cite{Aalbers:2016jon,Lang:2016zhv}) will observe a burst of coherent elastic neutrino-nucleus scattering events (see Sec.~\ref{sec:WG3_coherent}). Neutral-current elastic scattering channels are not affected by uncertainties related to flavor conversion physics; therefore detectors sensitive to these channels will offer a complementary view with respect to other detection technologies.   The observed energy distribution of recoils also gives information on the all-flavor neutrino spectrum as a function of time.

Nearly all supernova-neutrino detectors are located underground, in order to reduce the cosmic-ray backgrounds. However, some detectors on the surface will have sensitivity as well, e.g.\ NOvA~\cite{NOvA:2020dll}.

% Scaling with distance

The supernova neutrino burst is emitted promptly after core collapse, and therefore enables a potential early warning for the impending supernova, given that the first electromagnetic observations may not be possible for hours or longer. The Supernova Early Warning System (SNEWS)~\cite{Antonioli:2004zb, Scholberg:2008fa} is a world-wide network of neutrino detectors which will provide a fast warning to astronomers for a reported neutrino burst.  Some pointing information may be available from the observed neutrino signal.  The highest-quality pointing will likely be from anisotropic interactions; elastic scattering on electrons is the most promising~\cite{Beacom:1998fj}, although other channels have some potential anisotropies as well~\cite{Tomas:2001dh,Fischer:2015oma,Abi:2020evt}.  Triangulation holds some promise for fast information, as well~\cite{Linzer:2019swe,Brdar:2018tce,Brdar:2018zds}.
The prospects for multi-messenger astronomy with supernova neutrinos are excellent,  too.   Notably, the gravitational wave signal from a core-collapse signal will be prompt, and coincident detection with neutrinos should be possible~\cite{Nakamura:2016kkl} (and see Sec.~\ref{sec:numass_SN} for a discussion on neutrino mass limits from supernova observations).  An upgrade to SNEWS, SNEWS~2.0~\cite{Kharusi:2020ovw} is currently underway, which will enhance multimessenger capabilities.  High joint up-time and long-timescale running of large future supernova-burst-sensitive detectors will be of importance to increase the likelihood of capturing maximum information from a supernova burst.

Up to two or three supernovae may occur in our Galaxy per century, however, a supernova explodes every second somewhere in the Universe. 
Hence, another possible observable is the flux of all past explosions, the Diffuse Supernova Background (DSNB). Measuring the DSNB is challenging, but may give valuable information on the star formation rate in the Universe and on fundamental physics~\cite{Beacom:2010kk,deGouvea:2020eqq}. Recent developments point towards possible important effects on the DSNB signal coming from the presence of binaries and large theoretical uncertainties are currently linked to the supernova rate~\cite{Kresse:2020nto, Horiuchi:2020jnc, Moller:2018kpn}. The Super-Kamiokande experiment has been enriched with gadolinium~\cite{Beacom:2003nk} and is expected to observe the DSNB flux~\cite{Simpson:2018snj}. Also, future experiments like DUNE, JUNO or Hyper-Kamiokande have interesting prospects for DSNB observation~\cite{Moller:2018kpn}.\label{sec:source_SN}

%\subsection{Astrophysical Neutrinos (K. Scholberg, M. Kowalski)\label{sec:astroneutrino}}

\subsection{Atmospheric Neutrinos }
Atmospheric neutrinos provide a natural beam of high-energy neutrinos that can be used to probe neutrino properties. 
They are  produced when primary cosmic ray protons and heavier nuclei interact with atoms from Earth's atmosphere, resulting in  atmospheric air showers. These air showers  contain large numbers of energetic charged pions and kaons, as well as heavier mesons, which produce neutrinos in their decay (for a review see e.g.\ \cite{Gaisser:2002jj}).

The pioneering experiments to observe atmospheric neutrinos were started in the mid-1960s. These experiments were carried out in the Kolar Gold Field mines in India  and the East Rand Proprietary mine in South Africa. They were performed in extremely deep underground laboratories at the depth of about 8000 meters water equivalent to 
shield the experiments from the background of atmospheric muons. The next generation of atmospheric neutrino experiments began in the mid-1980s in Europe (NUSEX \cite{Aglietta:1988be} and Frejus \cite{Berger:1990rd} detectors), USA (IMB-3 detector \cite{Casper:1990ac}), and Japan (Kamiokande \cite{Hirata:1992ku} and later in 1990s Super-Kamiokande \cite{Ashie:2005ik})

The so-called {\it conventional} atmospheric neutrinos arise from pion and kaon decay. The atmospheric neutrino spectrum peaks around a GeV, and at higher energies can be described by a power-law, which is steeper by $1/E_\nu$ than that of the primary cosmic ray spectrum (this is because above the critical energies of 115 (850) GeV, the pions (kaons) more likely interact before they decay). Current uncertainties arise from the less well constrained pion-to-kaon ratio, as well as uncertainties in the primary cosmic ray spectrum and composition. The flux of conventional atmospheric neutrinos has been observed over a large energy range, from sub-GeV to $\sim 100$ TeV \cite{ANTARES:2010izk, Richard:2015aua}, above which cosmic neutrinos start dominating the flux \cite{Aartsen:2016xlq}. 

Neutrinos resulting from the decay of heavier mesons, containing charm or heavier quarks are called {\it prompt} neutrinos, as they originate from a prompt decay, and as a result the flux follows the primary cosmic ray spectrum more closely. The onset of the prompt component depends on the poorly constrained  cross-section for forward charm production, which is  being constrained from above by observing the high-energy part of the spectrum \cite{Aartsen:2016xlq}.  Interestingly, these uncertainties can be reduced through future accelerator experiments (see e.g.\  \cite{Ahdida:2020evc}).

Today, atmospheric neutrinos are being observed in large quantities, e.g.\ Super-Kamiokande has observed $40,000$ atmospheric neutrino events above 100 MeV \cite{Richard:2015aua}. Above several tens of GeV, ANTARES and IceCube have also detected both the muon and electron atmospheric neutrino components \cite{ANTARES:2021cwc}, e.g.\ IceCube at the South Pole has already detected $800,000$ muon-neutrino events \cite{Aartsen:2019fau} (with a contamination from cosmic neutrinos of not more than 1\%). 

Because the flux of atmospheric neutrinos is observable over five orders of magnitude in energy, and over a range of baselines, it provides an essential probe for the study of standard and non-standard neutrino properties. A milestone for neutrino physics was the detection of atmospheric neutrino mixing by Super-Kamiokande in 1998 \cite{Fukuda:1998mi}, while today the mixing parameters $\theta_{23}$ and $\Delta m_{32}^2$ are being constrained also by neutrino beam experiments. In addition, tau neutrinos, which are not directly produced in the atmosphere in significant amounts, appear due to  oscillations of atmospheric muon neutrinos. They have been recently observed by Super-Kamiokande \cite{Li:2017dbe} and IceCube-DeepCore \cite{Aartsen:2019tjl}. The neutrino mass ordering also has an imprint on the atmospheric neutrino flux  (most notable around 10 GeV for vertical directions). A new generation of atmospheric neutrino detectors, Hyper-Kamiokande \cite{Abe:2018uyc}, ORCA \cite{KM3NeT:2021ozk} and IceCube-Upgrade \cite{Bezerra:2019dao}  is  capable to observe mixing parameters, as well as the unique signature of the neutrino mass-ordering, due to significantly improved sensitivity. \label{sec:source_atm}

\subsection{High-Energy Astrophysical Neutrinos}

\begin{figure}[t]
\centering\includegraphics[trim={0cm 2.7cm 0cm 2.0cm},clip,width=1\linewidth]{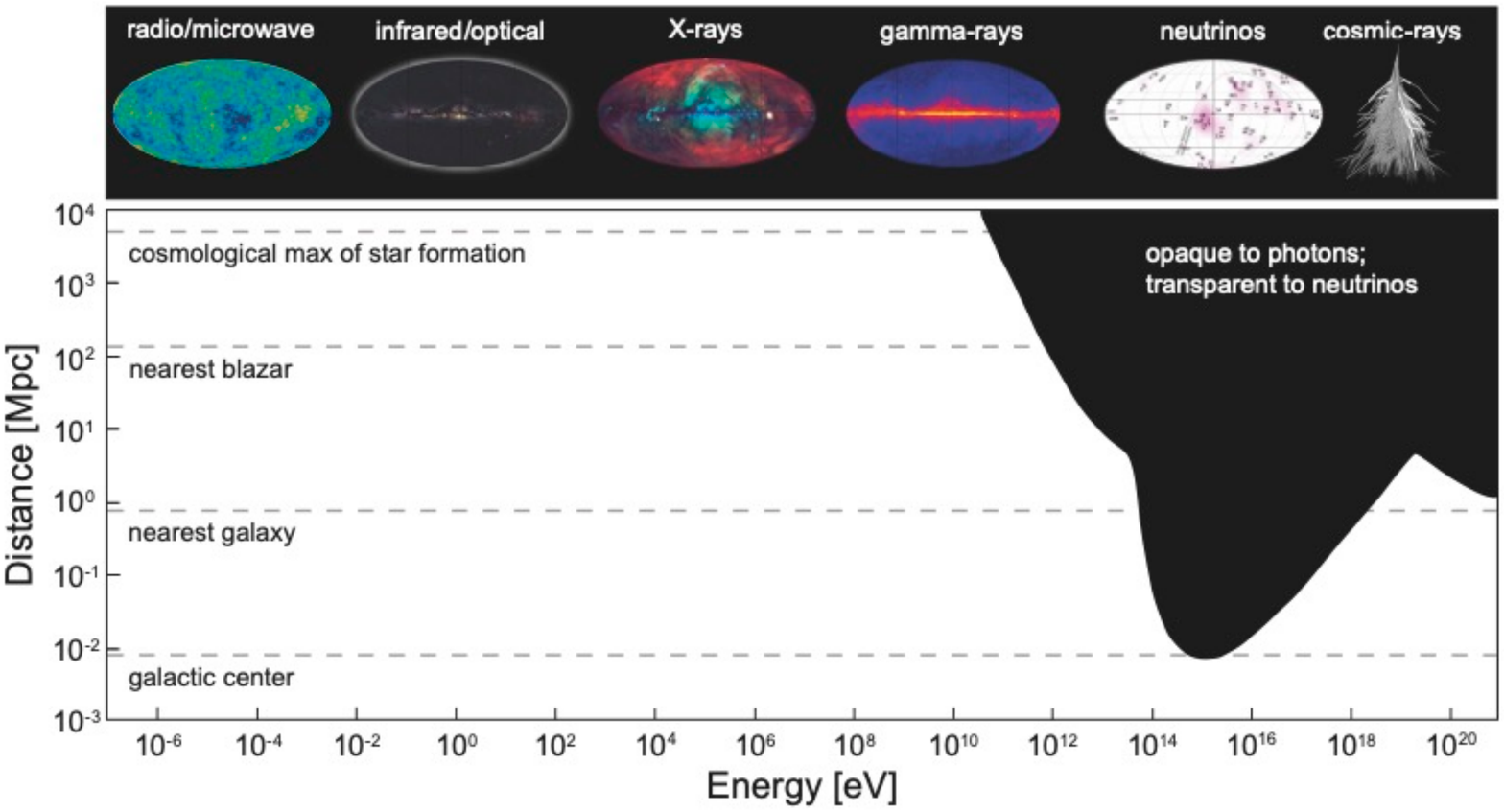}
\caption{Distance horizon at which the Universe becomes intransparent to electromagnetic radiation \cite{10.1088/978-0-7503-1369-8}.}\label{fig:horizons}
\end{figure}
 High-energy neutrinos escape energetic and dense astrophysical environments that are opaque to electromagnetic radiation. In addition, at PeV (10$^{15}$~eV) energies, extragalactic space becomes opaque to electromagnetic radiation due to the scattering of high-energy photons ($\gamma$ rays) on the cosmic microwave background and other radiation fields (see Fig.\ \ref{fig:horizons}). This leaves neutrinos as the only messengers to search for the most extreme particle accelerators in the cosmos -- the sources of the ultra-high-energy cosmic rays (UHECRs). These UHECRs reach energies of more than 10$^{20}$~eV, which is a factor of 10$^{7}$ times higher than the most powerful man-made particle accelerators. 

\subsubsection{Introduction}
\label{sec:WG_4_he_intro}
Astrophysical neutrinos are produced from the interactions between cosmic rays and matter or radiation; they therefore trace the origin of cosmic rays. The dominant neutrino production modes are $pp$ and $p\gamma$ interactions, where relativistic protons (or nuclei) interact with gas and radiation, respectively. The relative importance of these production modes depends on the gas or radiation (target) density, as well as the energy spectrum of the radiation. For $pp$ interactions, one roughly obtains $\pi^+$, $\pi^-$ and $\pi^0$ in equal fractions, whereas $p\gamma$ interactions are at threshold dominated by the $\Delta(1232)$-resonance 
\begin{equation}
	p + \gamma \rightarrow \Delta^+ \rightarrow \left\{\begin{array}{lc} n + \pi^+ & \frac{1}{3} \text{ of all cases} \\[0.2cm]  p + \pi^0 & \frac{2}{3} \text{ of all cases} \end{array} \right.  . \label{equ:Delta}
\end{equation}
The pions decay via the usual weak decay chains such as 
\begin{eqnarray}
\pi^+ & \rightarrow & \mu^+ + \nu_\mu \, ,\nonumber \\
& & \mu^+ \rightarrow e^+ + \nu_e + \bar{\nu}_\mu \, , \label{equ:piplusdec}  
\end{eqnarray}
where in this standard picture $\nu_e:\nu_\mu:\nu_\tau$ are produced in the ratio $1:2:0$ if neutrinos and antineutrinos are not distinguished. Flavor mixing (averaged neutrino oscillations), described by $P_{\alpha \beta}=\sum_i |U_{\alpha i}|^2 |U_{\beta i}|^2$, then is widely believed to lead to a flavor composition close to $1:1:1$ at detection; see~\cite{Farzan:2008eg} for a critical discussion. 
Unlike the charged cosmic rays, neutrinos are not deflected by magnetic fields on the way from the source to the Earth, but point back to their origin, thus providing for a smoking-gun signature of cosmic-ray acceleration. The physics implications regarding the relationship to multiple messengers are discussed in \Sec~\ref{sec:heneutrinos}.

The scientific potential of using high-energy neutrinos for astronomy has been obvious for many decades, and yet, the path towards discovery was a stony one \cite{Katz:2011ke}. However, the technical problems were overcome and by now, the concept of open water/ice neutrino telescopes, that are sensitive from tens of GeV to beyond PeV energies, has been successfully demonstrated in several places world-wide: the Baikal collaboration deployed a first functional detector in lake-water\cite{BAIKAL:1997iok}, the ANTARES collaboration deployed the first successful undersea detector \cite{ANTARES:2011hfw} and the AMANDA collaboration installed the first in-ice neutrino detector \cite{Andres:1999hm}. The detection principle  is similar for all detectors: Cherenkov-light produced by charged particles -- either background muons produced in air showers above the detector, or particles produced in a neutrino interaction -- is recorded by a three-dimensional array of photomultipler tubes (PMTs) contained in appropriate pressure-resisting glass housings. The arrival time allows reconstructing the direction of the particles (to better than a degree for muons), while the total number of photons recorded is used to reconstruct  its energy. Arrival direction, energy and topology allows to distinguish  background from neutrino-induced events. A more refined analysis is then needed to distinguish astrophysical neutrinos from atmospheric neutrinos.

In addition, the energy range accessible with neutrino observatories is being expanded into the EHE 
(Extremely High Energy) region ($10^{18}$~eV) through a diverse range of technologies. Particle showers developing in the ice or the atmosphere produce a coherent signal at radio frequencies, that because of the longer attenuation length, can be detected over large distances. Through instrumenting natural ice with radio antennas, larger detection volumes compared to the optical regime are achievable, providing  sensitivity beyond tens of PeV. Monitoring the atmosphere for Earth skimming atmospheric air showers using a range of giant air shower detection techniques, including radio but also Cherenkov radiation in the optical, is another cost effective method to expand the sensitivity at EHE energies.   

Key scientific goals for current and future projects include:
\begin{enumerate}
  
\item Resolving the high-energy sky from TeV to EeV energies: What are the  sources of high energy neutrinos detected by IceCube? 

\item Understanding cosmic particle acceleration through multimessenger observation: This implies studying particle acceleration and neutrino emission from a range of multimessenger sources (e.g.\ AGN, GRBs, TDEs, SNe or kilonovae, see below and Sec.\  \ref{sec:heneutrinos} for explanations). Constraints on the physics within these sources can also come from measurements of spectrum and flavor composition of the astrophysical neutrino flux.

\item Revealing the sources and propagation of the highest energy particles in the Universe: This includes extragalactic cosmic ray sources and their neutrino emission, as well as the propagation of cosmic rays through the measurement of cosmogenic neutrinos, extending well into the EHE range. 

\item Identifying hadronic sources of cosmic rays in our galaxy, as well as detecting the high energy emission from hadronic cosmic rays propagation in our galaxy.

\item Probing fundamental physics with high-energy neutrinos: This entails the measurement of neutrino cross sections at high energies, searching for new physics  affecting neutrino flavor mixing on cosmic baselines, and searches for heavy dark matter.
\end{enumerate}

\subsubsection{Current Status}

The South Pole is home to the  currently largest operating neutrino detector. The IceCube detector, which was deployed between 2005 and 2010, consists of 86 strings with 5160 PMTs in total. The instrumented volume comprises a cubic kilometer and it is deployed between 1450 and 2450 m depth.  It has collected 
neutrino induced events with up to 10 PeV in energy, corresponding to the highest energy elementary particles ever observed and opening new scientific avenues not just for astronomy but also for probing physics beyond the Standard Model of particle physics (see, e.g., \cite{Aartsen:2017kpd}).   

With the first detection of high-energy neutrinos of extraterrestrial origin in 2013 by the IceCube Neutrino Observatory~\cite{Aartsen:2013jdh},  a new window to some of the most extreme parts of our Universe was opened.
In the Northern Hemisphere, ANTARES, has been taking data since 2006. Despite its modest size (12 strings of 75 PMTs), and thanks to its excellent angular accuracy, it provides constraints on the origin of the origin of the high-energy cosmic neutrino flux measured by IceCube. Due to its location, ANTARES has also good visibility over a large part of the Galactic plane, e.g.\ see \cite{ANTARES:2020srt} for a joint search for point-like sources in the Southern sky.

The most compelling evidence for a neutrino point source to date is the detection of one neutrino event (IC-170922A) in spatial and temporal coincidence with an enhanced $\gamma$-ray emission state of the blazar TXS~0506+056 \cite{IceCube:2018dnn}. Evidence for another period of enhanced neutrino emission from this source, in 2014/15, was revealed in a dedicated search in the IceCube archival data \cite{IceCube:2018cha}. The individual chance probabilities of the blazar-neutrino association and the observed excess in the IceCube data alone are each at a significance level of 3~--~3.5$\sigma$.  

Additional events of a similar nature are required to provide definitive statements about the production mechanism of neutrinos in blazars. At the same time, it is becoming increasingly clear that $\gamma$-ray blazars can not explain the majority of astrophysical neutrinos observed by IceCube: the number of observed coincidences  is  smaller than expected if compared to the total number of cosmic neutrino events~\cite{IceCube:2018dnn, Aartsen:2019gxs}. Further, a comparison of the full set of IceCube neutrinos with a catalog of $\gamma$-ray blazars does not produce evidence of a correlation and results in an upper bound of $\sim$ 30\% as the maximum contribution from these blazars to the diffuse astrophysical neutrino flux below 100 TeV~\cite{Glusenkamp:2015jca}. Accordingly, a blazar population responsible for the whole astrophysical neutrino flux would have to be appropriately dim in gamma-rays (see, e.g.~\cite{Halzen:2018iak, Palladino:2018lov, Neronov:2018wuo}). Empirical correlations between some astrophysical neutrinos and specific blazar populations have, for example, been proposed for radio-bright blazars \cite{Plavin:2020emb, Plavin:2020mkf} and intermediate/high frequency-peaked BL Lacs~\cite{Giommi:2020hbx}.

\begin{figure}[t!]
\centering\includegraphics[width=1.0\linewidth]{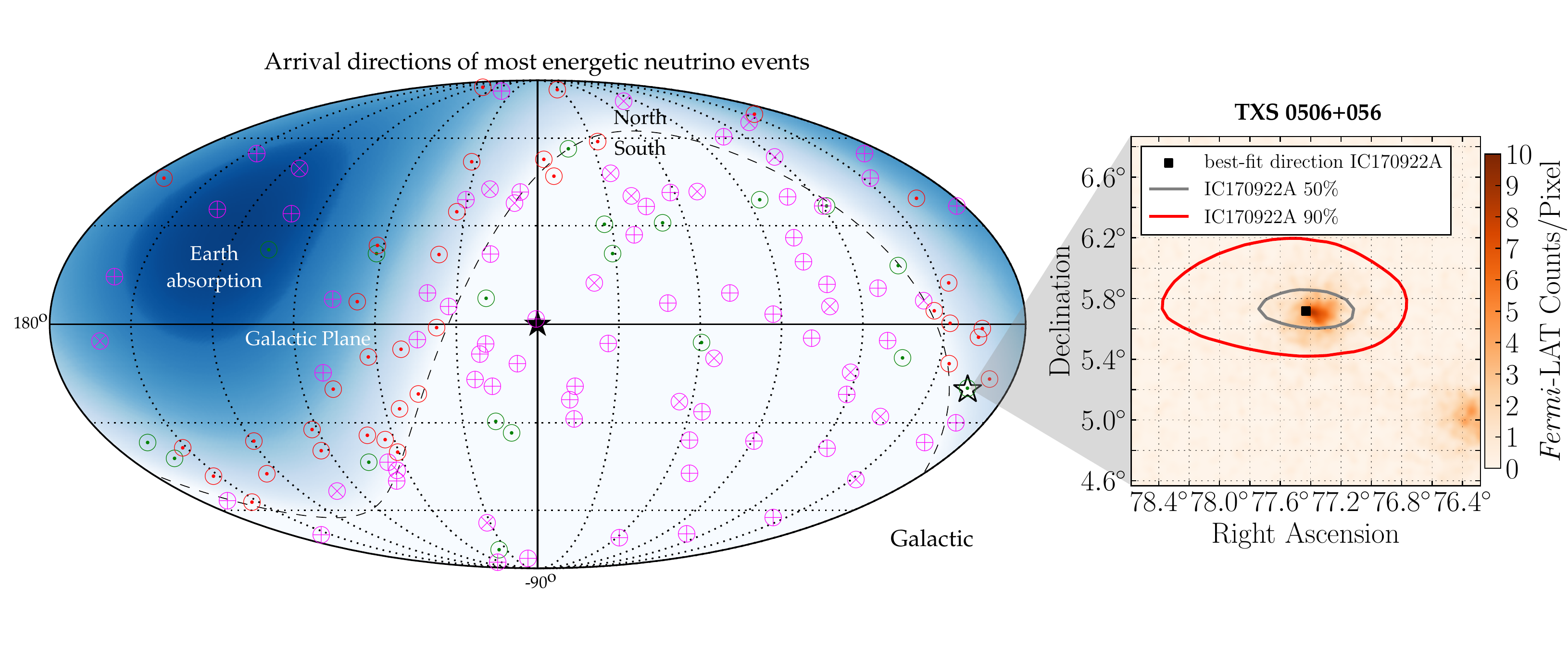}
\caption{The current sky map of highly energetic neutrino events detected by IceCube. Shown are upgoing track events~\cite{Aartsen:2016xlq,Haack:2017dxi}, the high-energy starting events (HESE) and cascades~\cite{Aartsen:2014gkd,Kopper:2015vzf,Kopper:2017zzm}, and additional track events published as public alerts~\cite{Smith:2012eu}. The distribution of the events is largely isotropic. The location of the first compelling neutrino source, blazar TXS~0506+056, is marked with a star.  Shown in the inset are the related \emph{Fermi Large Area Telescope}~(LAT) measurements of the region centred on TXS~0506+056 from September 2017~\cite{IceCube:2018dnn}.  The uncertainty ellipses of the IceCube neutrino event IC-170922A are shown for reference.}
\label{fig:skymap}
\end{figure}

Apart from the association with a $\gamma$-ray blazar, a neutrino from the Tidal Disruption Event (TDE) AT2019dsg has been observed very recently~\cite{Stein:2020xhk}, which points towards another (probably sub-dominant) source population producing astrophysical neutrinos; it is therefore likely that several source populations contribute to the diffuse astrophysical neutrino flux. Another widely considered candidate source of extragalactic neutrinos are $\gamma$-ray bursts (GRBs); because of their short duration atmospheric backgrounds can be efficiently reduced for this transient population, yielding perhaps the best sensitivity in IceCube. Similar to blazars, the non-detection of neutrinos in spatial and temporal coincidence with GRBs over many years has placed a strict upper bound of 1\% for the maximum contribution from observed GRBs to the diffuse flux observed by IceCube~\cite{Aartsen:2014aqy}. 
In spite of individual source associations, the distribution of astrophysical neutrinos in the sky is largely consistent with isotropy to date (see Fig.~\ref{fig:skymap}), implying that the dominant contribution to the astrophysical neutrino flux is of extragalactic origin. However, a contribution of neutrinos from Galactic sources, such as supernova remnants, or a diffuse component from interactions between Galactic cosmic rays and gas will lead to anisotropies as interesting targets for instruments more sensitive to the Southern hemisphere.

Independent evidence for astrophysical neutrinos comes from different detection channels, including shower-like events~\cite{Niederhausen:2017mjk}, events that start inside the instrumented volume~\cite{Aartsen:2014gkd}, through-going events~\cite{Aartsen:2015rwa}, as well as first candidates for ``double-bang'' tau-neutrino events \cite{Stachurska:2019srh} that are not expected to be produced in the atmosphere through conventional channels. The ANTARES Collaboration also has reported a mild excess of high-energy neutrinos that is not significant by its own \cite{ANTARES:2017srd}.

While the collective significance for the cosmic origin of the neutrinos has reached a level that is beyond any doubt, a decade of IceCube data taking has demonstrated the rarity of the measurements; e.g., only two tau neutrino candidates \cite{Abbasi:2020zmr} and one electron antineutrino candidate at the Glashow resonance of 6.3 PeV~\cite{Glashow:1960zz} have been observed to date. Clearly, much larger statistics are needed to exploit the full potential of all-flavor neutrino astronomy. 

At EeV-energies, a number of experiments has been searching for neutrinos from cosmic sources and neutrinos produced in the propagation of cosmic rays through the Universe. So far, only  upper-limits have been reported. The experiments include the ground based cosmic air shower observatory AUGER \cite{Aab:2019auo}, the balloon borne experiment ANITA \cite{Gorham:2019guw} (which has, however, observed interesting anomalous events \cite{Gorham:2018ydl}), or experiments operating on the surface or at shallow depth of the Antarctic ice sheet (ARA \cite{ARA:2019wcf} and ARIANNA \cite{Anker:2019rzo}). IceCube has also set limits in the energy range \cite{Aartsen:2018vtx}. Besides their scientific value in limiting the flux of highest-energy neutrinos, the experiments serve as important technology path-finder missions for a series of next generation detectors.

\subsubsection{Future Outlook}

Given the limited statistics that IceCube collects at the very highest energies, the identification of counterparts requires very long integration time. Furthermore, the moderate angular resolution of $\sim0.5^{\circ}$ for muon neutrinos and $\sim 10^{\circ}$ for electron and tau neutrinos (so-called cascade-like events) make identification of neutrino point sources currently very challenging. 
Consequently, the initial association of cosmic neutrinos with the first extragalactic objects has been an essential step, however, the sources for the bulk of the cosmic neutrino flux observed by IceCube remain to be resolved using instruments with much higher size and improved properties. The list of well motivated candidates is long: transient sources such as 
Gamma Ray Bursts (GRBs), TDEs, or steady sources, such as Active Galactic Nuclei (AGN) or Starburst galaxies, for example.

Furthermore, more elaborate multi-messenger studies which combine information from various observatories, ranging from $\gamma$~rays, via X-rays to the UV, optical and radio bands, and including gravitational waves, are pointing the way for more associations of high-energy neutrinos with their sources; see Sec.~\ref{sec:heneutrinos} for a more theoretical perspective. 

\noindent 
To reach the goals mentioned above and in  Sec.\ \ref{sec:WG_4_he_intro}, work for a new generation of instruments is ongoing. In the PeV energy range, the KM3NeT and Baikal-GVD detectors, under construction in the Mediterranean sea and in Lake Baikal %in Siberia 
 respectively, target a similar size as the one from IceCube and will complement IceCube in terms of sky coverage  \cite{Adrian-Martinez:2016fdl, Avrorin:2014vca}, and will  provide for comparable numbers of astrophysical neutrinos. Construction of KM3NeT has started. The effort is distributed over two sites, one on the French  coast named KM3NeT-ORCA,  focusing on lower energy neutrinos (10 GeV) and the other one called KM3NeT-ARCA focusing on the energy regime of IceCube. 
 As of today (summer 2021), 6 strings of ORCA and further 6 strings of ARCA have been deployed and are operational. Completion of KM3NeT is expected for 2026. 
 
 The construction of the Baikal-GVD detector was started in 2016. The current (summer 2021) effective volume of the detector for cascade events in the energy range 100 TeV -- 10 PeV is about 0.4 km$^3$. In this energy range, first cascade-events, candidates for neutrino events of astrophysical origin, have been detected already \cite{Belolaptikov:2021a1}. 
By 2024, the effective volume of the detector that is already funded is expected to reach about 0.7 km$^3$. The plan is to further extend the effective volume in the years after, reaching a volume of up to 1.5 km$^3$.

Another site in the northeast Pacific Ocean, 200 kilometers off the Canadian coast, was recently optically qualified \cite{Bailly:2021dxn}. The 2.6 kilometers deep site is being explored to host the Pacific Ocean Neutrino Experiment (P-ONE) \cite{Agostini:2020aar}, a neutrino telescope that will be based at an existing underwater facility (Ocean Network Canada). 

A number of projects are also being developed targeting the EeV-energy range. These include the detection of neutrino interactions in the ice using their radio signature (ARIANA200 \cite{Anker:2020lre}, RNO-G \cite{RNO-G:2020rmc}), or the search for tau neutrinos which are just skimming the Earth, interacting near the surface, so that the tau lepton can escape the dense environment to  decay in the atmosphere. Such upgoing air shower events can be observed using air Cherenkov  and flurescence telescopes (POEMMA \cite{Olinto:2019euf}, TRINITY \cite{Otte:2019knb}) or using again their radio signature (GRAND \cite{Alvarez-Muniz:2018bhp}, BEACOM \cite{Wissel:2020sec}). These project will have the sensitivity to probe the leading models. Finally, entirely new detection methods are being developed, such as the detection of EeV neutrinos via the radio echo signature \cite{Prohira:2019glh}.
 
IceCube-Gen2, a proposed wide-band neutrino observatory (MeV--EeV) that employs two complementary detection technologies for neutrinos -- optical and  radio, will provide  order of magnitude improved event rates of astrophysical neutrinos in the PeV range and five times the sensitivity to point sources compared to IceCube and for the first time provide a comparative sensitivity in the EHE range \cite{TheIceCube-Gen2:2016cap}. 
Construction of its low-energy core has already started as part of the IceCube Upgrade project~\cite{Ishihara:2019aao} that is a smaller realization of the PINGU concept~\cite{Aartsen:2014oha}. The completion of the IceCube-Gen2 construction is foreseen for 2032.

\label{sec:source_IC}

\subsection{Geoneutrinos }
\label{wg4_geo}

\subsubsection{Introduction}
Geoneutrinos are antineutrinos produced by natural radioactivity in the Earth \cite{fiorentini07}. Most geoneutrinos come from the decay of $^{40}$K and relatively smaller contributions come from the decays of $^{232}$Th and $^{238}$U. Together these three nuclear isotopes account for more than $99$\% of the heat generated by Earth's radioactivity. The distribution of radiogenic heating between the crust, mantle, and core gives unique insights about the formation and evolution of Earth. The direct connection between geoneutrinos and radiogenic heating makes antineutrino detectors important instruments for geophysical research.

The geoneutrino generating elements (K, Th, U) are lithophilic, according to the Goldschmidt classification. They exhibit geochemical affinity for the Earth's outer rocky crust and mantle layers but not for the metallic core. Their distributions, which are not fully known, throughout these silicate layers smooths over the effect of neutrino oscillations on the geoneutrino fluxes. An average oscillation probability provides an accuracy at the level of a few percent \cite{mao19}, which is comparable to the uncertainties in the oscillation parameters. For comparison, these uncertainties are small compared with those introduced by the geological modeling. Assessments of the concentrations of K, Th, and U in the largely inaccessible rocky layers of the Earth typically come with non-Gaussian uncertainties at the level of tens of percent. These uncertainties carry through to the predicted geoneutrino fluxes. Geoneutrino flux measurements, with their statistical and systematic uncertainties presently at the level exceeding $10$\%, fail to inform on neutrino oscillations and other neutrino properties. 

Predictions of the fluxes of geoneutrinos, originating from terrestrial $^{40}$K, $^{232}$Th, $^{238}$U, along with other less important nuclear isotopes, began to appear in the scientific literature about a decade after the discovery of the neutrino \cite{eder66}. Successive refinements \cite{krauss84,raghavan98,roth98}, including geothermal, seismic, and geochemical constraints, converged on a reference model \cite{mantovani04, huang13}. The general conclusion is that geoneutrino fluxes are largest over continents and smallest over ocean basins. A less prominent result is a slightly softer energy spectrum over continents than over ocean basins. This is due to a higher ratio of Th to U in continental crust than in the mantle. Interestingly, the difference could be the imprint of biological activity \cite{sleep13}. Estimates of the geoneutrino fluxes from the crust take inputs from physics for decay rates, from seismology for mapping the densities and locations of the various Earth layers \cite{crust1.0}, and from geochemistry for assessing the concentrations of the nuclear isotopes in these Earth layers. The uncertainties are largest ($\sim20$\%) on the concentrations and smallest on the decay rates ($\sim1$\%). Typically, there are separate estimates of the fluxes from the near-field and the far-field crust. Improving the accuracy and uncertainty of these estimates is an area of active research \cite{takeuchi19}. 

The energy spectra of geoneutrinos emitted from K, Th, and U \cite{sanshiro_spec} are known at the level of $\sim1$\%. These spectra clearly show the endpoint energies of the many contributing nuclear beta decays. Only four of these decays, two each in the $^{232}$Th ($^{228}$Ac, $^{212}$Bi) and $^{238}$U ($^{234m}$Pa, $^{214}$Bi) series, have endpoint energy above the $1.806$ MeV threshold for inverse beta decay of the free proton. Antineutrinos from $^{40}$K and the $^{235}$U series, with the exception of a very rare decay ($^{215}$Bi), have maximum energy below this threshold. The cross sections for inverse beta decay of free protons \cite{strumia03} and elastic scattering off atomic electrons, are known at the percent level or better, see Sec.\  \ref{chap:interactions}. Electron elastic scattering has no threshold energy, providing sensitivity to geoneutrinos from $^{40}$K. Sensitivity through inverse beta decay of nuclear targets other than hydrogen is possible \cite{krauss84} but requires further detector development.

Information from geoneutrinos on the amounts and spatial distributions of K, Th, and U in Earth comes from comparing measurements at various locations. Comparisons of measured rates, spectral shapes, and directions of observed signals are possible. Time variation of geoneutrino measurements at a given location is discounted due to the constancy of decay rates and the exceedingly slow relative motion between Earth reservoirs. The challenges for geoneutrino observations include reducing measurement errors, detecting K, and developing sensitivity to the directions of the geoneutrinos, all of which contribute in constraining the geological models. Missing tests include measuring surface variation of the magnitudes (rate) and the relative contributions (spectral shape) of the geoneutrino fluxes, as well as assessing the roles of K, Th, and U in radiogenic heating in the mantle and the core.
\subsubsection{Current Status}
There are measurements of the geoneutrino fluxes from Th and U at Japan by KamLAND \cite{watanabe19} and at Italy by Borexino \cite{agostini20}. Both detectors efficiently ($\sim80\%$) record geoneutrino interactions by inverse beta decay on free protons in scintillating liquid. After subtraction of well studied sources of background, measurements resolve the energies, but not the directions, of the interacting geoneutrinos. The KamLAND measurement, which results from an exposure of $7.2\times10^{32}$ proton-years, rejects the zero signal hypothesis from Th and U at $1.68\sigma$ and $3.15\sigma$, respectively. The Borexino measurement, which results from an exposure of $1.3\times10^{32}$ proton-years, rejects the zero signal hypothesis from Th as well as U at $\sim2.4\sigma$. The measurements of KamLAND and Borexino, which clearly demonstrate the detection of geoneutrinos from Th and U, are compared with predictions from a reference model in Tab.~\ref{tab:gnu_status}. Although differences in the relative strengths of the fluxes from continental crust and the mantle at different locations lead to predicted surface variations in the geoneutrino fluxes, the measurements from KamLAND and Borexino agree within uncertainties.

\begin{table}
\begin{tabular}{l c c c c c c c}
\hline\hline\noalign{\smallskip}
                            &  & Japan &  &  &  & Italy & \\
\cline{2-4} \cline{6-8}
                             & $\phi_\mathrm{U}\pm 1\sigma$ & $\phi_\mathrm{Th}\pm 1\sigma$ & ${\rm Th/U}\pm 1\sigma$ & & $\phi_\mathrm{U}\pm 1\sigma$ & $\phi_\mathrm{Th}\pm 1\sigma$ & ${\rm Th/U}\pm 1\sigma$  \\
\hline\noalign{\smallskip}
Measurements     & $1.79^{+0.60}_{-0.57}$ & $2.00^{+1.19}_{-1.19}$ & $5.3^{+6.0}_{-3.6}$ & & $1.9^{+1.1}_{-0.8}$  & $4.7^{+2.2}_{-2.0}$  & $-$ \\
Model prediction  & $3.47^{+0.65}_{-0.53}$ & $3.03^{+0.67}_{-0.43}$  & $4.11^{+1.19}_{-0.86}$ & & $4.34^{+0.96}_{-0.75}$ & $4.23^{+1.26}_{-0.80}$ & $4.59^{+1.45}_{-1.17}$ \\
\hline\hline\noalign{\smallskip}
\end{tabular}
\caption{Geoneutrino measurements at Japan \cite{watanabe19} and Italy \cite{agostini20} compared with a model prediction \cite{huang13}.}
\label{tab:gnu_status}
\end{table}

Resolution of the geoneutrino fluxes from the mantle, leading to estimates of global heating due to radioactivity, follow from further analysis of the measurements \cite{fiorentini12}. Due to the lithophilic nature of K, Th, and U, the reference model specifies that the geoneutrino fluxes originate in the crust and mantle only. The standard analysis respects this guidance, making the mantle contributions simply the differences between the calculated total fluxes and the estimated fluxes from the crust. At underground locations the reference model predicts larger geoneutrino fluxes from the crust than from the mantle. For the present measurements the mantle fluxes are the differences of two larger fluxes with uncertainties that impede resolution. The standard analysis constrains the shape of the measured energy spectrum to conform to the cosmochemical value of Th/U  $=3.9$ \cite{Rocholl:1993}. This constraint, which is due to the limited statistics of the measurements, reduces the reported uncertainties and brings the calculated fluxes of the observed Th and U geoneutrinos into agreement with the reference model.

The present observations of the geoneutrino fluxes from Th and U by KamLAND and Borexino are outstanding scientific achievements. They represent decades-long efforts to build and operate detectors capable of real-time monitoring of the heat generated by global radioactivity. The initial assessments of radiogenic heating benefit from reference model constraints. Resolving the geoneutrino fluxes from the mantle with more model-independent analyses requires greater exposures and more favorable detector locations than afforded by the existing measurements. The challenges remaining for future geoneutrino observations include detecting the flux from K and gaining more direct sensitivity to fluxes from the mantle and possibly the core. Each of these may be met by measuring the directions of the geoneutrino fluxes \cite{leyton17}.

\subsubsection{Future Prospects}
The existing geoneutrino measurements come from detectors with an impressive record of advancing knowledge of neutrino oscillations and solar fusion. Several upcoming neutrino detectors, which are motivated by fundamental questions in physics and astrophysics, forecast sensitivity to geoneutrinos. The siting of these neutrino detectors is motivated by access to overburden to reduce background due to cosmic rays and often by proximity to nuclear power reactors for an intense source of antineutrinos. These underground locations optimize the ability to perform the physics and astrophysics experiments rather than those enabled by geoneutrino measurements. The deep mines and tunnels beneath high mountains are typically in geologically complex regions with corresponding challenges in estimating the geoneutrino fluxes from the crust. Nonetheless, the upcoming detectors are planning to perform geoneutrino measurements.

One detector nearing operation with reported sensitivity to geoneutrinos is SNO+ \cite{chen06}. This is the former Solar Neutrino Observatory in Sudbury, Canada, reconfigured to search for neutrinoless double beta decay, see Section \ref{sec:0vbb}. The detector site is one of the deepest in the world, greatly reducing cosmogenic sources of background. Unlike KamLAND and Borexino, it is located well inside a continent with the prediction of strong geoneutrino fluxes from the crust. Using a target mass of $780$ tons of scintillating liquid, the expected rate of recorded geoneutrino interactions is about $20$ per year, assuming $80$\% detection efficiency. Using these values, it would take about eight years of detector operation to accumulate the measurement precision afforded by the 165 recorded interactions already reported by KamLAND \cite{watanabe19}. It is not clear if the precision of the SNO+ geoneutrino flux measurements over this period would provide evidence of surface variation. The error bars of the projected SNO+ measurements could still overlap with those of the KamLAND and Borexino measurements. There is certainty that the geoneutrino Th/U measurement from SNO+ over this period finds no disparity with those from KamLAND and Borexino.    

Another detector, which plans to begin operation within the next several years and with reported sensitivity to geoneutrinos, is JUNO \cite{An:2015jdp}. This is a new detector, carefully situated about $53$ km from several nuclear reactor complexes to maximize sensitivity to the neutrino mass ordering. The detector site in South China is near the continental shelf and offers an overburden of $\sim2000$ meters of water equivalent (m.w.e.), making it similar in shielding to the  KamLAND site. The distinguishing characteristics of JUNO are its size and sensitivity. At $20$ kton of scintillating liquid it is about $25$ times larger than SNO+ and the projected detector resolutions are unprecedented. Without question, the world sample of recorded geoneutrino interactions more than doubles after one year of operation of the JUNO detector. The main obstacle in measuring the geoneutrino fluxes is accurate knowledge of the reactor antineutrino rate and spectrum. Even with perfect subtraction of the reactor antineutrinos and with the superb precision projected by the unprecedented size and sensitivity of JUNO, the site location may be geologically too similar to Japan to provide evidence of surface or spectral variation of the geoneutrino fluxes compared alongside the measurements of KamLAND, Borexino, and SNO+. 

There are several detectors under consideration that hold promise of measuring geoneutrino fluxes significantly ($>1\sigma$) different from the KamLAND and the high statistics future JUNO measurements. This promise stems from the proposed multi-kton target masses for the needed exposures and the deep continental locations for the predicted strong crust fluxes. It appears less likely that measurements of Th/U by these detectors would be significantly different from the JUNO measurement. A detector with $10$ kton of scintillating liquid is planned at Baksan, which is a very deep site ($4760$ m.w.e.) beneath the Caucasus mountains \cite{barabanov17}. The proponents predict a $10$\% measurement of Th/U. A detector with $3$ kton of scintillating liquid is planned at Jinping, which is an extremely deep site ($6720$ m.w.e.) in central China \cite{wan17}. The proponents predict a $\sim25$\% measurement of Th/U with an exposure of $\sim12$ kton-y. A detector with $50$ kton of water-based scintillating liquid is suggested for the Homestake Mine, which is a deep site, $4300$ m.w.e.,  in the Black Hills of South Dakota \cite{askins2019theia}. Assessment of the subdominant mantle fluxes expected at the Baksan, Jinping, and Homestake sites entails subtracting the model-dependent estimates of the larger crust fluxes from the measured total fluxes.

Resolving geoneutrino fluxes with different values of Th/U to study any variation across the planet requires very large detector exposures at distinct locations. Measurements with a precision of about $5$\% are desirable to constrain model predictions. Comparing measurements made near thick continental crust with those made over thin oceanic crust in the deep ocean would be a very favorable scenario \cite{enomoto07}. A high statistics assessment of the mantle fluxes at an oceanic site would be relieved of model dependencies associated with existing assessments at continental sites. While there are suggestions for deploying antineutrino observatories in the deep ocean, underwater locations do not offer compelling advantages to foreseeable physics and astrophysics experiments with sensitivity to geoneutrino fluxes.

Geoneutrino research anticipates benefits from advances in detector technology. The capability of resolving the geoneutrino fluxes through their directions is emerging from the joint physics and astrophysics quest to measure CNO solar neutrinos \cite{bonventre18}. Adding direction information reduces sources of isotropic background, gaining sensitivity to signal from the source. Several efforts are underway for selectively collecting the directional Cherenkov light in scintillating liquid \cite{Kaptanoglu:2018sus,wan17}. The potential for improving the identification of inverse beta decay and reducing background is apparent with a novel detector using a dense array of optical fibers immersed in opaque scintillating liquid \cite{Cabrera:2019kxi}.

There are excellent prospects for continued development of geoneutrino research, leading to greater understanding of the magnitude and distribution of Earth's radioactivity. New detectors in Canada and China are expected to soon begin augmenting the ongoing flux measurements from Japan and Italy. Proposed multi-kton detectors in central China and the Caucasus mountains hopefully move forward. They would contribute substantially to demonstrating surface variation of the geoneutrino fluxes from Th and U. Advanced detection techniques are poised to enable measurement of geoneutrino source directions, leading to the rich reward of resolving geoneutrino fluxes from K and the mantle.\label{sec:source_geo}

\subsection{Cosmological Neutrinos }
\label{wg4_cosmo}

\subsubsection{Introduction}

At a red-shift of roughly 10 billion, nearly half of the total energy density of the Universe was in the form of neutrino kinetic energy, according to the Standard Cosmology Model.  After decoupling from the thermal bath of the hot Big Bang at under 1 second, Big Bang neutrinos, also known as the Cosmic Neutrino Background (C$\nu$B) or relic neutrinos, have continued to influence the Hubble expansion and rates of large-scale structure formation during over 13.7 billion years to the present day.  Due to the finite mass splittings measured through flavor oscillations, an important transition from relativistic to non-relativistic energies is believed to have already transpired for at least two massive states of the neutrinos.  As a result, the C$\nu$B is the largest known source of non-relativistic neutrinos.

Despite the unequivocal importance of neutrinos in shaping the expansion of the Universe, there is no present-day evidence that the C$\nu$B neutrinos continue to pervade all of space with a predicted average number density of 336 particles per cubic centimeter, assuming three flavors of light neutrinos and making no assumptions on whether the neutrino is distinct from the antineutrino.  Indirect hints of its existence are discussed in Sec.\ \ref{sec:CNB2}. 
The experimental effort on detecting the C$\nu$B is being advanced on two fronts.  One is through a new generation of precision cosmology measurements to detect the sub-percent fraction of the critical energy density from massive neutrinos, and the second is through the development of direct detection methods based on neutrino capture on $\beta$-decaying nuclei.

\subsubsection{Direct Detection Experiments}

Several methods to detect relic neutrinos have been proposed. Most of them have estimated sensitivities many orders of magnitude too low to detect relic neutrinos. However, one of them -- neutrino capture on $\beta$-decaying nuclei, looks conceivable with major improvements in the detection techniques. The idea was suggested by Weinberg~\cite{Weinberg:1962zza} for massless neutrinos. Cocco  et al.~\cite{Cocco:2007za} have noticed that for massive neutrinos the energy of the electrons from the neutrino capture process exceeds the maximum energy in the $\beta$-decay by about two neutrino masses. Therefore, the separation of the neutrino capture process from the overwhelming background from usual $\beta$-decays would be feasible with an extremely good energy resolution of about 50 meV or even better. Effects of zero-point motion of tritium atoms absorbed to graphene or other materials were recently discussed in Refs.\ \cite{Cheipesh:2021fmg,Nussinov:2021zrj}, they may challenge the observation of light relic neutrinos. 

The capture rate of Majorana neutrinos is twice as large as for Dirac neutrinos \cite{Long:2014zva}, but this effect is degenerate with potential relic neutrino clustering, which however would enhance the rate \cite{Ringwald:2004np}. To illustrate the challenging nature of an observation, note that in the currently world-leading neutrino mass experiment KATRIN the rate of capture of relic neutrinos on tritium is of order $10^{-6}$ yr$^{-1}$. Nevertheless, the PTOLEMY collaboration~\cite{betti2019neutrino} performs active R\&D~\cite{baracchini2018ptolemy} in order to demonstrate the feasibility of such a goal. The possibility to detect the neutrino capture on $\beta$-decaying nuclei using correlations between the neutrino direction and the spin of the $\beta$-decaying nucleus was discussed recently~\cite{lisanti2014measuring,Akhmedov:2019oxm}.
Results of the relic neutrino searches could also be sensitive to physics beyond the SM, like sterile neutrinos or neutrino decays \cite{Long:2014zva}.
\label{sec:source_rel}

\subsection{Neutrino Sources: Summary}

There are many natural sources of neutrinos: the early Universe, Earth, the Sun, the atmosphere, supernovae, or other astrophysical sources  as violent and as far away as Active Galactic Nuclei. % WW: Neutrinos from GRBs have not been observed (replaced by AGN) 
Weak interactions imply that neutrinos can travel long distances and large densities, giving access to  environments which cannot be tested otherwise. If other probes are accessible, neutrinos provide valuable complementary information.  One particular field where this clearly shows is high-energy  astrophysics where a 
combination of various cosmic ray messengers including neutrinos and recently gravitational waves as well allows to identify the sources where particles are accelerated to energies exceeding any terrestrial source. 

Using those sources we have also learned how our star and others produce energy. 
%, or started learning (together with other astrophysical messengers like photons or gravitational waves) what the origin of cosmic rays is.
Still many  open issues remain, like the role of neutrinos in cosmological structure formation or in supernova explosions, the amount of metalicity of the Sun, or the distribution of radioactive heat production within the Earth. 
Such studies are accompanied by human-made sources like nuclear reactors and  accelerators.  Neutrinos from such artificial sources have been, are, and will be used to unveil fundamental properties of the neutrino, but also for  more mundane applications like understanding nuclear fission, including  safe-guarding. 

%Together with other astrophysical messengers like photons, cosmic-rays, and/or gravitational waves, our understanding of the Universe would be greatly enhanced. Neutrinos from artificial sources can be used to unveil fundamental properties of the neutrino themselves such as leptonic CP violation, absolute neutrino mass, and Majorana vs Dirac nature of neutrinos, etc. 
While over more than 70 years great progress has been made in neutrino physics, from the discovery of the neutrinos from a reactor over understanding the basic of lepton mixing to the discovery of astrophysical neutrino sources, we have not still fully understood physics behind all these sources. Many  fundamental properties of neutrinos remain unclear, too. Some guaranteed sources of neutrinos such as relic neutrinos from the early Universe or past supernovae have never been detected mainly due to technical challenges. Recent R\&D developments make us confident that those will be overcome, and the resulting discovery of additional neutrino sources will further complement our understanding of the Universe on various scales, and in the future may be used to learn further about fundamental physics. 

%d/or bigger-scale improved detectors currently under construction or planned would lead to the discovery of these neutrinos as well as more precise studies on already known physics, which might lead to the discovery of new physics, within about two decades. 

%\subsection{WG4 Summary and Conclusion}
%\input{WG4/WG4_summary.tex}

%\begin{thebibliography}{99}
%\bibitem{}

%\end{thebibliography}

%\end{document}
 
%\input{WG4/WG4_outline.tex}

%\end{document}

\clearpage

\section{Neutrino Oscillations}
{Contributing additional authors: Silvia Pascoli (Durham U.), Raymond R.\ Volkas (ARC, CoEPP), Roger A.\ Wendell (Kyoto U.\ and Kavli IPMU)}
%KL: 19Jun2020-11:15
\label{WG1}
\subsection{Introduction}
\label{sec:osc}
If neutrinos have mass it is conceivable that their flavor changes periodically with distance over energy while they propagate.  The huge mass differences of charged fermions and quarks (which in addition almost immediately hadronize) makes neutrinos the only elementary fermions where such oscillations can be observed. 
%The Standard Model neutrino has weak quantum numbers but no electric
%charge. 
%It is therefore possible that the weak flavour of a neutrino
%changes, or oscillates, as it propagates through space and time.
Such neutrino-flavor oscillations are triggered by non-zero masses as well as by a non-trivial Pontecorvo-Maki-Nakagawa-Sakata (PMNS) lepton mixing matrix~\cite{Pontecorvo:1957qd,Maki:1962mu,Pontecorvo:1967fh}, which is the analogue of the Cabibbo-Kobayashi-Maskawa (CKM) 
matrix of the quark 
sector~\cite{Cabibbo:1963yz,Kobayashi:1973fv,10.1093/ptep/ptaa104}. 
Neutrino oscillations were discovered by the Super-Kamiokande collaboration in
1998~\cite{Fukuda:1998mi} and by the SNO collaboration in
2001~\cite{Ahmad:2001an}.
Thus, the existence of neutrino oscillations reveals that neutrinos have
mass and that lepton flavors mix. 

Oscillations among the three Standard Model neutrino flavors are 
readily described in terms of the mixing of three mass eigenstates, 
$\nu_i$, $i=1,2,3$.
The probability, $P(\nu_\alpha \rightarrow \nu_\beta)$, that a neutrino
created in an eigenstate of flavor $\alpha$ and which travels through
a vacuum is detected in flavor state $\beta$ is given by
\cite{Olive:2016xmw}:
\begin{equation}
  P(\nu_\alpha \rightarrow \nu_\beta) =
  \sum_{i,j} U_{\alpha i} U^*_{\beta i} U^*_{\alpha j} U_{\beta j}
  \exp \left[
       -i \frac{\Delta m_{ji}^2}{2} \frac{L}{E}
       \right]  .
  \label{Eq:VacOsc}
\end{equation}
Here $E$ is the neutrino energy, $L$ is the distance between source
and detector, and $\Delta m_{ji}^2 = m^2_j - m^2_i$.
The first observations of neutrino oscillations exploited muon neutrinos
produced in the cosmic-ray bombardment of the Earth's atmosphere and the 
electron neutrinos produced in nuclear processes in the Sun.
These observations, subsequently confirmed using neutrinos produced in nuclear
reactors and at accelerator facilities, established that the three-flavor
oscillations of Eq.\ (\ref{Eq:VacOsc}) can be described to a good approximation by two, decoupled oscillations.
The first, describing the oscillations of atmospheric muon neutrinos, is characterized by a large mass-squared difference and a mixing angle that is approximately $45^\circ$.
The second, describing the oscillations of solar electron neutrinos, is
characterized by a small mass-squared splitting and a large ($\sim 35^\circ$)
mixing angle.
These observations allow the unitary PMNS matrix, $U$, to be parameterized
in terms of three mixing angles, $\theta_{ij}$ and one phase parameter $\delta_{\rm CP}$:
\begin{equation}
  U =
    \left(
      \begin{array}{c c c}
        1 &  0     & 0      \\
        0 &  c_{23} & s_{23} \\
        0 & -s_{23} & c_{23}
      \end{array}
    \right)
    \left(
      \begin{array}{c c c}
         c_{13}             &  0 & s_{13} e^{-i\delta_{\rm CP}} \\
         0                 &  1 & 0                  \\
        -s_{13} e^{i\delta_{\rm CP}} & 0 & c_{13}
      \end{array}
    \right)
    \left(
      \begin{array}{c c c}
        c_{12} & s_{12} & 0 \\
       -s_{12} & c_{12} & 0 \\
        0     & 0     & 1
      \end{array}
    \right).
  \label{Eq:SnuM}
\end{equation}
Here $c_{ij}=\cos\theta_{ij}$ and $s_{ij}=\sin\theta_{ij}$; 
$\theta_{23}$ is referred to as the ``atmospheric mixing angle'' as it
determines at leading order the oscillations of atmospheric muon neutrinos
while $\theta_{12}$ is referred to as the solar mixing angle as it is used, 
at leading order, to describe the oscillations of solar electron neutrinos.
The mixing angle $\theta_{13}$ is small, accounting for the approximate
decoupling of the atmospheric and solar oscillations. For Majorana neutrinos there are two additional phases (``Majorana phases''), which an be put in a diagonal phase matrix to the right\footnote{The original ``symmetrical''  parameterization gives each individual rotation a phase \cite{Schechter:1980gr} and provides slightly more insight when discussing lepton number violating processes \cite{Rodejohann:2011vc}. } of Eq.\ (\ref{Eq:SnuM}). See Sec.\ \ref{sec:0vbb} for a discussion.

The evaluation of the oscillation probabilities requires that the product
of $U$ with its Hermitian conjugate be evaluated.
Such a calculation yields terms in the expression for the oscillation formul\ae
that depend on $\sin \delta_{\rm CP}$ and for which the sign differs depending on whether the expression is for the oscillation of neutrinos or antineutrinos.
Therefore, if $\sin \delta_{\rm CP} \ne 0$, CP invariance is violated in
neutrino oscillations.
Two additional phases that might arise if neutrinos are Majorana particles 
can not be measured in neutrino-oscillation experiments and 
are omitted from Eq.~(\ref{Eq:SnuM}). We will discuss them in Sec.\ \ref{sec:0vbb}.  Since a hierarchy in the mass-squared splittings $\Delta m_{21}^2 \ll |\Delta m_{31}^2|$ is present,
which neutrino oscillation observations have revealed, the two-flavor case is in many cases a reasonable approximation and nicely illustrates the features of oscillations. That probability reads for flavor changes
\begin{equation}\label{eq:2osc}
P(\nu_\alpha \rightarrow \nu_{\beta\neq\alpha}) = \sin^2 2 \theta \sin^2 \frac{\Delta m^2 L}{4 E} \,,
\end{equation}
where $\theta$ and $\Delta m^2 = m_2^2 - m_1^2$ are the only mixing angle and mass-squared difference, respectively, in this case. 

Neutrinos that pass through matter may interact with its
constituents.
The probability for incoherent inelastic scattering is very small.
However, coherent scattering is dominated by events in which very
little energy is transferred between the incident neutrino and the
target particle.
As a result, the coherent-scattering amplitude is strongly peaked for
neutrinos that continue to propagate in the forward direction.
Since the scatter is coherent, i.e.\ the quantum numbers of the final
state are the same as those of the initial state, the scattered
neutrino wave can interfere with the unscattered wave.
The effect of this interference may be expressed in the form of an
effective matter potential that causes observable variations in
the rate of neutrino oscillations.
The oscillation probabilities for neutrinos passing through matter are
therefore modified from the vacuum probabilities given by
Eqs.\ (\ref{Eq:VacOsc}, \ref{eq:2osc}). In the two-flavor limit from Eq.\ (\ref{eq:2osc}) the probability takes the same form, but with parameters $\theta$ and $\Delta m^2 $ changed to $\theta_m$ and $\Delta m^2_m $, taking into account the matter effects. In particular, for constant matter density one finds 
\begin{equation}\label{eq:theta_matter}
\sin^2 2 \theta_m = \frac{\sin^2 2\theta}{(A/\Delta m^2 - \cos 2 \theta)^2 + \sin^2 2\theta }, 
\end{equation}
where $A = 2\sqrt{2} G_F N_e E$ with $N_e$ the electron number density. 
The size of the matter effect depends on the density and composition
of the medium and on the oscillation parameters.
In particular, the matter effect may be exploited to determine the octant of the mixing angles  and the sign of the mass-squared differences.
Moreover, the matter effect is different for neutrinos and
antineutrinos because for the latter $A$ changes sign. 
The difference between the oscillation probabilities of neutrinos and
antineutrinos that arises from the matter effect must be taken into
account in searches for CP-invariance violation.

Neutrino oscillations are studied using both terrestrial and
astrophysical sources of neutrinos.
The oscillation channels that can be studied using a particular source
are determined by the neutrino-energy spectrum.
Atmospheric neutrinos are produced in the decay of mesons created
when cosmic rays strike the upper atmosphere, see Sec.\ \ref{sec:source_atm}.
The atmospheric-neutrino flux contains $\parenbar{\nu}_\mu$ and
$\parenbar{\nu}_e$ in a ratio of approximately $2:1$, which decreases with energy as the muons do not decay. 
The effective baseline at which the effect of oscillation is observed
is a function of the zenith angle of the neutrino-production point.
The atmospheric neutrino energy spectrum falls with energy, providing
a measurable flux at energies in excess of 10\,GeV.
As a result, atmospheric neutrinos allow oscillation effects to be
studied over a wide range of energies and baselines.
The dominant oscillation channel, $\nu_{\mu} \rightarrow \nu_{\tau}$ -- indirectly tested by the disappearance of muon neutrinos -- was key to the discovery of neutrino oscillations. 
This channel constrains the parameters $\theta_{23}$ and $|\Delta m^2_{31}|$. 

Solar neutrinos are produced through fusion processes in the core
of the Sun, see Sec.~\ref{sec:source_sol}. 
A variety of fusion and decay processes result in an electron-neutrino
energy spectrum that drops sharply for energies above $\sim 20$\,MeV. 
The measurement of the solar neutrino flux and its energy dependence
has been critical for the establishment of the three-neutrino mixing
paradigm, where the adiabatic neutrino flavor conversions between the Sun's core and ``surface'' are governed by the so-called MSW-effect~\cite{Wolfenstein:1977ue,Mikheev:1986gs}. The parameters that are constrained are mainly $\theta_{12}$ and $\Delta m^2_{21}$, where for the latter the sign was established to be positive. 

The decay of unstable fission products produced in the core of a
nuclear reactor, see Sec.\ \ref{sec:source_rea}, produces an intense flux of reactor antineutrinos
with an energy spectrum that runs from a few keV to several MeV. 
A detector placed at an appropriate baseline (from $\sim 1$~km to $\sim 100$~km) from the core is able to make precise measurements of the oscillation parameters $\theta_{13}$, $\theta_{12}$ and $|\Delta m^2_{31}|$.
Reactor neutrino experiments provide the most precise measurement of
the smallest mixing angle, $\theta_{13}$.

Accelerator neutrino beams are produced from the decay of mesons
generated when high-energy proton beams strike nuclear targets, see Sec.~\ref{sec:source_acc}. 
Magnetic focusing at the target is used to select the sign of the
secondary meson beam and to direct it to a decay channel.
If negative mesons are selected a neutrino beam dominated by muon
neutrinos is produced; a beam dominated by muon antineutrinos is
produced if positive mesons are selected.
Such beams have been used to constrain a variety of the parameters
that determine the mixing matrix $U$. 

Three-flavor mixing allows neutrino oscillations to be described
using six parameters: three mixing angles ($\theta_{12}$,
$\theta_{13}$ and $\theta_{23}$), two independent mass splittings 
($\Delta m^2_{21}$ and $\Delta m^2_{32}$, or $\Delta m^2_{31}$), and
one CP phase ($\delta_{\rm CP}$).
The measurements made using astrophysical and terrestrial sources have
been combined in global fits (see Section~\ref{WG1:SubSect:Glob})  to
determine the values for all the mixing angles and $\Delta m^2_{21}$
as well as the magnitude of $\Delta m^2_{32}$.
The value of the CP-invariance violating phase, $\delta_{\rm CP}$, and
the sign of $\Delta m^2_{32}$ are not known, though first hints have 
emerged.

The formalism outlined above is able to describe the majority of data
on neutrino oscillations.
Looking beyond the determination of the parameters, it will be
important to establish whether the model is correct as a description
of nature.
To do this requires redundant and precise measurements of $\theta_{23}$,
the degree to which it differs from $\pi/4$, $\theta_{13}$, and $\theta_{12}$.
Ideally, the precision of these measurements will approach that with
which the CKM matrix elements are known.
Such measurements will be important to establish deviations from the 
three-neutrino-mixing paradigm, test the unitarity of the neutrino-mixing 
matrix and other new physics effects, and to seek relationships between
the parameters that govern neutrino oscillations and those that govern 
quark mixing.

The quantum mechanical treatment of neutrino oscillations induces correction terms to the standard probability formula (\ref{Eq:VacOsc}), see e.g.\  \cite{Giunti:1997wq}. This is associated to the separation of the wave packets associated to the mass eigenstates. In the correct treatment (reviews on the subtle quantum mechanical issues are \cite{Akhmedov:2009rb,Akhmedov:2010ms}), one takes into account that a produced flavor neutrino is described by  neutrinos having different masses, thus their corresponding wave packets move with different speed. If those two packets do not overlap anymore, the oscillation pattern is lost. The correction term is given by  exp$\{-(L \Delta m^2/(4\sqrt{2} \, E^2 \sigma))^2 \}$, where $\sigma$ is the width of the wave packets, defined by properties of the source and detector. The correction term suppresses the oscillatory pattern. For reactor neutrinos with $E \sim$ MeV, the relevant leading mass-squared difference is  $2 \cdot 10^{-3}$ eV$^2$. Assuming a wave packet uncertainty governed by the size of the decaying nucleus of 10 fm, implies that $\Delta m^2/(4\sqrt{2} \, E^2 \sigma) \simeq 30$ m, which is very close to the distance in actual reactor neutrino experiments. Therefore,  current experiments are  close to observing effects caused by decoherence, and therefore test fundamental quantum mechanics with neutrinos, or test our understanding of neutrino oscillations \cite{An:2016pvi,deGouvea:2020hfl}. 

\subsection{Atmospheric Neutrino Experiments}

\subsubsection{Introduction}
The wide variety of both energies and baselines available in the atmospheric neutrino data enables sensitivity to a variety of oscillation effects. This variability was key to the discovery of oscillations with this source, which was best described by two-neutrino mixing ($\nu_{\mu} \rightarrow \nu_{\tau}$)  at $L/E \approx 450\,  \mbox{km/GeV}$.
Indeed, atmospheric neutrinos provided the first indication of disappearance consistent with the $L/E$ dependence characteristic of neutrino oscillations~\cite{Fukuda:1998mi,Super-Kamiokande:2004orf}.
With the subsequent discovery of mixing between all active neutrino flavors it is now clear that atmospheric neutrinos are also sensitive to the neutrino mass ordering via matter effects at low energies and to a lesser extent to $\delta_{\rm CP}$.
Furthermore, the large range of $L/E$ available to atmospheric neutrinos has made them a useful probe of exotic scenarios (classified by a different oscillatory behavior, e.g.\ $L E^{n}$ dependence) and the ability to observe oscillation-induced $\nu_{\tau}$ in their flux gives unique access to deviations from unitarity in the mixing matrix. The status of each of these topics is reviewed in the following pages.

\subsubsection{Measurement of \texorpdfstring{$\theta_{23}$}{theta23} and \texorpdfstring{$\Delta m_{32}^{2}$}{Dm32}}
\begin{figure}
\centering
\includegraphics[width = 0.8\textwidth]{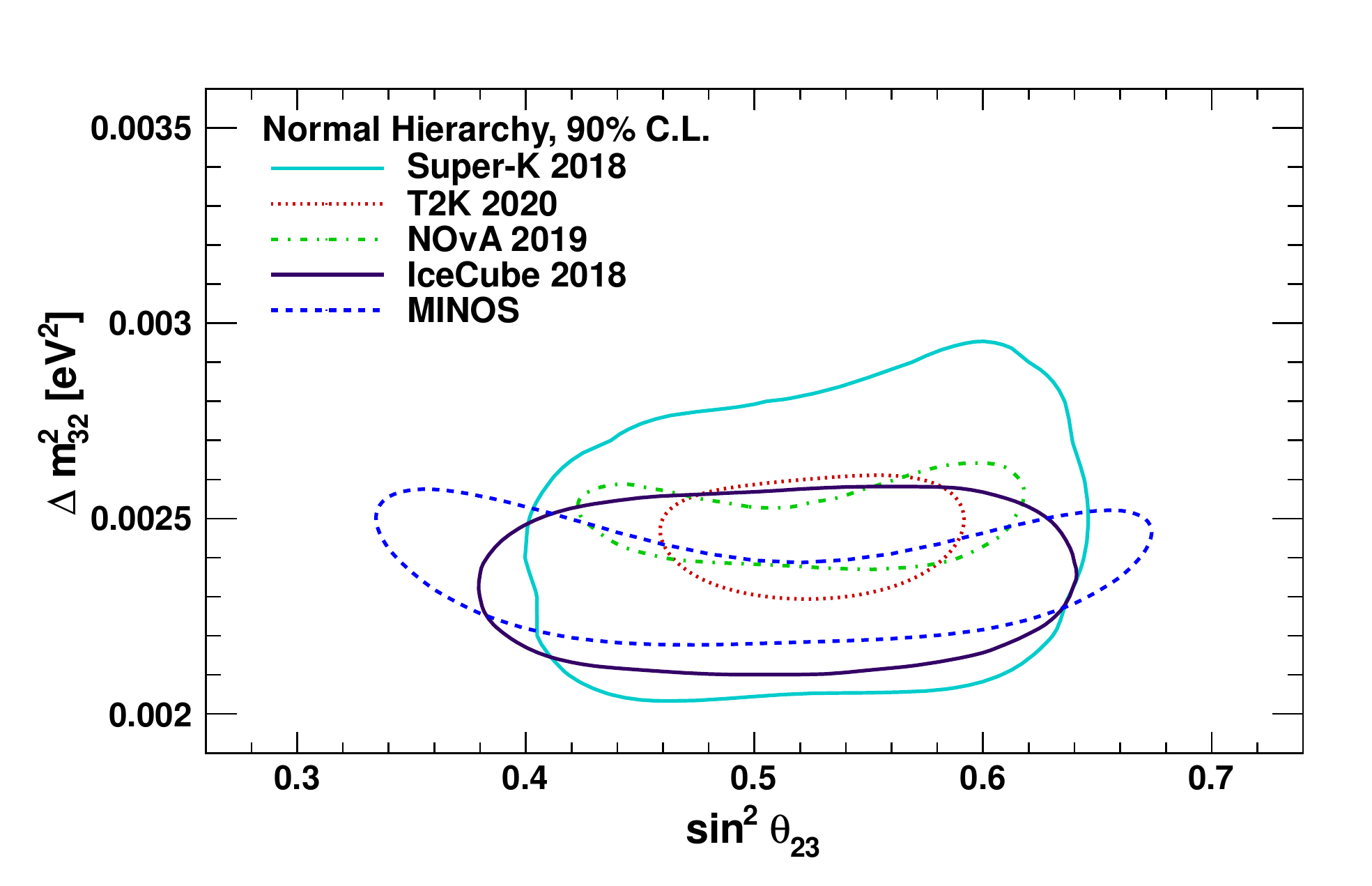}
    \caption{Neutrino mixing parameter measurements from both atmospheric (Super-Kamiokande~\cite{Abe:2017aap} and IceCube~\cite{Aartsen:2017nmd}) and accelerator (T2K~\cite{Abe:2019vii}, NOvA~\cite{Acero:2019ksn}, and MINOS~\cite{Adamson:2014vgd}) neutrino experiments.}
    \label{fig:atm_mixing}
\end{figure}

Atmospheric neutrino oscillations are dominated by transitions driven by a phase and amplitude governed by $\Delta m_{32}^{2}$ and $\theta_{23}$, respectively. The signal manifests as the disappearance of upward-going muon-like interactions and the subsequent appearance of tau-like interactions when the latter can be reconstructed.
Figure~\ref{fig:atm_mixing} shows current constraints on these mixing parameters for both atmospheric and long-baseline accelerator neutrino measurements.
Since atmospheric measurements have no a priori knowledge of the incoming neutrino direction, one must infer it using an event's interaction products.
The parameters extracted from atmospheric data are consistent with accelerator neutrino measurements which nowadays provide better  precision.
A summary of oscillation parameter measurements from atmospheric neutrinos is shown in Tab.~\ref{tbl:atm_mixing}.

\begin{table}[t]
\begin{center}
\begin{tabular}{l|c|c}
\hline\hline
Experiment & $\mbox{sin}^2\theta_{23}$  & $|\Delta m^{2}_{32}| \, [ 10^{-3}
\mbox{eV}^{2}]$ \\
\hline
Antares \cite{Albert:2018mnz} & $0.50^{+0.2}_{-0.19}$ & $2.0^{+0.4}_{-0.3}$ \\
IceCube \cite{Aartsen:2017nmd}   & $0.51^{+0.07}_{-0.09}$ & $2.31^{+0.11}_{-0.13} $ \\
Super-Kamiokande (Super-K) \cite{Abe:2017aap}  & $0.588^{+0.031}_{-0.064}$ & $2.50^{+0.13}_{-0.20}$ \\
\hline\hline
\end{tabular}
\end{center}
\caption{Summary of atmospheric neutrino mixing measurements from atmospheric neutrino experiments. }
\label{tbl:atm_mixing}
\end{table}

Two important points should be noted. First, atmospheric neutrino oscillations have been observed at the neutrino telescopes Antares and IceCube. Not only does this provide additional constraints on mixing with a higher energy threshold than Super-K and beam experiments, but it also serves as a proof-of-concept for oscillation studies proposed at upgrades of these facilities.  Second, due to enhanced oscillation effects for neutrinos traversing the Earth, atmospheric neutrino measurements bring additional sensitivity to the $\theta_{23}$ octant.  At present all measurements are consistent with maximal mixing, though Super-K has a weak ($\approx 1\sigma$) preference for the second octant.

\subsubsection{Measurement of the Mass Ordering}
At energies between two and ten GeV, upward-going neutrinos that traverse the Earth's core and mantle experience enhanced oscillation effects due to their interaction with matter along their trajectory. However, the enhancement is present only for neutrinos if the mass ordering is normal and only for antineutrinos if the ordering is inverted. Assuming normal ordering, an increase in the $\nu_{\mu} \rightarrow \nu_{e}$ appearance probability and a suppression of the $\nu_{\mu} \rightarrow \nu_{\mu}$ survival probability are expected for these neutrinos.

Since there are both neutrinos and antineutrinos in the atmospheric flux, experiments are sensitive to the ordering via modulations in the rate of both upward-going electron-like and muon-like events.
For Super-K, although its energy threshold is sufficiently small ($\approx$ 100 MeV) and it has  a 22.5~kton fiducial volume, it suffers for statistics at ${\cal O}(1)$ GeV energies and above.
Indeed, the flux at these higher energies is nearly three orders of magnitude smaller than that at 600~MeV. IceCube, in contrast, has a considerably larger volume, but a higher threshold (5 GeV in DeepCore~\cite{Aartsen:2019eht}) that cuts into the region between two and 10 GeV, where mass ordering-sensitive matter effects are largest.
There are additional challenges in separating $\nu_{e}$ charged current interactions from $\nu_{x}$ neutral current interactions, meaning there are larger backgrounds in its appearance sample. Both experiments suffer from an inability to cleanly distinguish neutrinos from antineutrinos, though Super-K has demonstrated some ability to do so statistically~\cite{Abe:2017aap}.

In spite of these challenges both experiments have attempted measurements of the mass ordering.
Super-K data indicate a weak preference for the normal ordering, rejecting the inverted ordering by between 81.2\% and 96.7\%, depending upon
assumptions about the other oscillation parameters~\cite{Abe:2017aap}, with an expected sensitivity of between $1\sim 1.8\sigma$.
On the other hand, IceCube measurements with three years of DeepCore data showed a similarly mild preference (53.3\%) for the normal ordering.
Though its sensitivity is only $0.45-0.65\sigma$~\cite{Aartsen:2019eht}, this  represents an important proof-of-principle for IceCube's future physics searches.
Due to weak degeneracies between $\theta_{23}$ and the mass ordering, the sensitivity of these experiments is expected to improve with stronger constraints on the  mixing angle, in particular.
Measurements of atmospheric $\nu_{\mu}$ disappearance serve this purpose, though accelerator neutrino experiments place tighter bounds on the range of $\theta_{23}$.
Indeed, the Super-K sensitivity was shown to improve by $0.2\sigma$ when constrained with only a fraction of the currently-available T2K data~\cite{Abe:2017aap}.

\subsubsection{Projected Oscillation Measurements with Atmospheric Neutrinos}\label{sec:proj_atm}
Though each of the measurements presented above will continue at existing facilities, they will likely be superseded by next-generation experiments.
In the following only constraints from atmospheric neutrinos are presented. Several experiments are anticipating combined measurements with beam or reactor data to improve sensitivity overall, but they are not considered here.

In terms of atmospheric neutrino mixing, IceCube Upgrade is expected to achieve roughly 20\% precision on the value of $\Delta m^2_{31}$ \cite{Stuttard:2020zsj}, making it comparable to current long-baseline experiments. 
After three years of operation the KM3NeT/ORCA project can determine this parameter to better than 4\% and distinguish the $\theta_{23}$ octant at $2\sigma$ if $| \mbox{sin}^{2}\theta_{23} - 0.5 | > 0.06 $.
Hyper-Kamiokande (Hyper-K), the next-generation water Cherenkov experiment in Kamioka,  will be able to resolve the $\theta_{23}$ octant at the same level within 10 years for $| \mbox{sin}^{2}\theta_{23} - 0.5 | > 0.07 $ ~\cite{Abe:2018uyc}.
Both Super-K~\cite{Li:2017dbe} and IceCube~\cite{Aartsen:2019tjl} have observed oscillation-induced $\nu_{\tau}$ events and these measurements will be extended at their successor experiments.
In terms of the normalization of the $\nu_{\tau}$ cross section, Hyper-K expects better than 15\%~\cite{Abe:2018uyc} sensitivity and IceCube Upgrade better than 10\%~\cite{Stuttard:2020zsj} after one year of operation. 
KM3NeT/ORCA can place a $7\%$ constraint on the normalization with three years of data~\cite{KM3NeT:2021ozk}.

Atmospheric neutrino measurements will a have strong impact on our understanding of the neutrino mass ordering.
These measurements are expected to include both muon-like and electron-like interactions and are summarized in Tab.~\ref{tbl:atm_hier}. 
In the table $\sin^2 \theta_{23}$ is mostly assumed to be near 0.5 for comparison purposes, though the sensitivity of these experiments typically improves (degrades) for larger (smaller) values. 
Combining atmospheric measurements with other approaches to the mass ordering such as JUNO are expected to improve the situation further~\cite{Bezerra:2019dao,KM3NeT:2021rkn}.

\begin{table}[t]
\begin{center}
\begin{tabular}{l|c|c}
\hline\hline
Experiment & MO Sensitivity ($\sigma$)  & Years  \\
\hline
DUNE~\cite{Abi:2020evt}          &  4.0             & 10  \\
Hyper-Kamiokande~\cite{Abe:2018uyc} &  3.0             & 10  \\
ICAL-INO~\cite{ICAL:2015stm}         &  3.0             &  10  \\
KM3NeT/ORCA~\cite{KM3NeT:2021ozk}   &  4.4 NO (2.3 IO)            &  3  \\
IceCube Upgrade~\cite{Bezerra:2019dao}       &  3.8 NO (1.8 IO) &  6  \\
\hline\hline
\end{tabular}
\end{center}
\caption{Summary of sensitivity to the neutrino mass ordering (MO) with atmospheric neutrino experiments. Here values assuming $\sin^{2} \theta_{23}$ is close to $0.5$ are used for all experiments other than ORCA (ICAL-INO), which assumes 0.56 (marginalized over its $3\sigma$ range). The number of years of operation to achieve the listed sensitivity is presented in the column headed ``Years''.}
\label{tbl:atm_hier}
\end{table}

\subsubsection{Possibilities with Atmospheric Neutrinos}
Atmospheric neutrinos have been used successfully to test a variety of oscillation scenarios and have placed stringent limits on several
types of mixing beyond the standard PMNS framework.
Indeed, ANTARES~\cite{Albert:2018mnz}, IceCube~\cite{Aartsen:2020iky}, and Super-Kamiokande~\cite{Abe:2014gda}, have all searched for sterile neutrinos with this source and the latter two have placed tight constraints on Lorentz-violating oscillations (cf.\   \cite{Aartsen:2017ibm} and \cite{Abe:2014wla}, respectively).
These searches are expected to be continued and expanded upon with the next-generation of experiments.

Besides observations of oscillation effects from known neutrinos, the large span of energies and distances (thus a varying matter profile) of atmospheric neutrinos make them ideal probes of non-standard neutrino physics, such as sterile neutrinos or non-standard neutrino interactions, see Secs.\ \ref{sec:atm_sol} and \ref{sec:np}. Often their effects are  mostly independent from those expected from the mass ordering and CP-violation. These measurements (e.g.\ \cite{Aartsen:2017xtt,PhysRevD.84.113008,Salvado:2016uqu}) are therefore complimentary to the future long-baseline neutrino program where these effects can be largely degenerate \cite{Abi:2020kei,PhysRevD.60.119905}.

\subsection{Solar Neutrino Experiments}
\label{WG1:Sect:Solars}

 Solar neutrinos are powerful probes of both the Sun and the properties of the neutrino. 
 The energy spectra and fluxes of solar neutrinos, as predicted by the Standard Solar Model (SSM) as well as measured by neutrino experiments, are discussed in Sec.~\ref{subsec:WG4solar}. In this section we present the solar neutrino measurements from the point of view of oscillation physics. These measurements have the best sensitivity to constrain the so-called solar mixing angle $\theta_{12}$ and, to a lesser degree, the $\Delta m^2_{21}$ mass splitting, as it is shown in Sec.~\ref{WG1:SubSec:Solar:ThetaDm}.
 Assuming the validity of the SSM predictions for solar neutrino fluxes, the electron-neutrino survival probability ($P_{ee}$) of solar neutrinos can be measured for different solar neutrino species and energy ranges from below 1\,MeV up to about 15\,MeV. In the dense solar matter, solar neutrinos undergo the process of adiabatic flavor conversion described by the MSW mechanism~\cite{PhysRevD.17.2369, Mikheev:1986gs} predicting a strong energy dependence for $P_{ee}$ and a transition from the so-called vacuum to the matter-dominated region. Deviations from this model, especially in the transition region at around 3\,MeV, could indicate the presence of new physics beyond the Standard Model. The current status of measurements of  $P_{ee}$ for solar neutrinos as well as direct observation of the Earth's matter effects will be discussed in Sec.~\ref{WG1:SubSec:Solar:Matter}. The latter effect leads to the regeneration of the electron flavor during the passage of neutrinos through the Earth during the night. This effect has a particular sensitivity to $\Delta m^2_{21}$.

\subsubsection{Measurement of \texorpdfstring{$\theta_{12}$}{theta12} and \texorpdfstring{$\Delta m_{21}^{2}$}{Dm21} }
%  \texorpdfstring{
%  $\Delta m_{21}^{2}$}{Dm21}
%  }
  %$\mathbf{\Delta \it m^2_{21}}$ 

\label{WG1:SubSec:Solar:ThetaDm}

\begin{figure}[t]
     \subfigure[]{\includegraphics[width = 0.5\textwidth]{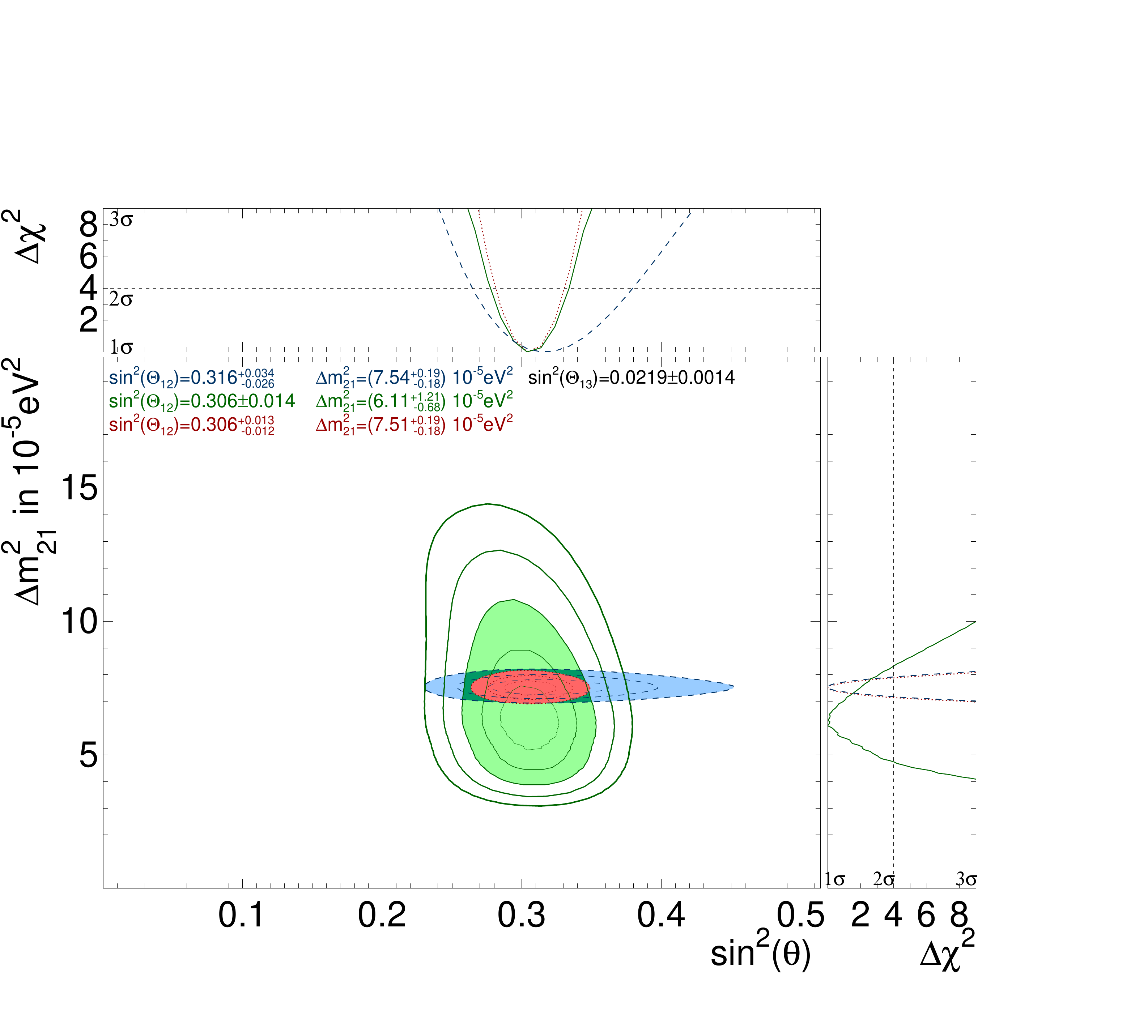}
     \label{fig:theta12-m12}
    }
    \subfigure[]{\includegraphics[width = 0.45\textwidth]{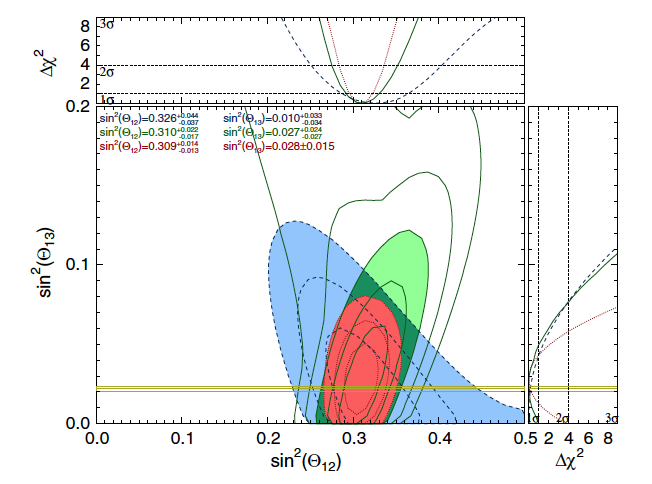}
    \label{fig:theta13-12}
    }
    \caption{Allowed regions for the neutrino oscillation parameters from all solar neutrino data (green), KamLAND (Kamioka, Japan) reactor antineutrino data (blue), and the combined result (red). The filled regions give the 3$\sigma$ C.L.\ results, the other contours shown are at the 1 and 2$\sigma$ C.L.\ (for the solar analyses also 4 and 5$\sigma$ C.L.). (a) Allowed regions for $\Delta m^2_{21}$ vs $\sin ^2 \theta_{12}$~\cite{SuperK:nu2020}. 
    (b) Allowed contours of $\sin^2 \theta_{13}$ vs $\sin^2 \theta_{12}$~\cite{Abe:2016nxk}. The yellow band is the $\sin^2 \theta_{13}$ measurement from reactor neutrino data~\cite{An:2013uza}.}
    \label{fig:solar_osci}
\end{figure}

The currently allowed regions for the oscillation parameters based on the solar-neutrino data compared to that based on the KamLAND reactor antineutrino data, as well as the respective combined result, are shown in Fig.\ \ref{fig:solar_osci}.
The high-precision measurements of $^8$B solar neutrinos by Super-K~\cite{Abe:2016nxk,SuperK:nu2020}  and SNO~\cite{SNO+B8}  dominate the combined fit to all solar neutrino data. 
In part (a), the figure shows the parameter space for $\Delta m^2_{21}$ versus $\sin^2 \theta_{12}$.
Some tension at the level of 2$\sigma$ between the solar neutrino and reactor antineutrino measurements of the solar mass splitting $\Delta m^2_{21}$ was previously reported~\cite{Abe:2016nxk}, stemming from the Super-Kamiokande measurement of the day/night asymmetry for $^8$B neutrinos (Sec.~\ref{WG1:SubSec:Solar:Matter}). This tension has recently strongly reduced (Fig.~\ref{fig:theta12-m12}) thanks to the updated Super-Kamiokande analysis, as it was reported at the Neutrino 2020 conference~\cite{SuperK:nu2020}. 
Solar neutrinos have only a very mild sensitivity to the $\theta_{13}$ mixing angle, as it is shown in Fig.~\ref{fig:theta13-12}.

\subsubsection{Matter Effects in Solar Neutrino Oscillations}
\label{WG1:SubSec:Solar:Matter}

\begin{figure}[t]
    \centering
     \subfigure[]{\includegraphics[width = 0.53\textwidth]{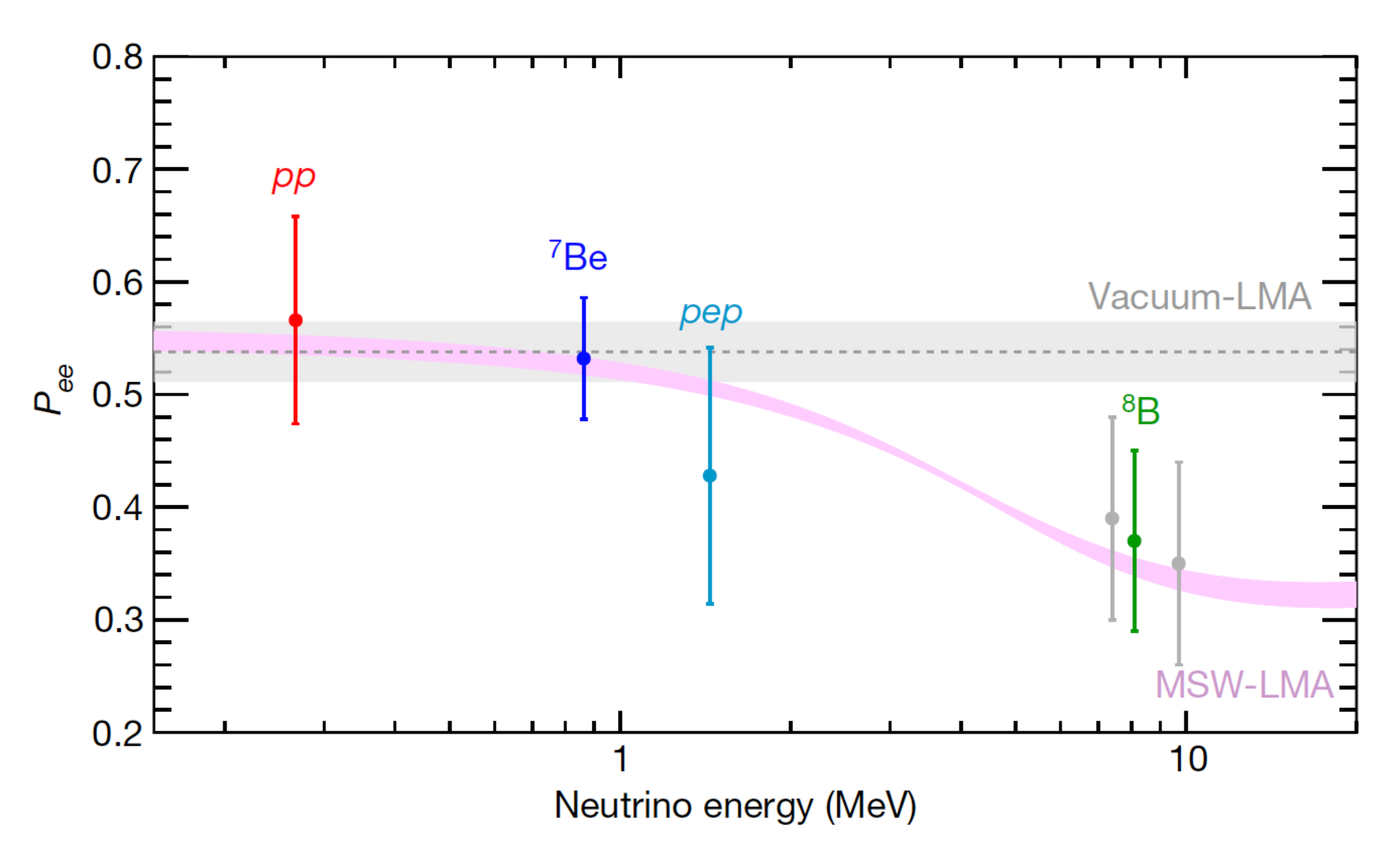}
     \label{fig:Pee_BX}
}
    \subfigure[]{\includegraphics[width = 0.44\textwidth]{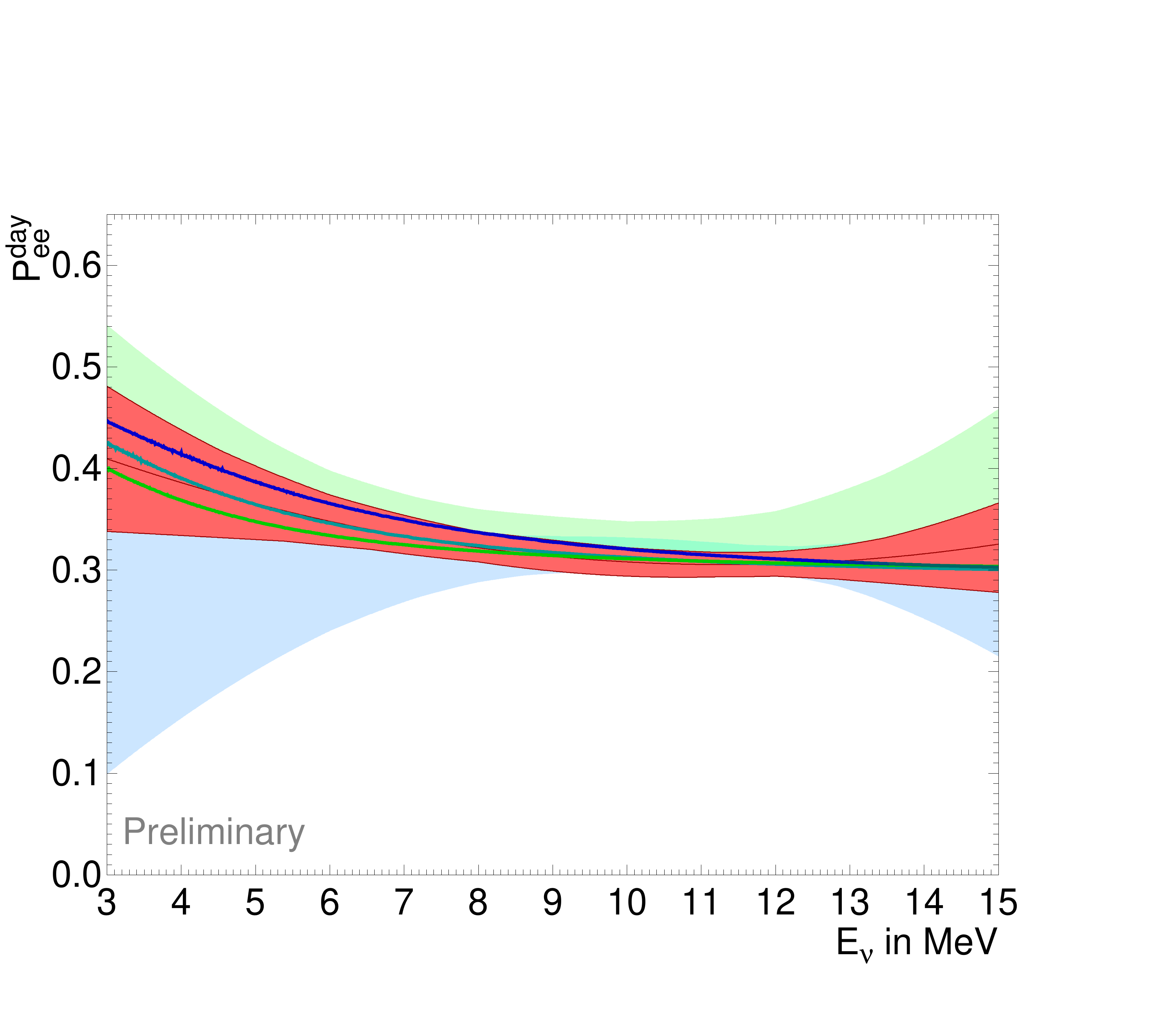}
    \label{fig:Pee_SuperK}
    }
    \caption{Electron neutrino survival probability $P_{ee}$ as a function of their energy. (a) Data points represent the Borexino results, in which the error bars include experimental and theoretical uncertainties. The pink band is the $\pm 1\sigma$ prediction of the MSW mechanism in matter, while the grey band is the case of vacuum-only oscillation. LMA label in the figure stands for the Large Mixing Angle, the current best-fit solution for solar oscillation parameters. From~\cite{Agostini:2018uly}. (b) Allowed survival probability 1$\sigma$ band from the combined $^8$B solar neutrino data of Super-Kamiokande and SNO (red). The pastel colored bands are the separate Super-Kamiokande (green) and SNO (blue) fits. The solid lines are the MSW predictions using particular oscillation parameters resulting from the fit: to all solar data (green), all solar $+$ KamLAND data (blue), to  Super-Kamiokande and SNO data (robin egg blue). Taken from~\cite{SuperK:nu2020}.}
\end{figure}

The measured interaction rates of $pp$, $^7$Be, $pep$, and $^8$B solar
neutrinos (see Sec.\ \ref{sec:source_sol}) can be used to infer the electron neutrino survival
probability at different energies. 
Assuming the high-metallicity\footnote{We recall that metallicity means the relative abundance of heavy elements to that of hydrogen in the Sun.} SSM fluxes (see~\cite{Vinyoles_2017} and Tab.\ \ref{tab:SolarFluxes}), Borexino obtained the electron-neutrino survival
probabilities for each solar-neutrino component, as it is shown in
Fig.~\ref{fig:Pee_BX}~\cite{Agostini:2018uly}. 
Borexino provides the most precise measurement of  $P_{ee}$ in the
low energy region, below 1.5 MeV, where flavor conversion is
vacuum-dominated.
At higher energies above 5 MeV, where flavor conversion is dominated by the matter effects in the Sun, the Borexino results are in agreement with the high-precision measurements performed by Super-Kamiokande~\cite{Abe:2016nxk, SuperK:nu2020} and SNO~\cite{Aharmim:2011vm}.
Borexino is the only experiment that can simultaneously test neutrino
flavor conversion both in the vacuum and in the matter-dominated regime and disfavors the vacuum only oscillation hypothesis at 98.2\% C.L. Figure~\ref{fig:Pee_SuperK} shows the recently updated~\cite{SuperK:nu2020} survival probability above 3 MeV for $^8$B solar neutrinos as measured by Super-Kamiokande and SNO.

During the night, while the Sun is below the horizon, solar neutrinos are crossing the Earth on their passage towards the detector. Thus the matter density of the Earth affects solar neutrino oscillations through the MSW mechanism and leads to an enhancement of the $\nu_e$ flavor content during the nighttime. As a consequence, the rate of events measured via the neutrino elastic scattering off an electron, predominantly sensitive to electron flavor, increases at night. This is often called a “day/night effect” resulting in an asymmetry between the experimental rates observed during the day and at night. This effect is energy dependent and according to the current knowledge of the oscillation parameters, expected to be of some importance only for the high energy part of $^8$B solar neutrinos. Defining $\Phi_D$ $(\Phi_N)$ as the day (night) flux with zenith angle $\cos\theta_z <0$ $(>0)$, the asymmetry is defined as $(\Phi_D - \Phi_N)/\frac 12 (\Phi_D + \Phi_N)$. An extended maximum likelihood fit to the amplitude of the solar zenith angle variation of the neutrino-electron elastic scattering rate in Super-Kamiokande results in a day/night asymmetry of ($-3.3 \pm 1.0$ (stat) $\pm$ 0.5 (syst))\%~\cite{Abe:2016nxk}, where the SK-IV phase contributes ($-3.6 \pm 1.6$ (stat) $\pm$ 0.6 (syst))\%~\cite{Abe:2016nxk}. 
At Neutrino 2020, Super-Kamiokande reported an updated value for SK-IV of ($-2.1 \pm 1.1$ (stat))\%~\cite{SuperK:nu2020}, with the systematic error and combination with other SK phases still ongoing.
Borexino excluded the day-night asymmetry for 0.867\,MeV $^7$Be solar
neutrinos~\cite{Bellini:2011yj} ($A_{}$ = 0.001 $\pm$ 0.012 (stat) $\pm$ 0.007 (syst)), in agreement with the prediction of the MSW solution for neutrino oscillations.

\subsubsection{Projected Oscillation Measurements with
Solar Neutrinos}
\label{WG1:SubSec:Solar:Projected}

Solar neutrinos have a well established position among the scientific
goals of the running, future, as well as next generation experiments.
Borexino has recently reported the first observation of neutrinos from the CNO fusion cycle~\cite{BXCNO}.
This experimentally confirms the existence of this process in
nature, which is extremely important for our understanding of
stellar physics.
Super-Kamiokande is working on the final analysis of all SK-IV phase data, that could improve the precision concerning the observed day-night
asymmetry and the low-energy part of the $^8$B spectrum.
SNO+, a successor of the SNO experiment in Sudbury, Canada,  has the
main goal to measure neutrinoless double beta ($0\nu\beta\beta$)
decay using the liquid scintillator loaded with $^{130}$Te.
Thanks to its depth (nearly 6000 m.w.e.) and a relatively
big volume (780 ton, 3 times more than Borexino), it has a large
potential in solar neutrino physics~\cite{Andringa:2015tza}.
The experiment is currently filling the detector with liquid
scintillator, after a phase with pure water during during which 
$^8$B neutrinos above 5 MeV have been observed~\cite{SNO+B8}.
Before loading the scintillator with $^{130}$Te, a period of several
months of data taking has the potential to provide precise low-energy 
solar neutrino spectroscopy, the extent of which will be
dictated by the final levels of the radiopurity reached.
During the several years long $^{130}$Te phase, only the $^8$B
neutrinos will be measured, possibly down to about 2.5 MeV. Coming back to low-energy solar neutrino physics after the $0\nu\beta\beta$ phase, is among the open possibilities.

JUNO is the next generation liquid scintillator detector under construction in Jiangmen, China that has its main goal determination of the neutrino mass ordering with reactor antineutrinos (Sec.~\ref{WG1:SubSect:Reactor:projection}).
In spite of its relatively shallow depth of about 700 m, thanks to a huge mass of 20 kton and extremely high energy resolution of 3\% at 1 MeV, it has also a potential in solar neutrino physics~\cite{An:2015jdp}. In particular, measurement of $^8$B neutrinos down to a unprecedented 2\,MeV energy threshold might be feasible~\cite{JUNO_B8}. Combined with high achievable statistics, JUNO will be able to perform precision tests of the transition region of $P_{ee}$, a measurement of the day-night asymmetry, and consequently, determination of the $\theta_{12}$ and $\Delta m^2_{21}$ parameters. JUNO will be the only experiment able to determine these parameters both with solar as well as reactor neutrinos (Sec.~\ref{WG1:Sect:Reactor}). The future Jinping neutrino experiment~\cite{Beacom_2017} plans to deploy a 2 kton target for precision solar neutrino physics in the world's deepest underground laboratory located in China. The project aims to use new detection techniques, for example the slow liquid scintillator. This could enable a separation of the Cherenkov and scintillation light, what would significantly help in background suppression utilising directionality of the Cherenkov light, while keeping the high light yield of about 500 photoelectrons per MeV. Further tests of the $P_{ee}$ transition region with high statistics would
however require a larger target mass, possibly using multi-modular neutrino detectors. The proposed THEIA~\cite{askins2019theia} detector collaboration also plans to exploit Cherenkov light to
observe particle direction while having  excellent energy
resolution and low threshold with a next generation,
few-tens-of-kton-scale detector using water-based liquid
scintillator.
Additional improvement might be possible by loading the scintillator
with, for example $^7$Li, that would enable the measurement of solar neutrinos
in addition to elastic scattering using charged current interactions.
Another possible technique for precision solar spectroscopy could
be based on two-phase liquid argon time projection
chambers~\cite{Franco_2016}.
This technique is under development for direct Dark Matter WIMP
searches within the DarkSide-20k~\cite{DS20k} collaboration at LNGS in Italy.
Argo, its long-time scale successor planned to be located at
SNOLAB, is conceived to accumulate an exposure of 1000 ton$\cdot$yr, free of
backgrounds other than that induced by coherent scattering of
neutrinos.
Thus, Argo would also enable precision measurements of solar neutrino
fluxes, representing the ``neutrino floor'' for the Dark Matter
searches. DARWIN as a two-phase xenon detector will also observe neutrinos from the Sun \cite{Aalbers:2016jon}. Precise observations of $pp$ and $^7$Be neutrinos will be possible, allowing the measurement of the $\nu_e$ survival oscillation probability at low energies and contributing to distinguishing the low and high metallicity scenarios \cite{Aalbers:2020gsn}.

\subsection{Reactor Neutrino Experiments}
\label{WG1:Sect:Reactor}

The flux and spectrum of reactor antineutrinos have been extensively studied, and are described in Sec.~\ref{sec:source_rea}. The inverse $\beta$-decay (IBD) reaction, $\overline{\nu}_e + p \rightarrow n + e^+$, which has the largest cross section in the few-MeV range and incomparable power to reject backgrounds with coincidence of prompt-delayed signals, is the classical channel to detect reactor antineutrinos with liquid scintillator (LS). Roughly, the event rate without oscillation is $\sim1$ (ton$\cdot$GW$_{\rm th}\cdot$day)$^{-1}$ at 1\,km distance from the reactor, where ton is the unit of the target mass of the liquid scintillator and GW$_{\rm th}$  is the unit of the thermal power of the reactors. The reactor antineutrino survival probability in vacuum can be written as \cite{Petcov:2001sy}
\begin{eqnarray}
\label{eq:P_nue2nue}
		P_{\bar{\nu}_e\to\bar{\nu}_e}=1 -\sin^22\theta_{13} \left(\cos^2\theta_{12}\sin^2\Delta_{31}+\sin^2\theta_{12}\sin^2\Delta_{32}\right)
		 -\cos^4\theta_{13}\sin^22\theta_{12}\sin^2\Delta_{21},
\end{eqnarray}
where $\Delta_{ij}=\Delta m_{ij}^2L/(4E)=(m_i^2-m_j^2)L/(4E)$, in which $L$ is the baseline and $E$ is the antineutrino energy. Note that from this expression one can define a mass splitting  $|\Delta m^2_{ee}|$, which can be safely approximated as $\cos^2\theta_{12} \Delta m^2_{31} + \sin^2\theta_{12} \Delta m^2_{32}$ at baselines of $\mathcal{O}$(1 km). 

Due to the low energy, the oscillation effect can only be observed via the disappearance of electron antineutrinos. CP-invariance violation, which is only present in the appearance channel, cannot be measured directly with reactor neutrinos. For the same reason, it does not rely on unknown parameters thus has advantages in precision measurements of relevant oscillation parameters. Therefore, reactor experiments are complementary to accelerator, atmospheric and solar neutrino oscillation experiments. The combination of these measurements can significantly improve our knowledge of physics of neutrino oscillation.

\subsubsection{Measurement of \texorpdfstring{$\theta_{12}$}{theta12} and \texorpdfstring{$\Delta m^2_{21}$}{Dm21}}
\label{WG1:SubSect:Reactor:t12}

KamLAND observed neutrino oscillation with reactors for the first time in 2002~\cite{Eguchi:2002dm}. It detects antineutrinos from more than 50 reactors at an average baseline of $\sim 180$~km with a 1-kton LS detector. The measurement allowed the determination of the Large Mixing Angle (LMA) MSW solution of the solar neutrino problem. In particular, the solar mass-splitting $\Delta m^2_{21}$ was determined to high precision. The latest results from three-flavor neutrino oscillation analyses with constraints from solar and short-baseline reactor neutrino experiments are~\cite{Gando:2013nba}
$$ \tan^2\theta_{12}=0.436^{+0.029}_{-0.025}\,, \quad \Delta m^2_{21}=7.53 \pm 0.18 \times 10^{-5} \ {\rm eV}^2\,.$$
Reactor neutrino flux and spectrum models have been found to deviate in recent measurements, i.e.\ the rate deficit and the shape anomalies discussed in Sec.\ \ref{sec:source_rea}. However, the impact to the above measurements is found to be very small. 
The value of $\theta_{12}$ is consistent with the solar neutrino results, while $\Delta m^2_{21}$ differs from the Super-K measurement $\Delta m^2_{21}=4.8^{+1.5}_{-0.8}\times 10^{-5} \ {\rm eV}^2$~\cite{Abe:2016nxk} and the SNO measurement $\Delta m^2_{21}=5.6^{+1.9}_{-1.4}\times 10^{-5}\ {\rm eV}^2$~\cite{Aharmim:2011vm} by about 2$\sigma$. The tension on $\Delta m^2_{21}$ between solar and reactor measurements is an interesting topic for future oscillation experiments and could in fact be explained by new physics. Recent Super-Kamiokande data seems however to weaken this tension considerably \cite{SuperK:nu2020}.

\subsubsection{Measurement of \texorpdfstring{$\theta_{13}$}{theta13} and \texorpdfstring{$\Delta m^2_{ee}$}{Dmee}}
\label{WG1:SubSect:Reactor:t13}

The negative results of the CHOOZ~\cite{Apollonio:2002gd} and Palo Verde~\cite{Boehm:2000vp} experiments, at a baseline of $\sim1$~km from reactors, demonstrated that atmospheric neutrino oscillations do not involve electron neutrinos and set an upper limit of $\sin^22\theta_{13}<0.12$ at 90\% C.L. Proposed in the early 2000s, Daya Bay~\cite{An:2012eh}, Double Chooz~\cite{Abe:2012tg}, and RENO~\cite{Ahn:2012nd} determined that $\theta_{13}$ is non-zero in 2012. All three experiments detect reactor antineutrinos with LS detectors of fiducial masses of tens of tons by near-far relative measurements, with the far detector(s) at a baseline of $\sim1$~km. 
The measurements of $\theta_{13}$ and $|\Delta m^2_{ee}|$ (see comment after Eq.\ (\ref{eq:P_nue2nue}))
by Daya Bay, Double Chooz, and RENO are listed in Tab.\ \ref{Tab:WG1:reactor:current}. 
The fit results on $\Delta m^2_{32}$ are also reported by Daya Bay and RENO for normal and inverted mass ordering.

Thanks to the near-far relative measurement, the deviation of reactor neutrino flux and spectrum model has negligible impact on the $\theta_{13}$ and $|\Delta m^2_{ee}|$ measurements.

\begin{table}[t]
\begin{center}
\begin{tabular}{c|c|c|c|c}
\hline\hline
 & \multicolumn{2}{c|}{$\sin^22\theta_{13}$} & \multicolumn{2}{c|}{$|\Delta m^2_{ee}|$ [$\times10^{-3} {\rm eV}^2$]} \\ \hline
	& Current & Ultimate & Current  & Ultimate  \\ \hline
Daya Bay & $0.0856\pm0.0029$ & $\sim2.7\%$ & $2.522^{+0.068}_{-0.070}$ & $\sim2.1\%$ \\\hline
Double Chooz & $0.102\pm0.012$ & $\sim 10\%$ &  NA & NA \\\hline
RENO & $0.0892\pm0.0063$ & $\sim 6.9\%$ & $2.74\pm0.12$ & $\sim4.5\%$ \\\hline\hline
\end{tabular}
\caption{Measurements and projected ultimate precision on $\sin^22\theta_{13}$ and $|\Delta m^2_{ee}|$. Data are taken from Ref.~\cite{Adey:2018zwh,DoubleChooz:nu2020,RENO:nu2020}. The ultimate precision is estimated with current systematics and increased statistics to the end of the operation.
\label{Tab:WG1:reactor:current}}
\end{center}
\end{table}

\subsubsection{Projected Oscillation Measurements with Reactors}
\label{WG1:SubSect:Reactor:projection}

The JUNO experiment~\cite{JUNO:2021vlw} is located in Jiangmen in southern China, at equal distance of $\sim 53$~km to the Yangjiang power plant (six 2.9 GW$_{\rm th}$ cores) and Taishan power plant (two 4.6 GW$_{\rm th}$ cores and another two to be built). The detector is located at 700~m underground and consists of 20~kton liquid scintillator viewed by 17,612 20-inch photomultiplier tubes (PMTs) and 25,600 3-inch PMTs. The energy resolution is designed to be $<3$\% at 1 MeV, driven by its main physics goal to determine the neutrino mass ordering~\cite{Zhan:2008id,Zhan:2009rs}.

\begin{figure}[t]
\begin{center}
  \includegraphics[width=0.65\textwidth]{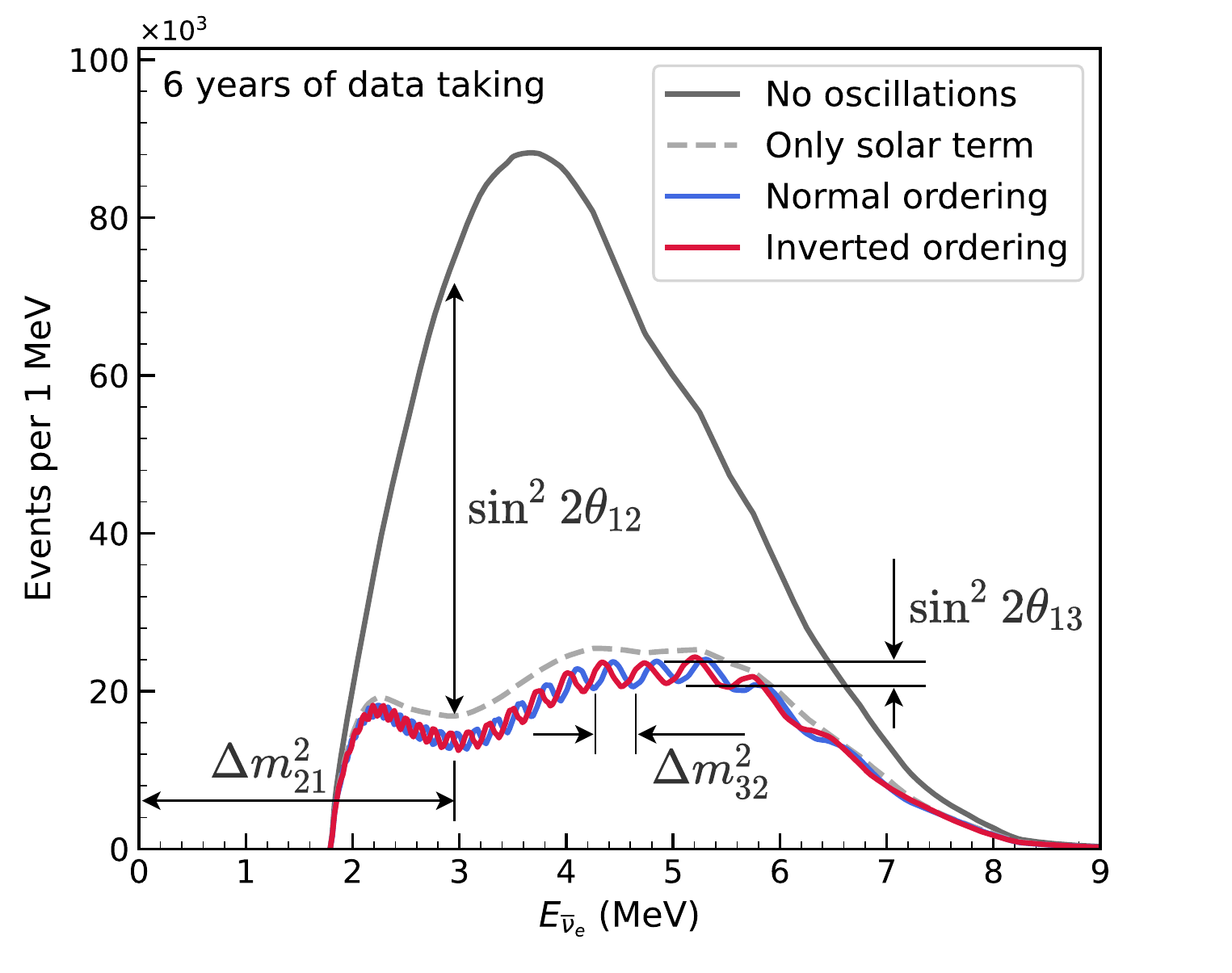}
  \caption{Expected energy spectrum of reactor antineutrinos in JUNO. The black curve shows the un-oscillated spectrum. The dashed curve shows the spectrum assuming $\theta_{13}=0$. The blue and red curves correspond to expected spectra for normal and inverted mass ordering, respectively. The features in the spectrum reflecting the sensitivity in oscillation parameters are demonstrated.    Taken from Ref.\ \cite{JUNO:2021vlw}. \label{fig:juno_oscillation}}
\end{center}
\end{figure}

The neutrino mass ordering can be revealed using the oscillation interplay between $\Delta m^2_{31}$ and $\Delta m^2_{32}$~\cite{Petcov:2001sy}. As shown in Fig.~\ref{fig:juno_oscillation}, see also 
Eq.\ (\ref{eq:P_nue2nue}), the difference in the multiple oscillation cycles in the oscillated spectra can be used to determine the neutrino mass ordering, with a sensitivity of $3-4 \sigma$ in 6 years by JUNO~\cite{An:2015jdp}. Note that the mass ordering is measured with vacuum oscillation with reactor neutrinos, while accelerator and atmospheric experiments measure it by matter effects. When combining the measurements from the reactor, accelerator, and atmospheric experiments, the interplay in the vacuum oscillation, the matter effects, and the difference in the $\Delta m^2$ measurements will provide a robust determination and a significantly boosted sensitivity~\cite{Li:2013zyd,IceCube-Gen2:2019fet,KM3NeT:2021rkn}. JUNO will also measure 3 out of 6 neutrino mixing parameters to a precision of better than 1\% and a 4th to 10\%, as demonstrated in Fig.~\ref{fig:juno_oscillation}, while its science endeavor will extend beyond particle physics, covering astrophysics, Earth science, and cosmology. JUNO started the civil construction in 2015 and expects to start data-taking in 2023.

\begin{figure}[htb]
\begin{center}
\includegraphics[width=0.65\textwidth]{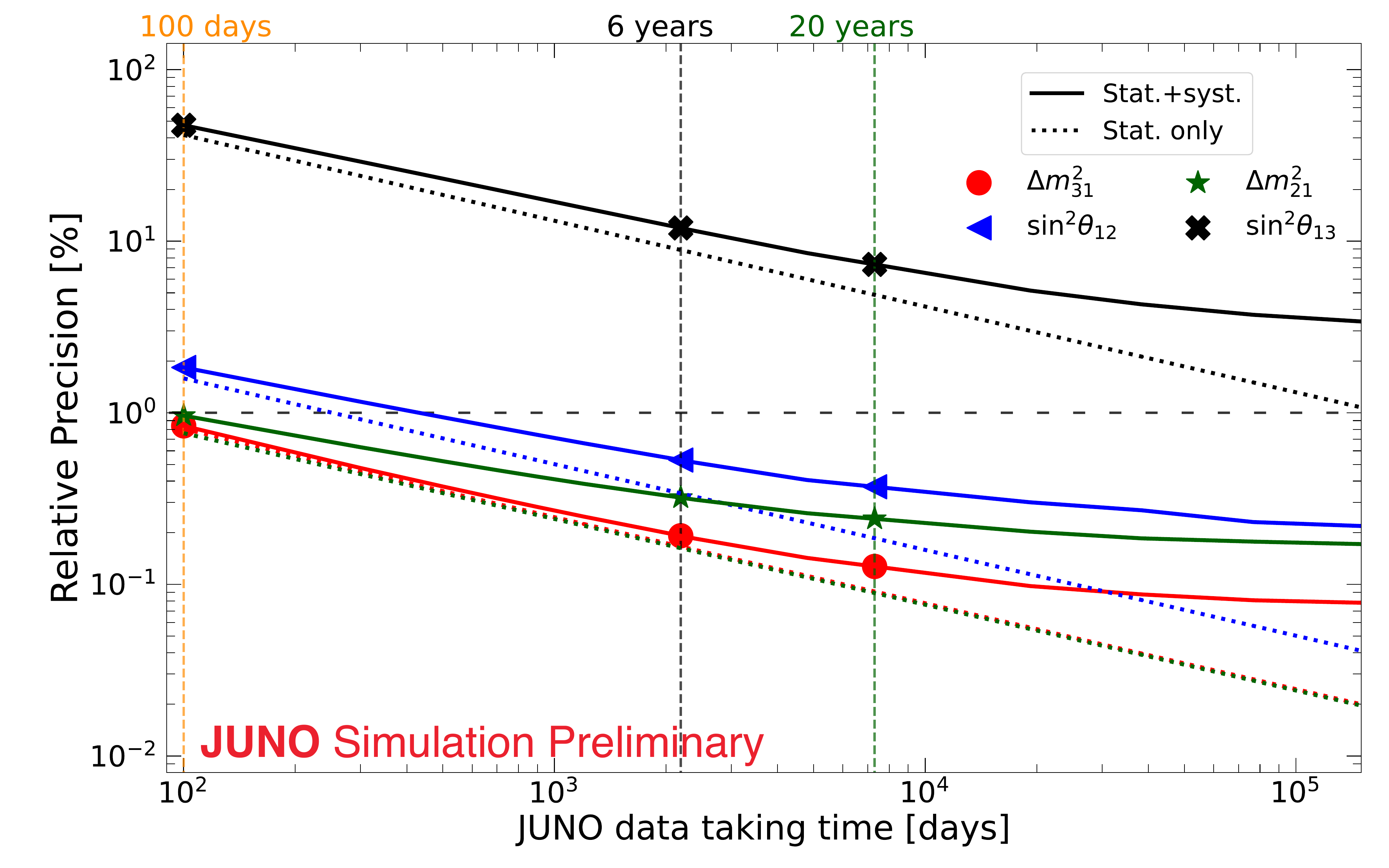}
    \caption{Sensitivity of oscillation parameters by JUNO. The vertical orange, black, and blue dashed lines correspond to 100 days, 6 years, and 20 years of JUNO data taking, respectively. The dotted curves show the statistics-only sensitivities and the solid lines show the sensitivities with projected JUNO systematics and backgrounds. Taken from~\cite{junoprojection}. \label{WG1:fig:par3reactor} }
\end{center}
\end{figure}

JUNO is anticipated to operate for more than 20 years. The relative precision of $\sin^2\theta_{12}$, $|\Delta m^2_{32}|$ and $|\Delta m^2_{21}|$ is shown in Fig.~\ref{WG1:fig:par3reactor}, where the vertical orange, black, and blue dashed lines correspond to 100 days, 6 years, and 20 years of JUNO data taking, respectively. The dotted curves show the statistics-only sensitivities and the solid lines show the sensitivities with projected JUNO systematics and backgrounds. At 53~km baseline, JUNO also has certain sensitivity to $\sin^2\theta_{13}$, as shown in Fig.~\ref{WG1:fig:par3reactor}.
The projected relative precision of oscillation parameters is also listed in Tab.\  \ref{Tab:WG1:reactor:JUNO}, comparing with the current knowledge~\cite{Zyla:2020zbs}. Inputs from JUNO-TAO~\cite{junotaocdr} have been considered in these evaluations to avoid the model dependence due to the reactor antineutrino flux and spectrum anomalies and lack of knowledge on the fine structure in the spectrum (see Sec.~\ref{sec:source_rea}).

\begin{table}[t]
\begin{center}
\begin{tabular}{c|c|c|c|c|c}
\hline\hline
 & Mass Ordering & $|\Delta m^2_{32}|$ & $\Delta m^2_{21}$ & $\sin^2\theta_{12}$ &$\sin^2\theta_{13}$ \\ \hline
6 years of data & $3-4\sigma$	& $\sim 0.2$\% & $\sim 0.3$\% & $\sim 0.5$\% & $\sim 12$\% \\ \hline
PDG2020 &  & 1.4\% & 2.4\% & 4.2\% & 3.2\% \\ \hline\hline
\end{tabular}
\caption{Projected relative precision of oscillation parameter measurements by JUNO~\cite{junoprojection}.
\label{Tab:WG1:reactor:JUNO}}
\end{center}
\end{table}

Strong motivation might emerge to further improve the precision of oscillation parameters, e.g.\ if hints of broken unitarity are found by future experiments. Reactor neutrino experiments could continue playing important roles. In a JUNO-like detector, statistics dominates the precision of $|\Delta m^2_{32}|$ since the sensitivity comes from the multi-cycle oscillation pattern in the observed spectrum. The precision can be improved to below 0.1\% with a moderate increase of exposure.  Measurement of the solar oscillation parameters $\sin^2\theta_{12}$ and $|\Delta m^2_{21}|$ relies on precise understanding of the reactor neutrino flux and spectrum shape, in addition to the statistics.  A better model prediction of the spectrum including the spent fuel and non-equilibrium contributions, or implementing dedicated near detector(s), could significantly improve the precision. Precision of these 3 parameters can be improved by a factor of $2-3$ to below 0.1\% with current technology. Daya Bay has a target mass of 80~ton at the oscillation maximum. At larger exposure, shape distortion will dominate the $\sin^2\theta_{13}$ sensitivity. The precision could be improved by a factor of 3 or more with a kton-scale detector. The precision foreseen by JUNO will allow unitarity tests on the first row of the PMNS matrix on the percent-level \cite{Ellis:2020hus}. Since the upper limit on the half-life of neutrinoless double beta decay for the inverted mass ordering depends strongly on $\theta_{12}$ \cite{Dueck:2011hu}, the experiment will have ramifications for this process as well, see Sec.\ \ref{sec:0vbb}.

\graphicspath{ {WG1/WG1_figs/} }

\subsection{Accelerator Neutrino Experiments}

Neutrino oscillations cause the flavor-composition of a neutrino beam
to evolve as it travels from source to detector.
Muon neutrinos dominate the flux of neutrino beams produced from meson
decay at proton-accelerator facilities, see Sec.\ \ref{sec:source_acc}.
The evolution of the flavor composition of the flux may therefore be
described in terms of the ``survival'' probability, 
$P_{{\nu}_\mu \rightarrow {\nu}_\mu}$, that a
muon-neutrino produced at the source is detected as a muon-neutrino,
and the ``appearance'' probability,
$P_{{\nu}_\mu \rightarrow {\nu}_X}$ that a neutrino
undergoes the transition
${\nu}_\mu \rightarrow {\nu}_X$.
The ``atmospheric parameters'', $\theta_{23}$ and $\Delta m^2_{32}$,
may be extracted from measurements of the disappearance channel.
The search for CP-invariance violation and the determination of the
mixing angle $\theta_{13}$ requires the measurement of the
${\nu}_e$ appearance channel.

Important contributions to the measurement of the atmospheric
parameters have been made by the MINOS experiment.
The MINOS experiment was a magnetized iron-scintillator tracking
calorimeter placed on the axis of the neutrino beam produced by the
120 GeV Main Injector at Fermilab.
The ``NuMI'' beam line is able to produce neutrino beams over a wide
range of energies.
Most of the MINOS data were taken in a low-energy configuration which
delivered a relatively broad neutrino spectrum peaked at 3 GeV.
Today, the accelerator-based experiments most sensitive to the
parameters $\theta_{23}$ and $\Delta m^2_{32}$ are NOvA and T2K.
The T2K experiment exploits the 30 GeV proton beam from the J-PARC
Main Ring to create a $\parenbar{\nu}_\mu$ beam that illuminates the
SK water-Cherenkov detector.
The distance from the source of the J-PARC neutrino beam to SK is
295 km.
The beam is directed such that the SK detector samples the flux at an
angle of 2.5$^\circ$ from the beam axis.
This arrangement was chosen to position the peak of the
neutrino-energy spectrum at 0.6 GeV, which corresponds to the position
of the first oscillation maximum at the 295 km baseline.
The NOvA experiment operates at a distance of 810 km from the
source of the NuMI beam.
The NuMI beamline is configured such that the NOvA detector
samples the beam at an angle of $0.84^\circ$ from the beam axis.
This produces a peak in the neutrino-energy spectrum at approximately
2 GeV, which corresponds to the first oscillation maximum at
810 km.
Each long-baseline (LBL) neutrino-oscillation experiment exploits a
detector placed close enough to the source to measure the neutrino
flux before the flavor composition of the beam has been affected by
oscillations. 
The near detector is required to measure the neutrino energy spectrum
and to constrain the flavor composition of the beam since the
(anti)muon-neutrino beam produced from meson decay contains a small
contamination of other neutrino flavors, mostly (anti)electron
neutrinos.

\begin{figure}[t!]
  \begin{center}
    \includegraphics[width=0.65\textwidth]{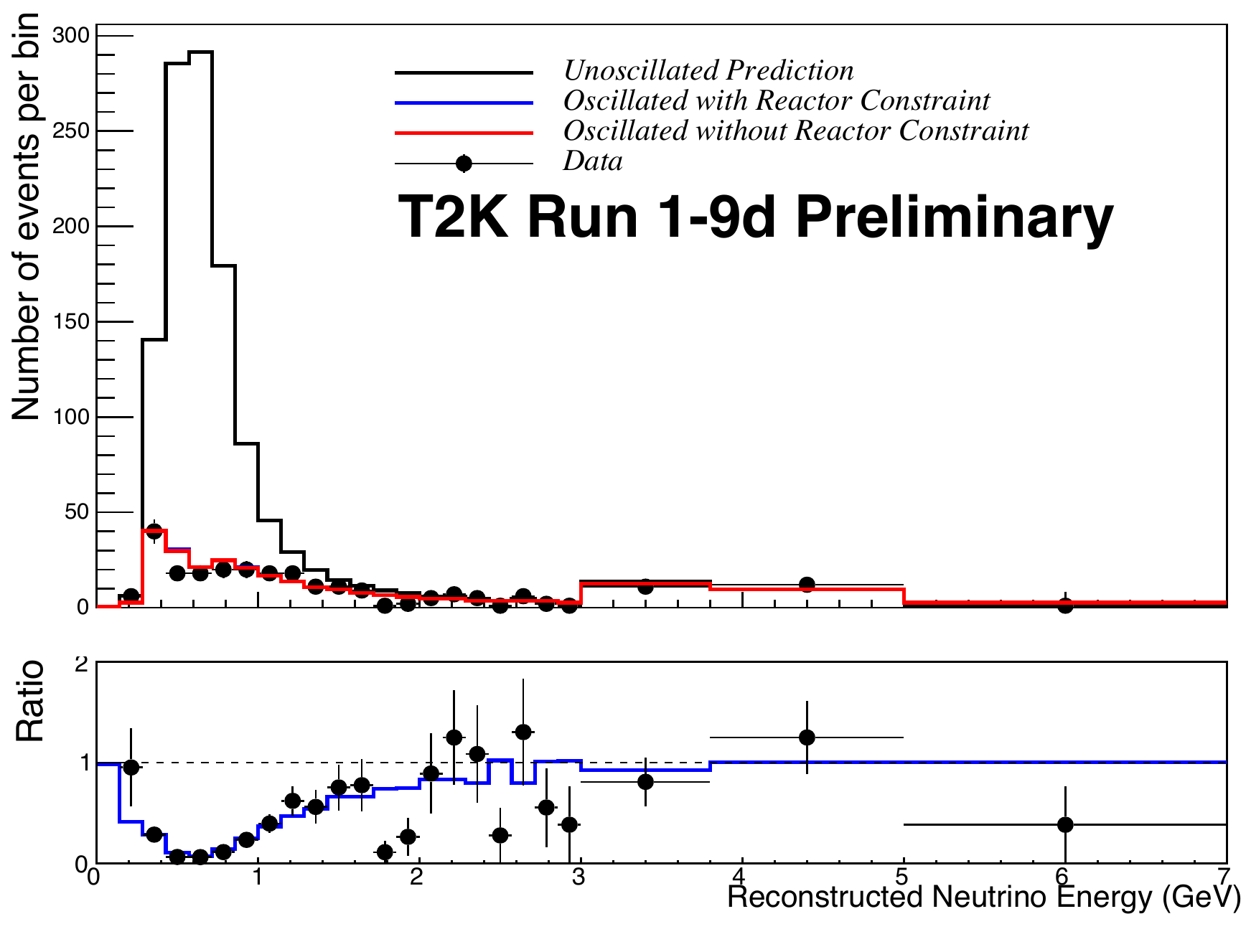}
  \end{center}
  \caption{
    %Left panel: 
    Neutrino energy spectrum measured using the Super-K
    detector (top).
    The data (solid points) are compared to the expected distribution
    in the absence of oscillations (black histogram) and the 
    distribution that results when the best-fit values for
    $\theta_{23}$ and $\Delta m^2_{32}$ are used.
    The ratio of the measured spectrum to the unoscillated spectrum is
    shown in the lower panel.
    The blue line shows the expectations of the best-fit spectrum. Taken from Ref.\ \cite{Zyla:2020zbs}.
      }
  \label{WG1:Sect:Accel:AtmPrm:Fig}
\end{figure}

\subsubsection{Measurement of \texorpdfstring{${\theta_{23}}$}{theta23} and
  \texorpdfstring{$\Delta m^2_{32}$}{Dm32}
  }
\label{WG1:Sect:Accel:AtmPrm}

The parameters $\theta_{23}$ and $\Delta m^2_{32}$ may be extracted
from the shape of the neutrino energy spectrum measured at the far
detector in LBL experiments for which the baseline exceeds $\sim 250$\,km.
An example of such a spectrum, taken from the T2K experiment, is shown
in Fig.\ \ref{WG1:Sect:Accel:AtmPrm:Fig}.
The position of the oscillation minimum on the ``Reconstructed
neutrino energy'' axis is sensitive to $|\Delta m^2_{32}|$, while the
depth of the minimum is sensitive to $\theta_{23}$.
The parameters $\theta_{23}$ and $\Delta m^2_{32}$ extracted from
fits to the data obtained by the T2K and NOvA collaborations are
compared with those obtained by Super-K and IceCube in
Fig.\ \ref{fig:atm_mixing}.
The various determinations are broadly consistent.

\subsubsection{Measurement of Mass Ordering and \texorpdfstring{$\mathbf{\delta_{\rm CP}}$}{deltaCP}}

\begin{figure}[h!]
  \begin{center}
    \includegraphics[width=0.485\textwidth]{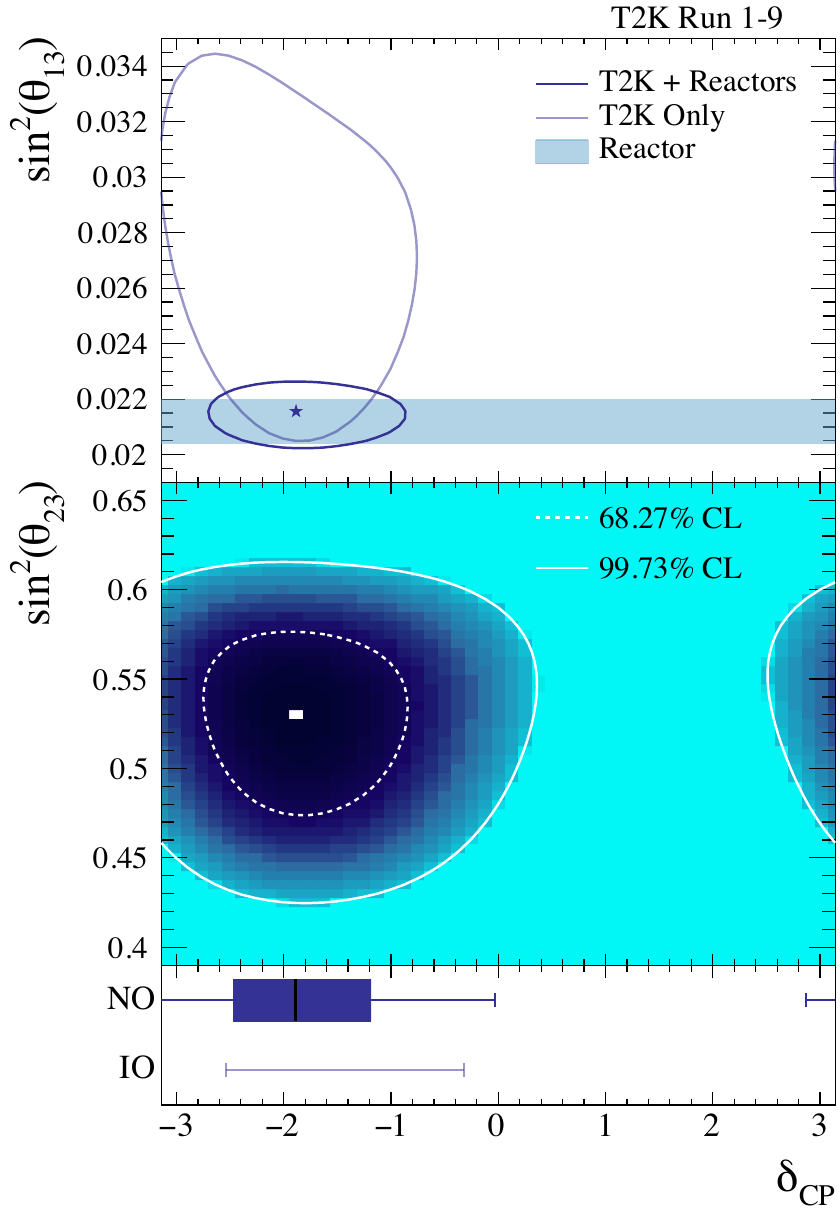}
    \quad \quad
    \includegraphics[width=0.448\textwidth]{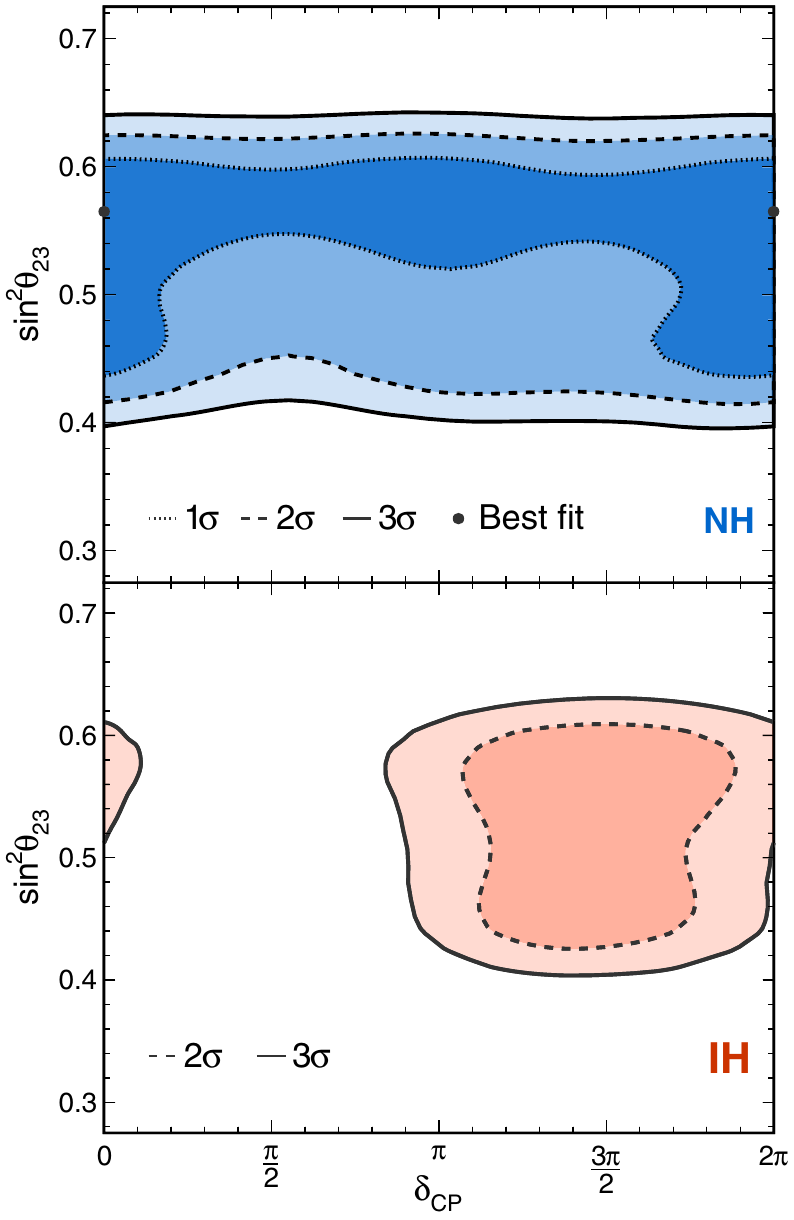}
  \end{center}
  \caption{
    Left panel: The top plot shows 68.27\% confidence-level allowed
    regions in the $\sin^2\theta_{13} - \delta_{\rm CP}$ plane
    assuming normal ordering.
    The result using T2K data only is shown using the solid light blue
    line.
    When the T2K data are combined with the reactor-neutrino
    constraint on $\theta_{13}$ (shown as the light-blue shaded band)
    the allowed region is delineated by the solid dark blue line and the
    best-fit point is shown as the dark-blue star.
    The middle plot shows the 68.27\% (dashed white line) and 99.73\%
    (solid white line) confidence-level allowed regions in the
    $\sin^2\theta_{23} - \delta_{\rm CP}$ plane assuming normal
    ordering for T2K data combined with the reactor-neutrino
    constraint on $\theta_{13}$.
    The bottom plot shows 68.27\% (dark-blue shaded region) and
    99.73\% (horizontal error bar) $\delta_{\rm CP}$ confidence
    intervals using T2K data combined with the reactor $\theta_{13}$
    constraint for both normal and inverted ordering. Taken from Ref.\ \cite{Abe:2019vii} \\
    Right panel: The top plot shows the 1$\sigma$, 2$\sigma$, and
    3$\sigma$ allowed regions in the
    $\sin^2\theta_{23} - \delta_{\rm CP}$ plane assuming normal
    ordering using NOvA data.
    The best-fit point is shown by the black markers lying on the
    $\delta_{\rm CP}=0, 2\pi$ axes.
    The bottom plot shows the 2$\sigma$ and 3$\sigma$ allowed regions
    in the $\sin^2\theta_{23} - \delta_{\rm CP}$ plane assuming
    inverted ordering. Taken from Ref.\ \cite{Acero:2019ksn}.
  }
  \label{WG1:Sect:Accel:MHtheta13del:Fig}
\end{figure}

The determination of the parameters $\theta_{13}$ and
$\delta_{\rm CP}$ requires the measurement of electron-neutrino
appearance in a muon-neutrino beam.
The CP-invariance violation arising from $\delta_{\rm CP}$ must be
distinguished from that which arises due to the matter effect.
This can be accomplished by exploiting the differences in the
modulations of the four oscillation probabilities:
$P_{{\nu}_\mu \rightarrow {\nu}_e}$, 
$P_{\bar{\nu}_\mu \rightarrow \bar{\nu}_e}$, 
$P_{{\nu}_e \rightarrow {\nu}_\mu}$ and
$P_{\bar{\nu}_e \rightarrow \bar{\nu}_\mu}$.
Both the NOvA and T2K collaborations exploit the
particle-identification capabilities of their detectors to partition
their data into four samples enriched in events corresponding to the
four disappearance and appearance channels.
The oscillation parameters $\theta_{13}$ and $\delta_{\rm CP}$ are
determined in a likelihood fit that takes into account the constraints
imposed by near-detector measurements, the far-detector simulation,
and matter effects. 
Two fits are performed; one assuming normal mass ordering; the second
assuming inverted ordering.
The values of the oscillation parameters extracted in this way are 
summarized in Fig.\ \ref{WG1:Sect:Accel:MHtheta13del:Fig}.
Both the T2K and the NOvA collaborations present their results as
allowed regions in the $\sin^2\theta_{23} - \delta_{\rm CP}$ plane.
The T2K collaboration presents its results for $\delta_{\rm CP}$ in the
range $-\pi \le \delta_{\rm CP} < \pi$ while the NOvA
collaboration presents its data over the range
$0 \le \delta_{\rm CP} < 2 \pi$.
When the inverted ordering is assumed, neither collaboration finds
allowed solutions inside the 1$\sigma$ confidence level; at the
2$\sigma$ and 3$\sigma$ confidence levels, the regions allowed by the
two collaborations overlap\footnote{Also new data by NOvA with a 50\% increase in neutrino data does not show any significant difference to T2K \cite{NOvA:2021nfi}.}.
When normal ordering is assumed, both collaborations find solutions
within the 1$\sigma$ confidence level.
The T2K data show a clear preference for
$\delta_{\rm CP} \approx -\frac{\pi}{2}$ (maximal CP-invariance
violation); CP-conserving values of $\delta_{\rm CP} = 0, \pi$ being
ruled out at the 95\% confidence level~\cite{Abe:2019vii}.

\subsubsection{Projected Oscillation Measurements with Accelerators}

Data taking at T2K will benefit from the incremental upgrades of the
J-PARC Main Ring.
The upgrade of the main magnet power supply will reduce the cycle time
by almost a factor of two by 2022.
Second harmonic RF will also be installed so that the beam power on
target will increase above 700 kW in 2022.
Incremental upgrades to the RF system will bring the power on target
to 1~MW by 2025 and 1.3~MW by 2028.
Improvements to the beam will be complemented by upgrades to the near
detectors designed to reduce the systematic uncertainties in the
oscillation analysis below 4\%. 
Gadolinium salts will be added to the water in the Super-K detector to
enhance its neutron-detection capability. 
%and improve the background
%rejection in the electron-neutrino samples.
A second phase of the experiment (T2K-II) will exploit the upgraded
beam and detectors until 2027 to search for CP-invariance violation with a sensitivity of $3\sigma$ in case of maximal CP-invariance violation.
%by accumulating $\sim3\times10^{22}$ protons on target (POT) by around 2028.

NOvA will continue to take data until 2026.
Improvements to the NuMI beam will allow the accumulation of equal 
proton-on-target exposures using neutrino and antineutrino beams.
The results of test beam measurements will be combined with improved
analysis techniques at the near and far detectors to enhance the
oscillation analysis.
In the absence of CP-invariance violation, NOvA data alone will
provide sensitivity to the mass ordering at the $2-3\sigma$
confidence level by 2025.
If the present preference for maximal CP-invariance violation  with
$\delta_{\rm CP} \simeq -\frac{\pi}{2}$ is confirmed, NOvA data will
allow the mass ordering to be determined at the $4-5\sigma$
level. The sensitivity to the mass ordering is below the $1\sigma$ level
if $\delta_{\rm CP} \sim \frac{\pi}{2}$.

To take the programme beyond the reach of the T2K and NOvA
experiments requires large, high-precision data sets.
Two experiments have been initiated to deliver such measurements: the
Deep Underground Neutrino Experiment (DUNE) served by the Fermilab Long
Baseline Neutrino Facility (LBNF) and the Hyper-K experiment served by
the J-PARC neutrino beam.
The primary oscillation-physics goals and projected timescales for the
two experiments are similar.
The complementary of the two experiments \cite{Cao:2015ita} rests on
key differences in their specification.
Hyper-K will be sited 295 km from J-PARC, while DUNE will be located
1300 km from Fermilab. 
With these baselines, the energy at which the first oscillation
maximum occurs is different: $\sim 600$ MeV for Hyper-K and
$\sim 3$ GeV for DUNE.
Hyper-K will be located at an off-axis angle of $2.5^\circ$, yielding
a narrow neutrino-energy spectrum peaked at $\sim 600$ MeV with a
high signal-to-background ratio in the critical
${\nu}_\mu\to{\nu}_e$ channel.
DUNE will be located on-axis so that the beam with which it will be
illuminated will have a broad energy spectrum, peaked at
$\sim 3$ GeV, which will allow the second oscillation maximum to be
studied.

The Hyper-K and DUNE detectors are each designed to achieve optimal
performance given their beams.
Hyper-K will use a water Cherenkov detector since the technique is
proven for the detection of neutrino
interactions up to $\sim 1$ GeV where low multiplicity channels such as
quasi-elastic and resonant single-pion production dominate.  Thanks to scalability and cost effectiveness of the water Cherenkov technique, Hyper-K will feature a far detector of 260 ktons, more than 8 times larger than its predecessor, SuperKamiokande.
DUNE will exploit the high granularity and fine tracking capabilities of the liquid-argon time-projection chamber (LAr-TPC) technology, which allows the reconstruction of the more complex events resulting
from neutrino interactions at energies $\gsim 2$ GeV. The fully exclusive reconstruction of the final state will enable enhanced resolution on neutrino energy. DUNE will deploy four LAr-TPCs of 10 ktons each.

Matter effects in the long-baseline programme at Hyper-K will be
small and neutrino-oscillation effects such as asymmetries in the
neutrino and antineutrino oscillations will be dominated by ``vacuum''
effects such as CP-invariance violation. Indeed, Hyper-K features very large sensitivity to CP-asymmetry thanks to the huge mass of the far detector, enabling very large statistics for the electron-(anti)neutrino appearance. Matter effects will be instead significant for DUNE, allowing a detailed study
of related phenomena and the resolution of the mass ordering at $5\sigma$ for all possible values of $\delta_{\rm CP}$ after 7 years of running. 
The deep underground location of both experiments permits detailed
studies of atmospheric neutrinos to be made over a large range of
energies and baselines.
The study of the atmospheric-neutrino sample is a complementary probe of the oscillation physics, as discussed in Sec.~\ref{sec:proj_atm}. 

\begin{figure}
  \begin{center}
    \includegraphics[width=0.44\textwidth]{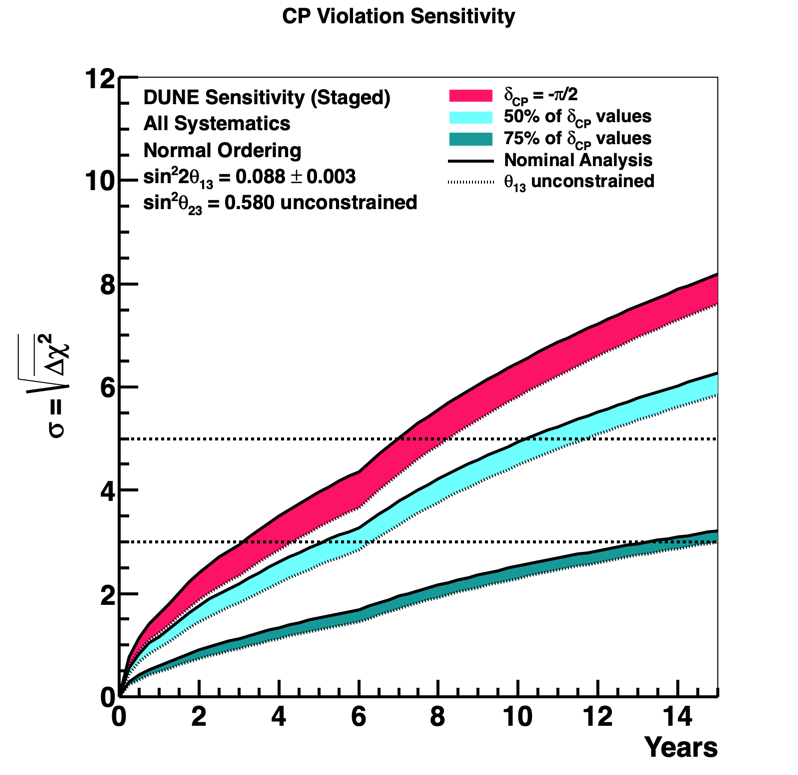}
    \quad \quad
    \includegraphics[width=0.50\textwidth]{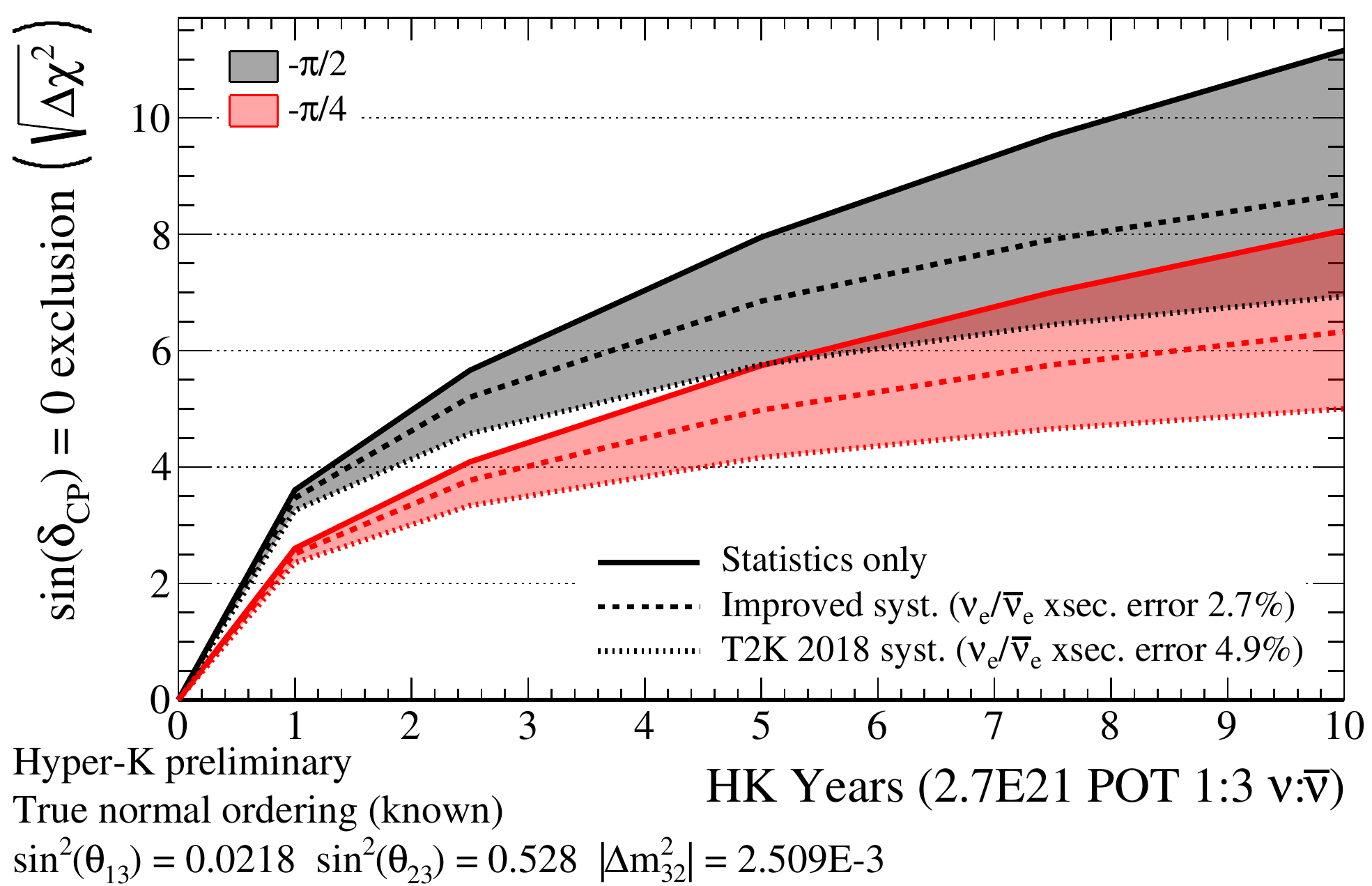}
  \end{center}
  \caption{
    Left panel:
    DUNE sensitivity to CP-invariance violation (i.e.\ 
    $\delta_{\rm CP} \ne 0$ or $\pi$) for the case when
    $\delta_{\rm CP} = -\frac{\pi}{2}$ and for 50\% and 75\% of all
    possible true $\delta_{\rm CP}$ values, as a function of time.
    The normal mass ordering is assumed.
    The width of the band shows the impact of applying an external
    constraint on $\sin^2 2 \theta_{13}$. Taken from Ref.\ \cite{Abi:2020evt}.\\
    Right panel:
    Hyper-K sensitivity to CP-invariance violation for $\delta_{\rm CP} = -\frac{\pi}{2}$ and $\delta_{\rm CP} = -\frac{\pi}{4}$ as a function of time. 
    The normal mass ordering is assumed.
    The width of the band show the impact of systematic uncertainties.
%Taken from Ref.\ \cite{Scott:2020gng}.
}
  \label{WG1:Sect:Accel:DUNEHK:Fig}  
\end{figure}

The sensitivity to CP-invariance violation of the DUNE and Hyper-K
experiments is summarized in Fig.\ \ref{WG1:Sect:Accel:DUNEHK:Fig}, assuming known mass ordering (and normal ordering).
The evolution of DUNE sensitivity takes into account the staging in three years of the installation of the far detector modules and an initial beam of 1.2 MW, upgraded to 2.4 MW after 6 years of data-taking. A combined uptime and efficiency of the the accelerator complex and beamline of 56\% is assumed. The evolution of Hyper-K sensitivity considers a beam of 1.3~MW as a function of 'Snowmass years', corresponding to 32\% availability of the beam.

In case of maximal CP-asymmetry ($\delta_{\rm CP} = -\frac{\pi}{2}$) Hyper-K can establish CP-invariance violation at the $3\sigma$ ($5\sigma$) confidence level after less than 1 (3) years of operation. In case of non-maximal violation, Hyper-K reaches in 5 years $5\sigma$ sensitivity to CP-invariance violation for 50\% of all values of $\delta_{\rm CP}$ and $3\sigma$ sensitivity to CP-invariance violation for 70\% of all values of $\delta_{\rm CP}$. 
In case of maximal CP-asymmetry DUNE can establish CP-invariance violation at the $3\sigma$ ($5\sigma$) confidence level after 4 (8) years of operation.
In case of non-maximal violation, $5\sigma$ sensitivity to CP-invariance violation is obtained with DUNE for 50\% of $\delta_{\rm CP}$ values after approximately 10-years of running; after 13 years $3\sigma$ sensitivity is reached over 70\% of all values of $\delta_{\rm CP}$.

Beyond the establishment of mass ordering and CP-invariance violation, DUNE and Hyper-K feature a long-term physics program of precision measurements of oscillation parameters, inside and beyond the standard PMNS paradigm. An unprecedented control of systematic uncertainties, due to detector effects and modeling of neutrino-nucleus interactions, will be needed. For instance, in case of maximal CP-invariance violation, in order to meet the target precision on $\delta_{\rm CP}$ of better than 20 degrees, a control of about 1\% on the scale of neutrino energy reconstruction is needed. The comparison and, eventually, combination of DUNE and Hyper-K will be crucial to meet this challenge and to have a robust cross-check of possible biases due to systematic effects. This is made possible by the different detector technology, energy reconstruction technique and nuclear effects at play in the two experiments. Finally, it is worth mentioning that the standard parametrization of neutrino oscillations assumes minimal three-flavour scenario, standard matter effects and standard neutrino production and detection. Establishing the fundamental properties of neutrino oscillation in a more general (model-independent) paradigm, will require measurements on a large range of the ratio baseline over energy ($L/E$), as can be provided only by the combination of multiple experiments.

\subsubsection{Possibilities with Accelerators}\label{sec:acc_fut}

An upgrade of the 2 GeV ESS linac has been proposed to deliver an
average power of 10 MW to be shared between neutrino and neutron
production~\cite{Wildner:2015yaa,ESSnuSB:2021lre}.
A neutrino beam with a mean energy of 0.4 GeV could be produced to
illuminate a megaton-scale underground water-Cherenkov detector
located 540 km from the ESS at the second oscillation maximum 
where the effect of CP-invariance violation is approximately three
times larger than at the first oscillation maximum.
The low neutrino-beam energy reduces the background from inelastic
scattering. 
Assuming a ten-year exposure with five years running time in neutrino mode and  five years in antineutrino mode, 
CP-invariance violation could be established with a significance of $5\sigma$ over more 
than 70\% of all values of $\delta_{\rm CP} $ and with an error in the measurement of the phase  of less than 8 degrees for all values of  $\delta_{\rm CP} $.

The next generation of accelerator-based neutrino-oscillation
experiments, DUNE and Hyper-K, will demonstrate detection techniques
that are at once extremely precise and capable of instrumenting
enormous sensitive volumes.
These techniques represent the culmination of many years of innovation
and development. 
By contrast, the neutrino-beam production techniques that will serve
DUNE and Hyper-K are incremental developments of that pioneered at
CERN in the 1960s.
Each exploits a Van der Meer horn which was first used to focus
pions produced using protons extracted from the Proton Synchroton.
Such horn-focused beams have been used at CERN, ANL, BNL, FNAL, IHEP,
KEK, and J-PARC, first to establish the quark-parton model and the 
Standard Model, and then to study neutrino oscillations and to search 
for new phenomena such as the existence of sterile neutrinos.

The neutrino flux produced by conventional, horn-focused, meson-decay
beams is contaminated with neutrino flavors other than the dominant
$\parenbar{\nu}_\mu$ contribution.
The presence of such contamination produces systematic uncertainties
and systematic biases in the extraction of the oscillation
parameters.
LBL experiments manage these systematic effects using sophisticated
near detectors to constrain the unoscillated neutrino flux and detailed
measurements of the particle spectra produced in the proton-target
interaction. 

To reach the precision required to determine whether the
three-neutrino-mixing model is a good description of nature and to
study the unitarity of the neutrino-mixing matrix is likely to require
novel neutrino beams in which the composition of the neutrino flux and
its energy spectrum are both precisely known. 
Two possible routes to the production of such beams are presently
under study.
The first exploits intense muon beams of low emittance to produce
neutrino beams with equal fluxes of electron- and muon-neutrinos.
The charge-to-mass ratio of the muon makes it possible to optimize
such a ``neutrino factory'' so that the neutrino-beam energy is
matched to a particular choice of detector technology and
source-detector distance.
The feasibility of the implementation of the ``Neutrinos from Stored
Muons'', $\nu$STORM, facility at CERN has been established in the context
of the Physics Beyond Colliders study group~\cite{Ahdida:2020whw}. 
$\nu$STORM has the potential to provide definitive, \%-level measurements
of neutrino-nucleus scattering, exquisitely sensitive searches for
light sterile neutrinos, and to provide a test bed for the development
of the technologies required to deliver a multi-TeV muon collider.
An alternative approach, being studied by the ENUBET collaboration, is
to tag the electron or positron produced in kaon decay to produce a
tagged electron-neutrino beam in which the decay kinematics is used to
estimate the energy of each $\overline{\nu}_e$.
ENUBET is the subject of an EU-funded design study.

An interesting proposal is the use of tagged neutrino beams, where the muon from the meson decay is detected in coincidence with the distant neutrino interaction. This technique has been explored many times but has historically run into the technical challenge of detecting the right meson decay out of the billions needed to produce a single neutrino interaction. A letter of interest has been submitted for the P2O experiment \cite{Akindinov:2019flp} which would use a tagged neutrino beam originating at Protvino at the KM3Net/ORCA detector.  The extremely large size of the KM3Net detector would allow a lower beam intensity, which combined with advances in tracking technology could allow efficient tagging of the initial decays.

\graphicspath{ {WG1/WG1_figs/} }

\subsection{Global Fits}
\label{WG1:SubSect:Glob}

The combination of data from different experiments plays a crucial
role in constraining the oscillation parameters. 
Solar neutrino experiments, namely radio-chemical ones,
Super-Kamiokande, SNO and Borexino, and the long-baseline reactor
neutrino experiment KamLAND probe the $\nu_e \rightarrow \nu_e$
disappearance channel, whose oscillation probability is controlled by
$\Delta m^2_{21}$, $\theta_{12}$ and $\theta_{13}$.
In particular, they provide the most precise determination of
$\Delta m^2_{21}$ and $\theta_{12}$. 
Reactor neutrino experiments, Daya Bay and at
sub-leading level RENO and Double CHOOZ, test the electron antineutrino
disappearance channel with a probability dependent on 
$\Delta m^2_{31}$, and, sub-dominantly, on $\Delta m^2_{21}$.
These experiments also give the best measurements of $\theta_{13}$.
Long baseline accelerator neutrino experiments, mainly T2K, NOvA
and MINOS, measure both the $\nu_\mu \rightarrow \nu_\mu$
disappearance and the $\nu_\mu \rightarrow \nu_e$ appearance
oscillation probabilities in the neutrino and antineutrino modes.
These oscillations are driven by $\Delta m^2_{31}$, $\theta_{32}$ and
$\theta_{13}$, with sub-leading but important effects due to the mass
ordering (normal or inverted, NO or IO) and $\delta_{\rm CP}$.
This is the main source of information on the mass ordering (MO) and on the violation of
the leptonic CP symmetry, especially when combined with the reactor-neutrino constraint on $\theta_{13}$. 
Atmospheric neutrino data is sensitive to the same parameters but in
different combinations, providing a synergy in extracting their
values.
Super-Kamiokande and IceCube-DeepCore are the most important sources
of information, with some contribution by ANTARES. 
\begin{table}
  \begin{center}
    \includegraphics[width=0.99\textwidth]{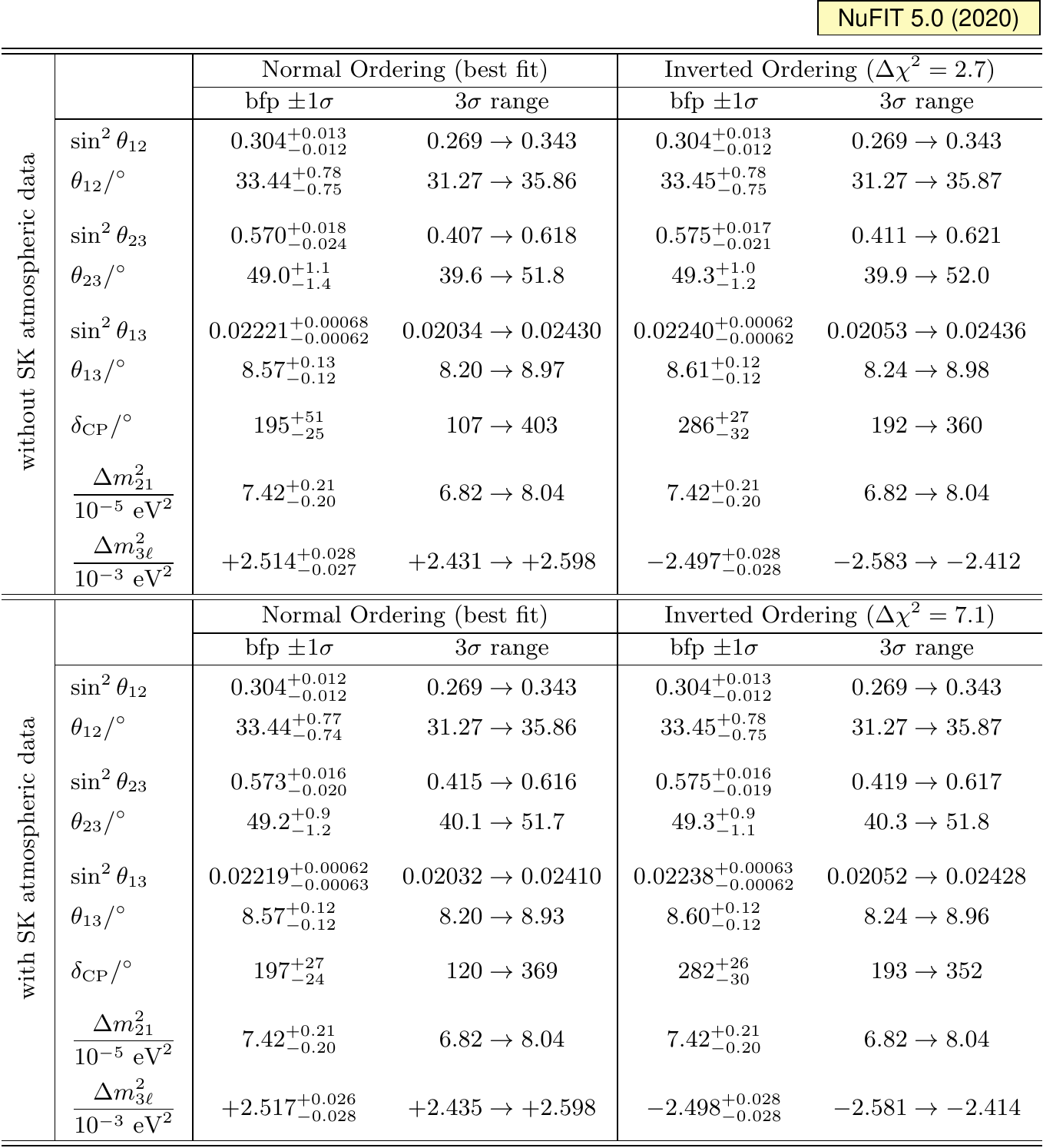}
  \end{center}
  \caption{
    Oscillation parameters from a fit of the global data as of July
    2020, version NuFit-5.0.
    The results in the lower (upper) sections are obtained (without)
    including atmospheric neutrino data from Super-Kamiokande.
    Note that $\Delta m^2_{3 \ell} = \Delta m^2_{31}> 0$ for NO and
    $\Delta m^2_{3 \ell} = \Delta m^2_{32} < 0$  for IO.
    Taken from Ref.~\protect\cite{Esteban:2020cvm,nufit}.
  }
  \label{WG1:fig:globalfit_1}
\end{table} 
\begin{figure}
  \begin{center}
    \includegraphics[width=0.75\textwidth]{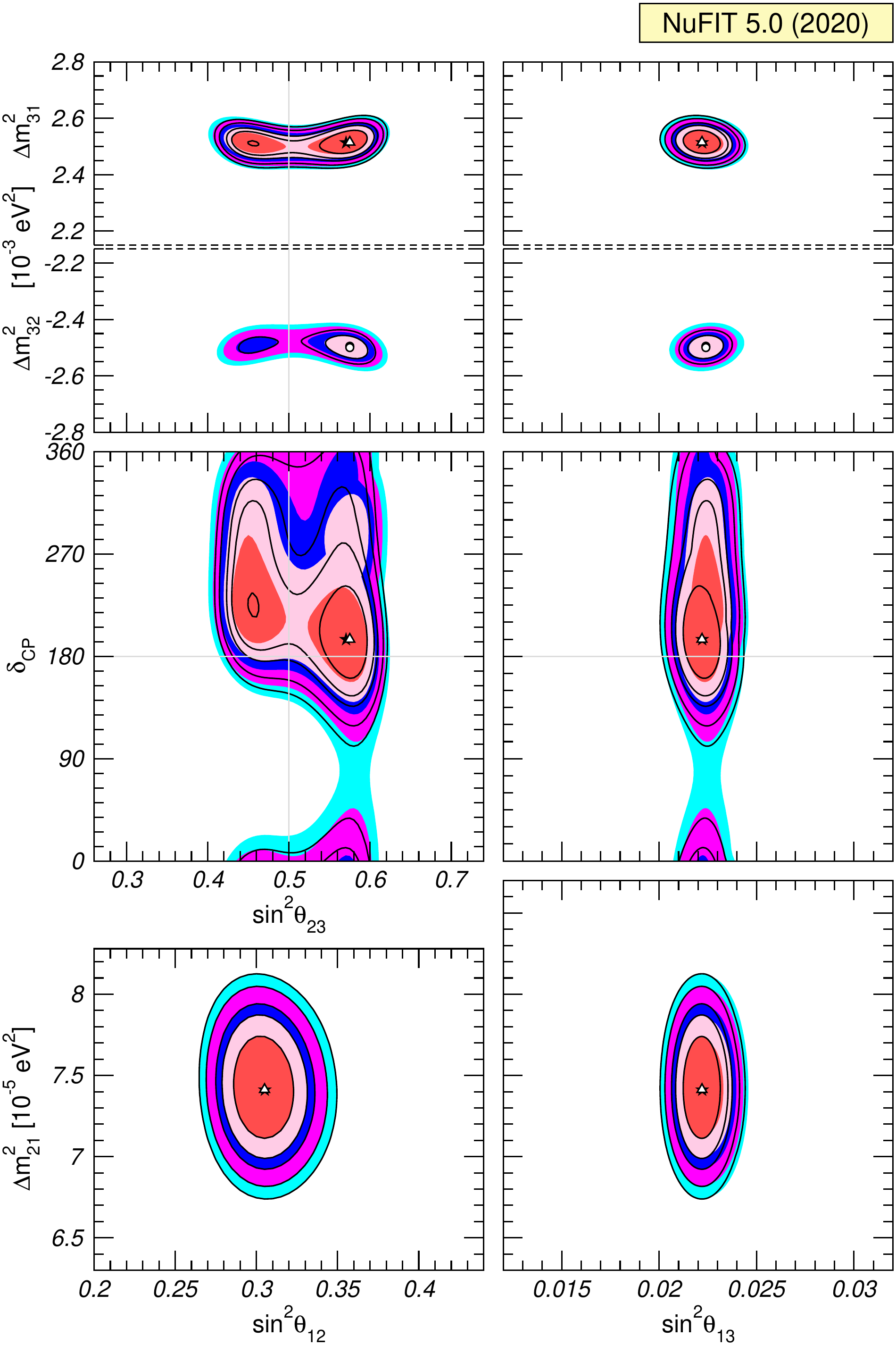}
  \end{center}
  \caption{
    Allowed values of the oscillation parameters at $1\sigma$, 90\%,
    $2\sigma$, 99\%, $3\sigma$ C.L.\ (2 dof).
    Each panel shows the two dimensional projection after
    marginalisation with respect to the undisplayed parameters.
    In the lower 4 panels, the results are minimized with respect to
    the mass ordering.
    Colored regions (black contour curves) do not (do) include
    Super-Kamiokande atmospheric neutrino data.
    Taken from~\protect\cite{Esteban:2020cvm,nufit}, version
    NuFit-5.0.
  }
  \label{WG1:fig:globalfit_2}
\end{figure} 

The main results on the oscillation parameters are reported in
Tab.\ \ref{WG1:fig:globalfit_1} and Fig.\ \ref{WG1:fig:globalfit_2},
taken from~\cite{Esteban:2020cvm,nufit}.
Two independent analyses by
Ref.~\cite{Capozzi:2019vbz,Capozzi:2018ubv} and
Ref.~\cite{deSalas:2020pgw,deSalas:2017kay} find similar results,
using data available up to May 2020.
In the solar sector, the mass-squared difference $\Delta m^2_{21}$  is
known at better than 3\% (at 1$\sigma$) mainly thanks to KamLAND
reactor antineutrino data.
$\theta_{12}$ is also precisely determined at around 2\%.
As discussed in Sec.\ \ref{WG1:SubSec:Solar:ThetaDm}, a mild tension
between the value of $\Delta m^2_{21}$ obtained by KamLAND
and other solar-neutrino experiments has been reduced with the inclusion of the latest
Super-Kamiokande Phase IV results~\cite{SuperK:nu2020,Esteban:2020cvm,nufit}. 
Overall, the data show a very good consistency and allow a very
precise determination of the oscillation parameters in the solar
sector. 

The best-known angle is $\theta_{13}$ whose 1$\sigma$ error is smaller
than 2\%, thanks to the reactor  data and in particular Daya Bay. The angle  
$\theta_{23}$ is the least known.
Its value is constrained to be very close to maximal (see
Tab.\ \ref{WG1:fig:globalfit_1}) but the octant is not yet determined at high
significance.
There is a mild preference for the second octant, arising from some
tension in the values of $\theta_{13}$ between accelerator and reactor
neutrino data, that is reduced for relatively large $\theta_{23}$. 

The value of $\Delta m^2_{31}$ ($\Delta m^2_{32}$) for NO (IO) is known to better than 2\%  (at 1$\sigma$) but its sign remains undetermined. All experiments, both using accelerator and atmospheric neutrinos, provide consistent results for the value of the mass-squared difference.
The preference for NO, which was around the $\sim 3\sigma$ level prior to the summer 2020 NOvA and T2K data, has significantly decreased to around $\sim 1.6\sigma$ (without Super-Kamiokande atmospheric neutrino results)~\cite{Esteban:2020cvm,nufit}. Exploiting matter effects in 
long-baseline neutrino oscillations, T2K is better compatible with NO and $\delta_{\rm CP} \sim 3 \pi/2$ while NOvA data is better fitted with  NO and $\delta_{\rm CP} \sim \pi/2$ or IO and $\delta_{\rm CP}$ close to $3 \pi/2$. The combination of the two would point towards IO and $\delta_{\rm CP} \sim 3 \pi/2$. Once reactor neutrino experiments are included, using the complementarity between the electron and muon neutrino disappearance channels, a preference for NO arises. It is expected that atmospheric neutrino data can somewhat strengthen this conclusions, but the latest Super-Kamiokande results have not been included yet in a global fit~\cite{Esteban:2020cvm,nufit}.

Data also show some hints in favor of leptonic CP violation with $\delta_{\rm CP}>180^o$. The key information comes from the T2K and NOvA appearance channels combined with the precise measurement of $\theta_{13}$ from reactors. Global analyses disfavor CP conservation for $\delta_{\rm CP}=0$ at above $2\sigma$ for both mass orderings and for $\delta_{\rm CP}=\pi$ at a smaller significance for NO. In particular, recent T2K results have provided further hints in favor of CP violation, with both CP conserving values $\delta_{\rm CP}=0$ and $\delta_{\rm CP}=\pi$ excluded at 95\%~C.L.~\cite{Abe:2019vii}. For IO, both T2K and NOvA point towards maximal CP violation at $\delta_{\rm CP} \sim 3 \pi/2$. For NO, which is mildly preferred as discussed above, the tension between these two experiments broadens the allowed range of $\delta_{\rm CP}$ and shifts its best-fit value towards $\delta_{\rm CP}\sim \pi$, weakening the previous hints of CP-violation. These conclusions are evolving as new data from accelerator neutrino experiments becomes available. 

It should be noted that new physics beyond the 3-neutrino mixing scheme, e.g.\ non-standard interactions and non-unitarity of the mixing matrix, could affect the results discussed above, see Sec.\ \ref{sec:np}. Future even more precise data will be able to shed further light on these issues.

\subsection{Neutrino Oscillations: Summary}
Neutrinos oscillate which implies that leptons mix in analogy to quarks. This means that neutrinos have non-vanishing rest masses, which requires at least one new particle species beyond the ones in the standard model. 

The large mixing angles and the tiny mass-squared differences made it possible that neutrino oscillations were observed. At the same time, these properties are surprising from a theoretical point of view, as the charged leptons and the quarks have huge mass differences, and quarks mix with small to tiny angles. While the leading aspects of lepton mixing are clear by now, one close-to-maximal, one large, one small mixing angles, there is much to do. 
The mass ordering and the CP phase remain to be determined, though first interesting hints have emerged. 
Experimentally, large scale facilities are required to determine unknown and precisely measure known neutrino parameters. 
The physics potential of these large experiments reaches beyond pure neutrino physics.  

The fact that the different neutrino experiments, ranging from measurements of solar or atmospheric neutrinos to reactor or accelerator neutrinos, and spanning many orders of magnitude in energy and distance, can be combined in a common framework, is far from trivial. Further checking whether the three-neutrino paradigm can indeed describe all available and future data, or of new physics modifies the parameters at some level, is crucial. This requires in particular different and complementary approaches using different energies and baselines. 
Answering those open questions has many ramifications in neutrino physics, particle physics, and beyond. It will contribute to understanding what is beyond the standard model. 

\label{chap:osc}

\newpage
\section{Absolute Masses}
Contributing additional author: Kathrin Valerius (KIT)
\label{WG2}
\subsection{Introduction}
\label{sec:mass_intro}
Neutrinos are massless particles in the Standard Model. The straightforward extension of the SM  to introduce neutrino masses similar to the charged lepton masses is the addition of  right-handed (SM singlet) neutrino fields; Yukawa interactions will then lead to  Dirac neutrino masses after electroweak symmetry breaking (EWSB). This ansatz is, however, perceived to be unsatisfactory by the neutrino theory community  for two reasons: a) it does not explain why the absolute neutrino mass scale is at least  a factor one million smaller than that of the other SM fermion masses, and b) the SM symmetries do not forbid other, so-called Majorana mass terms for the newly introduced  right--handed neutrino fields. These masses are not bounded from above by the Higgs vacuum expectation value and thus expected to take values much larger than the top quark mass. 
Taking the Majorana mass terms into account, leads (after integrating out the heavy masses) to effective light Majorana neutrino masses at an absolute neutrino mass scale $m_\nu \simeq m_D^2/M_R$; here $m_D$ is the scale of the electroweak symmetry breaking and $M_R \gtrsim 10
^{14} \, \mathrm{GeV}$ the scale of the heavy Majorana neutrinos (where the constraint on $M_R$ is inferred from the current bounds on the absolute neutrino mass scale). This mechanism is established as the see-saw mechanism (type~I)~\cite{Minkowski:1977sc,Yanagida:1979as,GellMann:1980vs,Mohapatra:1979ia}; it is  attractive because it  describes the smallness of neutrino mass, has a  potential connection to leptogenesis (cf.\  Sec.\ \ref{WG1:SubSec:PMNS-Theory:CPV}), and may even imply a relationship to a scale unifying the electroweak and strong forces. Light neutrino masses then emerge as the mass eigenstates of the effective light Majorana mass matrix, and the PMNS matrix $U$ is obtained as the relative rotation between the left-handed charged lepton and neutrino fields (which the charged current couples to); see Sec.\ \ref{WG1:SubSec:PMNS-Theory:flavorSymmetries} for theoretical implications of the PMNS matrix. In the meanwhile, many different versions of the see-saw mechanism have been studied which include one or more new fields; the simplest  alternatives are the type-II~\cite{Magg:1980ut,Schechter:1980gr,Wetterich:1981bx,Lazarides:1980nt,Mohapatra:1980yp,Cheng:1980qt} and type-III~\cite{Foot:1988aq} see-saw mechanism including a triplet scalar and a triplet fermion, respectively; most of these lead to effective light Majorana neutrino masses. There are however many more mechanisms, Sec.\ \ref{sec:D_or_M} and Sec.\ \ref{sec:origin} discuss general aspects and implications.

Recall that the absolute neutrino masses emerge from the theory as mass eigenstates of the effective light neutrino mass matrix. While neutrino oscillation experiments can measure the mass-squared splittings among these and even the ordering of the masses, they cannot access the absolute neutrino mass scale, which corresponds to the overall normalization of that mass matrix. Neutrino oscillations imply lower bounds for the sum of the neutrino masses of 0.06~eV and 0.10~eV  for the normal and inverted orderings, respectively, while the current upper bounds are $\lesssim$~1~eV using different methods. If the sum of the neutrino masses is close to the lower bound, we speak of a hierarchical mass scheme with the lightest neutrino mass closer to (or equal to) zero compared to both mass splittings. If it is close to the upper bound, we speak of degenerate neutrino masses, because the splittings $|\Delta m^2| \ll m^2$ are small compared to the masses.  Neutrino mass ordering and mass scale are important indicators for theoretical models, because the underlying structure of flavor in the Lagrangian describing neutrino mass will be very different in the normal hierarchical, inverted hierarchical, and degenerate cases, see Sec.\ \ref{WG1:SubSec:PMNS-Theory:flavorSymmetries}. 

\begin{figure}
    \centering
    \includegraphics[width=0.32\textwidth]{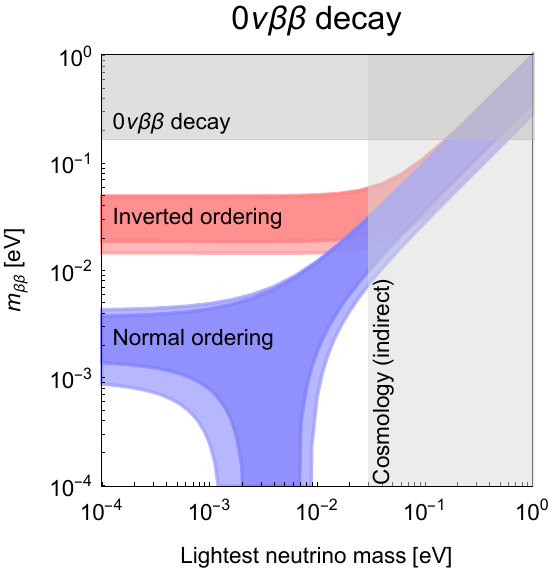}     
    \includegraphics[width=0.32\textwidth]{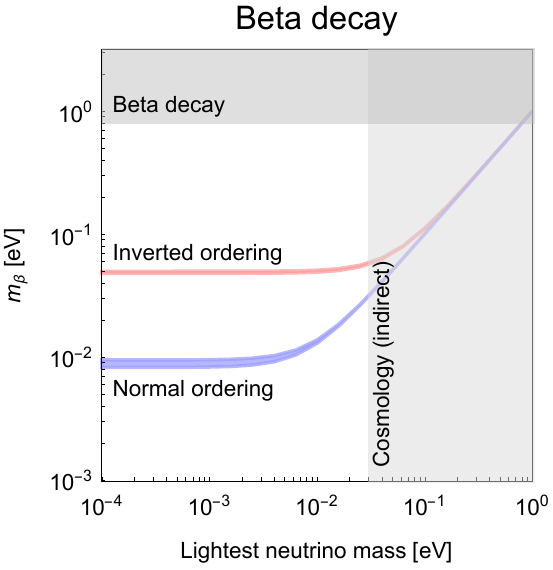} 
    \includegraphics[width=0.32\textwidth]{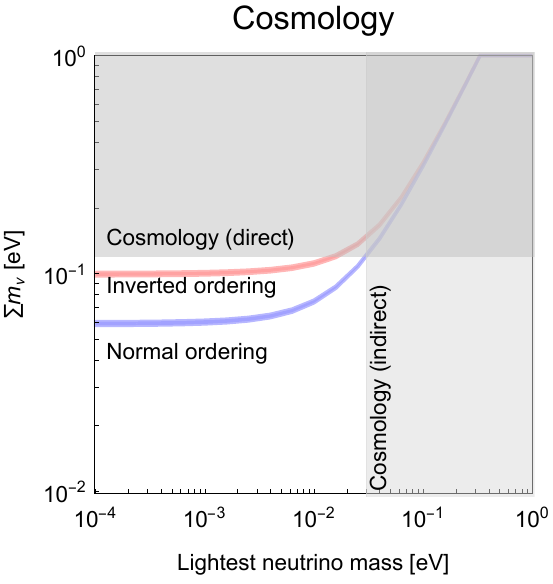}
    \caption{Experimentally observable combinations of neutrino mass in neutrinoless double beta decay ($0\nu\beta\beta$), beta decay, and cosmology (panels) as a function of the lightest neutrino mass. Dark-colored shadings refer to the uncertainties of the Majorana phases only, whereas light colored shadings include the current oscillation parameter uncertainties (parameters varied within their $3\sigma$ regions taken from Ref.~\cite{Esteban:2020cvm}, NuFit 5.0). Current bounds (upper grey-shaded regions) are taken from KamLAND-Zen~\cite{KamLAND-Zen:2016pfg} (90\% CL, conservative nuclear matrix element), KATRIN~\cite{Aker:2021gma} (90\% CL), and Planck~\cite{Aghanim:2018eyx} (95\% C.L., CMB lensing and galaxy clustering, incl.\ BAO) for $0\nu\beta\beta$, beta decay, and cosmology, respectively. The right grey-shaded areas are indirect bounds from cosmology, derived from the direct bound in the right panel (for the normal ordering).}
    \label{fig:nmass}
\end{figure}

If the neutrino mixing matrix, mass ordering and mass-squared splittings are fixed, the absolute neutrino mass scale can be parameterized by one remaining free parameter. A frequent choice is the lightest neutrino mass $m$, which can be either $m_1$ (normal ordering) or $m_3$ (inverted ordering). This parameter is, however, not directly accessible to experiments or observations, as discussed in detail in this section. Focusing on three active neutrinos, the three most prominent experimentally accessible combinations of neutrino mass are described as follows:  
\begin{description}
\item[Cosmological tests of neutrino mass\!] are sensitive to the sum of the neutrino masses
\begin{equation}
 \sum m_\nu = m_1 + m_2 + m_3 \, . \label{equ:mcosmo}
\end{equation}
\item[Beta decay experiments sensitive to the endpoint\!] probe the incoherent sum of the neutrino masses coupling to the electron flavor
\begin{equation}
 m_\beta \equiv \sqrt{ |U_{e1}|^2 m_1^2 + |U_{e2}|^2 m_2^2 + |U_{e3}|^2 m_3^2} \, . \label{equ:mbeta}
\end{equation}
\item[Neutrinoless double beta decay experiments\!] test the combination
\begin{equation}
 m_{\beta \beta} \equiv \left| U_{e1}^2 m_1 + U_{e2}^2 m_2 + U_{e3}^2 m_3 \right| \, \label{equ:mbetabeta}
\end{equation}
if the light neutrinos have Majorana masses. 
 Note that for Majorana neutrinos, the mixing matrix carries two additional phases, the so-called Majorana phases often denoted by $\phi_1$ and $\phi_2$ (or $\alpha_1$ and $\alpha_2$). They do not influence neutrino oscillations \cite{Bilenky:1980cx,Langacker:1986jv}, and can for instance be included by writing $U_{e2}^2 = |U_{e2}|^2 e^{i \phi_1}$ and $U_{e3}^2 = |U_{e3}|^2 e^{i \phi_2}$, where  $|U_{e2,3}|^2$ can be parametrized as in Eq.\ (\ref{Eq:SnuM}). They can have a strong impact on the prediction of $m_{\beta\beta}$. 
\end{description}

If the existing information on masses and mixings from neutrino oscillations is applied to Eqs.~(\ref{equ:mcosmo}) to (\ref{equ:mbetabeta}), these combinations can be expressed in terms of the lightest neutrino mass $m$.
We therefore show the experimentally observable combinations of neutrino mass in $0\nu\beta\beta$, beta decay, and cosmology  as a function of the lightest neutrino mass in \figu{nmass}. The current oscillation parameter uncertainties (denoted by the light-colored shadings) are already so small that extremely good predictions for the observables as a function of $m$ can be made. Only for $0\nu\beta\beta$  the uncertainties are much larger and come from the (unknown) Majorana phases (dark colors). 

The results depend in all cases on the mass ordering, as  is most obvious in the right panel for $m=0$, where the limits discussed earlier are recovered. 
The most complicated case is $0\nu\beta\beta$, where  cancellations are possible for the normal ordering in the appropriate range of $m$. Note that the horizontal grey-shaded regions in \figu{nmass} are current direct experimental bounds on the observables on the vertical axes, while the vertical bound on $m$ is indirectly derived from the right panel from the bound on $\sum m_\nu$ (where the gray-shaded areas meet); it will obviously depend somewhat on the mass ordering. From any panel, it can be read off that $m  \gg 0.1 \, \mathrm{eV}$ corresponds to the degenerate regime in which the observable is proportional to $m$ and the impact of the mass ordering is negligible. For $m<0.1 \, \mathrm{eV}$, there are substantial differences in the observables;  we will discuss the direct correlations between the observables at the end of this section after explaining in detail  the various approaches to gain insight into the mass of neutrinos.

\subsection{Kinematic Measurements of Neutrino Mass}

\subsubsection{Direct Mass Measurements with \texorpdfstring{$\nu_e$}{nue}}
\label{sec:nu_e}

A largely model-independent way to access the absolute neutrino mass scale in a laboratory measurement is offered by precision kinematics measurements of weak decays. The imprint of the  effective electron-based neutrino mass $m_\beta$, see \equ{mbeta}, consists both of a reduction of the kinematic endpoint ($E_0 - m_\beta$) and of a spectral shape modification close to the endpoint. The latter is the signature more readily exploited by experiments.

Current experimental efforts are focused on two nuclides which are particularly suitable in terms of their half-life (accessible event rate), spectral range (low kinematic endpoint), and isotopic availability: the $\beta^{-}$ emitter tritium (\textsuperscript{3}H) and the electron capture isotope \textsuperscript{163}Ho (see Tab.~\ref{tab:kinematical_exp}). The two approaches are highly complementary due to the widely different experimental techniques they rely on. Interestingly, they allow to address $m_\beta(\nu_\mathrm{e})$ and $m_\beta(\overline{\nu}_\mathrm{e})$ independently, which are expected to be identical in the case of CPT conservation. Since both methods are based on relativistic energy-momentum conservation, the experimentally accessible observable in either case is the squared mass $m_\beta^2$, which illustrates the difficulty in gaining an extra order of magnitude in sensitivity on the neutrino mass.
We note here that in principle, such kinematic experiments can also be used for a lab-based direct search for sterile neutrinos (essentially by expanding \equ{mbeta} to a fourth neutrino mass state and its corresponding mixing matrix entry). This will be discussed in Sec.\ \ref{sec:sterile_mass}. 

\begin{table}[t]
\resizebox{\textwidth}{!}{%
	\centering
	\begin{tabular}{lll}\hline\hline
{\bfseries Decay type}	& {\bfseries Project} & {\bfseries Staged goals and projected timelines} \\
\hline
$\beta^-$-decay of \textsuperscript{3}H\textsubscript{2}	&	KATRIN	& First 2 science runs (2019), $m_\beta < 0.8$~eV (90\% C.L.) \cite{Aker:2021gma}\\[0.2ex]
	&		& Data-taking for $0.2$~eV sensitivity ($2019- 2024$)\\[0.2ex]
	&       & R\&D on differential read-out and novel source concepts\\[0.2ex]
	&		& Detector upgrade for keV-sterile $\nu$ search ($>$~2025)\\
\midrule
IC of \textsuperscript{83m}Kr	&	Project 8	& Phase I: demonstration of CRES technique ($2014 - 2016$) \cite{Asner:2014cwa}\\[0.2ex]
$\beta^-$-decay of \textsuperscript{3}H\textsubscript{2}	&		& Phase II: tritium demonstrator ($2015-2020$)\\[0.2ex]
$\beta^-$-decay of \textsuperscript{3}H\textsubscript{2}	&  		& Phase III: {\small large-volume CRES demonstrator,  atomic source develop.} \\
& & (R\&D ongoing,  starting 2023 for  $3-5$~eV sensitivity)\\[0.2ex]
$\beta^-$-decay of \textsuperscript{3}H								&  		& Phase IV: atomic source \\
& & (R\&D ongoing,  starting 2024 for  $40$~meV sensitivity)\\
\midrule
$\nu_e$ capture on \textsuperscript{3}H	& PTOLEMY & R\&D stage, target $m_\beta$ sensitivity $<100$~meV*\\
\midrule
EC of \textsuperscript{163}Ho		& ECHo		& ECHo-1k: Medium-sized array ($\sim$ 100 detectors) \\
& 	& First science run (2018), $m_\beta < 150$~eV (95\% C.L.) \cite{Velte:2019jvx}\\
&   & Data-taking for $10 - 20$~eV sensitivity ($2019 - 2020$)\\ [0.2ex]
& 	& ECHo-100k: Large array ($\sim$ 12000 detectors)\\ 
&	& Production phase, starting 2021 for $1 - 2$~eV sensitivity \\
\midrule
EC of \textsuperscript{163}Ho		& HOLMES	& Short-term ($2020 - 2021$): Medium-sized array ($\sim$100 ch.)\\
& & for $10 - 20$~eV sensitivity \\[0.2ex]
& & Medium-term ($2021 - 2023$): Increase no.\ of deployed arrays \\
& & ($\sim$ 1000 channels) for $1 - 2$~eV sensitivity \\
\hline\hline
\end{tabular}}
\caption{Overview of current and upcoming experimental approaches in direct kinematic mass measurements using $\beta^-$-decay of molecular and atomic tritium or electron capture in \textsuperscript{163}Ho. Most experiments are taking a phased approach towards improving sensitivity on the neutrino mass. --- *) The PTOLEMY project is intended to become a tritium-based observatory for cosmic relic neutrinos. Its potential to probe neutrino masses is currently being investigated.
	\label{tab:kinematical_exp} }
\end{table}

\paragraph{Tritium Beta-Decay} For tritium-based experiments, the spectroscopic method utilizing an electrostatic retardation spectrometer with magnetic adiabatic collimation to maximize angular acceptance (MAC-E filter) \cite{Lobashev:1985mu,PICARD1992345} has yielded the most stringent bounds obtained thus far (see \cite{Kraus:2004zw,Aseev:2011dq} and   \cite{Otten:2008zz,Drexlin:2013lha} for a comprehensive review). The KATRIN experiment exploits the full reach of this technology both in terms of its high-luminosity gaseous tritium source ($10^{11}$ decays per second) and dimensions of the high-resolution spectrometer (10~m diameter), which allows a highly precise measurement of the $\beta$-decay spectrum with low systematic uncertainties. In its four-week first neutrino-mass run, KATRIN has  achieved an upper limit of $1.1$ eV (90\% C.L.) \cite{Aker:2019uuj}, followed by a second measurement campaign  that was the first ever to have sub-eV sensitivity. The limit of 0.9 eV was combined with the first neutrino-mass run to achieve an upper limit of $0.8$ eV (90\% C.L.) \cite{Aker:2021gma}. 
KATRIN will continue data-taking for a total of $\sim 1000$ measurement days to reach its design sensitivity of $0.2\,\mathrm{eV}$ (90\% C.L.) \cite{Angrik:2005ep}. Because of its unprecedented statistics measuring a super-allowed decay, the KATRIN experiment is also sensitive to eV-scale light sterile neutrinos \cite{Aker:2020vrf} (see Sec.\ \ref{sec:sterile_mass}) and other BSM physics. A program to search for keV-scale sterile neutrinos with the TRISTAN detector upgrade \cite{Mertens:2018vuu} is in the R\&D phase. The option of turning the integral spectrum measurement of the high-pass MAC-E filter into a differential measurement by adding time-of-flight information is also being investigated \cite{1999NIMPA.421..256B,Steinbrink:2013ska}. 
Another possible improvement is related to the fact that a MAC-E filter collects low-energy electrons, released by several background
processes within its large volume, and accelerates them to the focal-plane detector. Such
low-energy electrons, which cannot be differentiated energy-wise by the detector, typically
have much smaller transverse energies or incident angles. R\&D on developing a novel
“active transverse energy filter” (aTEF) detector for KATRIN is on-going. 

The technology used in KATRIN cannot easily be pushed further since the windowless gaseous tritium source is already nearly opaque to electrons. This limitation is circumvented by a new approach proposed in \cite{Monreal:2009za} which employs the cyclotron radiation emitted by electrons stored in a magnetic trap for a non-destructive measurement of the $\beta$-decay spectrum via radiofrequency detection. A successful proof-of-principle of the cyclotron radiation emission spectroscopy (CRES) technique has been performed by Project~8 using monoenergetic internal conversion (IC) electrons from \textsuperscript{83m}Kr \cite{Asner:2014cwa}, and, subsequently, spectroscopic information was acquired for first $\beta$-electrons from a small volume containing gaseous tritium. In future project phases, an open receiver array encompassing a large source volume is foreseen in order to obtain neutrino-mass sensitivity at the $3 - 5$~eV scale  \cite{Esfahani:2017dmu}. Moreover, R\&D work has started towards the development of an atomic tritium source which holds the promise to surpass the fundamental sensitivity limitation (around $100$ meV) due to the population of excited molecular final states. An effort to combine a large-scale CRES detector with an atomic tritium source to reach eV neutrino mass sensitivity is under investigation. Explorative studies of techniques to approach the hierarchical mass scale are under way, either using trapped tritium atoms (as proposed by Project~8) or quasi-atomic sources on substrates (as discussed, for instance, for the PTOLEMY study \cite{Betti:2019ouf}). 
A sensitivity of $\sim40$ meV, as envisioned for Project~8, would cover the inverted mass ordering range (see Fig.\ \ref{fig:nmass}), and a null result of $m_\beta$ at that level would point towards the normal hierarchy. 

\paragraph{Electron Capture in $^{163}$Ho} 
An alternative path towards kinematic neutrino-mass determination through electron capture (EC) in $^{163}$Ho was  opened in the 1980s \cite{DeRujula:1982qt}. 
More recently, substantial progress in the development of cryogenic microcalorimeters has kindled major incentives leading to a new generation of holmium-based neutrino-mass experiments: ECHo \cite{Gastaldo:2017edk} and HOLMES \cite{Giachero:2016xnn}.
The required amount of \textsuperscript{163}Ho nuclei is produced through neutron irradiation and subsequent purification of the source material. In the two experiments, the source nuclei are implanted into two different types of cryogenic detectors: ECHo has already successfully demonstrated the acquisition of high-resolution EC spectra with arrays of holmium-implanted Metallic Magnetic Calorimeters (MMC) \cite{Fleischmann2005,Gastaldo:2009ovy}, whereas the HOLMES technology is based on Transition Edge Sensors (TES) \cite{Faverzani:2012zgs} which have been shown to exhibit excellent detector properties in characterization measurements prior to implantation with holmium. Among the challenges in setting up a large-scale neutrino-mass experiment of eV- to sub-eV sensitivity are the control of the pile-up fraction, key detector characteristics such as energy resolution and signal rise time of implanted calorimeters, and the operation and multiplexed read-out of large detector arrays on the order of tens of thousands of individual detectors to acquire sufficient event statistics. These challenges are being addressed by current experiments, with near-term goals targeting a sensitivity around $10 - 20$~eV. Towards the medium-term goal of approaching the $1 - 2$~eV benchmark, the two collaborations have developed different concepts for leveraging array size (i.e., number of channels) against activity load of individual detectors. New ideas with the aim of reaching sub-eV sensitivities of 200~meV or even beyond are being investigated for future stages of calorimetric arrays. Next to further developments on the experimental methods, obtaining an improved model of the calorimetric EC spectrum, including both the theoretical spectral shape \cite{Faessler:2014xpa,Brass:2017kov,Velte:2019jvx} and the detector response, is of vital importance for inference of the neutrino mass.

\paragraph{Outlook} With several experiments now operational and more projects in preparation (cf.\  Tab.\ \ref{tab:kinematical_exp}), direct neutrino mass search is entering a decisive phase. The next years are about to bring the scientific return of long-standing development work. The reach of current, proven technologies extending present-day sensitivity by a factor of 5 down to $\sim 200$ meV forms the foundation for progressive, novel ideas which promise even further improvement by another order of magnitude.

\subsubsection{Direct Mass Measurements with \texorpdfstring{$\nu_{\mu}$}{numu} and \texorpdfstring{$\nu_{\tau}$}{nutau}}
%(Expertise: \underline{Mikhail Danilov})
%
Direct mass measurements of $\nu_{\mu}$ and $\nu_{\tau}$ are performed using $\pi \rightarrow \mu + \nu_{\mu}$ and $\tau \rightarrow n\pi + \nu_{\tau}$ decays. These decays are sensitive to the incoherent sum of the neutrino masses coupling to the muon and tau flavors like in \equ{mbeta}. The best limits on the masses of $\nu_{\mu}$ and $\nu_{\tau}$ are 0.19~MeV \cite{Zyla:2020zbs}%0.17~MeV \cite{Assamagan:1995wb}
and 18.2~MeV \cite{Barate:1997zg}, respectively. In the case of $\nu_{\mu}$, a near surface muon beam was used in order to reduce the energy losses in the target material and to determine very precisely the ${\mu}^+$ momentum in the ${\pi}^+$ decay. The limit on the $\nu_{\mu}$ mass was obtained using the measured value of the muon momentum and known masses of ${\pi}^+$ and ${\mu}^+$. In the case of $\nu_{\tau}$, the 5 and 6 pion $\tau$ decays with large invariant masses of the multi-pion system provided the strongest limits on the $\nu_{\tau}$ mass. Considerable improvements on the $\nu_{\tau}$  mass accuracy can be made by the Belle II and BES III experiments mainly because of larger statistics, better multi-pion invariant mass resolution especially in case of BES III, and good energy-scale determination. It should be noted that in all known scenarios those masses are very close to the 
electron neutrino mass from beta-decay.

\subsubsection{Neutrino Mass from Supernova Neutrino Detection}\label{sec:numass_SN}

Another possible kinematic method for learning about the neutrino absolute mass scale is via detection of a burst of neutrinos from a core-collapse supernova.  More detailed information about supernova neutrinos can be found in Sec.\ \ref{sec:source_SN}; % elsewhere in this report~\ref{sec:sn};
in short, the core collapse of a massive star will yield an intense flash of neutrinos of all flavors with energies from a few to a few tens of MeV, over a few tens of seconds.  The detectable range for current large neutrino detectors is approximately the Milky Way (tens of kpc); next-generation detectors such as Hyper-K will observe a handful of neutrino events from Andromeda, $\sim 700$ kpc away.
The idea for constraining neutrino mass is that neutrino propagation over the very long distance from a supernova will result in an energy-dependent time delay with respect to propagation at $c$ according to 
$\Delta t \, [\mathrm{s}]= 0.515 \left( m_\nu \, [\mathrm{eV}]/E_\nu \, [\mathrm{MeV}] \right)^2 \, D \, [\mathrm{kpc}]$ for distance to the supernova $D$.  
Over Galactic distances, given current experimental limits on the absolute neutrino mass scale, the expected kinematic delay is much smaller than the spread of emission times from the supernova, making it a challenge to extract information about neutrino mass.   The time spread of the observed $\bar{\nu}_e$ burst from SN1987A 55~kpc away, about 13 seconds, enabled some of the best mass limits of the era, $m_\nu<20$~eV%$^2$
~\cite{Schramm:1990pf} (improved with updated analysis to around 6~eV%/$^2$
~\cite{Loredo:2001rx,Pagliaroli:2010ik}). 
Although it will be hard to compete with KATRIN's final expected sensitivity, nevertheless some information may be extracted with high-statistics and low energy thresholds of the next core-collapse supernova burst, with expected sensitivity down to around the eV scale~\cite{Beacom:1998ya,Beacom:1998yb,Lu:2014zma}.  Possible improvements to this measurement could result from a sharp emission cutoff of neutrino luminosity due to formation of a black hole~\cite{Beacom:2000qy} or the coincident observation of a gravitational wave burst~\cite{Arnaud:2001gt}; both of these would help set a $t_0$ for relative delay and would improve the constraints.

\subsection{Neutrinoless Double Beta Decay}
\label{sec:0vbb}

One of the most interesting questions in neutrino physics concerns the character of the neutrinos, i.e.\ whether these fermions are  Majorana or Dirac particles. The necessity of performing  sophisticated experiments to determine the Majorana nature of neutrinos lies in the $V-A$ structure of weak interactions. This implies that the difference of Dirac and Majorana neutrinos is of order $m_\nu/E$ on the amplitude level, where $m_\nu$ is the neutrino mass and $E$ the energy scale of the process. Indeed, all other processes in principle sensitive to the Majorana nature of light neutrinos, like $Z$ decays in neutrino pairs or neutrino-antineutrino oscillations, are not suited to probe the Majorana nature. Another difference is given by the fact that for Dirac neutrinos the right-handed neutrinos are independent particles (in contrast to Majorana neutrinos, where they are related). Hence, these species would contribute to the relativistic number of degrees of freedom, see Sec.\ \ref{sec:neff}. However, the smallness of neutrino mass implies that they do not thermalize in the early Universe and thus contribute in negligible amounts.

\subsubsection{General Aspects}
Neutrinoless double beta decay ($0\nu\beta\beta$) is one of the most sensitive probes for physics beyond the Standard Model of Particle Physics. In this decay mode, two beta particles are emitted in the final state but no neutrino,
\begin{equation}
    \left(Z, A\right)\ \rightarrow\ \left(Z+2,A\right)+2e^-\,. \label{equ:0nbb-eq}
    %^N_ZA_{\beta\beta}\ \rightarrow\ ^{N-2}_{Z+2}A+2e^-\,.
\end{equation}
Here, the atomic number $Z$ changes by two units while the mass number $A$ remains unchanged. Such a decay violates lepton number conservation, which is an accidental symmetry in the SM, 
by two units and would imply that the neutrino is a Majorana particle \cite{Schechter:1981bd}. Equivalent decay modes in neutron-deficient nuclei are $0\nu\beta^+\beta^+$, $0\nu EC\,EC$ (double Electron Capture), and $0\nu\beta^+EC$. However, in $\beta^+\beta^+$ and $\beta^+\,EC$ decays the $Q$-value is reduced by $4m_e$ and $2m_e$, respectively, which results in slower decay rates. The reduced $Q$-value, i.e., emitted energy, further limits the sensitivity of experiments since the background rate typically increases for decreasing energies (see, e.g., \cite{Blaum:2020ogl} for a review of $EC\,EC$, including the resonant option). Therefore, the decay mode presented in \equ{0nbb-eq} is generally preferred in searches for lepton number violation.

An observation of $0\nu\beta\beta$ would have far-reaching implications for our understanding of the Universe: it would be the first observation of a fundamental particle with properties that are completely different from all other known fermions. Lepton number violation in weak decays could help explain the observed matter-antimatter imbalance in our Universe \cite{Fukugita:1986hr}, as most ideas to suppress neutrino mass or explain the matter-antimatter asymmetry of the Universe predict lepton number violation, see Secs.\ \ref{sec:D_or_M}, \ref{sec:origin} and \ref{WG1:SubSec:PMNS-Theory:CPV}.  Furthermore, an observation of \textbf{$0\nu\beta\beta$} could, depending on the underlying physics, allow the extraction of the effective neutrino mass $m_{\beta\beta}$ (see~\equ{mbetabeta}) and has ramifications on particle physics and possibly cosmology. 
%and help explain why the mass of the neutrino is so much smaller than that of any other fundamental particle. %Status and prospects of neutrinoless double beta decay are discussed in detail in \cite{Dolinski:2019nrj}.
\begin{figure}
    \centering
    \includegraphics[angle=-90,width=0.65\textwidth]{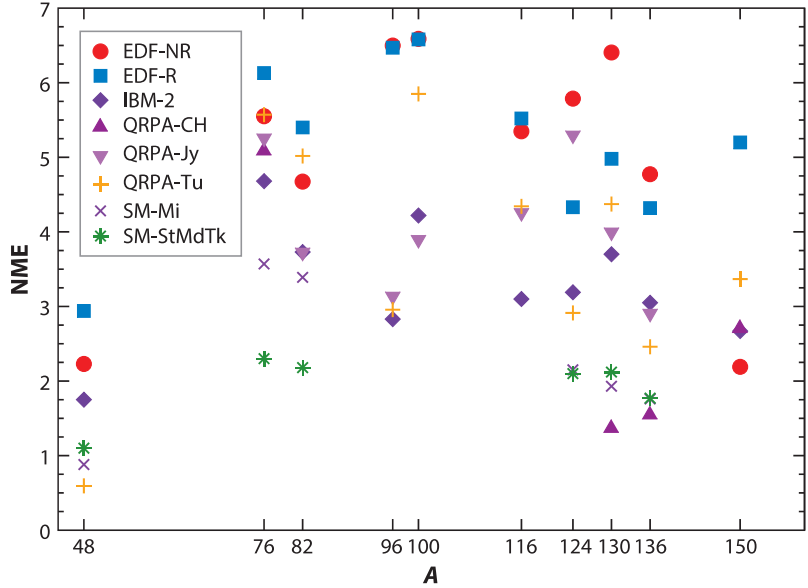}     
    \caption{Nuclear Matrix Elements (NMEs) calculated in different theoretical models for $0\nu\beta\beta$  isotopes under investigation by current experiments. For these calculations an unquenched $g_A = 1.27$ has been used. Figure reproduced from \cite{Dolinski:2019nrj}, original figure and references can be found in \cite{Engel:2016xgb}. Abbreviations: EDF -- energy-density functional; IBM -- interacting boson model; QRPA -- quasi-particle  random-phase approximation; SM -- Shell Model. }
    \label{fig:nme}
\end{figure}

Neutrinoless double beta decay can only occur in isotopes that  undergo Standard-Model allowed $2\nu\beta\beta$ decay. %The latter 
This is a second-order weak process that can only occur if single-$\beta$ decay is energetically forbidden or highly suppressed as in $^{48}$Ca. Thirty-five $\beta\beta$ decay isotopes have been identified \cite{Tretyak:2002dx}, of which $\approx 10$ are considered for $0\nu\beta\beta$ searches due to their availability and $\beta\beta$ endpoint energies  above 2 MeV. The latter is essential for a potential signal-of-interest to lie above most naturally occurring backgrounds. 

The arguably best-motivated scenario studied in the literature %preferred by the community 
is the light-Majorana exchange mechanism (see, e.g., \cite{Dolinski:2019nrj} and references therein). In this theoretical model the effective Majorana neutrino mass $m_{\beta\beta}$ can be extracted from the decay rate of the transition
\begin{equation}
    \Gamma^{0\nu} = g_A^4\ G_{0\nu}|M_{0\nu}|^2 \left( m_{\beta\beta} \right)^2,
     %\Gamma^{0\nu} = G_{0\nu}|M_{0\nu}|^2\langle m_{\beta\beta}\rangle^2,
\end{equation}
where $G_{0\nu}$ is the phase-space factor which is calculated using Dirac wave functions \cite{Kotila:2012zza,Stoica:2013lka} and $g_A$ is the axial-vector coupling for free nucleons. The crucial  nuclear matrix element (NME) is given by $M_{0\nu}$; it is calculated in various theoretical frameworks. 
Reviews of such frameworks and calculations of the associated NMEs are presented, e.g., in  \cite{Avignone:2007fu,Vergados:2012xy,Engel:2016xgb}. 
However, when comparing NMEs calculated in different frameworks their values vary by factors of a few depending on isotope and method. The situation is illustrated in Fig.\ \ref{fig:nme} and discussed in detail in, e.g., \cite{Engel:2016xgb}. The relatively large spread in values suggests that underlying models are incomplete and missing certain features. The discrepancy between different NMEs directly translates into a spread in limits on the effective neutrino mass extracted from experimental half-life limits making a direct comparison between experiments based on their reached mass limits problematic. Furthermore, following an observation of $0\nu\beta\beta$, the uncertainties in NMEs will limit the precision at which $m_{\beta\beta}$ can be known. However, with the increase in computational power and progress in nuclear theory, the situation is expected to improve. Especially recent progress in first-principle calculations in light- and medium-mass nuclei \cite{Gysbers:2019uyb} offers an interesting novel approach to calculating $0\nu\beta\beta$ decay NMEs for all isotopes of interest. Increasingly helpful are also the approaches undertaken by the lattice community \cite{Cirigliano:2020yhp}.

Another open question in nuclear theories describing $\beta\beta$ is whether the weak axial-vector coupling strength $g_A$ is quenched as has been observed in $\beta$ decays, or not. In the description of $\beta$ and $2\nu\beta\beta$ decays, $g_A$ is quenched by a factor $q$ to calculate an \textit{effective} value of $g_A$ as $g_A^{\text{eff}}=q\ g_A^{\text{free}}$ in order to reproduce experimental values, with $g_A^{\text{free}}=1.27$ being the free-nucleon axial-vector coupling measured in neutron beta decay. The quenching factor $q$ significantly impacts the reach of experimental $0\nu\beta\beta$ searches as $g_A$ enters \equ{mbetabeta} at the fourth power. The situation is discussed in detail in \cite{Ejiri:2019ezh} and references therein. 
There are arguments that the quenching factor in $0\nu\beta\beta$, which has a larger momentum scale as $\beta$ and $2\nu\beta\beta$ decays, is closer to $1$. 

\subsubsection{Experimental Aspects}
In direct $0\nu\beta\beta$ searches, the energy of the emitted electrons is measured in calorimeter-type detectors. No neutrinos are emitted in the process and all energy is carried away by the two electrons. A positive $0\nu\beta\beta$ signal will manifest itself as a peak in the energy spectrum at the $\beta\beta$ decay endpoint energy $Q_{\beta\beta}$.  The region of interest (ROI) where experiments search for an excess of events is chosen based on the detector energy resolution $\Delta E$ at $Q_{\beta\beta}$. 
The sensitivity of an experiment depends on the actual number of $0\nu\beta\beta$ events detected in this ROI. Background events in the ROI limit a detector sensitivity, which scales according to 
\begin{equation}
    T_{1/2}^{0\nu} \propto 
\begin{cases}
a\, M\, \epsilon\, t & \text{with no background, } \\
a\, \epsilon\, \sqrt{\frac{M\,t}{B\,\Delta E}} & \text{with background}.
\end{cases}
\label{eq:sensitivity}
\end{equation}
Here $M$ is the mass of source deployed, $a$ is the abundance of the $\beta\beta$ decaying isotope, $t$ is the measuring time, $\epsilon$ is the detection efficiency, and $B$ is the background index, which is typically quoted in events/keV/kg/yr in the ROI. The sensitivity in conventional counting experiments scales linearly with observation time $t$ in an experiment with no background in the ROI, while it improves with $\sqrt{t/B}$ in experiments with background events in the ROI. However, in large monolithic detectors the spectrum is extracted by a simultaneous fit of the data in energy, event multiplicity, and event location within the detector volume. This approach studies the signal as a function of the depth within the detector \cite{Albert:2017hjq}. As a result the background index is no longer a good measure of sensitivity in these detectors.

A worldwide search for $0\nu\beta\beta$ is ongoing in several isotopes. Groups are pursuing different detector technologies which are influenced by the isotope under investigation and its availability. Natural abundance and the ability to enrich an isotope are crucial factors in the selection of the target isotope, along with achievable background levels and  energy resolution. While there is no obvious isotope of choice to search for $0\nu\beta\beta$, experiments aim to maximize their discovery potential by optimizing the parameters in \eq~(\ref{eq:sensitivity}). 
Current experiments typically reach half-life sensitivities on the order of $10^{25}-10^{26}$ years depending on the isotope under investigation as well as the detection technology (see \Tab~\ref{tab:0nbbexperiments}). The most stringent limits on $T_{1/2}^{0\nu}$ are $1.8 \times 10^{26}$ years \cite{Agostini:2020xta} and $1.1 \times 10^{26}$ years \cite{KamLAND-Zen:2016pfg} in $^{76}$Ge and $^{136}$Xe, respectively, translating into limits on the effective Majorana neutrino mass of $0.07-0.16$ eV ($^{76}$Ge) and $0.06 - 0.17$ eV ($^{136}$Xe). 
The global limit as well as parameter space disfavored by $\beta\beta$ experiments is shown in Fig.\ \ref{fig:nmass}. In the parametrization of $m_{\beta\beta}$ versus lightest mass $m$, current experiments exclude the degenerate region. Planned upgrades to current experiments, such as SNO+ Phase I, SuperNEMO, LEGEND-200 and KamLAND-Zen800, aim to probe into the horizontal parameter band of the inverted ordering in Fig.\ \ref{fig:nmass}. However, in order to completely probe the parameter space allowed in the inverted ordering, next-generation detectors with sensitivities close to or exceeding $10^{28}$ years are required. Several experiments are being prepared in order to probe the parameter space allowed in the inverted ordering including  LEGEND-1000 ($^{76}$Ge), CUPID ($^{100}$Mo), AMoRE ($^{100}$Mo), SNO+ Phase II  ($^{130}$Te), JUNO ($^{136}$Xe), KamLAND2-Zen  ($^{136}$Xe), nEXO ($^{136}$Xe), NEXT HD  ($^{136}$Xe), and PANDAX 1k  ($^{136}$Xe). 

Future multi-ton dark matter direct detection experiments using xenon as target material automatically contain a considerable amount of $^{136}$Xe (natural isotopic abundance $\approx8.9\%$), which allows for  $0\nu\beta\beta$ searches with predicted sensitivities on the order of $10^{26}$ years (LUX \cite{Akerib:2019dgs} and PandaX \cite{Chen:2016qcd}) to $10^{27}$ years (DARWIN \cite{Agostini:2020adk}). This is one of the interesting connections of dark matter and neutrino experiments. However, it should be pointed out that the latter experiments are designed to maximize the sensitivity to dark matter interactions. These experiments focus on reducing backgrounds that would mimic dark matter interactions. Thus, the predicted $0\nu\beta\beta$ half life sensitivity is less than that of next-generation $\beta\beta$ decay experiments, e.g., DARWIN deploying 50 tons of natural xenon has a predicted sensitivity of $2.4\times10^{27}$ years \cite{Agostini:2020adk}, while the projected sensitivity of nEXO deploying 5 tons of xenon enriched in $^{136}$Xe is $9.2\times10^{27}$ years (90\% C.L.) \cite{Albert:2017hjq}. In addition, the timescale to deploy DARWIN is different from next-generation $\beta\beta$ decay experiments, which are anticipated to start construction within the next years. 
While the most sensitive next-generation experiments will probe the complete parameter space allowed in the inverted mass ordering, they will also explore a large fraction of the parameter space in a normal mass ordering scenario. In \cite{Agostini:2017jim} the discovery probabilities of several experiments are calculated for both mass orderings. The authors predict a discovery probability of more than $50-60$\% for normal ordering and searches in the isotopes $^{76}$Ge (LEGEND-1000), $^{130}$Te, and $^{136}$Xe (nEXO). The predicted discovery potential for inverted ordering is more than 80\% in the aforementioned isotopes.

\begin{table}[t]\centering
\begin{tabular}{cccccc}
    \hline\hline
    \textbf{} & \textbf{Exposure}& \textbf{Sensitivity} & $T_{1/2}^{0\nu}$ & $ m_{\beta\beta}$& \textbf{Experiment}\\
    & (kg $\times$ yr)&($\times 10^{25}$ yr)&($\times 10^{25}$ yr)&(eV)&\\
    \midrule
    $^{48}$Ca  &13.5&$1.8\times10^{-3}$  & $>5.8\times10^{-3}$ & $<3.5 - 22$ &  ELEGANT VI \cite{Umehara:2008zz}\\
    \addlinespace[0.5em]
    \multirow{3}{*}{$^{76}$Ge}  &127.2&18  &$>18$  &$<0.08 - 0.18$  & GERDA \cite{Agostini:2020xta}\\
    &\multirow{2}{*}{26.0}  &\multirow{2}{*}{4.8} &\multirow{2}{*}{$>2.7$} & \multirow{2}{*}{$<0.20 - 0.43$} & Majorana\\ &&&&&Demonstrator \cite{Alvis:2019sil} \\
    \addlinespace[0.5em]
    $^{82}$Se   & 5.29 &$5.0\times10^{-1}$& $>3.5\times10^{-1}$ & $<0.31 - 0.64$ & CUPID-0 \cite{Azzolini:2019tta}\\
    \addlinespace[0.5em]
    $^{96}$Zr   & $(-)$ &$(-)$& $>9.2\times 10^{-4}$ & $< 7.2 - 19.5$ & NEMO-3 \cite{Argyriades:2009ph}\\
    \addlinespace[0.5em]
    $^{100}$Mo   & 1.17 &$(-)$& $>1.5\times10^{-1}$ & $<0.31 - 0.54$ & CUPID-Mo \cite{Armengaud:2020luj}\\
    \addlinespace[0.5em]
    $^{116}$Cd   &$(-)$  &$(-)$& $>2.2 \times 10^{-2}$ & $<1.0 - 1.7$ & Aurora \cite{Barabash:2018yjq}\\
    \addlinespace[0.5em]
    $^{128}$Te   & $(-)$ &$(-)$&$>1.1\times10^{-2}$  & $(-)$ &Arnaboldi et al. \cite{Arnaboldi:2002te} \\
    \addlinespace[0.5em]
    $^{130}$Te   &  1038.4$^{\diamond}$  &2.8&$>2.2$  & $<0.09 - 0.31$ & CUORE \cite{Adams:2021rbc} \\
    \addlinespace[0.5em]
    \multirow{2}{*}{$^{136}$Xe}   &504$^{\dagger}$& 5.6 & $>10.7$ & $<0.06 - 0.17$  & KamLAND-Zen \cite{KamLAND-Zen:2016pfg} \\  
      & 234.1  &5.0& $>3.5$ & $<0.09  - 0.29$ & EXO-200 \cite{Anton:2019wmi}\\ 
    \addlinespace[0.5em]
    $^{150}$Nd   &0.19 & $(-)$& $>2.0\times10^{-3}$ & $<1.6 - 5.3$ & NEMO-3 \cite{Arnold:2016qyg} \\
    %\multicolumn{2}{c}{Multi-column}\\
    %X&X\\
    \midrule
    $^3$H&\multicolumn{3}{l}{$\beta$-endpoint measurement }&$m_{\beta}<0.8$&KATRIN \cite{Aker:2021gma}\\
    \hline\hline
\end{tabular}
\caption{Comparison of current experimental limits of $0\nu\beta\beta$ searches in different isotopes. Limits are given at the 90\% C.L. Fields with $(-)$ indicate where values are not provided in the listed reference. $^{\diamond}$ is the $^{\textnormal{nat}}$TeO$_2$ exposure and $^{\dagger}$ lists the complete xenon dissolved in scintillator. For comparison, the limit from the direct measurement of $m_{\beta}$ is shown as well. \label{tab:0nbbexperiments}}
\end{table}

Neutrinoless double beta decay experiments are inherently difficult. Backgrounds must  be reduced as much as possible and irreducible backgrounds must be thoroughly studied and understood. While the sensitivity of an experiment improves with increased energy resolution, energy alone will not be sufficient in next-generation experiments to claim a discovery. It will be crucial to also demonstrate that the observed signal at $Q_{\beta\beta}$ is inconsistent with observed backgrounds. At least two experiments are required to unambiguously demonstrate observation of $0\nu\beta\beta$ and several technological developments should be pursued in parallel by the global community. While current and next-generation experiments are focusing on maximizing their discovery potential,  tracking calorimeters are required to study the underlying physical principles driving $0\nu\beta\beta$ once an observation has been made. 

The purity and low background rates of double beta decay experiments also allow to search for a variety of exotic processes, including axions, Majorons, low mass dark matter, exotic nuclear decay or electron decay \cite{Majorana:2016hop}. Also precision studies of $2\nu\beta\beta$ are sensitive to new physics \cite{Deppisch:2020mxv,Deppisch:2020sqh,Agostini:2020cpz}.

The projected final sensitivity of next generation experiments is $10^{27} - 10^{28}$ years. Depending on isotope and selected NME this sensitivity calculates to effective Majorana neutrino masses ranging from $m_{\beta\beta} \approx 5 - 20$ meV. Experiments will reach their final sensitivity in the second half of the decade $2030 - 2040$. However, despite the discovery potential of next generation experiments, even larger multi-tonnes detectors and technological breakthroughs (see Sec.\ \ref{wg5_tech}) may be required to observe $0\nu\beta\beta$ in less favorable scenarios. These experiments should aim for sensitivities of $m_{\beta\beta} \approx 1$ meV, as discussed, e.g., in \cite{Agostini:2020oiv,Cao:2019hli}.

\subsection{Cosmology}

Cosmology provides one of the most promising avenues to constrain the sum of the neutrino masses in the next decade, thanks to the impact of neutrinos  on cosmological observables.  Massive neutrinos alter the expansion history of the Universe in a peculiar way.   At early times, they are relativistic and  contribute   as a radiation term. They later become non-relativistic, at a redshift that depends on their mass, and they  contribute to the Universe expansion as a hot dark matter component. The most notable feature of massive neutrinos, however, is the imprint they leave on the evolution of matter perturbation at late times. On scales smaller than their free-streaming length, a size comparable to the particle horizon at the epoch when the neutrinos become non-relativistic, the large thermal velocities of neutrinos prevent them from falling into the gravitational potential wells of overdensities.
The growth of matter perturbations is therefore suppressed and  matter clustering  exhibits a step-like power suppression  proportionally to the sum of the neutrino masses. On larger scales, in contrast, neutrinos cluster just like cold dark matter  would. This distinctive imprint allows cosmological probes to constrain the sum of the neutrino masses, in particular when combining information from large and small scales. Hence, cosmic microwave background (CMB) data used jointly with large-scale structure (LSS) data offer a unique approach on the sum of the neutrino masses. 
Current cosmological observations already provide the tightest bounds on $\sum m_\nu$ and next-generation surveys are aiming at moving from upper bounds to the first clear measurement of $\sum m_\nu$. 

As the upper bounds are now approaching the 0.1 eV mass range, thus strongly disfavoring quasi-degenerate neutrinos \cite{Lattanzi:2020iik}, it could seem relevant to account for  neutrino mass hierarchies allowed by neutrino oscillation results~\cite{deSalas:2020pgw}. As mentioned in Sec.\ \ref{sec:mass_intro}, three different mass hierarchy scenarios can be considered: the normal hierarchy, the inverted hierarchy, and the degenerate case where the three neutrino species equally share the total mass $\sum m_\nu$.
It has been shown that the amplitude of the matter power spectrum  shifts by 0.2\% at most between these three cases~\cite{Lesgourgues:2006nd}, below the level of sensitivity of current and next-generation experiments. Massive neutrinos have a larger impact on smaller scales. These are probed with Lyman-$\alpha$ forest flux power-spectrum measurements, where the various mass-hierarchy scenarios produce even smaller differences~\cite{Palanque-Delabrouille:2015pga}. Neutrino hierarchy can therefore be neglected in cosmological studies, and assuming that cosmology only constraints $\sum m_\nu$ is a good approximation.

\subsubsection{Main Experimental Approaches to Constrain \texorpdfstring{$\sum m_\nu$}{MNU}}\label{sec:neff}

Promising experimental approaches to a detection of $\sum m_\nu$ in the next decade are summarized below (from~\cite{Dvorkin:2019jgs} and~\cite{Brinckmann:2018owf}). Most current bounds are obtained by allowing for massive neutrinos on top of the six-parameter  $\Lambda$CDM model parameterized by the  densities in baryons $\omega_b$ and in cold dark matter $\omega_c$,  the angular scale of the sound horizon at electron-proton recombination $\theta_s$, the  amplitude $A_s$ and scalar index $n_s$ of primordial fluctuations, and the redshift of reionization $z_{\rm reio}$. We discuss more general models in Sec.\ \ref{sec:cosmo_degen}.

{\bf CMB alone:} CMB alone is not an ideal probe of neutrino masses since all three neutrino flavors are still relativistic at the time of neutrino decoupling. Including polarization information, notably on small scales, in addition to temperature helps in tightening the constraints. Current 95\% C.L.\ upper bounds on $\sum m_{\nu}$ are in the $0.26-0.54$~eV range depending on the actual choice of CMB data set~\cite{Aghanim:2018eyx}.

{\bf CMB lensing:} CMB photons are deflected by intervening structures. The deflection angle is proportional to the integrated distribution of matter along the line of sight. CMB lensing is thus probing scales in the quasi-linear regime, where structure growth is still linear and  easier to model than for smaller scales where one needs to resort to higher-order perturbation theory or numerical simulations. Crucial information  lies in large angular scale $E$-mode polarization data, which today  constraint the sum of the neutrino masses at the level of  $\sum m_\nu<0.24\,{\rm eV}$~\cite{Aghanim:2018eyx}.

{\bf Galaxy clustering:} Galaxies reside in massive halos and can  therefore probe the scale-dependent impact of neutrinos on structure formation~\cite{Alam:2016hwk}. Galaxy surveys measure clustering  at smaller redshifts than achievable with CMB lensing, making these two approaches  highly complementary. Combined, CMB lensing and galaxy clustering tighten the limit to $\sum m_\nu<0.12\,{\rm eV}$~\cite{Aghanim:2018eyx}. A complementary approach is to consider the combination of galaxy clustering, type IA supernaovae and CMB temperature and polarization data, which pushes the limit to $\sum m_\nu<0.09\,{\rm eV}$~\cite{DiValentino:2021hoh}.
To fully benefit from next-generation LSS spectroscopic surveys, however, we must improve our modelling of the effect of neutrinos on non-linear scales with dedicated $N$-body simulations. 

{\bf Cluster counts:} Massive neutrinos impact the abundance and properties of  dark matter halos, well correlated with massive clusters. Cluster counts as a function of mass and redshift can therefore  constrain $\sum m_\nu$~\cite{Ade:2015fva}. The main drawback of this approach is due to the complex nature of clusters. The uncertain halo mass function  prevents this method from being competitive today. If systematics can be reduced, however, it provides an independent handle on neutrino mass.

{\bf Lyman-$\alpha$ forest:} The absorption of the light from background quasars by intervening hydrogen produces a characteristic absorption feature in quasar spectra, dubbed Lyman-$\alpha$ forest. It carries unique information on small scales, where the matter power spectrum is maximally suppressed by massive neutrinos. Combined with CMB temperature and polarization data, this method currently provides one of the most stringent upper bounds, $\sum m_\nu<0.11\,{\rm eV}$~\cite{Palanque-Delabrouille:2019iyz}, although tightening it further with similar data  is a challenge because of the extensive hydrodynamical simulations that are required for a proper modeling of the intergalactic medium. The tightest limit to date, $\sum m_\nu<0.09\,{\rm eV}$, is obtained when considering in addition CMB lensing and galaxy clustering data~\cite{Palanque-Delabrouille:2019iyz}.

\subsubsection{Parameter Degeneracies} \label{sec:cosmo_degen}

The sum of the neutrino masses has non-trivial correlations with several of the parameters describing the cosmological model, even in the simple $\Lambda$CDM scenario. The main correlation is between $\sum m_\nu$ and the optical depth of reionization $\tau$ (or equivalently the redshift of reionization, assuming quasi-instantaneous reionization). It affects all results derived from a combination of CMB and LSS data, since the constraint relies upon a measurement of the primordial fluctuation amplitude $A_s$ in order to infer the power  suppression induced by massive neutrinos. Because CMB surveys constrain $A_s e^{-2\tau}$,  the improved determination of $\tau$ that next-generation surveys will provide, thanks to a better measurement of  large-scale $E$-modes,  is crucial. Given current sensitivity on the optical depth $\sigma(\tau)= 0.007$ from the Planck CMB data~\cite{Aghanim:2018eyx}, an optimal combination of next-generation CMB and LSS measurements can potentially reach a $4\sigma$ detection of neutrino mass, assuming minimal normal mass ordering, with $\sigma(\sum m_\nu)\sim15$~meV. 

Another source of uncertainty is the degeneracy between $\sum m_\nu$ and parameters that govern the evolution of the Universe. The best example is  the dark energy equation-of-state parameter $w$. Except for early dark-energy models, most extensions to  $\Lambda$CDM  can be parameterized as $w=w_0+(1-a)w_a$, where $a$ is the scale factor, related to redshift  by $a = 1/(1+z)$. Because most of the effect of dark energy occurs at small redshift, LSS data and tomographic measurements, both relevant in the $0<z<3$ redshift range in particular, will help in disentangling the effects of $\sum m_\nu$ and $w(z)$. Varying dark energy is expected to loosen the determination of $\sum m_\nu$ by a factor about 1.5 to 2.0. For exotic modifications of cosmology or general relativity it is often not known yet how the neutrino mass limits would change.

The degeneracy between $\sum m_\nu$ and the effective number of relativistic species $N_{\rm eff}$ has long been a source of concern. With the advent of the Planck survey, however, this issue is now solved~\cite{Aghanim:2018eyx}. With Planck data alone, $N_{\rm eff}$ is constrained to $N_{\rm eff}=2.92\pm 0.37$, in agreement with the standard model prediction $N_{\rm eff}=3.044$ \cite{Froustey:2020mcq,Bennett:2020zkv}. Including lensing and BAO measurements only slightly modifies the limit to $N_{\rm eff}=2.99\pm 0.34$.  Allowing for $N_{\rm eff}$ to float at the same time as the sum of the neutrino masses does not alter the determination of $\sum m_\nu$ by more than a few percent, and the combined result is very close to the one on either $N_{\rm eff}$ or $\sum m_\nu$ alone. Big Bang Nucleosynthesis (BBN), the generation of light elements in the early Universe, is also sensitive to the number of relativistic species \cite{Pitrou:2018cgg}. 
Current constraints from are in the $N_{\rm eff} = 3-4$ range, and thus consistent with values obtained from late-Universe observables. 

We should note that besides degeneracies, also new physics related in particular to neutrinos can modify limits on their mass. Examples are neutrino decay~\cite{Beacom:2004yd,Escudero:2020ped} or exotic scenarios in which neutrino mass varies with time \cite{Dvali:2016uhn}.

\subsubsection{Upcoming and Proposed Experiments }

The next decade is rich in surveys aiming at unveiling the nature of dark energy. These same projects will provide the ingredients to obtain a detection of $\sum m_\nu$ at a 3 to 4$\sigma$ level. 

On the LSS side, we highlight three main surveys. The Dark Energy Spectroscopic Instrument (DESI) will measure BAO over at sub-percent-level precision out to $z=1.85$, providing unique data on galaxy clustering at low redshift~\cite{Levi:2019ggs, Aghamousa:2016zmz}. DESI began operations in December 2020 and is expected to run its main survey from mid 2021 to mid 2026. It is the first survey running that should reach this precision. The Euclid satellite, with a launch scheduled for 2022,  is designed to provide 1\% accuracy on galaxy clustering and weak shear observables. 

On the CMB side, the  LiteBIRD experiment, currently in phase A, will provide  large-scale polarization information~\cite{Matsumura:2013aja}. The ground-based CMB-S4 project, with a smaller sky coverage but a much better resolution than LiteBIRD, should reach an unprecedented sensitivity to CMB lensing~\cite{Abazajian:2019eic}. Other projects, more ambitious but less advanced in the approval process, are already under discussion. This is the case for projects like the CORE-M5 satellite project that would reach both goals (polarisation and lensing) at once~\cite{Delabrouille:2017rct}. None of those projects has been  approved  yet.

\subsection{Theoretical Interpretation and Complementarity of Approaches}
\label{sec:mass_compl}
\begin{figure}[tp]
    \centering
    \includegraphics[width=0.32\textwidth]{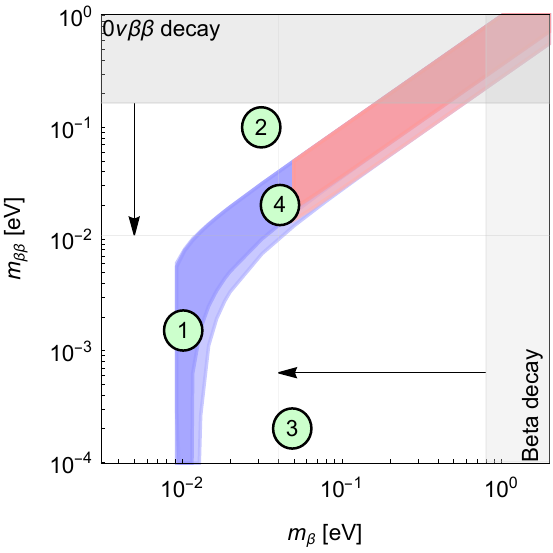}    
    \includegraphics[width=0.33\textwidth]{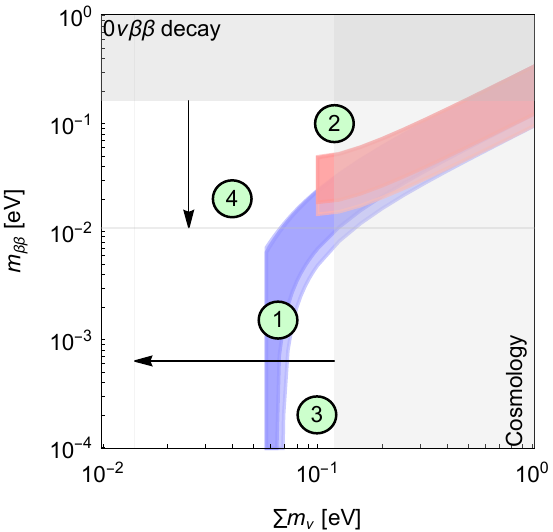}
    \includegraphics[width=0.32\textwidth]{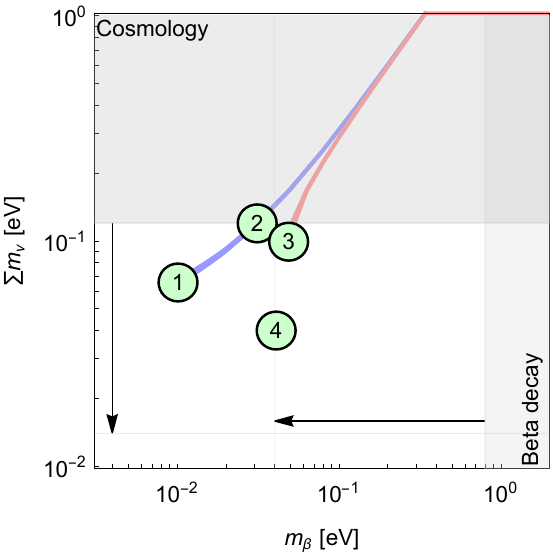}
    \caption{Correlations between experimentally observable combinations of neutrino mass in $0\nu\beta\beta$ $|m_{ee}|$, beta decay $m_\beta$, and cosmology $\sum m_\nu$. Dark--colored shadings refer to the uncertainties of the Majorana phases only, whereas light colored shadings include the current oscillation parameter uncertainties; see \figu{nmass} for references and descriptions of bounds. Future projected exemplary bounds are indicated with arrows and taken from nEXO~\cite{Albert:2017hjq} (90\% C.L., one middle case of nuclear matrix element, 10 years), Project~8~\cite{Esfahani:2017dmu} (90\% C.L., phase IV), and an optimal combination of next-generation CMB and LSS surveys~\cite{Dvorkin:2019jgs} (95\% C.L.) for $0\nu\beta\beta$, beta decay, and cosmology, respectively. The different scenarios corresponding to the marked disks are discussed in the main text.}
    \label{fig:masscorr}
\end{figure}

A direct comparison of the experimental observables is shown in \figu{masscorr} together with the corresponding bounds on the respective quantities (grey-shaded regions). If a signal is found, it will lead to a fit region in this parameter space which can be directly inferred from the observables. In the standard picture (effective light Majorana neutrino masses and self-consistent measurements) the fit has to lie  within the colored regions; see discussion below. Currently the strongest bound on neutrino mass comes from cosmology, with a sensitivity close to be able to rule out the inverted mass ordering. Here we highlight the complementarity of the different approaches, especially pointing out that the cosmological measurements need to be confirmed by direct tests of neutrino mass.

\begin{figure}[tp]
    \centering
    \includegraphics[width=0.6\textwidth]{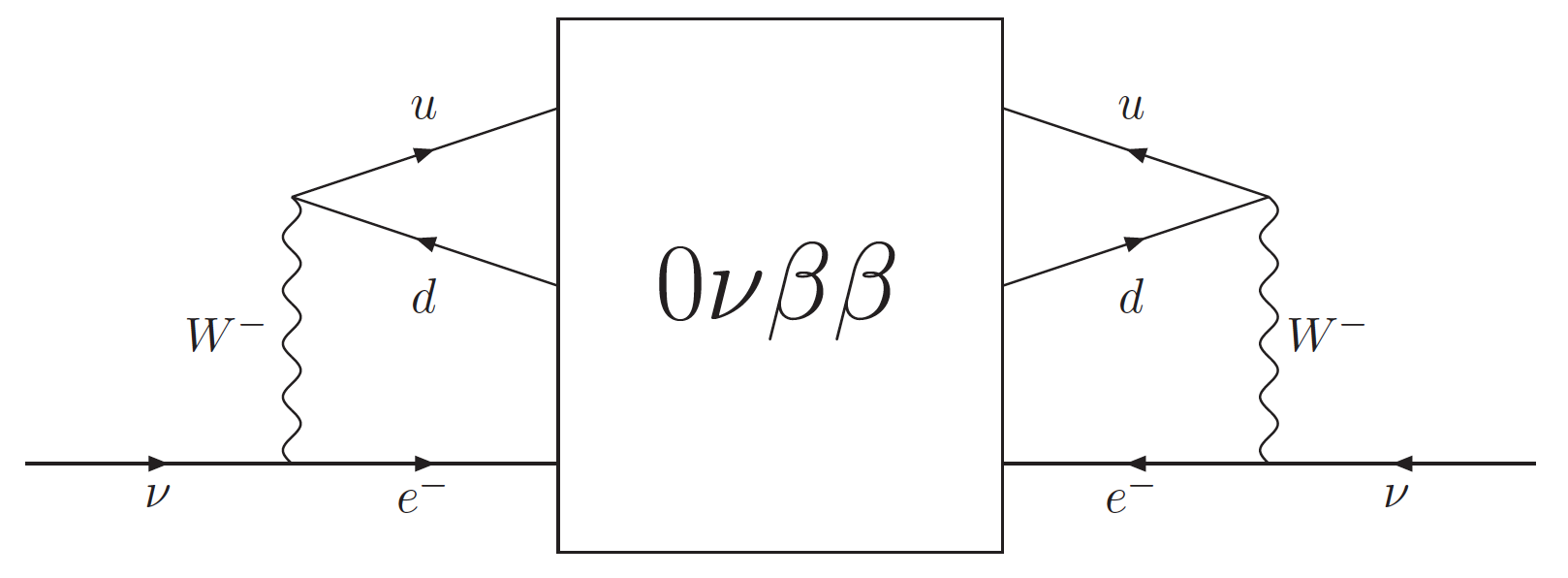} \hspace{0.05\textwidth} \includegraphics[width=0.28\textwidth]{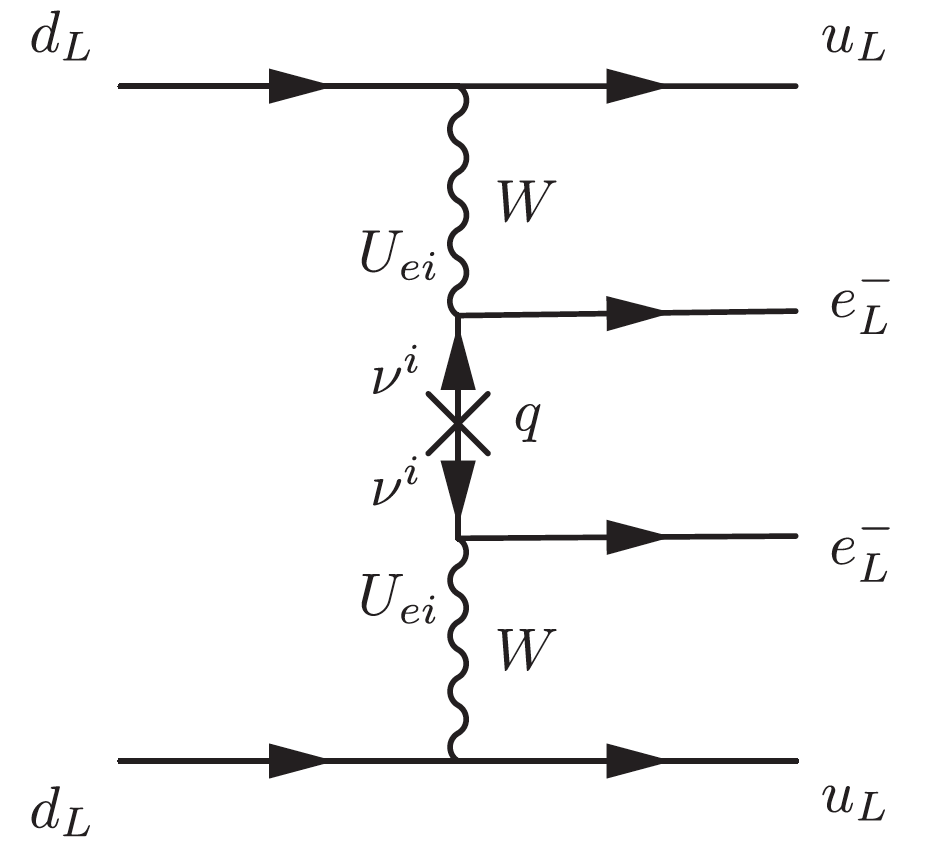}      
    \caption{Left panel: Black box theorem, illustrated. The existence of a (lepton number) effective operator leading to $0\nu\beta\beta$  implies loop-generated Majorana neutrino masses from this diagram.  Right panel: the  standard mass mechanism for Majorana neutrino masses leads to $0\nu\beta\beta$ decay; it  is one possible decomposition of the effective operator in \equ{effoperatorbetabeta}. Figures taken from \Refs~\cite{Bonnet:2012kh,Rodejohann:2011mu}.}
    \label{fig:blackbox}
\end{figure}

As far as neutrinoless double beta decay is concerned, its measurement is mainly driven by the question about the nature of neutrino mass and whether lepton number is violated, see  \Sec~\ref{sec:D_or_M} for a detailed discussion. In this respect, the interpretation of the decay in terms of neutrino mass is less straightforward than suggested so far. The observation of $0\nu\beta\beta$ decay
can be interpreted in terms of an effective lepton-number-violating operator of dimension nine
\begin{equation}
  \mathcal{O} \propto \bar{u} \bar{u} \, d d \, \bar{e} \bar{e} \, ,
    \label{equ:effoperatorbetabeta}
\end{equation}
whose completion (fundamental theory) at high energies is a priori unknown.

It has been demonstrated that  this operator always implies  Majorana neutrino masses (``black box theorem'' \cite{Schechter:1981bd}), generated by the diagram shown in the left panel of \figu{blackbox} by inserting \equ{effoperatorbetabeta} into the box. However, as discussed in  Ref.~\cite{Duerr:2011zd}, this four-loop black-box diagram itself generates radiatively only mass terms which are many orders of magnitude too small to explain neutrino masses, which means that translating  an observed rate of neutrinoless double beta decay into neutrino mass would then be potentially misleading. 

On the other hand, if the neutrinos are Majorana particles, the diagram in the right panel of  \figu{blackbox} will inevitably lead to  $0\nu\beta\beta$ decay (if there are no cancellations from the Majorana phases). The standard interpretations of $0\nu\beta\beta$ decay in terms of neutrino mass typically rely on the  assumption of its {\em exclusiveness} driving it, meaning that this diagram is the leading contribution to \equ{effoperatorbetabeta}.

So what other mechanisms could lead to $0\nu\beta\beta$ decay, i.e., the operator in \equ{effoperatorbetabeta}? For example, a systematic analysis has been performed in Ref.~\cite{Bonnet:2012kh}, where all tree-level decompositions (fundamental theories) leading to \equ{effoperatorbetabeta} have been identified\footnote{Note that BSM loop contributions may also have a larger contribution to $0\nu\beta\beta$ decay than \figu{blackbox}, left panel.}. While more than ten options have been discussed in well-motivated models in the literature before (such as $R$-parity violating SUSY, left-right-symmetric, and leptoquark models), even more possibilities exist; some of these may be tested at the LHC~\cite{Helo:2013dla}. 
The fact that lepton-number-violating TeV-scale physics is currently testable can be understood by the simple estimate of the $0\nu\beta\beta$ amplitude being $\propto G_F^2 m_{\beta\beta}/q^2$, with $q^2 \simeq (100 \, \,\rm MeV)^2 $ the energy scale of the process (virtual neutrino momentum). For heavy particle exchange at energy scales $\Lambda$ much higher, the amplitude of the $d=9$ operator-induced transition must be proportional to $G_F^2 v^4/\Lambda^5$. This  numerically corresponds to the $0\nu\beta\beta$ amplitude order of magnitude-wise for $\Lambda = {\cal O}(\rm TeV)$, which is in the range currently getting testable by the LHC. Note that realistic extensions of the Standard Model typically have several possible diagrams for $0\nu\beta\beta$, left-right symmetric theories being one example \cite{Rodejohann:2011mu}. In this sense, searches for $0\nu\beta\beta $ provide constraints on a large number of models and parameters. 
Moreover, TeV-scale lepton number violation has dramatic consequences for leptogenesis, as any lepton asymmetry generated at high energies is washed out by such interactions  in the early Universe \cite{Deppisch:2013jxa,Deppisch:2012nb}.

All possibilities (including the standard mass mechanism) leading to $0\nu\beta\beta$ decay have in common that they require new fields and lepton number violation, i.e., physics beyond the Standard Model. As a consequence, while the  discovery of $0\nu\beta\beta$ decay will imply the discovery of physics beyond the Standard Model, the interpretation in terms of neutrino mass is only one (arguably the best-motivated) possibility. Only the combination with other direct and indirect mass measurements will reveal a self-consistent picture, which means that the detected neutrino mass is compatible with the colored regions in \figu{masscorr} to establish credibility in the Majorana mass mechanism.

We highlight the importance of different complementary approaches to measure neutrino mass in \figu{masscorr}, where also selected experimental bounds of future potential experiments are shown for illustration (arrows), and we assume for this discussion that the discovery reach is similar. In all cases, in principle, the inverted mass ordering can be excluded with a high enough sensitivity in the future. Several exemplary scenarios are depicted, see marked disks in the different panels:
\begin{enumerate}
    \item {\bf Majorana neutrinos with normal hierarchy and small $\boldsymbol{0\nu\beta\beta}$ mass.} Neutrino mass is probably found by cosmology; the  $0\nu\beta\beta$ lifetime is too long to be measured because of cancellations from the Majorana phases.
    \item {\bf Non-standard $\boldsymbol{0\nu\beta\beta}$ signal, normal ordering.} Will lead to an inconsistency between cosmological neutrino mass measurement and $0\nu\beta\beta$ (middle panel), whereas the scenario is consistent with the colored region in the right panel (beta decay-cosmology). Points towards discovery of new physics driving $0\nu\beta\beta$.
    \item {\bf Dirac neutrino masses with inverted hierarchy.} No signal in $0\nu\beta\beta$, consistent measurement between cosmology and beta decay (right panel). Here the Dirac nature of neutrino mass can be inferred, as Majorana neutrinos would be seen in $0\nu\beta\beta$.
    \item {\bf Majorana neutrinos, unknown systematics in cosmology, normal ordering.} Here the scenario leads to a consistent result between $0\nu\beta\beta$ and beta decay (left panel), whereas the cosmology result does not match (other panels).  This would point to non-standard cosmology beyond $\Lambda$CDM.
\end{enumerate}
From these examples it is clear that different complementary techniques are needed to probe the absolute mass scale of the neutrinos and the nature of neutrino mass, as the most important discoveries will be made by inconsistencies in these measurements. In most of the above chosen examples (except for example 1) two of the techniques produce a consistent result, whereas one measurement is in contradiction. Without a ``tie-breaker'', it will not be possible to identify the origin of such an inconsistency. While cosmological tests of neutrino mass appear to have extremely good sensitivity, the direct test of neutrino mass and 0$\nu\beta\beta$ decay provide a straightforward path to neutrino mass and its nature. Finally, note that the case of one massless neutrino can be falsified by future experiments.  

In summary, the determination of neutrino mass from neutrinoless double beta decay relies on the assumption of Majorana neutrinos and the dominance of the diagram in which those are mediating the decay. The neutrino mass from cosmology depends on the validity of the underlying model. Direct searches are the most model-independent way to determine neutrino mass. All methods need to be pursued. Consistency of different measurements would be a spectacular confirmation of the standard neutrino paradigm, whereas inconsistencies may  have dramatic consequences for particle physics or cosmology.

\subsection{Neutrino Mass: Summary}

Tests of the absolute neutrino mass scale and the nature of neutrino mass are currently being pursued by three avenues: kinematic measurements, 
neutrinoless double beta decay experiments, and cosmological measurements. These approaches are complementary in the sense of testing different  combinations of neutrino mass eigenstates depending on the parameter space they can access. Kinematic measurements are the most direct test of neutrino mass regardless of its nature; however, testing very small mass scales is challenging and requires detectors with sub-eV energy resolution. Neutrinoless double beta decay experiments are searching for lepton-number violation in weak interactions to determine whether neutrinos are so-called Majorana particles, i.e., their own antiparticles. An observation of this decay 
may test the absolute scale of the neutrino masses and the nature of the neutrino mass term. Nuclear matrix element uncertainties affect the translation of lifetime into neutrino mass, which implies that measurements with different isotopes are required. Cosmological tests of neutrino mass currently provide the most stringent bounds on neutrino mass; the extraction depends on astrophysical and cosmological models and is sensitive to different types of systematics and parameter degeneracies. Establishing a self-consistent picture of neutrino mass will therefore require all three techniques. 

Significant progress has been achieved in improving current experiments and the development of next generation experiments. Kinematic measurements are at the verge of pushing the sensitivity below 1 eV with the goal of achieving sub-100-meV sensitivities. Current neutrinoless double beta decay experiments reach halflife limits of $10^{25}-10^{26}$ years which convert to limits on the effective Majorana neutrino mass on the order of ~ 100s meV, depending on isotope and nuclear matrix element. Next generation experiments are being developed with the goal of reaching sensitivities of $10^{28}$ years or down to the few-meV level in mass space. Neutrinoless double beta decay and kinematic measurements are extremely challenging experiments. However, the community has been developing technological solutions to push sensitivities beyond the 100-meV level. Over the next $10-15$ years experiments will probe the complete parameter space of the inverted mass ordering and a large fraction of the parameter space in a normal ordering scenario. On a similar time scale, limits from cosmology are expected to improve with the availability of larger observation times and new telescopes coming online, so that they are expected to reach sensitivities in which a measurement is guaranteed, if our standard theories on neutrinos and cosmology are correct. 

\label{chap:mass}

\clearpage

\section{Neutrino Interactions}

\subsection{Introduction}
%\subsection{Neutrino interactions}
\newcommand{\hms}[1]{{\color{blue} #1 \color{black}}}
\newcommand{\gt}{\rightarrow}
\newcommand{\nubar}{\overline{\nu}}

\begin{figure}[h]
\begin{center}
  \includegraphics[height=8 cm] {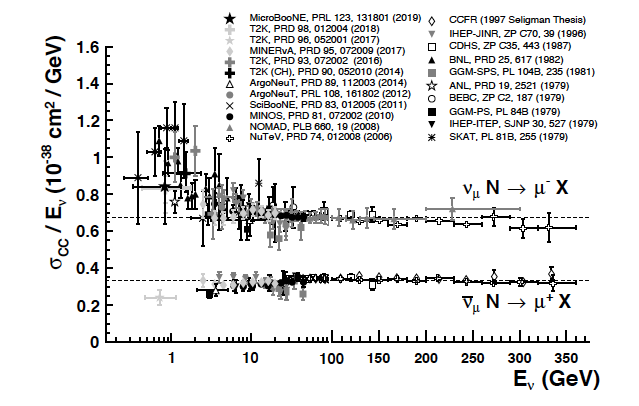}
\end{center}
\caption{Neutrino inclusive cross sections scaled by neutrino energy from \cite{Zyla:2020zbs}.  Quasi-elastic processes dominate at very low energies while partonic processes take over at higher energies. 
}\label{fig:WG3_pdgcross}
\end{figure}

Neutrino interactions span a very wide energy range, from elastic scattering of very low energy neutrinos off electrons, nucleons, and nuclei, which can become a significant background for dark matter searches, to ultra-high energy neutrinos which can scatter off  cosmic microwave background neutrinos. It is a very rich field of study and there are several distinct kinematic regions, defined by energy thresholds and resolving power. Based on this, the various processes are classified as:

\begin{itemize}
\item{ \bf Coherent elastic neutrino-nucleus scattering:} This process was recently observed experimentally and can occur at any energy but dominates at lower energies. Its hallmark is the detection of a very low energy nuclear recoil. See Section \ref{sec:WG3_coherent}. 
\item{\bf Elastic scattering from atomic electrons:} Such processes have a much smaller but more well-known cross section than that for scattering off nucleons or nuclei,  hence are commonly used as a standard candle for neutrino flux determinations and/or BSM searches. See Section \ref{sec:WG3_electrons}.
\item{\bf Scattering from individual nucleons:} Neutrinos can also scatter  off individual nucleons, either elastically, quasi-elastically, or leading to the production of a resonant state. There are multiple processes possible with numerous final state topologies.  See Section \ref{sec:WG3_nucleons}.
\item{\bf Scattering from partons:} Such deep inelastic scattering (DIS) processes begin to dominate when the  momentum transfer  $Q$ reaches the strong interaction scale $\lambda  \simeq 300$ MeV.  
\end{itemize}
These neutrino cross sections generally scale with lab energy as:
\begin{eqnarray}
\sigma(E_\nu) \propto \frac{s_{CM}}{(M_{W/Z}^2  \pm Q^2)^2} , 
\end{eqnarray}
where $s_{CM}$ is the center-of-mass energy squared ($\sim 2m E_\nu$ if the target mass is neglected,  where $m$ the mass of the target, $E_\nu$ is the neutrino energy), $M_{W/Z}$ is the mass of the exchanged boson, and $Q^2$ is the momentum-transfer squared. The $\pm$ corresponds to $t$-channel scattering and $s$-channel annihilation.  For $|Q|\ll  M_W$, the cross section grows with energy but at very high energies it falls as $|Q^2|$ grows. Fig.~\ref{fig:WG3_pdgcross} illustrates the charged current (CC) inclusive neutrino cross section ($\sigma_{\rm CC}/E_\nu$) as a function of neutrino energy $E_\nu$~\cite{Zyla:2020zbs}. It should be noted that at the very highest energies (not shown), neutrino absorption by the cosmic neutrino background via %the Glashow resonance \cite{PhysRev.118.316} $\nubar_e + e^- \gt W^-$ or through 
the process $\nu + \nubar \gt Z $ is possible. See \cite{Barenboim:2004di} for a discussion of the relevant energy scale $\sim 10^{12}$ GeV (depending on the neutrino masses), where these effects become important.

\subsubsection{Electron, Muon and  Tau Neutrinos}\label{sec:WG3_threshold}

Electron-type neutrinos are mainly produced in nuclear beta decay in reactors, in the Earth’s core, and  in fusion reactions in the Sun, while most accelerator neutrino beams originate from light meson decays and are produced as muon neutrinos. Neutrino mixing effects lead to transformations between species but our ability to tag flavor depends on the presence of a charged-current neutrino interaction where the charged lepton in the final state can be identified.  This leads to limitations on detectability due to leptonic mass thresholds.  For the simplest charged current scattering process $\nu_\ell + n \rightarrow \ell^- + p$, the requirement that the center-of-mass energy squared accommodate the final state particles, 
\begin{eqnarray}
s &=& (m_\ell + m_p)^2 \\
E_\nu &\ge& \frac{(m_p^2 -m_n^2) + m_\ell^2 + 2 m_p m_\ell}{{2 m_n}} , 
\end{eqnarray}
leads to an energy threshold of $\sim 100$ MeV  for muon-neutrino interactions and $\sim 3.5 $ GeV for tau neutrinos. Due to these thresholds, only electron-neutrino appearance can be detected in solar and reactor experiments. Appropriate oscillation lengths for detecting the transformations of muon and tau neutrinos are therefore higher, with $L \ge 50-100$ km necessary to reach an oscillation minimum.  Despite the difficulties imposed by the high energy threshold, DONUT was the first experiment to observe tau neutrinos \cite{Kodama:2007aa} and more recently, the OPERA collaboration~\cite{PhysRevLett.120.211801} reported their final results on the appearance of 10 tau neutrinos in a muon-neutrino beam with energies ranging from 5 to 30 GeV in 2018. Very high-energy astrophysical tau-neutrino interactions are detectable in the IceCube experiment \cite{Aartsen:2019tjl} as extended cascade from the neutrino scatter and subsequent decay of the tau lepton. Tau-neutrino interactions and oscillations remain,  however, a largely unexplored topic for future experiments.

We begin our discussion of neutrino   cross sections with the cross sections for scattering off free electrons (Sec.\ \ref {sec:WG3_electrons}), then free nucleons  (Sec.\ \ref{sec:WG3_nucleons}), followed by the complications that arise due to nuclear effects (Sec.\ \ref{sec:WG3_nuclei}).

\subsection{Scattering from Atomic Electrons}
\label{sec:WG3_electrons}

\begin{figure}[t]
\begin{center}
  \includegraphics[height=9 cm] {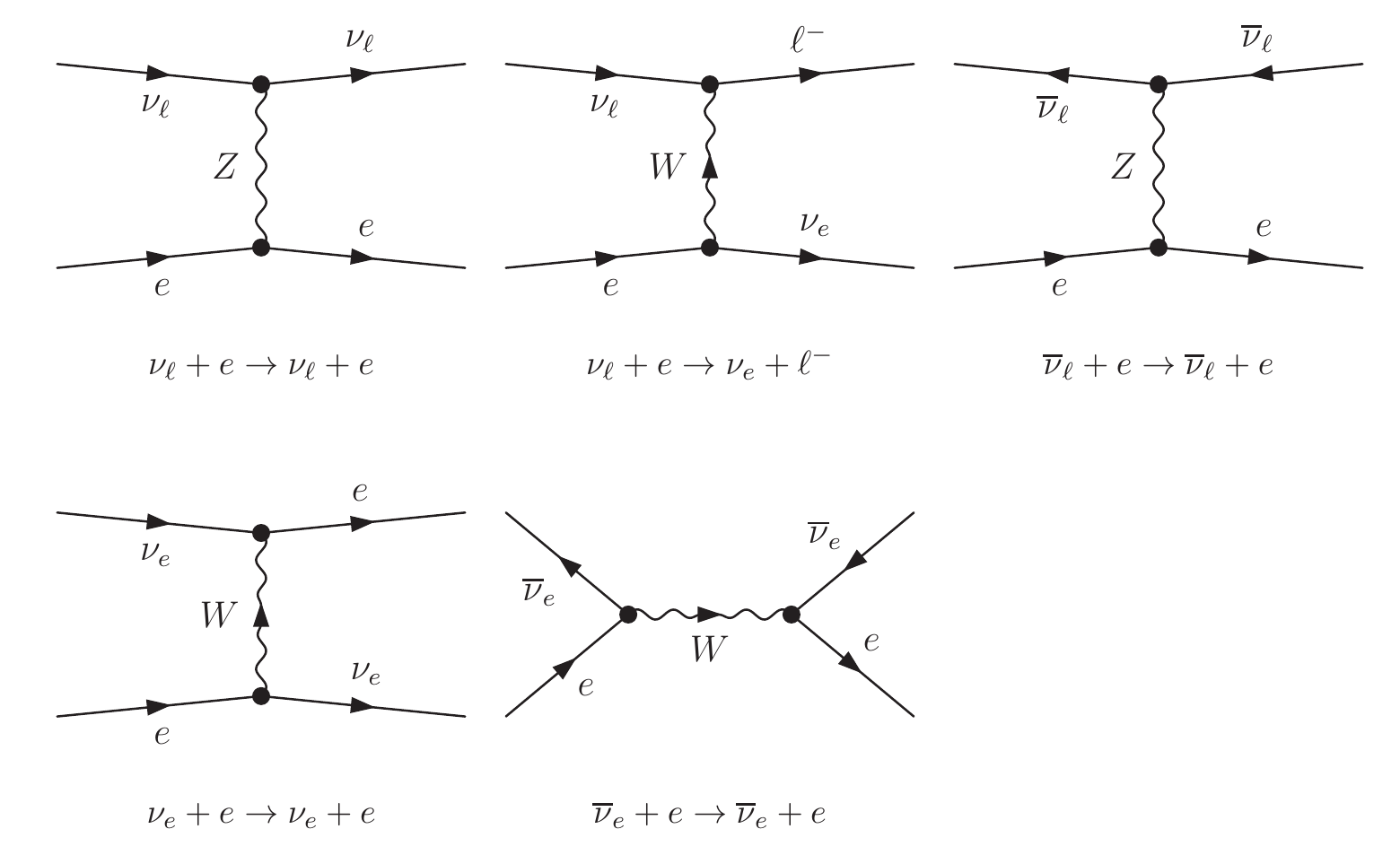}
\end{center}
\caption{Feynman diagrams for neutrino electron scattering.  The top row shows diagrams that are possible (although possibly kinematically not allowed) for all neutrino species while the bottom row shows the different diagrams for electron neutrinos and antineutrinos that lead to differential matter effects. }\label{fig:WG3_fig0}
\end{figure}

All flavors of neutrinos can interact with atomic electrons via neutral current interactions, while electron neutrinos have additional charged current diagrams (shown in Figure \ref{fig:WG3_fig0}) which lead to matter effects when neutrinos traverse dense media.  This leads to the matter effects observed in neutrino mixing as discussed in Chapter \ref{WG1}. 

With the advent of high intensity neutrino beams such as the Fermilab NuMI beam, event rates for the process $\nu_\mu + e^- \gt \nu_\mu + e^-$ are now high enough to provide a statistically significant standard candle based solely on this pure electroweak process. As an example, the MINER$\nu$A experiment recently reported a measurement of this process in the $1-20$ GeV region which resulted in a cross section measurement with 3\% accuracy and considerable improvement in the neutrino flux prediction for the experiment~\cite{PhysRevD.100.092001}.

\subsection{Neutrino Interactions with Nucleons}\label{sec:WG3_nucleons}
 %\label{sec:WG3_nucleons}
{Due to the need for higher event statistics, modern neutrino experiments use heavier nuclear targets (C, O, Ar, Fe), although there are historical low-statistics data on hydrogen and deuterium~\cite{Zyla:2020zbs} that provide information on single nucleon interactions.  Interactions on heavy nuclei then combine the physics of single nucleon interactions with nuclear effects that will be discussed in Section \ref{sec:WG3_nuclei}.}

Neutrinos and antineutrinos interact with free nucleons through the following processes induced by charged currents (CC) and neutral 
currents (NC):
\begin{itemize}
 \item Quasi-elastic (QE) and elastic scattering: Neutrinos and antineutrinos (where $\ell=e,\mu, \tau$),
interact with a nucleon through:
\begin{eqnarray}\label{Ch-4:process1_nu}
 {\nu}_\ell/\bar{\nu}_\ell (k) + N (p) &\longrightarrow& \ell^-/\ell^{+} (k^\prime) + N^{\prime} (p^\prime), \quad \quad N, N^{\prime} = n,p \;\; ({\rm CC}),\\
 \label{Ch-4:process5_nu1}
 \text{and}~~~{\nu}_\ell/\bar{\nu}_\ell (k) + N (p) &\longrightarrow& {\nu}_\ell/\bar{\nu}_\ell (k^\prime) + N (p^\prime) \;\; ({\rm NC}),
\end{eqnarray}
leading to a nucleon final state in the $\Delta S=0$ sector (see Fig.\ \ref{fig:WG3_fig1}a).  In the strangeness sector, meson final states are possible. Such reactions are constrained by the $\Delta S = \Delta Q$  and flavor-changing neutral current rules leading to CC reactions induced by antineutrinos  (Fig.\ \ref{fig:WG3_fig1}b), i.e.:
\begin{eqnarray}
\label{Ch-4:process5_nu}
  \bar{\nu}_\ell (k) + N (p) &\longrightarrow& \ell^+ (k^\prime) + Y (p^\prime), \quad ~~~~~\qquad \quad Y=\Lambda, \Sigma^{0},
 \Sigma^{-} .
\end{eqnarray}
In each case, the quantities in parentheses represent the four momenta of the corresponding particles.
 
\begin{figure}[t]
 \begin{center}
	\includegraphics[height=3.3 cm, width=2.5 cm]{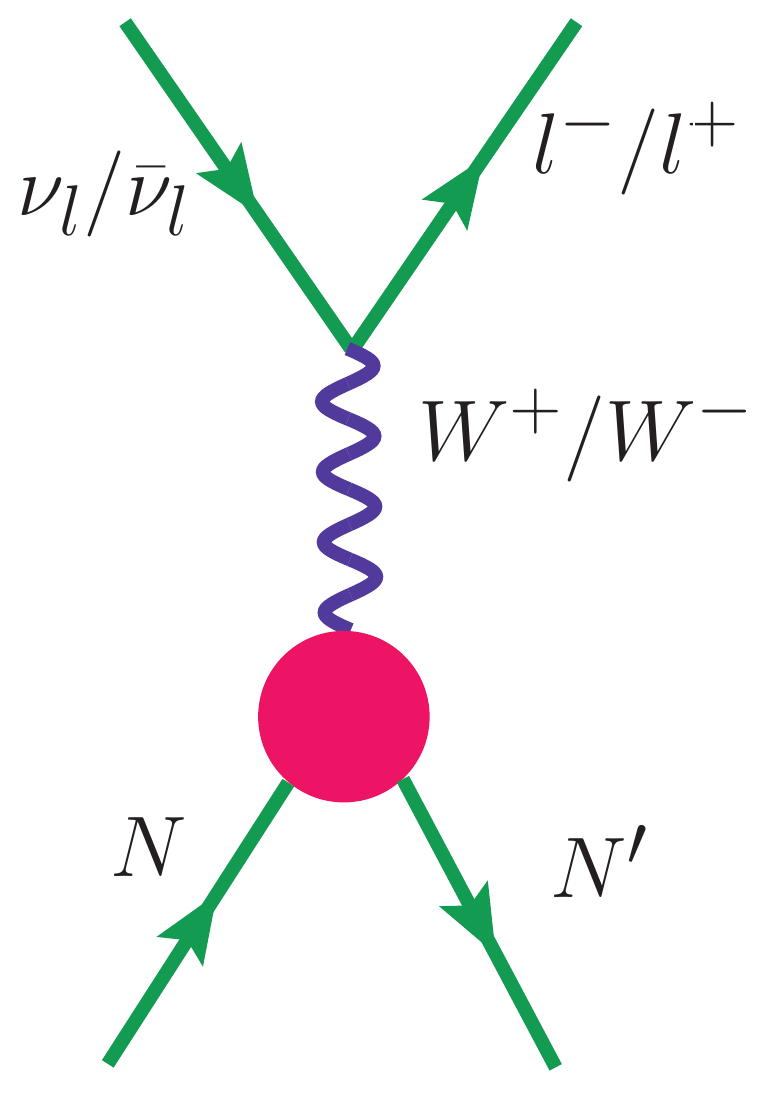}\hspace*{5mm}
	\includegraphics[height=3.3 cm, width=2.5 cm]{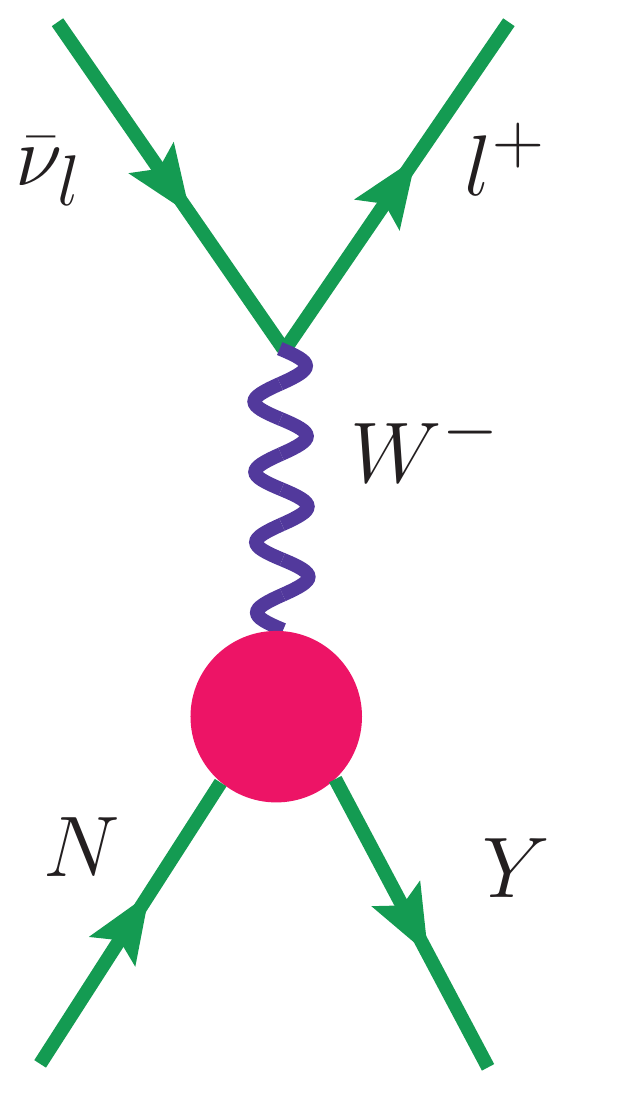}\hspace*{5mm}
     	\includegraphics[height=3.3 cm, width=2.5 cm]{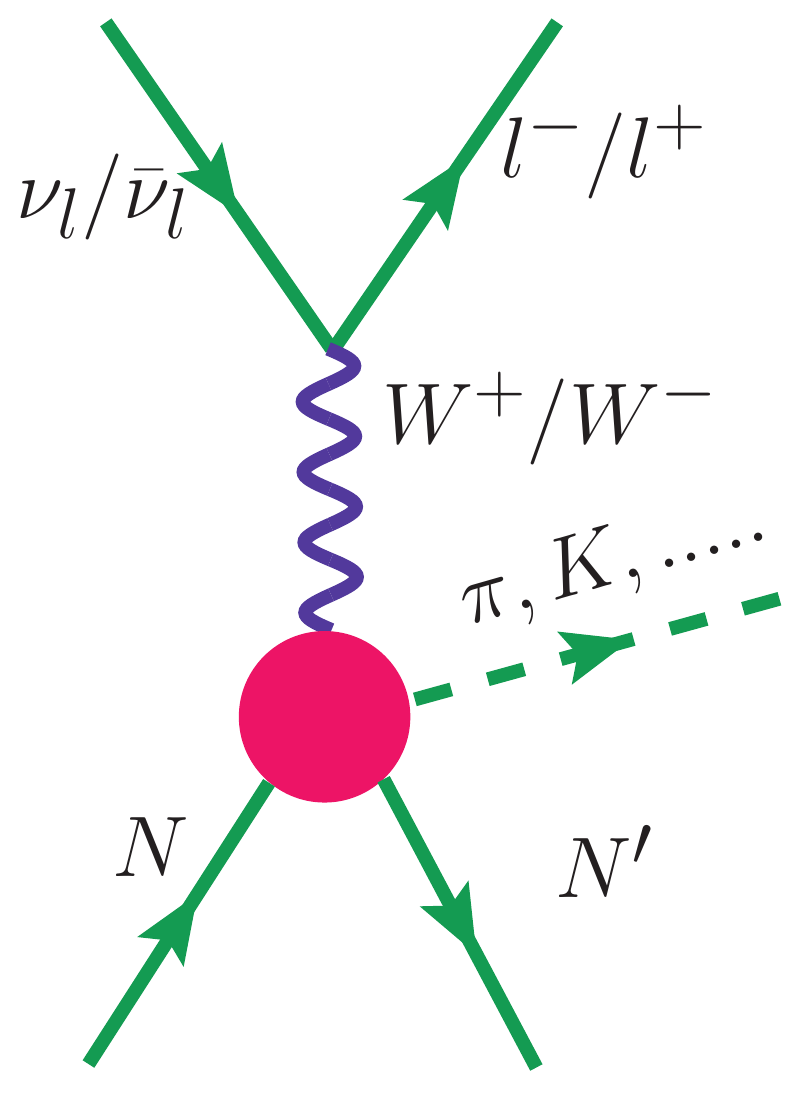}\hspace*{5mm}
       	\includegraphics[height=3.3 cm, width=2.5 cm]{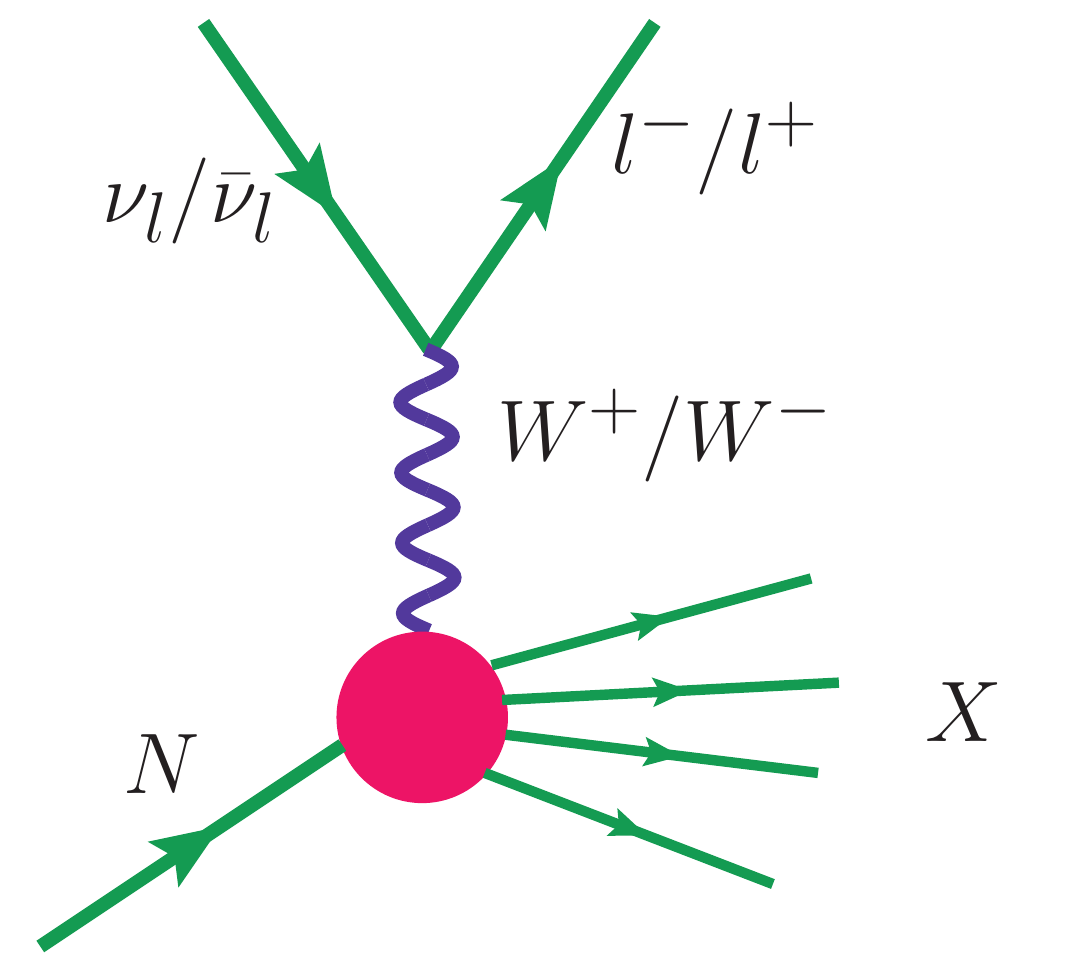}
 \end{center}
\caption{Feynman diagrams representing (from left to right) (a) QE, (b) CC meson, (c) CC pion and kaon production,  and (d) DIS processes. In the case of NC-induced processes, the final state lepton $\ell^-(\ell^+)$ and exchange boson $W^\pm$ are replaced by ${\nu}_\ell(\bar{\nu}_\ell)$ and 
$Z$, respectively.}\label{fig:WG3_fig1}
\end{figure}

\item Inelastic scattering (IE): In CC and NC inelastic processes, single (Fig.\ \ref{fig:WG3_fig1}c) and multiple mesons are produced in the 
reactions subject to the absence of flavor-changing neutral currents. A list of such reactions is given in Tab.\  \ref{Table1}. 
 
 \begin{table}[t]
  %\begin{center}
%     \vspace{1cm}
    \begin{tabular}{c c c }
%       \noalign{\vspace{-8pt}}
      \hline \hline
      S.\ No.\               & CC induced $\nu(\bar{\nu})$ reactions & NC induced $\nu(\bar{\nu})$ reactions
        \\ \hline
      
      1. & $\nu_{\ell} (\bar{\nu}_{\ell}) + N \longrightarrow \ell^{-} (\ell^{+})+ N^{\prime} + \pi$ & $\nu_{\ell} (\bar{\nu}_{\ell}) + N 
      \longrightarrow \nu_{\ell}(\bar{\nu_{\ell}})+ N^{\prime} + \pi$ \\ 
      
      2. & $\nu_{\ell} (\bar{\nu}_{\ell}) + N \longrightarrow \ell^{-} (\ell^{+})+ N^{\prime} + n \pi$ & $\nu_{\ell} (\bar{\nu}_{\ell}) + N 
      \longrightarrow \nu_{\ell} (\bar{\nu}_{\ell})+ N^{\prime} + n\pi$ \\ 
      
      3. & $\nu_{\ell} (\bar{\nu}_{\ell}) + N \longrightarrow \ell^{-} (\ell^{+})+ N^{\prime} + \eta$ &  $\nu_{\ell} (\bar{\nu}_{\ell}) + N 
      \longrightarrow \nu_{\ell}(\bar{\nu}_{\ell}) + N^{\prime} + \eta$\\ 
      
      4. & $\nu_{\ell} (\bar{\nu}_{\ell}) + N \longrightarrow \ell^{-} (\ell^{+})+ Y + K$ & $\nu_{l} (\bar{\nu}_{\ell}) + N \longrightarrow 
      \nu_{\ell} (\bar{\nu}_{\ell})+ Y + K$ \\ 
      
      5. & $\nu_{\ell} (\bar{\nu}_{\ell}) + N \longrightarrow \ell^{-} (\ell^{+})+ N^{\prime} + K(\bar{K})$ & $\bar\nu_l+N \longrightarrow l^{+}+Y+\pi$ \\ \hline \hline
    \end{tabular}
 %     \end{center}
 %%changehere
\caption{Charged- and neutral-current-induced inelastic processes. Here $N,N^{\prime}$ represent proton and neutron, $Y = \Lambda, 
\Sigma$ represents the hyperons, $K=K^{+}, K^{0}$ represents the kaons, $\bar{K}=K^-, \bar{K}^{0}$ represents the antikaons 
and $\ell=e,\mu$ represents the leptons.}\label{Table1}
%    \vspace{15mm}
\end{table}
\item Deep inelastic scattering (DIS): CC and NC DIS processes (Fig.\  \ref{fig:WG3_fig1}d) are represented by the same reactions as in Eq.\ (\ref{Ch-4:process5_nu}) 
except with the replacement that $N, N^\prime$ are now instead a jet of hadrons rather than a single nucleon in the final state. 

\end{itemize}

We will next go into more detail on each of these possible nucleon-level scattering processes.

\subsubsection{Quasi-Elastic Scattering}
Quasi-elastic scattering events are commonly used in the analysis of accelerator-based and atmospheric neutrino oscillation measurements. The transition matrix element for such processes as given in Eqs.\ (\ref{Ch-4:process1_nu})--(\ref{Ch-4:process5_nu}), can be simply written as 
\begin{eqnarray}
 \label{Ch-4:matrixelement}
 {\cal{M}} = a\frac{G_F}{\sqrt{2}} ~ \left[\bar{u} (k^\prime) \gamma_\mu (1 \pm \gamma_5) u (k)\right]~
 \left[\bar{u} (p^\prime) \left({V^\mu - A^\mu}\right) u (p)\right],
\end{eqnarray}
where the Cabibbo angle $\theta_C$ enters through  factors $a=\cos\theta_{C}\, (\sin\theta_{C})$ in the strangeness conserving (changing) processes, and  
\begin{eqnarray}\label{Ch-4:vx}
 \bar{u}(p^\prime)  V^\mu u(p) &=& \bar{u}(p^\prime) \left[\gamma^\mu f_1^{NB}(Q^2)+i\sigma^{\mu \nu} 
 \frac{q_\nu}{M+M_B} f_2^{NB}(Q^2) \right.\nonumber \\
 &+& \left.\frac{2 ~q^\mu}{M+M_B} f_3^{NB}(Q^2) \right] u(p),~~~~\nonumber\\
 \label{Ch-4:vy}
  \bar{u}(p^\prime)A^\mu u(p)&=&\bar{u}(p^\prime)\left[\gamma^\mu \gamma_5 g_1^{NB} (Q^2) + i \sigma^{\mu 
  \nu} \frac{q_\nu}{M+M_B} \gamma_5 g_2^{NB}(Q^2) \right.\nonumber \\
  &+& \left. \frac{2 ~q^\mu} {M+M_B} g_3^{NB}(Q^2) \gamma_5 \right] u(p),~~~~
\end{eqnarray}
where $B$ represents a nucleon $N$ or a hyperon $Y$, $M$ and $M_B$ are the masses of the initial nucleon and final baryon; $q_\mu$ 
is the four-momentum transfer with $Q^2(=-q^2) \ge 0$, while $f_1^{NB}(Q^2)$, $f_2^{NB}(Q^2)$ and $f_3^{NB}(Q^2)$ are the weak vector, 
magnetic and induced scalar form factors and $g_1^{NB}(Q^2)$, $g_2^{NB} (Q^2)$ and $g_3^{NB}(Q^2)$ are the axial vector, induced 
 tensor (also known as weak electric) and pseudoscalar form factors, respectively \cite{LlewellynSmith:1971uhs}.
 $T$ invariance implies that $f_{1-3} (Q^2)$ and $g_{1-3} (Q^2)$ are real.
 In the absence of second class currents, i.e., assuming 
 $T$- and $G$-invariance, $f_3^{NB}(Q^2)=0$ and $g_2^{NB} (Q^2)=0$. The hypothesis of a conserved vector current (CVC), which follows from the assumption that weak vector currents along 
with the EM current form an isotriplet, implies that $f_{1,2}^{np} (Q^2) = f_{1,2}^p (Q^2) - f_{1,2}^n (Q^2)$.
 The vector form factors for the nucleons $f_{1,2}$ are given in terms of EM form factors $f_{1,2}^p (Q^2)$ and $f_{1,2}^n (Q^2)$ which 
in turn are expressed in terms of the Sachs electric ($G_E^{p,n} (Q^2)$) and magnetic ($G_M^{p,n} (Q^2)$) form factors of the nucleons. {Information from charged lepton scattering can be used to constrain the vector form factors while the axial form factors are more easily
accessible in neutrino data.} For details, please see Ref.~\cite{Fatima:2018wsy,Akbar:2016awk}.

{ Historically, the axial vector form factor $g_1(q^2)$ has been  parameterized as a dipole} given by  
 \begin{equation}
 g_{1}^{np}(Q^2)= \frac{g_A (0)}{\left( 1 + \frac{Q^2}{M_A^2} \right)^2},
 \end{equation}
  where the axial charge $g_A(0)= 1.267 \pm 0.003$ and the axial dipole mass
  $M_A = 1.026 \pm 0.021$ GeV   is the world average value \cite{Bernard:2001rs}. Even if its asymptotic behaviour at high $Q^2$ is the one predicted by perturbative QCD, the dipole ansatz is not theoretically well founded. Alternative  representations have been therefore developed such as the z-expansion, based on the analytic properties of strong interactions~\cite{Bhattacharya:2011ah}. The partially conserved axial current (PCAC) assumes that the divergence of the axial current is given in terms of a pion field, i.e.\ 
$\partial_\mu A^\mu(x)=C_\pi \phi_\pi(x)$ 
implying that $g_3 (Q^2)=\frac{2Mg_1 (Q^2)}{m_\pi^2 + Q^2}$.

\subsubsection{Cross Sections and Polarization Observables in Quasi-Elastic Processes}
The cross sections associated with quasi-elastic scattering are calculated using the matrix element given in Eq.\ (\ref{Ch-4:matrixelement}). There are many resulting observations 
made: total cross sections ($\sigma$), the energy and angular distributions of the final state particles, as well as the double differential cross 
sections for the charged lepton and the nucleon in the final state. If the processes given in Eq.\ (\ref{Ch-4:process1_nu}) take place with a nucleon bound 
inside a nucleus, note that nuclear medium effects are then very important. These
 nuclear medium effects play an important role in the final state composition, in interpreting the experimental results, and in the determination of $M_A$ in nuclear target data. It has been also observed that the quasi-elastic hyperon production induced by antineutrino scattering can give significant contribution to the pion production arising due to the hyperon decay and is comparable to the pions arising due to $\Delta$ decay in nuclear targets in the sub-GeV
energy region relevant for atmospheric and accelerator experiments being performed to study neutrino oscillation phenomenona~\cite{Alam:2014bya,Fatima:2018wsy,Alam:2013cra, Fatima:2021ctt}.

It has also been suggested~\cite{Fatima:2018wsy,Akbar:2016awk} that polarisation measurements of the baryon in the final state can give important information about 
the various form factors as has been found in the case of electron-proton scattering. Moreover, a determination of all the form 
factors including $f_3(q^2)$ and $g_2(q^2)$ would also help to study the status of $G$ and $T$-invariance in the weak 
interaction. Notwithstanding the experimental difficulties in measuring the nucleon polarisation in quasi-elastic reactions as it 
would involve a double scattering experiment, it is possible to make such polarisation measurements in the case of $\bar{\nu} + 
N\rightarrow e^++Y,~Y\rightarrow N\pi$ reactions. Those are self analysing by performing measurements on the polarisation observables  
and asymmetries in the angular distribution of final nucleons and pions. 
The taus produced in charged current $\nu_\tau({\bar\nu}_\tau)$ interactions are polarized and this tau polarization affects the distributions of its decay products used in identifying charged current and neutral current events.
Theoretical estimates and the feasibility of measuring 
these observables have been recently discussed in the literature along with the implications in determining the various form 
factors~\cite{Fatima:2018wsy,Akbar:2016awk}.

\begin{figure}[t]
\begin{center}
\includegraphics[height=7 cm, width=8cm]{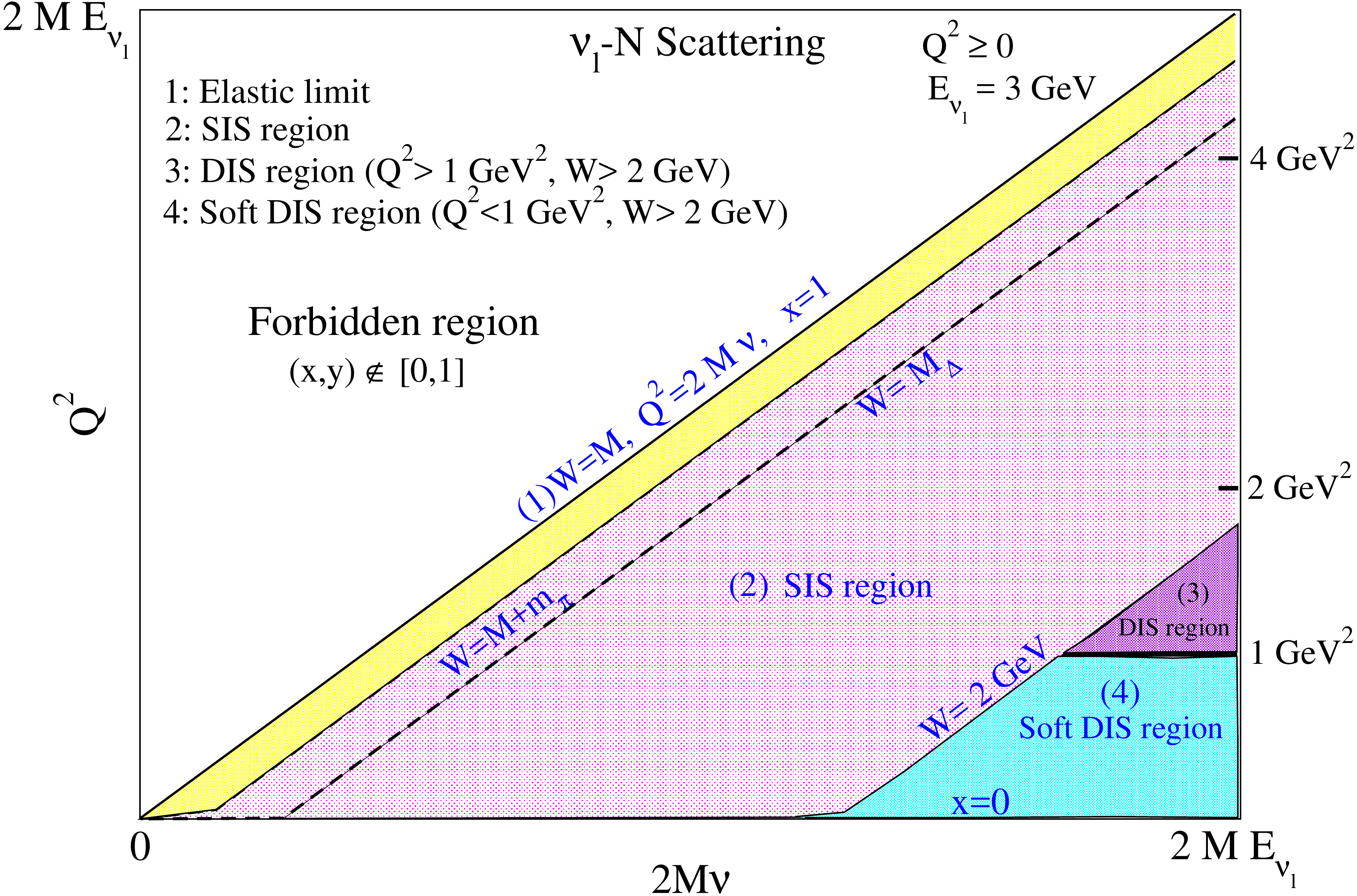}
\includegraphics[height=7 cm, width=8cm]{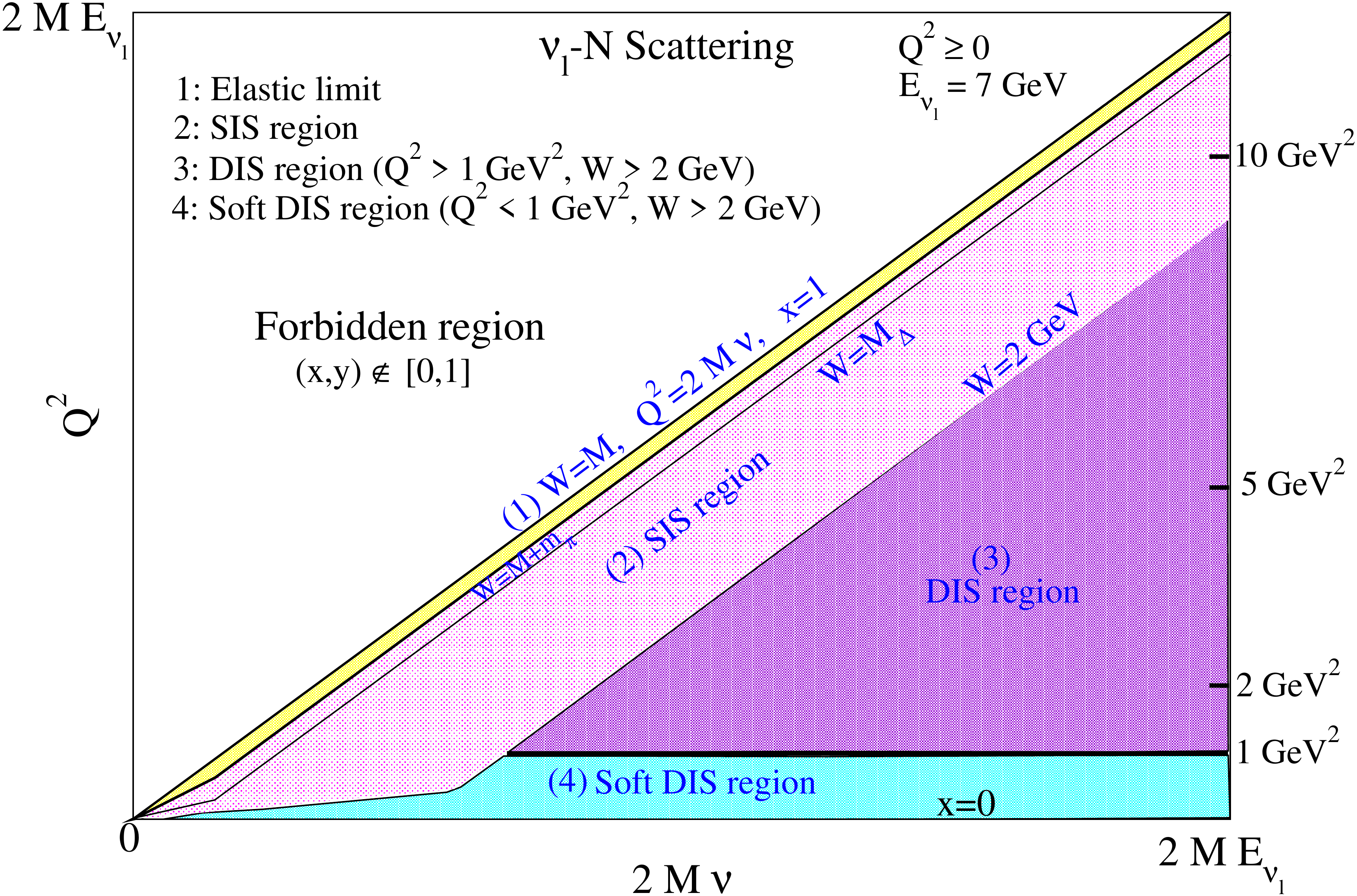}
 \end{center}
\caption{Allowed kinematical region for neutrino-nucleon scattering in the ($Q^2, \nu$) plane for 
$E_\nu=3$ GeV (left panel) and $E_\nu=7$ GeV (right panel). The square of the invariant mass  is 
defined as $W^2=M_N^2+2M_N\nu-Q^2$ with  nucleon mass $M_N$ and  energy transfer $\nu$. The inelasticity is defined as $y=\frac{\nu}{E_\nu}=\frac{(E_\nu - E_l)}{E_\nu}$ and the forbidden region in terms of
$x$ and $y$ is then defined as $x,y~\notin~[0,1]$.
The elastic limit is $x=\frac{Q^2}{2M_N\nu}=1$  and the shallow inelastic scattering (SIS) region is defined as the
region for which $M_N+M_\pi \le W \le 2$ GeV and $Q^2 \ge 0$ covering both non-resonant and resonant meson production. The DIS region is defined as the region for which $Q^2 \ge 1$ GeV$^2$ and $W \ge 2$ GeV, and the 
soft DIS region is defined as $Q^2 < 1$ GeV$^2$ and $W \ge 2$ GeV.
%(also part of SIS). 
Notice the yellow  band ($M_N < W < M_N+M_\pi$), where we do not 
expect anything from neutrino-nucleon scattering. However, this region becomes important when the scattering takes place with a nucleon within a nucleus due to the multi-nucleon correlation effect.
In the yellow band process like photon emission is possible. Soft DIS region is also nothing but the SIS region. The boundaries between regions are not sharply established and are indicative only.
} \label{fig:WG3_fig2}
\end{figure}

\subsubsection{Inelastic Scattering}
\begin{figure}[t]
 \begin{center}
\includegraphics[height=4.5cm, width=11.5cm]{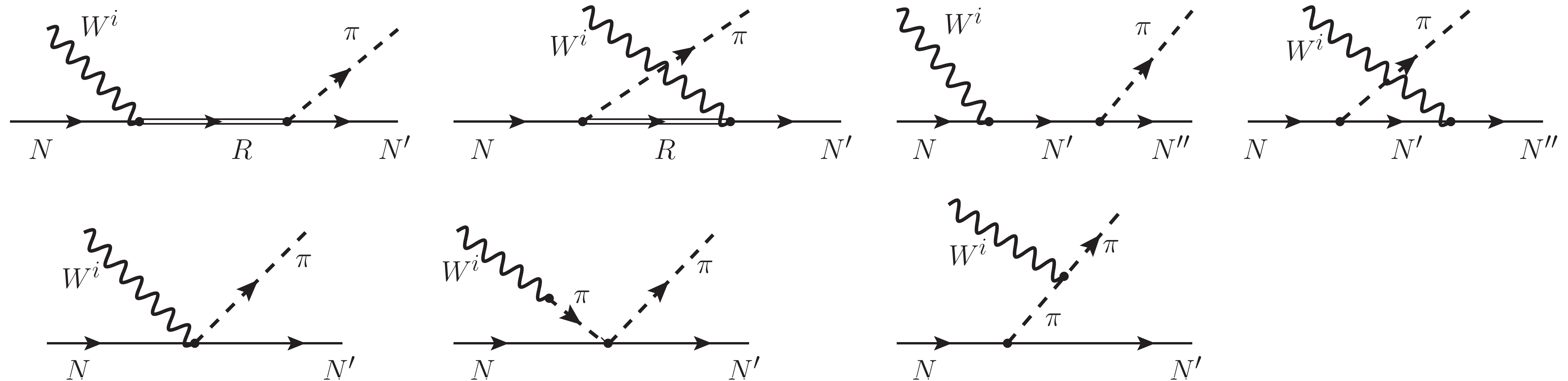}
\caption{Feynman diagrams contributing to the hadronic current corresponding to $W^{i} N \to N^{\prime}(N^{\prime\prime}) 
\pi^{\pm,0}$, where $(i=\pm)$ for charged-current processes and $(W^i \equiv Z \; ; i=0)$ for neutral 
current processes with $N,N^{\prime},N^{\prime\prime}=p$ or $n$. The first row (left to right) represents $s$- and $u$-channel diagrams for 
the resonance production, and the nucleon pole terms and the second row shows the contact, pion pole and pion-in-flight term.}\label{fig:WG3_feynmann}
\end{center}
\end{figure}

Above the QE scattering region %, in effective hadronic mass~($W$), 
the region of inelastic scattering starts with the excitation of the $\Delta$ resonance 
followed by excitation of increasingly higher mass resonant states. These resonances sit atop a continuum of non-resonant $\pi$ 
production that starts at hadronic system mass-squared $W^2 = p^\prime\cdot p^\prime = (M + m_{\pi})^2$ (Fig.\ \ref{fig:WG3_fig2}). The generic Feynman diagrams describing these reactions are shown 
in Fig.\ \ref{fig:WG3_feynmann}, where one vertex is the weak vertex describing the weak interactions of leptons with 
$W^{\pm}(Z)$ bosons, while the second vertex is a mixed vertex describing the weak interaction of nucleons
and the strong interactions of the meson-baryon system described by a phenomenological Lagrangian consistent with the symmetries of the 
strong interaction or by an effective Lagrangian motivated by the symmetries of QCD like the chiral symmetry. This resonant plus 
non-resonant $\pi$ production region transitions leads directly into the DIS region (Fig.\  \ref{fig:WG3_fig2}), where the interactions occur on 
quarks, at a kinematically defined regions for most experiments as $W \ge$ 2.0 GeV and $Q^2 \ge 1$ GeV$^2$ and this kinematical cut is adhoc and in most of the experimental analyses the region $W \ge$ 2.0 GeV and $Q^2 \ge 1$ GeV$^2$ are considered to be the safe DIS region. The non-resonant pion 
production region as well as the region with   $W \ge$ 2.0 GeV and $Q^2 \le 1$ GeV$^2$ is the 
intriguing kinematic region now referred to as the shallow-inelastic scattering (SIS) region. In this transition region, the 
principle of quark-hadron duality can be very effectively used to connect the DIS cross section to the cross section in the 
resonance region, which states that the nucleon structure functions 
%~(which are used to describe the cross section) 
at low $Q^2$, 
averaged over a certain energy are similar to the structure functions of DIS processes at higher $Q^2$ averaged over the same energy. 
Thus, for a given energy, the structure functions in the transition region can be equivalently described either by the inelastic 
or DIS structure functions with appropriate evolution from low to high $Q^2$ \cite{SajjadAthar:2020nvy}. 

\begin{figure}[t]
 \begin{center}
\includegraphics[height=3.5cm, width=4.9cm]{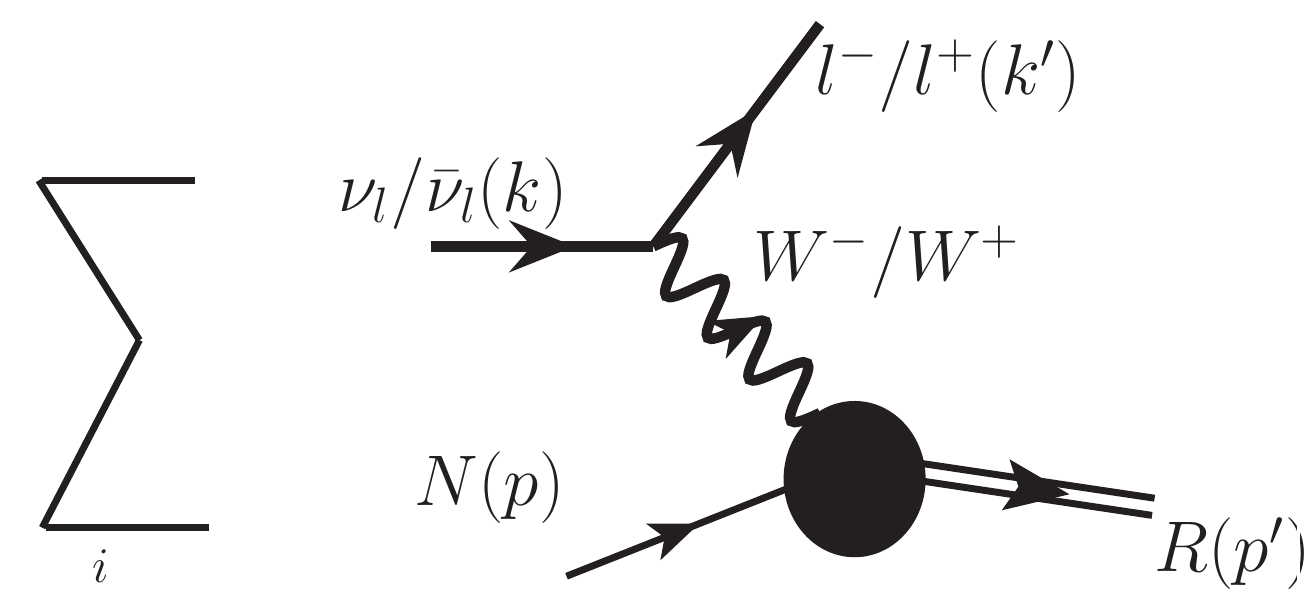}
\caption{Diagrammatic representation of resonance excitations for the charged current induced $\nu_\ell({\bar\nu}_\ell) + N \rightarrow \ell^-(\ell^+) + R$ process, where $R$ 
represents the different resonances contributing to the hadronic current.}\label{fig:reswk}
\end{center}
\end{figure}

The nucleon resonances which are excited in the inelastic reactions (shown in Fig.\ \ref{fig:reswk}) are characterized by their mass, parity, spin and isospin and 
are represented by the symbol $R_{IJ}(M_{R})$, where $R$ is the name of the resonance given on the basis of its orbital angular 
momentum, i.e., $L=0,1,2$ and named $S,~ P,~ D$, etc., showing its parity, $M_{R}$ is the mass while $I$ and $J$ specify their 
isospin and spin quantum numbers. Near the threshold region of single pion production, the $P_{33}(1232)$ resonance is dominant while at higher energies the resonances such as the  $P_{11} (1440)$, $S_{11}(1535)$, $D_{13}(1520)$ and $P_{13} 
(1720)$, lying in the second and third resonance regions, become increasingly important~\cite{SajjadAthar:2020nvy, Hernandez:2007qq, Alam:2015gaa}.

Weak pion production has been studied for a long time and the various calculations are based on (i) dynamical models with 
dispersion theory, (ii) quark models with higher symmetry such as $SU(6)$, and (iii) phenomenological Lagrangians for the interaction of mesons with nucleons and higher resonances. In most of the models, the vector form factors in the resonance sector are determined in terms of the helicity amplitudes, which 
are extracted from real and/or virtual photon scattering experiments. Information on the axial vector form factors is scarce, 
therefore, the PCAC hypothesis is generally used to obtain this contribution. The strong coupling at the 
meson-baryon-resonance vertex is obtained using the branching ratio and the partial decay width of the resonance decaying into the meson-baryon mode. 

In recent times, many calculations have been performed in the isobar model with or without explicitly taking into account the final 
state interaction of the pion-nucleon state. The existing experimental data on the single pion production process from (almost) 
free nucleons are available only from the bubble chamber experiments performed almost 40 years ago at ANL and BNL with 
deuteron and hydrogen targets. The ANL and BNL results differ with each other by about $30-40$\%. Reanalysis of these data 
has resulted a better agreement between these two data sets \cite{Wilkinson:2014yfa}. 
Nevertheless, it has been strongly felt that it is important to have a good understanding of the basic inelastic processes on 
nucleon targets as current and future neutrino oscillation experiments are being performed using medium to heavy nuclear targets and neutrino-nucleon
scattering cross sections serve as an important input in all the Monte Carlo generators. 
The (anti)neutrino induced single kaon production~\cite{RafiAlam:2010kf, Alam:2011vwg}, eta production~\cite{RafiAlam:2013jcs} and associated particle production~\cite{Adera:2010zz, RafiAlam:2013tgm}
have also been studied recently. 

In recent times, many calculations have been performed in the isobar model with or without taking into account non-resonant contributions. Unitarity, which is absent in tree level amplitudes, can be accounted for perturbatively using chiral perturbation theory~\cite{Yao:2018pzc} but this approach is applicable only close to threshold. It can also be approximately restored by imposing Watson’s theorem \cite{Alvarez-Ruso:2015eva}. Ultimately, it can be implemented dynamically by solving the Lippmann-Schwinger equation in coupled channels for the meson-baryon system. This is the approach followed by the dynamically coupled channel~(DCC) model to achieve a unified description of weak production of $N \pi$, $N \pi\pi$, $N \eta$ and $Y K$ final states with invariant masses $W \lesssim 2$~GeV \cite{Nakamura:2015rta}.

\subsubsection{Deep Inelastic Scattering }
For inclusive neutrino and antineutrino induced deep inelastic scattering (DIS) processes on free nucleons (Fig.\ \ref{fig:WG3_fig1}), 
the differential cross section is calculated in a quark parton model using the assumption of {approximate} Bjorken scaling. 
 The differential scattering cross section in terms of the Bjorken scaling variables $x$ and $y$ 
is given by:
\begin{eqnarray}\label{dsig_xy} \frac{ d^2\sigma_N^{} }{ dx dy }&=
&\frac{G_F^2 M E_\nu}{\pi} 
           \bigg [y\Big(xy + \frac{m_\ell^2}{2 E_\nu M}\Big)F_1 + 
            \Big(1-y -\frac{Mxy}{2 E_\nu} - \frac{m_\ell^2}{4 E_\nu^2}\Big) F_2 \pm \nonumber\\
      &   & \Big(xy(1-\frac{y}{2})-y\frac{m_\ell^2}{4 M E_\nu}\Big) F_3 + 
            \Big(x y \frac{m_\ell^2}{2 M E_\nu} + \frac{m_\ell^4}{4 M^2 E_\nu^2}\Big) F_4 -
            \frac{m_\ell^2}{2 M E_\nu} F_5\bigg],~~~\;\;\;
            \end{eqnarray}    
           where $+(-)$ corresponds to neutrino (antineutrino)-nucleon scattering. In the limit that the lepton mass $m_\ell \rightarrow 0$, only the $F_{1-3}$ structure functions contribute. In the limit of high $Q^2$ and energy transfer $\nu$, such that the Bjorken variable $x=\frac{Q^2}{2M\nu} \rightarrow$ constant, the nucleon structure functions become a function of the
dimensionless variable $x$ only, and $F_{1}(x)$ and $F_{2}(x)$ satisfy the Callan-Gross relation $F_2(x)=2xF_1(x)$. 
            Through the explicit evaluation of the nucleon structure functions, one may
write them in terms of the parton distribution functions
(PDFs) viz.\ $u(x)$, $d(x)$, etc., which provide information about the momentum
distribution of the partons within the nucleon, and are given by:
\begin{eqnarray}
F_2^{\rm CC}(\nu p)& = & 2 x [d + s + \bar{u} +\bar{c}]; \;\;x F_3^{\rm CC}(\nu p) =  2 x [d + s - \bar{u} -\bar{c}]\,,\nonumber\\
F_2^{\rm CC}(\bar{\nu} p)& = & 2 x [u + c + \bar{d} +\bar{s}]; \;\; x F_3^{\rm CC}(\bar{\nu} p) =  2 x [u + c - \bar{d} -\bar{s}]\,.
\end{eqnarray}
If one assumes isospin symmetry,  $u$ and $d$ quark distributions are swapped for a neutron target and expressions of the cross section for neutrino and antineutrino DIS are obtained. The effect of various perturbative and nonperturbative QCD corrections on the free nucleon structure functions $F_{i=1-3}$ 
 have been studied in the literature such as target mass corrections, higher twist effects, etc.\  and emphasis has been made to understand their implications in the determination of nuclear structure functions~\cite{Zaidi:2019asc}. 
{Historically, thanks to the ability to separate quark and antiquark flavors, neutrino interactions on nuclear targets played an important role in the early development of parton distributions.  Limitations due to the presence of nuclear effects at the $10-20$\% level have led to neutrino data being deemphasized in modern fits intended for proton-proton scattering experiments; $pA$ data sets from the LHC at CERN have led to renewed interest in these nuclear parton distributions. See Ref.~\cite{AbdulKhalek:2020yuc} for a recent comparison of $pA$ and neutrino data.}

 An important feature is that through neutrino (antineutrino) scattering on nucleons, quarks and antiquarks can be directly probed which is not possible in the case of electromagnetic interactions. In this case, the cross 
sections for neutrino and antineutrino scattering off  free protons in the four flavour scheme are obtained as:
\begin{eqnarray}
\label{xeq1}
  \frac{d^2\sigma^{\nu p}}{dx dy}
  &=& \frac{G^2_F\;sx}{\pi}\;(d(x)+s(x)+(1-y^2)(\bar u(x)+\bar c(x))\,,\\
  \label{xeq2}
   \frac{d^2\sigma^{\bar\nu p}}{dx dy} 
   &=& \frac{G^2_F\;sx}{\pi}\;(\bar d(x)+\bar s(x)+(1-y^2)(u(x)+c(x))\,~~~
\end{eqnarray}
\noindent
whereby neutrino data can be used, for example, to extract the strange and charm quark PDFs.

\subsection{Neutrino Interactions with Nuclei}\label{sec:WG3_nuclei}
%\subsection{Nuclear effects in Neutrino-Nucleus Scattering}
%\label{sec:WG3_nuclei}
Most neutrino oscillation experiments rely on massive detectors to achieve the target mass necessary to detect sizable statistics of neutrino interactions over large distances.  Although there is some data from hydrogen bubble chambers~\cite{Zeller:2003ey}, most detectors use heavier materials such as scintillator (CH), water (H$_2$O), iron (Fe) or noble liquids (Ar).  As a result, the nucleon phenomenology described above must be expanded to include the effects of interactions within a complex nucleus.  To perform precision neutrino oscillation measurements, we must also understand the flavor ($e ,\mu,\tau$) and weak-charge ($\nu ,\nubar$) dependence of neutrino interaction rates and how the determination of the neutrino energy and the measured scattering rates depends on those parameters.

\subsubsection{Neutrino Energy Determination}
For neutrino oscillation measurements, a good understanding of the initial neutrino energy in charged current interactions is required.  This can be estimated in two ways: first by summing the total energy (leptonic and hadronic) exciting the nucleus and second, in the special case of quasi-elastic scattering, by using the final state lepton kinematics to estimate the incoming lepton energy via two-body kinematics. This presumes that the initial nucleon is at rest. 

In the first case, most of the final state hadrons need to be detected and a good physics model is needed to correct for any missed particle (in particular neutrons and $K_L$) and additional  intrinsic differences between neutrino  and antineutrino scattering, where the final state has differing fractions of easily detected  particles such as protons and charged pions and harder to detect particles such as $K_L$ and neutrons. At low energies, where final state multiplicities are low, these effects can be very large, while at very high energies, most of these differences in the final state can be expected to cancel.

For a subset of events, namely those with a quasi-elastic signature ($\nu + n \gt \ell + p$ or $\overline{\nu} + p \gt \ell + n$)  with no recoil energy aside from a proton or neutron, a kinematic estimate of the neutrino energy can be made, namely 
\begin{eqnarray}
 E^{\rm QE}_{\nu(\nubar)} \simeq \frac{M^2_{n(p)} - (M_{p(n)}-E_b)^2 - M_\ell^2+2(M_{p(n)}-E_b)E_\ell}{2(M_{p(n)}-E_b-E_\mu+P_\ell \cos\theta_\ell)},
\end{eqnarray}
where $E^{\rm QE}_{\nu}$ is the estimated neutrino energy, $M_n$ and $M_p$ are the neutron and proton masses, $M_\ell$, $p_\ell$, $E_\ell$ and $\theta_\ell$ represent the final state lepton kinematics  and $E_b$ is a binding energy of order 10's of MeV. Both of these methods of neutrino energy estimation rely on detailed nuclear models to obtain the precise predictions needed for modern oscillation experiments, as detailed below. 

\subsubsection{Neutrino Interaction Rate Determination}
In order to determine neutrino oscillation parameters, the final state neutrino flavor must be identified.  In theory, this relies only on detecting the final state charged lepton, but for each target nucleus, detector technology and neutrino flavor, the final state signatures require unique selection criteria which depend on a large number of model parameters.  Reference \cite{Abi:2020qib} lists some of the factors that must be considered in optimizing experimental sensitivity to oscillations. 
They include nuclear corrections to quasi-elastic scattering (1p1h), the presence of correlated nucleon effects (2p2h), resonance production of final state pions, high-$W$ inelastic scatters in which the nucleon breaks up, and final state interactions (FSI) in which the outgoing hadronic particles interact in the nucleus and have their type or kinematics altered. For recent summaries of the impact of nuclear effects on neutrino interaction predictions and hence neutrino oscillation measurements, please see~\cite{Alvarez-Ruso:2017oui, Katori:2016yel}.

%\subsubsection{Nuclear Models}

\begin{figure}[t]
\begin{center}
  \includegraphics[width=.7\textwidth]{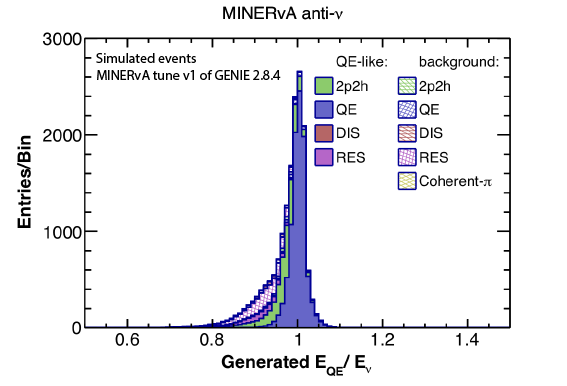}
    
\end{center}
\caption{Ratio of reconstructed to true antineutrino  energy in quasi-elastic interactions, illustrating the effects of resonance production and 2p2h effects.  Pure quasi-elastic scattering is spread out by Fermi motion while most other effects decrease the reconstructed energy.  The solid regions represent interactions where the final state contains no additional pions while the hatched regions contain additional low energy pions that were below detection threshold. This shows the extent to which such effects need to be accounted for in order to obtain an accurate neutrino energy estimate. The effects can be sizable. }\label{fig:WG3_Enudist}
\end{figure}

Over the past two decades, improvements in detector technology and increased statistical power have been accompanied by neutrino-nucleus interaction  event generators of increasing sophistication and accuracy, including GENIE \cite{Andreopoulos:2009rq}, NuWro \cite{Golan:2012wx}, NEUT \cite{Hayato:2009zz}, and 
GiBUU \cite{Buss:2011mx}.  These models now incorporate a wide range of nuclear effects, including short range interactions, long range screening effects, and final state interactions. The simplest models start with a Relativistic Fermi Gas  (RFG) treatment where the nucleus is modeled as a simple potential well populated by neutrons and protons \cite{Smith:1972xh}.  RFG models predict a velocity distribution for the nucleons inside the nucleus and hence an intrinsic spread $E_\nu^{\rm QE}$ around the true $E_\nu$. Fig.\  \ref{fig:WG3_Enudist} shows a GENIE \cite{Andreopoulos:2009rq} prediction at particle level for the ratio of the kinematic estimate to true neutrino energy $E_\nu^{\rm QE}/E_\nu$ in antineutrino events in the $2-6$ GeV region. Pure quasi-elastic events are smeared by Fermi motion but centered on the correct value of $E_\nu$.  Almost all other nuclear effects result in an underestimate of the neutrino energy if $E^{\rm QE}$ is used to estimate the energy for  candidate quasi-elastic scattering events.

Continued improvement in the underlying interaction models, for both nuclei and nucleons, and in event generators which implement those models will remain an extremely important component of all neutrino measurements, from oscillations to astrophysics.

\subsubsection{Multi-Nucleon Correlation Effects}
\begin{figure}[t]
\begin{center}
  \includegraphics[width=.7\textwidth]{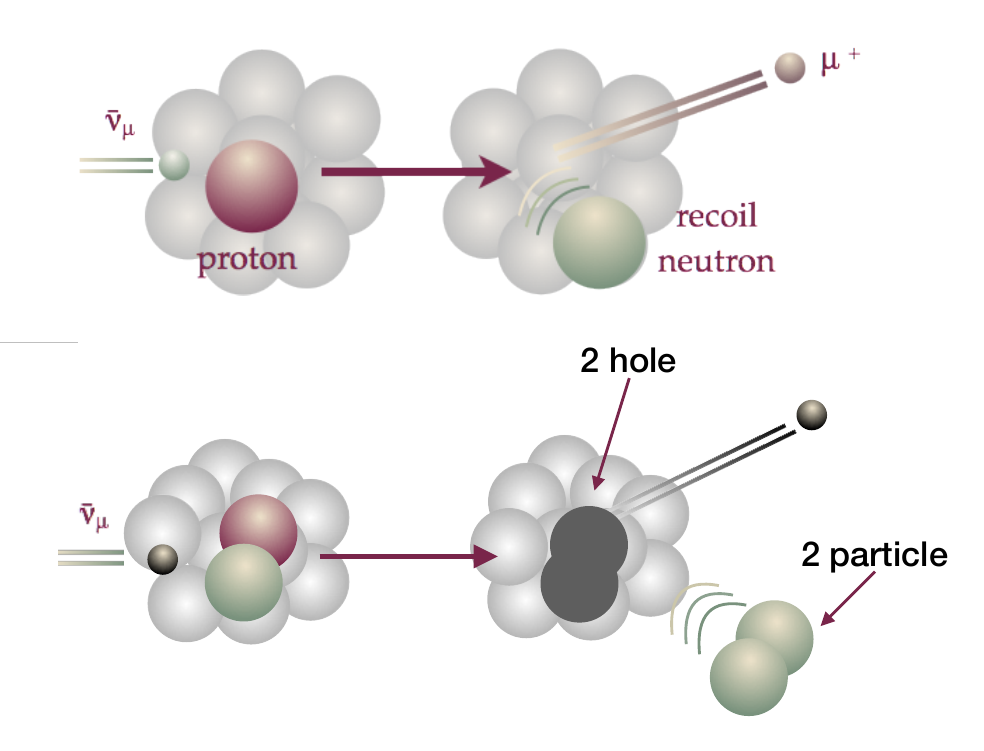}
\end{center}
\caption{Illustration of a standard charged current process where an antineutrino interacts with a single proton and produces a recoil neutron (top) and  a 2p2h process where the antineutrino interacts quasi-elastically with an $np$ pair in the nucleus producing a two nucleon final state (bottom). Credit C.\ Patrick.}\label{fig:WG3_2p2h}
\end{figure}

\begin{figure}[t]
\begin{center}
  \includegraphics[width=.9\textwidth]{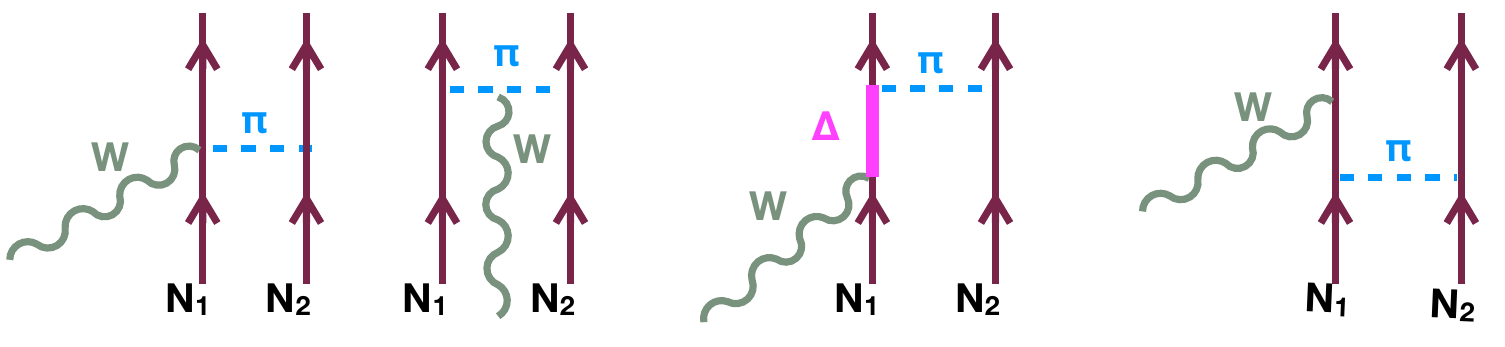}
\end{center}
\caption{Correlated exchange processes. }\label{fig:WG3_MEC}
\end{figure}

There is now substantial evidence from both elastic electron scattering \cite{Subedi:2008zz} 
and quasi-elastic  neutrino scattering \cite{minibooneccqe,t2kccqe,laura,arturo} that a significant proportion of neutrino interactions involve scattering from correlated nucleon pairs.    This process is alternately described as a 2p2h (2-particle 2-hole) or MEC (Meson Exchange Current) process. Figure~\ref{fig:WG3_2p2h} illustrates this process schematically while Fig.\ \ref{fig:WG3_MEC} shows contributing diagrams. In Fig.\ \ref{fig:WG3_Enudist} events from this process are shown to give an underestimate of the true neutrino energy due to the failure of the 2-body scattering assumption. A significant body of theoretical work in recent years has led to improved calculations of such multi-nucleon effects~\cite{Alvarez-Ruso:2017oui,Carlson:2001mp,Martini:2011ui,Benhar:2015ula,Nieves:2016sma,Megias:2016fjk,Barbaro:2016hrt,Dolan:2018sbb,Rocco:2020jlx,Barbaro:2021psv}.

\subsubsection{Final State Interactions }
\begin{figure}[th]
    \centering
    \includegraphics[width=.6\textwidth]{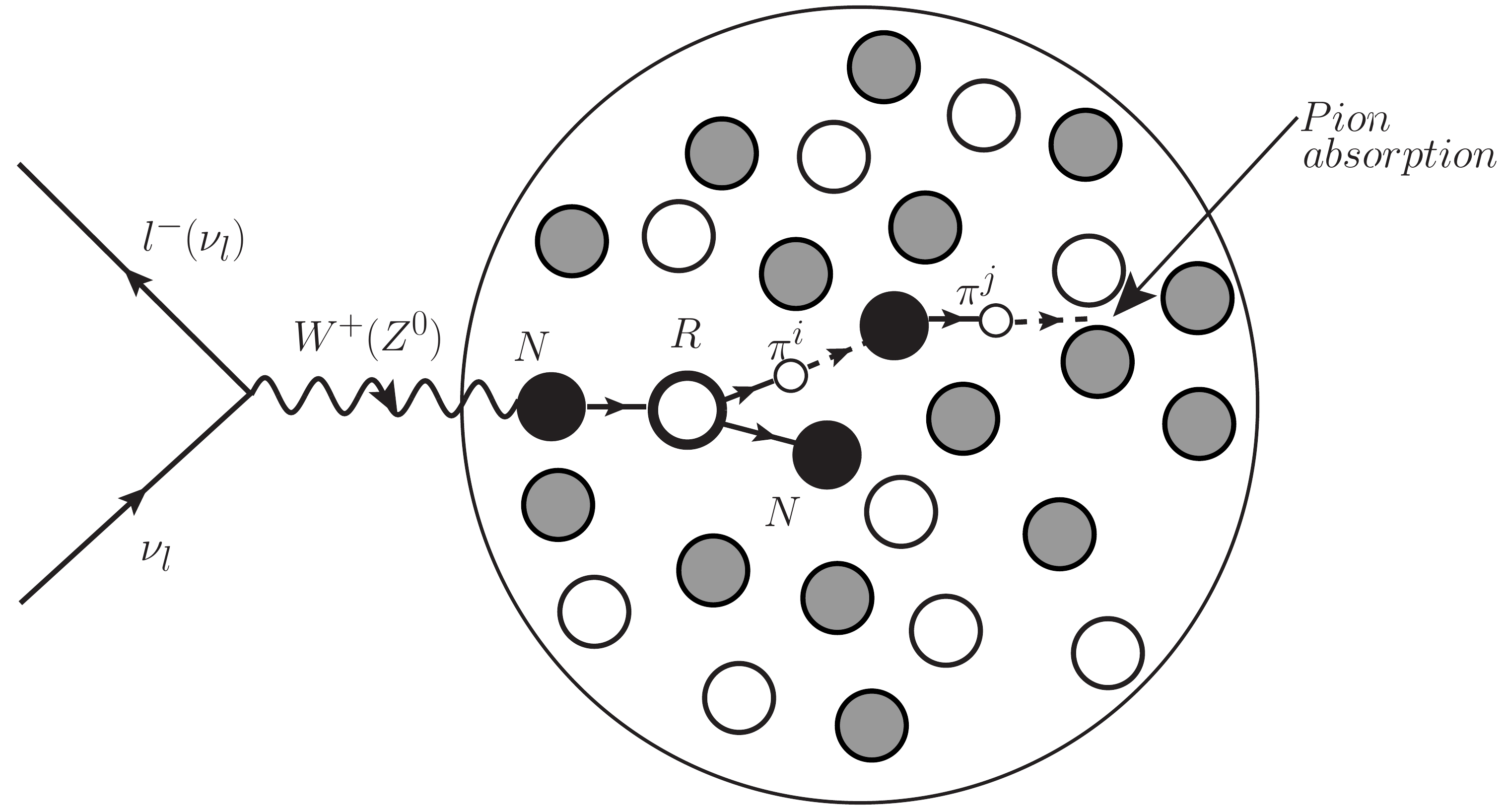}
    \caption{Illustration of the effects of final state interactions.  In this case, a neutrino interaction has produced a heavy resonance, which decays to a nucleon and pion, the pion is absorbed within the nucleus, mimicking a quasi-elastic interaction signature. }
    \label{fig:WG3_fig6}
\end{figure}

\begin{figure}[th]
    \centering
    \includegraphics[width=.45\textwidth]{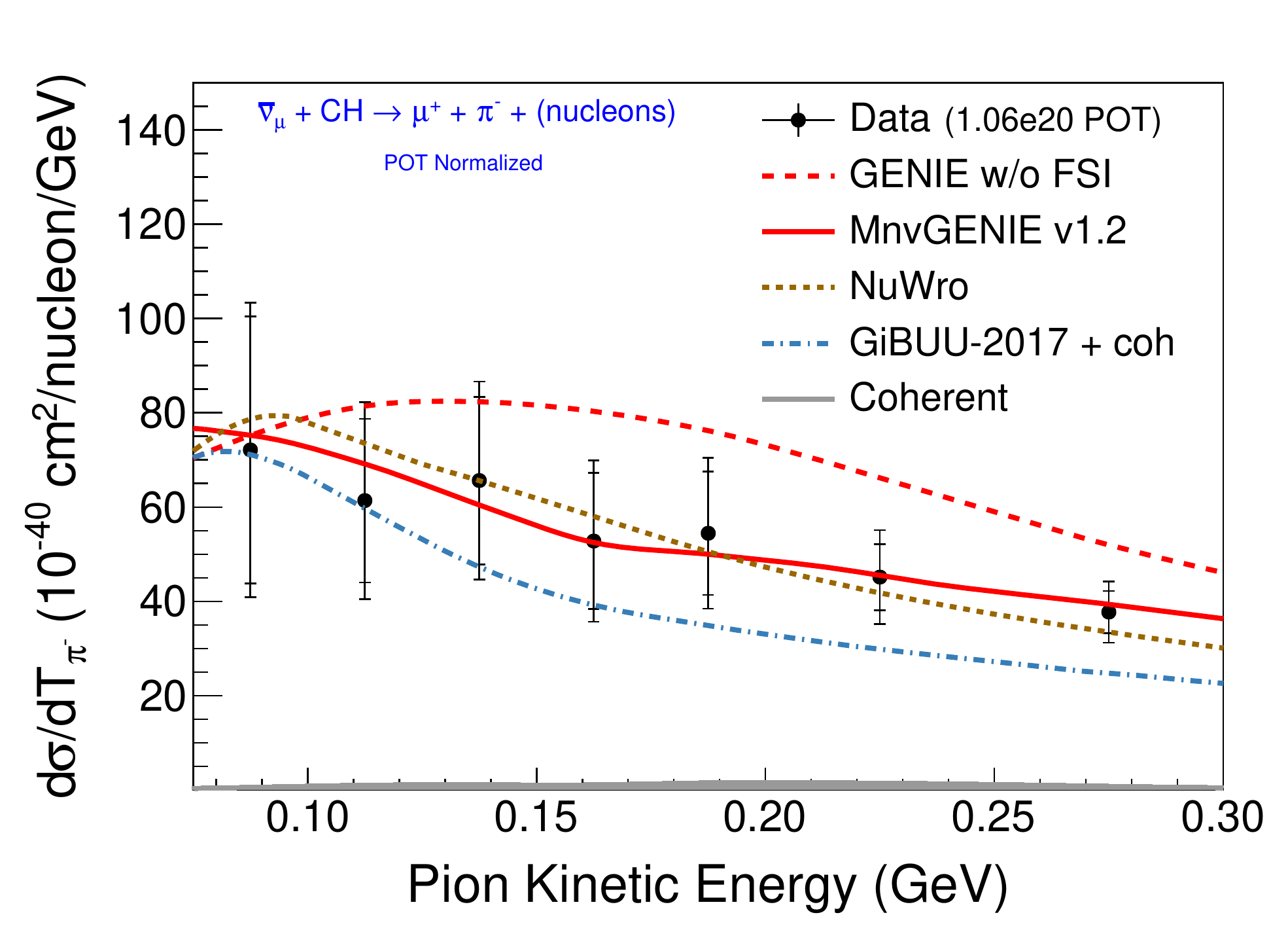}
     \includegraphics[width=.45\textwidth]{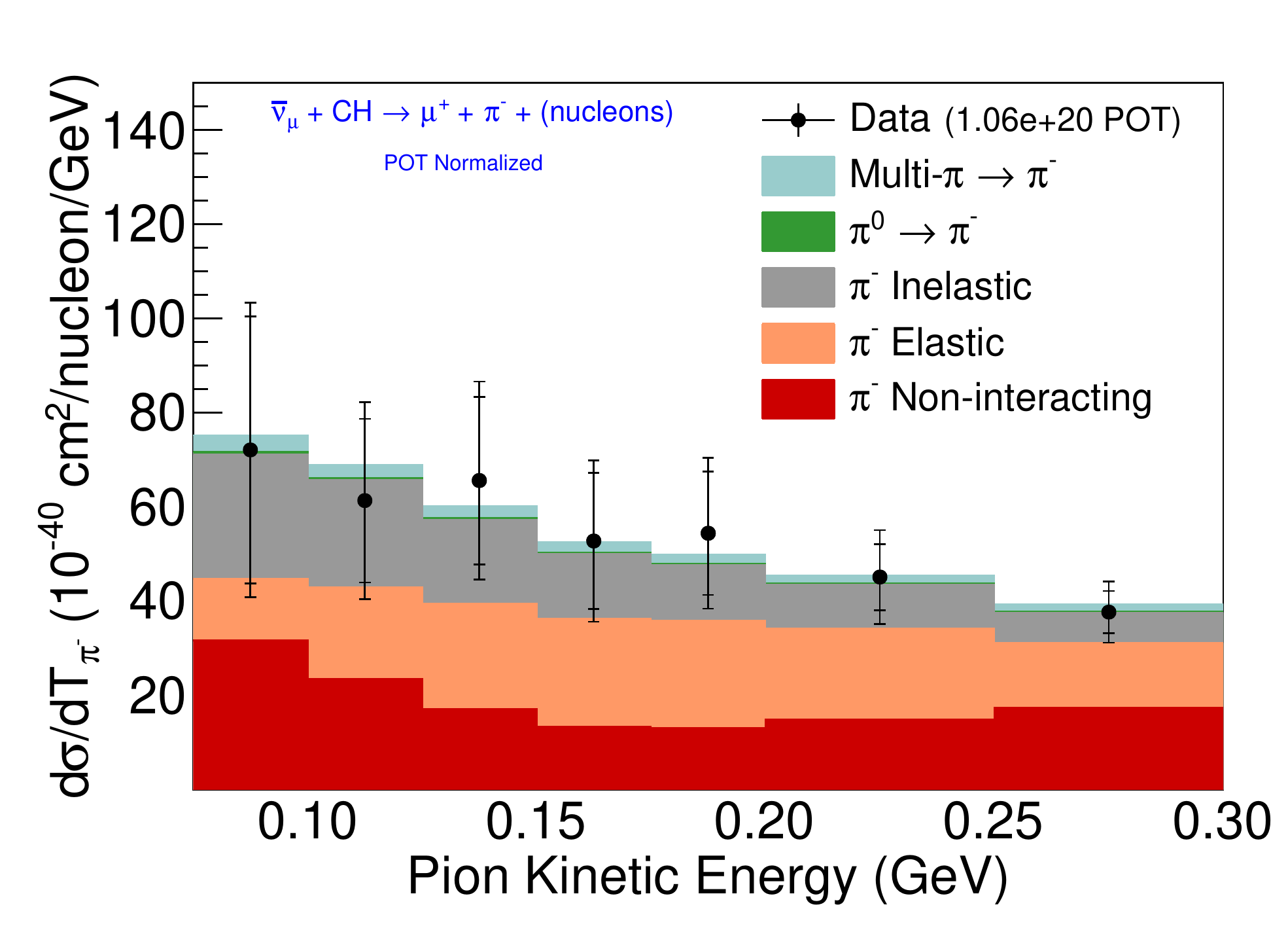}
    \caption{Differential cross sections of charged pions from the MINER$\nu$A experiment at neutrino energies $\sim4$ GeV. The plot on the left compares the data (solid points) to predictions from GENIE, NuWro, and GiBUU. The  plot on the right shows the components in the GENIE model compared to the data. The difference between the red-dashed (no-FSI) and solid red (MINER$\nu$A) curves suggests a substantial FSI effect. Taken from Ref.\ \cite{Le:2019jfy}.}
    \label{fig:WG3_pion}
\end{figure}

In nuclei, the products of the initial scatter must traverse the nuclear material.  This can lead to rescattering, production, or absorption of final state hadrons, and results in a final state that differs from the one that would be expected from the initial neutrino-nucleon interaction. As one example, Fig.\  \ref{fig:WG3_fig6} illustrates the way in which final state interactions of pions from neutrino-induced resonance production can mimic the 2-body signature of quasi-elastic scattering if a pion is absorbed in the target nucleus. In addition, Fig.\ \ref{fig:WG3_pion} shows a comparison of charged pion production data from the MINER$\nu$A experiment \cite{Le:2019jfy} with   modern models \cite{Andreopoulos:2009rq, Golan:2012wx, Buss:2011mx} that highlights the need to include final state interaction effects.

\subsubsection{Random Phase Approximation and Spectral Functions}

Other effects include screening effects at low momentum transfer squared, where the neutrino probe fails to fully resolve the individual nuclei.  These can be modeled using a Random Phase Approximation (RPA)~\cite{Nieves:2004wx, Pandey:2014tza} which provides a more accurate prediction of accelerator and atmospheric neutrino scattering at low $Q^2$ and is also of importance in double beta-decay matrix element calculations~\cite{Vogel:1986nj}.

\subsubsection{Neutrino Interactions in the few Tens of MeV Range}

The neutrino energy regime from a few MeV up to around 100~MeV is relevant for detection of solar and supernova burst neutrinos and for the low-energy tail of the atmospheric neutrino flux.  Reactor neutrinos and neutrinos from pion decay at rest fall in this range as well.  In this regime, neutrinos will interact with nuclei via charged and neutral current interactions, perhaps changing $Z$ but not leading to a nuclear breakup.  Below 100 MeV, only charged current interactions of $\nu_e$ and $\bar{\nu}_e$ flavors are kinematically accessible.  The final-state $e^\pm$ are observable for CC interactions, and if the final-state nucleus is left in an excited state, then there is potentially observable de-excitation debris (e.g., gamma rays or ejected nucleons) for both CC and NC interactions.  The inverse beta decay CC interaction on the simplest nucleus, a free proton ($\bar{\nu}_e + p \rightarrow n + e^+$), is very well understood theoretically. %, as is NC elastic scattering on the proton.  
However, for other nuclei, the interaction cross sections and differential final-state distributions tend to be highly dependent on the nuclear structure of initial and final states; there are considerable theoretical uncertainties.  There are very few measurements of cross sections in this energy range in the literature.  There are existing measurements of CC interactions of $\nu_e$ on  $^{12}$C~\cite{Auerbach:2001hz,Auerbach:2002iy}, $^{127}$I~\cite{Distel:2002ch} and Fe~\cite{Maschuw:1998jf}, and measurement of CC and NC $d$ breakup (reviewed in~\cite{Formaggio:2013kya}) and NC excitation of $^{12}$C~\cite{Armbruster:1998gk} in this energy regime, but little else.  A program of measurements on nuclei relevant for supernova neutrino detectors -- in particular, $^{16}$O, $^{12}$C, $^{40}$Ar, and Pb -- is  needed.  Knowledge of neutrino interaction cross sections on other nuclei is also of value for understanding of nuclear structure and BSM searches.

Neutrinos from pion decay at rest are well suited to measurements of these cross sections.  If positive pions produced by proton collisions on a nuclear target are stopped in a dense material, they decay at rest, producing, per pion, a $\nu_\mu$, a $\bar{\nu}_\mu$ and a $\nu_e$ with a well understood energy spectrum extending to about 50~MeV (half the mass of the muon).  If the pions are produced by a pulsed beam, the time distribution can also be well known.  The stopped-pion neutrino spectrum overlaps significantly with the expected neutrino spectrum from a supernova burst.   Existing and future sources have excellent potential for improved understanding of neutrino interactions in the few tens of MeV range.  Current stopped-pion sources in use for neutrino physics include the SNS~\cite{Bolozdynya:2012xv}, JSNS~\cite{Ajimura:2017ul}, and Lujan~\cite{ccm}; possible sources in future are the ESS~\cite{Baxter:2019mcx}, CSNS~\cite{Wang:2013aka} and DAE$\delta$ALUS~\cite{Alonso:2010fs}.

Neutrino cross sections in the energy range from about 50 MeV to a few hundred MeV are perhaps even more poorly understood than those for few-tens-of-MeV interactions.  There are few near-term prospects for well-understood neutrino sources.  Beta beams are a possibility, although the necessary technology does not currently exist.

\subsubsection{Coherent Elastic Neutrino-Nucleus Scattering}\label{sec:WG3_coherent}

Coherent elastic neutrino-nucleus scattering (CE$\nu$NS) is a process in which a neutrino scatters elastically with an entire nucleus, leaving it intact; the only significant observable from this process is the tiny recoil energy imparted to the nucleus~\cite{PhysRevD.9.1389,Freedman:1977xn,Drukier:1983gj}. 
The differential cross section with respect to nuclear recoil $T$ is 
\begin{equation}
\frac{d\sigma}{dT} = \frac{G_F^2 M}{4\pi} \left(1 - \frac{MT}{2E_\nu^2} \right) \left((1 - 4 \sin^2 \theta_W) Z - N \right)^2,  
\end{equation}
where $M$ is the nuclear mass. If the coherence condition is not fulfilled, both $Z$ and $N$ are modified with nuclear form factors. For reactor neutrinos, coherency is near-exact and those are $\sim1$.
The cross section is relatively large, scaling as the square of the weak charge of the nucleus, which is proportional to $(N-4\sin^2\theta_W Z)^2$.  Because the weak mixing angle is $\sim 1/4$, the proton contribution $Z$ to the weak coupling is small, and so the interaction rate scales as $ N^2$, where $N$ is the number of neutrons in the nucleus.  A nuclear form factor $F^2(Q)$, which is a function of the 4-momentum transfer $Q$, modulates the rate, suppressing it for $Q \gg 1/R$, where $R$ is the nuclear radius.  For neutrino energies less than about 50--100 MeV for medium-size nuclei, this low-$Q$ CE$\nu$NS process dominates.  The maximum recoil energy of the nucleus is $T\sim 2 E_\nu^2/M$, where $M$ is the nuclear mass; this translates to keV-scale recoils for reactor neutrino energies, and to tens of keV scale recoils for stopped-pion neutrino energies.  Detecting these tiny recoils is a technical challenge.  
% Vector dominates

Studies of CE$\nu$NS offer a window into (beyond-the)-Standard-Model physics. 
If high enough precision can be reached (percent level), the Weinberg angle could be determined at low momentum transfer~\cite{Scholberg:2005qs}, or the nuclear neutron distribution and neutron rms radius could be measured~\cite{Amanik:2009zz}.  The difference of the latter to the proton rms radius is called  neutron skin and influences for instance the neutron star equation of state, which links CE$\nu$NS to gravitational waves. 
Because the uncertainties on the nuclear form factor are small (at the few percent level), any deviation from the Standard Model expectation in the expected observables (rate, recoil spectrum and angular distribution as a function of $N$ and $Z$) could point to new interactions, mediators, or new particles in the final state \cite{Barranco:2005yy,deNiverville:2015mwa,Lindner:2016wff,Brdar:2018qqj}.  An anomalous neutrino magnetic moment could turn up in the spectrum as an upturn at low nuclear recoil energy~\cite{Scholberg:2005qs,Kosmas:2015vsa}.  
 The low energy scale of the process allows to distinguish several new physics scenarios which give the same effect in higher-energy oscillation experiments.   Furthermore, it offers a new probe of sterile neutrino oscillations~\cite{Formaggio:2011jt}.

CE$\nu$NS is also a new flavor-blind tool for observations of natural neutrinos -- the Sun, supernovae, geoneutrinos.  It is a background for searches for both natural and accelerator-produced dark matter, so understanding of its cross section matters for these searches.  Finally, thanks to its relatively large cross section, CE$\nu$NS is conceivably useful for nuclear reactor monitoring.
% Need some references for these

\begin{figure}[t]
\begin{center}
\includegraphics[width=.9\textwidth]{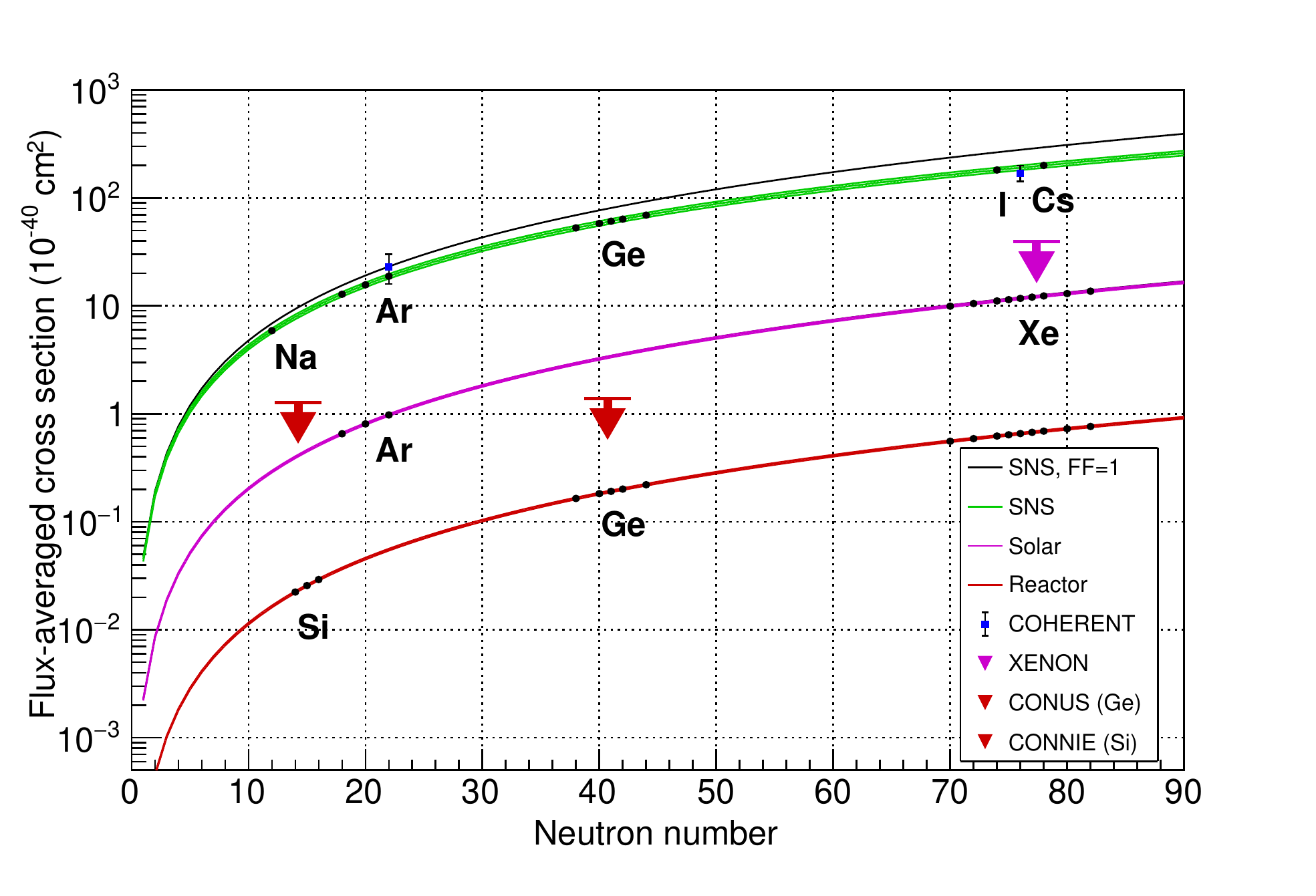}
    
\end{center}
\caption{CE$\nu$NS cross section as a function of the number of neutrons in a nucleus, averaged over the stopped-pion neutrino spectrum (green, above), the solar $^8$B spectrum (magenta) and the reactor spectrum $>0.1$~MeV (red, below). The black line represents the flux-averaged cross section for unity form factor; colored lines include the effect of the nuclear form factor. Indicated also by black dots are several isotopes for which there are existing experiments.  The blue points with error bars indicate measurements by COHERENT~\cite{Akimov:2017ade, Akimov:2020pdx}. The magenta arrow shows the upper limit from XENON1T~\cite{Aprile:2020thb} and the red arrows show the upper limits from CONNIE~\cite{Aguilar-Arevalo:2019jlr} and CONUS~\cite{Bonet:2020awv}.  }\label{fig:cevns-xscn}
\end{figure}

In 2017, the COHERENT experiment made the first measurement of the CE$\nu$NS process in Na-doped CsI crystals using the neutrinos from the Spallation Neutron Source~\cite{Akimov:2017ade}. In 2020, COHERENT made the first CE$\nu$NS measurement in Ar~\cite{Akimov:2020pdx}. COHERENT plans measurements of CE$\nu$NS on Ge and NaI in the near future, and possibly additional targets.  The Coherent Captain Mills experiment is planning to make use of the Los Alamos Lujan facility~\cite{CCM:2021leg}, and the European Spallation Source~\cite{Baxter:2019mcx} offers high power but less sharply pulsed beam.  Fig.~\ref{fig:cevns-xscn} summarizes measurements.

The frontier for future CE$\nu$NS experiments is at low recoil energy.  For the $\sim 30$ MeV neutrinos from stopped-pion decay, tens-of-keV nuclear recoils are relatively accessible. However for reactor neutrinos, the required thresholds are less than 1~keV, and in that experimental regime, achieving good signal to background is  technically very challenging in spite of high reactor fluxes.  Nevertheless, a number of experimental collaborations deploying diverse low-thresholds technologies are taking on this challenge.  These include CONUS~\cite{Hakenmuller:2019ecb, Bonet:2020awv}, CONNIE~\cite{Aguilar-Arevalo:2019jlr}, Ricochet~\cite{Billard:2016giu}, RED~\cite{RED-100:2019rpf}, MINER~\cite{Agnolet:2016zir}, Nu-Cleus~\cite{Rothe:2019aii},  NuGEN~\cite{Belov:2015ufh}, and NEON~\cite{Choi:2020gkm}.  As for dark matter experiments, precise knowledge of the detector response for nuclear recoils is important.

\subsubsection{Ultra High Energy Neutrino Cross Sections}
At energies above $\sim 1$ TeV, neutrino interactions are dominated by the DIS contribution and thew $Q^2$ dependence of both the weak boson propagator and the structure functions come into play yielding a decrease in the neutrino cross section $\sigma_{\rm CC}/E_\nu$ as $E_\nu$ increases. Predictions of the neutrino cross section above 10 GeV rely on structure functions evaluated using global fits to parton distribution functions. While there are no direct measurements of the neutrino-nucleon cross section above $E_\nu\sim 350$ GeV, the parton distribution functions in the relevant $(x,Q^2)$ ranges are well constrained and there is widespread agreement in predictions of the cross section for $E_\nu\lsim 10^7-10^8$ GeV \cite{Gandhi:1998ri,Reno:2004eb,Jeong:2010za,Gluck:2010rw,Connolly:2011vc,CooperSarkar:2011pa,Block:2014kza,Arguelles:2015wba,Gauld:2019pgt}. At even higher neutrino energies, small-$x$ extrapolations of the parton distribution functions are less constrained. For example, with Standard Model perturbative QCD evolution of the parton distribution functions and $E_\nu=10^{11}$ GeV, the neutrino-nucleon scattering cross section is uncertain by a factor of $\sim 4$ \cite{Connolly:2011vc,CooperSarkar:2011pa}.  Allowing non-perturbative QCD evolution \cite{Henley:2005ms} can  increase the range of predictions at the same neutrino energy to a factor of $\sim 10$, as shown in Ref.\ 
\cite{Ackermann:2019cxh}, for example.

What are the prospects for measurements of ultra high energy neutrino cross sections? The angular-dependent atmospheric and astrophysical neutrino-induced event rates in the IceCube Neutrino Observatory \cite{Aartsen:2016nxy} depend on the cross section through neutrino flux attenuation in the Earth and neutrino interactions in or near the detector \cite{Klein:2019nbu}. These data allow an extraction of the neutrino-nucleon cross section 
\cite{Aartsen:2017kpd,Bustamante:2017xuy} and inelasticity distribution \cite{Aartsen:2018vez} in the TeV-PeV neutrino energy range. Cosmic sources of ultra high energy neutrinos potentially provide fluxes of neutrinos with energies up to $E_\nu=10^{12}$ GeV, where neutrino-induced horizontal air showers and tau-neutrino-induced upward air showers may allow for neutrino cross section measurements \cite{Feng:2001ue,Kusenko:2001gj}.  A first Glashow resonance electron-antineutrino candidate at 6.3~PeV has been reported by IceCube~\cite{IceCube:2021rpz}.   A summary of current and proposed ground-based and space-based neutrino detectors for ultra high energy neutrinos appears in Ref.~\cite{Ackermann:2019cxh}. At the LHC, experiments proposed to be positioned along the beam line for BSM searches will intercept large numbers of muon and electron neutrinos and antineutrinos from pions and kaons \cite{Abreu:2019yak, XSEN:2019bel, SHiP:2020sos}, and all three neutrino flavors from charm hadron decays \cite{Abreu:2019yak,Bai:2020ukz}. The FASER$\nu$ experiment \cite{Abreu:2020ddv}, for example, would produce neutrino cross section measurements that bridge the energy gap between accelerator-based neutrino and IceCube measurements.

\subsubsection{Present and Future Experiments}

Neutrino cross section on protons (and deuterium) were historically measured in bubble chamber experiments prior to 1990~\cite{Formaggio:2013kya,Zyla:2020zbs}. %Data samples on hydrogen ranged up to 20,000 fully reconstructed interactions and beam energies ranged from 1 to 200 GeV.  
These measurements have been highly influential on understanding the basic physics of neutrino-nucleon scattering due to their ability to fully reconstruct most of the final state.  Since 1990, most neutrino experiments have emphasized cross section measurements relevant to neutrino oscillations;  concentrating on high statistics, energies in the $0.2-10$ GeV energy range and on target nuclei  best suited for short and  long-distance neutrino oscillation experiments. Table \ref{WG3_expts} summarizes the cross section experiments that have been published recently. Modern neutrino cross section experiments such as ArgoNeuT, MicroBooNE, MINER$\nu$A, MiniBooNE, NOvA, and T2K now have data samples with hundreds of thousands of neutrino interactions in quasi-elastic and meson final state channels. These high statistics measurements have led to increasing refinement of model parameters. In the future, additional high statistics neutrino-nucleus scattering data on argon is expected from ICARUS, SBND, and the DUNE near detector.  
Precise determination of the cross sections of neutrinos with target material are absolutely necessary to precisely determine neutrino parameters, and to probe physics beyond the standard neutrino picture. 

\begin{table}
\begin{tabular}{l|c|l|l}
\hline\hline
Experiment& Neutrino Energies & Neutrino Target & Websites\\
\hline
  ArgoNeuT   & $1-20$ GeV &Ar & \href{https://t962.fnal.gov/}{t962.fnal.gov} \\%cite websites for each \\ %\cite{Acciarri:2020lhp,Acciarri:2018ahy,Acciarri:2014isz,Anderson:2011ce,Acciarri:2018ahy,Acciarri:2015ncl,Acciarri:2014eit}\\
  COHERENT &  $<$ 50 MeV & Na, Ar, Ge, Cs, I  &\href{ https://sites.duke.edu/coherent/}{COHERENT}\\ %\url{https://coherent.phy.duke.edu}\\
%\cite{Akimov:2017ade,Akimov:2020pdx}\\
  K2K & $0.2-3$ GeV & water, scintillator  & \href{https://neutrino.kek.jp/}{K2K}\\
  MiniBooNE     & 0.2-3 GeV &liquid scintillator & \href{https://www-boone.fnal.gov/}{MiniBooNE}\\ %\cite{Grange:2013iwa,
% Aguilar-Arevalo:2013nkf,
% Aguilar-Arevalo:2013dva,
% AguilarArevalo:2010bm,
% AguilarArevalo:2010xt,
% AguilarArevalo:2010cx,
% AguilarArevalo:2010zc,
% AguilarArevalo:2009ww,
% AguilarArevalo:2009eb,
% AguilarArevalo:2010bm,
% AguilarArevalo:2010xt,
% AguilarArevalo:2009ww}\\
  MicroBooNE &  $0.2-3$ GeV & Ar &\href{https://microboone.fnal.gov/}{MicroBooNE} \\
  MINER$\nu$A & $1.5-20$ GeV & scintillator, He, C, Fe, Pb &\href{https://minerva.fnal.gov/}{MINERvA} \\
  MINOS & $3-50$ GeV & Fe & \href{https://www-numi.fnal.gov/}{MINOS}\\
  NOvA & $0.8-4$ GeV & liquid scintillator &\href{https://novaexperiment.fnal.gov/}{NOvA} \\
  SciBooNE & $0.2-3$ GeV &liquid scintillator &\href{https://www-sciboone.fnal.gov/}{SciBooNE}\\
  T2K INGRID & $0.2-10$ GeV & scintillator &\href{https://t2k-experiment.org/}{T2K} \\
  T2K ND280 & $0.2-10$ GeV & water and scintillator & \href{https://t2k-experiment.org/}{T2K} \\
  \hline\hline
\end{tabular}
\caption{Recent neutrino cross section experiments, their beam energies, and nuclear targets.}
\label{WG3_expts}
\end{table}

\subsection{Neutrino Interactions: Summary}

Predictions for the rates and topologies of neutrino interactions with matter are a crucial component in many current investigations within nuclear and astroparticle physics. Ultimately, we need to precisely understand neutrino-matter interactions to enable progress in high priority physics including neutrino oscillations, supernova dynamics, and BSM searches. Such improved understanding must in tandem include theoretical calculations of neutrino processes within a nuclear environment as well as dedicated experimental measurements to verify such predictions across wide energy ranges and varying nuclei. 

Over the past decade, notable advances have been made on both the theoretical calculations and experimental measurements of neutrino-nucleus scattering. However, neutrino-nucleus interaction uncertainties remain a limiting factor in many neutrino oscillation searches at both short and long distances. Experiments using heavier nuclear targets to increase their signal yields have to contend with the presence of significant nuclear effects impacting neutrino interaction rates, particle kinematics, and observed final states. Uncertainties in both the neutrino interaction cross sections and associated nuclear effects must be understood to maximize the sensitivity of an experiment to neutrino oscillations, interpret a supernova neutrino burst observation, and to uncover possible BSM physics in neutrinos. Continued progress on both the theoretical and experimental fronts is crucial for the success of our endeavors in all of these areas. The future is bright with such efforts underway using electron and neutrino scattering probes, encompassing multiple theoretical approaches, and spanning a wide kinematic range with a variety of nuclear targets.

\label{chap:interactions}

\clearpage

\section{The Number of Neutrinos}
\label{WG0}
% change to how many neutrino species are there? 
%

\subsection{Introduction }

Sterile neutrinos are singlets of the SM gauge group. Since  they do not couple directly to the gauge bosons they can only participate in  weak interactions  
through mixing with active neutrinos.
The mass and number of these hypothetical particles are unconstrained by theory, so in spite of any theoretical prejudice, they can be fairly light. 

The quest for light sterile neutrinos, in the sub-eV--eV mass scale range, has been motivated by a series of low energy anomalies which cannot be accounted for 
by the standard three-neutrino framework.
These involve SBL experiments with neutrinos from accelerators, nuclear reactors and radioactive sources. 

The {\em LSND anomaly},  is the 3.8$\sigma$ excess of events compatible to $\bar \nu_e$ appearance in a $\bar \nu_\mu$ beam observed by the LSND experiment \cite{Aguilar:2001ty}, 
the {\em MiniBooNE anomaly}, is the $4.8\sigma$ excess of electron-like events in the MiniBooNE experiment observed in both $\nu_\mu$ and $\bar \nu_\mu$ beams~\cite{Aguilar-Arevalo:2018gpe}, 
the {\em reactor antineutrino anomaly}, is the $\sim 6$\% deficit (a $3\sigma$ effect)  of $\bar \nu_e$ in reactor SBL ($<  100$ m) experiments resulting from the re-evaluation of the reactor antineutrino flux~\cite{Mention:2011rk,Huber:2011wv}  (although there are some doubts about the validity of this hint, see Sec.\ \ref{sec:reactorflux}) and, finally, the {\em gallium anomaly}, is the deficit of about $15$\%  (a $2.5\sigma$ effect) in the observed neutrino count rate  in the calibration runs with radioactive sources of the Ga solar neutrino experiments GALLEX/GNO and SAGE~\cite{Giunti:2010zu}. Very recently BEST \cite{ Barinov:2021asz } confirmed the gallium anomaly. The deficit becomes larger ($20\pm 5\%$) and has a larger significance. 

Regardless of the different sources, baselines and energy ranges of these experiments, all of the above results can be understood individually via short-baseline neutrino oscillations driven by 
$\Delta m^2_{\rm SBL}\sim 1$~eV$^2$, a substantially higher scale than the solar ($\Delta m^2_{21}$) and 
atmospheric ($|\Delta m^2_{32}|$) ones. So  the oscillation interpretation requires a fourth neutrino mass  eigenstate to account for this new (higher) mass-squared difference. 
  
On the other hand, the Large Electron-Positron collider result on the $Z$ invisible decay width established that there are only three light neutrinos, with masses lower than $m_Z/2$ and
 SM couplings to this particle~\cite{ALEPH:2005ab}. This implies that if there is a fourth light neutrino, it must be sterile in nature having no direct couplings to the SM bosons.
 
The simplest implementation of this idea of adding an extra massive state to the neutrino mass spectrum to solve the SBL anomalies and still provide a good  flavor oscillation solution to the solar and atmospheric data.  The PMNS matrix is now a $4\times4$ matrix, new parameters of interest are a new mass denoted $m_4$, and PMNS matrix elements $U_{\alpha 4}$, with $\alpha = e, \mu, \tau$. In the usual parametrization, in particular, $|U_{e 4}| = \sin \theta_{14}$. 
It comes down to two basic schemes: the (2+2) and the (3+1) schemes.  The (2+2) scheme, where two groups of mass eigenstates, one  accounting for the solar and one for the atmospheric mass-squared difference, are separated by a gap, has been  since long discarded by global fits of solar and atmospheric neutrino data~\cite{Maltoni:2002xd}. The (3+1) scheme, however, remains a viable possibility.
 In the (3+1) scheme there are four possible mass spectra, depending on the mass ordering.  Here the  three (mostly) active mass eigenstates  can 
 account for the solar and atmospheric neutrino data in the usual way and the  (mostly) sterile mass eigenstate can be used to explain  the anomalies.
It turns out that due to the dominance of the large $\Delta m^2_{\rm SBL} \approx \Delta m^2_{41} \gg \Delta m^2_{21}, \vert \Delta m^2_{32} \vert$ the  survival and the oscillation probabilities  for SBL experiments become effectively two-neutrino highly 
correlated probabilities, 
   which in particular implies  that $\nu_\mu \to \nu_e$ appearance will result in both $\nu_e \to \nu_e$ and $\nu_\mu \to \nu_\mu$ disappearance,  making it non-trivial to both satisfy the 
   SBL anomalies and  the negative results from  disappearance experiments; see Sec.~\ref{sec:sterile_global}. That is why different global analyses including the   available data, while they may differ on the conclusions about 
   the degree of disagreement,    seem to indicate  tension between appearance and disappearance SBL data in this scenario.

   Non-oscillation experiments that aim to determine the  
   absolute neutrino mass scale~\cite{Aker:2020vrf} and whether neutrinos are Dirac or Majorana particles~\cite{Rodejohann:2011mu}, can also independently help to constrain the parameter space spanned by $\Delta m^2_{41}$ and $U_{e4}$.   
   
   Neutrino properties leave also observable imprints in cosmological observations. Cosmological data can be complementary to laboratory
   experiments by placing limits on the sum of neutrino masses and the number of relativistic light degrees of freedom. Both seem to currently challenge the sterile neutrino needed to solve the  anomalies. Non-standard effects~\cite{Dasgupta:2013zpn,Hannestad:2013ana} may, nevertheless, reduce sterile neutrino production in the early Universe evading some of these limits.
   
   The existence of light sterile neutrinos is still an open 
   question and some of the aforementioned  SBL anomalies may have some other physical origin\footnote{Moreover, statistical issues may lead to an over-interpretation of the results \cite{Agostini:2019jup,Giunti:2020uhv}.}, thus they 
   must  continue to be vigorously scrutinized by experiments.
\label{sec:sterile_intro}

\subsection{Sterile Neutrinos and Accelerators }
\subsubsection{Current Status }

Across history, there have been multiple searches for the existence of sterile neutrinos using accelerator-based neutrino sources and studying both $\nu_e$ appearance and $\nu_\mu$ disappearance oscillation signatures (see Tab.\ \ref{Tab:sterile:accelerator-neutrino}). Two experiments have observed anomalous signals. The first is the LSND experiment which searched for neutrino oscillations using neutrinos from a stopped pion source at Los Alamos National Laboratory in the mid to late 1990s. Using the inverse beta decay process, LSND observed a $3.8\sigma$ excess of $\overline{\nu}_e$ events that to this day remains unexplained~\cite{Aguilar:2001ty}. One should note that the KARMEN experiment, using a similar setup, found no evidence for $\overline{\nu}_e$ appearance \cite{Armbruster:2002mp}, though a combination with LSND still allows parameter space \cite{Church:2002tc}.

More recently, the MiniBooNE experiment has likewise reported $4.7\sigma$ $\nu_e$ and $2.7\sigma$ $\overline{\nu}_e$ event excesses after analyzing its complete data set after $\sim15$ years of operations~\cite{Aguilar_Arevalo_2021,Aguilar-Arevalo:2018gpe}. The MicroBooNE~\cite{Acciarri_2017} experiment at Fermilab uses the same neutrino source as MiniBooNE, but a different and more capable detector technology, a liquid argon TPC, will be probing the source of the excess of events seen in MiniBooNE. Results from MicroBooNE using a variety of analysis strategies and multiple neutrino interaction modes are expected soon.

\begin{table}[h]
\begin{center}
\begin{tabular}{c|c|c|c}
\hline\hline
Experiment & Baseline [m]  & Target Material  & Mode \\ \hline \hline
CCFR & 715, 1116 & iron, scintillator & $\nu_\mu$, $\overline{\nu}_\mu$ disappearance \\ \hline
CDHS & 130 & solid scintillator & $\nu_\mu$ disappearance \\ \hline
KARMEN & 17.7 & liquid scintillator  & $\overline{\nu}_e$ appearance \\ \hline
LSND   &  30   & liquid scintillator  & $\overline{\nu}_e$ appearance  \\ \hline
MicroBooNE & 470 & liquid argon & $\nu_e$ appearance  \\ \hline
MiniBooNE & 541 & mineral oil & $\nu_e$, $\overline{\nu}_e$ appearance  \\ \hline
MiniBooNE/SciBooNE & 100, 540 & solid scintillator, mineral oil & $\nu_\mu$, $\overline{\nu}_\mu$ disappearance \\ \hline
MINOS, MINOS+ & 1040, 734 k  & solid scintillator & $\nu_\mu$, $\overline{\nu}_\mu$ disappearance \\ \hline
NOMAD  & 625 & solid scintillator  & $\nu_e$ appearance \\ \hline
NOvA & 1000, 809 k & solid scintillator & NC disappearance \\ \hline
T2K & 280, & multiple sub-detectors, & NC disappearance,   \\ 
    & 295 k & water Cherenkov & $\nu_e$ \& $\nu_\mu$ disappearance \\ \hline\hline
\end{tabular}
\caption{Past and currently operating accelerator-based short-baseline neutrino experiments. In this list, LSND and MiniBooNE are the two accelerator-based neutrino experiments which have observed signals, both in appearance mode.
\label{Tab:sterile:accelerator-neutrino}}
\end{center}
\end{table}

Both the LSND and MiniBooNE observations were made in the appearance mode. Interestingly, none of the accelerator-based short-baseline neutrino experiments (including MiniBooNE itself~\cite{Cheng_2012}) have observed the $\nu_\mu$ disappearance signature that one would expect to observe as the muon neutrinos oscillate to electron neutrinos through a sterile neutrino state. This includes the non-observation of $\nu_\mu$ disappearance from CCFR, CDHS, MiniBooNE, MINOS, MINOS+, NOvA, and T2K (see Tab.\ \ref{Tab:sterile:accelerator-neutrino}). 
Recently, there has been a large body of theoretical work that attempts to collectively explain the experimental measurements by invoking additional BSM physics including heavy sterile neutrinos, dark portals, new scalar bosons, and hidden sector sector physics~\cite{Ballett:2018ynz,deGouvea:2019qre}. Such mechanisms tend to lead to more complex final states (for example, $e^+e^-$ pairs as opposed to a single electron or positron), some of which could be observed in current and future liquid argon-based neutrino experiments.

\subsubsection{Future Prospects}
There is more to come on the accelerator-based sterile neutrino front, as summarized in Tab.\ \ref{Tab:sterile:accelerator-neutrino-future}. While MicroBooNE is currently operating, it will soon be accompanied by two additional liquid argon detectors (SBND and ICARUS) as part of the Short-Baseline Neutrino (SBN) program at Fermilab~\cite{acciarri2015proposal}. This is the first time that a series of liquid argon TPCs will have been positioned on the same beamline to study neutrino oscillations. The ICARUS detector is currently being commissioned at Fermilab following an extensive refurbishment at CERN. The SBND detector is currently under construction and will sit closest to the Fermilab Booster neutrino source. Data from the SBND near detector to constrain the un-oscillated neutrino flux, MicroBooNE (with its head start), and ICARUS (with its large mass and longer baseline) will work together to fully address the sterile neutrino phase space suggested by the LSND and MiniBooNE anomalies.

\begin{table}[t]
\begin{center}
\begin{tabular}{c|c|c|c}
\hline\hline
Experiment & Baseline [m]  & Target Material  & Mode \\ \hline \hline
IsoDAR & 16 & liquid scintillator & $\overline{\nu}_e$ interactions \\ \hline
JSNS$^2$ & 24 & liquid scintillator & $\overline{\nu}_e$ appearance \\ \hline
$\nu$STORM & TBD & TBD & $\nu_e$, $\overline{\nu}_e$ appearance \& $\nu_\mu$, $\overline{\nu}_\mu$ appearance\\ \hline
SBN & 110, 470, 600 & liquid argon & $\nu_e$ appearance, $\nu_\mu$ disappearance  \\ \hline\hline
\end{tabular}
\caption{Future planned and proposed accelerator-based short-baseline neutrino experiments that will search for sterile neutrinos.
\label{Tab:sterile:accelerator-neutrino-future}}
\end{center}
\end{table}

In addition, the JSNS$^2$~\cite{Maruyama:2016mqj} experiment, which is aiming to provide a direct test of LSND, is about to start operations at J-PARC in Japan using a detector filled with gadolinium loaded liquid scintillator exposed to a beam of neutrinos from muon decay at rest. Future probes of sterile neutrinos will also be possible in the DUNE~\cite{Abi_2020} and Hyper-K~\cite{hyperk-loi} near detectors with extremely high statistics given the intensities of their planned neutrino beams. Next-generation sterile neutrino searches are also in the planning and include concepts using a high intensity $^8$Li beta-decay-at-rest antineutrino source  (IsoDAR~\cite{Bungau:2012ys,Conrad:2013sqa}) and neutrinos uniquely created from the decay of muons confined within a storage ring ($\nu$STORM~\cite{Adey:2013pio}).
\label{sec:sterile_acc}

\subsection{Sterile Neutrinos and Reactors }

As discussed in Sec.\ \ref{sec:reactorflux}, 
the 6\% flux deficit compared to the new prediction in 2011 \cite{Mueller:2011nm} is called reactor antineutrino anomaly (RAA), and was suggested to be due to active-to-sterile neutrino oscillation at an eV-scale with  best-fit values of $\Delta m^2_{41} = 2.4$ eV$^2$ and $\sin^2 2\theta_{14}$ = 0.14 \cite{Mention:2011rk}. 
Recent developments of this particular hint and the  current discussion on its validity can be found in Sec.\  \ref{sec:FluxAnomaly}. 
As discussed in previous sections, evidence of the eV-scale sterile neutrinos was also observed in accelerator-based (LSND~\cite{Athanassopoulos:1996jb}, MiniBooNE~\cite{Aguilar-Arevalo:2013pmq}) experiments and also in calibration measurements of radio-chemical solar neutrino experiments (GALLEX, SAGE). 

\subsubsection{Current Status}\label{sec:sterile_rea0}

Uncertainties from each VSBL experiment from the '80s and the '90s were large. 
To reduce these uncertainties in measurements and to have better understanding of the unexpected ``5~MeV excess'', many VSBL experiments have been created to take data.
Table~\ref{t:VSBL_comp} summarizes the VSBL experiments currently operating or being prepared. 
NuLat and CHANDLER~\cite{Huber:2019mro,Subedi:2019shd} are mainly for  nuclear non-proliferation and currently in R\&D. 
Among these VSBL reactor neutrino experiments, NEOS, STEREO, Neutrino-4 and PROSPECT  are based on liquid scintillators while the others on plastic scintillator. NEOS and DANSS detect neutrinos from commercial reactors while the others from research reactors, i.e.\ they use a $^{235}$U enriched neutrino source. PROSPECT and NEOS have the best energy resolution, 4.5\% and 5\%, respectively while the others have values larger  than 10\%. Only NEOS has a  homogeneous detector while the others are segmented (2D or 3D), i.e.\ better for background rejection.

\begin{table}[t]
\centering
\begin{tabular}{l|r|r|c|c|c}
\hline\hline
Experiment & Power & Baseline & Target mass & Target & Segmentation \\
           & [MW$_{\mathrm{th}}$] & [m] & or volume & material & \\
\hline
NEOS       & 2800     & 24    & $\sim$1~m$^3$ & GdLS                    & No \\
DANSS      & 3100     & 11-13 & 1~m$^3$       & PS (Gd layer)           & quasi-3D \\
Neutrino-4 & 100      & 6-12  & 1.8 ton       & GdLS                    & 2D \\
PROSPECT   & 85       & 7-12  & 4 ton         & $^{6}\rm{LiLS}$         & 2D \\
SoLid      & 72       & 6-9   & 1.6 ton       & PS ($^{6}\rm{Li}$ layer)& 3D \\
STEREO     & 57       & 9-11  & 2.4~m$^3$     & GdLS                    & 2D \\
NuLat*     & any      & any   & 0.9 ton       & $^{6}\rm{LiPS}$         & 3D \\
CHANDLER*  & any      & any   & $\sim$1 ton   & PS ($^{6}$Li layer)     & 3D \\
iDREAM*    & 3100     &  20   & 1~m$^3$       & GdLS                    & No \\
\hline\hline
\end{tabular}
\caption{
Comparison of the current VSBL reactor neutrino experiments~\cite{Seo:2020ehv}. 
--- *) The main purpose of NuLat, CHANDLER and iDREAM is to monitor reactors by observing neutrinos from the reactors. 
}
\label{t:VSBL_comp}
\end{table}

All the VSBL experiments perform model-independent analyses, 
and so far no significant evidence for eV-scale sterile neutrino was observed from these experiments. 
NEOS'~\cite{Ko:2016owz} spectral shape divided by Daya Bay's showed an oscillation pattern indicating possibly a  sterile neutrino at 2.5$\sigma$ but further reduction of systematic uncertainty is needed to claim it.

Neutrino-4~\cite{Serebrov:2018vdw} claimed observation of sterile neutrinos with best-fit values of $\Delta m^2_{41} \simeq 7$ eV$^2$ and $\sin^2 2\theta_{14} \simeq 0.4$. However, the statistical significance of the result was only 2.8$\sigma$. Moreover, the Neutrino-4 analysis method was questioned in Refs.\ \cite{Danilov:2018dme,Danilov:2020rax,Almazan:2020drb,Giunti:2021iti}.
Neutrino-4 replied to these critical comments on their analysis~\cite{Serebrov:2020yvp,Serebrov:2020wny} 
  and addressed recently two of them~\cite{Serebrov:2020kmd}. A more accurate treatment of energy resolution resulted in the reduction of the significance from 2.8$\sigma$ to 2.5$\sigma$. With increased data sample the significance of the signal reached 2.9$\sigma$. The  MC-based statistical analysis employed now by Neutrino-4 reduced the significance of the signal from 2.9$\sigma$ to 2.7$\sigma$.
The obtained  fit results of $\Delta m^2_{41} = 7.3\pm 1.17$ eV$^2$ and $\sin^2 2\theta_{14} = 0.36\pm0.12_{\rm stat}$ are in tension with limits obtained by Daya Bay, Bugey-3 and RENO (see for example~\cite{MINOS:2020iqj}), where the limits are obtained by taking into account the large uncertainties of the predictions for the antineutrino flux from reactors. However, the very recent BEST results  \cite{Barinov:2021asz} favor large $\Delta m_{41}^2$ and large  $\sin^2 2\theta_{14}$ in agreement with the Neutrino-4 best-fit values, see Sec.\ \ref{sec:sterile_radio}. 
The Neutrino-4 results are also in tension with limits obtained by PROSPECT~\cite{Andriamirado:2020erz}. 
A comparison of the Neutrino-4 result with other experiments was done in Ref.~\cite{Serebrov:2020rhy}.
The Neutrino-4 claim can be tested by upgraded DANSS~\cite{Svirida:2020zpk } and PROSPECT~\cite{ Andriamirado:2021qjc}.
Currently Neutrino-4 is constructing a new detector with 3 times better sensitivity in comparison with the existing one~\cite{Serebrov:2020kmd }. Therefore, situation with the Neutrino-4 claim will be clarified in $3-4$ years.

Both DANSS~\cite{Alekseev:2018efk,Danilov:2019aef} and NEOS excluded the RAA best-fit values at 5$\sigma$ and 4.6$\sigma$, respectively. Recent PROSPECT results using 96 calendar days of reactor-ON data \cite{Andriamirado:2020erz} showed no evidence of sterile neutrinos and also disfavored the RAA best-fit at 2.5$\sigma$. 
Recent STEREO~\cite{AlmazanMolina:2019qul} result using 179-day reactor-ON data rejects the RAA at more than 99.9\% C.L.  \\

\subsubsection{Future Prospects}

Recently, NEOS has finished its phase-II data-taking in October 2020, and their new results would be available in 2021 or 2022 with 500-day reactor-ON data, covering a full fuel cycle, with two sets of reactor-OFF data (before and after the reactor-ON period) for background subtraction. 
Neutrino-4 will upgrade their detector (Neutrino-6) with a pulse shape discrimination  capability. 
DANSS will continue to take data until Spring 2022 to cover one more reactor-OFF period and will upgrade the spectrometer in 2022 in order to improve the energy resolution. 
PROSPECT plans to de-construct, repair and upgrade their detector because of an unacceptable number of non-functioning PMTs. PROSPECT is expected to redeploy the detector after its upgrade (PROSPECT-II) for additional years of data taking. 
STEREO 
collected 334-days reactor-ON data and finished data taking. 
SoLid has collected 196 (146)-day reactor-ON (OFF) data since 2018 and is expected to release new results hopefully soon. 
\label{sec:sterile_rea}

\subsection{Sterile Neutrinos and other Experiments}

\subsubsection{Radioactive Sources }\label{sec:sterile_radio}
Radioactive $^{51}$Cr and $^{37}$Ar neutrino sources have been used for the calibration of the Ga-Ge solar neutrino experiments GALLEX \cite{Anselmann:1994ar,Hampel:1997fc} and SAGE \cite{PhysRevC.59.2246,Abdurashitov:2005tb}. Using  $^{71}$Ge  production by neutrinos from the sources through the charged current reaction $^{71}$Ga$(\nu_e,e^-)^{71}$Ge, the observed event rate was only $0.88 \pm 0.05$ of the expected one \cite{Abdurashitov:2005tb}. This so-called gallium anomaly could be explained by oscillations of electron neutrinos into sterile ones with mass-squared difference around eV$^2$  \cite{GA}. In order to test this hypothesis, the BEST experiment used recently a $^{51}$Cr source of a huge activity of 3~MCi placed inside a 50-ton liquid Ga target split into two nested volumes~\cite{Kozlova:2018cfx,Kozlova:2019zee}. 
The ratios of the measured  and expected rate were $R_{\rm in}=0.791\pm0.05$ and  $R_{\rm out}=0.766\pm0.05$ for the inner and outer detector volumes, respectively \cite{Barinov:2021asz}. These results are consistent with the gallium anomaly, and fitting all source experiments gives $R = 0.80 \pm 0.05$, implying a $4\sigma$ hint. This could be interpreted as oscillation of electron neutrinos into sterile ones with large mixing and $\Delta m_{41}^2 > 1$ eV$^2$ (the BEST  best-fit values are $\Delta m_{41}^2= 3.3$ eV$^2$ and $\sin^2 2\theta_{14} =0.42$). 
 A large fraction of sterile neutrino parameter space preferred by BEST including the best-fit point is already excluded by the DANSS, NEOS, PROSPECT and STEREO experiments. 
%However, for $\Delta m_{41}^2>4eV^2 $ the disagreement becomes weaker and disappears  above $10eV^2$. 
The BEST results are also in tension with the Daya Bay, Bugey-3 and RENO limits~\cite{MINOS:2020iqj} even at large $\Delta m_{41}^2$. 
On the other hand, the BEST results are in a perfect agreement with the Neutrino-4 best-fit point~\cite{Serebrov:2020kmd}, see Sec.\ \ref{sec:sterile_rea0}.

A $^{65}$Zn source was proposed for the next round of sterile neutrino searches at the BEST-2 experiment \cite{Gavrin:2018zmf}.
A $^{144}$Ce~source \cite{Cribier:2011fv,Kornoukhov:1994zq} was considered for sterile neutrino searches at the CeLAND \cite{Gando:2013zla} and SOX \cite{Borexino:2013xxa} experiments. Unfortunately both experiments have not been performed. However the technology of the source production was developed and can be used in the  future. For instance, a possibility to use a $^{144}$Ce~source for sterile neutrino searches at Jinping laboratory is considered \cite{Smirnov:2020bcr}.   Radioactive sources can be used not only for the sterile neutrino searches but also for other goals. For example a 5~MCi $^{51}$Cr source was proposed \cite{Bellenghi:2019vtc} for studies of the coherent elastic neutrino-nucleus scattering that can be sensitive to new physics. see Sec.\ \ref{sec:WG3_coherent}.  Another proposal was to put a $^{51}$Cr source in a liquid xenon dark matter detector in order to test for new physics in neutrino-electron scattering \cite{Link:2019pbm}.  Several experiments study the electron capture process for measurements of the electron neutrino mass and the search for sterile neutrinos in the keV range using $^{163}$Ho, $^{131}$Cs, and $^7$Be  isotopes~\cite{DeGerone:2019xce, Gastaldo:2017edk,Martoff:2021vxp, Friedrich:2020nze}.

\subsubsection{Neutrino Mass Experiments}\label{sec:sterile_mass}
\begin{figure}[t]
\centering
\includegraphics[width=11cm]{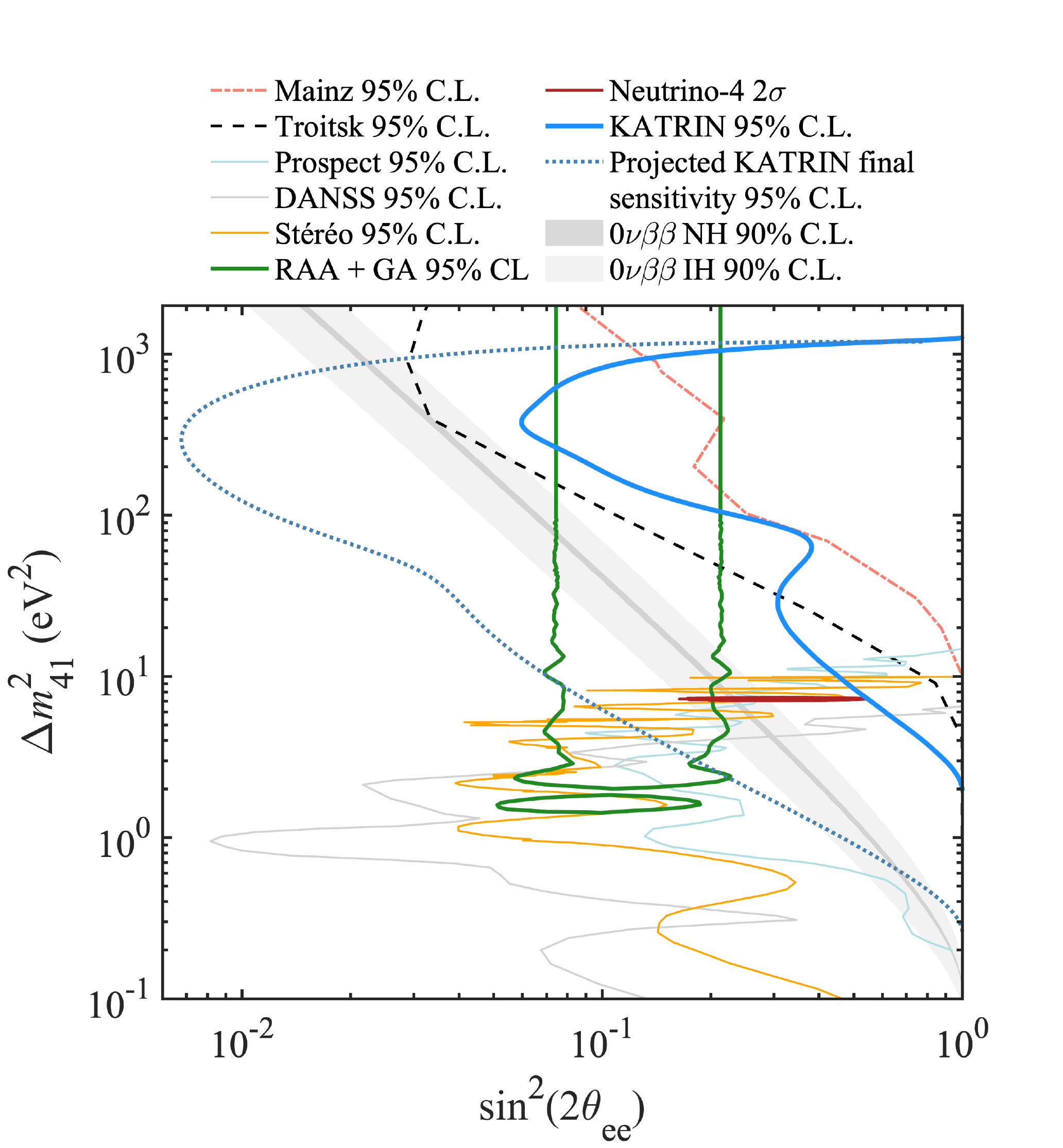}
\caption{KATRIN's 95\% C.L.\ constraints on sterile neutrino parameters in comparison with fits to the reactor and gallium anomalies, as well as with dedicated reactor neutrino experiments looking for sterile neutrinos. A comparison to neutrinoless double beta decay is also made. Taken from Ref.\ \cite{Aker:2020vrf}.}
\label{fig:katrin_sterile}
\end{figure}

Sterile neutrinos of eV-scale also influence observables related to neutrino mass. The interplay with cosmological neutrino mass determinations is discussed in Sec.\ \ref{sec:sterile_cosmo}. For direct mass experiments, the relevant quantity is now a sum of four terms 
\begin{equation}
    m_\beta^2 = \sum\limits_{i=1}^4 |U_{ei}|^2 m_i^2   \,,
\end{equation}
that is, a sterile neutrino mixing with the electron neutrino would
manifest itself as a distortion of the spectrum of the $\beta$-decay electrons, leading to a  kink-like signature. It is important that the same sterile neutrino mixing as the one responsible for the reactor and gallium anomalies is probed here. 
The KATRIN experiment has used the same data set that lead to the groundbreaking neutrino mass limit of 1.1 eV \cite{Aker:2019uuj} to look for this feature \cite{Aker:2020vrf} (see also \cite{Giunti:2019fcj} for a phenomenological study). The result is seen in Fig.\ \ref{fig:katrin_sterile}, which compares the 95\% C.L.\ limits with various other sterile neutrino probes.  KATRIN improves the exclusion of DANSS, PROSPECT, and  STEREO reactor spectral ratio measurements for mass-squared differences larger than 
10 eV$^2$; reactor and gallium anomalies are constrained for $ 100~<~\Delta{m}^2_{41}~<~1000$~eV$^2$. The Neutrino-4 hint of large active-sterile mixing is at the edge of the current  exclusion. Also shown in the figure is the expected 5-year sensitivity, which will improve the global sensitivity further. Assuming the smallest neutrino mass to be close to zero, a comparison with neutrinoless double beta decay is also made in the figure. 
Indeed, if neutrinos are Majorana particles, sterile neutrinos modify the effective mass (see Sec.\ \ref{sec:0vbb}) to 
\begin{equation}
    m_{\beta\beta} = \left| \sum\limits_{i=1}^4 U_{ei}^2 m_i  \right| . 
\end{equation}
The limit of about 0.2 eV on the effective mass in the three-neutrino picture applies directly to the case of eV-scale sterile neutrinos. Both approaches, direct neutrino mass searches and neutrinoless double beta decay, are again testing quite similar parameter ranges. 

Interestingly, the ``sterile contribution'' $m_4 |U_{e4}|^2$ to the effective mass is about the same order as the 
one of the active neutrinos in the inverted hierarchy case. Therefore,  both terms could cancel each other due to the presence of various Majorana phases \cite{Rodejohann:2011mu}. This could imply that for the inverted hierarchy the effective mass can be very small, while for the normal hierarchy it is large. This is the opposite situation compared to the standard 3-neutrino interpretation.

These considerations and limits assume of course that the sterile neutrino is heavier than the active ones. The opposite case, a very light sterile neutrinos separated from  three eV-scale active neutrinos is in principle also possible, but highly unlikely given the cosmological and  terrestrial limits.

\subsubsection{Solar and Atmospheric Neutrinos }\label{sec:atm_sol}

The (2+2) scheme can explain the SBL anomalies and 
does not suffer from the appearance-disappearance tension present in the (3+1), but it is incompatible with the solar and atmospheric neutrino data~\cite{Maltoni:2002xd}, so it has been discarded.
The (3+1) scheme can be easily made compatible with the solar and 
atmospheric data, as long as $\nu_4$ is essentially sterile 
and the other three mass eigenstates behave like in the standard three neutrino framework and are essentially active,  i.e.\ $\vert U_{\alpha 4}\vert^2 \ll 1, \alpha =e,\mu, \tau$.

Solar neutrino experiments may, however, be sensitive  to an 
even lighter sterile state.
Although the standard LMA-MSW~\cite{Wolfenstein:1977ue,Mikheev:1986gs} has been believed to be for almost 20 years the solution to the solar neutrino problem, some tensions remain. 
The best-fit value of $\Delta m^2_{21}$ preferred by KamLAND~\cite{Abe:2008aa},  which controls the value of the global fits for this parameter, has been consistently somewhat  higher than the one 
preferred by all the solar neutrino oscillation experiments combined~\cite{Esteban:2020cvm}. The LMA-MSW solution for the value of $\Delta m^2_{21}$ indicated by KamLAND 
predicts a low energy upturn in the $^8$B energy spectrum not observed by SNO~\cite{Aharmim:2011vm}, Borexino~\cite{Bellini:2008mr} or Super-Kamiokande~\cite{Abe:2016nxk}, as well as a 
slightly smaller day-night asymmetry than observed by Super-Kamiokande. 
This tension has decreased the level of significance from 
$2.2\sigma$ to $1.14\sigma$, after the latest Super-Kamiokande solar neutrino result, 
which indicates a smaller day-night asymmetry and a slightly more pronounced upturn~\cite{Abe:2020tyy}. This problem is still, however, not completely settled yet. 
It has been suggested that a super-light sterile neutrino at the $\Delta m^2$ scale of ${\cal O}(10^{-5})$ eV$^2$ could explain the suppression of the upturn~\cite{deHolanda:2003tx,deHolanda:2010am}. 
A precise measurement of the solar neutrino spectrum in the transition region could help to address this problem. 
The JUNO experiment, which will soon measure $\Delta m^2_{21}$ with a precision of a few per mill,   
can also search for sterile neutrinos at this $\Delta m^2$ scale using reactor neutrinos~\cite{An:2015jdp}.

The search for sterile neutrinos with eV-scale mass splittings can be effectively carried out using  atmospheric neutrinos in the GeV to TeV range. The power to constrain sterile mixing parameter space comes from the large amount of matter that neutrinos traverse when propagating through the Earth. Resonant disappearance of muon antineutrinos targets primarily $\Delta m_{41}^2$ (position of resonance) vs.\ $\theta_{24}$ (depth of deficit). Recent results by ANTARES, IceCube and Super-Kaomikande have provided constraints on a large range of parameters utlizing both GeV \cite{PhysRevD.95.112002, PhysRevD.91.052019, Albert:2018mnz} and TeV energies \cite{Aartsen:2020iky,PhysRevD.102.052009}.  There is also limited sensitivity to one of the three CP violating phases that are present if one sterile neutrino exists \cite{Albert:2018mnz}. 

Studying atmospheric neutrinos does not just allow for a sensitive probe of the parameter space, it also comes with very different systematic uncertainties compared to other probes.  The signature of sterile neutrinos in atmospheric neutrinos is generally at higher energies, reducing the impact of uncertainties due to the cross-sections and other nuclear effects/backgrounds. Therefore, for a convincing sterile neutrino discovery one would really want to see it in multiple probes, including atmospheric neutrinos (in a similar way as the historic solar neutrino deficit was not sufficient to establish neutrino oscillations).

\subsection{eV-Scale Sterile Neutrinos and Cosmology }
Cosmology data is sensitive to a possible existence of sterile neutrinos through the effective number of neutrino families $N_{\rm eff}$. In the standard case of  three species only, corresponding to the three active left-handed neutrinos, $N_{\rm eff} =  3.044$, where the excess from a value of 3 comes from a small  reheating of the cosmic neutrino background by the $e^+e^-$ annihilations that occur sufficiently close in time to neutrino decoupling to slightly affect the neutrino temperature. The neutrino temperature $T_\nu$ is commonly defined relative to the photon temperature $T_\gamma$ by the relation 
\begin{equation}
T_\nu = \left(\frac{4}{11}\right)^{1/3}\,T_\gamma\,. 
\end{equation} 
The extra energy density is instead absorbed into the definition of $N_{\rm eff}$, via 
\begin{equation}
\rho_\nu = N_{\rm eff}\frac{7\pi^2}{120}T_\nu^4 \,. 
\label{eq:neff}
\end{equation} 
where $\rho_\nu$ is the total neutrino energy density in the radiation-dominated era. Any value of $N_{\rm eff}$ in excess of the standard model value would indicate the contribution of additional relativistic relics such as sterile neutrinos. 

The total energy density $\rho_{\rm tot}$ of the Universe at any given time dictates the expansion rate $H(t)$ through $ H^2(t)= 8\pi G_N\rho_{\rm tot}/3$, where $G_N$ is the Newton constant. In the radiation-dominated era, this translates into 
\begin{equation}
    H^2(t)= \frac{8\pi G_N}{3}(\rho_{\gamma}+\rho_\nu)\,.
    \label{eq:H_rhonu}
\end{equation}
Thus any observation that is affected by the expansion rate at early times will directly constrain $N_{\rm eff}$. The main two cosmological epochs that are sensitive to sterile neutrinos are therefore  Big-Bang Nucleosynthesis (BBN) and the emission of the cosmic microwave background (CMB). 

At later epochs, the exact timing of which depends on the particle mass, sterile neutrinos eventually become non-relativistic. They will then contribute to the expansion rate of the Universe but not to the clustering on small scales since they will have free-streamed out of gravitational potential wells during the fraction of the Universe history when they were relativistic. The existence of sterile neutrinos can therefore also be constrained at late times by studying the clustering properties of the Universe on small scales. These constraints, however, are mostly relevant in the case of keV-sterile neutrinos, that we discuss in Sec.\ \ref{sec:dm} on neutrinos as dark matter. 

To contribute as $\Delta N_{\rm eff}=1$ to the effective number of neutrinos, the additional species would have to be thermally produced in the early Universe through oscillations with active neutrinos, such that both active and sterile neutrinos end up with the same temperature $T_\nu$, although with a different normalization of the phase-space distribution~\cite{Gariazzo:2019gyi}. The oscillation production is enhanced with larger mixing angles or shorter oscillation periods due to larger $\Delta m^2$. Additional radiation does not need to be fully thermalized, however, in which case it could contribute with $\Delta N_{\rm eff}<1$. 

Given the current best estimates of their mass ($\sim 1~\rm eV$) and mixing angle between the active and the sterile neutrino states ($\sin^2 2\theta \gtrsim 10^{-3}$),  sterile neutrinos associated with the SBL anomalies would be fully thermalized, hence contributing $\Delta N_{\rm eff}= 1$. The expansion rate of the Universe prior to BBN would thus be increased (cf.\ Eqs.\ (\ref{eq:neff}) and (\ref{eq:H_rhonu})), enhancing the neutron-to-proton ratio at the onset of BBN. Since this ratio fixes the abundances of the light elements, the measurement in particular of the abundances of $^4\rm He$ and deuterium  impose interesting constraints on the effective number of neutrino families. Because BBN predictions for fixed $N_{\rm eff}$  depend on a single parameter, the baryon-to-photon ratio, the determination of $N_{\rm eff}$ from BBN exhibits some degeneracy with the baryon density. Current constraints from BBN are in the $N_{\rm eff} = 3-4$ range, and thus can not rule out fully thermalized sterile neutrinos. 

The speed-up of the expansion rate of the Universe caused by additional radiation also impacts CMB measurements via the determination of the sound horizon at recombination. The data from Planck therefore also constrain $N_{\rm eff}$. Using its most conservative choice of priors, the Planck collaboration obtained the following bounds, $N_{\rm eff}<3.3$ and $m_{\rm eff}<0.65$~eV (95\% C.L.), using the ``TT,TE,EE+lowE'' Planck data combined with lensing and BAO~\cite{Aghanim:2018eyx}. Here, $m_{\rm eff}$ is an effective sterile neutrino mass defined as $m_{\rm eff}=94.1\,\Omega_{\rm sterile}\, h^2$ eV, with $\Omega_{\rm sterile}$ the contribution to the energy density relative to the critical one and $h=H_0/100\,{\rm km \, s^{-1}\,  Mpc^{-1}}$ the re-scaled Hubble parameter. In the case of a thermally-distributed sterile neutrino, this parameter is related to the true mass by $m_{\rm eff} = (\Delta N_{\rm eff})^{3/4} \, m_4$.

The tightest result comes from the combination of BBN and CMB that provides a constraint on $N_{\rm eff}$ in the $2.9-3.0$ range with an uncertainty $\sigma(N_{\rm eff}) \sim 0.3$ (see the white paper~\cite{Abazajian:2012ys} for a detailed review on the cosmological impact of eV sterile neutrinos). Such a constraint on $N_{\rm eff}$ is incompatible with a fully thermalized light sterile neutrino as currently favored by the best-fit parameters of short-baseline experiments. In addition, a thermalized light sterile neutrino in the eV mass range is too heavy to be compatible with the CMB constraint. These cosmological constraints can be alleviated in some BSM models in which, for instance, the sterile neutrino couples to a new light pseudoscalar degree of freedom~\cite{Archidiacono:2020yey}. Such models provide a good fit to BBN and CMB temperature data, even reducing the tension on the value of the expansion rate $H_0$ between early and late-time measurements. The inclusion of CMB polarisation  data, however,  constrains the sterile neutrino mass to be less than $1~\rm eV$ in this scenario. A global fit to cosmology data and short-baseline experiments only allows for a narrow window around 1~eV.

\label{sec:sterile_cosmo}

\subsection{Global Fits }
We have seen in the previous sections that a variety of hints towards the existence of light eV-scale sterile neutrinos exists. All those hints need to be incorporated in a global fit, in analogy to the standard three-flavor pattern discussed in Sec.\ \ref{WG1:SubSect:Glob}. 
The fits by different groups largely agree with each other \cite{Gariazzo:2017fdh,Dentler:2018sju,Berryman:2020agd}. 

With a fourth neutrino, the PMNS matrix is a $4\times 4$ matrix that contains six mixing angles and three CP phases relevant for oscillations (three additional phases appear for Majorana neutrinos). Obviously, determining all those parameters would be a huge undertaking which would require a large number of different experiments.  

\begin{figure}[t]
\centering
\includegraphics[width=11cm]{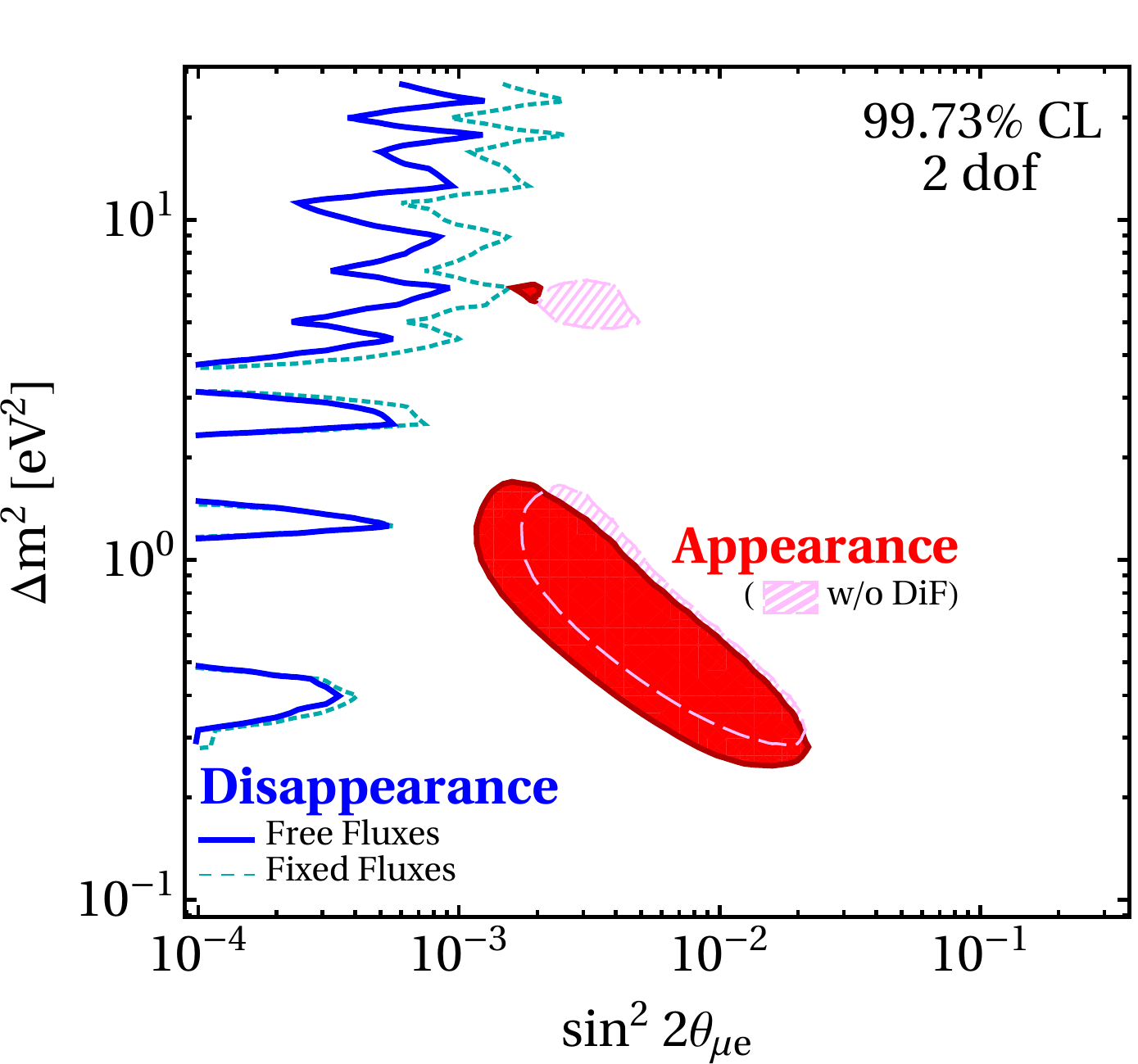}
\caption{Appearance versus disappearance data for the relevant mixing parameters $\sin^2 2 \theta_{e\mu} = 4 |U_{e4} U_{\mu 4}|$ and $\Delta m^2_{ 41}$ at 99.73\% C.L.\ for 2 dof.  Disappearance data using free reactor fluxes (solid) or fixed reactor fluxes (dashed) are used. The allowed parameters are to the left of the dark and light blue lines. The shaded contours in red (pink hatched) are the allowed parameters from appearance data including (excluding) decay-in-flight LSND results. Taken from Ref.\ \cite{Dentler:2018sju}.}
\label{fig:ster_global}
\end{figure}

\begin{figure}[t]
\centering
\includegraphics[width=0.45\textwidth]{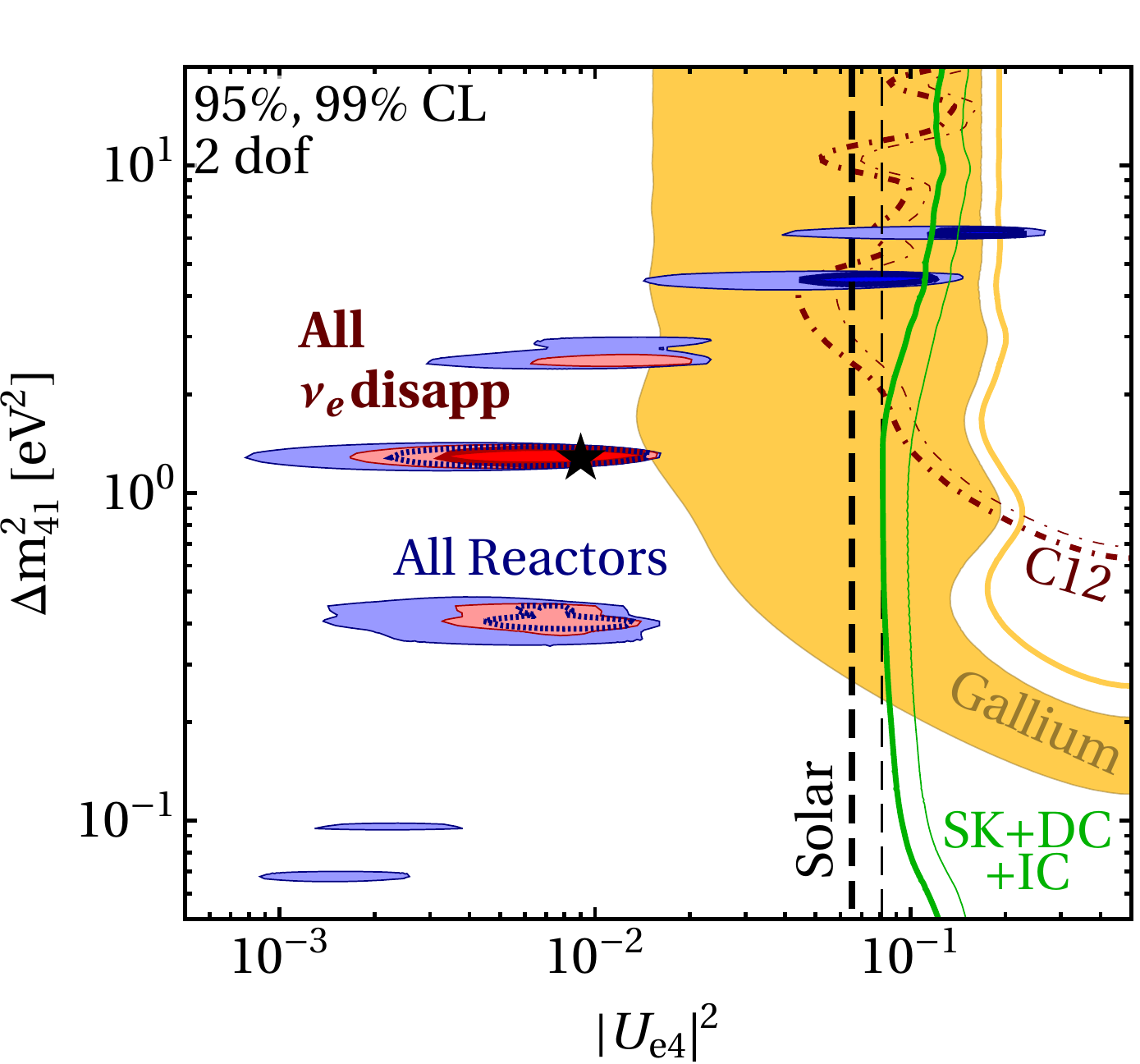}
\includegraphics[width=0.45\textwidth]{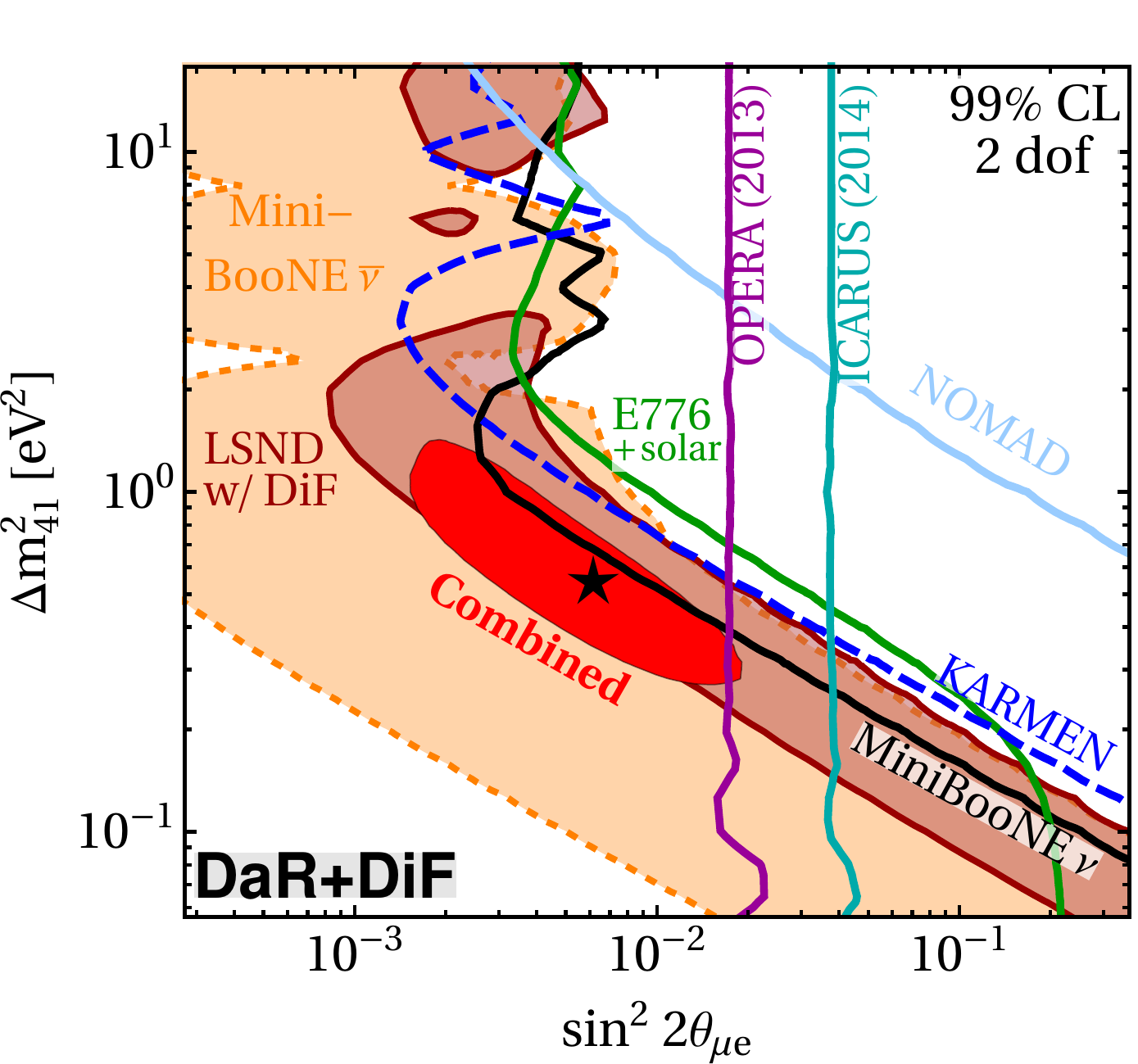}
\caption{Fit results on sterile neutrinos using only $\nu_e \to \nu_e$ disappearance (left) and  $\nu_\mu \to \nu_e$ appearance (right). Taken from Ref.\ \cite{Dentler:2018sju}.}
\label{fig:ster_global2}
\end{figure}

The main hints stem from $\nu_e \to \nu_e$ disappearance and  $\nu_\mu \to \nu_e$ appearance, as well as many constraints exist on  $\nu_\mu \to \nu_\mu$ disappearance. In the relevant $L/E$ regime for an eV-scale $\Delta m^2_{\rm st} = \Delta m^2_{ 41}$, unitarity implies that if $\nu_e \to \nu_e$ disappearance and  $\nu_\mu \to \nu_e$ appearance exists, then $\nu_\mu \to \nu_\mu$ disappearance must exist. This argument is independent of the number of sterile neutrinos (there could be more than one), and leads to strong tensions in the global fits. The probabilities, in the relevant limit in which the large sterile mass-squared difference dominates, read 
\begin{equation}
    P(\nu_\alpha \to \nu_\beta)= 1- \sin^2 2\theta_{\alpha \alpha} \sin^2 \Delta_{\rm 41}  \, , \quad \quad P(\nu_\alpha \to \nu_\beta) = \sin^2 2\theta_{\alpha \beta} \sin^2\Delta_{\rm 41}\, ,
 \label{eq:probsterile}
\end{equation}
 where $\Delta_{\rm 41} \equiv \Delta m^2_{\rm 41}L/4E$,  $\sin^2 2\theta_{\alpha \alpha} \equiv 4 \vert U_{\alpha 4}\vert^2 (1-\vert U_{\alpha 4}\vert^2)$
  and $\sin^2 2 \theta_{\alpha \beta} \equiv 4 \vert U_{\alpha 4} \vert^2 \vert U_{\beta 4} \vert^2$. 
To be more specific, $P(\nu_e \to \nu_e)$ depends on 
$|U_{e4}|$, $P(\nu_\mu \to \nu_\mu)$ on 
$|U_{\mu 4}|$, and $P(\nu_\mu \to \nu_e)$ on 
$|U_{e4} U_{\mu 4}|$. The tension is displayed in Fig.\ \ref{fig:ster_global}, where the two regions according to appearance and disappearance have very little overlap. It is independent of the assumed reactor flux, and also on which data set one excludes in the fit: the $p$-value that appearance and disappearance data agrees never exceeds $10^{-5}$ \cite{Dentler:2018sju} (see also \cite{Gariazzo:2017fdh}). The inclusion of extra sterile states may alleviate some of this 
   tension~\cite{Dentler:2018sju,Diaz:2019fwt,Boser:2019rta}, but
    the latest result from IceCube further discards some regions of these solutions~\cite{PhysRevD.102.052009}, see Sec.\ \ref{sec:atm_sol}. 
The obvious solution to this issue is to conjecture that one (or both) sets of results are not reliable. Individual fits to appearance and disappearance data are therefore useful. The results of such a fit are seen in Fig.\ \ref{fig:ster_global2}. 

\begin{figure}[t]
\centering
\includegraphics[width=0.45\textwidth]{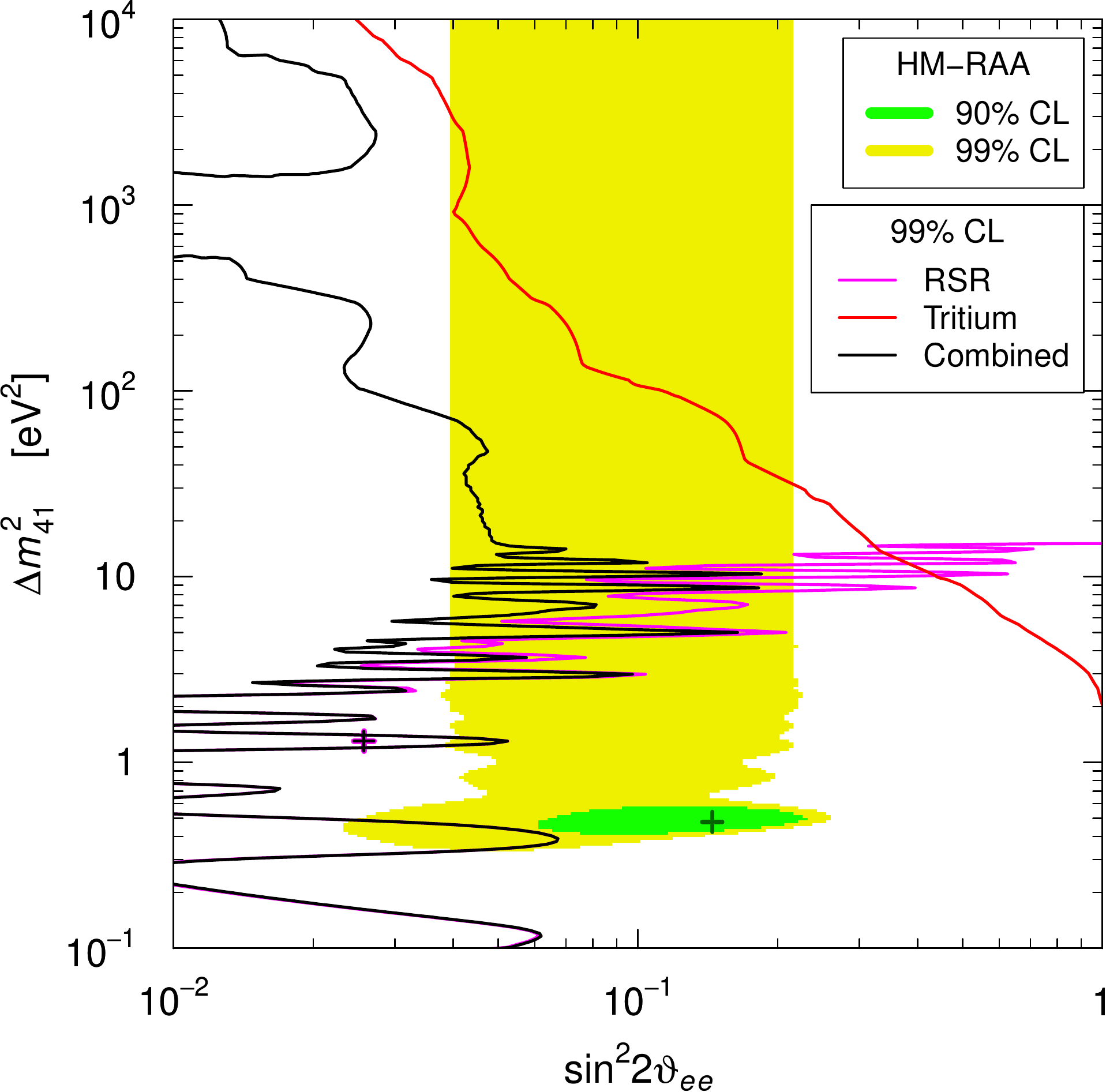}
\includegraphics[width=0.45\textwidth]{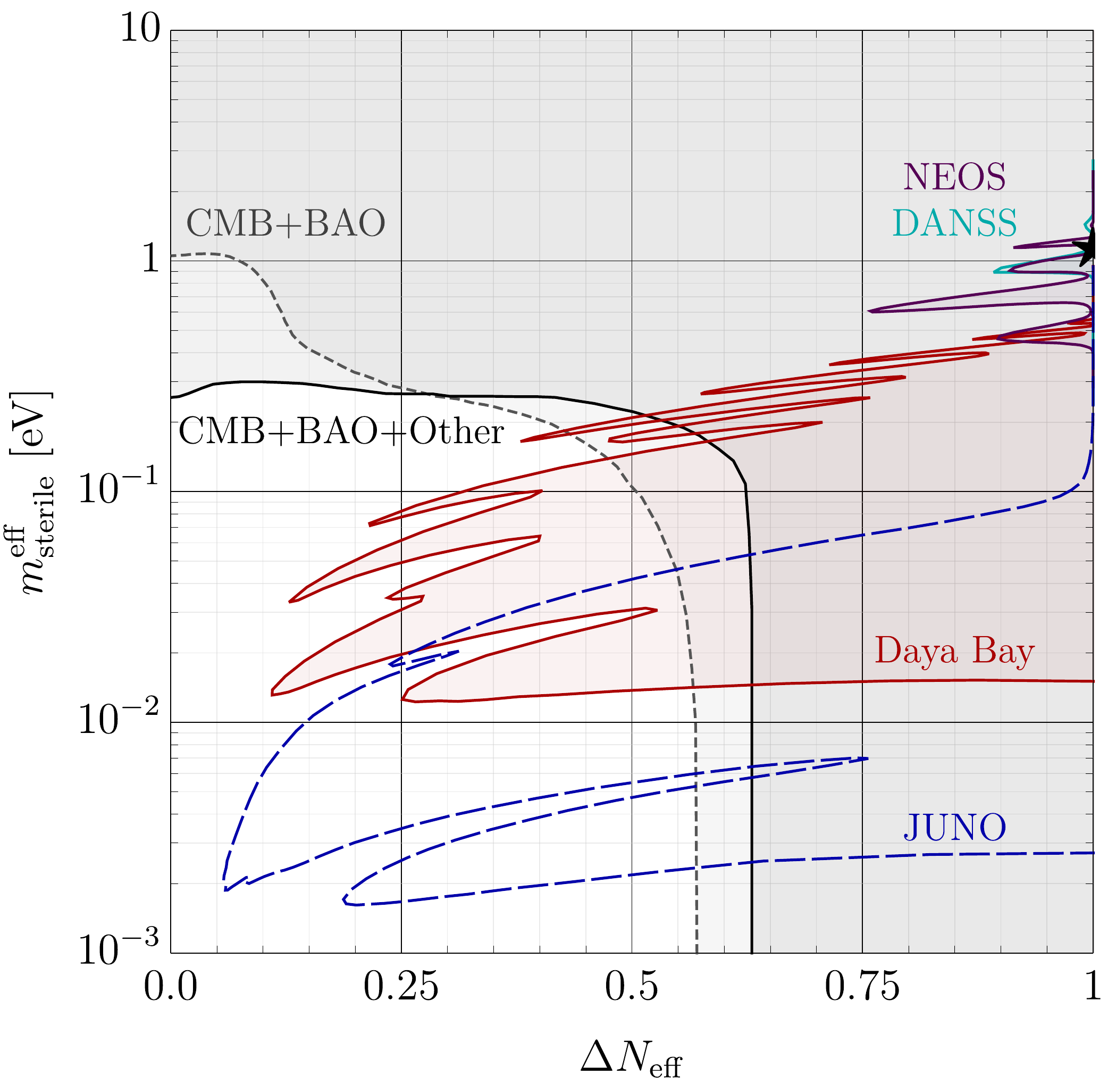}
\caption{The left plot shows in green and yellow the 90\% and 99\% C.L.\ areas coming from the reactor flux deficit solution, while the pink line shows the result of an analysis of reactor experiments with spectral ratio measurements  (RSR). A combination of Mainz, Troitsk and KATRIN tritium experiments is seen in red, which can be combined with the RSR fit to the black line. Taken from Ref.\ \cite{Giunti:2019fcj}. 
Here $\sin^2 2 \theta_{ee} =  4 |U_{e4}|^2 (1 - |U_{e4}|^2)$. The right plot shows constraints 
in the plane of $\Delta N_{\rm eff}$ and sterile neutrino mass from reactor experiments, including prospects of the future JUNO project. Taken from \cite{Berryman:2019nvr}. 
}
\label{fig:ster_comb}
\end{figure}

It is also possible to fit the oscillation experiments including  cosmology data \cite{Archidiacono:2013xxa,Hagstotz:2020ukm}, or with $\beta$ decay data, which is shown in Fig.\ \ref{fig:ster_comb}. 
One identifies a tension between the active-sterile oscillations indicated by the reactor flux deficit and the combined tritium and reactor spectral-ratio 
measurements. Since all these aspects are in flux, it is too early to take this very seriously. However, it shows the interplay of different approaches to the problem, which will be central in the coming years, when the issue of sterile neutrinos will be hopefully settled. 
Regarding cosmology, the analysis of data should keep the effective number of relativistic degrees of freedom $N_{\rm eff}$, or its difference $\Delta N_{\rm eff} = N_{\rm eff} - 3.044$, into account. 
As expected, cosmology data does not allow an eV-scale sterile neutrino (recall though the possibility to allow this via new physics mentioned in Sec.\ \ref{sec:sterile_cosmo}), though for smaller masses (which cannot explain the reactor flux deficit) present experiments reach parts of parameter space to which cosmology data is currently not sensitive.

\label{sec:sterile_global}

\subsection{Searches for Sterile Neutrinos beyond eV}
\label{sec:sterile_other}

\begin{figure}[t]
\centering
\includegraphics[width=11cm]{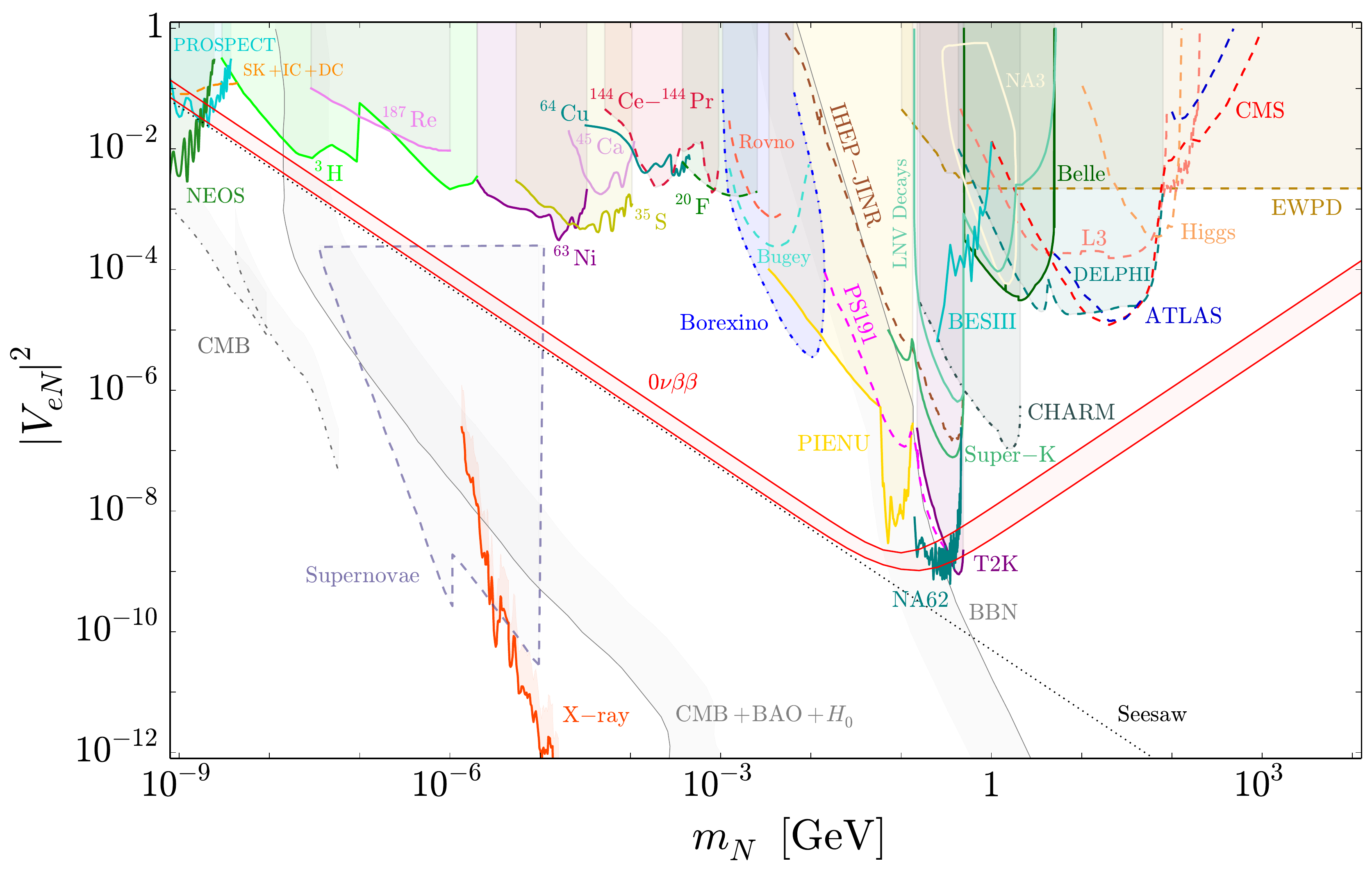}
\caption{Constraints on the mass of a sterile neutrino and its mixing with electron neutrinos from a variety of astroparticle and particle physics searches. Taken from \cite{Bolton:2020ncv}. 
}
\label{fig:sterile_other}
\end{figure}

Sterile neutrinos may exist at various scales, see Sec.\ \ref{sec:ster}, and one can search for them in a variety of ways. 
Their mass may lie below or above the eV-scale, with most activity dealing with the case of heavier sterile neutrino masses. This includes even masses around the TeV-scale, where they would be produced  at colliders \cite{Deppisch:2015qwa}. Lower mass neutrinos below GeV are usually searched for in decays of mesons, where instead of the usual active neutrinos, sterile neutrinos are emitted. In all searches, the mass of the sterile neutrino and its mixing with the active neutrinos of flavor $e$, $\mu$ or $\tau$ is constrained. The search for those particles also includes astrophysical observations, cosmological constraints, electroweak precision tests, or neutrinoless double beta decay, if the sterile neutrinos are Majorana particles. 
A summary of various limits, taken from \cite{Bolton:2020ncv}, can be found in Fig.\ \ref{fig:sterile_other}.
 
\begin{figure}[t]
\centering
\includegraphics[width=11cm]{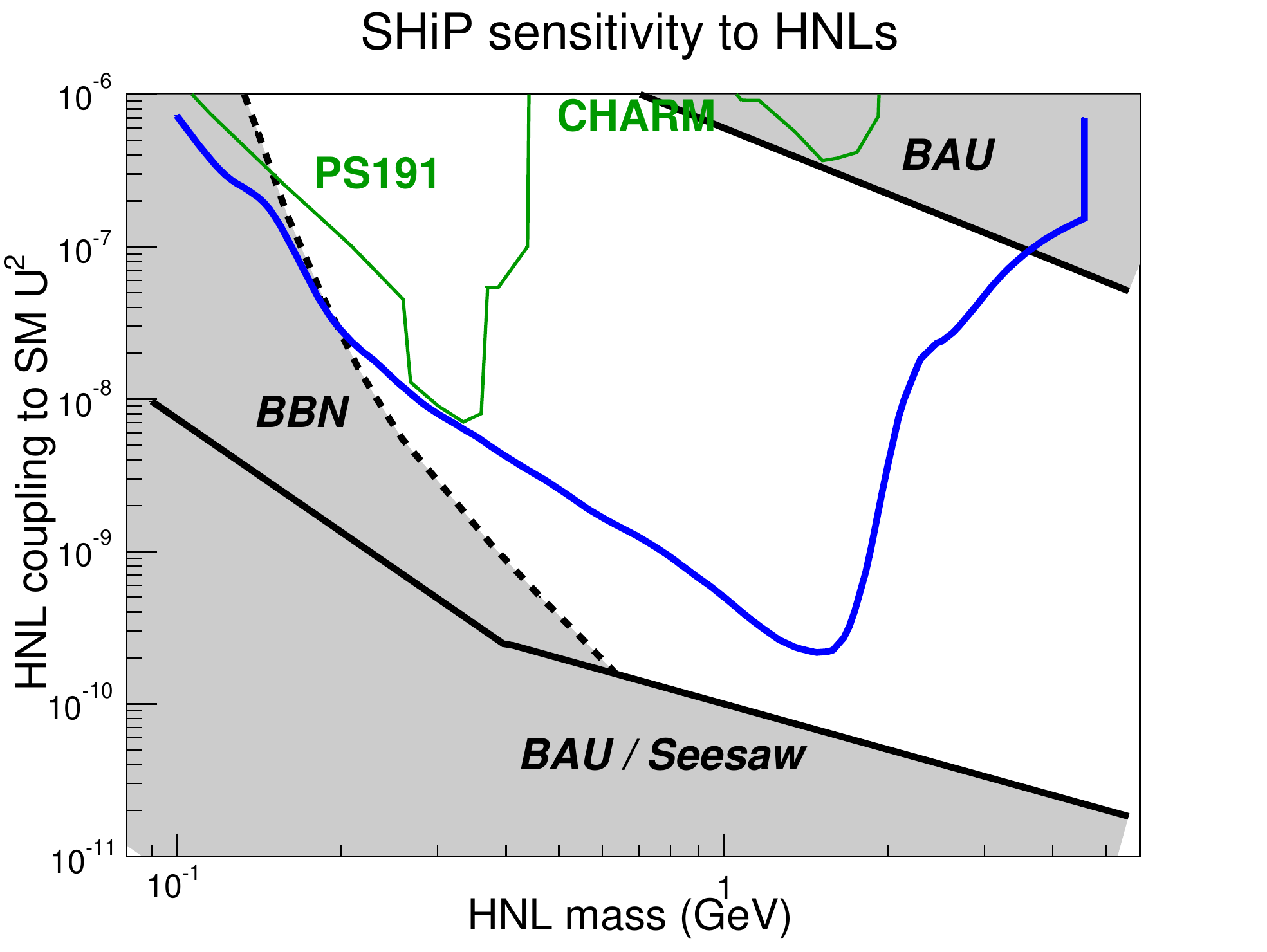}
\caption{The white region displays mass and mixing of sterile neutrinos, here called HNL (Heavy Neutral Lepton), which can lead to  successful resonant leptogenesis. Green lines are constraints from past experiments, the blue line is the expected sensitivity of SHiP. Taken from \cite{Anelli:2015pba}. 
}
\label{fig:res_lep}
\end{figure}

 An interesting regime is the one of masses around a few keV, where neutrinos become attractive warm dark matter candidates, see Sec.\ \ref{sec:dm}. Noteworthy approaches to search for those particles are the TRISTAN detector upgrade for KATRIN ($\beta$ decay of tritium) \cite{Mertens:2018vuu}, the BeEST experiment (electron capture of $^7$Be implanted into  superconducting tunnel junction quantum sensors) \cite{Friedrich:2020nze} or HUNTER (electron capture of $^{137}$Cs in a magneto-optical trap) \cite{Smith:2016vku}. Indeed,  BeEST has already obtained the best laboratory mixing limits  in the range between $100-800$ keV, planing to improve these limits by 3 orders of magnitude in the next 5 years. 
 The use of quantum sensors and atom/ion traps for searches of new physics related to neutrino is promising, see also Sec.\ \ref{wg5_tech}, and may eventually reach the cosmologically interesting range of mixing angles below $ \sim 10^{-5}$. 
 
 Another noteworthy scenario is related to ``resonant leptogenesis'', see 
 Sec.\ \ref{WG1:SubSec:PMNS-Theory:CPV}. In principle one can arrange sterile neutrinos to be almost degenerate in mass, which enhances the CP asymmetry in their decay and arrange leptogenesis at GeV-scale temperatures. Fig.\ \ref{fig:res_lep} shows the possible parameter space of the ``$\nu$MSM'' \cite{Gorbunov:2007ak} and the experimental sensitivity of the SHiP experiment \cite{Anelli:2015pba}.

\subsection{The Number of Neutrinos: Summary}
Various hints for the existence of light eV-scale sterile neutrinos exist, originating from neutrinos from nuclear reactors, accelerators and radioactive sources. It is likely that not all of those hints are correct. Nevertheless, the long-standing nature of the hints and the lack of fully convincing other explanations of the experimental anomalies has lead to an active pursue of dedicated experiments looking for sterile neutrinos. Furthermore, every neutrino (not necessarily oscillation) experiment has sensitivity on such particles as well, providing complementary information. Moreover, renewed interest in the calculation of nuclear reactor fluxes has arisen. 
Sterile neutrinos can easily be motivated by theoretical considerations, but essentially always their mass is much larger in such scenarios. Hence,  a confirmation of their existence would have strong consequences for model-building.

If those particles are verified, the overall mass of neutrinos is increased by at least the square-root of the $\Delta m^2$ associated to the apparent sterile neutrino oscillations. This has impact on neutrino mass experiments and measurements. Direct searches such as KATRIN start to provide competitive constraints on the scenario, as well as neutrinoless double  beta decay experiments, in case the sterile neutrinos are Majorana particles. Cosmology strongly disfavors the existence of sterile neutrinos, both from the overall mass-scale, as well as the number-of-neutrino point of view. If confirmed in terrestrial experiments, sterile neutrinos as motivated by the above mentioned anomalies would require rather non-trivial modifications of early Universe cosmology. 

Within the next few years the question on whether eV-scale sterile neutrinos exist, is expected to be answered. 

\newpage

%\subsection{Non-standard $\nu$ Interactions}

%\subsection{Neutrino Magnetic Moment}
%\input{BSM/BSM_mm.tex}
 
\label{sec:Neff}

\clearpage

\section{New Technologies, Cross-over to other Science, and Frameworks for Neutrino Physics}\label{sec:tech}
\subsection{Introduction and Community}

Neutrino physics covers a huge spectrum of experimental, technological and theoretical aspects. It is therefore not surprising that one can identify sub-communities with connections to many other scientific and technological topics in neutrino physics and to other fields of science. In particle physics this are groups coming from high energy physics, doing experiments with beams and using corresponding detector technologies. Groups with a nuclear physics background tend to work with neutrinos at lower energies where very different methods and technologies are used. A third sub-group comes from astrophysics, where neutrinos are important, for example for the evolution of stars and for supernovae. Yet another sub-group comes from cosmology, clearly again with very different techniques and requirements. A final sub-group are theorists who often tie some or all of the other groups together into a global picture of neutrino physics. 

We list in this section a number of the these diverse technologies and methods to indicate directions which are important for the future of neutrino physics. We touch on many relevant topics, but note that we do not claim that this list is complete. Note also that the ordering or length of the text does not express any preference. Instead this compilation should be seen as directions which all are very interesting and deserve in principle to be further improved due to their valuable potential. Some technologies are more mature or exist and can be/are more or less realized, while others need more R\&D. These topics are listed in the {\bf ``Technologies and Capabilities''} subsection. Another direction can be called scientific {\bf ``Infrastruture''} which is important for the often spacial environments where experiments are performed. 

It is important to keep in mind that -- as in the past -- new ideas will emerge which can change the landscape significantly. We would like to emphasize therefore the importance of an adequate level of flexible R\&D funds. Similarly theory support is very important since it unites the vastly different experimental efforts and since it comes up with new ideas which trigger exciting new experimental projects. 

Decisions for the most promising big or small experimental projects requires continuous careful scientific discussions and a balancing of the effort to gain ratio by the world-wide community. At the same time one must keep the lead times of big projects in mind. Good international cooperation and coordination is therefore very important for best science results by making best use of resources for developments, for R\&D efforts, for experiments with similar goals and for shared infrastructure facilities.

\subsection{Technologies and Capabilities}
%\subsection{Technologies}
\label{wg5_tech}

\begin{itemize}

\item Accelerator neutrinos: In 1961 the first true neutrino beam was created at CERN using the Van der Meer horn to focus pions produced in the bombardment of a solid target by protons extracted from the PS. Such horn-focused beams have been used at CERN, ANL, BNL, FNAL, IHEP, KEK, and J-PARC, first to establish the quark-parton model and the Standard Model, and then to study neutrino oscillations and to search for new phenomena such as the existence of sterile neutrinos. Future exploration of the nature of neutrino flavor using neutrino-beams produced by accelerator facilities will continue to exploit the technique pioneered at CERN.

The Deep Underground Neutrino Experiment (DUNE)~\cite{Abi:2018dnh,Acciarri:2016crz,Acciarri:2015uup,Strait:2016mof,Acciarri:2016ooe} in the US and the Tokai-to-Hyper-Kamio\-kande (Hyper-K)~\cite{Abe:2014oxa,Abe:2015zbg,Abe:2016ero,Hyper-Kamiokande:2018ofw} experiment in Japan will use horn-focused pion beams produced using proton-beam powers in excess of 1\,MW to search for the violation  of CP invariance in the neutrino sector. The high-flux beams illuminating the large DUNE and Hyper-K detectors will allow very large data sets to be accumulated. Projections of the rate at which data will be collected indicate that the statistical error will be reduced to the percent level by 2028--30.

The Short Baseline Neutrino (SBN) programme at Fermilab will be served by the horn-focused Booster Neutrino beam. Over the coming half decade, the SBND, MicroBooNE, and ICARUS experiments will place make stringent tests of the consistency of the three-neutrino mixing paradigm and explore the anomalous results reported in the study of short-baseline neutrino oscillations reported elsewhere in this report.

\label{SubSect:Accel}

\item Systematic uncertainties on beam neutrino fluxes, their flavor composition and neutrino cross-sections impact future neutrino experiments, Hyper-K and DUNE. Thus,  precise measurements of the beam neutrino fluxes and neutrino cross-sections (1\% level) are necessary to maximize the sensitivity of the next generation of long-baseline experiments, and to take the next step in sensitivity. 

Precision CP phase measurements would, in particular, be greatly improved by  facilities and detectors to directly measure the interaction cross sections of electron-neutrinos and electron antineutrinos.  Existing experiments mainly use muon neutrino beams with small electron neutrino components and electron neutrino cross section predictions rely on low-statistics measurements and extrapolation from muon neutrino scattering results. 
These measurements can be done in muon decay rings by $\nu$STORM~\cite{Kyberd:2012iz} or in instrumented decay tunnels by ENUBET~\cite{Longhin:2014yta}. 

The muon decay ring is a facility to provide a muon beam and a test bed for accelerator and detector technology. 
At $\nu$STORM, the flavor composition of the neutrino beam is known and its energy spectrum may be determined precisely, courtesy of advanced detector techniques, using the storage-ring instrumentation. It would be the first neutrino-beam facility to be based on a stored muon beam and will provide a test-bed for the development of the  technologies required for a multi-TeV muon collider and/or a neutrino factory. It will also serve the nuclear physics community by providing a unique 100\% polarized probe of flavor-dependent collective effects in nuclei.

Optimization for electron neutrino measurements will require different beam tune and detectors designed to sign analyze electrons. 
ENUBET will be the first ``monitored neutrino beams'' by  monitoring the leptons in the decay tunnel at the single particle level. In particular, the use of an instrumented decay channel to constrain the energy of electron-neutrinos produced in kaon decays will be studied.

\item Detectors for precision neutrino oscillation: Neutrino oscillation experiments, for both measurement of three-flavor oscillation parameters, including a CP phase, as well as searches for sterile neutrinos, require a neutrino source and one or more detectors. Detector technologies for existing and planned neutrino oscillation experiments are diverse, ranging from fine-grained low-threshold detectors for very-short baseline reactor experiments, to hundred-ton-scale tracking time projection chambers for short-baseline oscillation experiments, to large, homogeneous liquid volumes of tens-of-kiloton scale, for long-baseline beams.  For oscillation experiments, one needs to measure neutrino energy and flavor with high statistics and resolution.   This requires generically a well-understood, high-intensity source, coupled with a high-mass, high-performance detector.

In general, one needs fine granularity that enables precision reconstruction of the neutrino interaction final state, in order to tag the interacting neutrino's flavor and determine neutrino energy with high resolution. Especially above the hundreds of MeV range, systematic uncertainties in oscillation parameter determination may eventually be dominated by understanding of neutrino interaction cross sections.  Therefore, well-understood neutrino beams and detectors that enable measurements of these cross sections will be a vital component of a future program.

\item Novel techniques to search for sterile neutrino searches are also tried using beam-neutrinos produced by muon decay at rest (DAR) in JSNS$^2$~\cite{Harada:2013yaa} and by $^{8}$Li beta decay in IsoDAR~\cite{Alonso:2017fci} where $^{8}$Li is formed mostly by $^{7}$Li capturing neutron produced by proton beam striking a beryllium target. Neutrinos produced from the world's most intense proton source would allow to explore leptonic CP violation and the neutrino mass ordering in ESSnuSB~\cite{Baussan:2013zcy}.

\item Improved detectors for MeV events: For rare phenomena searches such as neutrinoless double beta decay, low radioactivity material for shielding and sensor apparatus material have been important. In addition,  high-sensitivity underground searches now require thorough understanding of muon spallation and its products. These experiments require measurements of detailed production cross section and shower propagation for a useful background estimation. Muon beam experiments should be conducted systematically for relevant nuclei for this purpose. Isotope enrichment and purification of detector materials have been vigorously studied for better signal to noise ratio, and these should be extended for much larger mass aiming at several tens of ton target material. Future large mass detectors will naturally require large area and high quantum efficiency photon sensors for good energy resolution. This will also require capability of particle identification 
(e.g.\ Timepix-based detectors)
and directional information for a reliable discovery and for understanding underlying physics. 
 
\item Improvement in photon sensor technologies: For several decades, photomultiplier tubes (PMTs) have been the working horse in neutrino- and dark matter detectors, e.g., in aforementioned large-scale water-Cherenkov detectors. PMTs have been tremendously improved over the years and have potential for future improvements to further increase their performance in specialized applications such as ultra-low-background experiments or cryogenic environments. PMTs are being developed where the dynode structure is replaced by either a multi-channel plate detector or by a silicon-based electron-to-digital converter. These developments aim at increasing the performance and reducing the cost of the integrated detection system. However, to date conventional PMTs remain the photosensors of choice for large volume neutrino detectors. This is not just because of the cost advantages per photo-sensitive area but also due to lower noise associated with PMTs and their demonstrated long-term stability. At the same time, the trend to ever larger PMTs has stopped and instead applications with larger numbers of medium sized PMTs ($3-4$~inch), that can be produced in large quantities for modest costs, have become more frequent. In water-Cherenkov neutrino detectors, several of these smaller PMTs are typically arranged in  a transparent pressure sphere pointing in different directions, providing the advantage of additional spacial or directional information. With an increase in target volume in future neutrino detectors, and thus an increase in sensitivity, large areas within the detector volume must be covered with PMTs. Thus, further development of low-cost photo-sensors and corresponding readout electronics is required. 
    
In certain applications, such as noble-gas neutrino detectors, Silicon PhotoMultipliers (SiPMs) are becoming the devices of choice. Their intrinsic low radioactivity, low bias voltage of typically less than 100~V, high gain of $10^5-10^6$, and gain stability towards temperature and bias fluctuations gives them a competitive edge compared to PMTs and avalanche photodiodes. Especially when operated inside cryogenic, liquid noble-gas detectors, their dark count rate and correlated noise are at sufficiently low levels, effectively enabling single-photon counting. The sensitivity of SiPMs to vacuum-ultra-violet photons has improved over the years to photon-detection efficiencies (PDEs) of more than 15\% at 175~nm \cite{ Jamil:2018tkx}, the scintillation wavelength in liquid xenon \cite{FUJII2015293}. Their PDE to 128~nm scintillation photons in liquid argon still remains insufficient for direct photon detection, thus wavelength shifters are required to shift the wavelength towards the visible range where SiPM PDE peaks. Further development to increase the PDE in the ultra-violet spectral range should be pursued for future application of SiPMs in liquid noble-gas detectors.  Next-generation experiments require SiPM coverages of several square meters to be covered by SiPMs. This necessitates the integration of SiPMs, which are on the order of 1~cm$^2$ in area, into larger modules and development of readout electronics that can be placed in close proximity to the SiPM modules in order to reduce cabling. The placement of SiPMs and readout electronics in liquid noble-gas detectors puts constraints on the acceptable power consumption in order to prevent boiling and the creation of bubbles. Due to the placement inside the detector volume, low radioactivity levels are required for these cryogenic electronics. Readout electronics, such as low radioactive cryogenic ASICS, and integrated SiPM modules should be developed further. Low radioactive SiPM modules including cryogenic readout electronics will transform light detection in noble-gas neutrino and dark matter detectors if large areas can be covered at a moderate cost. Development of fast timing electronics may enable additional topological suppression of background events in liquid noble gas detectors, such as in xenon \cite{ Brodsky:2018abk}, and should be pursued.

\item Better and novel neutrino sources: Radioactive sources are very compact and powerful sources of (anti)neutrinos with well known spectra. They can be used in various studies for example in sterile neutrino searches, investigations of coherent neutrino-nucleus scattering, and searches for new phenomena in neutrino-electron scattering. Radioactive $^{51}$Cr and $^{37}$Ar neutrino sources have already been used for the calibration of the Ga-Ge solar neutrino experiments GALLEX and SAGE. A $^{51}$Cr source of huge activity of 3 MCi was successfully produced and used by the BEST experiment in searches for sterile neutrinos. The technology of a  $^{144}$Ce~source production was developed for sterile neutrino searches. High activity $^{65}$Zn and $^{170}$Tm sources are also being considered by the BEST collaboration~\cite{Barinov:2021asz}. The development and production of radioactive sources require close cooperation between fundamental science and the nuclear power industry.

\item Reactors as strong neutrino sources: Nuclear power reactors are one of the strongest neutrino sources which allow very interesting experiments. Access to very promising sites close to strong reactor cores requires close cooperation with the industry operating nuclear power plants. Such sites have, however, very strong limitations for the allowed technologies and access (both safety). This requires special low background techniques which is acceptable for a given site. Proximity to the reactor core gives more flux, while it often excludes most liquids, cryogenic equipment and special gases. Technologies which are compatible with the requirements should be further developed.

\item Nuclear safeguarding:
neutrino detectors are potentially useful for nuclear safeguarding applications.  A recent study~\cite{nutools} has evaluated the potential, coming to the conclusion that while for many existing reactors, current IAEA safeguards are sufficient, there are possible use cases for neutrino monitoring associated with advanced reactors. Other possibilities  identified in Ref.~\cite{nutools} include: future nuclear deals involving cooperative monitoring or verification, non-destructive assay of spent nuclear fuel~\cite{Brdar:2016swo}, and post-accident response.  Technology development is desirable for these use cases.

\item Continued development of noble liquid detection technologies: Liquid Argon Time projection chambers (LArTPC's) have come of age as large scale detectors for accelerator based neutrino experiments.  Current devices rely upon wire or pad based multi-view 1-dimensional readout for the coordinates perpendicular to the drift direction. Truly 3-D detectors using full 2-D pixel readout  have been developed and now being deployed, for example in ArgonCube\cite{Asaadi:2018xfh} and the future DUNE near detector\cite{AbedAbud:2021hpb}.  

Large LArTPC's pose significant technical challenges due to the need for high electric fields (500 V/cm) over many meters in a high purity cryogenic environment.  These experiments rely on sophisticated high volume filtration systems, which lead to bulk motion in the active medium and requires constant monitoring.  High voltage systems need to be robust against sparking and capable of dealing with very large stored energies.    Detector electronics may need to be placed in high voltage regions, which is leading to the development of optical  methods for power delivery and signal transmission.   See \cite{Majumdar:2021llu} for a very recent overview of current LArTPC technology.  Continue development of these vital engineering technologies will be required as LArTPC detectors grow from the 1 kT (ProtoDUNE) to the $10-20$ kT (DUNE) scale. 

%%%%%%%%%%%%

Noble liquids allow also very powerful low background neutrino experiments. The operation and construction of such detectors employs technologies which have significant overlap with direct dark matter detection experiments based on noble liquids. In neutrino physics these experiments aim especially at neutrinoless double beta decay of xenon $^{136}$Xe, either enriched or with a natural abundance of 9\% in xenon TPCs, both liquid and gaseous. 3-D event reconstruction in the TPC allows to select the cleanest inner part of the liquid xenon as active volume, while an outer layer acts as further shielding aiming at completely suppressing background events. 

This has been demonstrated in the liquid xenon TPC of the EXO-200 experiment where 2-D drifted electrons were read out by two wire planes \cite{Albert:2013gpz}. Ionization readout tiles with orthogonal metal charge-collection strips \cite{Jewell:2017dzi, Li:2019qgs} are being developed for full 3-D event reconstruction in the nEXO liquid xenon TPC. The $2\nu\beta\beta$ background is, however, intrinsic and cannot be removed by 3-D event reconstruction. Therefore R\&D is being performed to identify the $^{136}$Ba daughter isotope as a clear indicator for the $^{136}$Xe-decay \cite{Moe:1991ik}. Recent progress has been made in the identification of individual Ba atoms \cite{Chambers:2018srx}, and Ba$^+$ \cite{Green:2007rc} and Ba$^{++}$ \cite{McDonald:2017izm} ions. These developments should be pursued along with the challenging developments of techniques to extract barium from the detector volume. In a nEXO-style detector with only $2\nu\beta\beta$ decays contributing counts to the region of interest, the sensitivity would increase by a factor of $3-4$ compared to the projected nEXO sensitivity \cite{Albert:2017hjq}.
   
Another route is to look for neutrinoless double beta decay of natural xenon $^{136}$Xe in the DARWIN project \cite{Aalbers_2016}. The purpose of this detector which uses a dual phase liquid xenon TPC is primarily to look for dark matter. The 50 tonnes of natural xenon contain about 4.5 tonnes of $^{136}$Xe which allows to search in addition to dark matter for neutrinoless double beta decay with an interesting sensitivity \cite{Aalbers:2016jon}. Such a detector would also be sensitive to solar neutrinos \cite{Aalbers:2020gsn} and neutrinos from galactic supernovae.

Scintillation light from noble gases can provide fast timing and enhanced low energy trigger capabilities. An extensive program studying light emission from Ar and Xe, with Xe doping of large volume Ar detectors, or H2 (or D2) doping of large Xe detectors are promising technologies to get more out of a given size of these experiments.    

\item New scintillator technologies: Liquid scintillator detectors with photomultipliers (PMTs) on the outside have been a work-horse technology in neutrino physics for several decades. This can and should be further improved by developments of improved scintillators with better optical properties, better radiopurity, improved stability and light yield combined with further improved PMTs or novel optical sensors.    Growing detector sizes lead, however, to a number of challenges which warrant in addition new developments to meet the functional, cost, reliability and environmental requirements. An example is a water based liquid scintillator detector \cite{Alonso:2014fwf}, currently in a R\&D phase. Another example is wax-like scintillator~\cite{Buck:2019tsa} which allows high and stable loading without being an environmentally more difficult liquid. This may be important for future experiments on neutrinoless double beta decay aiming at the normal mass hierarchy. The wax-like scintillator additionally warrants the development of new readout technologies, e.g.\ based on optical fibers~\cite{Cabrera:2019kxi}. 

\item Low energy low threshold detectors: Low energy neutrinos lead to events with low recoil energy (electron or nucleus scatters) which requires detectors with lowest possible threshold. Coherent scattering at low energies and very high neutrino fluxes at reactors allow interesting experimenst to test coherent scattering and to search for new physics. An example is germanium PPC detector technology, for which the threshold has already been demonstrated for kilogram-size ultra low background detectors at a few hundred eV levels (see e.g. \cite{Bonet_2021}). These detectors can and should be further improved. Even lower thresholds are studied for for other materials, such as cryogenic bolometers \cite{Strauss_2017}.  

\item Water Cherenkov detectors have a long history of success and are able to scale to very large sizes.  Here improvements in photon detector technology, for low cost, high quantum efficiency, and fast timing, could have a major impact on the cost and efficiency of future larger detectors such as Hyper-K. Adding at sub-percent level gadolinium to the water will improve searching for proton decay and supernova relic neutrinos. Adding liquid scintillator to water will improve energy resolution and threshold, and R\&D studies on water-based liquid scintillator techniques are in progress. Water Cherenkov technology will continue to play an important role from exploring fundamental properties of neutrinos and neutrino astronomy to its application to nuclear monitoring at a remote distance. The potential of water Cherenkov technology would be maximized when combined with advanced technologies in photon detection, background tagging and light yield increase.

\item Much larger detector for cosmogenic and astrophysical neutrinos (including radio detection techniques): Cubic-kilometer sized neutrino detectors are required to observe cosmic neutrino flux at PeV ($10^{15}$ eV) energies, however, for exploring cosmic neutrinos with EHE ($10^{18}$ eV) energies, the required sensitive volume needs to increase by $1-2$ orders of magnitudes. For these highest energies, open water/ice Cherenkov neutrino detectors are surpassed by radio detectors as a more economic technological choice. 
    
\item Advancing R\&D on high resolution techniques for separating relic neutrino capture from $\beta$-decay endpoint electrons:  Development of RF tracking methods to dynamically select endpoint electrons through cyclotron emission radiation, new electromagnetic filter methods to transport endpoint electrons to cryogenic microcalorimeters and target substrates that maintain the intrinsic energy separation at the endpoint due to neutrino mass.

\item Metallic magnetic calorimeters (MMCs) are low temperature detectors being operated at milli-kelvin temperatures. They are characterized by very good energy resolution, excellent linearity and a fast detector response and they can be further improved for future experiments. They are composed by an absorber, suitable for a particular application, which is tightly connected to a paramagnetic temperature sensor, typically Au:Er or Ag:Er sitting in a static magnetic field. The sensor, in turn, is weakly connected to a thermal bath kept at constant temperature. When a particle deposits energy in the absorber, the temperature of the detector slightly increases leading to a change of magnetization of the sensor which is then detected as a change of magnetic flux in a suitable pick-up coil. Low noise large bandwidth readout is achieved by using the two-stage SQUID scheme. Recently the concept of microwave SQUID multiplexing has been adapted for the readout of large MMC arrays, only slightly decreasing the single channel readout performance.
    
MMCs are already used in a large variety of experiments. 
In general, quantum sensing in neutrino experiments 
using also transition edge sensors (TES) or superconducting tunnel junctions (STJ) remains a promising technology. 
In the field of Neutrino Physics, MMCs have been selected for the ECHo experiment aiming at the neutrino mass scale by analyzing the endpoint region of the electron capture spectrum of $^{163}$Ho, and for the AMoRE experiment developed for the search of neutrinoless double beta decay in $^{100}$Mo. Thanks to the very good performance which was already achieved both for very tiny detectors, as the one used in ECHo, and for macroscopic detectors, as the one used in AMoRE, as well as to the possibility to adapt the design and particle absorber material, MMCs can be further optimized to meet the requirements of new applications in neutrino physics. For example, a natural extension of the technology developed for the AMoRE experiment, would be to apply small scintillating crystals of different chemical composition for the measurement of coherent neutrino nucleus scattering.

\item Atomic tritium: Experimental systematic effects and theoretical uncertainties associated with molecular
tritium limit the targeted sensitivity of current neutrino mass experiments to $0.1-0.2$ eV. The
beta decay of tritium molecules not only adds a significant broadening (of $ \sim 0.4$ eV) to the
measured endpoint, which corresponds to a limit of the possible energy resolution, but also
shifts it by about 8 eV \cite{Saenz:2000dul,Bodine:2015sma}. 
Since this constrains the neutrino mass sensitivity of any tritium-based direct measurement, 
it motivates the transition from molecular to atomic tritium sources. Technical challenges are
the generation of atomic tritium,  subsequent cooling it down to K to mK ranges to make it
accessible for spectroscopy, while preventing its recombination at any surfaces.
Future experimental approaches for  precision physics from tritium beta-decay investigations 
(e.g., determination of the neutrino mass with 40 meV sensitivity or the search for relic neutrinos) will depend on reliable infrastructures to supply atomic tritium sources and beams
with high throughput, while complying with the very stringent upper limits on remaining 
traces of tritium molecules as well as maintaining purity, and long-term stability. 
Indeed, for future experiments based on atomic tritium the processing will be even more challenging
as the generation of a purely atomic tritium beam requires additional stages such as
dissociation and beam cooling which reduces the ratio of fiducial tritium activity in the source
vs.\ the amount of employed tritium. Furthermore, atomic tritium is more chemically active 
than the molecular form which also increases the impurity generation rates. Therefore,
specialized tritium processing facilities are required to cope with the expected high total throughput of tritium. Thinking further, 
future experiments involving high-intensity neutrino sources from several 100 g of bound tritium need facilities for the safe
preparation of these ultra-strong sources. The scientific measurements may then take place
at a different location with a tritium handling license -- which therefore reduces the demand
for permanent tritium processing compared to neutrino mass experiments.

\item Simulation tools for future projects: Larger and more complex neutrino projects tend to be more costly than previous experiments and require decades to build and operate. It is therefore very important to develop tools which are able to simulate and assess a project's potential as realistically as possible. An example is the GLoBES\cite{Huber:2008zz} simulation package which was developed for neutrino beam experiments and where all relevant properties of the source and of a realistic detector are encoded in a general language. This allows to vary assumed parameters in order to optimize projects. This and similar tools for other applications become more important and should be systematically supported. 
    
Detector simulation codes such as Geant4 \cite{Allison:2016lfl}, FLUKA \cite{Ferrari:2005zk} and MARS \cite{Mokhov:2017klc}, will continue to play an important role, as they allow precision simulation of particle  interactions at energies ranging from eV to TeV.  These codes are used at all phases of an experiment, from initial beam line and detector design to final extraction of precision parameters.  Additional simulated physics processes will continue to be needed as well as continuous efficiency improvements in the codes themselves.   These codes were mainly developed for particle physics and now have very broad impact in fields ranging from Mars exploration to proton therapy. 
    
Neutrino interaction simulation codes, as discussed in Sec.\ \ref{chap:interactions} will also continue to improve as more data constrains the existing models. 
    
\item Reconstruction tools, machine learning:  the reconstruction of neutrino properties from complex interaction final states recorded by fine-grained detectors requires sophisticated algorithms.  Neutrino experiments also often face the the challenging problem of sifting subtle neutrino signals from overwhelmingly large backgrounds.  In both of these cases, machine-learning algorithms can be effectively deployed.  For some real-time applications, such as for triggering, such fast machine-learning algorithms can be implemented on FPGAs.

\item Improved data management and readout technologies: As neutrino detectors have grown in size and in spatial resolution, the data volumes they generate have grown. Interest in low energy-threshold physics such as supernovae and solar neutrinos means that aggressive zero-suppression is unwise.  For example, the existing ProtoDUNE and MicroBooNE LArTPC detectors generate $100-200$ MB of data for a single readout, with lossless compression only gaining a factor of three reduction. A single 5 ms readout  of a DUNE far detector module is $2-6$ GB in size while a full supernova readout over 100 s would generate $100-400$ TB of data. Data volumes from large water-based detectors are smaller but not by orders of magnitude. Raw signals from wires and photo-detectors need to be identified as energy depositions and then combined to form interactions.  The raw size of the data unfortunately requires that data be split up for processing and then combined once energy deposits have been found. At that point novel pattern recognition algorithms take over. 
  
These sophisticated high precision event reconstruction algorithms are CPU intensive. These problems are well suited to the effective use of the latest hardware (CPU, GPU and FPGA accelerators) and software technologies (deep learning, graph neural net, complex-valued neural net etc.) but will require substantial development due to the unique geometry and event size in neutrino interactions.   Overall the computational needs for an individual neutrino experiment are not as large as those for LHC experiments, but the novel computational ``shape'' of the problem, with large data volumes needing to be held in memory at the same time, or distributed and recombined after processing, requires new algorithmic development. 
    
Once events have been reconstructed, extraction of neutrino oscillation parameters in the presence of large numbers of uncertainties is equally challenging computationally.  For example, the NOvA parameter extraction \cite{Acero:2019ksn} has relied on the supercomputer facilities at NERSC. 
    
Neutrino experiments are now collaborating with other large experiments via the High Energy Physics Software Foundation (HSF) \cite{Alves:2017she} in the development of common tools for managing and reconstructing the data from HEP experiments. Examples include the adoption of Rucio \cite{Barisits:2019fyl} for file management, large scale databases for calibration and data description and the use of worldwide computing grid capabilities.   This collaborative work will have impact across the field of HEP and in nuclear physics and astrophysics. 

\end{itemize}

%\subsection{Capabilities}
%\input{WG5/WG5_capabilities.tex}

\subsection{Infrastructure}

    \subsubsection{Supporting Capabilities for the Science at Underground Facilities}
    Neutrino detectors that study natural processes and sources often require low radio-background construction, and ultra-low radioactive environments to observe the rare and weak signals from the neutrino interactions. To provide the latter ultra-quiet environment requires shielding from cosmic radiation and local radioactivity, which point to hosting such detectors deep underground. This is a similar requirement within the search for Galactic dark matter, and so the neutrino and dark matter communities share similar problems in creating ultra-quiet environments.
    
    A network of deep underground labs have been established around the world for both neutrino and dark matter studies in physics, and a growing list of additional science objectives that require this quiet environment. The greatest constraint on background levels is currently being placed by neutrinoless double-beta decay systems, to prevent cosmogenic activation of isotopes that might lead to background events in the region of interest. As tonne-scale, or larger, detectors are developed, either greater depth of better veto and shielding systems are required. Currently there are two facilities below 2 km depth in the world, and several shallower facilities developing shielding strategies. To shield local radiation also requires effective and usually active shielding, such as large water-Cherenkov or liquid scintillator veto systems, which can also be tasked as test facilities where full scale systems may need to be deployed for background assays.
    
    In addition to the physical environment, underground laboratories are enhancing their support for additional services and systems, such as liquid noble and cryogenic systems, safety and environmental control, accessibility and logistics, and project management.
    
    To facilitate the delivery of the scientific program, collaboration and sharing of best-practice between the facilities would be strongly encouraged. This should include an audit of available and accessible underground cavities and halls, which should be assessed against the community plans; to ensure the facilities can provide appropriate space for future projects. This will require an optimisation between depth, scale, location and capability.
    
    \subsubsection{Supporting Capabilities for Low Background Experiments} Many low energy neutrino experiments have very low event rates and it is therefore of out-most importance to further improve techniques or to develop new technologies to identify and mitigate natural radioactivity. Facilities for $\gamma$, $\alpha$ and Rn screening have already remarkable capabilities, but their sensitivity should be further improved. An example is the work towards a new generation of GeMPI detectors with even better sensitivity as the existing ones. The growing number of bigger neutrino detectors (together with larger direct dark matter detection experiments) require also an enlarged capacity of screening facilities. The gamma screening program must be accompanied by complementary direct radon emanation measurements, which are extremely sensitive to surface impurities. High sample throughput rates as well as strict reproducibility can be achieved by an automated system as pioneered in \cite{aprile2020222rn}. Inductively Coupled Plasma Mass Spectrometry (ICP-MS) is another important supporting technology since it is complimentary to gamma screening and can achieve excellent sensitivity for long-lived isotopes like $^{238}$U, $^{232}$Th, and $^{40}$K. In addition promising new directions could be realized, like a counting facility with low background comparable to that of real neutrino detectors. One example is a large liquid-scintillator detector that counts radiations from test samples immersed in it. It can evaluate surface alpha- beta-, gamma-rays, and neutron emission from the samples. It is important to share measurement time and results inside the community. Another aspect concerns improved techniques to optimally avoid re-contamination by cosmogenic activation, Radon-plate-out or other contamination. This requires adequate underground storage capabilities at institutes where detector components are produced and in some cases a supply of Radon free air. Furthermore new low background shielding and vetoing technologies like optimized graded shieldings should be further developed. This allows inside the shield conditions which correspond to deeper underground locations. This added ``virtual depth'' leads to more flexibility for detector locations and allows to make optimally use of existing underground laboratory space and the available infrastructure in each location. 
    
 In addition to screening capabilities, underground production, manufacture and storage of low background materials are becoming increasingly important. Underground copper production and machining has been established by the neutrino community as a viable operation, and will be expanded by several facilities. Production of low background target material is also being developed in many underground laboratories, and extraction of low background material where viable, such as underground argon as a low $^{39}$Ar  background shield or dark matter target. Finally, once material has been produced or procured, underground storage is required to prevent additional cosmogenic activation -- this requires coordination between underground facilities, although does not need to be at great depths to facilitate shielding.
    
    \subsubsection{Test Beams for Detector Development and Calibration}
    Precision measurements of neutrino properties require the development of novel detector techniques and precision calibration of detector components.  Experiments worldwide depend crucially on test beam programs.  Examples include \begin{itemize}
    
    \item The neutrino-specific \href{https://home.cern/science/experiments/cern-neutrino-platform}{CERN Neutrino platform}, which has hosted the DUNE prototypes and performed testing of the BabyMIND detector for T2K.  
    \item  The use of the Linac Coherent Light Source at SLAC to demonstrate the scattering of radio waves  from the interactions of high energy particles\cite{Prohira:2019glh} as an alternative method of detecting neutrino interactions in very large volumes of ice. 
    \item The invaluable work at test facilities worldwide to characterize detectors with beams of known particles. Charged particle beams in the -- difficult to achieve -- sub-GeV energy range are especially valuable.   
    \end{itemize}
    Test beam studies benefit the neutrino program but  also benefit and contribute to efforts across multiple other fields.  Improved data on interactions is fed back into physics models such as Geant4 \cite{Allison:2016lfl} and FLUKA \cite{Ferrari:2005zk} with benefits to neutrino physics, collider experiments, nuclear physics, space science and medicine.

\subsection{Theory}

Revealing the secrets of nature, i.e., developing a coherent physics picture, requires an intense interaction between theory and experiment. The wide range and interdisciplinary character of neutrino physics is reflected in the theory community both topic-wise and methodology-wise. While parts of neutrino physics theory share common characteristics with particle physics or astrophysics, asking the right questions at the right time has been especially important in the past and to develop the field (including experimental approaches) further --  similar to other theoretical disciplines in astroparticle physics. The pillars of neutrino physics theory are:
\begin{description}
\item[Particle physics phenomenology:] Combination of the vastly different experimental information in a global picture, interpretation, and guidance of experiments into directions where especially exciting results may show up. This combined information does not only include neutrino experiments, but also other fields such as searches for charged lepton or quark flavor violation, collider or astroparticle physics and cosmology. Examples are neutrino oscillation studies and global fits, connections of neutrinoless double beta decay results with the baryon asymmetry of the Universe, the simultaneous explanation of neutrino mass and dark matter in concrete models, or collider tests of neutrino mass mechanisms.  
\item[Physics BSM model building:] Taking the results of neutrino physics and other experiments into account to construct new theoretical models for particle physics. This includes the connection with adjacent disciplines through new particles, interactions and energy scales. Examples are Grand Unified Theories being able to explain maximal atmospheric mixing, combining flavor with CP symmetries to explain near-maximal CP violation, linking the anomalous magnetic moment of the muon with neutrino mass models, or linking the stability of dark matter particles with the conservation of lepton number.  
\item[Interdisciplinary approaches:] Identification, exploitation and application of methods and results from nuclear physics, geophysics, astrophysics and other disciplines for neutrino physics. Examples are the prediction of precise reactor neutrino spectra, the computation of neutrino-nucleus cross sections, the interpretation of georeactor hypothesis in terms of expected geoneutrino fluxes, and the prediction of atmospheric neutrino fluxes from cosmic-ray interactions, interpretation of neutrino data in terms of source physics, such as solar models.
\item[Astrophysical multi-messenger modelling:] Study of the acceleration, propagation and interactions of the cosmic rays, which are the primaries for the astrophysical neutrino production. Astrophysical multi-wavelength models including neutrino production and hadronic signatures in the electromagnetic spectrum. Phenomenological interpretation of the astrophysical observed neutrino flux in terms of different source classes and its characteristics. Identification of the production sites of astrophysical neutrinos in individual cases and development of macroscopic source models.
\end{description}
The above examples illustrate that a development of the field of neutrino physics requires substantial theory support owing the complexity, range and interdisciplinary character of the field.

\subsection{Impact and Societal Benefits}

As often the case in physics, fundamental research can have surprising hands-on applications with societal benefits. Examples are the discovery and utilization of X-rays, things like nuclear energy, solar energy, computer tomography scans, magnetic resonance imaging, positron emission tomography, semiconductors, superconductors, low temperature and high vacuum technologies, the worldwide web, electronic communication, grid and cloud computing, data science and machine learning, etc.\ have emerged from fundamental research. It is difficult to envisage daily life and international business without those developments. Predicting which next future technological breakthrough will emerge from fundamental physics is nearly impossible, but it is likely to happen as we push to more and more sophisticated technologies for our research projects. Many smaller developments happen on a daily basis in the word-wide neutrino physics community. Thereby we, most often involuntarily, pay back to the public which funds our endeavours. 

An important output to society from neutrino physics is highly qualified personnel which are sought after by high-tech industry. Young scientists working in neutrino experiments are trained in cutting-edge technologies such as cryogenic engineering, photodetectors, electronic systems, mass spectronomy, firmware programming, micro-machining, nano-tools, clean-room technology, big data analysis or machine learning, to name a few. Those are highly marketable and transferable skills, leading to application in high-tech industries, data science and artificial intelligence, health care, education, finance, natural resources exploration and many more.  Often there is direct partnering with industry partners to develop the highly sensitive devices we need, and a large number of spin-off companies are founded. A survey of the over 1000 highly qualified personnel that were members of the SNO collaboration, as an example, including students, scientists, post-doctoral fellows, technicians, and engineers, showed that about 19\% have gone on to technical positions in industry, 23\% to academic positions in research fields other than underground science, 49\% to higher positions in underground science research, and 9\% to other positions.

Probing neutrinos requires highly sensitive detection techniques. This leads to several possibilities for applications. One can monitor the production of weapons-grade plutonium  in nuclear detectors,  which is obviously helpful in non-proliferation of nuclear weapons and  monitoring of nuclear reactors for nuclear safety. Understanding better the reactor neutrino flux will make possible a more efficient use of nuclear energy. 
Further developments of detectors for radioactivity have been used for better security measures at airports or freight terminals. 
A particular example, among the many, is given by SiPMs mentioned in Sec.\ \ref{wg5_tech} above.  Those allow for single photon sensitivity with sub-nanosecond timing, and this possible in a robust and compact packaging with little energy consumption. Applications range from quantum cryptography (which rely on single photon transmission), single photon emission computed tomography, or single photon measurements of UV scattering, fluorescence or absorption to detect traces of smoke or specific molecules to monitor environmental hazards or give early warning on forest fires. 
Applications of neutrino and astroparticle detectors to medical technology are also manifold, ranging from developments of semiconductor-based dosimeters to dynamical X-ray imaging with photon counting and particle tracking pixel detectors. 

Neutrino physics contributes also to many other fields of science. Improvements in detectors and capabilities needed for neutrino studies have made it possible to scan archeological artefacts ranging from Napoleon's hair (checking with neutron activation  if he was poisoned with arsenic), to dating wine through measurement of various nuclear-test created isotopes, to the study of ancient pyramids (probing cavities with muon tomography). 
Other scientific fields that benefited from neutrino physics are geology by the additional information from geoneutrinos, mining through the possible detection of ore bodies, oil and minerals with neutrino tomography, and marine biology by the presence of sensitive detectors at the bottom of the ocean. Atmospheric science benefited from capabilities to detect traces of miniscule amounts of radioactive materials, such as $^{133}$Xe, stemming from the  Fukushima Daiichi nuclear disaster. 
All this  comes on top of the obvious parallels with dark matter research such as low background techniques, which have been frequently mentioned in this document. The synergy between neutrino physics and deep underground science is also strong, with many deep underground facilities being developed primarily for the large neutrino (or dark matter) detector systems, which allows additional convergence research to be undertaken, such as low radiation genetics studies, astrobiology and sub-surface biosphere studies, and tests of fundamental physics properties. 

We should stress here as well the benefits of international cooperation and supporting science in regions other than Europe, North America or Asia. The broadening to a world-wide level is good tradition in particular in high-energy physics and has brought many fruitful results, and exceptional talents to the field. It has contributed to peaceful cooperation of scientists from many competing countries and cultures.  

The fascinating properties of neutrinos are furthermore an ideal example to interest students and pupils in fundamental physics, make them study physics and thus keep the field going for decades to come. The open questions remaining in neutrino studies act as an attractor for those interested in solving some of the most challenging questions in contemporary physics, the active nature of this field being evidenced by two Nobel Prize awards in the last 20 years. The intriguing quantum mechanical foundations of neutrino oscillations, the many connections the light neutrinos may have with the cosmos, the incredible number of hardly interacting neutrinos around us, the huge distances they travel basically unperturbed  
and the spectacular experimental facilities deep underground or in the sea and ice will continue to attract bright minds and will thus be beneficial for the whole fields of physics, cosmology and astrophysics. This will also keep the steady flux of technological breakthroughs from fundamental research constant. Moreover, in combination with the fast developments in our field we will keep attracting the brightest and motivated minds, who may prefer this over doing for instance R\&D for future colliders taking data decades from now. 

Our understanding of the Universe will increase with more and more understanding of neutrinos and their properties. 
They influence the creation of light and heavy elements in the early Universe or in stellar explosions, and they might even be responsible for the existence of matter as such in the Universe. The physics that generates neutrino mass will necessarily be part of whatever theory will be the next Standard Model of particle physics. 
Further understanding neutrinos, hopefully also in combination with other breakthroughs in physics and cosmology, will further clarify how the cosmos works and what our role in it is, thus answering the most fascinating questions humanity dealt with.

\clearpage

\section{Physics Implications}
Neutrino physics has already led to a number of remarkable discoveries.  By summarizing in this section the main topics for the future, we would like to express our vision that the field has  excellent potential for more exciting discoveries and advances. 

There are two main directions: One is the use of neutrinos as probes into sources which are otherwise not accessible. Especially in astrophysics and cosmology, neutrinos can convey unique information from a variety of dense and/or other hidden places of our Universe, where they were produced. Neutrinos will thus allow us to learn about these sources and will improve our understanding of the most extreme environments in which neutrinos can propagate.

The other direction is the great potential of neutrinos to study central questions of fundamental physics in a unique way. The fact that neutrinos are massive is indeed the first physics Beyond the Standard Model (BSM). Explaining neutrino mass and lepton mixing is thus expected to be deeply connected to whatever theory will eventually replace the SM of particle physics. 
These masses can be explained in different ways, and unraveling the correct mechanism and how it connects to many other topics will be a very exciting task for the future. This new understanding may also help to understand how fermion masses arise, why three generations of quarks and leptons exist or perhaps why more fermions such as sterile neutrinos should exist. The specific answers to these questions have many interesting connections to high energy physics and astroparticle physics. The important questions include the question of whether neutrinos are Dirac or Majorana particles, and more generally, whether lepton number is violated. Neutrinos are very sensitive to new effects, and can probe energy scales that are comparable to, or above, the reach of current and future colliders.  
Another route is the connection to dark matter, or more generally to dark sectors. There are also very important connections of neutrinos to cosmology, for example to the understanding of the baryon asymmetry or of the formation and development of structure in the Universe. 

The following sections expand on the connections of neutrinos to broader questions in fundamental physics, astrophysics and cosmology in more detail. Various possibilities of new neutrino physics beyond the usual standard paradigm are also discussed. 

\subsection{Learning about Sources}
\label{sec:sources}

Neutrinos can be used as messengers to learn about their sources. The production of heat in the Earth's interior by natural radioactivity can be tested by the radioactive decays from $^{232}$Th and $^{238}$U, which produce neutrinos with energy above the inverse beta decay threshold. The absorption of TeV neutrinos and the oscillations of GeV neutrinos depend on the density of matter, which means that information on the density and structure of the Earth can be obtained by, for example, measuring the fluxes of atmospheric neutrinos as a function of path through the Earth's matter. Neutrinos also probe the nature of processes in the interior of the Sun, such as the contribution of the CNO fusion cycle, which is sub-dominant for a solar-mass star but which dominates for the majority of stars in the Universe. Neutrinos also play an important role in the dynamics of core-collapse supernovae, which means that they can be used to test our understanding of the explosion mechanism. An interesting target of future detectors is the contribution to a diffuse neutrino flux from all supernovae over the history of the Universe, which depends on the distribution of the sources. Neutrinos can also be used to study the cosmos at extremely early times via the potential (and challenging) detection of primordial neutrinos.

\subsubsection{Better Understanding the Interior of the Earth}

The present observations of the geoneutrino fluxes from the $^{232}$Th and $^{238}$U chains by KamLAND and Borexino are outstanding scientific achievements. It is significant that the experimental results are compatible with geophysical estimates of the flux, which, however, have a large uncertainty. First constraints on the amount of  heating from the Earth's interior  caused by radioactivity has been established by detectors that originally aimed at probing fundamental neutrino properties. The heat caused by radioactivity drives plate tectonics, and neutrino physics can provide information on this interplay which would be otherwise inaccessible. Using the chondritic ratio for the $^{232}$Th and $^{238}$U abundances, one may even extract the age of Earth. 
Further verification that mantle and crust are the sources of geoneutrinos, as predicted by most models, will require more exposure, technological advances, and different sites than the present ones. Further measurements of the distribution and the overall magnitude of the heat provided by geoneutrinos are long-term goals of this area of research.

The Earth's interior can be also studied using externally produced neutrinos, such as atmospheric neutrinos. There are two approaches in the literature: {\em Neutrino absorption tomography} and {\em neutrino oscillation tomography}; see \cite{Winter:2006vg} for a review. Neutrino absorption tomography uses the fact that the neutrino cross section increases with energy; the absorption length becomes comparable to the Earth's diameter at about 40~TeV. Consequently, the absorption of neutrinos along their straight paths through Earth can be used to study the density profile of Earth in a manner to the similar to the X-ray tomography technique; see e.g.\  Ref.~\cite{Donini:2018tsg} for tomography using atmospheric neutrinos. The main limitations of absorption tomography are the relatively low statistics at these very high neutrino energies regardless of the source class, and the increase of cross-section uncertainties at high energies. 

In contrast, neutrino {\em oscillation} tomography uses matter effects in neutrino oscillations at energies in the GeV range (for atmospheric neutrinos); see e.g.\ \cite{Rott:2015kwa,Winter:2015zwx}. Matter effects are primarily sensitive to the electron density, which can be translated into the matter density if the composition (actually the ratio $Z/A$) is known, or can in fact be used to study the composition of matter traversed, such as of the Earth's core.  An interesting effect in neutrino oscillation tomography is that the Hamiltonians for different matter density layers do not commute, which means that additional information beyond the simple column density is imprinted in the energy spectrum. While the statistics in the GeV range for atmospheric neutrinos is  generally better for this approach, the main limitations are parameter degeneracies, detector threshold effects, and directional uncertainties (which are especially relevant if one wants to study the inner core of the Earth). 

In both tomography approaches, a precision competitive with seismic wave tomography or other geophysical approaches is not expected; however, neutrinos measure different quantities than these approaches. Apart from tomography using atmospheric neutrinos, many other sources have been proposed (such as solar neutrinos, astrophysical neutrinos or neutrino beams), see e.g.\  \cite{Winter:2006vg}. For example, a new dedicated neutrino beam experiment may even measure the density (or corresponding composition variable) of the Earth's inner core at the percent level~\cite{Winter:2005we}. Since such an experiment, however, requires significant new technology and dedicated investment, the approaches using atmospheric neutrinos seem to be the most promising at this point. 

\subsubsection{Learning about the Sun}

Solar neutrinos represent an invaluable means to study neutrino properties as well as to learn about the interior of our Sun. Historically, detection of solar neutrinos was the first to hint towards the existence of neutrino oscillations. Today, we study the effects of dense matter, both in the Sun as well as in the Earth, on the electron-flavor survival probability of neutrinos, searching for neutrino properties and interactions beyond those included in the SM. Solar neutrinos are the only direct probe of hydrogen fusion processes powering the Sun. Currently, precision spectroscopy of solar neutrinos from the $pp$ chain fusion has been performed by Borexino for all species and by Super-Kamiokande for $^8$B neutrinos. Recently, Borexino has verified the existence of the CNO fusion cycle, contributing with about 1\% of the total solar energy. 
The CNO cycle dominates for more massive stars, which are, in fact, more abundant in the Universe. Thus, this measurement allows for the study of the primary mechanism 
for the conversion of hydrogen into helium in stars.
 Future experiments aim to further improve precision of solar neutrino measurements, that in  the case of CNO neutrinos can help to solve the long-standing problem of solar metallicity, e.g.\ solar abundances of elements heavier than helium.

\subsubsection{Learning about the Death of Massive Stars}

Neutrinos are key particles in core-collapse supernovae, which mark the death of massive stars. With the dawn of the multi-messenger era, neutrinos offer very exciting prospects to learn about the yet mysterious supernova physics. At the same time, supernovae are unique laboratories to study particle physics under extreme conditions.

As described in Sec.\ \ref{sec:source_SN}, the flash of neutrinos accompanying a core-collapse supernova in the Milky Way or its immediate neighborhood will provide an unprecedented view in neutrinos of the hidden processes underlying the collapse, the formation of a compact remnant (neutron star or black hole) and the subsequent supernova explosion.  The energy, time and flavor profile of the neutrinos, observable in multiple detectors worldwide, contains signatures of astrophysical mechanisms underlying the dramatic event.  
Neutrinos, together with gravitational waves, carry information about the physics of the pre-explosion dynamics, such as hydrodynamical instabilities, as well as rotation and black hole formation~\cite{Tamborra:2013laa,Walk:2019miz, Walk:2018gaw}. The observed neutrino burst can also be used to optimize the time window for gravitational wave searches and locate to the supernova in the sky~\cite{Nakamura:2016kkl,Adams:2013ana}. 
In addition, the long timescale signal of neutrinos emitted during the cooling phase carries information about the neutron star physics and possibly its equation of state~\cite{Li:2020ujl, GalloRosso:2018ugl}. In order to maximize our chances to extract precious information from the next nearby supernova burst, these concepts will be better explored in the near future, as a growing sample of supernova simulations becomes available. 

Because of the uncertainties on the flavor conversion physics and degeneracies with the supernova properties itself (e.g., its mass, nuclear equation of state), the neutrino signal from the next nearby explosion may not provide clear insight on the neutrino mass ordering and mixing parameters. Non-standard physics scenarios could greatly modify the expected neutrino signal~\cite{Suliga:2019bsq,deGouvea:2019goq, Das:2017iuj,Nunokawa:1997ct,Tamborra:2011is,Shalgar:2019rqe, Carenza:2020cis, Stapleford:2016jgz}. 
Furthermore, the fact that the emergence of the neutrino burst from the stellar envelope precedes the electromagnetic signatures by hours or longer means that the detection of the neutrino burst can provide an early alert of a core-collapse signal in multiple messengers, increasing the astronomical community's ability to harvest data from the supernova's early photon signals.

Another unsolved problem concerns the nucleosynthesis occurring in supernovae~\cite{Arcones:2012wj, Cowan:2019pkx}. Current simulations report relatively proton-rich environments, leaving room for a light rapid neutron capture process only. As physics linked to magneto-hydrodynamics will be treated consistently with neutrino transport, the amount of heavy elements that can be produced in supernovae may need to be reassessed, also in the light of an improved understanding of neutrino mixing. 

Finally, high energy neutrinos (with ${\cal O}(10-100)$ TeV energy) are expected to be produced from freely expanding supernova ejecta interacting with the circumstellar medium. The diffuse emission of high-energy neutrinos should constitute about 10\% of the diffuse background currently observed by the IceCube Neutrino Observatory; however the non-detection of neutrinos from targeted searches already allows constraints on the fraction of shock energy channeled into protons~\cite{Petropoulou:2017ymv, Murase:2017pfe, Murase:2010cu, Zirakashvili:2015mua}. 

Coincident observation of supernova burst neutrinos with other signals on different time scales -- gravitational waves, later high-energy neutrinos, electromagnetic radiation in all wavelengths -- will provide rich information for both particle physics and astrophysics in the multimessenger community.

\subsubsection{Learning about the Cosmos}\label{sec:CNB2}

The cosmic neutrino background (C$\nu$B, see Sec.~\ref{sec:source_rel}) has only been studied indirectly so far, through the impact that this bath of relic neutrinos has on the cosmic microwave background or on the growth of late-time structures. Most of these efforts lead to a measurement of the effective number of relativistic neutrino species in agreement with the canonical value of $ 3.044$, with less than  10\% uncertainty, thus  confirming the existence of the C$\nu$B~\cite{Aghanim:2018eyx}. As is often the case for indirect approaches, these measurements, however, are model-dependent: the results and the uncertainties vary with the exact assumptions that are made. A different approach has recently led to the detection of a subtle phase shift of the acoustic oscillations caused by the C$\nu$B. This shift comes from the fact that in the early Universe  neutrinos  propagate  at nearly the speed of light, faster than sound waves in the hot plasma of baryons and photons~\cite{Follin:2015hya, Baumann:2018qnt}. This phase shift cannot be mimicked  by  other properties of the primordial plasma, and its detection confirms the existence of a cosmic neutrino background at the predicted temperature of $T=1.95$ K~\cite{Baumann:2018qnt}. 

The PTOLEMY project offers an interesting prospect for the first {\em direct} detection of the C$\nu$B~\cite{Betti:2019ouf}, providing possibly the first model-independent confirmation of its existence. Furthermore, while the properties of the C$\nu$B are theoretically expected to be very similar to those of the cosmic microwave background, a direct detection would provide a unique confirmation of our cosmological model at an epoch when the Universe was only about one second old (compared to about three minutes at the epoch of Big-Bang nucleosynthesis and 380,000 years  when the cosmic microwave background was emitted). 

On the issue of neutrino masses, a measurement of $\sum m_\nu$ would fix the amount of power suppression on small scales caused by the free-streaming of massive  neutrinos. As a consequence, studies of other particles that could have a similar impact on the growth of structures, such as warm dark matter particles (bosonic or fuzzy dark matter, thermal relics, keV sterile neutrinos), would be facilitated.

\subsection{High-Energy Neutrino Astrophysics}
\label{sec:heneutrinos}

The detection of astrophysical neutrinos beyond TeV energies has opened a new way to test the origin of cosmic rays, which are the primaries needed for the production of the neutrinos.  Recent discoveries of a diffuse flux from astrophysical neutrinos and by the association of neutrinos to individual astrophysical objects have lead to a new field, which is perhaps best described as ``high-energy neutrino astrophysics''.  Its nature is different from conventional particle-physics-oriented neutrino physics as astrophysical scenarios themselves carry a lot of freedom and uncertainty.  Its current mainstream therefore pertains to the domain of multi-messenger astrophysics, combining the information from neutrinos, cosmic rays, electromagnetic radiation and also gravitational waves to maximally exploit the information from  the sources, the Universe between the sources and Earth through the transport of these messengers, and BSM physics. 
Here we put a strong focus on the multi-messenger perspective, which is particularly important in this field. 

\subsubsection{Neutrino Production from Cosmic-Ray Interactions}

Cosmic rays (protons or nuclei) are frequently assumed to be accelerated to power law spectra $dN/dE \propto E^{-\alpha}$ by processes such as Fermi shock acceleration with a spectral index $\alpha \simeq 2$; these spectra are also called ``non-thermal'' spectra, in contrast to ``thermal'' spectra which are peaked at a characteristic energy. In $pp$ interactions, the target gas is typically non-relativistic, which leads to neutrino spectra described by the same power law as the primaries $\propto E^{-\alpha}$. In $p \gamma$ interactions, the target radiation is (always) relativistic and may follow a power law $dN/dE \propto E^{-\beta}$ itself if it is generated by non-thermal primaries, such as from synchrotron radiation off co-accelerated electrons. In that case, the neutrino spectrum $dN/dE \propto E^{-\alpha+\beta-1}$ emerges in the $\Delta$-resonance approximation, which only follows the primary spectrum if $\beta \simeq 1$. Therefore, the physics of $p \gamma$ sources is typically more complicated, and the frequently used assumption of an $E^{-2}$ neutrino spectrum does not hold; see \cite{Fiorillo:2021hty} for a more detailed discussion. 

In many practical cases (such as for AGN) one has a photon spectral index $\beta>1$ in the relevant energy range (see below), which means  that the neutrino spectrum in $E^2 dN/dE$ is strongly peaked with a peak determined by the maximal neutrino energy. Since to leading approximation the pion takes about 20\% of the primary proton energy in the above interactions, and each pion decays into four leptons in \equ{piplusdec}, one has $E_{\nu, \text{peak}} \simeq 1/4 \times 0.2 \, E_{p, \text{max}} \simeq 0.05 \, E_{p, \text{max}}$ as the neutrino peak energy. This simple example illustrates how the observed neutrino energy directly traces the primary cosmic ray energy. The astrophysical neutrino detection can therefore  be used a) as direct evidence for cosmic ray acceleration in the source, and b) as a tracer of the primary cosmic ray energy. An exception to this principle are sources with strong magnetic fields, in which the maximal neutrino energy does not follow the maximal primary energy because beyond a critical energy the pions and muons in \equ{piplusdec} lose energy by synchrotron radiation  faster than they can decay. In that case, the critical energy determines the maximal neutrino energy and can be only used as a lower limit for the primary cosmic ray energy; GRB neutrinos related to the prompt phase of the emission represent an example for which this is the case.

As indicated earlier, astrophysical neutrinos are typically expected to be produced 
in the ratio $\nu_e:\nu_\mu:\nu_\tau$ of $1:2:0$ from the pion decay chain; deviations are possible if different production modes (such as neutron decays or kaon decays) are at work, or if the secondary muons (e.g.\ the ones in \equ{piplusdec}) synchrotron-cool faster in magnetic fields than they can decay, see \cite{Winter:2012xq} for a review. Different ratios of $\pi^+$ and $\pi^-$ at the sources are expected to leading order for $pp$ and $p\gamma$ interactions (see \equ{Delta}), respectively, leading to different ratios of neutrinos and antineutrinos  at the source, and (after flavor mixing) at detection. This particular feature may be tested by the Glashow resonance $\bar{\nu}_e + e^- \to W^- \to \text{anything}$ at around $6.3 \, \text{PeV}$, which is sensitive to electron antineutrinos only; one such event has been recently discovered~\cite{IceCube:2021rpz}. Note, however, that in practice additional production processes have to be taken into account for the hadronic $p\gamma$ interactions, which lead to substantial $\pi^-$ production.

\subsubsection{Connection with Electromagnetic Radiation}

First of all, both $pp$ and $p\gamma$ interactions lead to $\pi^0$ production (see e.g.\ \equ{Delta}), which pre-dominantly decay by 
\begin{equation}
 \pi^0 \overset{98.8\%}{\longrightarrow} \gamma + \gamma \, . \label{equ:pizerodec}
\end{equation}
These gamma rays have energies $E_\gamma \simeq 1/2 \times 0.2 \, E_p \simeq 0.1 \, E_p$ because two gamma rays are produced from a pion which carries about 20\% of the initial proton energy. For TeV -- PeV neutrinos, these gamma-ray energies are of a similar magnitude, an important secondary indicator for the neutrino production typically limiting the models indirectly. For example, for $pp$ sources transparent to gamma rays, the emission feeds into the extragalactic gamma-ray background limiting the spectral index of the neutrino flux~\cite{Murase:2013rfa,Bechtol:2015uqb}.
If, on the other hand, the source is sufficiently compact, as it might be expected for efficient $p\gamma$ neutrino producers, the gamma rays will be re-processed inside the source, which means that gamma rays may not be expected from such a source~\cite{Murase:2015xka}. These conclusions, however, are derived using assumptions for the diffuse astrophysical neutrino flux and do not necessarily apply to individual sources. 

Another interesting question for $p \gamma$ interactions is that of the target photon energy the cosmic rays interact with; it can be shown that $E_X \text{\,[keV]} \simeq 0.01 \, \Gamma^2/(E_\nu \text{ [PeV]})$ is the relevant target photon energy at the $\Delta$-resonance for interactions with internal radiation in a source traveling with Doppler factor $\Gamma$ towards the observer. For sources such as AGN and TDE jets, $\Gamma \simeq 10$ is a typical estimate, which means that cosmic rays producing PeV neutrinos typically interact with X-rays in the keV-range; X-ray monitoring is therefore important to learn about the neutrino production. The origin of these photons depends on the source class and model: typical examples are synchrotron radiation from co-accelerated electrons, a more complicated combination of radiation processes, or external (re-processed) accretion disk radiation (in which case $\Gamma$ cancels and $E_X \text{\,[keV]} \simeq 0.01 /(E_\nu \text{ [PeV]})$).

\begin{figure}[t]
\begin{tabular}{ccc}
\includegraphics[width=0.3\textwidth]{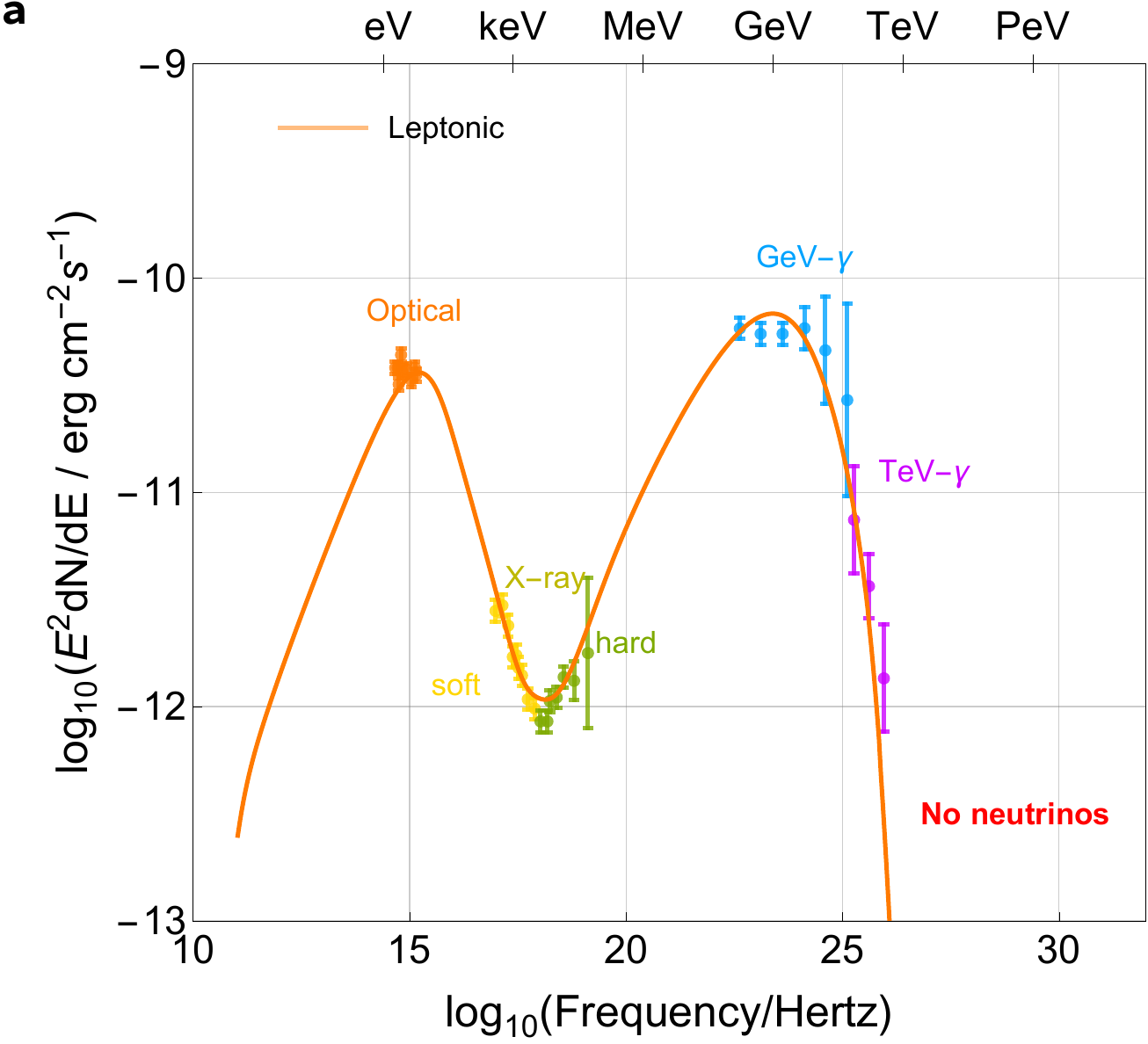} &
\includegraphics[width=0.3\textwidth]{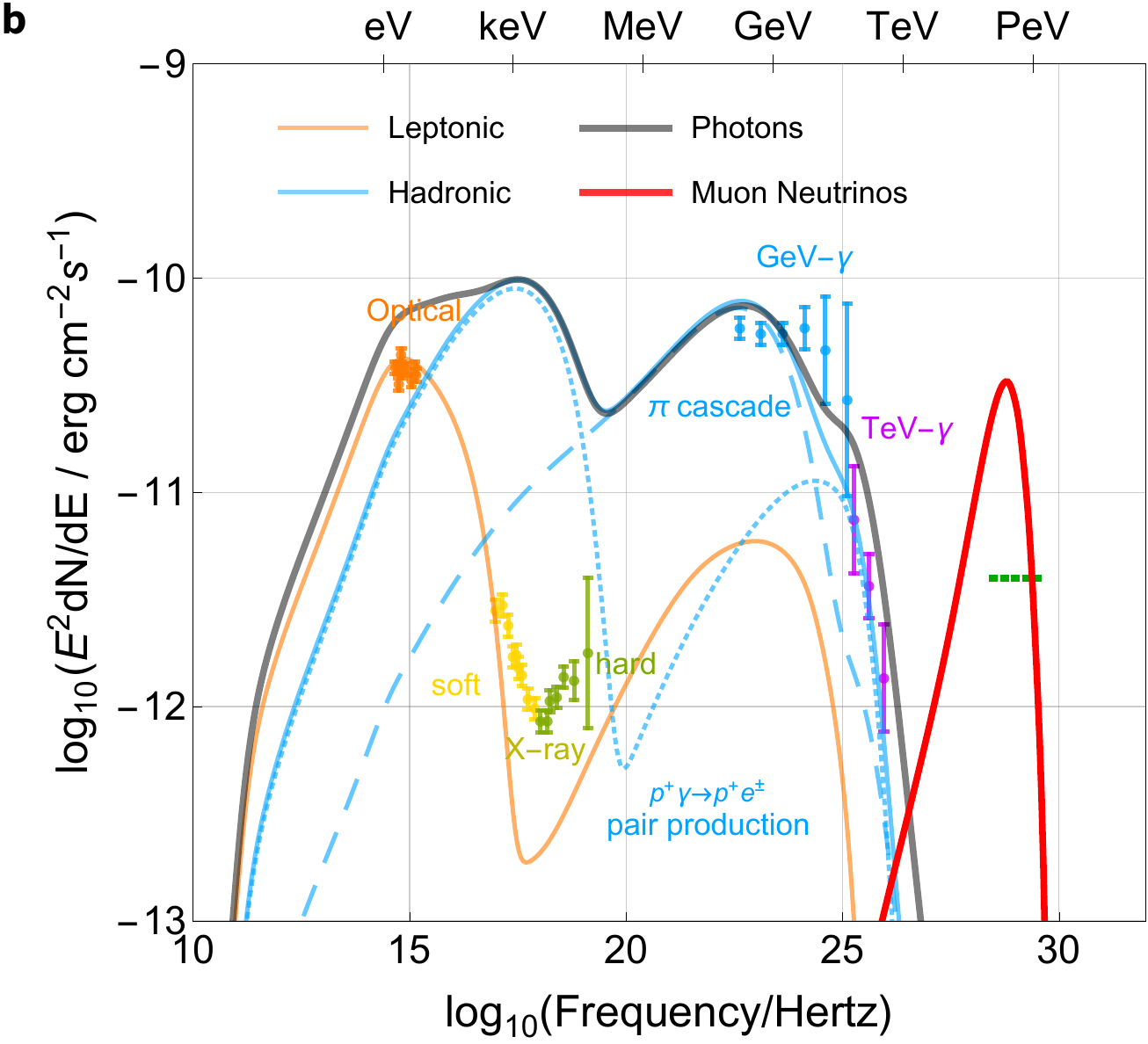} &
\includegraphics[width=0.3\textwidth]{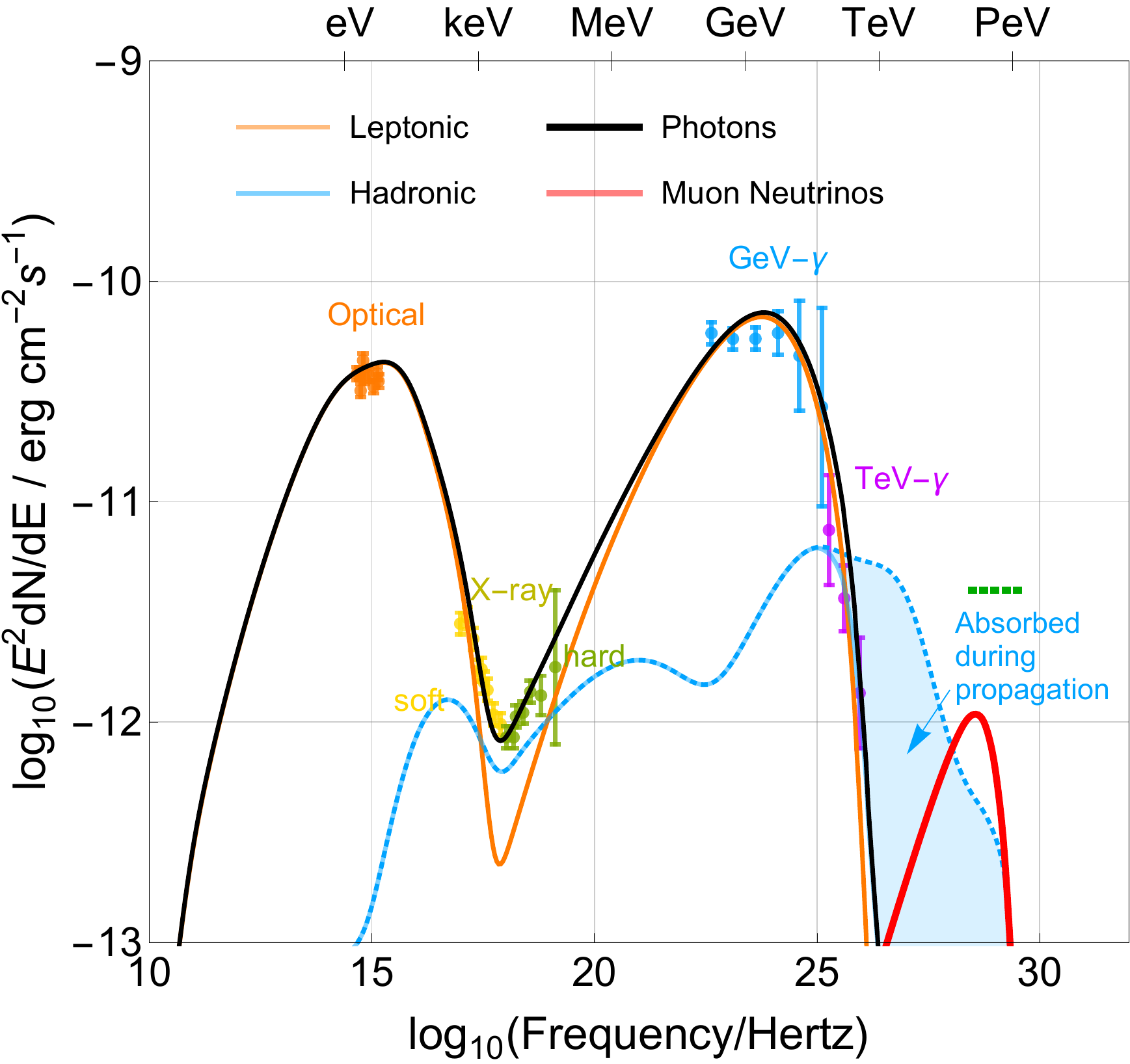} 
\end{tabular}
\caption{\label{fig:AGN} Examples for theoretical models describing the electromagnetic spectrum and neutrino associated with a gamma-ray flare of the AGN blazar TXS 0506+056 in  2017; observed data points from different wavelength bands are labeled. The left panel shows a purely leptonic model, which can describe the electromagnetic data but which does not produce neutrinos (orange curve). The middle panel shows a lepto-hadronic model in which the second hump is produced by the pion cascade accompanying the neutrinos (hadronic components: blue curves, leptonic components: orange curve). The neutrino flux (red curve) is significant and can describe the observation (estimated by  the green-dashed line), but the hadronic components overshoot the X-ray data. The right panel shows a  lepto-hadronic model which can describe both the electromagnetic and neutrino data: the two humps are dominated by the leptonic components (orange curve), whereas X-ray and TeV gamma ray-data constrain the hadronic contribution (blue curve), thus constraining the neutrino flux (red curve). Figure taken from \cite{2019NatAs...3...88G}.
}
\end{figure}

\subsubsection{Multi-Messenger Source Models}

So far, the most prominent detection of high energy neutrinos from a single source may be the AGN blazar TXS 0506+056; here we use the 2017 neutrino observation~\cite{IceCube:2018dnn} as an example to illustrate the multi-messenger physics implications of the neutrino observation, see \figu{AGN}. The neutrino arrived during a flaring state of the blazar, illustrated by the data available from different instruments (labeled data points). The more detailed physics question is in that case: what can we learn from the detection of the neutrino about the radiation processes in the source? 

Apparently, the electromagnetic spectrum needs to be described over many decades of energy, where data in certain regions are sparse. While a purely leptonic model (left panel) depends on few parameters only (such  as luminosity and the size of the region, which determine the target density, magnetic field, and properties of the injected electron spectrum), no neutrinos are produced.  
If, on the other hand, cosmic ray protons are accelerated in the source, additional processes are at work which may describe features of the electromagnetic spectrum~\cite{Mannheim:1993jg}, and which require additional parameters (such as the baryonic loading and the properties of the cosmic ray injection spectrum). For example, the gamma rays co-produced with the neutrinos in \equ{pizerodec} will feed into the electromagnetic cascade in the source and may describe the second hump (see \figu{AGN}, middle panel); the accompanying neutrino flux is at a similar level (red curve). 
It seems, however, difficult to reconcile this hypothesis without effects showing up in the dip at X-ray energies and at TeV energies (where, however, attenuation in the extragalactic background radiation is at play). It was therefore concluded that the hadronic processes must be sub-dominant in this case, see e.g.\ \cite{2019NatAs...3...88G,Cerruti:2018tmc,2018ApJ...864...84K} (and the right panel of \figu{AGN}), and that X-ray and TeV electromagnetic data are important indicators for the hadronic emission; see also \cite{Rodrigues:2020fbu} for the source PKS 1502+106. 

The  challenge of the electromagnetic cascade accompanying the neutrino production is also actively being discussed in the context of the 2014/15 neutrino flare of TXS 0506+056~\cite{IceCube:2018cha}. While efficient neutrino production typically comes together with gamma-ray suppression for compact sources and some re-processing~\cite{Halzen:2018iak,2019ApJ...881...46R}, details depend on the compactness and parameters of the source and require a full electromagnetic modeling~\cite{Rodrigues:2018tku}. 
It is, however, clear that re-processed gamma rays typically show up at lower energies, unless they leave the source and are absorbed in the background light. Astrophysical considerations also take into account the available power in the source, which in many cases exceeds the standard expectations from accretion theory. This can be an indicator either for more complicated multi-zone neutrino production models, or for new astrophysical processes such as AGN blazars undergoing super-Eddington accretion during hadronic flares.  

Note that beyond neutrinos from AGN, very recently neutrinos from TDE have been observed~\cite{Stein:2020xhk}. In that case,  a star is disrupted by the tidal forces of a black hole, and the remaining debris is partially accreted by the black hole; here the physics is very different and much less understood, see e.g.\  \cite{Hayasaki:2019kjy,Winter:2020ptf,Murase:2020lnu} for possible neutrino production sites.
These examples illustrate on the one hand the complexity of hadronic emission models and their tight interplay with astrophysics; on the other hand, they indicate that recent results in neutrino astronomy drive a newly emerging discipline of neutrino astrophysics.

\begin{figure*}[t!]
\centering
\includegraphics[width=0.4\textwidth]{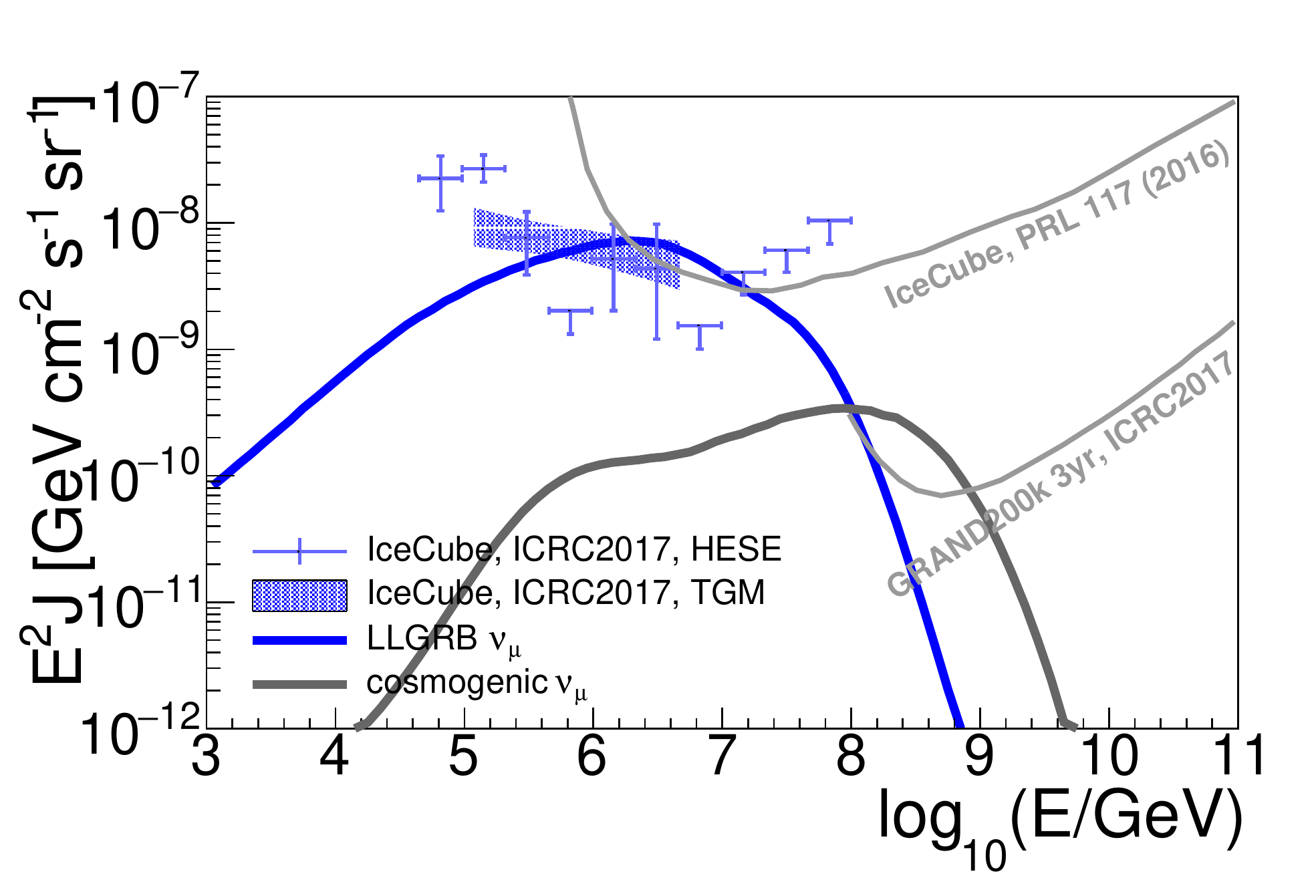}
\\
\includegraphics[width=0.3\textwidth]{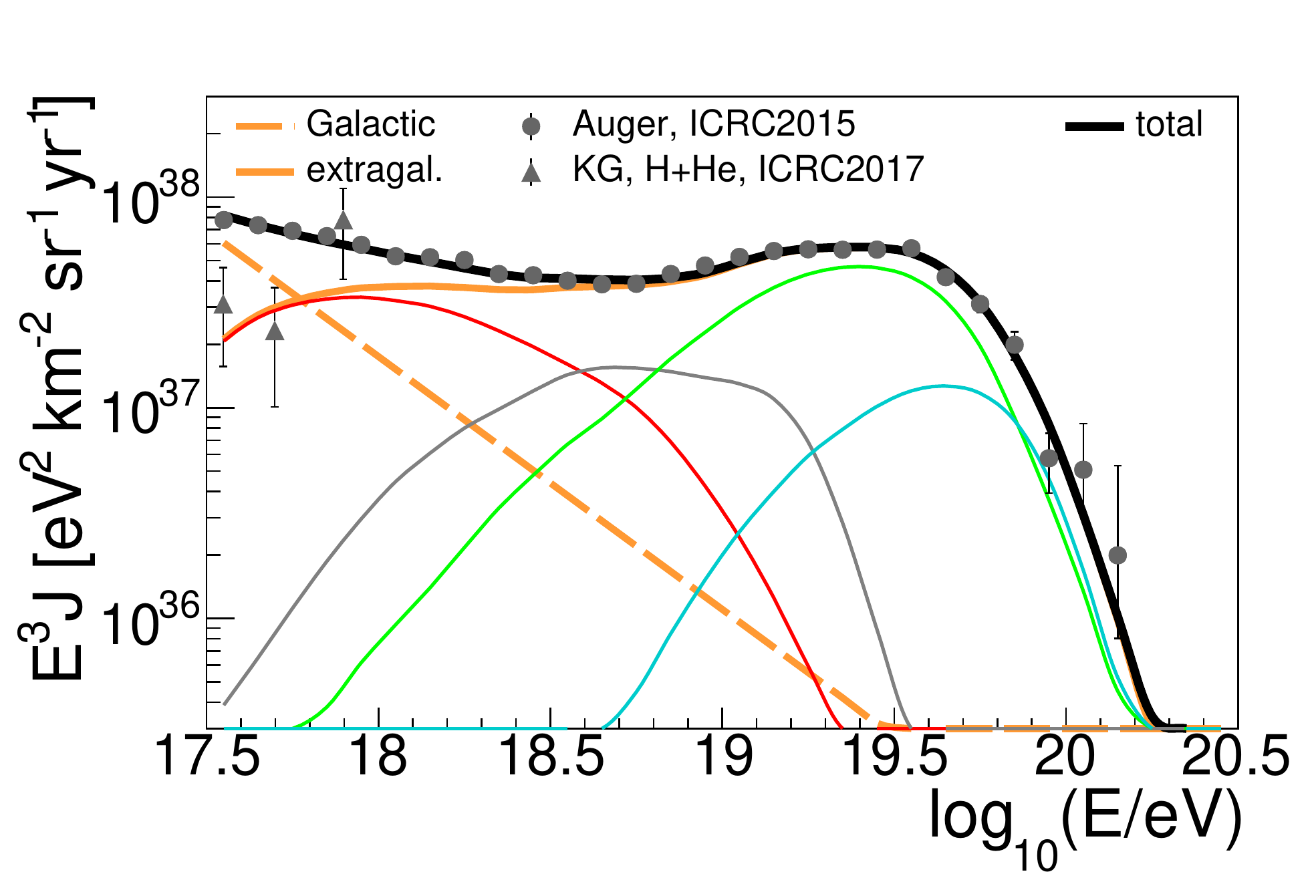}
\includegraphics[width=0.6\textwidth]{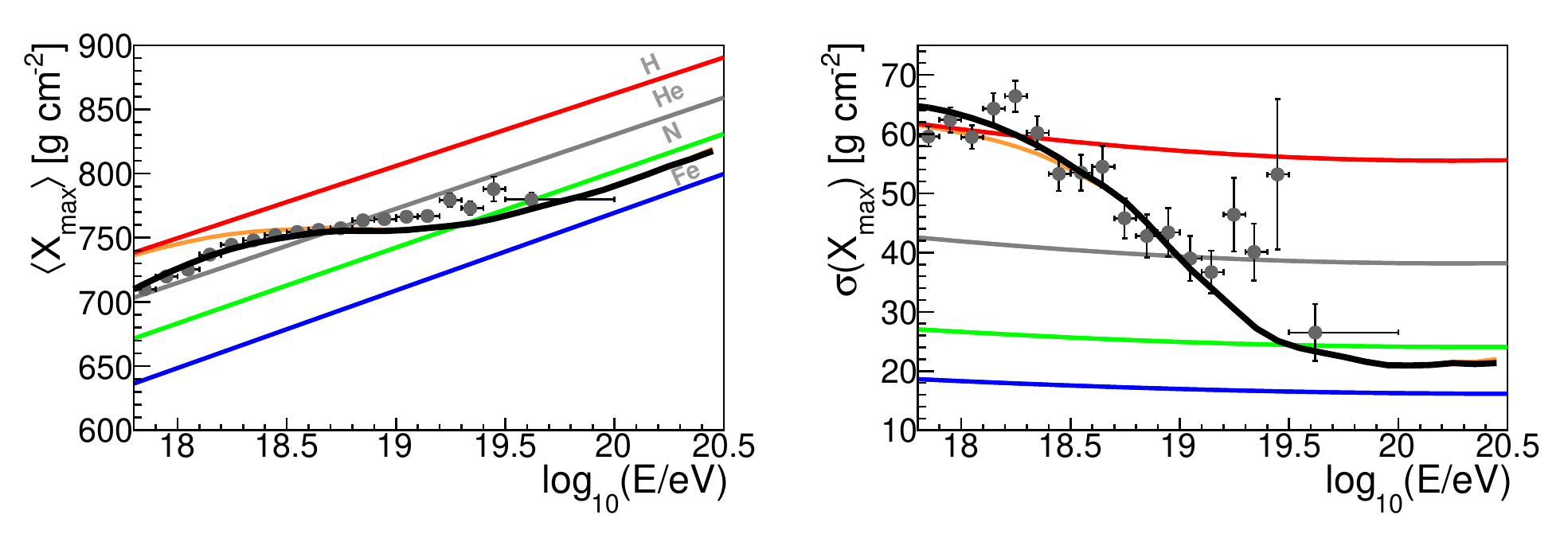}
\caption{Neutrino (upper panel) and cosmic-ray (lower panels) observables for a low-luminosity GRB model describing both neutrinos and cosmic rays at the highest energies. The neutrino panel shows both the neutrino fluxes from the source (blue curve) and the cosmic ray propagation (grey curve) for that model. The cosmic ray curves (colors in lower panels) corresponds to the mass groups indicated in the middle panel. Taken  from \cite{Boncioli:2018lrv}. \label{fig:CRfit}}
\end{figure*}

\subsubsection{Neutrinos and the Origin and Transport of the UHECRs}

In spite of the evidence for individual neutrino-source assocations with AGN and TDE, the observed diffuse neutrino flux may still not be dominated by these source classes since conceptual arguments (such as stacking searches, limits from the non-observation of multiplets, or the shape of the observed spectrum) point towards different dominant source classes or possibly even multiple contributions. A particular field of interest is the possible connection to UHECRs beyond about $10^9 \, \mathrm{GeV}$, as these are expected to be powered by very luminous or very abundant sources. 

A famous example is the Waxman-Bahcall bound~\cite{Waxman:1998yy}, asking the question about how high the neutrino flux would be if the UHECR energy was efficiently converted into astrophysical neutrinos. Interestingly, the derived neutrino flux is close to current observations in terms of magnitude; however, as discussed earlier, $E_{\nu} \simeq 0.05 \, E_{p}$, which means that the UHECR and neutrino energies differ by several orders of magnitude in energy. Current power-law fits of the astrophysical neutrino spectrum disfavor such a direct connection as the observed neutrino spectrum is much softer than $E^{-2}$~\cite{Abbasi:2020jmh}, whereas the UHECR extrapolation relies on an $E^{-2}$ spectrum. This may imply that the neutrino spectrum is not a simple power law, or that the source contains strong magnetic fields; see below. 

Apart from neutrinos produced in the astrophysical source, secondary neutrinos are produced during UHECR propagation by \equ{Delta} from interactions with the cosmic background radiation, such as the cosmic microwave background. This ``cosmogenic'' neutrino flux in the EeV range follows the UHECR energy directly and is expected at higher energies e.g. relevant for radio-detection experiments; its level is an indicator for the presence of light element fraction in UHECRs~\cite{vanVliet:2019nse} at the neutrino highest energies, and a possible indicator for high-redshift cosmic background light at lower energies. Note that for some source classes the source neutrino flux at EeV energies may even ``outshine'' the cosmogenic flux even if the source describes UHECR data; see e.g. \cite{Rodrigues:2020pli} for AGN.

In recent years, progress has been made to theoretically describe the observed UHECR spectrum and composition, while at the same time predicting the source and cosmogenic neutrino fluxes for many different source classes. Here we show one example for a population of low-luminosity GRBs in \figu{CRfit}: The upper panel shows the diffuse source neutrino flux (blue curve), which is peaked at around PeV energies -- as it is typical for GRB neutrino fluxes. Here the cosmogenic neutrino flux (grey curve) is comparatively low because the UHECR observables (lower panels) prefer a heavy composition at the highest energies.  In this example, the radiation density in the source controls the nuclear disintegration and the neutrino production inside the source at the same time; the appearance of light elements below the cosmic ray ankle ($E \lesssim 10^{18.6} \, \mathrm{eV}$), which are needed to describe UHECR data in that range, is therefore directly correlated with the neutrino flux.  This is just one modern example for the implementation of the Waxman-Bahcall paradigm, where in this case the magnetic field effects on the secondary pions and muons break the correlation between the UHECR and the neutrino energies; an example where the neutrino spectrum follows the UHECR spectrum more closely can be found in \cite{Fang:2017zjf}, more statistics on the neutrino spectrum can help to discriminate such options. 

\subsubsection{Neutrinos from Gravitational Wave Sources?}

Neutrinos produced in connection to gravitational wave sources, such as the binary neutron star merger  GW170817, which has been so far the best motivated source for neutrino detection, see~\cite{Ando:2012hna,Kimura:2017kan,Fang:2017tla}, have also been searched for~\cite{ANTARES:2017bia,Baikal-GVD:2018cya}. Dedicated computations however show that the expected neutrino fluence from the associated short GRB must not have been larger than about $10^{-4}$ of the instruments' sensitivities~\cite{Biehl:2017qen}. While gravitational wave sources are being monitored with neutrinos and an association would be ground-breaking, no neutrinos in coincidence with gravitational waves have been found so far \cite{ANTARES:2018bmu}.

\subsubsection{Astrophysical Neutrinos and BSM Physics}

\begin{figure}[t]
\begin{center}
\includegraphics[width=0.6\textwidth]{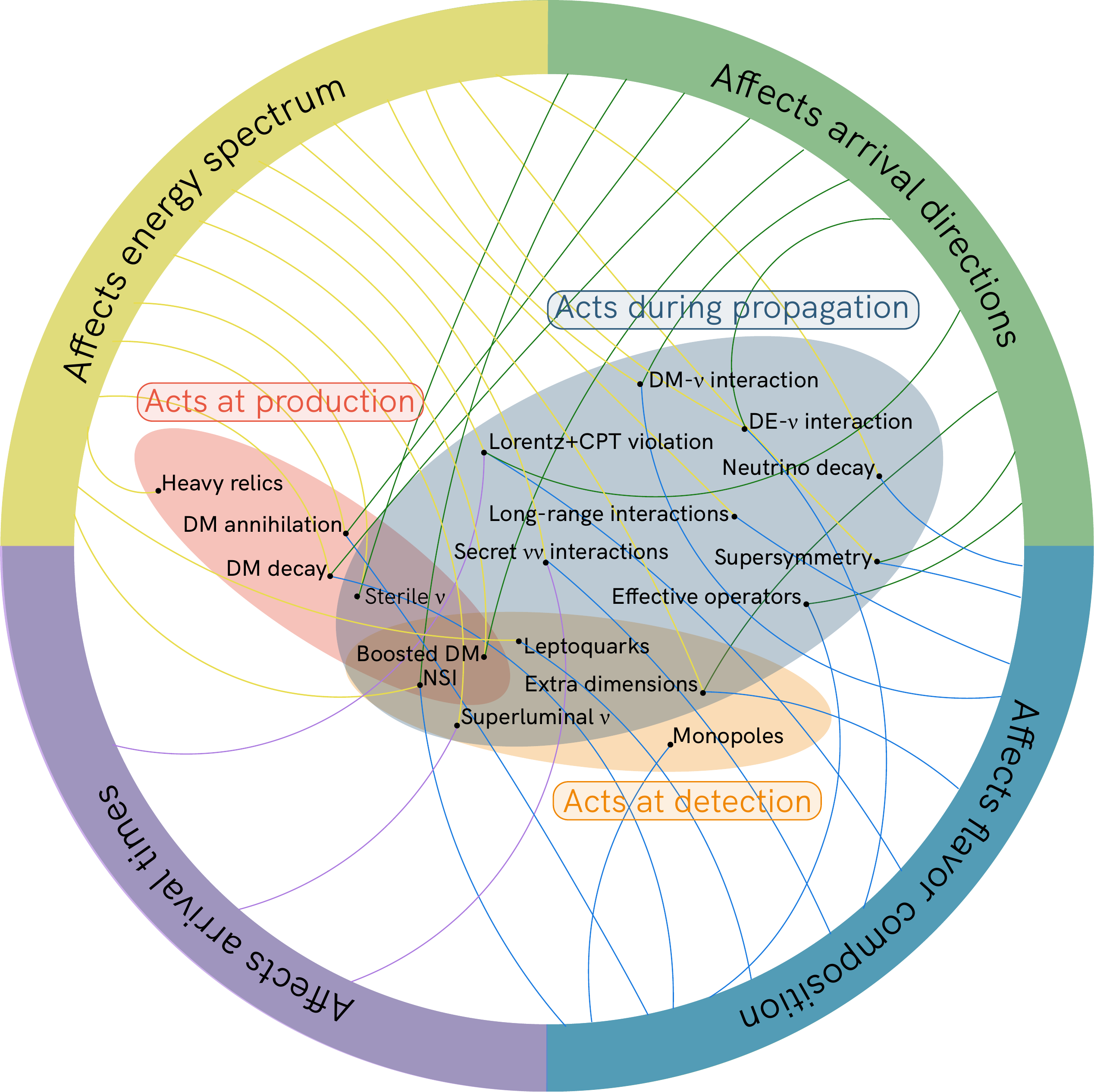}
\end{center}
\caption{\label{fig:astrobsm} Possible mechanisms for BSM physics testable with astrophysical neutrinos. Figure taken from \cite{Arguelles:2019rbn}.} 
\end{figure}

Finally, we would like to highlight that astrophysical neutrinos can also test BSM physics, thanks to the extreme distances, production environments, and energies; see \figu{astrobsm} for a summary. These mechanisms can be classified in two different dimensions: they may affect the neutrino arrival directions, neutrino spectrum, neutrino arrival times, or the flavor composition. The BSM effect may be at work  at production, during propagation or at detection. An example is neutrino lifetime (leading to neutrino decay, if short enough), which can also be regarded as one of the unknown properties of the neutrinos for which only lower limits exist, see Sec.\ \ref{sec:exotic}; over large distances, neutrino decay may be tested by changing the flavor composition, whereas it  may also  show up in the energy spectrum at low energies; it is questionable if changes in the arrival directions may be detectable in this case.

\subsection{Theoretical Implications of Neutrino Mass and Lepton Mixing}

Neutrino mass is tiny, which generically is interpreted as being caused by some suppression mechanism. Possibly connected to that is the question of whether neutrinos are Dirac or Majorana particles, i.e., whether lepton number is conserved or violated. Lepton number is an accidental global symmetry in the SM, which is hard to realize in theories beyond the SM. Along these lines, it is fair to say that the vast majority of models predicts neutrinos to be Majorana particles. 
This would mean that there is another source of fundamental mass generation in Nature besides the SM Higgs mechanism. The violation of lepton number that is implied by Majorana fermions will  have another set of deep consequences in terms of conservation of global symmetries or the creation of matter in the Universe.
Checking this prediction via observation of neutrinoless double beta decay is the only realistic possibility, but still very challenging. Apart from the character of neutrino mass, its value has a guaranteed impact on the global structure of the Universe, the magnitude of which remains to be explored by future measurements. 
Identifying the precise mechanism of neutrino mass generation is possible by identifying the various new features that it typically brings along, such as new particles or new energy scales. The most straightforward mechanism is the type I seesaw mechanism, which naively works at close-to-GUT scales, but can be arranged to work at more testable energies. Many other mechanisms have been proposed and can  typically be arranged in theories beyond the SM. Often the mechanisms work at low scales, allowing tests at colliders or with lepton flavor violation searches.

 The peculiar pattern observed for the three PMNS mixing angles, with $\theta_{12}$ and $\theta_{23}$ being large while $\theta_{13}$ is small (but not tiny), is in strong contrast to the CKM matrix pattern. This difference poses intriguing questions on the origin of the flavor structure of fermions. It remains an open question as to whether the apparent mixing patterns reflect some deeper theoretical principle such as a broken flavor symmetry, or whether the patterns are purely random or ``anarchic''. The relationship, if any, between the PMNS and CKM matrices is also  unknown. These issues are a major concern in the literature.  Among the countless examples in the literature, many flavor symmetry models predict the same mixing parameters. Precision beyond that expected for future experiments will not likely help much in distinguishing models.  Which models are favored can be influenced by other arguments, such as simplicity, minimality, the possibility of incorporating quark mixing or to compatibility with GUTs. 
The nature of CP-violating phases is a related theoretical challenge, with ramifications 
% through leptogenesis 
for cosmology, as well as for particle physics \emph{per se}. The current indications for a Dirac phase in the vicinity of $-\pi/2$  invite further thoughts of flavor symmetries, while the existence or not of CP-violating Majorana phases remains undetermined; see Sec.\ \ref{sec:D_or_M}. 
Observation of lepton number violation and CP violation would have fundamental cosmological impact, as those may lead to the generation of a baryon asymmetry.  In the most straightforward picture, the observations would provide circumstantial evidence of leptogenesis as the origin of the observed baryon asymmetry of the Universe. 

This subsection deals with details of these fundamental issues, which are triggered by the results of neutrino physics.

\subsubsection{Dirac or Majorana Neutrinos? \label{sec:D_or_M}}

As mentioned above, Dirac neutrinos imply the conservation of lepton number, which in the SM is an accidental global symmetry. Typically, additional symmetries are necessary in theories beyond the SM to keep lepton number conserved. The situation resembles that for dark matter, where often a symmetry that stabilizes the dark matter particle needs to be introduced. 
General considerations in quantum gravity lead to the claim that global symmetries are not conserved. 
An unbroken local symmetry could be chosen in analogy to QED, in which the electric charge is  conserved. However, the gauge coupling associated with lepton number would need to be extremely tiny in order to obey existing experimental limits. 
Of course, lepton number could be violated by $3$ or $4$ units, i.e.\ neutrinos would be Dirac particles, but processes which violate lepton number by $3$ or $4$ units would exist. In fact, non-perturbative SM processes relevant for leptogenesis (see Sec.\ \ref{WG1:SubSec:PMNS-Theory:CPV}), or more generally baryogenesis, do violate lepton (and baryon) number by three units. In general, the violation of lepton and/or baryon number is crucial for our ideas for the generation of the baryon asymmetry of the Universe. 
Moreover, Grand Unified Theories do generically predict Majorana neutrinos, e.g.\ the violation of lepton number. This is on equal footing with the prediction of baryon number violation (that is, proton decay); therefore, baryon number and lepton number violation are not separate questions. 

The Majorana nature of light neutrinos necessarily implies further terms in the overall particle physics Lagrangian,  and in particular implies new particles, parameters and energy scales. Determining those, also with the help of experiments beyond pure neutrino physics, such as direct searches at colliders or via lepton flavor violating processes, may teach us valuable lessons on the correct BSM approach.

Eventually, the question of the neutrino nature needs to be answered experimentally, via observation (or perhaps non-observation) of neutrinoless double beta decay ($0\nu\beta\beta$). 
Avogadro's number is the only way to beat the $(m_\nu/E)^2$  suppression of observables that can distinguish Majorana from Dirac neutrinos, leading to multi-100-kg experiments searching for $0\nu\beta\beta$. Nuclear physics uncertainties remain a potential problem in precision physics of $0\nu\beta\beta$. The nuclear community has been intensifying its attack on the problem, and indeed the field is understood much better than a decade ago. 
 An observation with at least two different isotopes, preferably with two different measurement techniques, is needed in order to provide convincing evidence for lepton number violation and Majorana neutrinos. 

All in all, the consequences of pinning down the neutrino nature would be fundamental, ranging from particle physics, to the fate of global symmetries, and to cosmology.

\subsubsection{Origin of Neutrino Mass}\label{sec:origin}

Where do tiny neutrino masses come from? The most simple possibility would be to add right-handed neutrinos to the SM particle content, and thus create a Dirac mass term $m_D$ for neutrinos in analogy to all other fermions of the SM. The Yukawa coupling would be at least six orders of magnitude smaller than the one for the electron. While this is the same hierarchy as for the third-generation top quark and the first-generation electron Yukawa, the point is that this strong hierarchy affects particles in the same $SU(2)_L$ doublet, for each generation: up- and down quarks have only a mild, if any, mass hierarchy, while electrons and electron neutrinos have a mass ratio of  $10^{-6}$ or below. 

Therefore, some suppression mechanism is required. Connected to that, the gauge symmetries of the SM allow a bare mass term for the right-handed neutrinos, $M_R$. ``Bare'' denotes here a mass term not connected to the SM Higgs mechanism, which gives mass to all other particles in the SM. This mass term is thus not bounded from above by perturbativity of couplings, thus can be arbitrarily high. Moreover, it is a Majorana mass term. Via the coupling of the right-handed neutrinos with the left-handed ones through the Dirac mass, the Majorana character is passed to the light neutrinos.  
In addition,  light neutrinos have a mass given by $m_D^2/M_R$ and thus are suppressed for all three generations. This is the type I seesaw mechanism \cite{Minkowski:1977sc,GellMann:1980vs,Yanagida:1979as,Glashow:1979nm,Mohapatra:1979ia}. 

The lessons of this most simple mechanism are that 
\begin{itemize} 
\item[(i)]  new particles exist, in this case, right-handed neutrinos. 
\item[(ii)] a new energy scale exists. Recall that the SM possesses only a single energy scale.
\item[(iii)] a new property exists, in this case the violation of lepton number due to the Majorana nature of the light neutrinos. 
\item[(iv)] the mass of the light neutrinos is inversely proportional to the energy scale related to their origin.  
\end{itemize}
In addition, the new particles often come with additional interactions of their own, for instance caused by a gauge symmetry related to the difference of baryon minus lepton number ($B-L$), left-right symmetry, etc. 
This and the above features are almost generic for the countless mechanisms that have been proposed to generate neutrino mass. These features allows testing and distinguishing of the mechanisms. 

For the type I seesaw, the naive picture implies that $M_R \sim m_D^2 /m_\nu \sim v^2/m_\nu \gtrsim 10^{14}$ GeV, and a mixing of the right-handed neutrinos with the charged current 
of order $m_D/M_R \simeq \sqrt{m_\nu/M_R}$, which implies little hope of testability. However, $m_D$ and $M_R$ are matrices, allowing for cancellations. In addition, simple variants and modifications of the type I seesaw exist, that allow even more flexibility. 
For instance the type II  \cite{Magg:1980ut,Schechter:1980gr,Cheng:1980qt,Lazarides:1980nt,Wetterich:1981bx,Mohapatra:1980yp} or III mechanisms \cite{Foot:1988aq} are other options, which introduce scalar and fermion triplets, respectively. More involved scenarios such as  inverse~\cite{Wyler:1982dd,Mohapatra:1986aw,Mohapatra:1986bd} or linear~\cite{Akhmedov:1995vm,Barr:2003nn,Malinsky:2005bi} seesaws have additional singlet fermions and more than one new energy scale. 

Loop mechanisms are the second-most popular way to generate neutrino masses. Examples are the Zee model \cite{Zee:1980ai} or the ``scotogenic'' model \cite{Ma:2006km}, 
which work at one-loop, or the Zee-Babu model \cite{Zee:1985id,Babu:1988ki} at two-loop order. Again, new particles are introduced, mostly scalars, but also fermions. The loop-suppression of neutrino mass allows for more easily testable scenarios at colliders or using lepton flavor violating processes, Higgs physics or anomalous magnetic moments of charged leptons.  

Another way to lower the scale of neutrino mass and of lepton number violation is to apply the  `t Hooft naturalness argument, which states that a parameter is small when the symmetry of the theory is enlarged in its absence. In our case, the symmetry is lepton number and the parameter is the scale of lepton number violation or neutrino mass generation. 

A different perspective to the origin of neutrino mass is the SM Effective Field Theory (SMEFT) framework, which treats BSM physics as a perturbation series of non-renormalizable effective operators made from SM fields and suppressed by powers of a new physics scale $\Lambda$. It turns out that the lowest-order perturbation is a unique, lepton number violating operator (called Weinberg operator \cite{Weinberg:1979sa}), suppressed by one power of $\Lambda$ and leading to Majorana neutrino masses after electroweak symmetry breaking. The theory producing this effective operator (the so-called ``ultraviolet completion'' at higher energies to make it a renormalizable theory) is, however, not unique: there are three possibilities to generate this operator from tree-level diagrams, which are the three types of seesaw mechanism mentioned earlier. 
In this language, neutrino mass can also be generated by operators of higher dimension than $5$ ($7, 9, 11,\ldots$), which lowers the suppression scale $\Lambda$, possibly to straightforwardly testable values. In the literature, different construction principles to more sophisticated neutrino mass models are being used, such as from lepton number violating effective operators (e.g.\ Refs.\  \cite{deGouvea:2007qla,Angel:2012ug})  or from a systematic decomposition of SMEFT operators (e.g.\ Refs.\ \cite{Cepedello:2017eqf,Anamiati:2018cuq}).\\

Thus, there are various ways for testable neutrino mass generation mechanisms. They can be distinguished by their different particle content, energy scales, couplings to SM particles and predictions for neutrino parameters. Identifying them will be of crucial importance to understand particle physics beyond the SM.

\subsubsection{Flavor Symmetries}
\label{WG1:SubSec:PMNS-Theory:flavorSymmetries}

Unless a flavor symmetry or some other constraining structure is imposed, the mixing angles and CP-violating phase(s) in the PMNS matrix $U$ are arbitrary parameters, in analogy to their CKM cousins. However, the measured values of the mixing angles in particular of the PMNS matrix with $\theta_{23}$ close to maximal, do encourage speculations about a deeper flavor structure, which possibly could eventually help to solve the long-standing mystery of quark-lepton family replication. In this sense, the information that oscillation experiments have been collecting so far has triggered an enormous amount of theoretical activity which brought new insights to flavor physics. See Refs.\  \cite{Altarelli:2010gt,King:2013eh,Xing:2019vks,Feruglio:2019ktm} for comprehensive reviews of flavor symmetries.

A flavor symmetry puts the weak fermion singlets or doublets of different generations in certain multiplets of a new symmetry group $G_f$. The latter is subsequently broken in order to explain the non-degenerate charged lepton masses. 
The breaking leads to different conserved subgroups of $G_f$ in the charged lepton and the neutrino sectors, which ultimately leads to non-trivial $U$. 
The understanding that fermion mixing may be caused by the conservation of different subgroups of a larger flavor symmetry group is one of the potential lessons learned. 

The value of $\theta_{12}$ can give insight on which flavor symmetry one could apply.  Commonly considered special cases are (a) \emph{tribimaximal mixing} 
which has $\sin^2 \theta_{12}=\frac 13$, (b) \emph{bimaximal mixing} which has 
$\sin^2 \theta_{12}=\frac 12$, (c) \emph{hexagonal mixing} 
which has  $\sin^2\theta_{12} = 1/4$, 
and (d) \emph{golden ratio mixings}, which have $\tan \theta_{12} = 1/\phi$ 
(or $\sin^2 \theta_{12}=0.28$) or $\cos\theta_{12} = \phi/2$ (or $\sin^2 \theta_{12}=0.35$), 
where $\phi = (1 + \sqrt{5})/2$. 
These patterns are typically derived from discrete or finite flavor symmetry groups, which may have their origins as discrete subgroups of continuous flavor groups such as $SO(3)$ and $SU(3)$ or products thereof, where the ``$3$'' is mandated by the observed threefold family replication. Examples include $S_4$, $A_4$, $A_5$, $D_N$, $\Sigma(2 N^2)$, $\Sigma(3 N^3)$, $\Delta(3 N^2)$, $\Delta(6 N^2)$. We refer to   the reviews \cite{Ishimori:2010au,King:2013eh,Xing:2019vks,Feruglio:2019ktm} for references to the large original literature.  It is also possible to use Abelian $U(1)$ symmetries. The most studied example is the  (anomaly-free) difference of muon and tau flavors  \cite{He:1990pn,Foot:1990mn,He:1991qd,Binetruy:1996cs}, 
$L_\mu - L_\tau$. Gauging and breaking the symmetry is necessary, leading to a massive $Z'$ boson of interest for the anomalous magnetic moment of the muon. Extending the symmetry to the quark sector is possible, which introduces the option to explain $B$-physics anomalies  \cite{Altmannshofer:2014cfa,Crivellin:2015mga}; see Sec.\ \ref{sec:unif}. 

The CKM matrix is famously not very different from the unit matrix, with the largest mixing angle (the Cabibbo angle $\theta_C$) being about $0.23$ and the others one to two orders of magnitude smaller. Its zeroth order form can be interpreted as the unit matrix, with higher-order corrections yielding in particular the Cabibbo angle. Its value is in fact close to $\theta_{13}$, which makes an overall picture, in which $\theta_C$ and $\theta_{13}$ appear as effects of corrections, quite appealing.

Given the expected precision on the oscillation parameters discussed in Sec.\  \ref{chap:osc},  it will be possible to rule out classes of models (e.g.\ hexagonal mixing from tribimaximal mixing). However, some of the schemes will not be distinguishable, e.g.\ tribimaximal mixing and the golden ratio scheme with $\cos \theta_{12} = \varphi/2$, see e.g.\ \cite{Petcov:2018snn}. This is true also when correlations to model parameters are included, which concerns the often-studied ``sum-rules'' (see e.g.\ Ref.\ \cite{Frampton:2004ud,King:2005bj,King:2013eh,Girardi:2015vha,Ballett:2013wya}) such as 
$\sin^2 \theta_{12}  = 1/3 + \zeta \sin^2 \theta_{13} \cos \delta_{\rm CP} $, with $\zeta$ a real parameter predicted by the model. Other sum-rules, in particular those relating neutrino masses and Majorana phases with each other \cite{Barry:2010yk,Dorame:2011eb,King:2013psa}, suffer in addition from the various ways neutrino mass observables and their correlations are modified by possible new physics or smeared by theoretical uncertainties, cf.\ Sec.\ \ref{sec:mass_compl}. Nevertheless, the 
determination of the absolute neutrino mass scale and the mass ordering is very important for ruling out models. The flavor structure of the neutrino mass matrix is very different for normal hierarchy, inverted hierarchy, or quasi-degeneracy. 
In addition, corrections to models are highly unlikely to change the normal hierarchy into the inverted one.

An interesting class of flavor symmetries are those that non-trivially combine transformations between families with CP transformations, the obvious motivation being to constrain the CP violating phases in the PMNS matrix, especially the Majorana phases. A generalized CP symmetry~\cite{Grimus:1995zi,Feruglio:2012cw,Holthausen:2012dk,Chen:2014tpa} is introduced, which is then combined with a given flavor symmetry. Typically, predictions for the CP phases, including the Majorana ones, are obtained, leading to testable signatures in neutrino oscillation and neutrinoless double beta decay experiments. A large number of special cases have been analysed in the literature, with useful summaries provided in the reviews of Refs.\ \cite{Coloma:2018ioo,Xing:2019vks,Feruglio:2019ktm}.

%It should be noted that the full models contain a variety of fields and parameters, typically exceeding the numbers of predictions they make. In view of this, a 
A more recent development~\cite{Feruglio:2017spp} has been the idea of using  non-linearly realized flavor symmetries known as \emph{modular symmetries}, a structure that has its origins in supersymmetry and string theory.  Modular symmetries are extended mathematical structures under which the coupling constants are functions of a chiral superfield $\tau$ called the \emph{modulus}. Requiring modular invariance leads to the coupling constants having to transform in a certain way, thus greatly constraining the form of the theory, and is able to constrain the neutrino masses in addition to the PMNS  parameters~\cite{Xing:2019vks,Feruglio:2019ktm}. A vigorous activity in this area akin to the conventional model building discussed above is ongoing. The number of free parameters in such models is much smaller, at least for simple ones, than for the conventional models, but this comes with the price of requiring  fundamental and hard-to-verify features like higher dimensions, high scale supersymmetry, and string theory. Nevertheless, it is an exciting development (see e.g.\ Refs.\ \cite{Penedo:2018nmg,Criado:2018thu,Kobayashi:2018vbk,Kobayashi:2018scp,Ding:2019zxk}), though the same comments regarding the possibilities to distinguish models as for conventional flavor symmetries apply here as well. \\

Overall, the field of flavor symmetries is still very active and boosted mainly by neutrino data. In the future, it will be quite possible to distinguish classes of models, but many different scenarios exist that lead to very similar, and thus experimentally indistinguishable, predictions. Various corrections to model predictions are possible (vacuum misalignment, renormalization, etc.), often depending on independent and inaccessible energy scales. Moreover, by slightly modifying models, the predictions can be adjusted to new data.  
Measuring neutrino parameters as precisely as possible is nevertheless a necessity in order to gain further insight. However, improving beyond the currently foreseen values is likely not to be helpful in flavor model building, but rather (in analogy to the CKM sector) for testing new physics scenarios, as discussed in Sec.\ \ref{sec:np}.

\subsection{Connections of Neutrinos to beyond the Standard Model Physics}
Neutrinos are linked to many aspects of particle physics. For each aspect, there are countless examples which cannot all be covered here. Instead, we give a few popular and illuminating examples.  More often than not, the various BSM aspects are interconnected with each other, so the separation made in this section is somewhat arbitrary. \\ 
 Within models, the violation of lepton flavor in the neutrino sector can directly or indirectly transfer to the charged-lepton sector. If models incorporate the quark sector, then the plethora of tests of meson decays can be linked. More often than not, dark matter candidates exist in a model responsible for neutrino mass. The baryon asymmetry of the Universe is another field with frequent connections to  neutrino mass. The parameters that we measure within neutrino physics provide another set of information on grand unified theories; in fact, neutrino mass is a typical prediction of such theories, even for less-unified theories such as left-right symmetry, and provides motivation for them. 
 All these theoretical connections are accompanied by experimental connections, such as the neutrino background to dark matter direct detection experiments, or searches for proton decay in neutrino oscillation experiments. Furthermore, astrophysical neutrinos can be used to test neutrino properties in extreme environments or over large distances.

\subsubsection{Flavor Physics}
\label{sec:flav}
Neutrino oscillations  show that neutral-lepton flavor violation exists. Via loops the charged-lepton sector violates flavor as well, though, due a GIM-suppression, only at negligible level \cite{Petcov:1976ff}. This means, however, that observation of charged-lepton flavor violation (cLFV) implies observation of new physics beyond neutrino mass. A plethora of possible processes exists. The limits (in particular on the $e$-$\mu$ sector) are so strong that they correspond to new energy scales exceeding the LHC center-of-mass energy \cite{deGouvea:2013zba,Calibbi:2017uvl}. 
All models generating neutrino mass imply cLFV at tree or loop level, and experimental limits are often the strongest constraints on neutrino mass models. While collider limits apply mostly to the energy scale, flavor observables provide information on the flavor structure in particular, thus  offering complementary information. 
Additional bounds to test models come from lepton flavor universality, Higgs decays, charged-lepton magnetic or electric dipole moments. There are models in which the rates of cLFV are given by the neutrino parameters (plus an energy scale) and models in which parameters not directly measurable by oscillations govern the rates. An example for a direct connection is the type II seesaw model, in which the rates for decays like $\mu \to e \gamma$ scale like 
$|(m_\nu m_\nu^\dagger)_{e\mu}|^2$ and for $\mu \to 3e$ like 
$|(m_\nu)_{ee} (m_\nu)_{e\mu}|^2$ \cite{Chun:2003ej}. Here $m_\nu$ is the neutrino mass matrix that can be reconstructed with  existing and future measurements of neutrino parameters.

Future limits are expected to further constrain in particular low scale neutrino mass models. Leading here will be MEG-2 (for $\mu \to e \gamma$ \cite{Baldini:2018nnn}), Mu3e (for $\mu \to 3e$ \cite{Blondel:2013ia}), Mu2e and COMET (for $\mu$-$e$ conversion \cite{Bartoszek:2014mya,Adamov:2018vin}), but also tau-sector decays will be very important. \\

Combining lepton with quark flavor is of interest for theories unifying quarks and leptons. Interestingly, long-standing anomalies in the $B$-meson sector may be connected to neutrino physics. This concerns mainly the decay $B \to K^\ast \mu\mu$ and the ratio of $B \to K \mu\mu$ and $B\to K e e$. It is possible to extend the neutrino-motivated $L_\mu - L_\tau$ gauge symmetry to the quark sector in order to explain such 
$b \to s \mu\mu$ anomalies \cite{Altmannshofer:2014cfa,Crivellin:2015mga}. Another explanation applies leptoquarks \cite{Hiller:2014yaa,Bauer:2015knc}, which can generate neutrino mass radiatively \cite{Cata:2019wbu}. This is just one example of the possible connections of neutrinos to more general flavor physics.

\subsubsection{Dark Matter}
\label{sec:dm}
Just as for neutrino mass, the presence of dark matter \cite{Aghanim:2018eyx} is a clear 
proof of physics beyond the standard model, though it is so far without direct verification in laboratories. 
The standard paradigm of Dark Matter (DM) is that a  Weakly Interacting Massive Particle (WIMP) interacted with the thermal SM plasma in the expanding Universe until the interaction could no longer keep up with the expansion: the DM particles ``freeze out''. 
Massive neutrinos  are WIMPs in the true sense of the word. However, their small mass would lead to freeze-out when they are highly relativistic, and their free-streaming would erase structures in a way that is incompatible with observations should they be the only constituent of the dark matter of the Universe. Neutrinos would thus be ``hot dark matter''.  Structure formation considerations require that DM be cold or warm, but rule out pure hot dark matter scenarios.

Heavier neutrinos with masses of around 100 GeV are good candidates, but their Yukawa  interaction with leptons and the Higgs would lead to immediate decay. Switching off this coupling with a symmetry would remove them from their role in seesaw neutrino mass generation, and also require additional interactions to produce them in the thermal plasma. 

An intermediate case is neutrinos of mass in the keV range (see \cite{Adhikari:2016bei} for a review), which could be produced via oscillations with active neutrinos early on in the history of the Universe. A keV neutrino can be part of the seesaw mechanism generating a very tiny active neutrino mass. In order not to exceed the DM abundance, keV sterile neutrinos should have a small mixing with the SM sector ($\theta^2 \sim 10^{-8}$). They never reach thermal equilibrium and hence contribute to the number of relativistic degrees of freedom as $\Delta N_{\rm eff}<1$, which allows them to evade the constraint on $N_{\rm eff}$ established by BBN and CMB measurements (see discussion in Sec.\ \ref{sec:sterile_cosmo}). An upper bound on the mass of a keV sterile neutrino is provided by the absence of detection in X-ray searches looking for a diffuse signal from the radiative decay of the sterile neutrino. For  non-resonantly produced sterile neutrinos~\cite{Dodelson:1993je}, this limit is of the order of 4~keV. For resonant production modes, where the neutrino density is enhanced in the presence of a lepton-antilepton asymmetry at the epoch of production~\cite{Shi:1998km}, this limit can be increased to about 50 keV depending on the active-sterile interaction strength~\cite{Boyarsky:2009ix,Laine:2008pg}. Becoming non-relativistic in the radiation-dominated era, keV neutrinos would behave as warm dark matter (see \cite{Abazajian:2017tcc,Boyarsky:2018tvu} for reviews), alleviating some of the drawbacks on galactic scales of the more commonly considered cold dark matter scenario (absence of a visible galactic cusp, small number of detected galactic satellites compared to simulations, etc.). 
This DM candidate has recently attracted renewed attention as an unidentified line at around $3.5$ keV in the X-ray spectra of galaxy clusters \cite{Bulbul:2014sua} and Andromeda \cite{Boyarsky:2014jta} may arise from the decay $\nu_s \to \nu_\alpha \gamma$ of a $7$-keV neutrino into an active one and a photon. The DM origin of the line is however subject to criticism \cite{Jeltema:2014qfa,Carlson:2014lla}. Furthermore, keV sterile neutrinos would smooth out density fluctuations below their free-streaming scale, a signature that small-scale clustering analyses use to constrain  the existence of such particles. The tightest bounds on keV neutrinos are currently provided by Lyman-$\alpha$ forest surveys. Constraints on non-resonantly produced sterile neutrinos are now in the $20-30$ keV mass range~\cite{Baur:2015jsy,Irsic:2017ixq,Yeche:2017upn}, closing the window for such particles as a major constituent of the  dark matter. The case for resonantly-produced keV sterile neutrinos is less clear-cut, with current  limits in light tension~\cite{Baur:2017stq} with the debated sterile neutrino interpretation of the  3.5~keV emission.

Within neutrino mass models, often a DM particle exists. A popular example is the ``scotogenic'' neutrino mass model, where right-handed neutrinos are prohibited with a $Z_2$ symmetry from having a Majorana mass term, but, with the help of additional scalar particles, can generate a Majorana neutrino mass via one-loop diagrams. The same $Z_2$ symmetry stabilizes the lightest particle of the scenario, which is therefore a DM candidate \cite{Ma:2006km}. There are also approaches where the DM-stabilizing $Z_2$ symmetry is used to forbid Majorana mass terms in general, and hence is connected to the Dirac nature of neutrinos \cite{Chulia:2016ngi}. 

Neutrinos are also helpful to indirectly probe dark matter, if it decays or annihilates into neutrinos. Atmospheric neutrino experiments or neutrino telescopes are sensitive to this signature, which above several GeV of energy is essentially background-free. Those experiments were not constructed with the aim of studying DM particles, but nevertheless provide highly valuable information on them.   
The DM-generated  neutrino flux can come from DM particles accumulated in the Earth or Sun, or from the galactic halo, which introduces uncertainties when integrating the DM density along the line of sight (the so-called $J$-factor). See Ref.\ \cite{losHeros:2020csi} for an overview.

Another connection of neutrinos to DM is provided by the ``neutrino floor'' \cite{Billard:2013qya}. Coherent elastic neutrino-nucleus scattering (CE$\nu$NS, see Sec.\ \ref{sec:WG3_coherent}) of unshieldable solar and atmospheric neutrinos with $\sim$ MeV energies will create a background of nuclear recoils in the keV-regime. This background is not distinguishable event by event from the direct detection signal that future  experiments such as DARWIN \cite{Aalbers:2016jon} aim to measure, caused by the scattering of non-relativistic DM particles with masses above $\sim$ GeV.   The neutrino floor  then corresponds approximately to a number of neutrino events larger than the number of DM events with similar recoil spectra. In principle, ways to get below the neutrino floor exist \cite{Davis:2014ama,Ruppin:2014bra,Grothaus:2014hja,OHare:2015utx,Dent:2016iht}, for example,  via direction-sensitive detectors; however, going beyond the neutrino floor will be challenging. Interestingly, new interactions of neutrinos can significantly enhance the neutrino floor \cite{Boehm:2018sux}. 
Thus, understanding CE$\nu$NS with dedicated experiments, as well as solar and atmospheric neutrino fluxes, is of great importance. Finally, we stress again the experimental similarities of dark matter direct detection and neutrino experiments, as discussed in Sec.\ \ref{sec:tech}.

\subsubsection{The Baryon Asymmetry of the Universe}
\label{WG1:SubSec:PMNS-Theory:CPV}

The PMNS matrix has up to three CP-violating phases. The Dirac phase is the analogue of the CKM CP-violating phase and it can be measured in neutrino oscillation experiments. If neutrinos are Majorana particles, the two additional CP phases cancel out in the oscillation probabilities, but affect other observables such as the neutrinoless double-beta-decay rate. For seesaw models that feature additional, usually very massive, neutral fermions, there will in general be additional CP-violating phases affecting the decays of the heavy neutral fermions into leptons and Higgs bosons. This is important for leptogenesis driven by out-of-equilibrium heavy neutral lepton decays above the electroweak phase transition~\cite{Fukugita:1986hr}.

Leptogenesis has received great attention because it can naturally take place in seesaw models and has been shown to be able to explain the observed baryon asymmetry of the Universe. It can be implemented in different ways, depending on the underlying neutrino mass model. In the simplest case of type I seesaw models, it readily satisfies the three Sakharov conditions:
(i) lepton number is violated by the heavy Majorana masses; (ii) several CP-violating phases can be present in the Yukawa coupling between heavy neutrinos and the Higgs; (iii) the departure from equilibrium is guaranteed by the expanding Universe. At very high temperatures, the heavy Majorana neutrinos are in thermal equilibrium with the rest of the plasma due to their Yukawa interactions and decouple after the temperature drops below their mass. In the presence of CP violation, their decays generate a lepton asymmetry which is converted into a baryon asymmetry by SM sphaleron processes. 
In fact, essentially all scenarios that generate neutrino mass have the option to generate the baryon asymmetry of the Universe \cite{Hambye:2012fh}. Leptogenesis is possible even if neutrinos are Dirac particles \cite{Dick:1999je}. 
Various aspects of leptogenesis in and beyond the standard scenario are summarized in Refs.\ \cite{Buchmuller:2004nz,Dev:2017wwc,Drewes:2017zyw,Chun:2017spz,Bodeker:2020ghk}.

The current experimental preference for a nonzero Dirac phase as outlined in Sec.\ \ref{sec:osc} is an encouraging sign for the existence of new sources of CP violation in nature, as is required for all theories of baryogenesis. The connection with leptogenesis is a very important one, as it could, in principle at least, shed light on the nature of leptonic CP-violation. Unless special flavor structures are imposed, all of these phases are free parameters, whose origin is as much of a mystery as are the fermion masses and lepton and quark mixing angles. As indicated above, most models of neutrino masses contain a larger number of parameters than those measurable, and in particular more CP-violating phases. Consequently, in a completely model-independent way, it is not possible to draw a direct link between the value of $\delta_{\rm CP}$ and the baryon asymmetry.

However, models which aim to explain the values of neutrino masses and of the mixing structure we observe have a reduced number of parameters and can present a direct connection between the two. 
It is therefore possible to make some more general statements in specific neutrino mass models. In the widely studied case of the type I seesaw at scales slightly below $10^{12}$~GeV, it can be shown (by setting to zero all other phases) that the $\delta_{\rm CP}$ phase can be the origin of the observed baryon asymmetry~\cite{Pascoli:2006ci,Anisimov:2007mw,Moffat:2018smo}, and it is also possible in extended see-saw models~\cite{Dolan:2018qpy}.  This would represent the terrestrial discovery of a parameter that is essential for very early-Universe cosmology. Alternatively, if the Dirac phase, depending on its value, could be proven to be inadequate, then that would point to the existence of additional phases, which could well have a leptonic origin. 

Within type I seesaw models, leptogenesis is not necessarily a high energy phenomenon. It is possible to bring down the scale of the heavy neutrino masses down to GeV-scale 
\cite{Akhmedov:1998qx, Canetti:2012kh,Canetti:2012vf}. This requires typically an extreme closeness of  right-handed neutrino masses (``resonant leptogenesis''), and  allows testability for instance in SHiP \cite{Alekhin:2015byh} (see also \cite{Agrawal:2021dbo}). As mentioned in Sec.\ \ref{sec:mass_compl}, high scale leptogenesis is not possible when there is TeV-scale lepton number violation on an observable level; hence it is falsifiable \cite{Deppisch:2013jxa}.\\

One can conclude that, generically, the observation of lepton number violation (e.g.\ neutrinoless double beta decay) and of CP violation in long-baseline neutrino oscillation experiments and/or possibly neutrinoless double beta decay would provide circumstantial evidence (not a proof!) in favor of thermal high scale leptogenesis as the origin of the baryon asymmetry of the Universe. Certain very constrained scenarios allow for testing all parameters related to leptogenesis in terrestrial experiments. Generically, mechanisms for neutrino mass generation come quite often together with new particles and CP phases, thus entertaining the possibility of generating the baryon asymmetry of the Universe. Testability requires model building input, but the search for low energy lepton number and CP violation is crucial to test our ideas on lepto- and baryogenesis, and will help in favoring neutrino-mass-related baryogenesis scenarios over other ones.

\subsubsection{Unification}
\label{sec:unif}
Neutrinos do not exist isolated from the other particles of the SM. They live with charged leptons in an $SU(2)$ doublet. In Grand Unified Theories (GUTs), neutrinos live additionally with quarks in GUT multiplets. Furthermore, mechanisms to generate small neutrino mass introduce new particles that can influence other particles of the SM as well. 
Among the many examples, we focus on the type I seesaw mechanism with right-handed singlet neutrinos, discussing several aspects of unification. 
Starting with supersymmetry, scalar partners of the singlet neutrinos are called sneutrinos, and inherit lepton number violating properties. If one of the sneutrinos is the lightest supersymmetric particle, it is a DM candidate, see e.g.\ \cite{Asaka:2005cn}.  In many supersymmetric frameworks, the Dirac mass matrix within the type I seesaw mechanism is the main source for lepton-flavor-violating decays mediated by SUSY particles, and is typically observable for not-too-large SUSY scales. In this case, one is able to reconstruct the seesaw parameter space, in contrast to the non-supersymmetric case where the heavy neutrinos and/or their small mixing suppress the rates. 

The Yukawa coupling of the right-handed neutrino with SM lepton doublets and the Higgs boson may have a remarkable impact on Higgs physics. First of all, it influences the running of the quartic Higgs coupling towards negative values at high energies, thus leading possibly to an unstable vacuum \cite{EliasMiro:2011aa}. Moroever, right-handed neutrinos provide a loop-induced contribution to the Higgs mass of order $\Delta m_h^2 \simeq m_D^2 \, M_N^2 /(16\pi^2  v^2) = m_\nu \, M_N^3 /(16\pi^2 v^2)
$, leading to a limit of $M_N \lesssim 10^7$ GeV if this contribution is not to be larger than the measured Higgs mass \cite{Vissani:1997ys,Casas:2004gh,Abada:2007ux,Clarke:2015gwa}. This ``naturalness'' mass limit is, interestingly,  in conflict with standard leptogenesis requirements (see Sec.\ \ref{WG1:SubSec:PMNS-Theory:CPV}). 

Theories that gauge the difference of baryon and lepton number $B-L$ are attractive as this charge is exactly conserved in the SM \cite{Wetterich:1981bx, Mohapatra:1980qe, Marshak:1979fm, Masiero:1982fi, Mohapatra:1982xz, Buchmuller:1991ce}. 
The symmetry is anomaly-free and can thus be consistently gauged, if three right-handed neutrinos are added to the particle content, providing therefore motivation for the seesaw mechanism. Furthermore, lepton and baryon number are connected here, linking neutrinoless double beta decay with proton decay.

The difference $B-L$ is also part of many other BSM theories. In particular, in left-right symmetric models the $SU(2)_L$ gauge group of the SM is extended by another $SU(2)_R$ that acts on right-handed fields only \cite{Pati:1974yy, Mohapatra:1974gc,Mohapatra:1974hk, Senjanovic:1975rk,Senjanovic:1978ev,Mohapatra:1979ia,
Mohapatra:1980yp}. 
Right-handed neutrinos are thus needed for the consistency of the theory. Their mass is generated by a Higgs mechanism  in analogy to all fermion masses of the SM (with a  Higgs multiplet different from the SM Higgs). The scale of their masses is the scale at which parity breaks down. 
Hence, the parity violation of weak interactions is connected to the smallness of neutrino mass. In this sense, left-right symmetric theories have gained additional support by determining that neutrinos have mass \cite{Senjanovic:2016bya}.

In GUTs the active and sterile neutrinos share a multiplet. In $SO(10)$ models, all 15 SM fermions (left- and right-handed up- and down-quarks of all three colors, left- and right-handed electrons and left-handed active neutrinos) of a single generation plus a right-handed neutrino, 
fit in the 16-dimensional spinor representation of the group. 
With a Higgs sector that breaks  the large GUT group down to the SM, the different Yukawa coupling matrices of quarks and leptons are connected. In the most-often applied scenario, one has all five relevant mass matrices given by a combination of only three Yukawa couplings \cite{Ross:1985ai,Dutta:2004zh,Dutta:2005ni,Senjanovic:2006nc}
\begin{equation}
\label{eqn:mrel}
\begin{aligned}
m_u &= v_{10}^u Y_{10} + v_{126}^u Y_{126} + v_{120}^u Y_{120} \, ,\\
m_d &= v_{10}^d Y_{10} + v_{126}^d Y_{126} + v_{120}^d Y_{120} \, ,\\
m_{D} &= v_{10}^u Y_{10} - 3 v_{126}^u Y_{126} + v_{120}^{D} Y_{120} \, ,\\
 m_\ell &= v_{10}^d Y_{10} - 3 v_{126}^d Y_{126} + v_{120}^\ell Y_{120} \, ,\\
M_R &= v_{126}^R Y_{126} \, ,
\end{aligned}
\end{equation}
where $m_u, m_d, m_D, m_l, M_R$ are the up-quark, down-quark, Dirac neutrino, charged lepton and right-handed Majorana neutrino mass matrices. 
Fitting this to the observed fermion masses is a check of the validity of GUT models. 
One particular result is that the case of only $Y_{10}$ and $Y_{126}$ being present is ruled out, since the value of the atmospheric neutrino mixing angle comes out too low \cite{Bertolini:2006pe}. Other choices are not compatible with an inverted mass ordering; hence neutrino data can seriously constrain grand unified theories.  
Furthermore, typically a hierarchical neutrino mass spectrum is predicted. Hence observing close-to-degenerate neutrino masses would lead to serious implications for GUTs. Moreover, while type I and II seesaw mechanisms, and to some extent the type III, can easily be implemented in such GUTs, many other mechanisms require non-minimal to exotic  extensions. 

One should mention that the obvious experimental connection of neutrino physics to GUTs, in the sense that present \cite{Takenaka:2020vqy} and upcoming \cite{Abi:2020kei,Abe:2018uyc} detectors can set strong limits on various proton decay modes, thereby testing GUTs. 

Finally, theories with additional gauge groups can have first-order phase transitions when one of the scalar multiplets necessary for the symmetry breaking obtains a vacuum expectation value. This can have testable consequences as it leads to a stochastic gravitational wave background, that can be probed in future experiments.

\subsection{Sterile Neutrinos}
\label{sec:ster}
Sterile, or right-handed, neutrinos $N_R$ have been encountered already a few times in this document. Some experiments discussed in Sec.\ \ref{WG0}, point towards the existence of eV-scale sterile neutrinos, having various possible consequences in particle physics and cosmology. This section is aiming at generalizing the notion of sterile neutrinos to arbitrary energy scales. Generically, sterile neutrinos, i.e.\ fermions without isospin or hypercharge, appear in many BSM frameworks and are thus an exciting window to new physics. Reviews on the interesting physics of sterile neutrinos can be found in Refs.\ \cite{Abazajian:2012ys,Drewes:2013gca,Drewes:2017zyw}.

There are two arguments for the mass scale of sterile neutrinos. In the limit of vanishing mass there is no Majorana mass term, hence the symmetry of the system is enhanced because lepton number is conserved. This is along 't Hooft's notion of naturalness: a parameter should be small if its absence enhances the symmetry of the system. On the other hand, right-handed neutrinos are singlets under the SM gauge group; hence their Majorana mass term is not protected by the gauge symmetry of the SM, thus can be arbitrarily large.  Good phenomenological arguments were made in the past for keV-scale neutrinos as warm dark matter (see Sec.\ \ref{sec:dm}), or for masses close to the GUT scale of $10^{15}$ GeV in order to explain the light neutrino mass scale via $v^2/M_R$ in the type I seesaw. Another argument is the slightly lower scale for the lightest sterile seesaw-neutrino of around $10^{10}$ GeV, a value at  which the simplest leptogenesis mechanism works (see Sec.\  \ref{WG1:SubSec:PMNS-Theory:CPV}). It is fair to say that no convincing argument has been made for the presence of an eV-scale sterile neutrino. All models are ``postdictions'' made after the LSND and reactor anomalies discussed in Sec.\ \ref{WG0}.

The number of sterile neutrinos is not constrained in general. Theories in which right-handed neutrinos are part of a gauge group require that they come in with the same number of SM generations, i.e.\ three.   In the seesaw context there should be one for each massive light active neutrino, which implies that two are enough (the lightest active neutrino could be massless). 
Two are also enough to explain the baryon asymmetry of the Universe via leptogenesis. 
This implies that sterile neutrinos of various energy scales may exist. Scenarios are conceivable in which one keV-scale neutrino is responsible for warm dark matter and two heavier ones are for leptogenesis. This is the spirit of the $\nu$MSM \cite{Asaka:2005pn,Canetti:2012vf}. 

While being called ``sterile'', these particle are not completely decoupled from the SM. They have the option to couple to the Higgs and a lepton doublet, which induces mixing with light neutrinos in the charged and neutral currents.  This mixing implies that they are produced in certain amounts whenever charged leptons or neutrinos take part in weak interactions. Therefore, one can look for their decay products and produce them at colliders or in decays. Indeed, limits on sterile neutrino mass and mixing have been obtained from beta decays, meson decays, or from LHC. Moreover,  sterile neutrinos have cosmological implications, see Sec.\  \ref{sec:sterile_cosmo}. 

Generally speaking, the vertex of Higgs-lepton-$N_R$ can  be viewed from different directions:  
(i) the Higgs couples to leptons and $N_R$, (ii) leptons couple to the Higgs and $N_R$, (iii) $N_R$ couples to leptons and Higgs. While this sounds trivial, it illustrates the rich phenomenology of sterile neutrinos. In the seesaw language, the above three directions of the vertex imply (i) Higgs-physics (naturalness, vacuum stability), (ii) lepton flavor violation, (iii) leptogenesis. 
In case the sterile neutrinos do have additional interaction (see Sec.\ \ref{sec:unif}), the discussion becomes even broader. Note that the seesaw connection of sterile neutrinos is not unique. Right-handed neutrinos could have left-handed Dirac partners and their mass comes from a different ``dark'' sector. This would modify the way one searches for them. 

 In summary, the notion of sterile neutrino is very broad and has rich phenomenology in particle physics and cosmology. While the presence of sterile neutrinos is expected from a theoretical point of view, its mass scale is unknown, and depending on its value has very different, but generically exciting, consequences.

\subsection{New Physics in Neutrino Experiments}
\label{sec:np}
Neutrino mass models often come with additional energy scales and interactions. This implies that neutrinos may show properties beyond the standard paradigm of three active neutrinos interacting via electroweak interactions. 
Indeed, new physics beyond the three-neutrino picture is expected in many well-motivated scenarios. 
There is even one long-standing hint of new neutrino physics, namely light sterile neutrinos, which is treated at various places in this report. Some possible new physics scenarios are well motivated, while others are speculative. The main point is that neutrinos offer completely new avenues to test for new physics. 
Here  neutrino experiments can probe TeV-scale new physics, which is on par with collider searches.  
Moreover, the presence of such effects may hinder the determination of unknown neutrino parameters, and therefore should be understood to avoid making wrong claims on the mass ordering or CP violation. The complementarity of different neutrino experiments to test the same parameters is important to avoid such wrong conclusions. 
More often than not, limits on non-standard neutrino features are obtained as a by-product of large-scale oscillation experiments, or by small dedicated experiments. This illustrates the variety and broadness of experimental approaches. 
  Popular examples on new physics are magnetic moments, unitarity  violation or non-standard interactions. 
Many more exotic possibilities such as long-range forces or CPT violation may show up, and strong sensitivities can be reached in many experiments. 

The present section tries to summarize the main theoretical ideas and features of various new physics scenarios.

\subsubsection{Non-Standard Interactions}
\label{sec:nsi}
Non-standard Interactions (NSIs) are a popular new physics option for neutrinos. They denote additional vector-like interactions of the left-handed SM neutrinos with other fermions $f$, described by the following neutral current Lagrangian \cite{Wolfenstein:1977ue}: 
\begin{equation}\label{eq:NSI}
    -{\cal L} = 2\sqrt{2} G_F\,  \epsilon_{\alpha \beta}^f 
    \left[ \bar \nu_\alpha \gamma_\mu P_L \nu_\beta \right]
    \left[ \bar f \gamma^\mu f \right]. 
\end{equation}
The strength of this new interaction is normalized to SM interactions and encoded in dimensionless (and complex) $\epsilon$ parameters. If $f$ are first-generation particles, these terms induce coherent forward scattering of neutrinos in matter, in analogy to the matter effects discussed in Sec.\  \ref{sec:osc}. In this way, they modify neutrino oscillation probabilities relevant for long-baseline experiments. Reviews on the effects of the above interaction can be found in Refs.\ \cite{Ohlsson:2012kf,Farzan:2017xzy,Dev:2019anc}.

The above interaction could stem from heavy particles that mediate interactions between neutrinos and other SM fermions. In this case one expects that $\epsilon \propto m_W^2/M_X^2$, where $m_W$ is the mass of the SM $W$-boson and $M_X$ the mass scale of the new particles. Therefore, probing percent-level $\epsilon$ implies testing TeV-scale new physics, that is, energy scales accessible at the LHC and beyond.  The interaction in Eq.\ (\ref{eq:NSI}) therefore has immediate collider phenomenology \cite{Babu:2020nna}. 
This demonstrates once more the exciting potential of neutrino physics. 
Indeed, many of the limits on the above $\epsilon_{\alpha \beta}^f $, as listed e.g.\ in Ref.\ \cite{Esteban:2018ppq} are around $0.1$ and partly below. Such limits also include results from 
coherent elastic neutrino-nucleon scattering (CE$\nu$NS, see Sec.\ \ref{sec:WG3_coherent}), where 
NSIs on up- and down quarks modify the cross section.  There is however a crucial difference. The matter effect in neutrino oscillations is caused at momentum transfer of 10 MeV and below. Therefore, the particles mediating the NSI can have a mass as small as this value \cite{Farzan:2015hkd}, and the smallness of the $\epsilon$ would be caused by a small coupling instead. Such a small value of the mediator mass will however modify the recoil spectrum observed in CE$\nu$NS, where the propagator of those particles is $1/(q^2 - m_X^2) \simeq 1/q^2 \propto 1/T$ and increases the cross section at low recoil $T$. 
Thus one can break the light/heavy mediator degeneracy by combining oscillation and scattering experiments. 
Regarding future neutrino oscillation experiments, those will further improve the bounds on NSIs or discover them. The presence of $\epsilon_{\alpha \beta}^f$ may hinder the determination of the neutrino mass ordering and the CP phase, or lead to a wrong measurement. Indeed, a recent result in this respect is the finding that CE$\nu$NS limits on certain $\epsilon$ parameters rule out the ``dark LMA'' solution $(\theta_{12}>0)$, which would be allowed by solar neutrino data  in the presence of non-zero values of those parameters \cite{Esteban:2018ppq}. 
This illustrates the necessity of different experiments and of over-determining the neutrino parameters. 

Note that the above interaction can be generalized to interactions different from vector, that is, scalar, pseudoscalar, axial vector, or tensor. These new interaction may stem from coupling neutrinos to dark matter. 
They would not influence neutrino oscillations \cite{Ge:2018uhz,Babu:2019iml}, but could also lead to effects in neutrinoless double beta decay \cite{Kovalenko:2013eba}, or observable  distortions in measurable recoil spectra e.g.\ in CE$\nu$NS \cite{Lindner:2016wff}. Also beta decays are used to constrain those extra interactions \cite{Cirgiliano:2019nyn}. The origin of those interactions often includes leptoquarks \cite{Bischer:2019ttk}. 
One can further generalize the new interactions in Eq.\ (\ref{eq:NSI}) from neutral currents to charged currents, which are however typically more constrained as the corresponding lepton flavor violation bounds are quite strong.

There have been several proposals for particles associated with NSIs. Examples are $Z'$ bosons, leptoquarks, scalar singlets, etc. Those typically induce not only a single set of NSI, and can thus be distinguished from each other by a global search in various experiments. In addition, the particles causing the NSI can be generated at colliders, if they are heavy. If they are light, they could be radiated off in processes in which the fermions taking part in Eq.\ (\ref{eq:NSI}) appear. It is noteworthy that this possibility includes also neutrino self-interactions, which are among the leading new physics solutions proposed to explain the long-standing discrepancy of Hubble parameter determinations in near- and far-distance cosmology \cite{Bernal:2016gxb}. Finally, one should mention the recent XENON1T excess of measured electron recoil $T$ \cite{Aprile:2020tmw}, which can be explained by solar neutrinos coupling to light mediators. The propagator of those particles is $\propto 1/T$ and is well-suited to explain an excess at low recoil.

\subsubsection{Unitarity Violation}
\label{WG1:SubSec:PMNS-Theory:Unitarity}

The PMNS matrix is exactly a $3 \times 3$ unitary matrix for two important neutrino mass scenarios: (1) the case of three Dirac neutrinos where their mass generation occurs in exactly the same way as for the charged fermions,  and (2) the case of three left-handed Majorana neutrinos with no additional fermionic states playing any role in the mass generation, as is the case for the type II seesaw model. 
%~\cite{Magg:1980ut,Schechter:1980gr,Cheng:1980qt,Lazarides:1980nt,Wetterich:1981bx,Mohapatra:1980yp}. 
However, if there are additional fermionic states, they will typically lead to unitarity violation. An example is type I seesaw right-handed neutrinos 
with Majorana masses, which generically couple with the three known neutrino flavors via a Yukawa interaction with the Higgs, unless a symmetry is invoked to forbid these terms. Once the Higgs acquires a vacuum expectation value, mixing arises and affects all the neutrino and neutrino-like states, rendering the $3 \times 3$ PMNS matrix only approximately unitary.  
Barring special flavor structures, one expects the deviations from unitarity to scale as powers of the ratio of the electroweak scale to the new physics scale. 
If the scale of the seesaw mechanism is low, or in the presence of additional states, as in inverse,  
linear 
or other extended see-saw models, the mixing can be sizable without 
contradicting the 
constraints from neutrino masses. Thus, tests of the unitarity of this matrix are also tests of the nature of the underlying mass generation mechanism. 
Note, however, that the presence of additional charged-lepton states would also imply unitarity violation. 

Underlying models put aside, we can describe a non-unitary lepton mixing matrix $N$ (as defined through charged-current interactions between charged leptons and neutrinos) as 
$N = (1 - \eta)\, U_\text{PMNS}$ with an exactly unitary matrix $U_\text{PMNS}$ via a Hermitian matrix $\eta$. This affects a raft of precision electroweak and flavor observables including the $W$-mass, the weak mixing angle, $Z$-decays, tests of flavor universality and many others, as was systematically studied for instance in Refs.~\cite{Antusch:2006vwa,Fernandez-Martinez:2016lgt}. A focus on non-unitarity in oscillation experiments only was made in Refs.\ \cite{Parke:2015goa,Ellis:2020hus}. 
Table IV of Ref.\ \cite{Fernandez-Martinez:2016lgt} presents a useful summary of the upper bounds on unitarity-violating deviations.  These results show that deviations up to the $2-7\%$ level are permitted by current data, depending on the matrix element. The $\eta_{\tau\tau}$ parameter is the least constrained, with $\eta_{\mu \mu}$ the most constrained. 
While in the CKM sector the unitarity of the mixing matrix is tested for all independent unitarity constraints, the prospects for the lepton sector are poor. Nevertheless, the precision foreseen in particular by JUNO will allow unitarity tests on the first row of the PMNS matrix on the percent-level \cite{Ellis:2020hus}. 
Studies of tau neutrino physics with atmospheric neutrinos \cite{Aartsen:2014oha,Adrian-Martinez:2016fdl}, long-baseline neutrinos \cite{deGouvea:2019ozk} or with SHiP \cite{Anelli:2015pba} 
are important to measure $U_{\alpha \tau}$ elements, which are currently hardly constrained.  Unitarity violation in long-baseline experiments may weaken the precision with which the currently unknown neutrino parameters can be measured \cite{Escrihuela:2016ube}.

It should be noted that also light sterile neutrinos would lead to an apparent non-unitarity of the PMNS matrix, if the oscillation experiment corresponds to $L/E \gg \Delta m^2_{\rm ster}$, i.e.\ the mass-squared difference  averages out. In addition, the ``zero-distance'' effect in non-unitary scenarios (flavor change even for vanishingly small baselines) is also present for non-standard interactions; see Sec.\ \ref{sec:nsi}. This demonstrates that distinguishing new physics requires a variety of experimental 
approaches.

\subsubsection{Magnetic Moments and other Electromagnetic Features}

Neutrino Magnetic Moments (NMM) are the most popular new physics scenario related to electromagnetic properties of neutrinos. The most general coupling of neutrinos to the photon field can be written in the form of charge, dipole magnetic, dipole electric, and anapole neutrino form factors, which display different Lorentz structures. 
In the limit of vanishing momentum exchange $q^2$, the charge form factor corresponds to a potential charge of the neutrino, while the dipole magnetic form factor is the magnetic moment. The latter is best understood and has attracted the most attention from experimentalists. We refer to the review in Ref.\ \cite{Giunti:2008ve} for a detailed summary of the physics of neutrino electromagnetic features, 
 in particular, the subtle connections of electromagnetic properties of neutrinos with the Dirac/Majorana nature, or with C, P, CP and CPT violation.  

In the SM extended with massive Dirac neutrinos, loop processes generate electromagnetic couplings, since the neutrino couples to charged $W$ bosons. The magnetic moment is of order $G_F e m_\nu/(16\pi^2)$, and takes a very small value of 
$\mu_\nu \simeq 3.2 \cdot 10^{-19} \, (m_\nu/{\rm eV}) \, \mu_B$, expressed in Bohr magnetons. Interestingly, for Majorana neutrinos, magnetic moments can only couple different  states with each other (``transition magnetic moments''); for Dirac neutrinos, only diagonal transitions are allowed. There is furthermore a relative factor of two between Majorana and Dirac neutrino NMM. 
As in the SM the NMM is very much suppressed, an observation would immediately imply new physics coupling to neutrinos. 
Indeed,  models in which particles with electroweak quantum numbers appear often lead to observable magnetic moments. A common issue in such models is that a magnetic moment generated at a scale $\Lambda$ corresponds to a neutrino mass proportional to $\mu_\nu \Lambda^2$, which typically is too small unless non-trivial model building is involved. 

An interaction caused by a magnetic moment adds incoherently to the SM cross section. For instance, neutrino-electron scattering, in the limit of electron energies $E_\nu$ much larger than recoil $T$, has a differential cross section with respect to recoil of order $G_F^2 m_e (a + b \, T/E_\nu)$ in the SM, while the magnetic moment contributes with $\alpha /m_{e}^2 (\mu_\nu/\mu_B)^2 \frac 1T$.  For  millicharged neutrinos, the differential cross section would be proportional to $\alpha /(m_e T^2)$. The increase at low recoil is a  characteristic feature of electromagnetic couplings, and is the main method to obtain limits in terrestrial experiments. Here experiments with reactor and solar (anti)neutrinos give the most stringent terrestrial limits, which are of order $10^{-11} \mu_B$, i.e.\ eight orders of magnitude below the prediction of the SM extended by massive neutrinos. Note that, again, the XENON1T excess can easily be explained by solar neutrino scattering involving millicharge or magnetic moments \cite{Aprile:2020tmw}, courtesy of the increase of the differential cross section at low recoil. Competing limits on electromagnetic couplings of neutrinos are also obtained by CE$\nu$NS experiments, which work at low recoil by definition.

Astrophysical limits are typically stronger than terrestrial ones, though they often depend on the modeling of the neutrino source under consideration. A NMM couples left-handed states to right-handed ones, the latter being sterile for Dirac neutrinos and thus able to leave a supernova core, thereby inducing energy loss. Confronting the observation of the SN1987A neutrino burst with this feature leads to limits of order $10^{-12} \mu_B$. 
Other processes that are of astrophysical interest are radiative decay $\nu_j \to \nu_i + \gamma$ (which would also influence the cosmic microwave background) or a resonant spin-flavor precession in the presence of a magnetic field. 
Such limits are summarized in Ref.\ \cite{Raffelt:1999gv}.

\subsubsection{Other Possibilities}
\label{sec:exotic}
The unusual features of neutrino mass and lepton mixing motivate speculation about the presence of new physics beyond the possibilities discussed so far. 
\begin{itemize}
    \item {\bf Long-range forces:} This is actually connected to NSIs: if the mass of a vector or scalar boson coupling to neutrinos has an extremely tiny mass, a long range force proportional to $1/r$ is felt by neutrinos. For instance, the Sun-Earth distance corresponds to about $10^{-18}$ eV, and mediators lighter than that value create a potential $V \simeq 1.0 \cdot 10^{-12} \, ({g'}^2/10^{-50})$ eV, which can be compared to quantities related to oscillations, namely $\Delta m^2/E$ or $G_F N_e$, where $N_e$ is the number density of neutrinos in the medium  \cite{Joshipura:2003jh,Grifols:2003gy,Heeck:2010pg,Bustamante:2018mzu,Smirnov:2019cae}.  
    If this force involves electrons, the limits are extremely strong 
    (${g}'^2$ much below $10^{-50}$), 
    as the Earth or the Sun contain an enormous amount of electrons. If muons or tau forces are involved, mixing of the bosons is required for observable effects. A list of current constraints in many models is given in Ref.\ \cite{Coloma:2020gfv}. 
    In principle the effect is distinguishable from the  vector NSIs discussed in Sec.\ \ref{sec:nsi} since long-range interactions depend differently on the distance traveled. However, in reality the difference is tiny.   
    
    \item {\bf Neutrino decay:} Massive neutrinos can decay radiatively $\nu_j \to \nu_i + \gamma$, though with only SM interactions the half-life is extremely long. This decay mode could however be enhanced via new physics, or the decay products could be one or more beyond the Standard Model particles, e.g.\ a sterile neutrino or a Majoron (``invisible decay''). A Majoron is a Goldstone boson associated with the broken global lepton number symmetry. This decay would have observable consequences for astrophysical (see e.g.\ \cite{Bustamante:2016ciw,deGouvea:2019ozk,Denton:2018aml}) neutrinos or for neutrino oscillation experiments (see e.g.\ \cite{Fogli:1999qt,Gago:2017zzy,Choubey:2018cfz}), where the classical oscillatory $L/E$ behavior would be modified by an exponential suppression exp$\{-m_\nu L/(E\tau)\}$ caused by the life-time $\tau$. A pure decay solution to the observed neutrino deficits was ruled out quite early \cite{Fogli:1999qt}. 
    Moreover, cosmological mass limits can be weakened if neutrinos decay \cite{Beacom:2004yd,Escudero:2020ped}. 
 Limits from atmospheric or long-baseline oscillation experiments are around $\tau \gtrsim 10^{-10} (m_\nu/{\rm eV})$ s \cite{Porto-Silva:2020gma} and solar neutrinos yield $\tau \gtrsim 10^{-4} (m_\nu/{\rm eV})$ \cite{Picoreti:2015ika},  whereas cosmological ones are much stronger, $\tau \gtrsim 10^{11} (m_\nu/{\rm eV})^5$ s \cite{Barenboim:2020vrr}. Observations from SN1987A give $\tau \gtrsim 10^{7} (m_\nu/{\rm eV})^3$ s \cite{Frieman:1987as}. Those numbers depend actually on which mass state decays and also on the mass ordering. Upcoming experiments, and a possible supernova, will further improve the limits. As usual, the fit results of oscillation experiments could change if neutrino decay were present. 
 
    \item {\bf Pseudo-Dirac neutrinos:} In cases where a Dirac and a Majorana mass term is present, but the former is much larger than the latter, neutrinos are formally Majorana particles. However, each mass state corresponding to an active neutrino is split in two states (one being sterile), the small splitting related to the ratio of the Majorana and Dirac mass terms. Several models have been found which realize such a scenario (see Ref.\ \cite{Anamiati:2019maf} for a discussion). Besides new mass-squared differences, also new mixing angles are introduced. All lepton-number-violating effects are  suppressed. Observable effects could nevertheless arise in oscillation experiments (see e.g.\ \cite{Joshipura:2000ts,Anamiati:2019maf}) 
    or in the context of astrophysical neutrinos \cite{Beacom:2003eu,deGouvea:2009fp}. 
    A further exotic modification is when certain mass states are Pseudo-Dirac while others are Majorana, which occurs when they are generated by a different source \cite{Allahverdi:2010us}.
    
    \item {\bf Anomalous decoherence:} 
    As mentioned in Sec.\ \ref{sec:osc} neutrino oscillations are closely connected to quantum-mechanical aspects such as decoherence: wave packets associated with different mass states need to overlap in order to observe oscillations. For large travel distances, coherence is lost. 
        One can envisage scenarios in which new physics induces additional decoherence, for instance caused by space-time foam within quantum gravity theories. For this case, one would fit the data with an arbitrary decoherence factor exp$\{-\xi(L,E)  \}$ multiplied with the oscillation probability. Those tests have been performed with various sources, see e.g.\ Refs.\  \cite{Fogli:2003th,Anchordoqui:2005gj,Mavromatos:2007hv,Stuttard:2020qfv}, and limits on anomalous decoherence have been obtained, testing those theories; a discovery would have obvious fundamental consequences.

    \item{\bf CPT and Lorentz invariance:}
    The final exotic possibility is the violation of holy principles in relativity and quantum field theory. An overview on different tests in the neutrino sector can be found in Ref.\ \cite {Diaz:2016xpw}. One should note that 
    Lorentz invariance violation leads to CPT violation, but not vice versa. CPT violation can be obtained by violating locality or causality, while keeping Lorentz invariance intact. As for anomalous decoherence caused by space-time foam, one would expect that the effect goes with energy scale over the Planck mass $M_{ \rm Pl} \simeq 10^{19}$ GeV. The likely high-scale origin of neutrino mass motivates the search for such effects with neutrinos, as the energy scale associated to their mass generation may not be far away from the Planck scale. 
    
    An obvious consequence of CPT violation would be that neutrino and antineutrino parameters are not the same. This is in principle distinguishable from matter effects, NSIs or long-range forces, which also generate this feature. Such explicit CPT violation has been applied to various cases in neutrino physics, see e.g.\  \cite{Murayama:2000hm,Barenboim:2001ac,Barenboim:2004wu,deGouvea:2017yvn,Liao:2017yuy}. It is worth noting that meson masses, in particular neutral kaons, are constrained to have a mass-squared difference of $m^2(K_0) - m^2(\bar K_0) \lesssim 0.25$ eV$^2$, and that for neutrinos the related difference of $\Delta m^2_{31}$ measured with neutrinos and antineutrinos is known to be much smaller. 
    
    Superluminal neutrinos are another candidate in this field. For tachyonic particles, processes are allowed that are otherwise kinematically forbidden, which was used to strongly refute \cite{Cohen:2011hx} the past (and later retracted) claim of faster-than-light neutrinos \cite{Adam:2011faa}. Strong limits on the amount of superluminality can also be given by such considerations.

    The so-called Standard Model Extension (SME) describes consequences of Lorentz invariance violation in a systematic way \cite{Kostelecky:2003cr,Kostelecky:2003fs}. Gauge-invariant operators violating Lorentz-invariance are constructed out of SM fields. In this framework, there are a great many free parameters that can be constrained in essentially any neutrino experiment.  Some of the constraints are obtained by the effect that the SME coefficients determine a fixed direction in space-time, around which the experiment rotates with sidereal frequency $2\pi$/(23 h 56 min). A variety of cases has been discussed in the literature, see e.g.\ \cite{Diaz:2009qk,Katori:2016eni,Agarwalla:2019rgv}.

\end{itemize}

All in all, there is some motivation that the properties of neutrinos make them more sensitive to exciting new physics than other particles. Searches for these effects can be done in every neutrino experiment (and meaningful limits set if the effects are not found) without additional costs, making these searches  attractive to perform. In the presence of such effects, typically the sensitivity to standard parameters is decreased.

\vspace{2cm}

{\bf Acknowledgments:} 
The members of the IUPAP neutrino panel would like to thank Mandy Lamarche for her continuous support towards the successful completion of this report. 
We thank 
Soud Al Kharusi, Summer Blot, Sara Bolognesi, Christian Buck, Giorgio Gratta, Jinhao Huang,  Mary Hall Reno, Pedro Ochoa-Ricoux,  Liang Zhan, Jinnan Zhang for clarifying comments and contributions,  
and are grateful to the international community to provide us with valuable feedback and criticism. 

\clearpage

\section{Executive Summary of the IUPAP Report}
Neutrino physics lies at the heart of many of the fundamental questions in contemporary physics: from why the Universe exists in its current form, to the mechanisms by which stars burn and explode thereby populating the galaxies with foundational elements, to the fundamental properties and interactions of sub-atomic elementary particles, the neutrino holds the key to our understanding. Initially thought to be impossible to observe, the dramatic progress over the last few decades in neutrino experiments and theory has led to eight individuals receiving the Nobel prize for studies into this fundamental particle. This impressive canon of work has already illuminated many of the secrets of the elusive neutrino, yet several essential questions remain about its intrinsic properties, its interactions with other families of sub-atomic particles, its role in astrophysical and cosmological processes, and possibilities for discovering new physics. Neutrino physics has already been an incredibly fruitful area of research, yet the future research programs foreseen in this field promise even greater rewards.

A comprehensive, extensive and globally collaborative research program is required to fully realize the bright scientific potential of neutrino physics and the opportunities afforded, including applications for societal benefit. The projects required to understand the properties and sources of neutrinos are generally growing in scale, cost and complexity, and are becoming international and interdisciplinary in nature.

\subsubsection*{Our charge}
It is in the context of promoting cooperation in neutrino physics that the International Union of Pure and Applied Physics (IUPAP) established an international panel on neutrino physics. This panel has the mandate ``to promote international cooperation in the development of an experimental program to study the properties of neutrinos and to promote international collaboration in the development of future neutrino experiments to establish the properties of neutrinos''. The specific objective defined was the creation of a community-informed {\it science-driven} white paper following the mandate of the panel: 

\begin{itemize}
\item To carry out a review of the present status of the global neutrino physics program and the development that can be expected on a 5 to 10-year timescale 
\item To give an overview of the measurements and R\&D (including software development) that are required for the near-term ($<$ 10-year) and medium- to long-term ($10 - 25$-year) programs to fulfill their potential
\item To identify opportunities within neutrino physics, mutual benefits of global connections within neutrino physics and other fields, as well as the synergies of an international program.
\end{itemize}

\subsubsection*{This report}
This report is the output from that process. It gives a scientific overview of the current status of the various research directions within neutrino physics, the opportunities that will arise in the medium and far term, and the challenges to realizing these opportunities. It describes the required developments to optimally deliver the  most promising and exciting physics program, by encouraging international cooperation and the most effective use of world-wide resources. Not intended as a road-map, the structure and length of this report is based on the scientific opportunities and concepts and is not an attempt to define the specific research projects required. Instead, this white paper aims at demonstrating that neutrinos are very special particles with unique features and many exciting opportunities.

Looking in more detail at the scientific questions that research into neutrinos can illuminate, the last two to three decades have been highly successful with the discovery that neutrinos must have mass, and that they must have quantum-mechanical ``mixing'' to explain the observed oscillations from one type (flavor) of neutrino to another. Laboratory measurements and cosmology have constrained the number of active neutrino species and their interactions. Progress within particle physics in recent years has also been augmented by studies of neutrinos from various astronomical sources  and in cosmology. Following these spectacular achievements, there remain several key questions which will guide this research field and the opportunities for the future:

\subsubsection*{The pattern of masses and couplings of elementary particles} One of the main unresolved questions of particle physics is the origin of flavor, i.e.\ why three generations of all elementary matter particles exist. Apparent regularities in the masses and mixings of the elementary particles strongly suggest some underlying principle. The fact that neutrino masses are tiny compared to quark and lepton masses suggests furthermore a special mechanism. Future neutrino physics can provide ever more precise measurements which will lead to stringent tests of theories and mechanisms aiming at describing their properties and this underlying principle. 

\subsubsection*{New physics beyond the Standard Model} Many of the well-motivated theoretical reasons for the incompleteness of the current Standard Model of particle physics imply new neutrino-like states and/or new interactions. Future neutrino experiments have a unique potential to explore these new sectors. An important illustrative example is lepton number violation, which is potentially connected to the mystery of why there is much more matter than antimatter in the Universe, i.e.\ why the Universe exists in its current form. Other examples of similar importance are the existence of light sterile neutrinos, enhanced neutrino magnetic moments, non-standard interactions and connections of neutrinos to dark matter. Understanding these threads will guide the development of a new understanding of Nature.

\subsubsection*{Neutrinos in Cosmology and Astrophysics}
Neutrinos play a very important role in cosmology and in astrophysics. The distribution of matter, the synthesis of elements, stellar evolution and their violent end in supernovae have strong connections with neutrino properties and interactions.   This connection provides a beneficial synergy between fields, with neutrino physics providing far reaching insights to astronomical models, and neutrino physics benefiting from cosmological and astronomical observations. 

 Solar, astronomical, cosmological, atmospheric, geological and artificially created neutrinos allow unique insights into the sources and production mechanisms of neutrinos. Neutrinos are unique messengers -- they allow us to peer into the inner cores of stars and through interstellar dust, revealing astrophysical systems that are otherwise invisible. Solar neutrinos allow us, for example, to test our model of the Sun as a template for stellar evolution and nuclear fusion; supernova neutrinos provide information about the processes and dynamics of the stellar progenitors. At high energies, neutrinos allow us to understand the origin of the most extreme particle accelerators in the cosmos and may eventually even lead to joint observations of sources of  gravitational waves. Finally, observing neutrinos from reactors opens routes to verifying nuclear safeguarding treaties.

\subsubsection*{Addressing the Future Challenge of Neutrino Physics}
To address even the sub-set of challenging questions detailed above requires a diverse program of research, with no single experiment or approach being able to address the full spectrum of opportunities in neutrino science. Collectively, a portfolio of various approaches, experiments and infrastructures is therefore required to maximize the potential of the field. 

An optimal global program requires a diverse set of experimental efforts aiming at different exciting questions. This requires balancing of many aspects, including the scale of the experimental projects and infrastructures; ensuring a diversity of technologies and timescales; the maturity of research R\&D and experimental techniques; mechanisms for nurturing new ideas and approaches; sustaining a vibrant theory community; and ensuring the availability of long-term research infrastructures. Such an optimal program will provide substantial scientific synergies between projects which study the neutrino from different perspectives, and provide synergies with other fields of research through the development of multi-purpose or multi-disciplinary experiments.

By ensuring a portfolio of projects with different objectives, scale, timescale, technologies and complexity, the community can ensure a continuous flow of new knowledge and opportunities. An optimal balance would ensure investments to occur in projects that improve measurements of known physics parameters, and also those which are exploring new physics, with commensurately higher risks and rewards. Ensuring a diversity of technologies reduces risk to the program, as does maintaining a vibrant R\&D program to exploit new technologies and new concepts. Ensuring support for a strong theory community provides the connections between different approaches and perspectives, putting all studies into a consistent global framework, whilst also ensuring new directions are illuminated. Many infrastructures utilized in neutrino research are large-scale, and therefore a balance also needs to be maintained between the local demands of operating these infrastructures and the ability to fully exploit their capabilities through international cooperation. Here one should also keep in mind that the same infrastructure helps also other fields. One example is low background conditions and measurement technologies required for both neutrino experiments and direct dark matter detection experiments.  

This optimization requires coordination between many stakeholders including the research community, national funding agencies and research platforms and institutes. Such coordination will be facilitated through theoretical advances which establish the physics connections between very different experimental methods or which even establish new directions with exciting opportunities. To illuminate the path forward, academic exchanges including conferences, workshops and activities such as this IUPAP-sponsored report are essential in order to refine and as much as possible agree on the scientific value and potential of research directions. Such a broad scientific process which includes the wide spectrum of neutrino physics is an essential basis for discussions and negotiations between national facilities and funding agencies. By evaluating the current status of the field and future potential opportunities, this report is intended to highlight potential areas of synergy and cooperation, especially those requiring international cooperation due to the scale, complexity, new idea/technology or location of research projects.

This report also highlights the potential direct societal benefits from research into neutrinos, either directly or through the technologies that are developed that find use in other fields such as medical imaging and national security.
Because neutrinos are so hard to detect, the field has already driven extensive and imaginative work in detection techniques, many  of which are synergistic with other fields of science.  In addition to these technological benefits, the training and development of highly skilled individuals whose skill-sets are applied more broadly is a core benefit -- this is facilitated due to the inspiring nature of neutrino research that attracts the next generation of researchers. It is thus important to ensure a healthy distribution of projects geographically to attract and educate early career researchers from many countries and regions.  \\

In conclusion, this report highlights  the strong, creative and dynamic neutrino science community addressing some of the most challenging questions in contemporary physics. Great progress has been made over the last few decades in understanding the intrinsic properties and interactions of the neutrino, and its influence on nuclear and particle physics, astronomy and cosmology. To address remaining challenges will require a coordinated and nimble global program of research, with a broad portfolio of experiments and theoretical approaches. International discussions and coordination will be essential to maximize synergies between communities and projects, using a science-driven approach to determine an optimal program and best use of resources. We hope this report can facilitate such discussion.

\clearpage
\bibliographystyle{JHEP}
\bibliography{references}

\providecommand{\href}[2]{#2}\begingroup\raggedright\begin{thebibliography}{100}

\bibitem{Vitagliano:2019yzm}
E.~Vitagliano, I.~Tamborra and G.~Raffelt, \emph{{Grand Unified Neutrino
  Spectrum at Earth: Sources and Spectral Components}},
  \href{https://doi.org/10.1103/RevModPhys.92.045006}{\emph{Rev. Mod. Phys.}
  {\bfseries 92} (2020) 45006}
  [\href{https://arxiv.org/abs/1910.11878}{{\ttfamily 1910.11878}}].

\bibitem{Kopeikin:2004cn}
V.~Kopeikin, L.~Mikaelyan and V.~Sinev, \emph{{Reactor as a source of
  antineutrinos: Thermal fission energy}},
  \href{https://doi.org/10.1134/1.1811196}{\emph{Phys. Atom. Nucl.} {\bfseries
  67} (2004) 1892} [\href{https://arxiv.org/abs/hep-ph/0410100}{{\ttfamily
  hep-ph/0410100}}].

\bibitem{Ma:2012bm}
X.B.~Ma, W.L.~Zhong, L.Z.~Wang, Y.X.~Chen and J.~Cao, \emph{{Improved
  calculation of the energy release in neutron-induced fission}},
  \href{https://doi.org/10.1103/PhysRevC.88.014605}{\emph{Phys. Rev. C}
  {\bfseries 88} (2013) 014605}
  [\href{https://arxiv.org/abs/1212.6625}{{\ttfamily 1212.6625}}].

\bibitem{Huber:2011wv}
P.~Huber, \emph{{On the determination of anti-neutrino spectra from nuclear
  reactors}}, \href{https://doi.org/10.1103/PhysRevC.85.029901}{\emph{Phys.
  Rev. C} {\bfseries 84} (2011) 024617}
  [\href{https://arxiv.org/abs/1106.0687}{{\ttfamily 1106.0687}}].

\bibitem{Mueller:2011nm}
T.~Mueller et~al., \emph{{Improved Predictions of Reactor Antineutrino
  Spectra}}, \href{https://doi.org/10.1103/PhysRevC.83.054615}{\emph{Phys. Rev.
  C} {\bfseries 83} (2011) 054615}
  [\href{https://arxiv.org/abs/1101.2663}{{\ttfamily 1101.2663}}].

\bibitem{Vogel:2015wua}
P.~Vogel, L.~Wen and C.~Zhang, \emph{{Neutrino Oscillation Studies with
  Reactors}}, \href{https://doi.org/10.1038/ncomms7935}{\emph{Nature Commun.}
  {\bfseries 6} (2015) 6935}
  [\href{https://arxiv.org/abs/1503.01059}{{\ttfamily 1503.01059}}].

\bibitem{Bemporad:2001qy}
C.~Bemporad, G.~Gratta and P.~Vogel, \emph{{Reactor Based Neutrino Oscillation
  Experiments}}, \href{https://doi.org/10.1103/RevModPhys.74.297}{\emph{Rev.
  Mod. Phys.} {\bfseries 74} (2002) 297}
  [\href{https://arxiv.org/abs/hep-ph/0107277}{{\ttfamily hep-ph/0107277}}].

\bibitem{An:2012eh}
{\scshape Daya Bay} collaboration, \emph{{Observation of electron-antineutrino
  disappearance at Daya Bay}},
  \href{https://doi.org/10.1103/PhysRevLett.108.171803}{\emph{Phys. Rev. Lett.}
  {\bfseries 108} (2012) 171803}
  [\href{https://arxiv.org/abs/1203.1669}{{\ttfamily 1203.1669}}].

\bibitem{Ahn:2012nd}
{\scshape RENO} collaboration, \emph{{Observation of Reactor Electron
  Antineutrino Disappearance in the RENO Experiment}},
  \href{https://doi.org/10.1103/PhysRevLett.108.191802}{\emph{Phys. Rev. Lett.}
  {\bfseries 108} (2012) 191802}
  [\href{https://arxiv.org/abs/1204.0626}{{\ttfamily 1204.0626}}].

\bibitem{Catanesi:2013fxa}
{\scshape T2K} collaboration, \emph{{T2K Results and Perspectives}},
  \href{https://doi.org/10.1016/j.nuclphysbps.2013.04.074}{\emph{Nucl. Phys. B
  Proc. Suppl.} {\bfseries 237-238} (2013) 129}.

\bibitem{Adamson:2011qu}
{\scshape MINOS} collaboration, \emph{{Improved search for muon-neutrino to
  electron-neutrino oscillations in MINOS}},
  \href{https://doi.org/10.1103/PhysRevLett.107.181802}{\emph{Phys. Rev. Lett.}
  {\bfseries 107} (2011) 181802}
  [\href{https://arxiv.org/abs/1108.0015}{{\ttfamily 1108.0015}}].

\bibitem{Abe:2011fz}
{\scshape Double Chooz} collaboration, \emph{{Indication of Reactor
  $\bar{\nu}_e$ Disappearance in the Double Chooz Experiment}},
  \href{https://doi.org/10.1103/PhysRevLett.108.131801}{\emph{Phys. Rev. Lett.}
  {\bfseries 108} (2012) 131801}
  [\href{https://arxiv.org/abs/1112.6353}{{\ttfamily 1112.6353}}].

\bibitem{Adey:2018qct}
{\scshape Daya Bay} collaboration, \emph{{Improved Measurement of the Reactor
  Antineutrino Flux at Daya Bay}},
  \href{https://doi.org/10.1103/PhysRevD.100.052004}{\emph{Phys. Rev. D}
  {\bfseries 100} (2019) 052004}
  [\href{https://arxiv.org/abs/1808.10836}{{\ttfamily 1808.10836}}].

\bibitem{DoubleChooz:2019qbj}
{\scshape Double Chooz} collaboration, \emph{{Double Chooz $\theta_{13}$
  measurement via total neutron capture detection}},
  \href{https://doi.org/10.1038/s41567-020-0831-y}{\emph{Nature Phys.}
  {\bfseries 16} (2020) 558}
  [\href{https://arxiv.org/abs/1901.09445}{{\ttfamily 1901.09445}}].

\bibitem{RENO:2020Nu}
{\scshape RENO} collaboration, \emph{{Recent Results from RENO Experiment}},
  \href{https://doi.org/https://doi.org/10.5281/zenodo.4123573}{\emph{Talk at
  Neutrino 2020\!\!} }.

\bibitem{An:2017osx}
{\scshape Daya Bay} collaboration, \emph{{Evolution of the Reactor Antineutrino
  Flux and Spectrum at Daya Bay}},
  \href{https://doi.org/10.1103/PhysRevLett.118.251801}{\emph{Phys. Rev. Lett.}
  {\bfseries 118} (2017) 251801}
  [\href{https://arxiv.org/abs/1704.01082}{{\ttfamily 1704.01082}}].

\bibitem{RENO:2018pwo}
{\scshape RENO} collaboration, \emph{{Fuel-composition dependent reactor
  antineutrino yield at RENO}},
  \href{https://doi.org/10.1103/PhysRevLett.122.232501}{\emph{Phys. Rev. Lett.}
  {\bfseries 122} (2019) 232501}
  [\href{https://arxiv.org/abs/1806.00574}{{\ttfamily 1806.00574}}].

\bibitem{Hayen:2018uyg}
L.~Hayen, J.~Kostensalo, N.~Severijns and J.~Suhonen, \emph{{First forbidden
  transitions in the reactor anomaly}},
  \href{https://doi.org/10.1103/PhysRevC.100.054323}{\emph{Phys. Rev. C}
  {\bfseries 100} (2019) 054323}
  [\href{https://arxiv.org/abs/1805.12259}{{\ttfamily 1805.12259}}].

\bibitem{Estienne:2019ujo}
M.~Estienne et~al., \emph{{Updated Summation Model: An Improved Agreement with
  the Daya Bay Antineutrino Fluxes}},
  \href{https://doi.org/10.1103/PhysRevLett.123.022502}{\emph{Phys. Rev. Lett.}
  {\bfseries 123} (2019) 022502}
  [\href{https://arxiv.org/abs/1904.09358}{{\ttfamily 1904.09358}}].

\bibitem{Kopeikin:2021ugh}
V.~Kopeikin, M.~Skorokhvatov and O.~Titov, \emph{{Reevaluating reactor
  antineutrino spectra with new measurements of the ratio between $^{235}$U and
  $^{239}$Pu $\beta$ spectra}},
  \href{https://arxiv.org/abs/2103.01684}{{\ttfamily 2103.01684}}.

\bibitem{Seo:2014xei}
{\scshape RENO} collaboration, \emph{{New Results from RENO and The 5 MeV
  Excess}}, \href{https://doi.org/10.1063/1.4915563}{\emph{AIP Conf. Proc.}
  {\bfseries 1666} (2015) 080002}
  [\href{https://arxiv.org/abs/1410.7987}{{\ttfamily 1410.7987}}].

\bibitem{Abe:2014bwa}
{\scshape Double Chooz} collaboration, \emph{{Improved measurements of the
  neutrino mixing angle $\theta_{13}$ with the Double Chooz detector}},
  \href{https://doi.org/10.1007/JHEP02(2015)074}{\emph{JHEP} {\bfseries 10}
  (2014) 086} [\href{https://arxiv.org/abs/1406.7763}{{\ttfamily 1406.7763}}].

\bibitem{AC}
A.~Cabrera, ``{Double Chooz III: First results}.''
  \url{https://indico.ijclab.in2p3.fr/event/2454/attachments/4626/5585/DCIIILAL_Anatael_140522.pdf}.

\bibitem{An:2015nua}
{\scshape Daya Bay} collaboration, \emph{{Measurement of the Reactor
  Antineutrino Flux and Spectrum at Daya Bay}},
  \href{https://doi.org/10.1103/PhysRevLett.116.061801,
  10.1103/PhysRevLett.118.099902}{\emph{Phys. Rev. Lett.} {\bfseries 116}
  (2016) 061801} [\href{https://arxiv.org/abs/1508.04233}{{\ttfamily
  1508.04233}}].

\bibitem{Adey:2019ywk}
{\scshape Daya Bay} collaboration, \emph{{Extraction of the $^{235}$U and
  $^{239}$Pu Antineutrino Spectra at Daya Bay}},
  \href{https://doi.org/10.1103/PhysRevLett.123.111801}{\emph{Phys. Rev. Lett.}
  {\bfseries 123} (2019) 111801}
  [\href{https://arxiv.org/abs/1904.07812}{{\ttfamily 1904.07812}}].

\bibitem{Andriamirado:2020erz}
{\scshape PROSPECT} collaboration, \emph{{Improved Short-Baseline Neutrino
  Oscillation Search and Energy Spectrum Measurement with the PROSPECT
  Experiment at HFIR}},  \href{https://arxiv.org/abs/2006.11210}{{\ttfamily
  2006.11210}}.

\bibitem{AlmazanMolina:2020jlh}
{\scshape STEREO} collaboration, \emph{{First antineutrino energy spectrum from
  $^{235}$U fissions with the STEREO detector at ILL}},
  \href{https://doi.org/10.1088/1361-6471/abd37a}{\emph{J. Phys. G} {\bfseries
  48} (2021) 075107} [\href{https://arxiv.org/abs/2010.01876}{{\ttfamily
  2010.01876}}].

\bibitem{Hayes:2015yka}
A.~Hayes, J.~Friar, G.~Garvey, D.~Ibeling, G.~Jungman, T.~Kawano et~al.,
  \emph{{Possible origins and implications of the shoulder in reactor neutrino
  spectra}}, \href{https://doi.org/10.1103/PhysRevD.92.033015}{\emph{Phys. Rev.
  D} {\bfseries 92} (2015) 033015}
  [\href{https://arxiv.org/abs/1506.00583}{{\ttfamily 1506.00583}}].

\bibitem{junotaocdr}
{\scshape JUNO} collaboration, A.~Abusleme et~al., \emph{{TAO Conceptual Design
  Report}},  2020.

\bibitem{Schwartz:1960hg}
M.~Schwartz, \emph{{Feasibility of using high-energy neutrinos to study the
  weak interactions}},
  \href{https://doi.org/10.1103/PhysRevLett.4.306}{\emph{Phys. Rev. Lett.}
  {\bfseries 4} (1960) 306}.

\bibitem{Dore:2018ldz}
U.~Dore, P.~Loverre and L.~Ludovici, \emph{{History of accelerator neutrino
  beams}}, \href{https://doi.org/10.1140/epjh/e2019-90032-x}{\emph{Eur. Phys.
  J. H} {\bfseries 44} (2019) 271}
  [\href{https://arxiv.org/abs/1805.01373}{{\ttfamily 1805.01373}}].

\bibitem{Adamson:2014vgd}
{\scshape MINOS} collaboration, \emph{{Combined analysis of $\nu_{\mu}$
  disappearance and $\nu_{\mu} \rightarrow \nu_{e}$ appearance in MINOS using
  accelerator and atmospheric neutrinos}},
  \href{https://doi.org/10.1103/PhysRevLett.112.191801}{\emph{Phys. Rev. Lett.}
  {\bfseries 112} (2014) 191801}
  [\href{https://arxiv.org/abs/1403.0867}{{\ttfamily 1403.0867}}].

\bibitem{Abe:2020iop}
{\scshape T2K} collaboration, \emph{{Measurements of $\overline{\nu}_{\mu}$ and
  $\overline{\nu}_{\mu} + \nu_{\mu}$ charged-current cross-sections without
  detected pions or protons on water and hydrocarbon at a mean anti-neutrino
  energy of 0.86 GeV}}, \href{https://doi.org/10.1093/ptep/ptab014}{\emph{PTEP}
  {\bfseries 2021} (2021) 043C01}
  [\href{https://arxiv.org/abs/2004.13989}{{\ttfamily 2004.13989}}].

\bibitem{Naples:2003fe}
{\scshape NuTeV} collaboration, \emph{{High energy neutrino scattering results
  from NuTeV}},
  \href{https://doi.org/10.1016/S0920-5632(03)01314-8}{\emph{Nucl. Phys. B
  Proc. Suppl.} {\bfseries 118} (2003) 164}.

\bibitem{Muether:2013gxa}
M.~Muether, \emph{{NOvA: Current Status and Future Reach}},
  \href{https://doi.org/10.1016/j.nuclphysbps.2013.04.075}{\emph{Nucl. Phys. B
  Proc. Suppl.} {\bfseries 237-238} (2013) 135}.

\bibitem{PhysRevD.87.012001}
{\scshape T2K Collaboration} collaboration, \emph{T2k neutrino flux
  prediction}, \href{https://doi.org/10.1103/PhysRevD.87.012001}{\emph{Phys.
  Rev. D} {\bfseries 87} (2013) 012001}.

\bibitem{Aguilar-Arevalo:2018ylq}
{\scshape MiniBooNE} collaboration, \emph{{First Measurement of Monoenergetic
  Muon Neutrino Charged Current Interactions}},
  \href{https://doi.org/10.1103/PhysRevLett.120.141802}{\emph{Phys. Rev. Lett.}
  {\bfseries 120} (2018) 141802}
  [\href{https://arxiv.org/abs/1801.03848}{{\ttfamily 1801.03848}}].

\bibitem{Albright:2000xi}
C.~Albright et~al., \emph{{Physics at a neutrino factory}},
  \href{https://arxiv.org/abs/hep-ex/0008064}{{\ttfamily hep-ex/0008064}}.

\bibitem{DeRujula:1998umv}
A.~De~Rujula, M.~Gavela and P.~Hernandez, \emph{{Neutrino oscillation physics
  with a neutrino factory}},
  \href{https://doi.org/10.1016/S0550-3213(99)00070-X}{\emph{Nucl. Phys. B}
  {\bfseries 547} (1999) 21}
  [\href{https://arxiv.org/abs/hep-ph/9811390}{{\ttfamily hep-ph/9811390}}].

\bibitem{Choubey:2011zzq}
{\scshape IDS-NF} collaboration, \emph{{International Design Study for the
  Neutrino Factory, Interim Design Report}},
  \href{https://arxiv.org/abs/1112.2853}{{\ttfamily 1112.2853}}.

\bibitem{Adey:2013pio}
{\scshape nuSTORM} collaboration, \emph{{nuSTORM - Neutrinos from STORed Muons:
  Proposal to the Fermilab PAC}},
  \href{https://arxiv.org/abs/1308.6822}{{\ttfamily 1308.6822}}.

\bibitem{Ahdida:2020whw}
C.~Ahdida et~al., \emph{{nuSTORM at CERN: Feasibility Study}}, .

\bibitem{Huber:2005jk}
P.~Huber, M.~Lindner, M.~Rolinec and W.~Winter, \emph{{Physics and optimization
  of beta-beams: From low to very high gamma}},
  \href{https://doi.org/10.1103/PhysRevD.73.053002}{\emph{Phys. Rev. D}
  {\bfseries 73} (2006) 053002}
  [\href{https://arxiv.org/abs/hep-ph/0506237}{{\ttfamily hep-ph/0506237}}].

\bibitem{Benedikt:2011za}
M.~Benedikt et~al., \emph{{Conceptual design report for a Beta-Beam facility}},
  \href{https://doi.org/10.1140/epja/i2011-11024-5}{\emph{Eur. Phys. J. A}
  {\bfseries 47} (2011) 24}.

\bibitem{Bungau:2012ys}
A.~Bungau et~al., \emph{{Proposal for an Electron Antineutrino Disappearance
  Search Using High-Rate $^{8}$Li Production and Decay}},
  \href{https://doi.org/10.1103/PhysRevLett.109.141802}{\emph{Phys. Rev. Lett.}
  {\bfseries 109} (2012) 141802}
  [\href{https://arxiv.org/abs/1205.4419}{{\ttfamily 1205.4419}}].

\bibitem{Conrad:2013sqa}
J.M.~Conrad, M.H.~Shaevitz, I.~Shimizu, J.~Spitz, M.~Toups and L.~Winslow,
  \emph{{Precision $\bar{\nu}_e$-electron scattering measurements with IsoDAR
  to search for new physics}},
  \href{https://doi.org/10.1103/PhysRevD.89.072010}{\emph{Phys. Rev. D}
  {\bfseries 89} (2014) 072010}
  [\href{https://arxiv.org/abs/1307.5081}{{\ttfamily 1307.5081}}].

\bibitem{Vitagliano:2017odj}
E.~Vitagliano, J.~Redondo and G.~Raffelt, \emph{{Solar neutrino flux at keV
  energies}}, \href{https://doi.org/10.1088/1475-7516/2017/12/010}{\emph{JCAP}
  {\bfseries 12} (2017) 010}
  [\href{https://arxiv.org/abs/1708.02248}{{\ttfamily 1708.02248}}].

\bibitem{Bahcall:2002ng}
J.N.~Bahcall, \emph{{Solar models: An Historical overview}},
  \href{https://doi.org/10.1016/S0920-5632(03)01306-9}{\emph{AAPPS Bull.}
  {\bfseries 12} (2002) 12}
  [\href{https://arxiv.org/abs/astro-ph/0209080}{{\ttfamily
  astro-ph/0209080}}].

\bibitem{Agostini:2018uly}
{\scshape Borexino} collaboration, \emph{{Comprehensive measurement of
  $pp$-chain solar neutrinos}},
  \href{https://doi.org/10.1038/s41586-018-0624-y}{\emph{Nature} {\bfseries
  562} (2018) 505}.

\bibitem{Gando:2014wjd}
{\scshape KamLAND} collaboration, \emph{{$^7$Be Solar Neutrino Measurement with
  KamLAND}}, \href{https://doi.org/10.1103/PhysRevC.92.055808}{\emph{Phys. Rev.
  C} {\bfseries 92} (2015) 055808}
  [\href{https://arxiv.org/abs/1405.6190}{{\ttfamily 1405.6190}}].

\bibitem{SuperK:nu2020}
Y.~Nakajima, \emph{{Recent results and future prospects from
  Super-Kamiokande}},
  \href{https://doi.org/https://doi.org/10.5281/zenodo.3959639}{\emph{Talk at
  Neutrino 2020\!\!} }.

\bibitem{Aharmim:2011vm}
{\scshape SNO} collaboration, \emph{{Combined Analysis of all Three Phases of
  Solar Neutrino Data from the Sudbury Neutrino Observatory}},
  \href{https://doi.org/10.1103/PhysRevC.88.025501}{\emph{Phys. Rev.}
  {\bfseries C88} (2013) 025501}
  [\href{https://arxiv.org/abs/1109.0763}{{\ttfamily 1109.0763}}].

\bibitem{SNO+B8}
{\scshape SNO+} collaboration, \emph{{Measurement of the $^8$B solar neutrino
  flux in SNO+ with very low backgrounds}},
  \href{https://doi.org/10.1103/PhysRevD.99.012012}{\emph{Phys. Rev. D}
  {\bfseries 99} (2019) 012012}
  [\href{https://arxiv.org/abs/1812.03355}{{\ttfamily 1812.03355}}].

\bibitem{Aharmim_2006}
{\scshape SNO} collaboration, \emph{{A Search for Neutrinos from the Solar hep
  Reaction and the Diffuse Supernova Neutrino Background with the Sudbury
  Neutrino Observatory}},
  \href{https://doi.org/10.1086/508768}{\emph{Astrophys. J.} {\bfseries 653}
  (2006) 1545} [\href{https://arxiv.org/abs/hep-ex/0607010}{{\ttfamily
  hep-ex/0607010}}].

\bibitem{PhysRevD.73.112001}
{\scshape Super-Kamiokande} collaboration, \emph{{Solar neutrino measurements
  in super-Kamiokande-I}},
  \href{https://doi.org/10.1103/PhysRevD.73.112001}{\emph{Phys. Rev. D}
  {\bfseries 73} (2006) 112001}
  [\href{https://arxiv.org/abs/hep-ex/0508053}{{\ttfamily hep-ex/0508053}}].

\bibitem{B8BX}
{\scshape Borexino} collaboration, \emph{{Improved measurement of $^8$B solar
  neutrinos with 1.5\,kt$\cdot$y of Borexino exposure}},
  \href{https://doi.org/10.1103/PhysRevD.101.062001}{\emph{Phys. Rev. D}
  {\bfseries 101} (2020) 062001}
  [\href{https://arxiv.org/abs/1709.00756}{{\ttfamily 1709.00756}}].

\bibitem{BXCNO}
{\scshape Borexino} collaboration, \emph{{Experimental evidence of neutrinos
  produced in the CNO fusion cycle in the Sun}}, {\emph{Nature} {\bfseries 587}
  (2020) 577–} [\href{https://arxiv.org/abs/2006.15115}{{\ttfamily
  2006.15115}}].

\bibitem{Vinyoles_2017}
N.~Vinyoles, A.M.~Serenelli, F.L.~Villante, S.~Basu, J.~Bergstr\"om,
  M.C.~Gonzalez-Garcia et~al., \emph{A new generation of standard solar
  models}, \href{https://doi.org/10.3847/1538-4357/835/2/202}{\emph{The
  Astrophysical Journal} {\bfseries 835} (2017) 202}.

\bibitem{RevModPhys.75.985}
R.~Davis, \emph{Nobel lecture: A half-century with solar neutrinos},
  \href{https://doi.org/10.1103/RevModPhys.75.985}{\emph{Rev. Mod. Phys.}
  {\bfseries 75} (2003) 985}.

\bibitem{ABDURASHITOV1994234}
{\scshape SAGE} collaboration, \emph{{Results from SAGE}},
  \href{https://doi.org/10.1016/0370-2693(94)90454-5}{\emph{Phys. Lett. B}
  {\bfseries 328} (1994) 234}.

\bibitem{ANSELMANN1992376}
{\scshape GALLEX} collaboration, \emph{{Solar neutrinos observed by GALLEX at
  Gran Sasso.}},
  \href{https://doi.org/10.1016/0370-2693(92)91521-A}{\emph{Phys. Lett. B}
  {\bfseries 285} (1992) 376}.

\bibitem{Abe:2016nxk}
{\scshape Super-Kamiokande} collaboration, \emph{{Solar Neutrino Measurements
  in Super-Kamiokande-IV}},
  \href{https://doi.org/10.1103/PhysRevD.94.052010}{\emph{Phys. Rev.}
  {\bfseries D94} (2016) 052010}
  [\href{https://arxiv.org/abs/1606.07538}{{\ttfamily 1606.07538}}].

\bibitem{Andringa:2015tza}
{\scshape SNO+} collaboration, \emph{{Current Status and Future Prospects of
  the SNO+ Experiment}}, \href{https://doi.org/10.1155/2016/6194250}{\emph{Adv.
  High Energy Phys.} {\bfseries 2016} (2016) 6194250}
  [\href{https://arxiv.org/abs/1508.05759}{{\ttfamily 1508.05759}}].

\bibitem{Beacom_2017}
{\scshape Jinping} collaboration, \emph{{Physics prospects of the Jinping
  neutrino experiment}},
  \href{https://doi.org/10.1088/1674-1137/41/2/023002}{\emph{Chin. Phys. C}
  {\bfseries 41} (2017) 023002}
  [\href{https://arxiv.org/abs/1602.01733}{{\ttfamily 1602.01733}}].

\bibitem{An:2015jdp}
{\scshape JUNO} collaboration, \emph{{Neutrino Physics with JUNO}},
  \href{https://doi.org/10.1088/0954-3899/43/3/030401}{\emph{J. Phys.}
  {\bfseries G43} (2016) 030401}
  [\href{https://arxiv.org/abs/1507.05613}{{\ttfamily 1507.05613}}].

\bibitem{JUNO_B8}
{\scshape JUNO} collaboration, \emph{{Feasibility and physics potential of
  detecting $^8$B solar neutrinos at JUNO}},
  \href{https://doi.org/10.1088/1674-1137/abd92a}{\emph{Chin. Phys. C}
  {\bfseries 45} (2021) 023004}
  [\href{https://arxiv.org/abs/2006.11760}{{\ttfamily 2006.11760}}].

\bibitem{Yano:2020aap}
{\scshape Hyper-Kamiokande Proto} collaboration, \emph{{Solar neutrino physics
  at Hyper-Kamiokande}}, \href{https://doi.org/10.22323/1.358.1037}{\emph{PoS}
  {\bfseries ICRC2019} (2020) 1037}.

\bibitem{askins2019theia}
{\scshape Theia} collaboration, \emph{{THEIA: an advanced optical neutrino
  detector}}, \href{https://doi.org/10.1140/epjc/s10052-020-7977-8}{\emph{Eur.
  Phys. J. C} {\bfseries 80} (2020) 416}
  [\href{https://arxiv.org/abs/1911.03501}{{\ttfamily 1911.03501}}].

\bibitem{DS20k}
{\scshape DarkSide-20k} collaboration, \emph{{DarkSide-20k: A 20 tonne
  two-phase LAr TPC for direct dark matter detection at LNGS}},
  \href{https://doi.org/10.1140/epjp/i2018-11973-4}{\emph{Eur. Phys. J. Plus}
  {\bfseries 133} (2018) 131}
  [\href{https://arxiv.org/abs/1707.08145}{{\ttfamily 1707.08145}}].

\bibitem{Franco_2016}
D.~Franco et~al., \emph{{Solar neutrino detection in a large volume
  double-phase liquid argon experiment}},
  \href{https://doi.org/10.1088/1475-7516/2016/08/017}{\emph{JCAP} {\bfseries
  08} (2016) 017} [\href{https://arxiv.org/abs/1510.04196}{{\ttfamily
  1510.04196}}].

\bibitem{Aalbers:2016jon}
{\scshape DARWIN} collaboration, \emph{{DARWIN: towards the ultimate dark
  matter detector}},
  \href{https://doi.org/10.1088/1475-7516/2016/11/017}{\emph{JCAP} {\bfseries
  1611} (2016) 017} [\href{https://arxiv.org/abs/1606.07001}{{\ttfamily
  1606.07001}}].

\bibitem{Aalbers:2020gsn}
{\scshape DARWIN} collaboration, \emph{{Solar neutrino detection sensitivity in
  DARWIN via electron scattering}},
  \href{https://doi.org/10.1140/epjc/s10052-020-08602-7}{\emph{Eur. Phys. J. C}
  {\bfseries 80} (2020) 1133}
  [\href{https://arxiv.org/abs/2006.03114}{{\ttfamily 2006.03114}}].

\bibitem{Bionta:1987qt}
R.M.~Bionta et~al., \emph{{Observation of a Neutrino Burst in Coincidence with
  Supernova SN 1987a in the Large Magellanic Cloud}},
  \href{https://doi.org/10.1103/PhysRevLett.58.1494}{\emph{Phys. Rev. Lett.}
  {\bfseries 58} (1987) 1494}.

\bibitem{Hirata:1987hu}
{\scshape Kamiokande-II} collaboration, \emph{{Observation of a Neutrino Burst
  from the Supernova SN 1987a}},
  \href{https://doi.org/10.1103/PhysRevLett.58.1490}{\emph{Phys. Rev. Lett.}
  {\bfseries 58} (1987) 1490}.

\bibitem{Alekseev:1987ej}
E.N.~Alekseev, L.N.~Alekseeva, V.I.~Volchenko and I.V.~Krivosheina,
  \emph{{Possible Detection of a Neutrino Signal on 23 February 1987 at the
  Baksan Underground Scintillation Telescope of the Institute of Nuclear
  Research}}, {\emph{JETP Lett.} {\bfseries 45} (1987) 589}.

\bibitem{Bethe:1984ux}
H.A.~Bethe and J.R.~Wilson, \emph{{Revival of a stalled supernova shock by
  neutrino heating}}, \href{https://doi.org/10.1086/163343}{\emph{Astrophys.
  J.} {\bfseries 295} (1985) 14}.

\bibitem{Mezzacappa:2020oyq}
A.~Mezzacappa, E.~Endeve, O.E.B.~Messer and S.W.~Bruenn, \emph{{Physical,
  numerical, and computational challenges of modeling neutrino transport in
  core-collapse supernovae}},
  \href{https://arxiv.org/abs/2010.09013}{{\ttfamily 2010.09013}}.

\bibitem{Lin:2019wwm}
Z.~Lin, C.~Lunardini, M.~Zanolin, K.~Kotake and C.~Richardson,
  \emph{{Detectability of standing accretion shock instabilities activity in
  supernova neutrino signals}},
  \href{https://doi.org/10.1103/PhysRevD.101.123028}{\emph{Phys. Rev. D}
  {\bfseries 101} (2020) 123028}
  [\href{https://arxiv.org/abs/1911.10656}{{\ttfamily 1911.10656}}].

\bibitem{Li:2020ujl}
S.W.~Li, L.F.~Roberts and J.F.~Beacom, \emph{{Exciting Prospects for Detecting
  Late-Time Neutrinos from Core-Collapse Supernovae}},
  \href{https://doi.org/10.1103/PhysRevD.103.023016}{\emph{Phys. Rev. D}
  {\bfseries 103} (2021) 023016}
  [\href{https://arxiv.org/abs/2008.04340}{{\ttfamily 2008.04340}}].

\bibitem{Mirizzi:2015eza}
A.~Mirizzi, I.~Tamborra, H.-T.~Janka, N.~Saviano, K.~Scholberg, R.~Bollig
  et~al., \emph{{Supernova Neutrinos: Production, Oscillations and Detection}},
  \href{https://doi.org/10.1393/ncr/i2016-10120-8}{\emph{Riv. Nuovo Cim.}
  {\bfseries 39} (2016) 1} [\href{https://arxiv.org/abs/1508.00785}{{\ttfamily
  1508.00785}}].

\bibitem{Duan:2010bg}
H.~Duan, G.M.~Fuller and Y.-Z.~Qian, \emph{{Collective Neutrino Oscillations}},
  \href{https://doi.org/10.1146/annurev.nucl.012809.104524}{\emph{Ann. Rev.
  Nucl. Part. Sci.} {\bfseries 60} (2010) 569}
  [\href{https://arxiv.org/abs/1001.2799}{{\ttfamily 1001.2799}}].

\bibitem{Tamborra:2020cul}
I.~Tamborra and S.~Shalgar, \emph{{New Developments in Flavor Evolution of a
  Dense Neutrino Gas}},  \href{https://arxiv.org/abs/2011.01948}{{\ttfamily
  2011.01948}}.

\bibitem{Chakraborty:2016yeg}
S.~Chakraborty, R.~Hansen, I.~Izaguirre and G.~Raffelt, \emph{{Collective
  neutrino flavor conversion: Recent developments}},
  \href{https://doi.org/10.1016/j.nuclphysb.2016.02.012}{\emph{Nucl. Phys. B}
  {\bfseries 908} (2016) 366}
  [\href{https://arxiv.org/abs/1602.02766}{{\ttfamily 1602.02766}}].

\bibitem{Scholberg:2017czd}
K.~Scholberg, \emph{{Supernova Signatures of Neutrino Mass Ordering}},
  \href{https://doi.org/10.1088/1361-6471/aa97be}{\emph{J. Phys. G} {\bfseries
  45} (2018) 014002} [\href{https://arxiv.org/abs/1707.06384}{{\ttfamily
  1707.06384}}].

\bibitem{Schramm:1990pf}
D.N.~Schramm and J.W.~Truran, \emph{{New physics from Supernova SN1987A}},
  \href{https://doi.org/10.1016/0370-1573(90)90020-3}{\emph{Phys. Rept.}
  {\bfseries 189} (1990) 89}.

\bibitem{Wright:2016gar}
W.P.~Wright, J.P.~Kneller, S.T.~Ohlmann, F.K.~Roepke, K.~Scholberg and
  I.R.~Seitenzahl, \emph{{Neutrinos from type Ia supernovae: The
  gravitationally confined detonation scenario}},
  \href{https://doi.org/10.1103/PhysRevD.95.043006}{\emph{Phys. Rev.}
  {\bfseries D95} (2017) 043006}
  [\href{https://arxiv.org/abs/1609.07403}{{\ttfamily 1609.07403}}].

\bibitem{Wright:2016xma}
W.P.~Wright, G.~Nagaraj, J.P.~Kneller, K.~Scholberg and I.R.~Seitenzahl,
  \emph{{Neutrinos from type Ia supernovae: The deflagration-to-detonation
  transition scenario}},
  \href{https://doi.org/10.1103/PhysRevD.94.025026}{\emph{Phys. Rev.}
  {\bfseries D94} (2016) 025026}
  [\href{https://arxiv.org/abs/1605.01408}{{\ttfamily 1605.01408}}].

\bibitem{Scholberg:2012id}
K.~Scholberg, \emph{{Supernova Neutrino Detection}},
  \href{https://doi.org/10.1146/annurev-nucl-102711-095006}{\emph{Ann. Rev.
  Nucl. Part. Sci.} {\bfseries 62} (2012) 81}
  [\href{https://arxiv.org/abs/1205.6003}{{\ttfamily 1205.6003}}].

\bibitem{Eguchi:2002dm}
{\scshape KamLAND} collaboration, \emph{{First results from KamLAND: Evidence
  for reactor anti-neutrino disappearance}},
  \href{https://doi.org/10.1103/PhysRevLett.90.021802}{\emph{Phys. Rev. Lett.}
  {\bfseries 90} (2003) 021802}
  [\href{https://arxiv.org/abs/hep-ex/0212021}{{\ttfamily hep-ex/0212021}}].

\bibitem{Agafonova:2014leu}
{\scshape LVD} collaboration, \emph{{Implication for the Core-collapse
  Supernova Rate From 21 Years of Data of the Large Volume Detector}},
  \href{https://doi.org/10.1088/0004-637X/802/1/47}{\emph{Astrophys. J.}
  {\bfseries 802} (2015) 47} [\href{https://arxiv.org/abs/1411.1709}{{\ttfamily
  1411.1709}}].

\bibitem{Monzani:2006jg}
M.E.~Monzani, \emph{{Supernova neutrino detection in Borexino}},
  \href{https://doi.org/10.1393/ncc/i2005-10230-2}{\emph{Nuovo Cim. C}
  {\bfseries 29} (2006) 269}.

\bibitem{Wei:2015qga}
H.~Wei, L.~Lebanowski, F.~Li, Z.~Wang and S.~Chen, \emph{{Design,
  characterization, and sensitivity of the supernova trigger system at Daya
  Bay}},
  \href{https://doi.org/10.1016/j.astropartphys.2015.10.011}{\emph{Astropart.
  Phys.} {\bfseries 75} (2016) 38}
  [\href{https://arxiv.org/abs/1505.02501}{{\ttfamily 1505.02501}}].

\bibitem{Ikeda:2007sa}
{\scshape Super-Kamiokande} collaboration, \emph{{Search for Supernova Neutrino
  Bursts at Super-Kamiokande}},
  \href{https://doi.org/10.1086/521547}{\emph{Astrophys. J.} {\bfseries 669}
  (2007) 519} [\href{https://arxiv.org/abs/0706.2283}{{\ttfamily 0706.2283}}].

\bibitem{Abe:2016waf}
{\scshape Super-Kamiokande} collaboration, \emph{{Real-Time Supernova Neutrino
  Burst Monitor at Super-Kamiokande}},
  \href{https://doi.org/10.1016/j.astropartphys.2016.04.003}{\emph{Astropart.
  Phys.} {\bfseries 81} (2016) 39}
  [\href{https://arxiv.org/abs/1601.04778}{{\ttfamily 1601.04778}}].

\bibitem{Abbasi:2011ss}
{\scshape IceCube} collaboration, \emph{{IceCube Sensitivity for Low-Energy
  Neutrinos from Nearby Supernovae}},
  \href{https://doi.org/10.1051/0004-6361/201117810e,
  10.1051/0004-6361/201117810}{\emph{Astron. Astrophys.} {\bfseries 535} (2011)
  A109} [\href{https://arxiv.org/abs/astro-ph/1108.0171}{{\ttfamily
  astro-ph/1108.0171}}].

\bibitem{Aartsen:2014njl}
{\scshape IceCube} collaboration, \emph{{IceCube-Gen2: A Vision for the Future
  of Neutrino Astronomy in Antarctica}},
  \href{https://arxiv.org/abs/1412.5106}{{\ttfamily 1412.5106}}.

\bibitem{Duba:2008zz}
C.A.~Duba et~al., \emph{{HALO: The helium and lead observatory for supernova
  neutrinos}}, \href{https://doi.org/10.1088/1742-6596/136/4/042077}{\emph{J.
  Phys. Conf. Ser.} {\bfseries 136} (2008) 042077}.

\bibitem{Abe:2018uyc}
{\scshape Hyper-Kamiokande} collaboration, \emph{{Hyper-Kamiokande Design
  Report}},  \href{https://arxiv.org/abs/1805.04163}{{\ttfamily 1805.04163}}.

\bibitem{Adrian-Martinez:2016fdl}
{\scshape KM3Net} collaboration, \emph{{Letter of intent for KM3NeT 2.0}},
  \href{https://doi.org/10.1088/0954-3899/43/8/084001}{\emph{J. Phys.}
  {\bfseries G43} (2016) 084001}
  [\href{https://arxiv.org/abs/1601.07459}{{\ttfamily 1601.07459}}].

\bibitem{Abi:2020lpk}
{\scshape DUNE} collaboration, \emph{{Supernova neutrino burst detection with
  the Deep Underground Neutrino Experiment}},
  \href{https://doi.org/10.1140/epjc/s10052-021-09166-w}{\emph{Eur. Phys. J. C}
  {\bfseries 81} (2021) 423}
  [\href{https://arxiv.org/abs/2008.06647}{{\ttfamily 2008.06647}}].

\bibitem{Abratenko:2020hfy}
{\scshape MicroBooNE} collaboration, \emph{{The continuous readout stream of
  the MicroBooNE liquid argon time projection chamber for detection of
  supernova burst neutrinos}},
  \href{https://doi.org/10.1088/1748-0221/16/02/P02008}{\emph{JINST} {\bfseries
  16} (2021) P02008} [\href{https://arxiv.org/abs/2008.13761}{{\ttfamily
  2008.13761}}].

\bibitem{Lang:2016zhv}
R.F.~Lang, C.~McCabe, S.~Reichard, M.~Selvi and I.~Tamborra, \emph{{Supernova
  neutrino physics with xenon dark matter detectors: A timely perspective}},
  \href{https://doi.org/10.1103/PhysRevD.94.103009}{\emph{Phys. Rev. D}
  {\bfseries 94} (2016) 103009}
  [\href{https://arxiv.org/abs/1606.09243}{{\ttfamily 1606.09243}}].

\bibitem{NOvA:2020dll}
{\scshape NOvA} collaboration, \emph{{Supernova neutrino detection in NOvA}},
  \href{https://doi.org/10.1088/1475-7516/2020/10/014}{\emph{JCAP} {\bfseries
  10} (2020) 014} [\href{https://arxiv.org/abs/2005.07155}{{\ttfamily
  2005.07155}}].

\bibitem{Antonioli:2004zb}
P.~Antonioli et~al., \emph{{SNEWS: The Supernova Early Warning System}},
  \href{https://doi.org/10.1088/1367-2630/6/1/114}{\emph{New J. Phys.}
  {\bfseries 6} (2004) 114}
  [\href{https://arxiv.org/abs/astro-ph/0406214}{{\ttfamily
  astro-ph/0406214}}].

\bibitem{Scholberg:2008fa}
K.~Scholberg, \emph{{The SuperNova Early Warning System}},
  \href{https://doi.org/10.1002/asna.200710934}{\emph{Astron. Nachr.}
  {\bfseries 329} (2008) 337}
  [\href{https://arxiv.org/abs/0803.0531}{{\ttfamily 0803.0531}}].

\bibitem{Beacom:1998fj}
J.F.~Beacom and P.~Vogel, \emph{{Can a supernova be located by its
  neutrinos?}}, \href{https://doi.org/10.1103/PhysRevD.60.033007}{\emph{Phys.
  Rev. D} {\bfseries 60} (1999) 033007}
  [\href{https://arxiv.org/abs/astro-ph/9811350}{{\ttfamily
  astro-ph/9811350}}].

\bibitem{Tomas:2001dh}
R.~Tomas, H.~Päs and J.W.F.~Valle, \emph{{Generalized bounds on Majoron -
  neutrino couplings}},
  \href{https://doi.org/10.1103/PhysRevD.64.095005}{\emph{Phys.\ Rev.\ D}
  {\bfseries 64} (2001) 095005}
  [\href{https://arxiv.org/abs/hep-ph/0103017}{{\ttfamily hep-ph/0103017}}].

\bibitem{Fischer:2015oma}
V.~Fischer et~al., \emph{{Prompt directional detection of galactic supernova by
  combining large liquid scintillator neutrino detectors}},
  \href{https://doi.org/10.1088/1475-7516/2015/08/032}{\emph{JCAP} {\bfseries
  08} (2015) 032} [\href{https://arxiv.org/abs/1504.05466}{{\ttfamily
  1504.05466}}].

\bibitem{Abi:2020evt}
{\scshape DUNE} collaboration, \emph{{Deep Underground Neutrino Experiment
  (DUNE), Far Detector Technical Design Report, Volume II DUNE Physics}},
  \href{https://arxiv.org/abs/2002.03005}{{\ttfamily 2002.03005}}.

\bibitem{Linzer:2019swe}
N.~Linzer and K.~Scholberg, \emph{{Triangulation Pointing to Core-Collapse
  Supernovae with Next-Generation Neutrino Detectors}},
  \href{https://doi.org/10.1103/PhysRevD.100.103005}{\emph{Phys. Rev. D}
  {\bfseries 100} (2019) 103005}
  [\href{https://arxiv.org/abs/1909.03151}{{\ttfamily 1909.03151}}].

\bibitem{Brdar:2018tce}
V.~Brdar and R.S.L.~Hansen, \emph{{IceCube Flavor Ratios with Identified
  Astrophysical Sources: Towards Improving New Physics Testability}},
  \href{https://doi.org/10.1088/1475-7516/2019/02/023}{\emph{JCAP} {\bfseries
  02} (2019) 023} [\href{https://arxiv.org/abs/1812.05541}{{\ttfamily
  1812.05541}}].

\bibitem{Brdar:2018zds}
V.~Brdar, M.~Lindner and X.-J.~Xu, \emph{{Neutrino astronomy with supernova
  neutrinos}}, \href{https://doi.org/10.1088/1475-7516/2018/04/025}{\emph{JCAP}
  {\bfseries 04} (2018) 025}
  [\href{https://arxiv.org/abs/1802.02577}{{\ttfamily 1802.02577}}].

\bibitem{Nakamura:2016kkl}
K.~Nakamura, S.~Horiuchi, M.~Tanaka, K.~Hayama, T.~Takiwaki and K.~Kotake,
  \emph{{Multimessenger signals of long-term core-collapse supernova
  simulations: synergetic observation strategies}},
  \href{https://doi.org/10.1093/mnras/stw1453}{\emph{Mon. Not. Roy. Astron.
  Soc.} {\bfseries 461} (2016) 3296}
  [\href{https://arxiv.org/abs/1602.03028}{{\ttfamily 1602.03028}}].

\bibitem{Kharusi:2020ovw}
{\scshape SNEWS} collaboration, \emph{{SNEWS 2.0: a next-generation supernova
  early warning system for multi-messenger astronomy}},
  \href{https://doi.org/10.1088/1367-2630/abde33}{\emph{New J. Phys.}
  {\bfseries 23} (2021) 031201}
  [\href{https://arxiv.org/abs/2011.00035}{{\ttfamily 2011.00035}}].

\bibitem{Beacom:2010kk}
J.F.~Beacom, \emph{{The Diffuse Supernova Neutrino Background}},
  \href{https://doi.org/10.1146/annurev.nucl.010909.083331}{\emph{Ann. Rev.
  Nucl. Part. Sci.} {\bfseries 60} (2010) 439}
  [\href{https://arxiv.org/abs/1004.3311}{{\ttfamily 1004.3311}}].

\bibitem{deGouvea:2020eqq}
A.~De~Gouv\^ea, I.~Martinez-Soler, Y.F.~Perez-Gonzalez and M.~Sen,
  \emph{{Fundamental physics with the diffuse supernova background neutrinos}},
  \href{https://doi.org/10.1103/PhysRevD.102.123012}{\emph{Phys. Rev. D}
  {\bfseries 102} (2020) 123012}
  [\href{https://arxiv.org/abs/2007.13748}{{\ttfamily 2007.13748}}].

\bibitem{Kresse:2020nto}
D.~Kresse, T.~Ertl and H.-T.~Janka, \emph{{Stellar Collapse Diversity and the
  Diffuse Supernova Neutrino Background}},
  \href{https://doi.org/10.3847/1538-4357/abd54e}{\emph{Astrophys. J.}
  {\bfseries 909} (2021) 169}
  [\href{https://arxiv.org/abs/2010.04728}{{\ttfamily 2010.04728}}].

\bibitem{Horiuchi:2020jnc}
S.~Horiuchi, T.~Kinugawa, T.~Takiwaki, K.~Takahashi and K.~Kotake,
  \emph{{Impact of binary interactions on the diffuse supernova neutrino
  background}}, \href{https://doi.org/10.1103/PhysRevD.103.043003}{\emph{Phys.
  Rev. D} {\bfseries 103} (2021) 043003}
  [\href{https://arxiv.org/abs/2012.08524}{{\ttfamily 2012.08524}}].

\bibitem{Moller:2018kpn}
K.~Moller, A.M.~Suliga, I.~Tamborra and P.B.~Denton, \emph{{Measuring the
  supernova unknowns at the next-generation neutrino telescopes through the
  diffuse neutrino background}},
  \href{https://doi.org/10.1088/1475-7516/2018/05/066}{\emph{JCAP} {\bfseries
  05} (2018) 066} [\href{https://arxiv.org/abs/1804.03157}{{\ttfamily
  1804.03157}}].

\bibitem{Beacom:2003nk}
J.F.~Beacom and M.R.~Vagins, \emph{{GADZOOKS! Anti-neutrino spectroscopy with
  large water Cherenkov detectors}},
  \href{https://doi.org/10.1103/PhysRevLett.93.171101}{\emph{Phys. Rev. Lett.}
  {\bfseries 93} (2004) 171101}
  [\href{https://arxiv.org/abs/hep-ph/0309300}{{\ttfamily hep-ph/0309300}}].

\bibitem{Simpson:2018snj}
{\scshape Super-Kamiokande} collaboration, \emph{{Physics Potential of Super-K
  Gd}}, \href{https://doi.org/10.22323/1.340.0008}{\emph{PoS} {\bfseries
  ICHEP2018} (2019) 008}.

\bibitem{Gaisser:2002jj}
T.~Gaisser and M.~Honda, \emph{{Flux of atmospheric neutrinos}},
  \href{https://doi.org/10.1146/annurev.nucl.52.050102.090645}{\emph{Ann. Rev.
  Nucl. Part. Sci.} {\bfseries 52} (2002) 153}
  [\href{https://arxiv.org/abs/hep-ph/0203272}{{\ttfamily hep-ph/0203272}}].

\bibitem{Aglietta:1988be}
{\scshape NUSEX} collaboration, \emph{{Experimental study of atmospheric
  neutrino flux in the NUSEX experiment}},
  \href{https://doi.org/10.1209/0295-5075/8/7/005}{\emph{Europhys. Lett.}
  {\bfseries 8} (1989) 611}.

\bibitem{Berger:1990rd}
{\scshape Frejus} collaboration, \emph{{A Study of atmospheric neutrino
  oscillations in the FREJUS experiment}},
  \href{https://doi.org/10.1016/0370-2693(90)90150-5}{\emph{Phys. Lett. B}
  {\bfseries 245} (1990) 305}.

\bibitem{Casper:1990ac}
D.~Casper et~al., \emph{{Measurement of atmospheric neutrino composition with
  IMB-3}}, \href{https://doi.org/10.1103/PhysRevLett.66.2561}{\emph{Phys. Rev.
  Lett.} {\bfseries 66} (1991) 2561}.

\bibitem{Hirata:1992ku}
{\scshape Kamiokande-II} collaboration, \emph{{Observation of a small
  atmospheric muon-neutrino / electron-neutrino ratio in Kamiokande}},
  \href{https://doi.org/10.1016/0370-2693(92)90788-6}{\emph{Phys. Lett. B}
  {\bfseries 280} (1992) 146}.

\bibitem{Ashie:2005ik}
{\scshape Super-Kamiokande} collaboration, \emph{{A Measurement of atmospheric
  neutrino oscillation parameters by SUPER-KAMIOKANDE I}},
  \href{https://doi.org/10.1103/PhysRevD.71.112005}{\emph{Phys. Rev. D}
  {\bfseries 71} (2005) 112005}
  [\href{https://arxiv.org/abs/hep-ex/0501064}{{\ttfamily hep-ex/0501064}}].

\bibitem{ANTARES:2010izk}
{\scshape ANTARES} collaboration, \emph{{Zenith distribution and flux of
  atmospheric muons measured with the 5-line ANTARES detector}},
  \href{https://doi.org/10.1016/j.astropartphys.2010.07.001}{\emph{Astropart.
  Phys.} {\bfseries 34} (2010) 179}
  [\href{https://arxiv.org/abs/1007.1777}{{\ttfamily 1007.1777}}].

\bibitem{Richard:2015aua}
{\scshape Super-Kamiokande} collaboration, \emph{{Measurements of the
  atmospheric neutrino flux by Super-Kamiokande: energy spectra, geomagnetic
  effects, and solar modulation}},
  \href{https://doi.org/10.1103/PhysRevD.94.052001}{\emph{Phys. Rev. D}
  {\bfseries 94} (2016) 052001}
  [\href{https://arxiv.org/abs/1510.08127}{{\ttfamily 1510.08127}}].

\bibitem{Aartsen:2016xlq}
{\scshape IceCube} collaboration, \emph{{Observation and Characterization of a
  Cosmic Muon Neutrino Flux from the Northern Hemisphere using six years of
  IceCube data}},
  \href{https://doi.org/10.3847/0004-637X/833/1/3}{\emph{Astrophys. J.}
  {\bfseries 833} (2016) 3} [\href{https://arxiv.org/abs/1607.08006}{{\ttfamily
  1607.08006}}].

\bibitem{Ahdida:2020evc}
{\scshape SHiP} collaboration, \emph{{SND@LHC}},
  \href{https://arxiv.org/abs/2002.08722}{{\ttfamily 2002.08722}}.

\bibitem{ANTARES:2021cwc}
{\scshape ANTARES} collaboration, \emph{{Measurement of the atmospheric $\nu_e$
  and $\nu_\mu$ energy spectra with the ANTARES neutrino telescope}},
  \href{https://doi.org/10.1016/j.physletb.2021.136228}{\emph{Phys. Lett. B}
  {\bfseries 816} (2021) 136228}
  [\href{https://arxiv.org/abs/2101.12170}{{\ttfamily 2101.12170}}].

\bibitem{Aartsen:2019fau}
{\scshape IceCube} collaboration, \emph{{Time-Integrated Neutrino Source
  Searches with 10 Years of IceCube Data}},
  \href{https://doi.org/10.1103/PhysRevLett.124.051103}{\emph{Phys. Rev. Lett.}
  {\bfseries 124} (2020) 051103}
  [\href{https://arxiv.org/abs/1910.08488}{{\ttfamily 1910.08488}}].

\bibitem{Fukuda:1998mi}
{\scshape Super-Kamiokande} collaboration, \emph{{Evidence for oscillation of
  atmospheric neutrinos}},
  \href{https://doi.org/10.1103/PhysRevLett.81.1562}{\emph{Phys. Rev. Lett.}
  {\bfseries 81} (1998) 1562}
  [\href{https://arxiv.org/abs/hep-ex/9807003}{{\ttfamily hep-ex/9807003}}].

\bibitem{Li:2017dbe}
{\scshape Super-Kamiokande} collaboration, \emph{{Measurement of the tau
  neutrino cross section in atmospheric neutrino oscillations with
  Super-Kamiokande}},
  \href{https://doi.org/10.1103/PhysRevD.98.052006}{\emph{Phys. Rev.}
  {\bfseries D98} (2018) 052006}
  [\href{https://arxiv.org/abs/1711.09436}{{\ttfamily 1711.09436}}].

\bibitem{Aartsen:2019tjl}
{\scshape IceCube} collaboration, \emph{{Measurement of Atmospheric Tau
  Neutrino Appearance with IceCube DeepCore}},
  \href{https://doi.org/10.1103/PhysRevD.99.032007}{\emph{Phys. Rev.}
  {\bfseries D99} (2019) 032007}
  [\href{https://arxiv.org/abs/1901.05366}{{\ttfamily 1901.05366}}].

\bibitem{KM3NeT:2021ozk}
{\scshape KM3NeT} collaboration, \emph{{Determining the Neutrino Mass Ordering
  and Oscillation Parameters with KM3NeT/ORCA}},
  \href{https://arxiv.org/abs/2103.09885}{{\ttfamily 2103.09885}}.

\bibitem{Bezerra:2019dao}
{\scshape IceCube-Gen2, JUNO} collaboration, \emph{{Combined sensitivity to the
  neutrino mass ordering with JUNO, the IceCube Upgrade, and PINGU}},
  \href{https://doi.org/10.1103/PhysRevD.101.032006}{\emph{Phys. Rev.}
  {\bfseries D101} (2020) 032006}
  [\href{https://arxiv.org/abs/1911.06745}{{\ttfamily 1911.06745}}].

\bibitem{10.1088/978-0-7503-1369-8}
I.~Bartos and M.~Kowalski, \emph{Multimessenger Astronomy}, 2399-2891, IOP
  Publishing (2017),
  \href{https://doi.org/10.1088/978-0-7503-1369-8}{10.1088/978-0-7503-1369-8}.

\bibitem{Farzan:2008eg}
Y.~Farzan and A.Y.~Smirnov, \emph{{Coherence and oscillations of cosmic
  neutrinos}},
  \href{https://doi.org/10.1016/j.nuclphysb.2008.07.028}{\emph{Nucl. Phys. B}
  {\bfseries 805} (2008) 356}
  [\href{https://arxiv.org/abs/0803.0495}{{\ttfamily 0803.0495}}].

\bibitem{Katz:2011ke}
U.~Katz and C.~Spiering, \emph{{High-Energy Neutrino Astrophysics: Status and
  Perspectives}},
  \href{https://doi.org/10.1016/j.ppnp.2011.12.001}{\emph{Prog.Part.Nucl.Phys.}
  {\bfseries 67} (2012) 651} [\href{https://arxiv.org/abs/1111.0507}{{\ttfamily
  1111.0507}}].

\bibitem{BAIKAL:1997iok}
{\scshape BAIKAL} collaboration, \emph{{The Baikal underwater neutrino
  telescope: Design, performance and first results}},
  \href{https://doi.org/10.1016/S0927-6505(97)00022-4}{\emph{Astropart. Phys.}
  {\bfseries 7} (1997) 263}.

\bibitem{ANTARES:2011hfw}
{\scshape ANTARES} collaboration, \emph{{ANTARES: the first undersea neutrino
  telescope}}, \href{https://doi.org/10.1016/j.nima.2011.06.103}{\emph{Nucl.
  Instrum. Meth. A} {\bfseries 656} (2011) 11}
  [\href{https://arxiv.org/abs/1104.1607}{{\ttfamily 1104.1607}}].

\bibitem{Andres:1999hm}
E.~Andres et~al., \emph{{The AMANDA neutrino telescope: Principle of operation
  and first results}},
  \href{https://doi.org/10.1016/S0927-6505(99)00092-4}{\emph{Astropart. Phys.}
  {\bfseries 13} (2000) 1}
  [\href{https://arxiv.org/abs/astro-ph/9906203}{{\ttfamily
  astro-ph/9906203}}].

\bibitem{Aartsen:2017kpd}
{\scshape IceCube} collaboration, \emph{{Measurement of the multi-TeV neutrino
  cross section with IceCube using Earth absorption}},
  \href{https://doi.org/10.1038/nature24459}{\emph{Nature} {\bfseries 551}
  (2017) 596} [\href{https://arxiv.org/abs/1711.08119}{{\ttfamily
  1711.08119}}].

\bibitem{Aartsen:2013jdh}
{\scshape IceCube} collaboration, \emph{{Evidence for High-Energy
  Extraterrestrial Neutrinos at the IceCube Detector}},
  \href{https://doi.org/10.1126/science.1242856}{\emph{Science} {\bfseries 342}
  (2013) 1242856} [\href{https://arxiv.org/abs/astro-ph/1311.5238}{{\ttfamily
  astro-ph/1311.5238}}].

\bibitem{ANTARES:2020srt}
{\scshape ANTARES, IceCube} collaboration, \emph{{ANTARES and IceCube Combined
  Search for Neutrino Point-like and Extended Sources in the Southern Sky}},
  \href{https://doi.org/10.3847/1538-4357/ab7afb}{\emph{Astrophys. J.}
  {\bfseries 892} (2020) 92}
  [\href{https://arxiv.org/abs/2001.04412}{{\ttfamily 2001.04412}}].

\bibitem{IceCube:2018dnn}
{\scshape IceCube, Fermi-LAT, MAGIC, AGILE, ASAS-SN, HAWC, H.E.S.S., INTEGRAL,
  Kanata, Kiso, Kapteyn, Liverpool Telescope, Subaru, Swift NuSTAR, VERITAS,
  VLA/17B-403} collaboration, \emph{{Multimessenger observations of a flaring
  blazar coincident with high-energy neutrino IceCube-170922A}},
  \href{https://doi.org/10.1126/science.aat1378}{\emph{Science} {\bfseries 361}
  (2018) eaat1378} [\href{https://arxiv.org/abs/1807.08816}{{\ttfamily
  1807.08816}}].

\bibitem{IceCube:2018cha}
{\scshape IceCube} collaboration, \emph{{Neutrino emission from the direction
  of the blazar TXS 0506+056 prior to the IceCube-170922A alert}},
  \href{https://doi.org/10.1126/science.aat2890}{\emph{Science} {\bfseries 361}
  (2018) 147} [\href{https://arxiv.org/abs/1807.08794}{{\ttfamily
  1807.08794}}].

\bibitem{Aartsen:2019gxs}
{\scshape Fermi-LAT, ASAS-SN, IceCube} collaboration, \emph{{Investigation of
  two Fermi-LAT gamma-ray blazars coincident with high-energy neutrinos
  detected by IceCube}},  \href{https://arxiv.org/abs/1901.10806}{{\ttfamily
  1901.10806}}.

\bibitem{Glusenkamp:2015jca}
{\scshape IceCube} collaboration, \emph{{Analysis of the cumulative neutrino
  flux from Fermi-LAT blazar populations using 3 years of IceCube data}},
  \href{https://doi.org/10.1051/epjconf/201612105006}{\emph{EPJ Web Conf.}
  {\bfseries 121} (2016) 05006}
  [\href{https://arxiv.org/abs/astro-ph/1502.03104}{{\ttfamily
  astro-ph/1502.03104}}].

\bibitem{Halzen:2018iak}
F.~Halzen, A.~Kheirandish, T.~Weisgarber and S.P.~Wakely, \emph{{On the
  Neutrino Flares from the Direction of TXS 0506+056}},
  \href{https://doi.org/10.3847/2041-8213/ab0d27}{\emph{Astrophys. J.}
  {\bfseries 874} (2019) L9}
  [\href{https://arxiv.org/abs/1811.07439}{{\ttfamily 1811.07439}}].

\bibitem{Palladino:2018lov}
A.~Palladino, X.~Rodrigues, S.~Gao and W.~Winter, \emph{{Interpretation of the
  diffuse astrophysical neutrino flux in terms of the blazar sequence}},
  \href{https://doi.org/10.3847/1538-4357/aaf507}{\emph{Astrophys. J.}
  {\bfseries 871} (2019) 41}
  [\href{https://arxiv.org/abs/1806.04769}{{\ttfamily 1806.04769}}].

\bibitem{Neronov:2018wuo}
A.~Neronov and D.V.~Semikoz, \emph{{Self-consistent model of extragalactic
  neutrino flux from evolving blazar population}},
  \href{https://doi.org/10.31857/S0044451020080064}{\emph{J. Exp. Theor. Phys.}
  {\bfseries 131} (2020) 265}
  [\href{https://arxiv.org/abs/1811.06356}{{\ttfamily 1811.06356}}].

\bibitem{Plavin:2020emb}
A.~Plavin, Y.Y.~Kovalev, Y.A.~Kovalev and S.~Troitsky, \emph{{Observational
  Evidence for the Origin of High-energy Neutrinos in Parsec-scale Nuclei of
  Radio-bright Active Galaxies}},
  \href{https://doi.org/10.3847/1538-4357/ab86bd}{\emph{Astrophys. J.}
  {\bfseries 894} (2020) 101}
  [\href{https://arxiv.org/abs/2001.00930}{{\ttfamily 2001.00930}}].

\bibitem{Plavin:2020mkf}
A.V.~Plavin, Y.Y.~Kovalev, Y.A.~Kovalev and S.V.~Troitsky, \emph{{Directional
  Association of TeV to PeV Astrophysical Neutrinos with Radio Blazars}},
  \href{https://doi.org/10.3847/1538-4357/abceb8}{\emph{Astrophys. J.}
  {\bfseries 908} (2021) 157}
  [\href{https://arxiv.org/abs/2009.08914}{{\ttfamily 2009.08914}}].

\bibitem{Giommi:2020hbx}
P.~Giommi, T.~Glauch, P.~Padovani, E.~Resconi, A.~Turcati and Y.L.~Chang,
  \emph{{Dissecting the regions around IceCube high-energy neutrinos: growing
  evidence for the blazar connection}},
  \href{https://doi.org/10.1093/mnras/staa2082}{\emph{Mon. Not. Roy. Astron.
  Soc.} {\bfseries 497} (2020) 865}
  [\href{https://arxiv.org/abs/2001.09355}{{\ttfamily 2001.09355}}].

\bibitem{Stein:2020xhk}
R.~Stein et~al., \emph{{A high-energy neutrino coincident with a tidal
  disruption event}},
  \href{https://doi.org/10.1038/s41550-020-01295-8}{\emph{Nature Astron.}
  (2021) } [\href{https://arxiv.org/abs/2005.05340}{{\ttfamily 2005.05340}}].

\bibitem{Aartsen:2014aqy}
{\scshape IceCube} collaboration, \emph{{Search for Prompt Neutrino Emission
  from Gamma-Ray Bursts with IceCube}},
  \href{https://doi.org/10.1088/2041-8205/805/1/L5}{\emph{Astrophys. J.}
  {\bfseries 805} (2015) L5}
  [\href{https://arxiv.org/abs/astro-ph/1412.6510}{{\ttfamily
  astro-ph/1412.6510}}].

\bibitem{Niederhausen:2017mjk}
{\scshape IceCube} collaboration, \emph{{High Energy Astrophysical Neutrino
  Flux Measurement Using Neutrino-induced Cascades Observed in 4 Years of
  IceCube Data}}, \href{https://doi.org/10.22323/1.301.0968}{\emph{PoS}
  {\bfseries ICRC2017} (2018) 968}.

\bibitem{Aartsen:2014gkd}
{\scshape IceCube} collaboration, \emph{{Observation of High-Energy
  Astrophysical Neutrinos in Three Years of IceCube Data}},
  \href{https://doi.org/10.1103/PhysRevLett.113.101101}{\emph{Phys. Rev. Lett.}
  {\bfseries 113} (2014) 101101}
  [\href{https://arxiv.org/abs/astro-ph/1405.5303}{{\ttfamily
  astro-ph/1405.5303}}].

\bibitem{Aartsen:2015rwa}
{\scshape IceCube} collaboration, \emph{{Evidence for Astrophysical Muon
  Neutrinos from the Northern Sky with IceCube}},
  \href{https://doi.org/10.1103/PhysRevLett.115.081102}{\emph{Phys. Rev. Lett.}
  {\bfseries 115} (2015) 081102}
  [\href{https://arxiv.org/abs/astro-ph/1507.04005}{{\ttfamily
  astro-ph/1507.04005}}].

\bibitem{Stachurska:2019srh}
{\scshape IceCube} collaboration, \emph{{IceCube High Energy Starting Events at
  7.5 Years -- New Measurements of Flux and Flavor}},
  \href{https://doi.org/10.1051/epjconf/201920702005}{\emph{EPJ Web Conf.}
  {\bfseries 207} (2019) 02005}
  [\href{https://arxiv.org/abs/1905.04237}{{\ttfamily 1905.04237}}].

\bibitem{ANTARES:2017srd}
{\scshape ANTARES} collaboration, \emph{{All-flavor Search for a Diffuse Flux
  of Cosmic Neutrinos with Nine Years of ANTARES Data}},
  \href{https://doi.org/10.3847/2041-8213/aaa4f6}{\emph{Astrophys. J. Lett.}
  {\bfseries 853} (2018) L7}
  [\href{https://arxiv.org/abs/1711.07212}{{\ttfamily 1711.07212}}].

\bibitem{Abbasi:2020zmr}
{\scshape IceCube} collaboration, \emph{{Measurement of Astrophysical Tau
  Neutrinos in IceCube's High-Energy Starting Events}},
  \href{https://arxiv.org/abs/2011.03561}{{\ttfamily 2011.03561}}.

\bibitem{Glashow:1960zz}
S.L.~Glashow, \emph{{Resonant Scattering of Antineutrinos}},
  \href{https://doi.org/10.1103/PhysRev.118.316}{\emph{Phys.Rev.} {\bfseries
  118} (1960) 316}.

\bibitem{Aab:2019auo}
{\scshape Pierre Auger} collaboration, \emph{{Probing the origin of
  ultra-high-energy cosmic rays with neutrinos in the EeV energy range using
  the Pierre Auger Observatory}},
  \href{https://doi.org/10.1088/1475-7516/2019/10/022}{\emph{JCAP} {\bfseries
  1910} (2019) 022} [\href{https://arxiv.org/abs/1906.07422}{{\ttfamily
  1906.07422}}].

\bibitem{Gorham:2019guw}
{\scshape ANITA} collaboration, \emph{{Constraints on the ultrahigh-energy
  cosmic neutrino flux from the fourth flight of ANITA}},
  \href{https://doi.org/10.1103/PhysRevD.99.122001}{\emph{Phys. Rev. D}
  {\bfseries 99} (2019) 122001}
  [\href{https://arxiv.org/abs/1902.04005}{{\ttfamily 1902.04005}}].

\bibitem{Gorham:2018ydl}
{\scshape ANITA} collaboration, \emph{{Observation of an Unusual Upward-going
  Cosmic-ray-like Event in the Third Flight of ANITA}},
  \href{https://doi.org/10.1103/PhysRevLett.121.161102}{\emph{Phys. Rev. Lett.}
  {\bfseries 121} (2018) 161102}
  [\href{https://arxiv.org/abs/1803.05088}{{\ttfamily 1803.05088}}].

\bibitem{ARA:2019wcf}
{\scshape ARA} collaboration, \emph{{Constraints on the diffuse flux of
  ultrahigh energy neutrinos from four years of Askaryan Radio Array data in
  two stations}},
  \href{https://doi.org/10.1103/PhysRevD.102.043021}{\emph{Phys. Rev. D}
  {\bfseries 102} (2020) 043021}
  [\href{https://arxiv.org/abs/1912.00987}{{\ttfamily 1912.00987}}].

\bibitem{Anker:2019rzo}
{\scshape ARIANNA} collaboration, \emph{{A search for cosmogenic neutrinos with
  the ARIANNA test bed using 4.5 years of data}},
  \href{https://doi.org/10.1088/1475-7516/2020/03/053}{\emph{JCAP} {\bfseries
  03} (2020) 053} [\href{https://arxiv.org/abs/1909.00840}{{\ttfamily
  1909.00840}}].

\bibitem{Aartsen:2018vtx}
{\scshape IceCube} collaboration, \emph{{Differential limit on the
  extremely-high-energy cosmic neutrino flux in the presence of astrophysical
  background from nine years of IceCube data}},
  \href{https://doi.org/10.1103/PhysRevD.98.062003}{\emph{Phys.\ Rev.\ D}
  {\bfseries 98} (2018) 062003}
  [\href{https://arxiv.org/abs/1807.01820}{{\ttfamily 1807.01820}}].

\bibitem{Avrorin:2014vca}
A.D.~Avrorin et~al., \emph{{Sensitivity of the Baikal-GVD neutrino telescope to
  neutrino emission toward the center of the galactic dark matter halo}},
  \href{https://doi.org/10.1134/S0021364015050021}{\emph{JETP Lett.} {\bfseries
  101} (2015) 289} [\href{https://arxiv.org/abs/1412.3672}{{\ttfamily
  1412.3672}}].

\bibitem{Belolaptikov:2021a1}
I.~Belolaptikov,  et~al., \emph{{Neutrino Telescope in Lake Baikal: Present and
  Nearest Future}}, \href{https://doi.org/10.22323/1.395.0002}{\emph{PoS}
  {\bfseries ICRC2021} (2021) 002}.

\bibitem{Bailly:2021dxn}
N.~Bailly et~al., \emph{{Two-Year Optical Site Characterization for the Pacific
  Ocean Neutrino Experiment P-ONE in the Cascadia Basin}},
  \href{https://arxiv.org/abs/2108.04961}{{\ttfamily 2108.04961}}.

\bibitem{Agostini:2020aar}
{\scshape P-ONE} collaboration, \emph{{The Pacific Ocean Neutrino Experiment}},
  \href{https://doi.org/10.1038/s41550-020-1182-4}{\emph{Nature Astron.}
  {\bfseries 4} (2020) 913} [\href{https://arxiv.org/abs/2005.09493}{{\ttfamily
  2005.09493}}].

\bibitem{Anker:2020lre}
A.~Anker et~al., \emph{{White Paper: ARIANNA-200 high energy neutrino
  telescope}},  \href{https://arxiv.org/abs/2004.09841}{{\ttfamily
  2004.09841}}.

\bibitem{RNO-G:2020rmc}
{\scshape RNO-G} collaboration, \emph{{Design and Sensitivity of the Radio
  Neutrino Observatory in Greenland (RNO-G)}},
  \href{https://doi.org/10.1088/1748-0221/16/03/P03025}{\emph{JINST} {\bfseries
  16} (2021) P03025} [\href{https://arxiv.org/abs/2010.12279}{{\ttfamily
  2010.12279}}].

\bibitem{Olinto:2019euf}
A.V.~Olinto et~al., \emph{{The POEMMA (Probe of Extreme Multi-Messenger
  Astrophysics) mission}},
  \href{https://doi.org/10.22323/1.358.0378}{\emph{PoS} {\bfseries ICRC2019}
  (2020) 378} [\href{https://arxiv.org/abs/1909.09466}{{\ttfamily
  1909.09466}}].

\bibitem{Otte:2019knb}
A.N.~Otte, A.M.~Brown, A.D.~Falcone, M.~Mariotti and I.~Taboada,
  \emph{{Trinity: An Air-Shower Imaging System for the Detection of Ultrahigh
  Energy Neutrinos}}, \href{https://doi.org/10.22323/1.358.0976}{\emph{PoS}
  {\bfseries ICRC2019} (2020) 976}
  [\href{https://arxiv.org/abs/1907.08732}{{\ttfamily 1907.08732}}].

\bibitem{Alvarez-Muniz:2018bhp}
{\scshape GRAND} collaboration, \emph{{The Giant Radio Array for Neutrino
  Detection (GRAND): Science and Design}},
  \href{https://doi.org/10.1007/s11433-018-9385-7}{\emph{Sci. China Phys. Mech.
  Astron.} {\bfseries 63} (2020) 219501}
  [\href{https://arxiv.org/abs/1810.09994}{{\ttfamily 1810.09994}}].

\bibitem{Wissel:2020sec}
S.~Wissel et~al., \emph{{Prospects for high-elevation radio detection of
  \ensuremath{>}100 PeV tau neutrinos}},
  \href{https://doi.org/10.1088/1475-7516/2020/11/065}{\emph{JCAP} {\bfseries
  11} (2020) 065} [\href{https://arxiv.org/abs/2004.12718}{{\ttfamily
  2004.12718}}].

\bibitem{Prohira:2019glh}
S.~Prohira et~al., \emph{{Observation of Radar Echoes From High-Energy Particle
  Cascades}}, \href{https://doi.org/10.1103/PhysRevLett.124.091101}{\emph{Phys.
  Rev. Lett.} {\bfseries 124} (2020) 091101}
  [\href{https://arxiv.org/abs/1910.12830}{{\ttfamily 1910.12830}}].

\bibitem{TheIceCube-Gen2:2016cap}
{\scshape IceCube} collaboration, \emph{{PINGU: A Vision for Neutrino and
  Particle Physics at the South Pole}},
  \href{https://arxiv.org/abs/hep-ex/1607.02671}{{\ttfamily
  hep-ex/1607.02671}}.

\bibitem{Ishihara:2019aao}
{\scshape IceCube} collaboration, \emph{{The IceCube Upgrade -- Design and
  Science Goals}}, \href{https://doi.org/10.22323/1.358.1031}{\emph{PoS}
  {\bfseries ICRC2019} (2020) 1031}
  [\href{https://arxiv.org/abs/1908.09441}{{\ttfamily 1908.09441}}].

\bibitem{Aartsen:2014oha}
{\scshape IceCube PINGU} collaboration, \emph{{Letter of Intent: The Precision
  IceCube Next Generation Upgrade (PINGU)}},
  \href{https://arxiv.org/abs/physics/1401.2046}{{\ttfamily
  physics/1401.2046}}.

\bibitem{Haack:2017dxi}
{\scshape IceCube} collaboration, \emph{{A measurement of the diffuse
  astrophysical muon neutrino flux using eight years of IceCube data.}},
  \href{https://doi.org/10.22323/1.301.1005}{\emph{PoS} {\bfseries ICRC2017}
  (2018) 1005}.

\bibitem{Kopper:2015vzf}
{\scshape IceCube} collaboration, \emph{{Observation of Astrophysical Neutrinos
  in Four Years of IceCube Data}},
  \href{https://doi.org/10.22323/1.236.1081}{\emph{PoS} {\bfseries ICRC2015}
  (2016) 1081}.

\bibitem{Kopper:2017zzm}
{\scshape IceCube} collaboration, \emph{{Observation of Astrophysical Neutrinos
  in Six Years of IceCube Data}},
  \href{https://doi.org/10.22323/1.301.0981}{\emph{PoS} {\bfseries ICRC2017}
  (2018) 981}.

\bibitem{Smith:2012eu}
M.W.E.~Smith et~al., \emph{{The Astrophysical Multimessenger Observatory
  Network (AMON)}},
  \href{https://doi.org/10.1016/j.astropartphys.2013.03.003}{\emph{Astropart.
  Phys.} {\bfseries 45} (2013) 56}
  [\href{https://arxiv.org/abs/1211.5602}{{\ttfamily 1211.5602}}].

\bibitem{fiorentini07}
G.~Fiorentini et~al., \emph{{Geoneutrinos and Earth's Interior}}, {\emph{Phys.
  Rep.} {\bfseries 453} (2007) 117}.

\bibitem{mao19}
X.~Mao, R.~Han and Y.-F.~Li, \emph{{Non-negligible oscillation effects in the
  crustal geoneutrino calculations}},
  \href{https://doi.org/10.1103/PhysRevD.100.113009}{\emph{Phys. Rev. D}
  {\bfseries 100} (2019) 113009}
  [\href{https://arxiv.org/abs/1911.12302}{{\ttfamily 1911.12302}}].

\bibitem{eder66}
G.~Eder, \emph{{Terrestrial neutrinos}},
  \href{https://doi.org/https://doi.org/10.1016/0029-5582(66)90903-5}{\emph{Nucl.
  Phys.} {\bfseries 78} (1966) 657}.

\bibitem{krauss84}
L.M.~Krauss, S.L.~Glashow and D.N.~Schramm, \emph{{Anti-neutrinos Astronomy and
  Geophysics}}, \href{https://doi.org/10.1038/310191a0}{\emph{Nature}
  {\bfseries 310} (1984) 191}.

\bibitem{raghavan98}
R.S.~Raghavan, S.~Schonert, S.~Enomoto, J.~Shirai, F.~Suekane and A.~Suzuki,
  \emph{{Measuring the global radioactivity in the earth by multidetector
  anti-neutrino spectroscopy}},
  \href{https://doi.org/10.1103/PhysRevLett.80.635}{\emph{Phys. Rev. Lett.}
  {\bfseries 80} (1998) 635}.

\bibitem{roth98}
C.G.~Rothschild, M.C.~Chen and F.P.~Calaprice, \emph{{Anti-neutrino geophysics
  with liquid scintillator detectors}},
  \href{https://doi.org/10.1029/98GL50667}{\emph{Geophys. Res. Lett.}
  {\bfseries 25} (1998) 1083}
  [\href{https://arxiv.org/abs/nucl-ex/9710001}{{\ttfamily nucl-ex/9710001}}].

\bibitem{mantovani04}
F.~Mantovani, L.~Carmignani, G.~Fiorentini and M.~Lissia, \emph{{Anti-neutrinos
  from the earth: The Reference model and its uncertainties}},
  \href{https://doi.org/10.1103/PhysRevD.69.013001}{\emph{Phys. Rev. D}
  {\bfseries 69} (2004) 013001}
  [\href{https://arxiv.org/abs/hep-ph/0309013}{{\ttfamily hep-ph/0309013}}].

\bibitem{huang13}
Y.~Huang et~al., \emph{{A reference Earth model for the heat producing elements
  and associated geoneutrino flux}},
  \href{https://doi.org/10.1002/ggge.20129}{\emph{Geochem., Geophys., Geosyst.}
  {\bfseries 14} (2013) 2003}
  [\href{https://arxiv.org/abs/1301.0365}{{\ttfamily 1301.0365}}].

\bibitem{sleep13}
N.H.~Sleep, D.K.~Bird and M.T.~Rosing, \emph{{Biological effects on the source
  of geoneutrinos}},
  \href{https://doi.org/10.1142/S0217751X13300470}{\emph{Int. J. Mod. Phys. A}
  {\bfseries 28} (2013) 1330047}.

\bibitem{crust1.0}
G.~Laske, A.~Dziewonski and G.~Masters, ``{The Reference Earth Model
  Website}.'' \url{https://igppweb.ucsd.edu/~gabi/rem.html}.

\bibitem{takeuchi19}
N.~Takeuchi, K.~Ueki, T.~Iizuka, J.~Nagao, A.~Tanaka, S.~Enomoto et~al.,
  \emph{{Stochastic modeling of 3-D compositional distribution in the crust
  with Bayesian inference and application to geoneutrino observation in
  Japan}}, \href{https://doi.org/10.1016/j.pepi.2019.01.002}{\emph{Phys. Earth
  Planet. Interiors} {\bfseries 288} (2019) 37}
  [\href{https://arxiv.org/abs/1901.01358}{{\ttfamily 1901.01358}}].

\bibitem{sanshiro_spec}
S.~Enomoto, ``{Geoneutrino Spectra and Luminosity}.''
  \url{https://www.awa.tohoku.ac.jp/~sanshiro/research/geoneutrino/spectrum/}.

\bibitem{strumia03}
A.~Strumia and F.~Vissani, \emph{{Precise quasielastic neutrino/nucleon
  cross-section}},
  \href{https://doi.org/10.1016/S0370-2693(03)00616-6}{\emph{Phys. Lett. B}
  {\bfseries 564} (2003) 42}
  [\href{https://arxiv.org/abs/astro-ph/0302055}{{\ttfamily
  astro-ph/0302055}}].

\bibitem{watanabe19}
H.~Watanabe, ``{Geoneutrino measurement with KamLAND}.''
  \url{https://indico.cern.ch/event/825708/contributions/3552210/attachments/1930535/3197332/HirokoWatanabe_NGS2019.pdf}.

\bibitem{agostini20}
{\scshape Borexino} collaboration, \emph{{Comprehensive geoneutrino analysis
  with Borexino}},
  \href{https://doi.org/10.1103/PhysRevD.101.012009}{\emph{Phys. Rev. D}
  {\bfseries 101} (2020) 012009}
  [\href{https://arxiv.org/abs/1909.02257}{{\ttfamily 1909.02257}}].

\bibitem{fiorentini12}
G.~Fiorentini, G.L.~Fogli, E.~Lisi, F.~Mantovani and A.M.~Rotunno,
  \emph{{Mantle geoneutrinos in KamLAND and Borexino}},
  \href{https://doi.org/10.1103/PhysRevD.86.033004}{\emph{Phys. Rev. D}
  {\bfseries 86} (2012) 033004}
  [\href{https://arxiv.org/abs/1204.1923}{{\ttfamily 1204.1923}}].

\bibitem{Rocholl:1993}
A.~Rocholl and K.~Jochum, \emph{{Th, U and other trace elements in carbonaceous
  chondrites: Implications for the terrestrial and solar-system Th/U ratios}},
  \href{https://doi.org/10.1016/0012-821X(93)90132-S}{\emph{Earth Planet. Sci.
  Lett.} {\bfseries 117} (1993) 265}.

\bibitem{leyton17}
M.~Leyton, S.~Dye and J.~Monroe, \emph{{Exploring the hidden interior of the
  Earth with directional neutrino measurements}},
  \href{https://doi.org/10.1038/ncomms15989}{\emph{Nature Commun.} {\bfseries
  8} (2017) 15989} [\href{https://arxiv.org/abs/1710.06724}{{\ttfamily
  1710.06724}}].

\bibitem{chen06}
M.C.~Chen, \emph{{Geo-neutrinos in SNO+}},
  \href{https://doi.org/10.1007/s11038-006-9116-4}{\emph{Earth Moon Planets}
  {\bfseries 99} (2006) 221}.

\bibitem{barabanov17}
I.R.~Barabanov et~al., \emph{{Large-volume detector at the Baksan Neutrino
  Observatory for studies of natural neutrino fluxes for purposes of geo- and
  astrophysics}}, \href{https://doi.org/10.1134/S1063778817030036}{\emph{Phys.
  Atom. Nucl.} {\bfseries 80} (2017) 446}.

\bibitem{wan17}
L.~Wan, G.~Hussain, Z.~Wang and S.~Chen, \emph{{Geoneutrinos at Jinping: Flux
  prediction and oscillation analysis}},
  \href{https://doi.org/10.1103/PhysRevD.95.053001}{\emph{Phys. Rev. D}
  {\bfseries 95} (2017) 053001}
  [\href{https://arxiv.org/abs/1612.00133}{{\ttfamily 1612.00133}}].

\bibitem{enomoto07}
S.~Enomoto, E.~Ohtani, K.~Inoue and A.~Suzuki, \emph{{Neutrino geophysics with
  KamLAND and future prospects}},
  \href{https://arxiv.org/abs/hep-ph/0508049}{{\ttfamily hep-ph/0508049}}.

\bibitem{bonventre18}
R.~Bonventre and G.D.~Orebi~Gann, \emph{{Sensitivity of a low threshold
  directional detector to CNO-cycle solar neutrinos}},
  \href{https://doi.org/10.1140/epjc/s10052-018-5925-7}{\emph{Eur. Phys. J. C}
  {\bfseries 78} (2018) 435}
  [\href{https://arxiv.org/abs/1803.07109}{{\ttfamily 1803.07109}}].

\bibitem{Kaptanoglu:2018sus}
T.~Kaptanoglu, M.~Luo and J.~Klein, \emph{{Cherenkov and Scintillation Light
  Separation Using Wavelength in LAB Based Liquid Scintillator}},
  \href{https://doi.org/10.1088/1748-0221/14/05/T05001}{\emph{JINST} {\bfseries
  14} (2019) T05001} [\href{https://arxiv.org/abs/1811.11587}{{\ttfamily
  1811.11587}}].

\bibitem{Cabrera:2019kxi}
A.~Cabrera et~al., \emph{{Neutrino Physics with an Opaque Detector}},
  \href{https://arxiv.org/abs/1908.02859}{{\ttfamily 1908.02859}}.

\bibitem{Weinberg:1962zza}
S.~Weinberg, \emph{{Universal Neutrino Degeneracy}},
  \href{https://doi.org/10.1103/PhysRev.128.1457}{\emph{Phys.\ Rev.} {\bfseries
  128} (1962) 1457}.

\bibitem{Cocco:2007za}
A.G.~Cocco, G.~Mangano and M.~Messina, \emph{{Probing low energy neutrino
  backgrounds with neutrino capture on beta decaying nuclei}},
  \href{https://doi.org/10.1088/1475-7516/2007/06/015}{\emph{JCAP} {\bfseries
  06} (2007) 015} [\href{https://arxiv.org/abs/hep-ph/0703075}{{\ttfamily
  hep-ph/0703075}}].

\bibitem{Cheipesh:2021fmg}
Y.~Cheipesh, V.~Cheianov and A.~Boyarsky, \emph{{Heisenberg's uncertainty as a
  limiting factor for neutrino mass detection in $\beta$-decay}},
  \href{https://arxiv.org/abs/2101.10069}{{\ttfamily 2101.10069}}.

\bibitem{Nussinov:2021zrj}
S.~Nussinov and Z.~Nussinov, \emph{{Quantum Induced Broadening- A Challenge For
  Cosmic Neutrino Background Discovery}},
  \href{https://arxiv.org/abs/2108.03695}{{\ttfamily 2108.03695}}.

\bibitem{Long:2014zva}
A.J.~Long, C.~Lunardini and E.~Sabancilar, \emph{{Detecting non-relativistic
  cosmic neutrinos by capture on tritium: phenomenology and physics
  potential}}, \href{https://doi.org/10.1088/1475-7516/2014/08/038}{\emph{JCAP}
  {\bfseries 08} (2014) 038} [\href{https://arxiv.org/abs/1405.7654}{{\ttfamily
  1405.7654}}].

\bibitem{Ringwald:2004np}
A.~Ringwald and Y.Y.~Wong, \emph{{Gravitational clustering of relic neutrinos
  and implications for their detection}},
  \href{https://doi.org/10.1088/1475-7516/2004/12/005}{\emph{JCAP} {\bfseries
  12} (2004) 005} [\href{https://arxiv.org/abs/hep-ph/0408241}{{\ttfamily
  hep-ph/0408241}}].

\bibitem{betti2019neutrino}
{\scshape PTOLEMY} collaboration, \emph{{Neutrino physics with the PTOLEMY
  project: active neutrino properties and the light sterile case}},
  \href{https://doi.org/10.1088/1475-7516/2019/07/047}{\emph{JCAP} {\bfseries
  07} (2019) 047} [\href{https://arxiv.org/abs/1902.05508}{{\ttfamily
  1902.05508}}].

\bibitem{baracchini2018ptolemy}
{\scshape PTOLEMY} collaboration, \emph{{PTOLEMY: A Proposal for Thermal Relic
  Detection of Massive Neutrinos and Directional Detection of MeV Dark
  Matter}},  \href{https://arxiv.org/abs/1808.01892}{{\ttfamily 1808.01892}}.

\bibitem{lisanti2014measuring}
M.~Lisanti, B.R.~Safdi and C.G.~Tully, \emph{{Measuring Anisotropies in the
  Cosmic Neutrino Background}},
  \href{https://doi.org/10.1103/PhysRevD.90.073006}{\emph{Phys. Rev. D}
  {\bfseries 90} (2014) 073006}
  [\href{https://arxiv.org/abs/1407.0393}{{\ttfamily 1407.0393}}].

\bibitem{Akhmedov:2019oxm}
E.~Akhmedov, \emph{{Relic neutrino detection through angular correlations in
  inverse $\beta$-decay}},
  \href{https://doi.org/10.1088/1475-7516/2019/09/031}{\emph{JCAP} {\bfseries
  09} (2019) 031} [\href{https://arxiv.org/abs/1905.10207}{{\ttfamily
  1905.10207}}].

\bibitem{Pontecorvo:1957qd}
B.~Pontecorvo, \emph{Inverse beta processes and nonconservation of lepton
  charge}, {\emph{Sov. Phys. JETP} {\bfseries 7} (1958) 172}.

\bibitem{Maki:1962mu}
Z.~Maki, M.~Nakagawa and S.~Sakata, \emph{{Remarks on the unified model of
  elementary particles}}, \href{https://doi.org/10.1143/PTP.28.870}{\emph{Prog.
  Theor. Phys.} {\bfseries 28} (1962) 870}.

\bibitem{Pontecorvo:1967fh}
B.~Pontecorvo, \emph{Neutrino experiments and the question of leptonic-charge
  conservation}, {\emph{Sov. Phys. JETP} {\bfseries 26} (1968) 984}.

\bibitem{Cabibbo:1963yz}
N.~Cabibbo, \emph{{Unitary Symmetry and Leptonic Decays}},
  \href{https://doi.org/10.1103/PhysRevLett.10.531}{\emph{Phys. Rev. Lett.}
  {\bfseries 10} (1963) 531}.

\bibitem{Kobayashi:1973fv}
M.~Kobayashi and T.~Maskawa, \emph{{CP Violation in the Renormalizable Theory
  of Weak Interaction}}, \href{https://doi.org/10.1143/PTP.49.652}{\emph{Prog.
  Theor. Phys.} {\bfseries 49} (1973) 652}.

\bibitem{10.1093/ptep/ptaa104}
{\scshape Particle Data Group} collaboration, \emph{{Review of Particle
  Physics}}, \href{https://doi.org/10.1093/ptep/ptaa104}{\emph{PTEP} {\bfseries
  2020} (2020) 083C01}.

\bibitem{Ahmad:2001an}
{\scshape SNO} collaboration, \emph{{Measurement of the rate of $\nu_e+d \to
  p+p+e^-$ interactions produced by $^8B$ solar neutrinos at the Sudbury
  Neutrino Observatory}},
  \href{https://doi.org/10.1103/PhysRevLett.87.071301}{\emph{Phys. Rev. Lett.}
  {\bfseries 87} (2001) 071301}
  [\href{https://arxiv.org/abs/nucl-ex/0106015}{{\ttfamily nucl-ex/0106015}}].

\bibitem{Olive:2016xmw}
{\scshape Particle Data Group} collaboration, \emph{{Review of Particle
  Physics}}, \href{https://doi.org/10.1088/1674-1137/40/10/100001}{\emph{Chin.
  Phys.} {\bfseries C40} (2016) 100001}.

\bibitem{Schechter:1980gr}
J.~Schechter and J.W.F.~Valle, \emph{{Neutrino Masses in $SU(2) \times U(1)$
  Theories}}, \href{https://doi.org/10.1103/PhysRevD.22.2227}{\emph{Phys. Rev.}
  {\bfseries D22} (1980) 2227}.

\bibitem{Rodejohann:2011vc}
W.~Rodejohann and J.W.F.~Valle, \emph{{Symmetrical Parametrizations of the
  Lepton Mixing Matrix}},
  \href{https://doi.org/10.1103/PhysRevD.84.073011}{\emph{Phys. Rev. D}
  {\bfseries 84} (2011) 073011}
  [\href{https://arxiv.org/abs/1108.3484}{{\ttfamily 1108.3484}}].

\bibitem{Wolfenstein:1977ue}
L.~Wolfenstein, \emph{{Neutrino Oscillations in Matter}},
  \href{https://doi.org/10.1103/PhysRevD.17.2369}{\emph{Phys. Rev. D}
  {\bfseries 17} (1978) 2369}.

\bibitem{Mikheev:1986gs}
S.P.~Mikheyev and A.Y.~Smirnov, \emph{{Resonance Amplification of Oscillations
  in Matter and Spectroscopy of Solar Neutrinos}}, {\emph{Sov. J. Nucl. Phys.}
  {\bfseries 42} (1985) 913}.

\bibitem{Giunti:1997wq}
C.~Giunti and C.W.~Kim, \emph{{Coherence of neutrino oscillations in the wave
  packet approach}},
  \href{https://doi.org/10.1103/PhysRevD.58.017301}{\emph{Phys. Rev. D}
  {\bfseries 58} (1998) 017301}
  [\href{https://arxiv.org/abs/hep-ph/9711363}{{\ttfamily hep-ph/9711363}}].

\bibitem{Akhmedov:2009rb}
E.K.~Akhmedov and A.Y.~Smirnov, \emph{{Paradoxes of neutrino oscillations}},
  \href{https://doi.org/10.1134/S1063778809080122}{\emph{Phys. Atom. Nucl.}
  {\bfseries 72} (2009) 1363}
  [\href{https://arxiv.org/abs/0905.1903}{{\ttfamily 0905.1903}}].

\bibitem{Akhmedov:2010ms}
E.K.~Akhmedov and J.~Kopp, \emph{{Neutrino Oscillations: Quantum Mechanics vs.
  Quantum Field Theory}},
  \href{https://doi.org/10.1007/JHEP04(2010)008}{\emph{JHEP} {\bfseries 04}
  (2010) 008} [\href{https://arxiv.org/abs/1001.4815}{{\ttfamily 1001.4815}}].

\bibitem{An:2016pvi}
{\scshape Daya Bay} collaboration, \emph{{Study of the wave packet treatment of
  neutrino oscillation at Daya Bay}},
  \href{https://doi.org/10.1140/epjc/s10052-017-4970-y}{\emph{Eur. Phys. J. C}
  {\bfseries 77} (2017) 606}
  [\href{https://arxiv.org/abs/1608.01661}{{\ttfamily 1608.01661}}].

\bibitem{deGouvea:2020hfl}
A.~de~Gouvea, V.~de~Romeri and C.A.~Ternes, \emph{{Probing neutrino quantum
  decoherence at reactor experiments}},
  \href{https://doi.org/10.1007/JHEP08(2020)049}{\emph{JHEP} {\bfseries 08}
  (2020) 018} [\href{https://arxiv.org/abs/2005.03022}{{\ttfamily
  2005.03022}}].

\bibitem{Super-Kamiokande:2004orf}
{\scshape Super-Kamiokande} collaboration, \emph{{Evidence for an oscillatory
  signature in atmospheric neutrino oscillation}},
  \href{https://doi.org/10.1103/PhysRevLett.93.101801}{\emph{Phys. Rev. Lett.}
  {\bfseries 93} (2004) 101801}
  [\href{https://arxiv.org/abs/hep-ex/0404034}{{\ttfamily hep-ex/0404034}}].

\bibitem{Albert:2018mnz}
{\scshape ANTARES} collaboration, \emph{{Measuring the atmospheric neutrino
  oscillation parameters and constraining the 3+1 neutrino model with ten years
  of ANTARES data}}, \href{https://doi.org/10.1007/JHEP06(2019)113}{\emph{JHEP}
  {\bfseries 06} (2019) 113}
  [\href{https://arxiv.org/abs/1812.08650}{{\ttfamily 1812.08650}}].

\bibitem{Aartsen:2017nmd}
{\scshape IceCube} collaboration, \emph{{Measurement of Atmospheric Neutrino
  Oscillations at 6–56 GeV with IceCube DeepCore}},
  \href{https://doi.org/10.1103/PhysRevLett.120.071801}{\emph{Phys. Rev. Lett.}
  {\bfseries 120} (2018) 071801}
  [\href{https://arxiv.org/abs/1707.07081}{{\ttfamily 1707.07081}}].

\bibitem{Abe:2017aap}
{\scshape Super-Kamiokande} collaboration, \emph{{Atmospheric neutrino
  oscillation analysis with external constraints in Super-Kamiokande I-IV}},
  \href{https://doi.org/10.1103/PhysRevD.97.072001}{\emph{Phys. Rev.}
  {\bfseries D97} (2018) 072001}
  [\href{https://arxiv.org/abs/1710.09126}{{\ttfamily 1710.09126}}].

\bibitem{Abe:2019vii}
{\scshape T2K} collaboration, \emph{{Constraint on the matter--antimatter
  symmetry-violating phase in neutrino oscillations}},
  \href{https://doi.org/10.1038/s41586-020-2177-0}{\emph{Nature} {\bfseries
  580} (2020) 339} [\href{https://arxiv.org/abs/1910.03887}{{\ttfamily
  1910.03887}}].

\bibitem{Acero:2019ksn}
{\scshape NOvA} collaboration, \emph{{First Measurement of Neutrino Oscillation
  Parameters using Neutrinos and Antineutrinos by NOvA}},
  \href{https://doi.org/10.1103/PhysRevLett.123.151803}{\emph{Phys. Rev. Lett.}
  {\bfseries 123} (2019) 151803}
  [\href{https://arxiv.org/abs/1906.04907}{{\ttfamily 1906.04907}}].

\bibitem{Aartsen:2019eht}
{\scshape IceCube} collaboration, \emph{{Development of an analysis to probe
  the neutrino mass ordering with atmospheric neutrinos using three years of
  IceCube DeepCore data}},
  \href{https://doi.org/10.1140/epjc/s10052-019-7555-0}{\emph{Eur. Phys. J.}
  {\bfseries C80} (2020) 9} [\href{https://arxiv.org/abs/1902.07771}{{\ttfamily
  1902.07771}}].

\bibitem{Stuttard:2020zsj}
{\scshape IceCube} collaboration, \emph{{Neutrino oscillations and PMNS
  unitarity with IceCube/DeepCore and the IceCube Upgrade}},
  \href{https://doi.org/10.22323/1.369.0099}{\emph{PoS} {\bfseries NuFact2019}
  (2020) 099}.

\bibitem{KM3NeT:2021rkn}
{\scshape KM3NeT} collaboration, \emph{{Combined sensitivity of JUNO and
  KM3NeT/ORCA to the neutrino mass ordering}},
  \href{https://arxiv.org/abs/2108.06293}{{\ttfamily 2108.06293}}.

\bibitem{ICAL:2015stm}
{\scshape ICAL} collaboration, \emph{{Physics Potential of the ICAL detector at
  the India-based Neutrino Observatory (INO)}},
  \href{https://doi.org/10.1007/s12043-017-1373-4}{\emph{Pramana} {\bfseries
  88} (2017) 79} [\href{https://arxiv.org/abs/1505.07380}{{\ttfamily
  1505.07380}}].

\bibitem{Aartsen:2020iky}
{\scshape IceCube} collaboration, \emph{{eV-Scale Sterile Neutrino Search Using
  Eight Years of Atmospheric Muon Neutrino Data from the IceCube Neutrino
  Observatory}},
  \href{https://doi.org/10.1103/PhysRevLett.125.141801}{\emph{Phys. Rev. Lett.}
  {\bfseries 125} (2020) 141801}
  [\href{https://arxiv.org/abs/2005.12942}{{\ttfamily 2005.12942}}].

\bibitem{Abe:2014gda}
{\scshape Super-Kamiokande} collaboration, \emph{{Limits on sterile neutrino
  mixing using atmospheric neutrinos in Super-Kamiokande}},
  \href{https://doi.org/10.1103/PhysRevD.91.052019}{\emph{Phys. Rev. D}
  {\bfseries 91} (2015) 052019}
  [\href{https://arxiv.org/abs/1410.2008}{{\ttfamily 1410.2008}}].

\bibitem{Aartsen:2017ibm}
{\scshape IceCube} collaboration, \emph{{Neutrino Interferometry for
  High-Precision Tests of Lorentz Symmetry with IceCube}},
  \href{https://doi.org/10.1038/s41567-018-0172-2}{\emph{Nature Phys.}
  {\bfseries 14} (2018) 961}
  [\href{https://arxiv.org/abs/1709.03434}{{\ttfamily 1709.03434}}].

\bibitem{Abe:2014wla}
{\scshape Super-Kamiokande} collaboration, \emph{{Test of Lorentz invariance
  with atmospheric neutrinos}},
  \href{https://doi.org/10.1103/PhysRevD.91.052003}{\emph{Phys. Rev. D}
  {\bfseries 91} (2015) 052003}
  [\href{https://arxiv.org/abs/1410.4267}{{\ttfamily 1410.4267}}].

\bibitem{Aartsen:2017xtt}
{\scshape IceCube} collaboration, \emph{{Search for Nonstandard Neutrino
  Interactions with IceCube DeepCore}},
  \href{https://doi.org/10.1103/PhysRevD.97.072009}{\emph{Phys. Rev. D}
  {\bfseries 97} (2018) 072009}
  [\href{https://arxiv.org/abs/1709.07079}{{\ttfamily 1709.07079}}].

\bibitem{PhysRevD.84.113008}
{\scshape Super-Kamiokande} collaboration, \emph{{Study of Non-Standard
  Neutrino Interactions with Atmospheric Neutrino Data in Super-Kamiokande I
  and II}}, \href{https://doi.org/10.1103/PhysRevD.84.113008}{\emph{Phys. Rev.
  D} {\bfseries 84} (2011) 113008}
  [\href{https://arxiv.org/abs/1109.1889}{{\ttfamily 1109.1889}}].

\bibitem{Salvado:2016uqu}
J.~Salvado, O.~Mena, S.~Palomares-Ruiz and N.~Rius, \emph{{Non-standard
  interactions with high-energy atmospheric neutrinos at IceCube}},
  \href{https://doi.org/10.1007/JHEP01(2017)141}{\emph{JHEP} {\bfseries 01}
  (2017) 141} [\href{https://arxiv.org/abs/1609.03450}{{\ttfamily
  1609.03450}}].

\bibitem{Abi:2020kei}
{\scshape DUNE} collaboration, \emph{{Prospects for beyond the Standard Model
  physics searches at the Deep Underground Neutrino Experiment}},
  \href{https://doi.org/10.1140/epjc/s10052-021-09007-w}{\emph{Eur. Phys. J. C}
  {\bfseries 81} (2021) 322}
  [\href{https://arxiv.org/abs/2008.12769}{{\ttfamily 2008.12769}}].

\bibitem{PhysRevD.60.119905}
J.~Arafune, M.~Koike and J.~Sato, \emph{{CP violation and matter effect in long
  baseline neutrino oscillation experiments}},
  \href{https://doi.org/10.1103/PhysRevD.60.119905}{\emph{Phys. Rev. D}
  {\bfseries 56} (1997) 3093}
  [\href{https://arxiv.org/abs/hep-ph/9703351}{{\ttfamily hep-ph/9703351}}].

\bibitem{PhysRevD.17.2369}
L.~Wolfenstein, \emph{Neutrino oscillations in matter},
  \href{https://doi.org/10.1103/PhysRevD.17.2369}{\emph{Phys. Rev. D}
  {\bfseries 17} (1978) 2369}.

\bibitem{An:2013uza}
{\scshape Daya Bay} collaboration, \emph{{Improved Measurement of Electron
  Antineutrino Disappearance at Daya Bay}},
  \href{https://doi.org/10.1088/1674-1137/37/1/011001}{\emph{Chin. Phys.}
  {\bfseries C37} (2013) 011001}
  [\href{https://arxiv.org/abs/1210.6327}{{\ttfamily 1210.6327}}].

\bibitem{Bellini:2011yj}
{\scshape Borexino} collaboration, \emph{{Absence of day--night asymmetry of
  862 keV $^7$Be solar neutrino rate in Borexino and MSW oscillation
  parameters}},
  \href{https://doi.org/10.1016/j.physletb.2011.11.025}{\emph{Phys. Lett.}
  {\bfseries B707} (2012) 22}
  [\href{https://arxiv.org/abs/1104.2150}{{\ttfamily 1104.2150}}].

\bibitem{Petcov:2001sy}
S.T.~Petcov and M.~Piai, \emph{{The LMA MSW solution of the solar neutrino
  problem, inverted neutrino mass hierarchy and reactor neutrino experiments}},
  \href{https://doi.org/10.1016/S0370-2693(02)01591-5}{\emph{Phys. Lett. B}
  {\bfseries 533} (2002) 94}
  [\href{https://arxiv.org/abs/hep-ph/0112074}{{\ttfamily hep-ph/0112074}}].

\bibitem{Gando:2013nba}
{\scshape KamLAND} collaboration, \emph{{Reactor On-Off Antineutrino
  Measurement with KamLAND}},
  \href{https://doi.org/10.1103/PhysRevD.88.033001}{\emph{Phys. Rev.}
  {\bfseries D88} (2013) 033001}
  [\href{https://arxiv.org/abs/1303.4667}{{\ttfamily 1303.4667}}].

\bibitem{Apollonio:2002gd}
{\scshape CHOOZ} collaboration, \emph{{Search for neutrino oscillations on a
  long baseline at the CHOOZ nuclear power station}},
  \href{https://doi.org/10.1140/epjc/s2002-01127-9}{\emph{Eur. Phys. J.}
  {\bfseries C27} (2003) 331}
  [\href{https://arxiv.org/abs/hep-ex/0301017}{{\ttfamily hep-ex/0301017}}].

\bibitem{Boehm:2000vp}
F.~Boehm et~al., \emph{{Results from the Palo Verde neutrino oscillation
  experiment}}, \href{https://doi.org/10.1103/PhysRevD.62.072002}{\emph{Phys.
  Rev.} {\bfseries D62} (2000) 072002}
  [\href{https://arxiv.org/abs/hep-ex/0003022}{{\ttfamily hep-ex/0003022}}].

\bibitem{Abe:2012tg}
{\scshape Double Chooz} collaboration, \emph{{Reactor electron antineutrino
  disappearance in the Double Chooz experiment}},
  \href{https://doi.org/10.1103/PhysRevD.86.052008}{\emph{Phys. Rev. D}
  {\bfseries 86} (2012) 052008}
  [\href{https://arxiv.org/abs/1207.6632}{{\ttfamily 1207.6632}}].

\bibitem{Adey:2018zwh}
{\scshape Daya Bay} collaboration, \emph{{Measurement of the Electron
  Antineutrino Oscillation with 1958 Days of Operation at Daya Bay}},
  \href{https://doi.org/10.1103/PhysRevLett.121.241805}{\emph{Phys. Rev. Lett.}
  {\bfseries 121} (2018) 241805}
  [\href{https://arxiv.org/abs/1809.02261}{{\ttfamily 1809.02261}}].

\bibitem{DoubleChooz:nu2020}
T.S.~Bezerra, \emph{{New Results from the Double Chooz Experiment}},
  \href{https://doi.org/https://doi.org/10.5281/zenodo.3959541}{\emph{Talk at
  Neutrino 2020\!\!} }.

\bibitem{RENO:nu2020}
J.~Yoo, \emph{Reno},
  \href{https://doi.org/https://doi.org/10.5281/zenodo.3959697}{\emph{Talk at
  Neutrino 2020\!\!} }.

\bibitem{JUNO:2021vlw}
{\scshape JUNO} collaboration, \emph{{JUNO Physics and Detector}},
  \href{https://arxiv.org/abs/2104.02565}{{\ttfamily 2104.02565}}.

\bibitem{Zhan:2008id}
L.~Zhan, Y.~Wang, J.~Cao and L.~Wen, \emph{{Determination of the Neutrino Mass
  Hierarchy at an Intermediate Baseline}},
  \href{https://doi.org/10.1103/PhysRevD.78.111103}{\emph{Phys. Rev. D}
  {\bfseries 78} (2008) 111103}
  [\href{https://arxiv.org/abs/0807.3203}{{\ttfamily 0807.3203}}].

\bibitem{Zhan:2009rs}
L.~Zhan, Y.~Wang, J.~Cao and L.~Wen, \emph{{Experimental Requirements to
  Determine the Neutrino Mass Hierarchy Using Reactor Neutrinos}},
  \href{https://doi.org/10.1103/PhysRevD.79.073007}{\emph{Phys. Rev. D}
  {\bfseries 79} (2009) 073007}
  [\href{https://arxiv.org/abs/0901.2976}{{\ttfamily 0901.2976}}].

\bibitem{Li:2013zyd}
Y.-F.~Li, J.~Cao, Y.~Wang and L.~Zhan, \emph{{Unambiguous Determination of the
  Neutrino Mass Hierarchy Using Reactor Neutrinos}},
  \href{https://doi.org/10.1103/PhysRevD.88.013008}{\emph{Phys. Rev. D}
  {\bfseries 88} (2013) 013008}
  [\href{https://arxiv.org/abs/1303.6733}{{\ttfamily 1303.6733}}].

\bibitem{IceCube-Gen2:2019fet}
{\scshape IceCube-Gen2} collaboration, \emph{{Combined sensitivity to the
  neutrino mass ordering with JUNO, the IceCube Upgrade, and PINGU}},
  \href{https://doi.org/10.1103/PhysRevD.101.032006}{\emph{Phys. Rev. D}
  {\bfseries 101} (2020) 032006}
  [\href{https://arxiv.org/abs/1911.06745}{{\ttfamily 1911.06745}}].

\bibitem{Zyla:2020zbs}
{\scshape Particle Data Group} collaboration, \emph{{Review of Particle
  Physics}}, \href{https://doi.org/10.1093/ptep/ptaa104}{\emph{PTEP} {\bfseries
  2020} (2020) 083C01}.

\bibitem{junoprojection}
``Preliminary results provided by {JUNO}, to appear\!\!.''

\bibitem{Ellis:2020hus}
S.A.R.~Ellis, K.J.~Kelly and S.W.~Li, \emph{{Current and Future Neutrino
  Oscillation Constraints on Leptonic Unitarity}},
  \href{https://doi.org/10.1007/JHEP12(2020)068}{\emph{JHEP} {\bfseries 12}
  (2020) 068} [\href{https://arxiv.org/abs/2008.01088}{{\ttfamily
  2008.01088}}].

\bibitem{Dueck:2011hu}
A.~Dueck, W.~Rodejohann and K.~Zuber, \emph{{Neutrinoless Double Beta Decay,
  the Inverted Hierarchy and Precision Determination of theta(12)}},
  \href{https://doi.org/10.1103/PhysRevD.83.113010}{\emph{Phys. Rev. D}
  {\bfseries 83} (2011) 113010}
  [\href{https://arxiv.org/abs/1103.4152}{{\ttfamily 1103.4152}}].

\bibitem{NOvA:2021nfi}
{\scshape NOvA} collaboration, \emph{{An Improved Measurement of Neutrino
  Oscillation Parameters by the NOvA Experiment}},
  \href{https://arxiv.org/abs/2108.08219}{{\ttfamily 2108.08219}}.

\bibitem{Cao:2015ita}
{\scshape ICFA Neutrino Panel} collaboration, \emph{{On the complementarity of
  Hyper-K and LBNF}},  \href{https://arxiv.org/abs/1501.03918}{{\ttfamily
  1501.03918}}.

\bibitem{Wildner:2015yaa}
E.~Wildner et~al., \emph{{The Opportunity Offered by the ESSnuSB Project to
  Exploit the Larger Leptonic CP Violation Signal at the Second Oscillation
  Maximum and the Requirements of This Project on the ESS Accelerator
  Complex}}, \href{https://doi.org/10.1155/2016/8640493}{\emph{Adv. High Energy
  Phys.} {\bfseries 2016} (2016) 8640493}
  [\href{https://arxiv.org/abs/1510.00493}{{\ttfamily 1510.00493}}].

\bibitem{ESSnuSB:2021lre}
{\scshape ESSnuSB} collaboration, \emph{{Updated physics performance of the
  ESSnuSB experiment}},  \href{https://arxiv.org/abs/2107.07585}{{\ttfamily
  2107.07585}}.

\bibitem{Akindinov:2019flp}
A.V.~Akindinov et~al., \emph{{Letter of Interest for a Neutrino Beam from
  Protvino to KM3NeT/ORCA}},
  \href{https://doi.org/10.1140/epjc/s10052-019-7259-5}{\emph{Eur. Phys. J. C}
  {\bfseries 79} (2019) 758}
  [\href{https://arxiv.org/abs/1902.06083}{{\ttfamily 1902.06083}}].

\bibitem{Esteban:2020cvm}
I.~Esteban, M.C.~Gonzalez-Garcia, M.~Maltoni, T.~Schwetz and A.~Zhou,
  \emph{{The fate of hints: updated global analysis of three-flavor neutrino
  oscillations}}, \href{https://doi.org/10.1007/JHEP09(2020)178}{\emph{JHEP}
  {\bfseries 09} (2020) 178}
  [\href{https://arxiv.org/abs/2007.14792}{{\ttfamily 2007.14792}}].

\bibitem{nufit}
NuFit, \url{http://www.nu-fit.org}.

\bibitem{Capozzi:2019vbz}
F.~Capozzi, E.~Lisi, A.~Marrone and A.~Palazzo, \emph{{Global analysis of
  oscillation parameters}},
  \href{https://doi.org/10.1088/1742-6596/1312/1/012005}{\emph{J. Phys. Conf.
  Ser.} {\bfseries 1312} (2019) 012005}.

\bibitem{Capozzi:2018ubv}
F.~Capozzi, E.~Lisi, A.~Marrone and A.~Palazzo, \emph{{Current unknowns in the
  three neutrino framework}},
  \href{https://doi.org/10.1016/j.ppnp.2018.05.005}{\emph{Prog. Part. Nucl.
  Phys.} {\bfseries 102} (2018) 48}
  [\href{https://arxiv.org/abs/1804.09678}{{\ttfamily 1804.09678}}].

\bibitem{deSalas:2020pgw}
P.F.~de~Salas, D.V.~Forero, S.~Gariazzo, P.~Mart\'\i{}nez-Mirav\'e, O.~Mena,
  C.A.~Ternes et~al., \emph{{2020 global reassessment of the neutrino
  oscillation picture}},
  \href{https://doi.org/10.1007/JHEP02(2021)071}{\emph{JHEP} {\bfseries 02}
  (2021) 071} [\href{https://arxiv.org/abs/2006.11237}{{\ttfamily
  2006.11237}}].

\bibitem{deSalas:2017kay}
P.~de~Salas, D.~Forero, C.~Ternes, M.~Tortola and J.~Valle, \emph{{Status of
  neutrino oscillations 2018: 3$\sigma$ hint for normal mass ordering and
  improved CP sensitivity}},
  \href{https://doi.org/10.1016/j.physletb.2018.06.019}{\emph{Phys. Lett. B}
  {\bfseries 782} (2018) 633}
  [\href{https://arxiv.org/abs/1708.01186}{{\ttfamily 1708.01186}}].

\bibitem{Minkowski:1977sc}
P.~Minkowski, \emph{{$\mu \to e\gamma$ at a Rate of One Out of $10^{9}$ Muon
  Decays?}}, \href{https://doi.org/10.1016/0370-2693(77)90435-X}{\emph{Phys.
  Lett.} {\bfseries B67} (1977) 421}.

\bibitem{Yanagida:1979as}
T.~Yanagida, \emph{{Horizontal Symmetry and Masses of Neutrinos}}, {\emph{Conf.
  Proc.} {\bfseries C7902131} (1979) 95}.

\bibitem{GellMann:1980vs}
M.~Gell-Mann, P.~Ramond and R.~Slansky, \emph{{Complex Spinors and Unified
  Theories}}, {\emph{Conf. Proc.} {\bfseries C790927} (1979) 315}
  [\href{https://arxiv.org/abs/1306.4669}{{\ttfamily 1306.4669}}].

\bibitem{Mohapatra:1979ia}
R.N.~Mohapatra and G.~Senjanovic, \emph{{Neutrino Mass and Spontaneous Parity
  Violation}}, \href{https://doi.org/10.1103/PhysRevLett.44.912}{\emph{Phys.
  Rev. Lett.} {\bfseries 44} (1980) 912}.

\bibitem{Magg:1980ut}
M.~Magg and C.~Wetterich, \emph{{Neutrino Mass Problem and Gauge Hierarchy}},
  \href{https://doi.org/10.1016/0370-2693(80)90825-4}{\emph{Phys. Lett.}
  {\bfseries B94} (1980) 61}.

\bibitem{Wetterich:1981bx}
C.~Wetterich, \emph{{Neutrino Masses and the Scale of B-L Violation}},
  \href{https://doi.org/10.1016/0550-3213(81)90279-0}{\emph{Nucl. Phys.}
  {\bfseries B187} (1981) 343}.

\bibitem{Lazarides:1980nt}
G.~Lazarides, Q.~Shafi and C.~Wetterich, \emph{{Proton Lifetime and Fermion
  Masses in an SO(10) Model}},
  \href{https://doi.org/10.1016/0550-3213(81)90354-0}{\emph{Nucl. Phys.}
  {\bfseries B181} (1981) 287}.

\bibitem{Mohapatra:1980yp}
R.N.~Mohapatra and G.~Senjanovic, \emph{{Neutrino Masses and Mixings in Gauge
  Models with Spontaneous Parity Violation}},
  \href{https://doi.org/10.1103/PhysRevD.23.165}{\emph{Phys. Rev.} {\bfseries
  D23} (1981) 165}.

\bibitem{Cheng:1980qt}
T.P.~Cheng and L.-F.~Li, \emph{{Neutrino Masses, Mixings and Oscillations in
  $SU(2) \times U(1)$ Models of Electroweak Interactions}},
  \href{https://doi.org/10.1103/PhysRevD.22.2860}{\emph{Phys. Rev.} {\bfseries
  D22} (1980) 2860}.

\bibitem{Foot:1988aq}
R.~Foot, H.~Lew, X.G.~He and G.C.~Joshi, \emph{{Seesaw Neutrino Masses Induced
  by a Triplet of Leptons}}, \href{https://doi.org/10.1007/BF01415558}{\emph{Z.
  Phys.} {\bfseries C44} (1989) 441}.

\bibitem{KamLAND-Zen:2016pfg}
{\scshape KamLAND-Zen} collaboration, \emph{{Search for Majorana Neutrinos near
  the Inverted Mass Hierarchy Region with KamLAND-Zen}},
  \href{https://doi.org/10.1103/PhysRevLett.117.109903,
  10.1103/PhysRevLett.117.082503}{\emph{Phys. Rev. Lett.} {\bfseries 117}
  (2016) 082503} [\href{https://arxiv.org/abs/1605.02889}{{\ttfamily
  1605.02889}}].

\bibitem{Aker:2021gma}
M.~Aker et~al., \emph{{First direct neutrino-mass measurement with sub-eV
  sensitivity}},  \href{https://arxiv.org/abs/2105.08533}{{\ttfamily
  2105.08533}}.

\bibitem{Aghanim:2018eyx}
{\scshape Planck} collaboration, \emph{{Planck 2018 results. VI. Cosmological
  parameters}},
  \href{https://doi.org/10.1051/0004-6361/201833910}{\emph{Astron. Astrophys.}
  {\bfseries 641} (2020) A6}
  [\href{https://arxiv.org/abs/1807.06209}{{\ttfamily 1807.06209}}].

\bibitem{Bilenky:1980cx}
S.M.~Bilenky, J.~Hosek and S.T.~Petcov, \emph{{On Oscillations of Neutrinos
  with Dirac and Majorana Masses}},
  \href{https://doi.org/10.1016/0370-2693(80)90927-2}{\emph{Phys. Lett. B}
  {\bfseries 94} (1980) 495}.

\bibitem{Langacker:1986jv}
P.~Langacker, S.T.~Petcov, G.~Steigman and S.~Toshev, \emph{{On the
  Mikheev-Smirnov-Wolfenstein (MSW) Mechanism of Amplification of Neutrino
  Oscillations in Matter}},
  \href{https://doi.org/10.1016/0550-3213(87)90699-7}{\emph{Nucl. Phys. B}
  {\bfseries 282} (1987) 589}.

\bibitem{Asner:2014cwa}
{\scshape Project 8} collaboration, \emph{{Single electron detection and
  spectroscopy via relativistic cyclotron radiation}},
  \href{https://doi.org/10.1103/PhysRevLett.114.162501}{\emph{Phys. Rev. Lett.}
  {\bfseries 114} (2015) 162501}
  [\href{https://arxiv.org/abs/1408.5362}{{\ttfamily 1408.5362}}].

\bibitem{Velte:2019jvx}
C.~Velte et~al., \emph{{High-resolution and low-background $^{163}$Ho spectrum:
  interpretation of the resonance tails}},
  \href{https://doi.org/10.1140/epjc/s10052-019-7513-x}{\emph{Eur. Phys. J. C}
  {\bfseries 79} (2019) 1026}.

\bibitem{Lobashev:1985mu}
V.~Lobashev and P.~Spivak, \emph{{A METHOD FOR MEASURING THE
  ANTI-ELECTRON-NEUTRINO REST MASS}},
  \href{https://doi.org/10.1016/0168-9002(85)90640-0}{\emph{Nucl.\ Instrum.\
  Meth.\ A} {\bfseries 240} (1985) 305}.

\bibitem{PICARD1992345}
A.~Picard et~al., \emph{A solenoid retarding spectrometer with high resolution
  and transmission for kev electrons},
  \href{https://doi.org/https://doi.org/10.1016/0168-583X(92)95119-C}{\emph{Nuclear
  Instruments and Methods in Physics Research Section B: Beam Interactions with
  Materials and Atoms} {\bfseries 63} (1992) 345 }.

\bibitem{Kraus:2004zw}
C.~Kraus et~al., \emph{{Final results from phase II of the Mainz neutrino mass
  search in tritium beta decay}},
  \href{https://doi.org/10.1140/epjc/s2005-02139-7}{\emph{Eur. Phys. J.}
  {\bfseries C40} (2005) 447}
  [\href{https://arxiv.org/abs/hep-ex/0412056}{{\ttfamily hep-ex/0412056}}].

\bibitem{Aseev:2011dq}
{\scshape Troitsk} collaboration, \emph{{An upper limit on electron
  antineutrino mass from Troitsk experiment}},
  \href{https://doi.org/10.1103/PhysRevD.84.112003}{\emph{Phys. Rev.}
  {\bfseries D84} (2011) 112003}
  [\href{https://arxiv.org/abs/1108.5034}{{\ttfamily 1108.5034}}].

\bibitem{Otten:2008zz}
E.W.~Otten and C.~Weinheimer, \emph{{Neutrino mass limit from tritium beta
  decay}}, \href{https://doi.org/10.1088/0034-4885/71/8/086201}{\emph{Rept.
  Prog. Phys.} {\bfseries 71} (2008) 086201}
  [\href{https://arxiv.org/abs/0909.2104}{{\ttfamily 0909.2104}}].

\bibitem{Drexlin:2013lha}
G.~Drexlin, V.~Hannen, S.~Mertens and C.~Weinheimer, \emph{{Current direct
  neutrino mass experiments}},
  \href{https://doi.org/10.1155/2013/293986}{\emph{Adv. High Energy Phys.}
  {\bfseries 2013} (2013) 293986}
  [\href{https://arxiv.org/abs/1307.0101}{{\ttfamily 1307.0101}}].

\bibitem{Aker:2019uuj}
{\scshape KATRIN} collaboration, \emph{{Improved upper limit on the neutrino
  mass from a direct kinematic method by KATRIN}},
  \href{https://doi.org/10.1103/PhysRevLett.123.221802}{\emph{Phys. Rev. Lett.}
  {\bfseries 123} (2019) 221802}
  [\href{https://arxiv.org/abs/1909.06048}{{\ttfamily 1909.06048}}].

\bibitem{Angrik:2005ep}
{\scshape KATRIN} collaboration, \emph{Katrin design report 2004\!\!}, .

\bibitem{Aker:2020vrf}
{\scshape KATRIN} collaboration, \emph{{Bound on 3+1 Active-Sterile Neutrino
  Mixing from the First Four-Week Science Run of KATRIN}},
  \href{https://doi.org/10.1103/PhysRevLett.126.091803}{\emph{Phys. Rev. Lett.}
  {\bfseries 126} (2021) 091803}
  [\href{https://arxiv.org/abs/2011.05087}{{\ttfamily 2011.05087}}].

\bibitem{Mertens:2018vuu}
{\scshape KATRIN} collaboration, \emph{{A novel detector system for KATRIN to
  search for keV-scale sterile neutrinos}},
  \href{https://doi.org/10.1088/1361-6471/ab12fe}{\emph{J. Phys. G} {\bfseries
  46} (2019) 065203} [\href{https://arxiv.org/abs/1810.06711}{{\ttfamily
  1810.06711}}].

\bibitem{1999NIMPA.421..256B}
J.~{Bonn}, L.~{Bornschein}, B.~{Degen}, E.W.~{Otten} and C.~{Weinheimer},
  \emph{{A high resolution electrostatic time-of-flight spectrometer with
  adiabatic magnetic collimation}},
  \href{https://doi.org/10.1016/S0168-9002(98)01263-7}{\emph{Nuclear
  Instruments and Methods in Physics Research A} {\bfseries 421} (1999) 256}.

\bibitem{Steinbrink:2013ska}
N.~Steinbrink, V.~Hannen, E.L.~Martin, R.G.H.~Robertson, M.~Zacher and
  C.~Weinheimer, \emph{{Neutrino mass sensitivity by MAC-E-Filter based
  time-of-flight spectroscopy with the example of KATRIN}},
  \href{https://doi.org/10.1088/1367-2630/15/11/113020}{\emph{New J. Phys.}
  {\bfseries 15} (2013) 113020}
  [\href{https://arxiv.org/abs/1308.0532}{{\ttfamily 1308.0532}}].

\bibitem{Monreal:2009za}
B.~Monreal and J.A.~Formaggio, \emph{{Relativistic Cyclotron Radiation
  Detection of Tritium Decay Electrons as a New Technique for Measuring the
  Neutrino Mass}},
  \href{https://doi.org/10.1103/PhysRevD.80.051301}{\emph{Phys. Rev.}
  {\bfseries D80} (2009) 051301}
  [\href{https://arxiv.org/abs/0904.2860}{{\ttfamily 0904.2860}}].

\bibitem{Esfahani:2017dmu}
{\scshape Project 8} collaboration, \emph{{Determining the neutrino mass with
  cyclotron radiation emission spectroscopy -- Project 8}},
  \href{https://doi.org/10.1088/1361-6471/aa5b4f}{\emph{J. Phys.} {\bfseries
  G44} (2017) 054004} [\href{https://arxiv.org/abs/1703.02037}{{\ttfamily
  1703.02037}}].

\bibitem{Betti:2019ouf}
{\scshape PTOLEMY} collaboration, \emph{{Neutrino physics with the PTOLEMY
  project: active neutrino properties and the light sterile case}},
  \href{https://doi.org/10.1088/1475-7516/2019/07/047}{\emph{JCAP} {\bfseries
  1907} (2019) 047} [\href{https://arxiv.org/abs/1902.05508}{{\ttfamily
  1902.05508}}].

\bibitem{DeRujula:1982qt}
A.~De~Rujula and M.~Lusignoli, \emph{{Calorimetric Measurements of $^{163}$Ho
  Decay as Tools to Determine the Electron Neutrino Mass}},
  \href{https://doi.org/10.1016/0370-2693(82)90218-0}{\emph{Phys. Lett.}
  {\bfseries 118B} (1982) 429}.

\bibitem{Gastaldo:2017edk}
L.~Gastaldo et~al., \emph{{The electron capture in $^{163}$Ho experiment -
  ECHo}}, \href{https://doi.org/10.1140/epjst/e2017-70071-y}{\emph{Eur. Phys.
  J. ST} {\bfseries 226} (2017) 1623}.

\bibitem{Giachero:2016xnn}
{\scshape HOLMES} collaboration, \emph{{Measuring the electron neutrino mass
  with improved sensitivity: the HOLMES experiment}},
  \href{https://doi.org/10.1088/1748-0221/12/02/C02046}{\emph{JINST} {\bfseries
  12} (2017) C02046} [\href{https://arxiv.org/abs/1612.03947}{{\ttfamily
  1612.03947}}].

\bibitem{Fleischmann2005}
A.~Fleischmann, C.~Enss and G.~Seidel, \emph{Metallic magnetic calorimeters},
  in \emph{Cryogenic Particle Detection}, C.~Enss, ed., (Berlin, Heidelberg),
  pp.~151--216, Springer Berlin Heidelberg (2005),
  \href{https://doi.org/10.1007/10933596_4}{DOI}.

\bibitem{Gastaldo:2009ovy}
L.~Gastaldo, J.~Porst, F.v.~Seggern, A.~Kirsch, P.~Ranitzsch, A.~Fleischmann
  et~al., \emph{{Low Temperature Magnetic Calorimeters For Neutrino Mass Direct
  Measurement}}, \href{https://doi.org/10.1063/1.3292415}{\emph{AIP Conf.\
  Proc.} {\bfseries 1185} (2009) 607}.

\bibitem{Faverzani:2012zgs}
M.~Faverzani, P.~Day, A.~Nucciotti and E.~Ferri, \emph{{Developments of
  Microresonators Detectors for Neutrino Physics in Milan}},
  \href{https://doi.org/10.1007/s10909-012-0538-2}{\emph{J.\ Low.\ Temp.\
  Phys.} {\bfseries 167} (2012) 1041}.

\bibitem{Faessler:2014xpa}
A.~Faessler, L.~Gastaldo and F.~Simkovic, \emph{{Electron capture in
  $^{163}$Ho, overlap plus exchange corrections and neutrino mass}},
  \href{https://doi.org/10.1088/0954-3899/42/1/015108}{\emph{J. Phys.}
  {\bfseries G42} (2015) 015108}
  [\href{https://arxiv.org/abs/1407.6504}{{\ttfamily 1407.6504}}].

\bibitem{Brass:2017kov}
M.~Braß, C.~Enss, L.~Gastaldo, M.~Haverkort and R.~Green, \emph{{$\textit{Ab
  initio}$ calculation of the calorimetric electron capture spectrum of
  $^{163}$Holmium: Intra-atomic decay into bound-states}},
  \href{https://doi.org/10.1103/PhysRevC.97.054620}{\emph{Phys.\ Rev.\ C}
  {\bfseries 97} (2018) 054620}
  [\href{https://arxiv.org/abs/1711.10309}{{\ttfamily 1711.10309}}].

\bibitem{Barate:1997zg}
{\scshape ALEPH} collaboration, \emph{{An Upper limit on the tau-neutrino mass
  from three-prong and five-prong tau decays}},
  \href{https://doi.org/10.1007/s100520050149}{\emph{Eur. Phys. J. C}
  {\bfseries 2} (1998) 395}.

\bibitem{Loredo:2001rx}
T.J.~Loredo and D.Q.~Lamb, \emph{{Bayesian analysis of neutrinos observed from
  supernova SN-1987A}},
  \href{https://doi.org/10.1103/PhysRevD.65.063002}{\emph{Phys. Rev.}
  {\bfseries D65} (2002) 063002}
  [\href{https://arxiv.org/abs/astro-ph/0107260}{{\ttfamily
  astro-ph/0107260}}].

\bibitem{Pagliaroli:2010ik}
G.~Pagliaroli, F.~Rossi-Torres and F.~Vissani, \emph{{Neutrino mass bound in
  the standard scenario for supernova electronic antineutrino emission}},
  \href{https://doi.org/10.1016/j.astropartphys.2010.02.007}{\emph{Astropart.
  Phys.} {\bfseries 33} (2010) 287}
  [\href{https://arxiv.org/abs/1002.3349}{{\ttfamily 1002.3349}}].

\bibitem{Beacom:1998ya}
J.F.~Beacom and P.~Vogel, \emph{{Mass signature of supernova muon-neutrino and
  tau-neutrino neutrinos in Super-Kamiokande}},
  \href{https://doi.org/10.1103/PhysRevD.58.053010}{\emph{Phys. Rev.}
  {\bfseries D58} (1998) 053010}
  [\href{https://arxiv.org/abs/hep-ph/9802424}{{\ttfamily hep-ph/9802424}}].

\bibitem{Beacom:1998yb}
J.F.~Beacom and P.~Vogel, \emph{{Mass signature of supernova muon-neutrino and
  tau-neutrino neutrinos in the Sudbury neutrino observatory}},
  \href{https://doi.org/10.1103/PhysRevD.58.093012}{\emph{Phys. Rev.}
  {\bfseries D58} (1998) 093012}
  [\href{https://arxiv.org/abs/hep-ph/9806311}{{\ttfamily hep-ph/9806311}}].

\bibitem{Lu:2014zma}
J.-S.~Lu, J.~Cao, Y.-F.~Li and S.~Zhou, \emph{{Constraining Absolute Neutrino
  Masses via Detection of Galactic Supernova Neutrinos at JUNO}},
  \href{https://doi.org/10.1088/1475-7516/2015/05/044}{\emph{JCAP} {\bfseries
  1505} (2015) 044} [\href{https://arxiv.org/abs/1412.7418}{{\ttfamily
  1412.7418}}].

\bibitem{Beacom:2000qy}
J.F.~Beacom, R.N.~Boyd and A.~Mezzacappa, \emph{{Black hole formation in core
  collapse supernovae and time-of-flight measurements of the neutrino masses}},
  \href{https://doi.org/10.1103/PhysRevD.63.073011}{\emph{Phys. Rev.}
  {\bfseries D63} (2001) 073011}
  [\href{https://arxiv.org/abs/astro-ph/0010398}{{\ttfamily
  astro-ph/0010398}}].

\bibitem{Arnaud:2001gt}
N.~Arnaud, M.~Barsuglia, M.A.~Bizouard, F.~Cavalier, M.~Davier, P.~Hello
  et~al., \emph{{Gravity wave and neutrino bursts from stellar collapse: A
  Sensitive test of neutrino masses}},
  \href{https://doi.org/10.1103/PhysRevD.65.033010}{\emph{Phys. Rev.}
  {\bfseries D65} (2002) 033010}
  [\href{https://arxiv.org/abs/hep-ph/0109027}{{\ttfamily hep-ph/0109027}}].

\bibitem{Schechter:1981bd}
J.~Schechter and J.W.F.~Valle, \emph{{Neutrinoless Double beta Decay in SU(2) x
  U(1) Theories}}, \href{https://doi.org/10.1103/PhysRevD.25.2951}{\emph{Phys.
  Rev.} {\bfseries D25} (1982) 2951}.

\bibitem{Blaum:2020ogl}
K.~Blaum, S.~Eliseev, F.A.~Danevich, V.I.~Tretyak, S.~Kovalenko,
  M.I.~Krivoruchenko et~al., \emph{{Neutrinoless Double-Electron Capture}},
  \href{https://doi.org/10.1103/RevModPhys.92.045007}{\emph{Rev. Mod. Phys.}
  {\bfseries 92} (2020) 045007}
  [\href{https://arxiv.org/abs/2007.14908}{{\ttfamily 2007.14908}}].

\bibitem{Fukugita:1986hr}
M.~Fukugita and T.~Yanagida, \emph{{Baryogenesis Without Grand Unification}},
  \href{https://doi.org/10.1016/0370-2693(86)91126-3}{\emph{Phys. Lett.}
  {\bfseries B174} (1986) 45}.

\bibitem{Dolinski:2019nrj}
M.J.~Dolinski, A.W.P.~Poon and W.~Rodejohann, \emph{{Neutrinoless Double-Beta
  Decay: Status and Prospects}},
  \href{https://doi.org/10.1146/annurev-nucl-101918-023407}{\emph{Ann. Rev.
  Nucl. Part. Sci.} {\bfseries 69} (2019) 219}
  [\href{https://arxiv.org/abs/1902.04097}{{\ttfamily 1902.04097}}].

\bibitem{Engel:2016xgb}
J.~Engel and J.~Men\'{e}ndez, \emph{{Status and Future of Nuclear Matrix
  Elements for Neutrinoless Double-Beta Decay: A Review}},
  \href{https://doi.org/10.1088/1361-6633/aa5bc5}{\emph{Rept. Prog. Phys.}
  {\bfseries 80} (2017) 046301}
  [\href{https://arxiv.org/abs/1610.06548}{{\ttfamily 1610.06548}}].

\bibitem{Tretyak:2002dx}
V.I.~Tretyak and Y.G.~Zdesenko, \emph{{Tables of double beta decay data: An
  update}}, \href{https://doi.org/10.1006/adnd.2001.0873}{\emph{Atom. Data
  Nucl. Data Tabl.} {\bfseries 80} (2002) 83}.

\bibitem{Kotila:2012zza}
J.~Kotila and F.~Iachello, \emph{{Phase space factors for double-$\beta$
  decay}}, \href{https://doi.org/10.1103/PhysRevC.85.034316}{\emph{Phys. Rev.}
  {\bfseries C85} (2012) 034316}
  [\href{https://arxiv.org/abs/1209.5722}{{\ttfamily 1209.5722}}].

\bibitem{Stoica:2013lka}
S.~Stoica and M.~Mirea, \emph{{New calculations for phase space factors
  involved in double-$\beta$ decay}},
  \href{https://doi.org/10.1103/PhysRevC.88.037303}{\emph{Phys. Rev.}
  {\bfseries C88} (2013) 037303}
  [\href{https://arxiv.org/abs/1307.0290}{{\ttfamily 1307.0290}}].

\bibitem{Avignone:2007fu}
F.T.~Avignone, III, S.R.~Elliott and J.~Engel, \emph{{Double Beta Decay,
  Majorana Neutrinos, and Neutrino Mass}},
  \href{https://doi.org/10.1103/RevModPhys.80.481}{\emph{Rev. Mod. Phys.}
  {\bfseries 80} (2008) 481} [\href{https://arxiv.org/abs/0708.1033}{{\ttfamily
  0708.1033}}].

\bibitem{Vergados:2012xy}
J.D.~Vergados, H.~Ejiri and F.~Simkovic, \emph{{Theory of Neutrinoless Double
  Beta Decay}},
  \href{https://doi.org/10.1088/0034-4885/75/10/106301}{\emph{Rept. Prog.
  Phys.} {\bfseries 75} (2012) 106301}
  [\href{https://arxiv.org/abs/1205.0649}{{\ttfamily 1205.0649}}].

\bibitem{Gysbers:2019uyb}
P.~Gysbers et~al., \emph{{Discrepancy between experimental and theoretical
  $\beta$-decay rates resolved from first principles}},
  \href{https://doi.org/10.1038/s41567-019-0450-7}{\emph{Nature Phys.}
  {\bfseries 15} (2019) 428}
  [\href{https://arxiv.org/abs/1903.00047}{{\ttfamily 1903.00047}}].

\bibitem{Cirigliano:2020yhp}
V.~Cirigliano, W.~Detmold, A.~Nicholson and P.~Shanahan, \emph{{Lattice QCD
  Inputs for Nuclear Double Beta Decay}},
  \href{https://doi.org/10.1016/j.ppnp.2020.103771}{\emph{Progress in Particle
  and Nuclear Physics} {\bfseries 112} (2020) 103771}
  [\href{https://arxiv.org/abs/2003.08493}{{\ttfamily 2003.08493}}].

\bibitem{Ejiri:2019ezh}
H.~Ejiri, J.~Suhonen and K.~Zuber, \emph{{Neutrino–nuclear responses for
  astro-neutrinos, single beta decays and double beta decays}},
  \href{https://doi.org/10.1016/j.physrep.2018.12.001}{\emph{Phys. Rept.}
  {\bfseries 797} (2019) 1}.

\bibitem{Albert:2017hjq}
{\scshape nEXO} collaboration, \emph{{Sensitivity and Discovery Potential of
  nEXO to Neutrinoless Double Beta Decay}},
  \href{https://doi.org/10.1103/PhysRevC.97.065503}{\emph{Phys. Rev.}
  {\bfseries C97} (2018) 065503}
  [\href{https://arxiv.org/abs/1710.05075}{{\ttfamily 1710.05075}}].

\bibitem{Agostini:2020xta}
{\scshape GERDA} collaboration, \emph{{Final Results of GERDA on the Search for
  Neutrinoless Double-$\beta$ Decay}},
  \href{https://doi.org/10.1103/PhysRevLett.125.252502}{\emph{Phys. Rev. Lett.}
  {\bfseries 125} (2020) 252502}
  [\href{https://arxiv.org/abs/2009.06079}{{\ttfamily 2009.06079}}].

\bibitem{Akerib:2019dgs}
{\scshape LZ} collaboration, \emph{{Projected sensitivity of the LUX-ZEPLIN
  experiment to the $0\nu\beta\beta$ decay of $^{136}Xe$}},
  \href{https://doi.org/10.1103/PhysRevC.102.014602}{\emph{Phys. Rev. C}
  {\bfseries 102} (2020) 014602}
  [\href{https://arxiv.org/abs/1912.04248}{{\ttfamily 1912.04248}}].

\bibitem{Chen:2016qcd}
X.~Chen et~al., \emph{{PandaX-III: Searching for neutrinoless double beta decay
  with high pressure$^{136}$Xe gas time projection chambers}},
  \href{https://doi.org/10.1007/s11433-017-9028-0}{\emph{Sci. China Phys. Mech.
  Astron.} {\bfseries 60} (2017) 061011}
  [\href{https://arxiv.org/abs/1610.08883}{{\ttfamily 1610.08883}}].

\bibitem{Agostini:2020adk}
{\scshape DARWIN} collaboration, \emph{{Sensitivity of the DARWIN observatory
  to the neutrinoless double beta decay of $^{136}$Xe}},
  \href{https://doi.org/10.1140/epjc/s10052-020-8196-z}{\emph{Eur. Phys. J. C}
  {\bfseries 80} (2020) 808}
  [\href{https://arxiv.org/abs/2003.13407}{{\ttfamily 2003.13407}}].

\bibitem{Agostini:2017jim}
M.~Agostini, G.~Benato and J.~Detwiler, \emph{{Discovery probability of
  next-generation neutrinoless double-$\beta$ decay experiments}},
  \href{https://doi.org/10.1103/PhysRevD.96.053001}{\emph{Phys. Rev.}
  {\bfseries D96} (2017) 053001}
  [\href{https://arxiv.org/abs/1705.02996}{{\ttfamily 1705.02996}}].

\bibitem{Umehara:2008zz}
S.~Umehara et~al., \emph{{Double beta decay of Ca-48 studied by CaF-2(Eu)
  scintillators}},
  \href{https://doi.org/10.1088/1742-6596/120/5/052058}{\emph{J.\ Phys.\ Conf.\
  Ser.} {\bfseries 120} (2008) 052058}.

\bibitem{Alvis:2019sil}
{\scshape Majorana} collaboration, \emph{{A Search for Neutrinoless Double-Beta
  Decay in $^{76}$Ge with 26 kg-yr of Exposure from the MAJORANA
  DEMONSTRATOR}},
  \href{https://doi.org/10.1103/PhysRevC.100.025501}{\emph{Phys.\ Rev.\ C}
  {\bfseries 100} (2019) 025501}
  [\href{https://arxiv.org/abs/1902.02299}{{\ttfamily 1902.02299}}].

\bibitem{Azzolini:2019tta}
{\scshape CUPID} collaboration, \emph{{Final result of CUPID-0 phase-I in the
  search for the $^{82}$Se Neutrinoless Double-$\beta$ Decay}},
  \href{https://doi.org/10.1103/PhysRevLett.123.032501}{\emph{Phys. Rev. Lett.}
  {\bfseries 123} (2019) 032501}
  [\href{https://arxiv.org/abs/1906.05001}{{\ttfamily 1906.05001}}].

\bibitem{Argyriades:2009ph}
{\scshape NEMO-3} collaboration, \emph{{Measurement of the two neutrino double
  beta decay half-life of Zr-96 with the NEMO-3 detector}},
  \href{https://doi.org/10.1016/j.nuclphysa.2010.07.009}{\emph{Nucl.\ Phys.\ A}
  {\bfseries 847} (2010) 168}
  [\href{https://arxiv.org/abs/0906.2694}{{\ttfamily 0906.2694}}].

\bibitem{Armengaud:2020luj}
{\scshape CUPID} collaboration, \emph{{New Limit for Neutrinoless Double-Beta
  Decay of $^{100}$Mo from the CUPID-Mo Experiment}},
  \href{https://doi.org/10.1103/PhysRevLett.126.181802}{\emph{Phys. Rev. Lett.}
  {\bfseries 126} (2021) 181802}
  [\href{https://arxiv.org/abs/2011.13243}{{\ttfamily 2011.13243}}].

\bibitem{Barabash:2018yjq}
A.~Barabash et~al., \emph{{Final results of the Aurora experiment to study
  $2\beta$ decay of $^{116}\mathrm{Cd}$ with enriched
  $^{116}\mathrm{Cd}{\mathrm{WO}}_{4}$ crystal scintillators}},
  \href{https://doi.org/10.1103/PhysRevD.98.092007}{\emph{Phys.\ Rev.\ D}
  {\bfseries 98} (2018) 092007}
  [\href{https://arxiv.org/abs/1811.06398}{{\ttfamily 1811.06398}}].

\bibitem{Arnaboldi:2002te}
C.~Arnaboldi et~al., \emph{{A Calorimetric search on double beta decay of
  Te-130}}, \href{https://doi.org/10.1016/S0370-2693(03)00212-0}{\emph{Phys.\
  Lett.\ B} {\bfseries 557} (2003) 167}
  [\href{https://arxiv.org/abs/hep-ex/0211071}{{\ttfamily hep-ex/0211071}}].

\bibitem{Adams:2021rbc}
{\scshape CUORE} collaboration, \emph{{High sensitivity neutrinoless
  double-beta decay search with one tonne-year of CUORE data}},
  \href{https://arxiv.org/abs/2104.06906}{{\ttfamily 2104.06906}}.

\bibitem{Anton:2019wmi}
{\scshape EXO-200} collaboration, \emph{{Search for Neutrinoless Double-$\beta$
  Decay with the Complete EXO-200 Dataset}},
  \href{https://doi.org/10.1103/PhysRevLett.123.161802}{\emph{Phys.\ Rev.\
  Lett.} {\bfseries 123} (2019) 161802}
  [\href{https://arxiv.org/abs/1906.02723}{{\ttfamily 1906.02723}}].

\bibitem{Arnold:2016qyg}
{\scshape NEMO-3} collaboration, \emph{{Measurement of the 2$\nu\beta\beta$
  decay half-life of $^{150}$Nd and a search for 0$\nu\beta\beta$ decay
  processes with the full exposure from the NEMO-3 detector}},
  \href{https://doi.org/10.1103/PhysRevD.94.072003}{\emph{Phys.\ Rev.\ D}
  {\bfseries 94} (2016) 072003}
  [\href{https://arxiv.org/abs/1606.08494}{{\ttfamily 1606.08494}}].

\bibitem{Majorana:2016hop}
{\scshape Majorana} collaboration, \emph{{New limits on Bosonic Dark Matter,
  Solar Axions, Pauli Exclusion Principle Violation, and Electron Decay from
  the Majorana Demonstrator}},
  \href{https://doi.org/10.1103/PhysRevLett.118.161801}{\emph{Phys. Rev. Lett.}
  {\bfseries 118} (2017) 161801}
  [\href{https://arxiv.org/abs/1612.00886}{{\ttfamily 1612.00886}}].

\bibitem{Deppisch:2020mxv}
F.F.~Deppisch, L.~Graf and F.~\v{S}imkovic, \emph{{Searching for New Physics in
  Two-Neutrino Double Beta Decay}},
  \href{https://doi.org/10.1103/PhysRevLett.125.171801}{\emph{Phys. Rev. Lett.}
  {\bfseries 125} (2020) 171801}
  [\href{https://arxiv.org/abs/2003.11836}{{\ttfamily 2003.11836}}].

\bibitem{Deppisch:2020sqh}
F.F.~Deppisch, L.~Graf, W.~Rodejohann and X.-J.~Xu, \emph{{Neutrino
  Self-Interactions and Double Beta Decay}},
  \href{https://doi.org/10.1103/PhysRevD.102.051701}{\emph{Phys. Rev. D}
  {\bfseries 102} (2020) 051701}
  [\href{https://arxiv.org/abs/2004.11919}{{\ttfamily 2004.11919}}].

\bibitem{Agostini:2020cpz}
M.~Agostini, E.~Bossio, A.~Ibarra and X.~Marcano, \emph{{Search for Light
  Exotic Fermions in Double-Beta Decays}},
  \href{https://doi.org/10.1016/j.physletb.2021.136127}{\emph{Phys. Lett. B}
  {\bfseries 815} (2021) 136127}
  [\href{https://arxiv.org/abs/2012.09281}{{\ttfamily 2012.09281}}].

\bibitem{Agostini:2020oiv}
M.~Agostini, G.~Benato, S.~Dell'Oro, S.~Pirro and F.~Vissani, \emph{{Discovery
  probabilities of Majorana neutrinos based on cosmological data}},
  \href{https://doi.org/10.1103/PhysRevD.103.033008}{\emph{Phys. Rev. D}
  {\bfseries 103} (2021) 033008}
  [\href{https://arxiv.org/abs/2012.13938}{{\ttfamily 2012.13938}}].

\bibitem{Cao:2019hli}
J.~Cao, G.-Y.~Huang, Y.-F.~Li, Y.~Wang, L.-J.~Wen, Z.-Z.~Xing et~al.,
  \emph{{Towards the meV limit of the effective neutrino mass in neutrinoless
  double-beta decays}},
  \href{https://doi.org/10.1088/1674-1137/44/3/031001}{\emph{Chin. Phys. C}
  {\bfseries 44} (2020) 031001}
  [\href{https://arxiv.org/abs/1908.08355}{{\ttfamily 1908.08355}}].

\bibitem{Lattanzi:2020iik}
M.~Lattanzi, M.~Gerbino, K.~Freese, G.~Kane and J.W.F.~Valle, \emph{{Cornering
  (quasi) degenerate neutrinos with cosmology}},
  \href{https://doi.org/10.1007/JHEP10(2020)213}{\emph{JHEP} {\bfseries 10}
  (2020) 213} [\href{https://arxiv.org/abs/2007.01650}{{\ttfamily
  2007.01650}}].

\bibitem{Lesgourgues:2006nd}
J.~Lesgourgues and S.~Pastor, \emph{{Massive neutrinos and cosmology}},
  \href{https://doi.org/10.1016/j.physrep.2006.04.001}{\emph{Phys. Rept.}
  {\bfseries 429} (2006) 307}
  [\href{https://arxiv.org/abs/astro-ph/0603494}{{\ttfamily
  astro-ph/0603494}}].

\bibitem{Palanque-Delabrouille:2015pga}
N.~Palanque-Delabrouille et~al., \emph{{Neutrino masses and cosmology with
  Lyman-alpha forest power spectrum}},
  \href{https://doi.org/10.1088/1475-7516/2015/11/011}{\emph{JCAP} {\bfseries
  1511} (2015) 011} [\href{https://arxiv.org/abs/1506.05976}{{\ttfamily
  1506.05976}}].

\bibitem{Dvorkin:2019jgs}
C.~Dvorkin et~al., \emph{{Neutrino Mass from Cosmology: Probing Physics Beyond
  the Standard Model}},  \href{https://arxiv.org/abs/1903.03689}{{\ttfamily
  1903.03689}}.

\bibitem{Brinckmann:2018owf}
T.~Brinckmann, D.C.~Hooper, M.~Archidiacono, J.~Lesgourgues and T.~Sprenger,
  \emph{{The promising future of a robust cosmological neutrino mass
  measurement}},
  \href{https://doi.org/10.1088/1475-7516/2019/01/059}{\emph{JCAP} {\bfseries
  1901} (2019) 059} [\href{https://arxiv.org/abs/1808.05955}{{\ttfamily
  1808.05955}}].

\bibitem{Alam:2016hwk}
{\scshape BOSS} collaboration, \emph{{The clustering of galaxies in the
  completed SDSS-III Baryon Oscillation Spectroscopic Survey: cosmological
  analysis of the DR12 galaxy sample}},
  \href{https://doi.org/10.1093/mnras/stx721}{\emph{Mon. Not. Roy. Astron.
  Soc.} {\bfseries 470} (2017) 2617}
  [\href{https://arxiv.org/abs/1607.03155}{{\ttfamily 1607.03155}}].

\bibitem{DiValentino:2021hoh}
E.~Di~Valentino, S.~Gariazzo and O.~Mena, \emph{{On the most constraining
  cosmological neutrino mass bounds}},
  \href{https://arxiv.org/abs/2106.15267}{{\ttfamily 2106.15267}}.

\bibitem{Ade:2015fva}
{\scshape Planck} collaboration, \emph{{Planck 2015 results. XXIV. Cosmology
  from Sunyaev-Zeldovich cluster counts}},
  \href{https://doi.org/10.1051/0004-6361/201525833}{\emph{Astron. Astrophys.}
  {\bfseries 594} (2016) A24}
  [\href{https://arxiv.org/abs/1502.01597}{{\ttfamily 1502.01597}}].

\bibitem{Palanque-Delabrouille:2019iyz}
N.~Palanque-Delabrouille, C.~Yèche, N.~Schöneberg, J.~Lesgourgues,
  M.~Walther, S.~Chabanier et~al., \emph{{Hints, neutrino bounds and WDM
  constraints from SDSS DR14 Lyman-$\alpha$ and Planck full-survey data}},
  \href{https://doi.org/10.1088/1475-7516/2020/04/038}{\emph{JCAP} {\bfseries
  04} (2020) 038} [\href{https://arxiv.org/abs/1911.09073}{{\ttfamily
  1911.09073}}].

\bibitem{Froustey:2020mcq}
J.~Froustey, C.~Pitrou and M.C.~Volpe, \emph{{Neutrino decoupling including
  flavour oscillations and primordial nucleosynthesis}},
  \href{https://doi.org/10.1088/1475-7516/2020/12/015}{\emph{JCAP} {\bfseries
  12} (2020) 015} [\href{https://arxiv.org/abs/2008.01074}{{\ttfamily
  2008.01074}}].

\bibitem{Bennett:2020zkv}
J.J.~Bennett, G.~Buldgen, P.F.~De~Salas, M.~Drewes, S.~Gariazzo, S.~Pastor
  et~al., \emph{{Towards a precision calculation of $N_{\rm eff}$ in the
  Standard Model II: Neutrino decoupling in the presence of flavour
  oscillations and finite-temperature QED}},
  \href{https://doi.org/10.1088/1475-7516/2021/04/073}{\emph{JCAP} {\bfseries
  04} (2021) 073} [\href{https://arxiv.org/abs/2012.02726}{{\ttfamily
  2012.02726}}].

\bibitem{Pitrou:2018cgg}
C.~Pitrou, A.~Coc, J.-P.~Uzan and E.~Vangioni, \emph{{Precision big bang
  nucleosynthesis with improved Helium-4 predictions}},
  \href{https://doi.org/10.1016/j.physrep.2018.04.005}{\emph{Phys. Rept.}
  {\bfseries 754} (2018) 1} [\href{https://arxiv.org/abs/1801.08023}{{\ttfamily
  1801.08023}}].

\bibitem{Beacom:2004yd}
J.F.~Beacom, N.F.~Bell and S.~Dodelson, \emph{{Neutrinoless universe}},
  \href{https://doi.org/10.1103/PhysRevLett.93.121302}{\emph{Phys. Rev. Lett.}
  {\bfseries 93} (2004) 121302}
  [\href{https://arxiv.org/abs/astro-ph/0404585}{{\ttfamily
  astro-ph/0404585}}].

\bibitem{Escudero:2020ped}
M.~Escudero, J.~Lopez-Pavon, N.~Rius and S.~Sandner, \emph{{Relaxing
  Cosmological Neutrino Mass Bounds with Unstable Neutrinos}},
  \href{https://doi.org/10.1007/JHEP12(2020)119}{\emph{JHEP} {\bfseries 12}
  (2020) 119} [\href{https://arxiv.org/abs/2007.04994}{{\ttfamily
  2007.04994}}].

\bibitem{Dvali:2016uhn}
G.~Dvali and L.~Funcke, \emph{{Small neutrino masses from gravitational
  \ensuremath{\theta}-term}},
  \href{https://doi.org/10.1103/PhysRevD.93.113002}{\emph{Phys. Rev. D}
  {\bfseries 93} (2016) 113002}
  [\href{https://arxiv.org/abs/1602.03191}{{\ttfamily 1602.03191}}].

\bibitem{Levi:2019ggs}
{\scshape DESI} collaboration, \emph{{The Dark Energy Spectroscopic Instrument
  (DESI)}},  \href{https://arxiv.org/abs/1907.10688}{{\ttfamily 1907.10688}}.

\bibitem{Aghamousa:2016zmz}
{\scshape DESI} collaboration, \emph{{The DESI Experiment Part I:
  Science,Targeting, and Survey Design}},
  \href{https://arxiv.org/abs/1611.00036}{{\ttfamily 1611.00036}}.

\bibitem{Matsumura:2013aja}
T.~Matsumura et~al., \emph{{Mission design of LiteBIRD}},
  \href{https://doi.org/10.1007/s10909-013-0996-1}{\emph{J. Low Temp. Phys.}
  {\bfseries 176} (2014) 733}
  [\href{https://arxiv.org/abs/1311.2847}{{\ttfamily 1311.2847}}].

\bibitem{Abazajian:2019eic}
K.~Abazajian et~al., \emph{{CMB-S4 Science Case, Reference Design, and Project
  Plan}},  \href{https://arxiv.org/abs/1907.04473}{{\ttfamily 1907.04473}}.

\bibitem{Delabrouille:2017rct}
{\scshape CORE} collaboration, \emph{{Exploring cosmic origins with CORE:
  Survey requirements and mission design}},
  \href{https://doi.org/10.1088/1475-7516/2018/04/014}{\emph{JCAP} {\bfseries
  04} (2018) 014} [\href{https://arxiv.org/abs/1706.04516}{{\ttfamily
  1706.04516}}].

\bibitem{Bonnet:2012kh}
F.~Bonnet, M.~Hirsch, T.~Ota and W.~Winter, \emph{{Systematic decomposition of
  the neutrinoless double beta decay operator}},
  \href{https://doi.org/10.1007/JHEP03(2013)055,
  10.1007/JHEP04(2014)090}{\emph{JHEP} {\bfseries 03} (2013) 055}
  [\href{https://arxiv.org/abs/1212.3045}{{\ttfamily 1212.3045}}].

\bibitem{Rodejohann:2011mu}
W.~Rodejohann, \emph{{Neutrino-less Double Beta Decay and Particle Physics}},
  \href{https://doi.org/10.1142/S0218301311020186}{\emph{Int. J. Mod. Phys. E}
  {\bfseries 20} (2011) 1833}
  [\href{https://arxiv.org/abs/1106.1334}{{\ttfamily 1106.1334}}].

\bibitem{Duerr:2011zd}
M.~Duerr, M.~Lindner and A.~Merle, \emph{{On the Quantitative Impact of the
  Schechter-Valle Theorem}},
  \href{https://doi.org/10.1007/JHEP06(2011)091}{\emph{JHEP} {\bfseries 06}
  (2011) 091} [\href{https://arxiv.org/abs/1105.0901}{{\ttfamily 1105.0901}}].

\bibitem{Helo:2013dla}
J.C.~Helo, M.~Hirsch, S.G.~Kovalenko and H.~Päs, \emph{{Neutrinoless double
  beta decay and lepton number violation at the LHC}},
  \href{https://doi.org/10.1103/PhysRevD.88.011901}{\emph{Phys. Rev.}
  {\bfseries D88} (2013) 011901}
  [\href{https://arxiv.org/abs/1303.0899}{{\ttfamily 1303.0899}}].

\bibitem{Deppisch:2013jxa}
F.F.~Deppisch, J.~Harz and M.~Hirsch, \emph{{Falsifying High-Scale Leptogenesis
  at the LHC}},
  \href{https://doi.org/10.1103/PhysRevLett.112.221601}{\emph{Phys. Rev. Lett.}
  {\bfseries 112} (2014) 221601}
  [\href{https://arxiv.org/abs/1312.4447}{{\ttfamily 1312.4447}}].

\bibitem{Deppisch:2012nb}
F.F.~Deppisch, M.~Hirsch and H.~Päs, \emph{{Neutrinoless Double Beta Decay and
  Physics Beyond the Standard Model}},
  \href{https://doi.org/10.1088/0954-3899/39/12/124007}{\emph{J. Phys. G}
  {\bfseries 39} (2012) 124007}
  [\href{https://arxiv.org/abs/1208.0727}{{\ttfamily 1208.0727}}].

\bibitem{Barenboim:2004di}
G.~Barenboim, O.~Mena~Requejo and C.~Quigg, \emph{{Diagnostic potential of
  cosmic-neutrino absorption spectroscopy}},
  \href{https://doi.org/10.1103/PhysRevD.71.083002}{\emph{Phys. Rev. D}
  {\bfseries 71} (2005) 083002}
  [\href{https://arxiv.org/abs/hep-ph/0412122}{{\ttfamily hep-ph/0412122}}].

\bibitem{Kodama:2007aa}
{\scshape DONuT} collaboration, \emph{{Final tau-neutrino results from the
  DONuT experiment}},
  \href{https://doi.org/10.1103/PhysRevD.78.052002}{\emph{Phys. Rev. D}
  {\bfseries 78} (2008) 052002}
  [\href{https://arxiv.org/abs/0711.0728}{{\ttfamily 0711.0728}}].

\bibitem{PhysRevLett.120.211801}
{\scshape OPERA} collaboration, \emph{{Final Results of the OPERA Experiment on
  $\nu_\tau$ Appearance in the CNGS Neutrino Beam}},
  \href{https://doi.org/10.1103/PhysRevLett.120.211801}{\emph{Phys. Rev. Lett.}
  {\bfseries 120} (2018) 211801}
  [\href{https://arxiv.org/abs/1804.04912}{{\ttfamily 1804.04912}}].

\bibitem{PhysRevD.100.092001}
{\scshape MINERvA} collaboration, \emph{{Constraint of the MINER$\nu$A medium
  energy neutrino flux using neutrino-electron elastic scattering}},
  \href{https://doi.org/10.1103/PhysRevD.100.092001}{\emph{Phys. Rev. D}
  {\bfseries 100} (2019) 092001}
  [\href{https://arxiv.org/abs/1906.00111}{{\ttfamily 1906.00111}}].

\bibitem{LlewellynSmith:1971uhs}
C.H.~Llewellyn~Smith, \emph{{Neutrino Reactions at Accelerator Energies}},
  \href{https://doi.org/10.1016/0370-1573(72)90010-5}{\emph{Phys. Rept.}
  {\bfseries 3} (1972) 261}.

\bibitem{Fatima:2018wsy}
A.~Fatima, M.S.~Athar and S.K.~Singh, \emph{{Weak Quasielastic Hyperon
  Production Leading to Pions in the Antineutrino-Nucleus Reactions}},
  \href{https://doi.org/10.3389/fphy.2019.00013}{\emph{Front.in Phys.}
  {\bfseries 7} (2019) 13} [\href{https://arxiv.org/abs/1807.08314}{{\ttfamily
  1807.08314}}].

\bibitem{Akbar:2016awk}
F.~Akbar, M.~Rafi~Alam, M.~Sajjad~Athar and S.~Singh, \emph{{Quasielastic
  production of polarized hyperons in antineutrino--nucleon reactions}},
  \href{https://doi.org/10.1103/PhysRevD.94.114031}{\emph{Phys. Rev. D}
  {\bfseries 94} (2016) 114031}
  [\href{https://arxiv.org/abs/1608.02103}{{\ttfamily 1608.02103}}].

\bibitem{Bernard:2001rs}
V.~Bernard, L.~Elouadrhiri and U.-G.~Meissner, \emph{{Axial structure of the
  nucleon: Topical Review}},
  \href{https://doi.org/10.1088/0954-3899/28/1/201}{\emph{J. Phys.} {\bfseries
  G28} (2002) R1} [\href{https://arxiv.org/abs/hep-ph/0107088}{{\ttfamily
  hep-ph/0107088}}].

\bibitem{Bhattacharya:2011ah}
B.~Bhattacharya, R.J.~Hill and G.~Paz, \emph{{Model independent determination
  of the axial mass parameter in quasielastic neutrino-nucleon scattering}},
  \href{https://doi.org/10.1103/PhysRevD.84.073006}{\emph{Phys. Rev. D}
  {\bfseries 84} (2011) 073006}
  [\href{https://arxiv.org/abs/1108.0423}{{\ttfamily 1108.0423}}].

\bibitem{Alam:2014bya}
M.R.~Alam, M.S.~Athar, S.~Chauhan and S.K.~Singh, \emph{{Quasielastic hyperon
  production in ${{\bar{\nu }}_{\mu }}$-nucleus interactions}},
  \href{https://doi.org/10.1088/0954-3899/42/5/055107}{\emph{J. Phys.}
  {\bfseries G42} (2015) 055107}
  [\href{https://arxiv.org/abs/1409.2145}{{\ttfamily 1409.2145}}].

\bibitem{Alam:2013cra}
M.~Rafi~Alam, S.~Chauhan, M.~Sajjad~Athar and S.K.~Singh, \emph{{$\bar{\nu}_l$
  induced pion production from nuclei at $\sim{1}$ GeV}},
  \href{https://doi.org/10.1103/PhysRevD.88.077301}{\emph{Phys. Rev.}
  {\bfseries D88} (2013) 077301}
  [\href{https://arxiv.org/abs/1310.7704}{{\ttfamily 1310.7704}}].

\bibitem{Fatima:2021ctt}
A.~Fatima, M.S.~Athar and S.K.~Singh, \emph{{$\bar{\nu}_{\mu}$ induced
  quasielastic production of hyperons leading to pions}},
  \href{https://arxiv.org/abs/2106.14590}{{\ttfamily 2106.14590}}.

\bibitem{SajjadAthar:2020nvy}
M.~Sajjad~Athar and J.G.~Morf\'\i{}n, \emph{{Neutrino(antineutrino)-nucleus
  interactions in the shallow- and deep-inelastic scattering regions}},
  \href{https://doi.org/10.1088/1361-6471/abbb11}{\emph{J. Phys. G} {\bfseries
  48} (2021) 034001} [\href{https://arxiv.org/abs/2006.08603}{{\ttfamily
  2006.08603}}].

\bibitem{Hernandez:2007qq}
E.~Hernandez, J.~Nieves and M.~Valverde, \emph{{Weak Pion Production off the
  Nucleon}}, \href{https://doi.org/10.1103/PhysRevD.76.033005}{\emph{Phys.
  Rev.} {\bfseries D76} (2007) 033005}
  [\href{https://arxiv.org/abs/hep-ph/0701149}{{\ttfamily hep-ph/0701149}}].

\bibitem{Alam:2015gaa}
M.~Rafi~Alam, M.~Sajjad~Athar, S.~Chauhan and S.K.~Singh, \emph{{Weak charged
  and neutral current induced one pion production off the nucleon}},
  \href{https://doi.org/10.1142/S0218301316500105}{\emph{Int. J. Mod. Phys.}
  {\bfseries E25} (2016) 1650010}
  [\href{https://arxiv.org/abs/1509.08622}{{\ttfamily 1509.08622}}].

\bibitem{Wilkinson:2014yfa}
C.~Wilkinson, P.~Rodrigues, S.~Cartwright, L.~Thompson and K.~McFarland,
  \emph{{Reanalysis of bubble chamber measurements of muon-neutrino induced
  single pion production}},
  \href{https://doi.org/10.1103/PhysRevD.90.112017}{\emph{Phys. Rev.}
  {\bfseries D90} (2014) 112017}
  [\href{https://arxiv.org/abs/1411.4482}{{\ttfamily 1411.4482}}].

\bibitem{RafiAlam:2010kf}
M.~Rafi~Alam, I.~Ruiz~Simo, M.~Sajjad~Athar and M.J.~Vicente~Vacas, \emph{{Weak
  Kaon Production off the Nucleon}},
  \href{https://doi.org/10.1103/PhysRevD.82.033001}{\emph{Phys. Rev. D}
  {\bfseries 82} (2010) 033001}
  [\href{https://arxiv.org/abs/1004.5484}{{\ttfamily 1004.5484}}].

\bibitem{Alam:2011vwg}
M.R.~Alam, I.R.~Simo, M.S.~Athar and M.J.~Vicente~Vacas, \emph{{$\bar{\nu}$
  induced $\bar{K}$ production off the nucleon}},
  \href{https://doi.org/10.1103/PhysRevD.85.013014}{\emph{Phys. Rev. D}
  {\bfseries 85} (2012) 013014}
  [\href{https://arxiv.org/abs/1111.0863}{{\ttfamily 1111.0863}}].

\bibitem{RafiAlam:2013jcs}
M.~Rafi~Alam, L.~Alvarez-Ruso, M.~Sajjad~Athar and M.J.~Vicente~Vacas,
  \emph{{Weak \ensuremath{\eta} production off the nucleon}},
  \href{https://doi.org/10.1063/1.4919520}{\emph{AIP Conf. Proc.} {\bfseries
  1663} (2015) 120014} [\href{https://arxiv.org/abs/1303.5951}{{\ttfamily
  1303.5951}}].

\bibitem{Adera:2010zz}
G.B.~Adera, B.I.S.~Van Der~Ventel, D.D.~van Niekerk and T.~Mart,
  \emph{{Strange-particle production via the weak interaction}},
  \href{https://doi.org/10.1103/PhysRevC.82.025501}{\emph{Phys. Rev. C}
  {\bfseries 82} (2010) 025501}
  [\href{https://arxiv.org/abs/1112.5748}{{\ttfamily 1112.5748}}].

\bibitem{RafiAlam:2013tgm}
M.~Rafi~Alam, M.~Sajjad~Athar, L.~Alvarez-Ruso, I.~Ruiz~Simo,
  M.J.~Vicente~Vacas and S.K.~Singh, \emph{{Weak Strangeness and Eta
  Production}},  in \emph{{15th International Workshop on Neutrino Factories,
  Super Beams and Beta Beams}}, 11, 2013
  [\href{https://arxiv.org/abs/1311.2293}{{\ttfamily 1311.2293}}].

\bibitem{Yao:2018pzc}
D.-L.~Yao, L.~Alvarez-Ruso, A.N.~Hiller~Blin and M.J.~Vicente~Vacas,
  \emph{{Weak pion production off the nucleon in covariant chiral perturbation
  theory}}, \href{https://doi.org/10.1103/PhysRevD.98.076004}{\emph{Phys. Rev.
  D} {\bfseries 98} (2018) 076004}
  [\href{https://arxiv.org/abs/1806.09364}{{\ttfamily 1806.09364}}].

\bibitem{Alvarez-Ruso:2015eva}
L.~Alvarez-Ruso, E.~Hern\'andez, J.~Nieves and M.J.~Vicente~Vacas,
  \emph{{Watson's theorem and the $N\Delta(1232)$ axial transition}},
  \href{https://doi.org/10.1103/PhysRevD.93.014016}{\emph{Phys. Rev. D}
  {\bfseries 93} (2016) 014016}
  [\href{https://arxiv.org/abs/1510.06266}{{\ttfamily 1510.06266}}].

\bibitem{Nakamura:2015rta}
S.X.~Nakamura, H.~Kamano and T.~Sato, \emph{{Dynamical coupled-channels model
  for neutrino-induced meson productions in resonance region}},
  \href{https://doi.org/10.1103/PhysRevD.92.074024}{\emph{Phys. Rev. D}
  {\bfseries 92} (2015) 074024}
  [\href{https://arxiv.org/abs/1506.03403}{{\ttfamily 1506.03403}}].

\bibitem{Zaidi:2019asc}
F.~Zaidi, H.~Haider, M.~Sajjad~Athar, S.K.~Singh and I.~Ruiz~Simo, \emph{{Weak
  structure functions in $\nu_l-N$ and $\nu_l-A$ scattering with
  nonperturbative and higher order perturbative QCD effects}},
  \href{https://doi.org/10.1103/PhysRevD.101.033001}{\emph{Phys. Rev.}
  {\bfseries D101} (2020) 033001}
  [\href{https://arxiv.org/abs/1911.12573}{{\ttfamily 1911.12573}}].

\bibitem{AbdulKhalek:2020yuc}
R.~Abdul~Khalek, J.J.~Ethier, J.~Rojo and G.~van Weelden, \emph{{nNNPDF2.0:
  quark flavor separation in nuclei from LHC data}},
  \href{https://doi.org/10.1007/JHEP09(2020)183}{\emph{JHEP} {\bfseries 09}
  (2020) 183} [\href{https://arxiv.org/abs/2006.14629}{{\ttfamily
  2006.14629}}].

\bibitem{Zeller:2003ey}
G.~Zeller, \emph{{Low-energy neutrino cross-sections: Comparison of various
  Monte Carlo predictions to experimental data}},  in \emph{{2nd International
  Workshop on Neutrino-Nucleus Interactions in the Few GeV Region}}, 12, 2003
  [\href{https://arxiv.org/abs/hep-ex/0312061}{{\ttfamily hep-ex/0312061}}].

\bibitem{Abi:2020qib}
{\scshape DUNE} collaboration, \emph{{Long-baseline neutrino oscillation
  physics potential of the DUNE experiment}},
  \href{https://arxiv.org/abs/2006.16043}{{\ttfamily 2006.16043}}.

\bibitem{Alvarez-Ruso:2017oui}
{\scshape NuSTEC} collaboration, \emph{{NuSTEC White Paper: Status and
  challenges of neutrino–nucleus scattering}},
  \href{https://doi.org/10.1016/j.ppnp.2018.01.006}{\emph{Prog. Part. Nucl.
  Phys.} {\bfseries 100} (2018) 1}
  [\href{https://arxiv.org/abs/1706.03621}{{\ttfamily 1706.03621}}].

\bibitem{Katori:2016yel}
T.~Katori and M.~Martini, \emph{{Neutrino–nucleus cross sections for
  oscillation experiments}},
  \href{https://doi.org/10.1088/1361-6471/aa8bf7}{\emph{J. Phys.} {\bfseries
  G45} (2018) 013001} [\href{https://arxiv.org/abs/1611.07770}{{\ttfamily
  1611.07770}}].

\bibitem{Andreopoulos:2009rq}
C.~Andreopoulos et~al., \emph{{The GENIE Neutrino Monte Carlo Generator}},
  \href{https://doi.org/10.1016/j.nima.2009.12.009}{\emph{Nucl. Instrum. Meth.
  A} {\bfseries 614} (2010) 87}
  [\href{https://arxiv.org/abs/0905.2517}{{\ttfamily 0905.2517}}].

\bibitem{Golan:2012wx}
T.~Golan, C.~Juszczak and J.T.~Sobczyk, \emph{{Final State Interactions Effects
  in Neutrino-Nucleus Interactions}},
  \href{https://doi.org/10.1103/PhysRevC.86.015505}{\emph{Phys. Rev. C}
  {\bfseries 86} (2012) 015505}
  [\href{https://arxiv.org/abs/1202.4197}{{\ttfamily 1202.4197}}].

\bibitem{Hayato:2009zz}
Y.~Hayato, \emph{{A neutrino interaction simulation program library NEUT}},
  {\emph{Acta Phys. Polon. B} {\bfseries 40} (2009) 2477}.

\bibitem{Buss:2011mx}
O.~Buss, T.~Gaitanos, K.~Gallmeister, H.~van Hees, M.~Kaskulov, O.~Lalakulich
  et~al., \emph{{Transport-theoretical Description of Nuclear Reactions}},
  \href{https://doi.org/10.1016/j.physrep.2011.12.001}{\emph{Phys. Rept.}
  {\bfseries 512} (2012) 1} [\href{https://arxiv.org/abs/1106.1344}{{\ttfamily
  1106.1344}}].

\bibitem{Smith:1972xh}
R.~Smith and E.~Moniz, \emph{{NEUTRINO REACTIONS ON NUCLEAR TARGETS}},
  \href{https://doi.org/10.1016/0550-3213(75)90612-4}{\emph{Nucl. Phys. B}
  {\bfseries 43} (1972) 605}.

\bibitem{Subedi:2008zz}
R.~Subedi et~al., \emph{{Probing Cold Dense Nuclear Matter}},
  \href{https://doi.org/10.1126/science.1156675}{\emph{Science} {\bfseries 320}
  (2008) 1476} [\href{https://arxiv.org/abs/0908.1514}{{\ttfamily 0908.1514}}].

\bibitem{minibooneccqe}
{\scshape MiniBooNE} collaboration, \emph{{First Measurement of the Muon
  Neutrino Charged Current Quasielastic Double Differential Cross Section}},
  \href{https://doi.org/10.1103/PhysRevD.81.092005}{\emph{Phys. Rev. D}
  {\bfseries 81} (2010) 092005}
  [\href{https://arxiv.org/abs/1002.2680}{{\ttfamily 1002.2680}}].

\bibitem{t2kccqe}
{\scshape T2K} collaboration, \emph{{Measurement of the $\nu_\mu$
  charged-current quasielastic cross section on carbon with the ND280 detector
  at T2K}}, \href{https://doi.org/10.1103/PhysRevD.92.112003}{\emph{Phys. Rev.
  D} {\bfseries 92} (2015) 112003}
  [\href{https://arxiv.org/abs/1411.6264}{{\ttfamily 1411.6264}}].

\bibitem{laura}
{\scshape MINERvA} collaboration, \emph{{Measurement of Muon Antineutrino
  Quasielastic Scattering on a Hydrocarbon Target at $E_{\nu} \sim 3.5$ GeV}},
  \href{https://doi.org/10.1103/PhysRevLett.111.022501}{\emph{Phys. Rev. Lett.}
  {\bfseries 111} (2013) 022501}
  [\href{https://arxiv.org/abs/1305.2234}{{\ttfamily 1305.2234}}].

\bibitem{arturo}
{\scshape MINERvA} collaboration, \emph{{Measurement of Muon Neutrino
  Quasielastic Scattering on a Hydrocarbon Target at $E_{\nu}\sim$ 3.5 GeV}},
  \href{https://doi.org/10.1103/PhysRevLett.111.022502}{\emph{Phys. Rev. Lett.}
  {\bfseries 111} (2013) 022502}
  [\href{https://arxiv.org/abs/1305.2243}{{\ttfamily 1305.2243}}].

\bibitem{Carlson:2001mp}
J.~Carlson, J.~Jourdan, R.~Schiavilla and I.~Sick, \emph{{Longitudinal and
  transverse quasielastic response functions of light nuclei}},
  \href{https://doi.org/10.1103/PhysRevC.65.024002}{\emph{Phys. Rev. C}
  {\bfseries 65} (2002) 024002}
  [\href{https://arxiv.org/abs/nucl-th/0106047}{{\ttfamily nucl-th/0106047}}].

\bibitem{Martini:2011ui}
M.~Martini, \emph{{Two Particle-Two Hole Excitations in Charged Current
  Quasielastic Neutrino-Nucleus Interactions}},
  \href{https://doi.org/10.1088/1742-6596/408/1/012041}{\emph{J. Phys. Conf.
  Ser.} {\bfseries 408} (2013) 012041}
  [\href{https://arxiv.org/abs/1110.5895}{{\ttfamily 1110.5895}}].

\bibitem{Benhar:2015ula}
O.~Benhar, A.~Lovato and N.~Rocco, \emph{{Contribution of two-particle-two-hole
  final states to the nuclear response}},
  \href{https://doi.org/10.1103/PhysRevC.92.024602}{\emph{Phys. Rev. C}
  {\bfseries 92} (2015) 024602}
  [\href{https://arxiv.org/abs/1502.00887}{{\ttfamily 1502.00887}}].

\bibitem{Nieves:2016sma}
J.~Nieves, I.R.~Simo, F.~S\'anchez and M.J.~Vicente~Vacas, \emph{{2p2h
  Excitations, MEC, Nucleon Correlations and Other Sources of QE-like Events}},
  \href{https://doi.org/10.7566/JPSCP.12.010002}{\emph{JPS Conf. Proc.}
  {\bfseries 12} (2016) 010002}.

\bibitem{Megias:2016fjk}
G.D.~Megias, J.E.~Amaro, M.B.~Barbaro, J.A.~Caballero, T.W.~Donnelly and
  I.~Ruiz~Simo, \emph{{Charged-current neutrino-nucleus reactions within the
  superscaling meson-exchange current approach}},
  \href{https://doi.org/10.1103/PhysRevD.94.093004}{\emph{Phys. Rev. D}
  {\bfseries 94} (2016) 093004}
  [\href{https://arxiv.org/abs/1607.08565}{{\ttfamily 1607.08565}}].

\bibitem{Barbaro:2016hrt}
M.B.~Barbaro, J.E.~Amaro, J.A.~Caballero, A.~De~Pace, T.W.~Donnelly,
  G.D.~Megias et~al., \emph{{The role of meson exchange currents in charged
  current (anti)neutrino-nucleus scattering}}, {\emph{Nucl. Theor.} {\bfseries
  35} (2016) 60} [\href{https://arxiv.org/abs/1610.02924}{{\ttfamily
  1610.02924}}].

\bibitem{Dolan:2018sbb}
S.~Dolan, U.~Mosel, K.~Gallmeister, L.~Pickering and S.~Bolognesi,
  \emph{{Sensitivity of Neutrino-Nucleus Interaction Measurements to 2p2h
  Excitations}}, \href{https://doi.org/10.1103/PhysRevC.98.045502}{\emph{Phys.
  Rev. C} {\bfseries 98} (2018) 045502}
  [\href{https://arxiv.org/abs/1804.09488}{{\ttfamily 1804.09488}}].

\bibitem{Rocco:2020jlx}
N.~Rocco, \emph{{Ab initio Calculations of Lepton-Nucleus Scattering}},
  \href{https://doi.org/10.3389/fphy.2020.00116}{\emph{Front. in Phys.}
  {\bfseries 8} (2020) 116}.

\bibitem{Barbaro:2021psv}
M.B.~Barbaro, A.~De~Pace and L.~Fiume, \emph{{The SuSA Model for Neutrino
  Oscillation Experiments: From Quasielastic Scattering to the Resonance
  Region}}, \href{https://doi.org/10.3390/universe7050140}{\emph{Universe}
  {\bfseries 7} (2021) 140} [\href{https://arxiv.org/abs/2104.10472}{{\ttfamily
  2104.10472}}].

\bibitem{Le:2019jfy}
{\scshape MINERvA} collaboration, \emph{{Measurement of $\bar{\nu}_{\mu}$
  Charged-Current Single $\pi^{-}$ Production on Hydrocarbon in the Few-GeV
  Region using MINERvA}},
  \href{https://doi.org/10.1103/PhysRevD.100.052008}{\emph{Phys. Rev. D}
  {\bfseries 100} (2019) 052008}
  [\href{https://arxiv.org/abs/1906.08300}{{\ttfamily 1906.08300}}].

\bibitem{Nieves:2004wx}
J.~Nieves, J.E.~Amaro and M.~Valverde, \emph{{Inclusive quasi-elastic neutrino
  reactions}}, \href{https://doi.org/10.1103/PhysRevC.70.055503}{\emph{Phys.
  Rev. C} {\bfseries 70} (2004) 055503}
  [\href{https://arxiv.org/abs/nucl-th/0408005}{{\ttfamily nucl-th/0408005}}].

\bibitem{Pandey:2014tza}
V.~Pandey, N.~Jachowicz, T.~Van~Cuyck, J.~Ryckebusch and M.~Martini,
  \emph{{Low-energy excitations and quasielastic contribution to
  electron-nucleus and neutrino-nucleus scattering in the continuum
  random-phase approximation}},
  \href{https://doi.org/10.1103/PhysRevC.92.024606}{\emph{Phys. Rev. C}
  {\bfseries 92} (2015) 024606}
  [\href{https://arxiv.org/abs/1412.4624}{{\ttfamily 1412.4624}}].

\bibitem{Vogel:1986nj}
P.~Vogel and M.~Zirnbauer, \emph{{Suppression of the Two Neutrino Double beta
  Decay by Nuclear Structure Effects}},
  \href{https://doi.org/10.1103/PhysRevLett.57.3148}{\emph{Phys. Rev. Lett.}
  {\bfseries 57} (1986) 3148}.

\bibitem{Auerbach:2001hz}
L.B.~Auerbach et~al., \emph{{Measurements of charged current reactions of nu/e
  on C- 12}}, \href{https://doi.org/10.1103/PhysRevC.64.065501}{\emph{Phys.
  Rev.} {\bfseries C64} (2001) 065501}
  [\href{https://arxiv.org/abs/hep-ex/0105068}{{\ttfamily hep-ex/0105068}}].

\bibitem{Auerbach:2002iy}
L.B.~Auerbach et~al., \emph{{Measurements of charged current reactions of nu/mu
  on C- 12}}, \href{https://doi.org/10.1103/PhysRevC.66.015501}{\emph{Phys.
  Rev.} {\bfseries C66} (2002) 015501}
  [\href{https://arxiv.org/abs/nucl-ex/0203011}{{\ttfamily nucl-ex/0203011}}].

\bibitem{Distel:2002ch}
J.~Distel, B.~Cleveland, K.~Lande, C.~Lee, P.~Wildenhain, G.~Allen et~al.,
  \emph{{Measurement of the cross-section for the reaction I-127 (nu(e), e-)
  Xe-127(bound states) with neutrinos from the decay of stopped muons}},
  \href{https://doi.org/10.1103/PhysRevC.68.054613}{\emph{Phys. Rev. C}
  {\bfseries 68} (2003) 054613}
  [\href{https://arxiv.org/abs/nucl-ex/0208012}{{\ttfamily nucl-ex/0208012}}].

\bibitem{Maschuw:1998jf}
{\scshape KARMEN} collaboration, \emph{{Neutrino spectroscopy with KARMEN}},
  \href{https://doi.org/10.1016/S0146-6410(98)00024-6}{\emph{Prog. Part. Nucl.
  Phys.} {\bfseries 40} (1998) 183}.

\bibitem{Formaggio:2013kya}
J.A.~Formaggio and G.P.~Zeller, \emph{{From eV to EeV: Neutrino Cross Sections
  Across Energy Scales}},
  \href{https://doi.org/10.1103/RevModPhys.84.1307}{\emph{Rev.\ Mod.\ Phys.}
  {\bfseries 84} (2012) 1307}
  [\href{https://arxiv.org/abs/1305.7513}{{\ttfamily 1305.7513}}].

\bibitem{Armbruster:1998gk}
B.~Armbruster et~al., \emph{{Measurement of the weak neutral current excitation
  C- 12(nu(mu) nu'(mu))C*-12(1+,1,15.1-MeV) at E(nu(mu)) = 29.8- MeV}},
  \href{https://doi.org/10.1016/S0370-2693(98)00087-2}{\emph{Phys. Lett.}
  {\bfseries B423} (1998) 15}.

\bibitem{Bolozdynya:2012xv}
A.~Bolozdynya et~al., \emph{{Opportunities for Neutrino Physics at the
  Spallation Neutron Source: A White Paper}},  11, 2012
  [\href{https://arxiv.org/abs/1211.5199}{{\ttfamily 1211.5199}}].

\bibitem{Ajimura:2017ul}
S.~Ajimura et~al., \emph{{Technical Design Report (TDR): Searching for a
  Sterile Neutrino at J-PARC MLF (E56, JSNS2)}},
  \href{https://arxiv.org/abs/1705.08629}{{\ttfamily 1705.08629}}.

\bibitem{ccm}
\url{https://p25ext.lanl.gov/lee/CaptainMills/Documentation/LDRD-DR-Proposal_2018.pdf}.

\bibitem{Baxter:2019mcx}
D.~Baxter et~al., \emph{{Coherent Elastic Neutrino-Nucleus Scattering at the
  European Spallation Source}},
  \href{https://doi.org/10.1007/JHEP02(2020)123}{\emph{JHEP} {\bfseries 02}
  (2020) 123} [\href{https://arxiv.org/abs/1911.00762}{{\ttfamily
  1911.00762}}].

\bibitem{Wang:2013aka}
F.~Wang, T.~Liang, W.~Yin, Q.~Yu, L.~He, J.~Tao et~al., \emph{{Physical design
  of target station and neutron instruments for China Spallation Neutron
  Source}}, \href{https://doi.org/10.1007/s11433-013-5345-5}{\emph{Sci. China
  Phys. Mech. Astron.} {\bfseries 56} (2013) 2410}.

\bibitem{Alonso:2010fs}
J.~Alonso, F.~Avignone, W.~Barletta, R.~Barlow, H.~Baumgartner et~al.,
  \emph{{Expression of Interest for a Novel Search for CP Violation in the
  Neutrino Sector: DAE$\delta$ALUS}},
  \href{https://arxiv.org/abs/1006.0260}{{\ttfamily 1006.0260}}.

\bibitem{PhysRevD.9.1389}
D.Z.~Freedman, \emph{Coherent neutrino nucleus scattering as a probe of the
  weak neutral current},
  \href{https://doi.org/10.1103/PhysRevD.9.1389}{\emph{Phys. Rev. D} {\bfseries
  9} (1974) 1389}.

\bibitem{Freedman:1977xn}
D.Z.~Freedman, D.N.~Schramm and D.L.~Tubbs, \emph{{The Weak Neutral Current and
  Its Effects in Stellar Collapse}},
  \href{https://doi.org/10.1146/annurev.ns.27.120177.001123}{\emph{Ann. Rev.
  Nucl. Part. Sci.} {\bfseries 27} (1977) 167}.

\bibitem{Drukier:1983gj}
A.~Drukier and L.~Stodolsky, \emph{{Principles and Applications of a Neutral
  Current Detector for Neutrino Physics and Astronomy}},
  \href{https://doi.org/10.1103/PhysRevD.30.2295}{\emph{Phys. Rev. D}
  {\bfseries 30} (1984) 2295}.

\bibitem{Scholberg:2005qs}
K.~Scholberg, \emph{{Prospects for measuring coherent neutrino-nucleus elastic
  scattering at a stopped-pion neutrino source}},
  \href{https://doi.org/10.1103/PhysRevD.73.033005}{\emph{Phys. Rev. D}
  {\bfseries 73} (2006) 033005}
  [\href{https://arxiv.org/abs/hep-ex/0511042}{{\ttfamily hep-ex/0511042}}].

\bibitem{Amanik:2009zz}
P.~Amanik and G.~McLaughlin, \emph{{Nuclear neutron form factor from neutrino
  nucleus coherent elastic scattering}},
  \href{https://doi.org/10.1088/0954-3899/36/1/015105}{\emph{J. Phys. G}
  {\bfseries 36} (2009) 015105}.

\bibitem{Barranco:2005yy}
J.~Barranco, O.~Miranda and T.~Rashba, \emph{{Probing new physics with coherent
  neutrino scattering off nuclei}},
  \href{https://doi.org/10.1088/1126-6708/2005/12/021}{\emph{JHEP} {\bfseries
  12} (2005) 021} [\href{https://arxiv.org/abs/hep-ph/0508299}{{\ttfamily
  hep-ph/0508299}}].

\bibitem{deNiverville:2015mwa}
P.~deNiverville, M.~Pospelov and A.~Ritz, \emph{{Light new physics in coherent
  neutrino-nucleus scattering experiments}},
  \href{https://doi.org/10.1103/PhysRevD.92.095005}{\emph{Phys. Rev. D}
  {\bfseries 92} (2015) 095005}
  [\href{https://arxiv.org/abs/1505.07805}{{\ttfamily 1505.07805}}].

\bibitem{Lindner:2016wff}
M.~Lindner, W.~Rodejohann and X.-J.~Xu, \emph{{Coherent Neutrino-Nucleus
  Scattering and new Neutrino Interactions}},
  \href{https://doi.org/10.1007/JHEP03(2017)097}{\emph{JHEP} {\bfseries 03}
  (2017) 097} [\href{https://arxiv.org/abs/1612.04150}{{\ttfamily
  1612.04150}}].

\bibitem{Brdar:2018qqj}
V.~Brdar, W.~Rodejohann and X.-J.~Xu, \emph{{Producing a new Fermion in
  Coherent Elastic Neutrino-Nucleus Scattering: from Neutrino Mass to Dark
  Matter}}, \href{https://doi.org/10.1007/JHEP12(2018)024}{\emph{JHEP}
  {\bfseries 12} (2018) 024}
  [\href{https://arxiv.org/abs/1810.03626}{{\ttfamily 1810.03626}}].

\bibitem{Kosmas:2015vsa}
T.~Kosmas, O.~Miranda, D.~Papoulias, M.~Tortola and J.~Valle,
  \emph{{Sensitivities to neutrino electromagnetic properties at the TEXONO
  experiment}},
  \href{https://doi.org/10.1016/j.physletb.2015.09.054}{\emph{Phys. Lett. B}
  {\bfseries 750} (2015) 459}
  [\href{https://arxiv.org/abs/1506.08377}{{\ttfamily 1506.08377}}].

\bibitem{Formaggio:2011jt}
J.A.~Formaggio, E.~Figueroa-Feliciano and A.~Anderson, \emph{{Sterile
  Neutrinos, Coherent Scattering and Oscillometry Measurements with
  Low-temperature Bolometers}},
  \href{https://doi.org/10.1103/PhysRevD.85.013009}{\emph{Phys. Rev. D}
  {\bfseries 85} (2012) 013009}
  [\href{https://arxiv.org/abs/1107.3512}{{\ttfamily 1107.3512}}].

\bibitem{Akimov:2017ade}
{\scshape COHERENT} collaboration, \emph{{Observation of Coherent Elastic
  Neutrino-Nucleus Scattering}},
  \href{https://doi.org/10.1126/science.aao0990}{\emph{Science} {\bfseries 357}
  (2017) 1123} [\href{https://arxiv.org/abs/1708.01294}{{\ttfamily
  1708.01294}}].

\bibitem{Akimov:2020pdx}
{\scshape COHERENT} collaboration, \emph{{First Measurement of Coherent Elastic
  Neutrino-Nucleus Scattering on Argon}},
  \href{https://doi.org/10.1103/PhysRevLett.126.012002}{\emph{Phys. Rev. Lett.}
  {\bfseries 126} (2021) 012002}
  [\href{https://arxiv.org/abs/2003.10630}{{\ttfamily 2003.10630}}].

\bibitem{CCM:2021leg}
{\scshape CCM} collaboration, \emph{{First Dark Matter Search Results From
  Coherent CAPTAIN-Mills}},  \href{https://arxiv.org/abs/2105.14020}{{\ttfamily
  2105.14020}}.

\bibitem{Hakenmuller:2019ecb}
J.~Hakenmüller et~al., \emph{{Neutron-induced background in the CONUS
  experiment}},
  \href{https://doi.org/10.1140/epjc/s10052-019-7160-2}{\emph{Eur. Phys. J. C}
  {\bfseries 79} (2019) 699}
  [\href{https://arxiv.org/abs/1903.09269}{{\ttfamily 1903.09269}}].

\bibitem{Bonet:2020awv}
{\scshape CONUS} collaboration, \emph{{Constraints on Elastic Neutrino Nucleus
  Scattering in the Fully Coherent Regime from the CONUS Experiment}},
  \href{https://doi.org/10.1103/PhysRevLett.126.041804}{\emph{Phys. Rev. Lett.}
  {\bfseries 126} (2021) 041804}
  [\href{https://arxiv.org/abs/2011.00210}{{\ttfamily 2011.00210}}].

\bibitem{Aguilar-Arevalo:2019jlr}
{\scshape CONNIE} collaboration, \emph{{Exploring low-energy neutrino physics
  with the Coherent Neutrino Nucleus Interaction Experiment}},
  \href{https://doi.org/10.1103/PhysRevD.100.092005}{\emph{Phys. Rev. D}
  {\bfseries 100} (2019) 092005}
  [\href{https://arxiv.org/abs/1906.02200}{{\ttfamily 1906.02200}}].

\bibitem{Billard:2016giu}
J.~Billard et~al., \emph{{Coherent Neutrino Scattering with Low Temperature
  Bolometers at Chooz Reactor Complex}},
  \href{https://doi.org/10.1088/1361-6471/aa83d0}{\emph{J. Phys. G} {\bfseries
  44} (2017) 105101} [\href{https://arxiv.org/abs/1612.09035}{{\ttfamily
  1612.09035}}].

\bibitem{RED-100:2019rpf}
{\scshape RED-100} collaboration, \emph{{First ground-level laboratory test of
  the two-phase xenon emission detector RED-100}},
  \href{https://doi.org/10.1088/1748-0221/15/02/P02020}{\emph{JINST} {\bfseries
  15} (2020) P02020} [\href{https://arxiv.org/abs/1910.06190}{{\ttfamily
  1910.06190}}].

\bibitem{Agnolet:2016zir}
{\scshape MINER} collaboration, \emph{{Background Studies for the MINER
  Coherent Neutrino Scattering Reactor Experiment}},
  \href{https://doi.org/10.1016/j.nima.2017.02.024}{\emph{Nucl. Instrum. Meth.
  A} {\bfseries 853} (2017) 53}
  [\href{https://arxiv.org/abs/1609.02066}{{\ttfamily 1609.02066}}].

\bibitem{Rothe:2019aii}
{\scshape NUCLEUS} collaboration, \emph{{NUCLEUS: Exploring Coherent
  Neutrino-Nucleus Scattering with Cryogenic Detectors}},
  \href{https://doi.org/10.1007/s10909-019-02283-7}{\emph{J. Low Temp. Phys.}
  {\bfseries 199} (2019) 433}.

\bibitem{Belov:2015ufh}
V.~Belov et~al., \emph{{The $\nu$GeN experiment at the Kalinin Nuclear Power
  Plant}}, \href{https://doi.org/10.1088/1748-0221/10/12/P12011}{\emph{JINST}
  {\bfseries 10} (2015) P12011}.

\bibitem{Choi:2020gkm}
J.J.~Choi, \emph{{Neutrino Elastic-scattering Observation with NaI[Tl](NEON)}},
  \href{https://doi.org/10.22323/1.369.0047}{\emph{PoS} {\bfseries NuFact2019}
  (2020) 047}.

\bibitem{Aprile:2020thb}
{\scshape XENON} collaboration, \emph{{Search for Coherent Elastic Scattering
  of Solar $^8$B Neutrinos in the XENON1T Dark Matter Experiment}},
  \href{https://doi.org/10.1103/PhysRevLett.126.091301}{\emph{Phys. Rev. Lett.}
  {\bfseries 126} (2021) 091301}
  [\href{https://arxiv.org/abs/2012.02846}{{\ttfamily 2012.02846}}].

\bibitem{Gandhi:1998ri}
R.~Gandhi, C.~Quigg, M.H.~Reno and I.~Sarcevic, \emph{{Neutrino interactions at
  ultrahigh-energies}},
  \href{https://doi.org/10.1103/PhysRevD.58.093009}{\emph{Phys. Rev.}
  {\bfseries D58} (1998) 093009}
  [\href{https://arxiv.org/abs/hep-ph/9807264}{{\ttfamily hep-ph/9807264}}].

\bibitem{Reno:2004eb}
M.H.~Reno, \emph{{Neutrino cross sections at HERA and beyond}},
  \href{https://doi.org/10.1016/j.nuclphysbps.2005.07.063}{\emph{Nucl. Phys. B
  Proc. Suppl.} {\bfseries 151} (2006) 255}
  [\href{https://arxiv.org/abs/hep-ph/0412412}{{\ttfamily hep-ph/0412412}}].

\bibitem{Jeong:2010za}
Y.S.~Jeong and M.~Reno, \emph{{Quark mass effects in high energy neutrino
  nucleon scattering}},
  \href{https://doi.org/10.1103/PhysRevD.81.114012}{\emph{Phys. Rev. D}
  {\bfseries 81} (2010) 114012}
  [\href{https://arxiv.org/abs/1001.4175}{{\ttfamily 1001.4175}}].

\bibitem{Gluck:2010rw}
M.~Glück, P.~Jimenez-Delgado and E.~Reya, \emph{{On the charged current
  neutrino-nucleon total cross section at high energies}},
  \href{https://doi.org/10.1103/PhysRevD.81.097501}{\emph{Phys. Rev. D}
  {\bfseries 81} (2010) 097501}
  [\href{https://arxiv.org/abs/1003.3168}{{\ttfamily 1003.3168}}].

\bibitem{Connolly:2011vc}
A.~Connolly, R.S.~Thorne and D.~Waters, \emph{{Calculation of High Energy
  Neutrino-Nucleon Cross Sections and Uncertainties Using the MSTW Parton
  Distribution Functions and Implications for Future Experiments}},
  \href{https://doi.org/10.1103/PhysRevD.83.113009}{\emph{Phys. Rev.}
  {\bfseries D83} (2011) 113009}
  [\href{https://arxiv.org/abs/astro-ph/1102.0691}{{\ttfamily
  astro-ph/1102.0691}}].

\bibitem{CooperSarkar:2011pa}
A.~Cooper-Sarkar, P.~Mertsch and S.~Sarkar, \emph{{The high energy neutrino
  cross-section in the Standard Model and its uncertainty}},
  \href{https://doi.org/10.1007/JHEP08(2011)042}{\emph{JHEP} {\bfseries 08}
  (2011) 042} [\href{https://arxiv.org/abs/1106.3723}{{\ttfamily 1106.3723}}].

\bibitem{Block:2014kza}
M.M.~Block, L.~Durand and P.~Ha, \emph{{Connection of the virtual $\gamma^*p$
  cross section of ep deep inelastic scattering to real $\gamma p$ scattering,
  and the implications for $\nu N$ and $ep$ total cross sections}},
  \href{https://doi.org/10.1103/PhysRevD.89.094027}{\emph{Phys.\ Rev.\ D}
  {\bfseries 89} (2014) 094027}
  [\href{https://arxiv.org/abs/1404.4530}{{\ttfamily 1404.4530}}].

\bibitem{Arguelles:2015wba}
C.A.~Arg{\"u}elles, F.~Halzen, L.~Wille, M.~Kroll and M.H.~Reno,
  \emph{{High-energy behavior of photon, neutrino, and proton cross sections}},
  \href{https://doi.org/10.1103/PhysRevD.92.074040}{\emph{Phys.\ Rev.\ D}
  {\bfseries 92} (2015) 074040}
  [\href{https://arxiv.org/abs/1504.06639}{{\ttfamily 1504.06639}}].

\bibitem{Gauld:2019pgt}
R.~Gauld, \emph{{Precise predictions for multi-TeV and PeV energy neutrino
  scattering rates}},
  \href{https://doi.org/10.1103/PhysRevD.100.091301}{\emph{Phys. Rev. D}
  {\bfseries 100} (2019) 091301}
  [\href{https://arxiv.org/abs/1905.03792}{{\ttfamily 1905.03792}}].

\bibitem{Henley:2005ms}
E.M.~Henley and J.~Jalilian-Marian, \emph{{Ultra-high energy neutrino-nucleon
  scattering and parton distributions at small x}},
  \href{https://doi.org/10.1103/PhysRevD.73.094004}{\emph{Phys.\ Rev.\ D}
  {\bfseries 73} (2006) 094004}
  [\href{https://arxiv.org/abs/hep-ph/0512220}{{\ttfamily hep-ph/0512220}}].

\bibitem{Ackermann:2019cxh}
M.~Ackermann et~al., \emph{{Fundamental Physics with High-Energy Cosmic
  Neutrinos}}, {\emph{Bull. Am. Astron. Soc.} {\bfseries 51} (2019) 215}
  [\href{https://arxiv.org/abs/1903.04333}{{\ttfamily 1903.04333}}].

\bibitem{Aartsen:2016nxy}
{\scshape IceCube} collaboration, \emph{{The IceCube Neutrino Observatory:
  Instrumentation and Online Systems}},
  \href{https://doi.org/10.1088/1748-0221/12/03/P03012}{\emph{JINST} {\bfseries
  12} (2017) P03012} [\href{https://arxiv.org/abs/1612.05093}{{\ttfamily
  1612.05093}}].

\bibitem{Klein:2019nbu}
S.R.~Klein, \emph{{Probing high-energy interactions of atmospheric and
  astrophysical neutrinos}},  pp.~75--107 (2020),
  \href{https://doi.org/10.1142/9789813275027\_0004}{DOI}
  [\href{https://arxiv.org/abs/1906.02221}{{\ttfamily 1906.02221}}].

\bibitem{Bustamante:2017xuy}
M.~Bustamante and A.~Connolly, \emph{{Extracting the Energy-Dependent
  Neutrino-Nucleon Cross Section Above 10 TeV Using IceCube Showers}},
  \href{https://doi.org/10.1103/PhysRevLett.122.041101}{\emph{Phys.\ Rev.\
  Lett.} {\bfseries 122} (2019) 041101}
  [\href{https://arxiv.org/abs/1711.11043}{{\ttfamily 1711.11043}}].

\bibitem{Aartsen:2018vez}
{\scshape IceCube} collaboration, \emph{{Measurements using the inelasticity
  distribution of multi-TeV neutrino interactions in IceCube}},
  \href{https://doi.org/10.1103/PhysRevD.99.032004}{\emph{Phys.\ Rev.\ D}
  {\bfseries 99} (2019) 032004}
  [\href{https://arxiv.org/abs/1808.07629}{{\ttfamily 1808.07629}}].

\bibitem{Feng:2001ue}
J.L.~Feng, P.~Fisher, F.~Wilczek and T.M.~Yu, \emph{{Observability of earth
  skimming ultrahigh-energy neutrinos}},
  \href{https://doi.org/10.1103/PhysRevLett.88.161102}{\emph{Phys. Rev. Lett.}
  {\bfseries 88} (2002) 161102}
  [\href{https://arxiv.org/abs/hep-ph/0105067}{{\ttfamily hep-ph/0105067}}].

\bibitem{Kusenko:2001gj}
A.~Kusenko and T.J.~Weiler, \emph{{Neutrino cross-sections at high-energies and
  the future observations of ultrahigh-energy cosmic rays}},
  \href{https://doi.org/10.1103/PhysRevLett.88.161101}{\emph{Phys.\ Rev.\
  Lett.} {\bfseries 88} (2002) 161101}
  [\href{https://arxiv.org/abs/hep-ph/0106071}{{\ttfamily hep-ph/0106071}}].

\bibitem{IceCube:2021rpz}
{\scshape IceCube} collaboration, \emph{{Detection of a particle shower at the
  Glashow resonance with IceCube}},
  \href{https://doi.org/10.1038/s41586-021-03256-1}{\emph{Nature} {\bfseries
  591} (2021) 220}.

\bibitem{Abreu:2019yak}
{\scshape FASER} collaboration, \emph{{Detecting and Studying High-Energy
  Collider Neutrinos with FASER at the LHC}},
  \href{https://doi.org/10.1140/epjc/s10052-020-7631-5}{\emph{Eur. Phys. J. C}
  {\bfseries 80} (2020) 61} [\href{https://arxiv.org/abs/1908.02310}{{\ttfamily
  1908.02310}}].

\bibitem{XSEN:2019bel}
{\scshape XSEN} collaboration, \emph{{XSEN: a $\nu$N Cross Section Measurement
  using High Energy Neutrinos from pp collisions at the LHC}},
  \href{https://arxiv.org/abs/1910.11340}{{\ttfamily 1910.11340}}.

\bibitem{SHiP:2020sos}
{\scshape SHiP} collaboration, \emph{{SND@LHC}},
  \href{https://arxiv.org/abs/2002.08722}{{\ttfamily 2002.08722}}.

\bibitem{Bai:2020ukz}
W.~Bai, M.~Diwan, M.V.~Garzelli, Y.S.~Jeong and M.H.~Reno, \emph{{Far-forward
  neutrinos at the Large Hadron Collider}},
  \href{https://doi.org/10.1007/JHEP06(2020)032}{\emph{JHEP} {\bfseries 06}
  (2020) 032} [\href{https://arxiv.org/abs/2002.03012}{{\ttfamily
  2002.03012}}].

\bibitem{Abreu:2020ddv}
{\scshape FASER} collaboration, \emph{{Technical Proposal: FASERnu}},
  \href{https://arxiv.org/abs/2001.03073}{{\ttfamily 2001.03073}}.

\bibitem{Aguilar:2001ty}
{\scshape LSND} collaboration, \emph{{Evidence for neutrino oscillations from
  the observation of $\bar{\nu}_e$ appearance in a $\bar{\nu}_\mu$ beam}},
  \href{https://doi.org/10.1103/PhysRevD.64.112007}{\emph{Phys. Rev. D}
  {\bfseries 64} (2001) 112007}
  [\href{https://arxiv.org/abs/hep-ex/0104049}{{\ttfamily hep-ex/0104049}}].

\bibitem{Aguilar-Arevalo:2018gpe}
{\scshape MiniBooNE} collaboration, \emph{{Significant Excess of ElectronLike
  Events in the MiniBooNE Short-Baseline Neutrino Experiment}},
  \href{https://doi.org/10.1103/PhysRevLett.121.221801}{\emph{Phys. Rev. Lett.}
  {\bfseries 121} (2018) 221801}
  [\href{https://arxiv.org/abs/1805.12028}{{\ttfamily 1805.12028}}].

\bibitem{Mention:2011rk}
G.~Mention, M.~Fechner, T.~Lasserre, T.~Mueller, D.~Lhuillier, M.~Cribier
  et~al., \emph{{The Reactor Antineutrino Anomaly}},
  \href{https://doi.org/10.1103/PhysRevD.83.073006}{\emph{Phys. Rev. D}
  {\bfseries 83} (2011) 073006}
  [\href{https://arxiv.org/abs/1101.2755}{{\ttfamily 1101.2755}}].

\bibitem{Giunti:2010zu}
C.~Giunti and M.~Laveder, \emph{{Statistical Significance of the Gallium
  Anomaly}}, \href{https://doi.org/10.1103/PhysRevC.83.065504}{\emph{Phys. Rev.
  C} {\bfseries 83} (2011) 065504}
  [\href{https://arxiv.org/abs/1006.3244}{{\ttfamily 1006.3244}}].

\bibitem{Barinov:2021asz}
V.V.~Barinov et~al., \emph{{Results from the Baksan Experiment on Sterile
  Transitions (BEST)}},  \href{https://arxiv.org/abs/2109.11482}{{\ttfamily
  2109.11482}}.

\bibitem{ALEPH:2005ab}
{\scshape ALEPH, DELPHI, L3, OPAL, SLD, LEP Electroweak Working Group, SLD
  Electroweak Group, SLD Heavy Flavour Group} collaboration, \emph{{Precision
  electroweak measurements on the $Z$ resonance}},
  \href{https://doi.org/10.1016/j.physrep.2005.12.006}{\emph{Phys. Rept.}
  {\bfseries 427} (2006) 257}
  [\href{https://arxiv.org/abs/hep-ex/0509008}{{\ttfamily hep-ex/0509008}}].

\bibitem{Maltoni:2002xd}
M.~Maltoni, T.~Schwetz, M.A.~Tortola and J.W.F.~Valle, \emph{{Ruling out four
  neutrino oscillation interpretations of the LSND anomaly?}},
  \href{https://doi.org/10.1016/S0550-3213(02)00747-2}{\emph{Nucl. Phys. B}
  {\bfseries 643} (2002) 321}
  [\href{https://arxiv.org/abs/hep-ph/0207157}{{\ttfamily hep-ph/0207157}}].

\bibitem{Dasgupta:2013zpn}
B.~Dasgupta and J.~Kopp, \emph{{Cosmologically Safe eV-Scale Sterile Neutrinos
  and Improved Dark Matter Structure}},
  \href{https://doi.org/10.1103/PhysRevLett.112.031803}{\emph{Phys. Rev. Lett.}
  {\bfseries 112} (2014) 031803}
  [\href{https://arxiv.org/abs/1310.6337}{{\ttfamily 1310.6337}}].

\bibitem{Hannestad:2013ana}
S.~Hannestad, R.S.~Hansen and T.~Tram, \emph{{How Self-Interactions can
  Reconcile Sterile Neutrinos with Cosmology}},
  \href{https://doi.org/10.1103/PhysRevLett.112.031802}{\emph{Phys. Rev. Lett.}
  {\bfseries 112} (2014) 031802}
  [\href{https://arxiv.org/abs/1310.5926}{{\ttfamily 1310.5926}}].

\bibitem{Agostini:2019jup}
M.~Agostini and B.~Neumair, \emph{{Statistical Methods Applied to the Search of
  Sterile Neutrinos}},
  \href{https://doi.org/10.1140/epjc/s10052-020-8279-x}{\emph{Eur. Phys. J. C}
  {\bfseries 80} (2020) 750}
  [\href{https://arxiv.org/abs/1906.11854}{{\ttfamily 1906.11854}}].

\bibitem{Giunti:2020uhv}
C.~Giunti, \emph{{Statistical Significance of Reactor Antineutrino
  Active-Sterile Oscillations}},
  \href{https://doi.org/10.1103/PhysRevD.101.095025}{\emph{Phys. Rev. D}
  {\bfseries 101} (2020) 095025}
  [\href{https://arxiv.org/abs/2004.07577}{{\ttfamily 2004.07577}}].

\bibitem{Armbruster:2002mp}
{\scshape KARMEN} collaboration, \emph{{Upper limits for neutrino oscillations
  muon-anti-neutrino ---\ensuremath{>} electron-anti-neutrino from muon decay
  at rest}}, \href{https://doi.org/10.1103/PhysRevD.65.112001}{\emph{Phys. Rev.
  D} {\bfseries 65} (2002) 112001}
  [\href{https://arxiv.org/abs/hep-ex/0203021}{{\ttfamily hep-ex/0203021}}].

\bibitem{Church:2002tc}
E.D.~Church, K.~Eitel, G.B.~Mills and M.~Steidl, \emph{{Statistical analysis of
  different muon-anti-neutrino ---\ensuremath{>} electron-anti-neutrino
  searches}}, \href{https://doi.org/10.1103/PhysRevD.66.013001}{\emph{Phys.
  Rev. D} {\bfseries 66} (2002) 013001}
  [\href{https://arxiv.org/abs/hep-ex/0203023}{{\ttfamily hep-ex/0203023}}].

\bibitem{Aguilar_Arevalo_2021}
{\scshape MiniBooNE} collaboration, \emph{{Updated MiniBooNE neutrino
  oscillation results with increased data and new background studies}},
  \href{https://doi.org/10.1103/PhysRevD.103.052002}{\emph{Phys. Rev. D}
  {\bfseries 103} (2021) 052002}
  [\href{https://arxiv.org/abs/2006.16883}{{\ttfamily 2006.16883}}].

\bibitem{Acciarri_2017}
R.~Acciarri and et~al., \emph{Design and construction of the {MicroBooNE}
  detector},
  \href{https://doi.org/10.1088/1748-0221/12/02/p02017}{\emph{Journal of
  Instrumentation} {\bfseries 12} (2017) P02017}.

\bibitem{Cheng_2012}
{\scshape MiniBooNE, SciBooNE} collaboration, \emph{{Dual baseline search for
  muon antineutrino disappearance at $0.1 {\rm eV}^2 < {\Delta}m^2 < 100 {\rm
  eV}^2$}}, \href{https://doi.org/10.1103/PhysRevD.86.052009}{\emph{Phys. Rev.
  D} {\bfseries 86} (2012) 052009}
  [\href{https://arxiv.org/abs/1208.0322}{{\ttfamily 1208.0322}}].

\bibitem{Ballett:2018ynz}
P.~Ballett, S.~Pascoli and M.~Ross-Lonergan, \emph{{U(1)' mediated decays of
  heavy sterile neutrinos in MiniBooNE}},
  \href{https://doi.org/10.1103/PhysRevD.99.071701}{\emph{Phys. Rev. D}
  {\bfseries 99} (2019) 071701}
  [\href{https://arxiv.org/abs/1808.02915}{{\ttfamily 1808.02915}}].

\bibitem{deGouvea:2019qre}
A.~de~Gouv\^ea, O.L.G.~Peres, S.~Prakash and G.V.~Stenico, \emph{{On The
  Decaying-Sterile Neutrino Solution to the Electron (Anti)Neutrino Appearance
  Anomalies}}, \href{https://doi.org/10.1007/JHEP07(2020)141}{\emph{JHEP}
  {\bfseries 07} (2020) 141}
  [\href{https://arxiv.org/abs/1911.01447}{{\ttfamily 1911.01447}}].

\bibitem{acciarri2015proposal}
{\scshape MicroBooNE, LAr1-ND, ICARUS-WA104} collaboration, \emph{{A Proposal
  for a Three Detector Short-Baseline Neutrino Oscillation Program in the
  Fermilab Booster Neutrino Beam}},
  \href{https://arxiv.org/abs/1503.01520}{{\ttfamily 1503.01520}}.

\bibitem{Maruyama:2016mqj}
{\scshape JSNS$^2$} collaboration, \emph{{Searching for a Sterile Neutrino at
  J-PARC MLF: JSNS$^2$ experiment}},
  \href{https://doi.org/10.22323/1.282.0482}{\emph{PoS} {\bfseries ICHEP2016}
  (2016) 482}.

\bibitem{Abi_2020}
{\scshape DUNE} collaboration, \emph{{Volume I. Introduction to DUNE}},
  \href{https://doi.org/10.1088/1748-0221/15/08/T08008}{\emph{JINST} {\bfseries
  15} (2020) T08008} [\href{https://arxiv.org/abs/2002.02967}{{\ttfamily
  2002.02967}}].

\bibitem{hyperk-loi}
K.~Abe et~al., \emph{{Letter of Intent: The Hyper-Kamiokande Experiment ---
  Detector Design and Physics Potential}},
  \href{https://arxiv.org/abs/1109.3262}{{\ttfamily 1109.3262}}.

\bibitem{Athanassopoulos:1996jb}
{\scshape LSND} collaboration, \emph{{Evidence for anti-muon-neutrino --->
  anti-electron-neutrino oscillations from the LSND experiment at LAMPF}},
  \href{https://doi.org/10.1103/PhysRevLett.77.3082}{\emph{Phys. Rev. Lett.}
  {\bfseries 77} (1996) 3082}
  [\href{https://arxiv.org/abs/nucl-ex/9605003}{{\ttfamily nucl-ex/9605003}}].

\bibitem{Aguilar-Arevalo:2013pmq}
{\scshape MiniBooNE} collaboration, \emph{{Improved Search for $\bar \nu_\mu
  \rightarrow \bar \nu_e$ Oscillations in the MiniBooNE Experiment}},
  \href{https://doi.org/10.1103/PhysRevLett.110.161801}{\emph{Phys. Rev. Lett.}
  {\bfseries 110} (2013) 161801}
  [\href{https://arxiv.org/abs/1303.2588}{{\ttfamily 1303.2588}}].

\bibitem{Huber:2019mro}
P.~Huber, J.~Link, C.~Mariani, S.~Pal and J.~Park, \emph{{CHANDLER R\&D
  status}}, \href{https://doi.org/10.1088/1742-6596/1216/1/012014}{\emph{J.
  Phys. Conf. Ser.} {\bfseries 1216} (2019) 012014}.

\bibitem{Subedi:2019shd}
T.~Subedi, J.M.~Link, S.~Li, J.~Park, P.~Huber, C.~Mariani et~al.,
  \emph{{Reactor Antineutrino Detection Using CHANDLER : A New Portable
  Neutrino Detector Tulasi Subedi Abstract CHANDLER}},
  \href{https://doi.org/10.22323/1.341.0059}{\emph{PoS} {\bfseries NuFACT2018}
  (2019) 059}.

\bibitem{Seo:2020ehv}
S.-H.~Seo, \emph{{Review of Sterile Neutrino Experiments}},  in \emph{{19th
  Lomonosov Conference on Elementary Particle Physics}}, 1, 2020
  [\href{https://arxiv.org/abs/2001.03349}{{\ttfamily 2001.03349}}].

\bibitem{Ko:2016owz}
{\scshape NEOS} collaboration, \emph{{Sterile Neutrino Search at the NEOS
  Experiment}},
  \href{https://doi.org/10.1103/PhysRevLett.118.121802}{\emph{Phys. Rev. Lett.}
  {\bfseries 118} (2017) 121802}
  [\href{https://arxiv.org/abs/1610.05134}{{\ttfamily 1610.05134}}].

\bibitem{Serebrov:2018vdw}
{\scshape NEUTRINO-4} collaboration, \emph{{First Observation of the
  Oscillation Effect in the Neutrino-4 Experiment on the Search for the Sterile
  Neutrino}}, \href{https://doi.org/10.1134/S0021364019040040}{\emph{Pisma Zh.
  Eksp. Teor. Fiz.} {\bfseries 109} (2019) 209}
  [\href{https://arxiv.org/abs/1809.10561}{{\ttfamily 1809.10561}}].

\bibitem{Danilov:2018dme}
M.~Danilov, \emph{{Searches for sterile neutrinos at very short baseline
  reactor experiments}},
  \href{https://doi.org/10.1088/1742-6596/1390/1/012049}{\emph{J. Phys. Conf.
  Ser.} {\bfseries 1390} (2019) 012049}
  [\href{https://arxiv.org/abs/1812.04085}{{\ttfamily 1812.04085}}].

\bibitem{Danilov:2020rax}
M.V.~Danilov and N.A.~Skrobova, \emph{{Comment on \textquotedblleft{}Analysis
  of the Results of the Neutrino-4 Experiment on the Search for the Sterile
  Neutrino and Comparison with Results of Other Experiments\textquotedblright{}
  (JETP Letters 112, 199 (2020))}},
  \href{https://doi.org/10.1134/S0021364020190066}{\emph{JETP Lett.} {\bfseries
  112} (2020) 452}.

\bibitem{Almazan:2020drb}
{\scshape PROSPECT, STEREO} collaboration, \emph{{Note on arXiv:2005.05301,
  'Preparation of the Neutrino-4 experiment on search for sterile neutrino and
  the obtained results of measurements'}},
  \href{https://arxiv.org/abs/2006.13147}{{\ttfamily 2006.13147}}.

\bibitem{Giunti:2021iti}
C.~Giunti, Y.F.~Li, C.A.~Ternes and Y.Y.~Zhang, \emph{{Neutrino-4 anomaly:
  oscillations or fluctuations?}},
  \href{https://doi.org/10.1016/j.physletb.2021.136214}{\emph{Phys. Lett. B}
  {\bfseries 816} (2021) 136214}
  [\href{https://arxiv.org/abs/2101.06785}{{\ttfamily 2101.06785}}].

\bibitem{Serebrov:2020yvp}
{\scshape Neutrino-4} collaboration, \emph{{A Comment on the note
  arXiv:2006.13147 on arXiv:2005.05301, ''Preparation of the Neutrino-4
  experiment on search for sterile neutrino and the obtained results of
  measurements''}},  \href{https://arxiv.org/abs/2006.13639}{{\ttfamily
  2006.13639}}.

\bibitem{Serebrov:2020wny}
A.P.~Serebrov and R.M.~Samoilov, \emph{{Reply to Comment on
  \textquotedblleft{}Analysis of the Results of the Neutrino-4 Experiment on
  the Search for the Sterile Neutrino and Comparison with Results of Other
  Experiments\textquotedblright{} (JETP Letters 112, 199 (2020))}},
  \href{https://doi.org/10.1134/S0021364020190108}{\emph{JETP Lett.} {\bfseries
  112} (2020) 455}.

\bibitem{Serebrov:2020kmd}
A.P.~Serebrov et~al., \emph{{Search for sterile neutrinos with the Neutrino-4
  experiment and measurement results}},
  \href{https://doi.org/10.1103/PhysRevD.104.032003}{\emph{Phys. Rev. D}
  {\bfseries 104} (2021) 032003}
  [\href{https://arxiv.org/abs/2005.05301}{{\ttfamily 2005.05301}}].

\bibitem{MINOS:2020iqj}
{\scshape MINOS+, Daya Bay} collaboration, \emph{{Improved Constraints on
  Sterile Neutrino Mixing from Disappearance Searches in the MINOS, MINOS+,
  Daya Bay, and Bugey-3 Experiments}},
  \href{https://doi.org/10.1103/PhysRevLett.125.071801}{\emph{Phys. Rev. Lett.}
  {\bfseries 125} (2020) 071801}
  [\href{https://arxiv.org/abs/2002.00301}{{\ttfamily 2002.00301}}].

\bibitem{Serebrov:2020rhy}
A.P.~Serebrov and R.M.~Samoilov, \emph{{Analysis of the Results of the
  Neutrino-4 Experiment on the Search for the Sterile Neutrino and Comparison
  with Results of Other Experiments}},
  \href{https://doi.org/10.31857/S1234567820160016}{\emph{JETP Lett.}
  {\bfseries 112} (2020) 199}
  [\href{https://arxiv.org/abs/2003.03199}{{\ttfamily 2003.03199}}].

\bibitem{Svirida:2020zpk}
{\scshape DANSS} collaboration, \emph{{DANSS experiment: current status and
  future plans}},
  \href{https://doi.org/10.1088/1742-6596/1690/1/012179}{\emph{J. Phys. Conf.
  Ser.} {\bfseries 1690} (2020) 012179}.

\bibitem{Andriamirado:2021qjc}
M.~Andriamirado et~al., \emph{{PROSPECT-II Physics Opportunities}},
  \href{https://arxiv.org/abs/2107.03934}{{\ttfamily 2107.03934}}.

\bibitem{Alekseev:2018efk}
{\scshape DANSS} collaboration, \emph{{Search for sterile neutrinos at the
  DANSS experiment}},
  \href{https://doi.org/10.1016/j.physletb.2018.10.038}{\emph{Phys. Lett. B}
  {\bfseries 787} (2018) 56}
  [\href{https://arxiv.org/abs/1804.04046}{{\ttfamily 1804.04046}}].

\bibitem{Danilov:2019aef}
{\scshape DANSS} collaboration, \emph{{Recent results of the DANSS
  experiment}},  in \emph{{2019 European Physical Society Conference on High
  Energy Physics}}, 11, 2019
  [\href{https://arxiv.org/abs/1911.10140}{{\ttfamily 1911.10140}}].

\bibitem{AlmazanMolina:2019qul}
{\scshape STEREO} collaboration, \emph{{Improved Sterile Neutrino Constraints
  from the STEREO Experiment with 179 Days of Reactor-On Data}},
  \href{https://arxiv.org/abs/1912.06582}{{\ttfamily 1912.06582}}.

\bibitem{Anselmann:1994ar}
{\scshape GALLEX} collaboration, \emph{{First results from the Cr-51 neutrino
  source experiment with the GALLEX detector}},
  \href{https://doi.org/10.1016/0370-2693(94)01586-2}{\emph{Phys. Lett. B}
  {\bfseries 342} (1995) 440}.

\bibitem{Hampel:1997fc}
{\scshape GALLEX} collaboration, \emph{{Final results of the Cr-51 neutrino
  source experiments in GALLEX}},
  \href{https://doi.org/10.1016/S0370-2693(97)01562-1}{\emph{Phys. Lett. B}
  {\bfseries 420} (1998) 114}.

\bibitem{PhysRevC.59.2246}
{\scshape SAGE} collaboration, \emph{{Measurement of the response of the
  Russian-American gallium experiment to neutrinos from a Cr-51 source}},
  \href{https://doi.org/10.1103/PhysRevC.59.2246}{\emph{Phys. Rev. C}
  {\bfseries 59} (1999) 2246}
  [\href{https://arxiv.org/abs/hep-ph/9803418}{{\ttfamily hep-ph/9803418}}].

\bibitem{Abdurashitov:2005tb}
J.~Abdurashitov et~al., \emph{{Measurement of the response of a Ga solar
  neutrino experiment to neutrinos from an Ar-37 source}},
  \href{https://doi.org/10.1103/PhysRevC.73.045805}{\emph{Phys. Rev. C}
  {\bfseries 73} (2006) 045805}
  [\href{https://arxiv.org/abs/nucl-ex/0512041}{{\ttfamily nucl-ex/0512041}}].

\bibitem{GA}
C.~Giunti and M.~Laveder, \emph{{Statistical Significance of the Gallium
  Anomaly}}, \href{https://doi.org/10.1103/PhysRevC.83.065504}{\emph{Phys. Rev.
  C} {\bfseries 83} (2011) 065504}
  [\href{https://arxiv.org/abs/1006.3244}{{\ttfamily 1006.3244}}].

\bibitem{Kozlova:2018cfx}
J.~Kozlova, E.~Veretenkin, V.~Gavrin, O.~Grekhov, T.~Ibragimova, A.~Kalikhov
  et~al., \emph{{Calorimetric System for Determination of Activity of a
  Neutrino Source Based on$^{51}$Cr}},
  \href{https://doi.org/10.1134/S1063779618040378}{\emph{Phys. Part. Nucl.}
  {\bfseries 49} (2018) 758}.

\bibitem{Kozlova:2019zee}
J.~Kozlova, E.~Veretenkin, V.~Gavrin, S.~Danshin, T.~Ibragimova and B.~Komarov,
  \emph{{Fabrication of reactor target from enriched $^{50}$Cr for artificial
  neutrino source}},
  \href{https://doi.org/10.1088/1742-6596/1390/1/012100}{\emph{J. Phys. Conf.
  Ser.} {\bfseries 1390} (2019) 012100}.

\bibitem{Gavrin:2018zmf}
V.~Gavrin et~al., \emph{{On the gallium experiment BEST-2 with a $^{65}$Zn
  source to search for neutrino oscillations on a short baseline}},
  \href{https://arxiv.org/abs/1807.02977}{{\ttfamily 1807.02977}}.

\bibitem{Cribier:2011fv}
M.~Cribier, M.~Fechner, T.~Lasserre, A.~Letourneau, D.~Lhuillier, G.~Mention
  et~al., \emph{{A proposed search for a fourth neutrino with a PBq
  antineutrino source}},
  \href{https://doi.org/10.1103/PhysRevLett.107.201801}{\emph{Phys. Rev. Lett.}
  {\bfseries 107} (2011) 201801}
  [\href{https://arxiv.org/abs/1107.2335}{{\ttfamily 1107.2335}}].

\bibitem{Kornoukhov:1994zq}
V.~Kornoukhov, \emph{{Some aspects of the creation and application of
  anti-neutrino artificial sources}}, {\emph{ITEP-90-94} (1994) }.

\bibitem{Gando:2013zla}
A.~Gando et~al., \emph{{White paper: CeLAND - Investigation of the reactor
  antineutrino anomaly with an intense $^{144}Ce-^{144}Pr$ antineutrino source
  in KamLAND}},  \href{https://arxiv.org/abs/1309.6805}{{\ttfamily 1309.6805}}.

\bibitem{Borexino:2013xxa}
{\scshape Borexino} collaboration, \emph{{SOX: Short distance neutrino
  Oscillations with BoreXino}},
  \href{https://doi.org/10.1007/JHEP08(2013)038}{\emph{JHEP} {\bfseries 08}
  (2013) 038} [\href{https://arxiv.org/abs/1304.7721}{{\ttfamily 1304.7721}}].

\bibitem{Smirnov:2020bcr}
M.~Smirnov, Z.~Hu, J.~Ling, Y.~Novikov, Z.~Wang and G.~Yang, \emph{{Sterile
  neutrino oscillometry with Jinping}},
  \href{https://doi.org/10.1140/epjc/s10052-020-8175-4}{\emph{Eur. Phys. J. C}
  {\bfseries 80} (2020) 609}
  [\href{https://arxiv.org/abs/2002.05246}{{\ttfamily 2002.05246}}].

\bibitem{Bellenghi:2019vtc}
C.~Bellenghi, D.~Chiesa, L.~Di~Noto, M.~Pallavicini, E.~Previtali and
  M.~Vignati, \emph{{Coherent elastic nuclear scattering of$^{51}$ Cr
  neutrinos}}, \href{https://doi.org/10.1140/epjc/s10052-019-7240-3}{\emph{Eur.
  Phys. J. C} {\bfseries 79} (2019) 727}
  [\href{https://arxiv.org/abs/1905.10611}{{\ttfamily 1905.10611}}].

\bibitem{Link:2019pbm}
J.M.~Link and X.-J.~Xu, \emph{{Searching for BSM neutrino interactions in dark
  matter detectors}},
  \href{https://doi.org/10.1007/JHEP08(2019)004}{\emph{JHEP} {\bfseries 08}
  (2019) 004} [\href{https://arxiv.org/abs/1903.09891}{{\ttfamily
  1903.09891}}].

\bibitem{DeGerone:2019xce}
M.~De~Gerone et~al., \emph{{Probing the absolute neutrino mass scale with
  $^{163}$Ho: The HOLMES project}},
  \href{https://doi.org/10.1016/j.nima.2018.10.108}{\emph{Nucl. Instrum. Meth.
  A} {\bfseries 936} (2019) 252}.

\bibitem{Martoff:2021vxp}
C.J.~Martoff et~al., \emph{{HUNTER: precision massive-neutrino search based on
  a laser cooled atomic source}},
  \href{https://doi.org/10.1088/2058-9565/abdb9b}{\emph{Quantum Sci. Technol.}
  {\bfseries 6} (2021) 024008}.

\bibitem{Friedrich:2020nze}
S.~Friedrich et~al., \emph{{Limits on the Existence of sub-MeV Sterile
  Neutrinos from the Decay of $^7$Be in Superconducting Quantum Sensors}},
  \href{https://doi.org/10.1103/PhysRevLett.126.021803}{\emph{Phys. Rev. Lett.}
  {\bfseries 126} (2021) 021803}
  [\href{https://arxiv.org/abs/2010.09603}{{\ttfamily 2010.09603}}].

\bibitem{Giunti:2019fcj}
C.~Giunti, Y.F.~Li and Y.Y.~Zhang, \emph{{KATRIN bound on 3+1 active-sterile
  neutrino mixing and the reactor antineutrino anomaly}},
  \href{https://doi.org/10.1007/JHEP05(2020)061}{\emph{JHEP} {\bfseries 05}
  (2020) 061} [\href{https://arxiv.org/abs/1912.12956}{{\ttfamily
  1912.12956}}].

\bibitem{Abe:2008aa}
{\scshape KamLAND} collaboration, \emph{{Precision Measurement of Neutrino
  Oscillation Parameters with KamLAND}},
  \href{https://doi.org/10.1103/PhysRevLett.100.221803}{\emph{Phys. Rev. Lett.}
  {\bfseries 100} (2008) 221803}
  [\href{https://arxiv.org/abs/0801.4589}{{\ttfamily 0801.4589}}].

\bibitem{Bellini:2008mr}
{\scshape Borexino} collaboration, \emph{{Measurement of the solar 8B neutrino
  rate with a liquid scintillator target and 3 MeV energy threshold in the
  Borexino detector}},
  \href{https://doi.org/10.1103/PhysRevD.82.033006}{\emph{Phys. Rev. D}
  {\bfseries 82} (2010) 033006}
  [\href{https://arxiv.org/abs/0808.2868}{{\ttfamily 0808.2868}}].

\bibitem{Abe:2020tyy}
{\scshape Super-Kamiokande} collaboration, \emph{{Search for solar electron
  anti-neutrinos due to spin-flavor precession in the Sun with
  Super-Kamiokande-IV}},  \href{https://arxiv.org/abs/2012.03807}{{\ttfamily
  2012.03807}}.

\bibitem{deHolanda:2003tx}
P.C.~de~Holanda and A.Y.~Smirnov, \emph{{Homestake result, sterile neutrinos
  and low-energy solar neutrino experiments}},
  \href{https://doi.org/10.1103/PhysRevD.69.113002}{\emph{Phys. Rev. D}
  {\bfseries 69} (2004) 113002}
  [\href{https://arxiv.org/abs/hep-ph/0307266}{{\ttfamily hep-ph/0307266}}].

\bibitem{deHolanda:2010am}
P.C.~de~Holanda and A.Y.~Smirnov, \emph{{Solar neutrino spectrum, sterile
  neutrinos and additional radiation in the Universe}},
  \href{https://doi.org/10.1103/PhysRevD.83.113011}{\emph{Phys. Rev. D}
  {\bfseries 83} (2011) 113011}
  [\href{https://arxiv.org/abs/1012.5627}{{\ttfamily 1012.5627}}].

\bibitem{PhysRevD.95.112002}
{\scshape IceCube} collaboration, \emph{{Search for sterile neutrino mixing
  using three years of IceCube DeepCore data}},
  \href{https://doi.org/10.1103/PhysRevD.95.112002}{\emph{Phys. Rev. D}
  {\bfseries 95} (2017) 112002}
  [\href{https://arxiv.org/abs/1702.05160}{{\ttfamily 1702.05160}}].

\bibitem{PhysRevD.91.052019}
{\scshape Super-Kamiokande} collaboration, \emph{{Limits on sterile neutrino
  mixing using atmospheric neutrinos in Super-Kamiokande}},
  \href{https://doi.org/10.1103/PhysRevD.91.052019}{\emph{Phys. Rev. D}
  {\bfseries 91} (2015) 052019}
  [\href{https://arxiv.org/abs/1410.2008}{{\ttfamily 1410.2008}}].

\bibitem{PhysRevD.102.052009}
{\scshape IceCube} collaboration, \emph{{Searching for eV-scale sterile
  neutrinos with eight years of atmospheric neutrinos at the IceCube Neutrino
  Telescope}}, \href{https://doi.org/10.1103/PhysRevD.102.052009}{\emph{Phys.
  Rev. D} {\bfseries 102} (2020) 052009}
  [\href{https://arxiv.org/abs/2005.12943}{{\ttfamily 2005.12943}}].

\bibitem{Gariazzo:2019gyi}
S.~Gariazzo, P.F.~de~Salas and S.~Pastor, \emph{{Thermalisation of sterile
  neutrinos in the early Universe in the 3+1 scheme with full mixing matrix}},
  \href{https://doi.org/10.1088/1475-7516/2019/07/014}{\emph{JCAP} {\bfseries
  07} (2019) 014} [\href{https://arxiv.org/abs/1905.11290}{{\ttfamily
  1905.11290}}].

\bibitem{Abazajian:2012ys}
K.~Abazajian et~al., \emph{{Light Sterile Neutrinos: A White Paper}},
  \href{https://arxiv.org/abs/1204.5379}{{\ttfamily 1204.5379}}.

\bibitem{Archidiacono:2020yey}
M.~Archidiacono, S.~Gariazzo, C.~Giunti, S.~Hannestad and T.~Tram,
  \emph{{Sterile neutrino self-interactions: $H_0$ tension and short-baseline
  anomalies}}, \href{https://doi.org/10.1088/1475-7516/2020/12/029}{\emph{JCAP}
  {\bfseries 12} (2020) 029}
  [\href{https://arxiv.org/abs/2006.12885}{{\ttfamily 2006.12885}}].

\bibitem{Gariazzo:2017fdh}
S.~Gariazzo, C.~Giunti, M.~Laveder and Y.F.~Li, \emph{{Updated Global 3+1
  Analysis of Short-BaseLine Neutrino Oscillations}},
  \href{https://doi.org/10.1007/JHEP06(2017)135}{\emph{JHEP} {\bfseries 06}
  (2017) 135} [\href{https://arxiv.org/abs/1703.00860}{{\ttfamily
  1703.00860}}].

\bibitem{Dentler:2018sju}
M.~Dentler, A.~Hern\'andez-Cabezudo, J.~Kopp, P.A.N.~Machado, M.~Maltoni,
  I.~Martinez-Soler et~al., \emph{{Updated Global Analysis of Neutrino
  Oscillations in the Presence of eV-Scale Sterile Neutrinos}},
  \href{https://doi.org/10.1007/JHEP08(2018)010}{\emph{JHEP} {\bfseries 08}
  (2018) 010} [\href{https://arxiv.org/abs/1803.10661}{{\ttfamily
  1803.10661}}].

\bibitem{Berryman:2020agd}
J.M.~Berryman and P.~Huber, \emph{{Sterile Neutrinos and the Global Reactor
  Antineutrino Dataset}},
  \href{https://doi.org/10.1007/JHEP01(2021)167}{\emph{JHEP} {\bfseries 01}
  (2021) 167} [\href{https://arxiv.org/abs/2005.01756}{{\ttfamily
  2005.01756}}].

\bibitem{Diaz:2019fwt}
A.~Diaz, C.A.~Arg\"uelles, G.H.~Collin, J.M.~Conrad and M.H.~Shaevitz,
  \emph{{Where Are We With Light Sterile Neutrinos?}},
  \href{https://doi.org/10.1016/j.physrep.2020.08.005}{\emph{Phys. Rept.}
  {\bfseries 884} (2020) 1} [\href{https://arxiv.org/abs/1906.00045}{{\ttfamily
  1906.00045}}].

\bibitem{Boser:2019rta}
S.~B\"oser, C.~Buck, C.~Giunti, J.~Lesgourgues, L.~Ludhova, S.~Mertens et~al.,
  \emph{{Status of Light Sterile Neutrino Searches}},
  \href{https://doi.org/10.1016/j.ppnp.2019.103736}{\emph{Prog. Part. Nucl.
  Phys.} {\bfseries 111} (2020) 103736}
  [\href{https://arxiv.org/abs/1906.01739}{{\ttfamily 1906.01739}}].

\bibitem{Berryman:2019nvr}
J.M.~Berryman, \emph{{Constraining Sterile Neutrino Cosmology with Terrestrial
  Oscillation Experiments}},
  \href{https://doi.org/10.1103/PhysRevD.100.023540}{\emph{Phys. Rev. D}
  {\bfseries 100} (2019) 023540}
  [\href{https://arxiv.org/abs/1905.03254}{{\ttfamily 1905.03254}}].

\bibitem{Archidiacono:2013xxa}
M.~Archidiacono, N.~Fornengo, C.~Giunti, S.~Hannestad and A.~Melchiorri,
  \emph{{Sterile neutrinos: Cosmology versus short-baseline experiments}},
  \href{https://doi.org/10.1103/PhysRevD.87.125034}{\emph{Phys. Rev. D}
  {\bfseries 87} (2013) 125034}
  [\href{https://arxiv.org/abs/1302.6720}{{\ttfamily 1302.6720}}].

\bibitem{Hagstotz:2020ukm}
S.~Hagstotz, P.F.~de~Salas, S.~Gariazzo, M.~Gerbino, M.~Lattanzi, S.~Vagnozzi
  et~al., \emph{{Bounds on light sterile neutrino mass and mixing from
  cosmology and laboratory searches}},
  \href{https://arxiv.org/abs/2003.02289}{{\ttfamily 2003.02289}}.

\bibitem{Deppisch:2015qwa}
F.F.~Deppisch, P.S.~Bhupal~Dev and A.~Pilaftsis, \emph{{Neutrinos and Collider
  Physics}}, \href{https://doi.org/10.1088/1367-2630/17/7/075019}{\emph{New J.
  Phys.} {\bfseries 17} (2015) 075019}
  [\href{https://arxiv.org/abs/1502.06541}{{\ttfamily 1502.06541}}].

\bibitem{Bolton:2020ncv}
P.D.~Bolton, F.F.~Deppisch, L.~Gr\'af and F.~\v{S}imkovic, \emph{{Two-Neutrino
  Double Beta Decay with Sterile Neutrinos}},
  \href{https://doi.org/10.1103/PhysRevD.103.055019}{\emph{Phys. Rev. D}
  {\bfseries 103} (2021) 055019}
  [\href{https://arxiv.org/abs/2011.13387}{{\ttfamily 2011.13387}}].

\bibitem{Anelli:2015pba}
{\scshape SHiP} collaboration, \emph{{A facility to Search for Hidden Particles
  (SHiP) at the CERN SPS}},  \href{https://arxiv.org/abs/1504.04956}{{\ttfamily
  1504.04956}}.

\bibitem{Smith:2016vku}
P.F.~Smith, \emph{{Proposed experiments to detect keV range sterile neutrinos
  using energy-momentum reconstruction of beta decay or K-capture events}},
  \href{https://doi.org/10.1088/1367-2630/ab1502}{\emph{New J. Phys.}
  {\bfseries 21} (2019) 053022}
  [\href{https://arxiv.org/abs/1607.06876}{{\ttfamily 1607.06876}}].

\bibitem{Gorbunov:2007ak}
D.~Gorbunov and M.~Shaposhnikov, \emph{{How to find neutral leptons of the
  $\nu$MSM?}}, \href{https://doi.org/10.1088/1126-6708/2007/10/015}{\emph{JHEP}
  {\bfseries 10} (2007) 015} [\href{https://arxiv.org/abs/0705.1729}{{\ttfamily
  0705.1729}}].

\bibitem{Abi:2018dnh}
{\scshape DUNE} collaboration, \emph{{The DUNE Far Detector Interim Design
  Report Volume 1: Physics, Technology and Strategies}},
  \href{https://arxiv.org/abs/1807.10334}{{\ttfamily 1807.10334}}.

\bibitem{Acciarri:2016crz}
{\scshape DUNE} collaboration, \emph{{Long-Baseline Neutrino Facility (LBNF)
  and Deep Underground Neutrino Experiment (DUNE)}: {Conceptual Design Report,
  Volume 1: The LBNF and DUNE Projects}},
  \href{https://arxiv.org/abs/1601.05471}{{\ttfamily 1601.05471}}.

\bibitem{Acciarri:2015uup}
{\scshape DUNE} collaboration, \emph{{Long-Baseline Neutrino Facility (LBNF)
  and Deep Underground Neutrino Experiment (DUNE)}: {Conceptual Design Report,
  Volume 2: The Physics Program for DUNE at LBNF}},
  \href{https://arxiv.org/abs/1512.06148}{{\ttfamily 1512.06148}}.

\bibitem{Strait:2016mof}
{\scshape DUNE} collaboration, \emph{{Long-Baseline Neutrino Facility (LBNF)
  and Deep Underground Neutrino Experiment (DUNE)}: {Conceptual Design Report,
  Volume 3: Long-Baseline Neutrino Facility for DUNE June 24, 2015}},
  \href{https://arxiv.org/abs/1601.05823}{{\ttfamily 1601.05823}}.

\bibitem{Acciarri:2016ooe}
{\scshape DUNE} collaboration, \emph{{Long-Baseline Neutrino Facility (LBNF)
  and Deep Underground Neutrino Experiment (DUNE)}: {Conceptual Design Report,
  Volume 4 The DUNE Detectors at LBNF}},
  \href{https://arxiv.org/abs/1601.02984}{{\ttfamily 1601.02984}}.

\bibitem{Abe:2014oxa}
{\scshape Hyper-Kamiokande Working Group} collaboration, \emph{{A Long Baseline
  Neutrino Oscillation Experiment Using J-PARC Neutrino Beam and
  Hyper-Kamiokande}},  12, 2014
  [\href{https://arxiv.org/abs/1412.4673}{{\ttfamily 1412.4673}}].

\bibitem{Abe:2015zbg}
{\scshape Hyper-Kamiokande Proto-} collaboration, \emph{{Physics potential of a
  long-baseline neutrino oscillation experiment using a J-PARC neutrino beam
  and Hyper-Kamiokande}},
  \href{https://doi.org/10.1093/ptep/ptv061}{\emph{PTEP} {\bfseries 2015}
  (2015) 053C02} [\href{https://arxiv.org/abs/1502.05199}{{\ttfamily
  1502.05199}}].

\bibitem{Abe:2016ero}
{\scshape Hyper-Kamiokande} collaboration, \emph{{Physics potentials with the
  second Hyper-Kamiokande detector in Korea}},
  \href{https://doi.org/10.1093/ptep/pty044}{\emph{PTEP} {\bfseries 2018}
  (2018) 063C01} [\href{https://arxiv.org/abs/1611.06118}{{\ttfamily
  1611.06118}}].

\bibitem{Hyper-Kamiokande:2018ofw}
{\scshape Hyper-Kamiokande} collaboration, \emph{{Hyper-Kamiokande Design
  Report}},  \href{https://arxiv.org/abs/1805.04163}{{\ttfamily 1805.04163}}.

\bibitem{Kyberd:2012iz}
{\scshape nuSTORM} collaboration, \emph{{nuSTORM - Neutrinos from STORed Muons:
  Letter of Intent to the Fermilab Physics Advisory Committee}},
  \href{https://arxiv.org/abs/1206.0294}{{\ttfamily 1206.0294}}.

\bibitem{Longhin:2014yta}
A.~Longhin, L.~Ludovici and F.~Terranova, \emph{{A novel technique for the
  measurement of the electron neutrino cross section}},
  \href{https://doi.org/10.1140/epjc/s10052-015-3378-9}{\emph{Eur. Phys. J. C}
  {\bfseries 75} (2015) 155} [\href{https://arxiv.org/abs/1412.5987}{{\ttfamily
  1412.5987}}].

\bibitem{Harada:2013yaa}
{\scshape JSNS2} collaboration, \emph{{Proposal: A Search for Sterile Neutrino
  at J-PARC Materials and Life Science Experimental Facility}},
  \href{https://arxiv.org/abs/1310.1437}{{\ttfamily 1310.1437}}.

\bibitem{Alonso:2017fci}
{\scshape IsoDAR} collaboration, \emph{{IsoDAR@KamLAND:A Conceptual Design
  Report for the Conventional Facilities}},
  \href{https://arxiv.org/abs/1710.09325}{{\ttfamily 1710.09325}}.

\bibitem{Baussan:2013zcy}
{\scshape ESSnuSB} collaboration, \emph{{A very intense neutrino super beam
  experiment for leptonic CP violation discovery based on the European
  spallation source linac}},
  \href{https://doi.org/10.1016/j.nuclphysb.2014.05.016}{\emph{Nucl. Phys. B}
  {\bfseries 885} (2014) 127}
  [\href{https://arxiv.org/abs/1309.7022}{{\ttfamily 1309.7022}}].

\bibitem{Jamil:2018tkx}
{\scshape nEXO} collaboration, \emph{{VUV-sensitive Silicon Photomultipliers
  for Xenon Scintillation Light Detection in nEXO}},
  \href{https://doi.org/10.1109/TNS.2018.2875668}{\emph{IEEE Trans. Nucl. Sci.}
  {\bfseries 65} (2018) 2823}
  [\href{https://arxiv.org/abs/1806.02220}{{\ttfamily 1806.02220}}].

\bibitem{FUJII2015293}
K.~Fujii, Y.~Endo, Y.~Torigoe, S.~Nakamura, T.~Haruyama, K.~Kasami et~al.,
  \emph{High-accuracy measurement of the emission spectrum of liquid xenon in
  the vacuum ultraviolet region},
  \href{https://doi.org/https://doi.org/10.1016/j.nima.2015.05.065}{\emph{Nuclear
  Instruments and Methods in Physics Research Section A: Accelerators,
  Spectrometers, Detectors and Associated Equipment} {\bfseries 795} (2015)
  293}.

\bibitem{Brodsky:2018abk}
J.P.~Brodsky, S.~Sangiorgio, M.~Heffner and T.~Stiegler, \emph{{Background
  Discrimination for Neutrinoless Double Beta Decay in Liquid Xenon Using
  Cherenkov Light}},
  \href{https://doi.org/10.1016/j.nima.2018.12.057}{\emph{Nucl. Instrum. Meth.
  A} {\bfseries 922} (2019) 76}
  [\href{https://arxiv.org/abs/1812.05694}{{\ttfamily 1812.05694}}].

\bibitem{nutools}
O.~Akindele et~al., \emph{{Nu Tools: Exploring Practical Roles for Neutrinos in
  Nuclear Energy and Security}},  Tech. Rep.
  \href{http://nutools.ornl.gov}{CERN-ESU-004}, Geneva (2021).

\bibitem{Brdar:2016swo}
V.~Brdar, P.~Huber and J.~Kopp, \emph{{Antineutrino monitoring of spent nuclear
  fuel}}, \href{https://doi.org/10.1103/PhysRevApplied.8.054050}{\emph{Phys.
  Rev. Applied} {\bfseries 8} (2017) 054050}
  [\href{https://arxiv.org/abs/1606.06309}{{\ttfamily 1606.06309}}].

\bibitem{Asaadi:2018xfh}
J.~Asaadi et~al., \emph{{A pixelated charge readout for Liquid Argon Time
  Projection Chambers}},
  \href{https://doi.org/10.1088/1748-0221/13/02/C02008}{\emph{JINST} {\bfseries
  13} (2018) C02008}.

\bibitem{AbedAbud:2021hpb}
{\scshape DUNE} collaboration, \emph{{Deep Underground Neutrino Experiment
  (DUNE) Near Detector Conceptual Design Report}},
  \href{https://arxiv.org/abs/2103.13910}{{\ttfamily 2103.13910}}.

\bibitem{Majumdar:2021llu}
K.~Majumdar and K.~Mavrokoridis, \emph{{Review of Liquid Argon Detector
  Technologies in the Neutrino Sector}},
  \href{https://doi.org/10.3390/app11062455}{\emph{Appl. Sciences} {\bfseries
  11} (2021) 2455} [\href{https://arxiv.org/abs/2103.06395}{{\ttfamily
  2103.06395}}].

\bibitem{Albert:2013gpz}
{\scshape EXO-200} collaboration, \emph{{Improved measurement of the
  $2\nu\beta\beta$ half-life of $^{136}$Xe with the EXO-200 detector}},
  \href{https://doi.org/10.1103/PhysRevC.89.015502}{\emph{Phys. Rev. C}
  {\bfseries 89} (2014) 015502}
  [\href{https://arxiv.org/abs/1306.6106}{{\ttfamily 1306.6106}}].

\bibitem{Jewell:2017dzi}
{\scshape nEXO} collaboration, \emph{{Characterization of an Ionization Readout
  Tile for nEXO}},
  \href{https://doi.org/10.1088/1748-0221/13/01/P01006}{\emph{JINST} {\bfseries
  13} (2018) P01006} [\href{https://arxiv.org/abs/1710.05109}{{\ttfamily
  1710.05109}}].

\bibitem{Li:2019qgs}
{\scshape nEXO} collaboration, \emph{{Simulation of charge readout with
  segmented tiles in nEXO}},
  \href{https://doi.org/10.1088/1748-0221/14/09/P09020}{\emph{JINST} {\bfseries
  14} (2019) P09020} [\href{https://arxiv.org/abs/1907.07512}{{\ttfamily
  1907.07512}}].

\bibitem{Moe:1991ik}
M.K.~Moe, \emph{{New approach to the detection of neutrinoless double beta
  decay}}, \href{https://doi.org/10.1103/PhysRevC.44.R931}{\emph{Phys. Rev. C}
  {\bfseries 44} (1991) 931}.

\bibitem{Chambers:2018srx}
{\scshape nEXO} collaboration, \emph{{Imaging individual barium atoms in solid
  xenon for barium tagging in nEXO}},
  \href{https://doi.org/10.1038/s41586-019-1169-4}{\emph{Nature} {\bfseries
  569} (2019) 203} [\href{https://arxiv.org/abs/1806.10694}{{\ttfamily
  1806.10694}}].

\bibitem{Green:2007rc}
M.~Green et~al., \emph{{Observation of single collisionally cooled trapped ions
  in a buffer gas}},
  \href{https://doi.org/10.1103/PhysRevA.76.023404}{\emph{Phys. Rev. A}
  {\bfseries 76} (2007) 023404}
  [\href{https://arxiv.org/abs/physics/0702122}{{\ttfamily physics/0702122}}].

\bibitem{McDonald:2017izm}
A.D.~McDonald et~al., \emph{{Demonstration of Single Barium Ion Sensitivity for
  Neutrinoless Double Beta Decay using Single Molecule Fluorescence Imaging}},
  \href{https://doi.org/10.1103/PhysRevLett.120.132504}{\emph{Phys. Rev. Lett.}
  {\bfseries 120} (2018) 132504}
  [\href{https://arxiv.org/abs/1711.04782}{{\ttfamily 1711.04782}}].

\bibitem{Aalbers_2016}
J.~Aalbers, F.~Agostini, M.~Alfonsi, F.~Amaro, C.~Amsler, E.~Aprile et~al.,
  \emph{Darwin: towards the ultimate dark matter detector},
  \href{https://doi.org/10.1088/1475-7516/2016/11/017}{\emph{Journal of
  Cosmology and Astroparticle Physics} {\bfseries 2016} (2016) 017–017}.

\bibitem{Alonso:2014fwf}
J.R.~Alonso et~al., \emph{{Advanced Scintillator Detector Concept (ASDC): A
  Concept Paper on the Physics Potential of Water-Based Liquid Scintillator}},
  \href{https://arxiv.org/abs/1409.5864}{{\ttfamily 1409.5864}}.

\bibitem{Buck:2019tsa}
C.~Buck, B.~Gramlich and S.~Schoppmann, \emph{{Novel Opaque Scintillator for
  Neutrino Detection}},
  \href{https://doi.org/10.1088/1748-0221/14/11/P11007}{\emph{JINST} {\bfseries
  14} (2019) P11007} [\href{https://arxiv.org/abs/1908.03334}{{\ttfamily
  1908.03334}}].

\bibitem{Bonet_2021}
{\scshape CONUS} collaboration, \emph{{Constraints on Elastic Neutrino Nucleus
  Scattering in the Fully Coherent Regime from the CONUS Experiment}},
  \href{https://doi.org/10.1103/PhysRevLett.126.041804}{\emph{Phys. Rev. Lett.}
  {\bfseries 126} (2021) 041804}
  [\href{https://arxiv.org/abs/2011.00210}{{\ttfamily 2011.00210}}].

\bibitem{Strauss_2017}
R.~Strauss et~al., \emph{{The $\nu$-cleus experiment: A gram-scale
  fiducial-volume cryogenic detector for the first detection of coherent
  neutrino-nucleus scattering}},
  \href{https://doi.org/10.1140/epjc/s10052-017-5068-2}{\emph{Eur. Phys. J. C}
  {\bfseries 77} (2017) 506}
  [\href{https://arxiv.org/abs/1704.04320}{{\ttfamily 1704.04320}}].

\bibitem{Saenz:2000dul}
A.~Saenz, S.~Jonsell and P.~Froelich, \emph{{Improved Molecular Final-State
  Distribution of HeT+ for the \ensuremath{\beta}-Decay Process of T2}},
  \href{https://doi.org/10.1103/PhysRevLett.84.242}{\emph{Phys. Rev. Lett.}
  {\bfseries 84} (2000) 242}.

\bibitem{Bodine:2015sma}
L.I.~Bodine, D.S.~Parno and R.G.H.~Robertson, \emph{{Assessment of molecular
  effects on neutrino mass measurements from tritium \ensuremath{\beta}
  decay}}, \href{https://doi.org/10.1103/PhysRevC.91.035505}{\emph{Phys. Rev.
  C} {\bfseries 91} (2015) 035505}
  [\href{https://arxiv.org/abs/1502.03497}{{\ttfamily 1502.03497}}].

\bibitem{Huber:2008zz}
P.~Huber, J.~Kopp, M.~Lindner and W.~Winter, \emph{{GLoBES: General long
  baseline experiment simulator}},
  \href{https://doi.org/10.22323/1.074.0145}{\emph{PoS} {\bfseries NUFACT08}
  (2008) 145}.

\bibitem{Allison:2016lfl}
J.~Allison et~al., \emph{{Recent developments in Geant4}},
  \href{https://doi.org/10.1016/j.nima.2016.06.125}{\emph{Nucl. Instrum. Meth.
  A} {\bfseries 835} (2016) 186}.

\bibitem{Ferrari:2005zk}
A.~Ferrari, P.R.~Sala, A.~Fasso and J.~Ranft, \emph{{FLUKA: A multi-particle
  transport code (Program version 2005)}}, .

\bibitem{Mokhov:2017klc}
N.V.~Mokhov and C.C.~James, \emph{{The MARS Code System User\textquoteright{}s
  Guide Version 15(2016)}}, .

\bibitem{Alves:2017she}
{\scshape HEP Software Foundation} collaboration, \emph{{A Roadmap for HEP
  Software and Computing R\&D for the 2020s}},
  \href{https://doi.org/10.1007/s41781-018-0018-8}{\emph{Comput. Softw. Big
  Sci.} {\bfseries 3} (2019) 7}
  [\href{https://arxiv.org/abs/1712.06982}{{\ttfamily 1712.06982}}].

\bibitem{Barisits:2019fyl}
M.~Barisits et~al., \emph{{Rucio - Scientific data management}},
  \href{https://doi.org/10.1007/s41781-019-0026-3}{\emph{Comput. Softw. Big
  Sci.} {\bfseries 3} (2019) 11}
  [\href{https://arxiv.org/abs/1902.09857}{{\ttfamily 1902.09857}}].

\bibitem{aprile2020222rn}
{\scshape XENON} collaboration, \emph{{$^{222}$Rn emanation measurements for
  the XENON1T experiment}},
  \href{https://doi.org/10.1140/epjc/s10052-020-08777-z}{\emph{Eur. Phys. J. C}
  {\bfseries 81} (2021) 337}
  [\href{https://arxiv.org/abs/2009.13981}{{\ttfamily 2009.13981}}].

\bibitem{Winter:2006vg}
W.~Winter, \emph{{Neutrino tomography: Learning about the earth's interior
  using the propagation of neutrinos}},
  \href{https://doi.org/10.1007/s11038-006-9101-y}{\emph{Earth Moon Planets}
  {\bfseries 99} (2006) 285}
  [\href{https://arxiv.org/abs/physics/0602049}{{\ttfamily physics/0602049}}].

\bibitem{Donini:2018tsg}
A.~Donini, S.~Palomares-Ruiz and J.~Salvado, \emph{{Neutrino tomography of
  Earth}}, \href{https://doi.org/10.1038/s41567-018-0319-1}{\emph{Nature Phys.}
  {\bfseries 15} (2019) 37} [\href{https://arxiv.org/abs/1803.05901}{{\ttfamily
  1803.05901}}].

\bibitem{Rott:2015kwa}
C.~Rott, A.~Taketa and D.~Bose, \emph{{Spectrometry of the Earth using Neutrino
  Oscillations}}, \href{https://doi.org/10.1038/srep15225}{\emph{Sci. Rep.}
  {\bfseries 5} (2015) 15225}
  [\href{https://arxiv.org/abs/1502.04930}{{\ttfamily 1502.04930}}].

\bibitem{Winter:2015zwx}
W.~Winter, \emph{{Atmospheric Neutrino Oscillations for Earth Tomography}},
  \href{https://doi.org/10.1016/j.nuclphysb.2016.03.033}{\emph{Nucl. Phys. B}
  {\bfseries 908} (2016) 250}
  [\href{https://arxiv.org/abs/1511.05154}{{\ttfamily 1511.05154}}].

\bibitem{Winter:2005we}
W.~Winter, \emph{{Probing the absolute density of the Earth's core using a
  vertical neutrino beam}},
  \href{https://doi.org/10.1103/PhysRevD.72.037302}{\emph{Phys. Rev. D}
  {\bfseries 72} (2005) 037302}
  [\href{https://arxiv.org/abs/hep-ph/0502097}{{\ttfamily hep-ph/0502097}}].

\bibitem{Tamborra:2013laa}
I.~Tamborra, F.~Hanke, B.~M\"uller, H.-T.~Janka and G.~Raffelt, \emph{{Neutrino
  signature of supernova hydrodynamical instabilities in three dimensions}},
  \href{https://doi.org/10.1103/PhysRevLett.111.121104}{\emph{Phys. Rev. Lett.}
  {\bfseries 111} (2013) 121104}
  [\href{https://arxiv.org/abs/1307.7936}{{\ttfamily 1307.7936}}].

\bibitem{Walk:2019miz}
L.~Walk, I.~Tamborra, H.-T.~Janka, A.~Summa and D.~Kresse, \emph{{Neutrino
  emission characteristics of black hole formation in three-dimensional
  simulations of stellar collapse}},
  \href{https://doi.org/10.1103/PhysRevD.101.123013}{\emph{Phys. Rev. D}
  {\bfseries 101} (2020) 123013}
  [\href{https://arxiv.org/abs/1910.12971}{{\ttfamily 1910.12971}}].

\bibitem{Walk:2018gaw}
L.~Walk, I.~Tamborra, H.-T.~Janka and A.~Summa, \emph{{Identifying rotation in
  SASI-dominated core-collapse supernovae with a neutrino gyroscope}},
  \href{https://doi.org/10.1103/PhysRevD.98.123001}{\emph{Phys. Rev. D}
  {\bfseries 98} (2018) 123001}
  [\href{https://arxiv.org/abs/1807.02366}{{\ttfamily 1807.02366}}].

\bibitem{Adams:2013ana}
S.M.~Adams, C.S.~Kochanek, J.F.~Beacom, M.R.~Vagins and K.Z.~Stanek,
  \emph{{Observing the Next Galactic Supernova}},
  \href{https://doi.org/10.1088/0004-637X/778/2/164}{\emph{Astrophys. J.}
  {\bfseries 778} (2013) 164}
  [\href{https://arxiv.org/abs/1306.0559}{{\ttfamily 1306.0559}}].

\bibitem{GalloRosso:2018ugl}
A.~Gallo~Rosso, S.~Abbar, F.~Vissani and M.C.~Volpe, \emph{{Late time supernova
  neutrino signal and proto-neutron star radius}},
  \href{https://doi.org/10.1088/1475-7516/2018/12/006}{\emph{JCAP} {\bfseries
  12} (2018) 006} [\href{https://arxiv.org/abs/1809.09074}{{\ttfamily
  1809.09074}}].

\bibitem{Suliga:2019bsq}
A.M.~Suliga, I.~Tamborra and M.-R.~Wu, \emph{{Tau lepton asymmetry by sterile
  neutrino emission -- Moving beyond one-zone supernova models}},
  \href{https://doi.org/10.1088/1475-7516/2019/12/019}{\emph{JCAP} {\bfseries
  12} (2019) 019} [\href{https://arxiv.org/abs/1908.11382}{{\ttfamily
  1908.11382}}].

\bibitem{deGouvea:2019goq}
A.~de~Gouv\^ea, I.~Martinez-Soler and M.~Sen, \emph{{Impact of neutrino decays
  on the supernova neutronization-burst flux}},
  \href{https://doi.org/10.1103/PhysRevD.101.043013}{\emph{Phys. Rev. D}
  {\bfseries 101} (2020) 043013}
  [\href{https://arxiv.org/abs/1910.01127}{{\ttfamily 1910.01127}}].

\bibitem{Das:2017iuj}
A.~Das, A.~Dighe and M.~Sen, \emph{{New effects of non-standard
  self-interactions of neutrinos in a supernova}},
  \href{https://doi.org/10.1088/1475-7516/2017/05/051}{\emph{JCAP} {\bfseries
  05} (2017) 051} [\href{https://arxiv.org/abs/1705.00468}{{\ttfamily
  1705.00468}}].

\bibitem{Nunokawa:1997ct}
H.~Nunokawa, J.T.~Peltoniemi, A.~Rossi and J.W.F.~Valle, \emph{{Supernova
  bounds on resonant active sterile neutrino conversions}},
  \href{https://doi.org/10.1103/PhysRevD.56.1704}{\emph{Phys. Rev. D}
  {\bfseries 56} (1997) 1704}
  [\href{https://arxiv.org/abs/hep-ph/9702372}{{\ttfamily hep-ph/9702372}}].

\bibitem{Tamborra:2011is}
I.~Tamborra, G.G.~Raffelt, L.~Hudepohl and H.-T.~Janka, \emph{{Impact of
  eV-mass sterile neutrinos on neutrino-driven supernova outflows}},
  \href{https://doi.org/10.1088/1475-7516/2012/01/013}{\emph{JCAP} {\bfseries
  01} (2012) 013} [\href{https://arxiv.org/abs/1110.2104}{{\ttfamily
  1110.2104}}].

\bibitem{Shalgar:2019rqe}
S.~Shalgar, I.~Tamborra and M.~Bustamante, \emph{{Core-collapse supernovae
  stymie secret neutrino interactions}},
  \href{https://doi.org/10.1103/PhysRevD.103.123008}{\emph{Phys. Rev. D}
  {\bfseries 103} (2021) 123008}
  [\href{https://arxiv.org/abs/1912.09115}{{\ttfamily 1912.09115}}].

\bibitem{Carenza:2020cis}
P.~Carenza, B.~Fore, M.~Giannotti, A.~Mirizzi and S.~Reddy, \emph{{Enhanced
  Supernova Axion Emission and its Implications}},
  \href{https://doi.org/10.1103/PhysRevLett.126.071102}{\emph{Phys. Rev. Lett.}
  {\bfseries 126} (2021) 071102}
  [\href{https://arxiv.org/abs/2010.02943}{{\ttfamily 2010.02943}}].

\bibitem{Stapleford:2016jgz}
C.J.~Stapleford, D.J.~V\"a\"an\"anen, J.P.~Kneller, G.C.~McLaughlin and
  B.T.~Shapiro, \emph{{Nonstandard Neutrino Interactions in Supernovae}},
  \href{https://doi.org/10.1103/PhysRevD.94.093007}{\emph{Phys. Rev. D}
  {\bfseries 94} (2016) 093007}
  [\href{https://arxiv.org/abs/1605.04903}{{\ttfamily 1605.04903}}].

\bibitem{Arcones:2012wj}
A.~Arcones and F.K.~Thielemann, \emph{{Neutrino-driven wind simulations and
  nucleosynthesis of heavy elements}},
  \href{https://doi.org/10.1088/0954-3899/40/1/013201}{\emph{J. Phys. G}
  {\bfseries 40} (2013) 013201}
  [\href{https://arxiv.org/abs/1207.2527}{{\ttfamily 1207.2527}}].

\bibitem{Cowan:2019pkx}
J.J.~Cowan, C.~Sneden, J.E.~Lawler, A.~Aprahamian, M.~Wiescher, K.~Langanke
  et~al., \emph{{Origin of the heaviest elements: The rapid neutron-capture
  process}}, \href{https://doi.org/10.1103/RevModPhys.93.015002}{\emph{Rev.
  Mod. Phys.} {\bfseries 93} (2021) 15002}
  [\href{https://arxiv.org/abs/1901.01410}{{\ttfamily 1901.01410}}].

\bibitem{Petropoulou:2017ymv}
M.~Petropoulou, S.~Coenders, G.~Vasilopoulos, A.~Kamble and L.~Sironi,
  \emph{{Point-source and diffuse high-energy neutrino emission from Type IIn
  supernovae}}, \href{https://doi.org/10.1093/mnras/stx1251}{\emph{Mon. Not.
  Roy. Astron. Soc.} {\bfseries 470} (2017) 1881}
  [\href{https://arxiv.org/abs/1705.06752}{{\ttfamily 1705.06752}}].

\bibitem{Murase:2017pfe}
K.~Murase, \emph{{New Prospects for Detecting High-Energy Neutrinos from Nearby
  Supernovae}}, \href{https://doi.org/10.1103/PhysRevD.97.081301}{\emph{Phys.
  Rev. D} {\bfseries 97} (2018) 081301}
  [\href{https://arxiv.org/abs/1705.04750}{{\ttfamily 1705.04750}}].

\bibitem{Murase:2010cu}
K.~Murase, T.A.~Thompson, B.C.~Lacki and J.F.~Beacom, \emph{{New Class of
  High-Energy Transients from Crashes of Supernova Ejecta with Massive
  Circumstellar Material Shells}},
  \href{https://doi.org/10.1103/PhysRevD.84.043003}{\emph{Phys. Rev.}
  {\bfseries D84} (2011) 043003}
  [\href{https://arxiv.org/abs/astro-ph/1012.2834}{{\ttfamily
  astro-ph/1012.2834}}].

\bibitem{Zirakashvili:2015mua}
V.N.~Zirakashvili and V.S.~Ptuskin, \emph{{Type IIn supernovae as sources of
  high energy astrophysical neutrinos}},
  \href{https://doi.org/10.1016/j.astropartphys.2016.02.004}{\emph{Astropart.
  Phys.} {\bfseries 78} (2016) 28}
  [\href{https://arxiv.org/abs/1510.08387}{{\ttfamily 1510.08387}}].

\bibitem{Follin:2015hya}
B.~Follin, L.~Knox, M.~Millea and Z.~Pan, \emph{{First Detection of the
  Acoustic Oscillation Phase Shift Expected from the Cosmic Neutrino
  Background}},
  \href{https://doi.org/10.1103/PhysRevLett.115.091301}{\emph{Phys. Rev. Lett.}
  {\bfseries 115} (2015) 091301}
  [\href{https://arxiv.org/abs/1503.07863}{{\ttfamily 1503.07863}}].

\bibitem{Baumann:2018qnt}
D.D.~Baumann, F.~Beutler, R.~Flauger, D.R.~Green, A.~Slosar, M.~Vargas-Maga\~na
  et~al., \emph{{First constraint on the neutrino-induced phase shift in the
  spectrum of baryon acoustic oscillations}},
  \href{https://doi.org/10.1038/s41567-019-0435-6}{\emph{Nature Phys.}
  {\bfseries 15} (2019) 465}
  [\href{https://arxiv.org/abs/1803.10741}{{\ttfamily 1803.10741}}].

\bibitem{Fiorillo:2021hty}
D.F.G.~Fiorillo, A.~Van~Vliet, S.~Morisi and W.~Winter, \emph{{Unified thermal
  model for photohadronic neutrino production in astrophysical sources}},
  \href{https://doi.org/10.1088/1475-7516/2021/07/028}{\emph{JCAP} {\bfseries
  07} (2021) 028} [\href{https://arxiv.org/abs/2103.16577}{{\ttfamily
  2103.16577}}].

\bibitem{Winter:2012xq}
W.~Winter, \emph{{Neutrinos from Cosmic Accelerators Including Magnetic Field
  and Flavor Effects}}, \href{https://doi.org/10.1155/2012/586413}{\emph{Adv.
  High Energy Phys.} {\bfseries 2012} (2012) 586413}
  [\href{https://arxiv.org/abs/1201.5462}{{\ttfamily 1201.5462}}].

\bibitem{Murase:2013rfa}
K.~Murase, M.~Ahlers and B.C.~Lacki, \emph{{Testing the Hadronuclear Origin of
  PeV Neutrinos Observed with IceCube}},
  \href{https://doi.org/10.1103/PhysRevD.88.121301}{\emph{Phys.Rev.} {\bfseries
  D88} (2013) 121301} [\href{https://arxiv.org/abs/1306.3417}{{\ttfamily
  1306.3417}}].

\bibitem{Bechtol:2015uqb}
K.~Bechtol, M.~Ahlers, M.~Di~Mauro, M.~Ajello and J.~Vandenbroucke,
  \emph{{Evidence against star-forming galaxies as the dominant source of
  IceCube neutrinos}},
  \href{https://doi.org/10.3847/1538-4357/836/1/47}{\emph{Astrophys. J.}
  {\bfseries 836} (2017) 47}
  [\href{https://arxiv.org/abs/1511.00688}{{\ttfamily 1511.00688}}].

\bibitem{Murase:2015xka}
K.~Murase, D.~Guetta and M.~Ahlers, \emph{{Hidden Cosmic-Ray Accelerators as an
  Origin of TeV-PeV Cosmic Neutrinos}},
  \href{https://doi.org/10.1103/PhysRevLett.116.071101}{\emph{Phys. Rev. Lett.}
  {\bfseries 116} (2016) 071101}
  [\href{https://arxiv.org/abs/astro-ph/1509.00805}{{\ttfamily
  astro-ph/1509.00805}}].

\bibitem{2019NatAs...3...88G}
S.~{Gao}, A.~{Fedynitch}, W.~{Winter} and M.~{Pohl}, \emph{{Modelling the
  coincident observation of a high-energy neutrino and a bright blazar flare}},
  \href{https://doi.org/10.1038/s41550-018-0610-1}{\emph{Nature Astronomy}
  {\bfseries 3} (2019) 88} [\href{https://arxiv.org/abs/1807.04275}{{\ttfamily
  1807.04275}}].

\bibitem{Mannheim:1993jg}
K.~Mannheim, \emph{{The Proton blazar}}, {\emph{Astron. Astrophys.} {\bfseries
  269} (1993) 67} [\href{https://arxiv.org/abs/astro-ph/9302006}{{\ttfamily
  astro-ph/9302006}}].

\bibitem{Cerruti:2018tmc}
M.~Cerruti, A.~Zech, C.~Boisson, G.~Emery, S.~Inoue and J.-P.~Lenain,
  \emph{{Leptohadronic single-zone models for the electromagnetic and neutrino
  emission of TXS 0506+056}},
  \href{https://doi.org/10.1093/mnrasl/sly210}{\emph{Mon. Not. Roy. Astron.
  Soc.} {\bfseries 483} (2019) L12}
  [\href{https://arxiv.org/abs/1807.04335}{{\ttfamily 1807.04335}}].

\bibitem{2018ApJ...864...84K}
A.~{Keivani}, K.~{Murase}, M.~{Petropoulou}, D.B.~{Fox}, S.B.~{Cenko},
  S.~{Chaty} et~al., \emph{{A Multimessenger Picture of the Flaring Blazar TXS
  0506+056: Implications for High-energy Neutrino Emission and Cosmic-Ray
  Acceleration}},
  \href{https://doi.org/10.3847/1538-4357/aad59a}{\emph{Astrophys. J.}
  {\bfseries 864} (2018) 84}
  [\href{https://arxiv.org/abs/1807.04537}{{\ttfamily 1807.04537}}].

\bibitem{Rodrigues:2020fbu}
X.~Rodrigues, S.~Garrappa, S.~Gao, V.S.~Paliya, A.~Franckowiak and W.~Winter,
  \emph{{Multiwavelength and Neutrino Emission from Blazar PKS 1502 + 106}},
  \href{https://doi.org/10.3847/1538-4357/abe87b}{\emph{Astrophys. J.}
  {\bfseries 912} (2021) 54}
  [\href{https://arxiv.org/abs/2009.04026}{{\ttfamily 2009.04026}}].

\bibitem{2019ApJ...881...46R}
A.~{Reimer}, M.~{B{\"o}ttcher} and S.~{Buson}, \emph{{Cascading Constraints
  from Neutrino-emitting Blazars: The Case of TXS 0506+056}},
  \href{https://doi.org/10.3847/1538-4357/ab2bff}{\emph{Astrophys. J.}
  {\bfseries 881} (2019) 46}
  [\href{https://arxiv.org/abs/1812.05654}{{\ttfamily 1812.05654}}].

\bibitem{Rodrigues:2018tku}
X.~Rodrigues, S.~Gao, A.~Fedynitch, A.~Palladino and W.~Winter,
  \emph{{Leptohadronic Blazar Models Applied to the 2014-2015 Flare of TXS
  0506+056}}, \href{https://doi.org/10.3847/2041-8213/ab1267}{\emph{Astrophys.
  J.} {\bfseries 874} (2019) L29}
  [\href{https://arxiv.org/abs/1812.05939}{{\ttfamily 1812.05939}}].

\bibitem{Hayasaki:2019kjy}
K.~Hayasaki and R.~Yamazaki, \emph{{Neutrino Emissions from Tidal Disruption
  Remnants}},  \href{https://arxiv.org/abs/1908.10882}{{\ttfamily 1908.10882}}.

\bibitem{Winter:2020ptf}
W.~Winter and C.~Lunardini, \emph{{A concordance scenario for the observation
  of a neutrino from the Tidal Disruption Event AT2019dsg}},
  \href{https://doi.org/10.1038/s41550-021-01305-3}{\emph{Nature Astron.}
  (2021) } [\href{https://arxiv.org/abs/2005.06097}{{\ttfamily 2005.06097}}].

\bibitem{Murase:2020lnu}
K.~Murase, S.S.~Kimura, B.T.~Zhang, F.~Oikonomou and M.~Petropoulou,
  \emph{{High-Energy Neutrino and Gamma-Ray Emission from Tidal Disruption
  Events}}, \href{https://doi.org/10.3847/1538-4357/abb3c0}{\emph{Astrophys.
  J.} {\bfseries 902} (2020) 108}
  [\href{https://arxiv.org/abs/2005.08937}{{\ttfamily 2005.08937}}].

\bibitem{Boncioli:2018lrv}
D.~Boncioli, D.~Biehl and W.~Winter, \emph{{On the common origin of cosmic rays
  across the ankle and diffuse neutrinos at the highest energies from
  low-luminosity Gamma-Ray Bursts}},
  \href{https://doi.org/10.3847/1538-4357/aafda7}{\emph{Astrophys. J.}
  {\bfseries 872} (2019) 110}
  [\href{https://arxiv.org/abs/1808.07481}{{\ttfamily 1808.07481}}].

\bibitem{Waxman:1998yy}
E.~Waxman and J.N.~Bahcall, \emph{{High-energy neutrinos from astrophysical
  sources: An Upper bound}},
  \href{https://doi.org/10.1103/PhysRevD.59.023002}{\emph{Phys. Rev.}
  {\bfseries D59} (1999) 023002}
  [\href{https://arxiv.org/abs/hep-ph/9807282}{{\ttfamily hep-ph/9807282}}].

\bibitem{Abbasi:2020jmh}
{\scshape IceCube} collaboration, \emph{{The IceCube high-energy starting event
  sample: Description and flux characterization with 7.5 years of data}},
  \href{https://arxiv.org/abs/2011.03545}{{\ttfamily 2011.03545}}.

\bibitem{vanVliet:2019nse}
A.~van Vliet, R.~Alves~Batista and J.R.~H\"orandel, \emph{{Determining the
  fraction of cosmic-ray protons at ultrahigh energies with cosmogenic
  neutrinos}}, \href{https://doi.org/10.1103/PhysRevD.100.021302}{\emph{Phys.
  Rev. D} {\bfseries 100} (2019) 021302}
  [\href{https://arxiv.org/abs/1901.01899}{{\ttfamily 1901.01899}}].

\bibitem{Rodrigues:2020pli}
X.~Rodrigues, J.~Heinze, A.~Palladino, A.~van Vliet and W.~Winter,
  \emph{{Active Galactic Nuclei Jets as the Origin of Ultrahigh-Energy Cosmic
  Rays and Perspectives for the Detection of Astrophysical Source Neutrinos at
  EeV Energies}},
  \href{https://doi.org/10.1103/PhysRevLett.126.191101}{\emph{Phys. Rev. Lett.}
  {\bfseries 126} (2021) 191101}
  [\href{https://arxiv.org/abs/2003.08392}{{\ttfamily 2003.08392}}].

\bibitem{Fang:2017zjf}
K.~Fang and K.~Murase, \emph{{Linking High-Energy Cosmic Particles by Black
  Hole Jets Embedded in Large-Scale Structures}},
  \href{https://doi.org/10.1038/s41567-017-0025-4}{\emph{Phys. Lett.}
  {\bfseries 14} (2018) 396}
  [\href{https://arxiv.org/abs/1704.00015}{{\ttfamily 1704.00015}}].

\bibitem{Ando:2012hna}
S.~Ando et~al., \emph{{Colloquium: Multimessenger astronomy with gravitational
  waves and high-energy neutrinos}},
  \href{https://doi.org/10.1103/RevModPhys.85.1401}{\emph{Rev. Mod. Phys.}
  {\bfseries 85} (2013) 1401}
  [\href{https://arxiv.org/abs/1203.5192}{{\ttfamily 1203.5192}}].

\bibitem{Kimura:2017kan}
S.S.~Kimura, K.~Murase, P.~M\'esz\'aros and K.~Kiuchi, \emph{{High-Energy
  Neutrino Emission from Short Gamma-Ray Bursts: Prospects for Coincident
  Detection with Gravitational Waves}},
  \href{https://doi.org/10.3847/2041-8213/aa8d14}{\emph{Astrophys. J. Lett.}
  {\bfseries 848} (2017) L4}
  [\href{https://arxiv.org/abs/1708.07075}{{\ttfamily 1708.07075}}].

\bibitem{Fang:2017tla}
K.~Fang and B.D.~Metzger, \emph{{High-Energy Neutrinos from Millisecond
  Magnetars formed from the Merger of Binary Neutron Stars}},
  \href{https://doi.org/10.3847/1538-4357/aa8b6a}{\emph{Astrophys. J.}
  {\bfseries 849} (2017) 153}
  [\href{https://arxiv.org/abs/1707.04263}{{\ttfamily 1707.04263}}].

\bibitem{ANTARES:2017bia}
{\scshape ANTARES, IceCube, Pierre Auger, LIGO Scientific, Virgo}
  collaboration, \emph{{Search for High-energy Neutrinos from Binary Neutron
  Star Merger GW170817 with ANTARES, IceCube, and the Pierre Auger
  Observatory}},
  \href{https://doi.org/10.3847/2041-8213/aa9aed}{\emph{Astrophys. J. Lett.}
  {\bfseries 850} (2017) L35}
  [\href{https://arxiv.org/abs/1710.05839}{{\ttfamily 1710.05839}}].

\bibitem{Baikal-GVD:2018cya}
{\scshape Baikal-GVD} collaboration, \emph{{Search for High-Energy Neutrinos
  from GW170817 with the Baikal-GVD Neutrino Telescope}},
  \href{https://doi.org/10.1134/S0021364018240025}{\emph{JETP Lett.} {\bfseries
  108} (2018) 787} [\href{https://arxiv.org/abs/1810.10966}{{\ttfamily
  1810.10966}}].

\bibitem{Biehl:2017qen}
D.~Biehl, J.~Heinze and W.~Winter, \emph{{Expected neutrino fluence from short
  Gamma-Ray Burst 170817A and off-axis angle constraints}},
  \href{https://doi.org/10.1093/mnras/sty285}{\emph{Mon. Not. Roy. Astron.
  Soc.} {\bfseries 476} (2018) 1191}
  [\href{https://arxiv.org/abs/1712.00449}{{\ttfamily 1712.00449}}].

\bibitem{ANTARES:2018bmu}
{\scshape ANTARES, IceCube, LIGO, Virgo} collaboration, \emph{{Search for
  Multimessenger Sources of Gravitational Waves and High-energy Neutrinos with
  Advanced LIGO during Its First Observing Run, ANTARES, and IceCube}},
  \href{https://doi.org/10.3847/1538-4357/aaf21d}{\emph{Astrophys. J.}
  {\bfseries 870} (2019) 134}
  [\href{https://arxiv.org/abs/1810.10693}{{\ttfamily 1810.10693}}].

\bibitem{Arguelles:2019rbn}
C.A.~Arguelles, M.~Bustamante, A.~Kheirandish, S.~Palomares-Ruiz, J.~Salvado
  and A.C.~Vincent, \emph{{Fundamental physics with high-energy cosmic
  neutrinos today and in the future}},
  \href{https://doi.org/10.22323/1.358.0849}{\emph{PoS} {\bfseries ICRC2019}
  (2020) 849} [\href{https://arxiv.org/abs/1907.08690}{{\ttfamily
  1907.08690}}].

\bibitem{Glashow:1979nm}
S.L.~Glashow, \emph{{The Future of Elementary Particle Physics}},
  \href{https://doi.org/10.1007/978-1-4684-7197-7_15}{\emph{NATO Sci. Ser. B}
  {\bfseries 61} (1980) 687}.

\bibitem{Wyler:1982dd}
D.~Wyler and L.~Wolfenstein, \emph{{Massless Neutrinos in Left-Right Symmetric
  Models}}, \href{https://doi.org/10.1016/0550-3213(83)90482-0}{\emph{Nucl.\
  Phys.\ B} {\bfseries 218} (1983) 205}.

\bibitem{Mohapatra:1986aw}
R.~Mohapatra, \emph{{Mechanism for Understanding Small Neutrino Mass in
  Superstring Theories}},
  \href{https://doi.org/10.1103/PhysRevLett.56.561}{\emph{Phys.\ Rev.\ Lett.}
  {\bfseries 56} (1986) 561}.

\bibitem{Mohapatra:1986bd}
R.~Mohapatra and J.~Valle, \emph{{Neutrino Mass and Baryon Number
  Nonconservation in Superstring Models}},
  \href{https://doi.org/10.1103/PhysRevD.34.1642}{\emph{Phys.\ Rev.\ D}
  {\bfseries 34} (1986) 1642}.

\bibitem{Akhmedov:1995vm}
E.K.~Akhmedov, M.~Lindner, E.~Schnapka and J.~Valle, \emph{{Dynamical
  left-right symmetry breaking}},
  \href{https://doi.org/10.1103/PhysRevD.53.2752}{\emph{Phys.\ Rev.\ D}
  {\bfseries 53} (1996) 2752}
  [\href{https://arxiv.org/abs/hep-ph/9509255}{{\ttfamily hep-ph/9509255}}].

\bibitem{Barr:2003nn}
S.~Barr, \emph{{A Different seesaw formula for neutrino masses}},
  \href{https://doi.org/10.1103/PhysRevLett.92.101601}{\emph{Phys.\ Rev.\
  Lett.} {\bfseries 92} (2004) 101601}
  [\href{https://arxiv.org/abs/hep-ph/0309152}{{\ttfamily hep-ph/0309152}}].

\bibitem{Malinsky:2005bi}
M.~Malinsky, J.~Romao and J.~Valle, \emph{{Novel supersymmetric SO(10) seesaw
  mechanism}},
  \href{https://doi.org/10.1103/PhysRevLett.95.161801}{\emph{Phys.\ Rev.\
  Lett.} {\bfseries 95} (2005) 161801}
  [\href{https://arxiv.org/abs/hep-ph/0506296}{{\ttfamily hep-ph/0506296}}].

\bibitem{Zee:1980ai}
A.~Zee, \emph{{A Theory of Lepton Number Violation, Neutrino Majorana Mass, and
  Oscillation}},
  \href{https://doi.org/10.1016/0370-2693(80)90349-4}{\emph{Phys. Lett. B}
  {\bfseries 93} (1980) 389}.

\bibitem{Ma:2006km}
E.~Ma, \emph{{Verifiable radiative seesaw mechanism of neutrino mass and dark
  matter}}, \href{https://doi.org/10.1103/PhysRevD.73.077301}{\emph{Phys. Rev.
  D} {\bfseries 73} (2006) 077301}
  [\href{https://arxiv.org/abs/hep-ph/0601225}{{\ttfamily hep-ph/0601225}}].

\bibitem{Zee:1985id}
A.~Zee, \emph{{Quantum Numbers of Majorana Neutrino Masses}},
  \href{https://doi.org/10.1016/0550-3213(86)90475-X}{\emph{Nucl. Phys. B}
  {\bfseries 264} (1986) 99}.

\bibitem{Babu:1988ki}
K.~Babu, \emph{{Model of 'Calculable' Majorana Neutrino Masses}},
  \href{https://doi.org/10.1016/0370-2693(88)91584-5}{\emph{Phys. Lett. B}
  {\bfseries 203} (1988) 132}.

\bibitem{Weinberg:1979sa}
S.~Weinberg, \emph{{Baryon and Lepton Nonconserving Processes}},
  \href{https://doi.org/10.1103/PhysRevLett.43.1566}{\emph{Phys. Rev. Lett.}
  {\bfseries 43} (1979) 1566}.

\bibitem{deGouvea:2007qla}
A.~de~Gouvea and J.~Jenkins, \emph{{A Survey of Lepton Number Violation Via
  Effective Operators}},
  \href{https://doi.org/10.1103/PhysRevD.77.013008}{\emph{Phys. Rev. D}
  {\bfseries 77} (2008) 013008}
  [\href{https://arxiv.org/abs/0708.1344}{{\ttfamily 0708.1344}}].

\bibitem{Angel:2012ug}
P.W.~Angel, N.L.~Rodd and R.R.~Volkas, \emph{{Origin of neutrino masses at the
  LHC: $\Delta L = 2$ effective operators and their ultraviolet completions}},
  \href{https://doi.org/10.1103/PhysRevD.87.073007}{\emph{Phys. Rev. D}
  {\bfseries 87} (2013) 073007}
  [\href{https://arxiv.org/abs/1212.6111}{{\ttfamily 1212.6111}}].

\bibitem{Cepedello:2017eqf}
R.~Cepedello, M.~Hirsch and J.~Helo, \emph{{Loop neutrino masses from $d = 7$
  operator}}, \href{https://doi.org/10.1007/JHEP07(2017)079}{\emph{JHEP}
  {\bfseries 07} (2017) 079}
  [\href{https://arxiv.org/abs/1705.01489}{{\ttfamily 1705.01489}}].

\bibitem{Anamiati:2018cuq}
G.~Anamiati, O.~Castillo-Felisola, R.M.~Fonseca, J.~Helo and M.~Hirsch,
  \emph{{High-dimensional neutrino masses}},
  \href{https://doi.org/10.1007/JHEP12(2018)066}{\emph{JHEP} {\bfseries 12}
  (2018) 066} [\href{https://arxiv.org/abs/1806.07264}{{\ttfamily
  1806.07264}}].

\bibitem{Altarelli:2010gt}
G.~Altarelli and F.~Feruglio, \emph{{Discrete Flavor Symmetries and Models of
  Neutrino Mixing}},
  \href{https://doi.org/10.1103/RevModPhys.82.2701}{\emph{Rev. Mod. Phys.}
  {\bfseries 82} (2010) 2701}
  [\href{https://arxiv.org/abs/1002.0211}{{\ttfamily 1002.0211}}].

\bibitem{King:2013eh}
S.F.~King and C.~Luhn, \emph{{Neutrino Mass and Mixing with Discrete
  Symmetry}}, \href{https://doi.org/10.1088/0034-4885/76/5/056201}{\emph{Rept.
  Prog. Phys.} {\bfseries 76} (2013) 056201}
  [\href{https://arxiv.org/abs/1301.1340}{{\ttfamily 1301.1340}}].

\bibitem{Xing:2019vks}
Z.-z.~Xing, \emph{{Flavor structures of charged fermions and massive
  neutrinos}}, \href{https://doi.org/10.1016/j.physrep.2020.02.001}{\emph{Phys.
  Rept.} {\bfseries 854} (2020) 1}
  [\href{https://arxiv.org/abs/1909.09610}{{\ttfamily 1909.09610}}].

\bibitem{Feruglio:2019ktm}
F.~Feruglio and A.~Romanino, \emph{{Lepton flavor symmetries}},
  \href{https://doi.org/10.1103/RevModPhys.93.015007}{\emph{Rev. Mod. Phys.}
  {\bfseries 93} (2021) 015007}
  [\href{https://arxiv.org/abs/1912.06028}{{\ttfamily 1912.06028}}].

\bibitem{Ishimori:2010au}
H.~Ishimori, T.~Kobayashi, H.~Ohki, Y.~Shimizu, H.~Okada and M.~Tanimoto,
  \emph{{Non-Abelian Discrete Symmetries in Particle Physics}},
  \href{https://doi.org/10.1143/PTPS.183.1}{\emph{Prog. Theor. Phys. Suppl.}
  {\bfseries 183} (2010) 1} [\href{https://arxiv.org/abs/1003.3552}{{\ttfamily
  1003.3552}}].

\bibitem{He:1990pn}
X.~He, G.C.~Joshi, H.~Lew and R.~Volkas, \emph{{NEW Z-prime PHENOMENOLOGY}},
  \href{https://doi.org/10.1103/PhysRevD.43.R22}{\emph{Phys. Rev. D} {\bfseries
  43} (1991) 22}.

\bibitem{Foot:1990mn}
R.~Foot, \emph{{New Physics From Electric Charge Quantization?}},
  \href{https://doi.org/10.1142/S0217732391000543}{\emph{Mod. Phys. Lett. A}
  {\bfseries 6} (1991) 527}.

\bibitem{He:1991qd}
X.-G.~He, G.C.~Joshi, H.~Lew and R.~Volkas, \emph{{Simplest Z-prime model}},
  \href{https://doi.org/10.1103/PhysRevD.44.2118}{\emph{Phys. Rev. D}
  {\bfseries 44} (1991) 2118}.

\bibitem{Binetruy:1996cs}
P.~Binetruy, S.~Lavignac, S.T.~Petcov and P.~Ramond, \emph{{Quasidegenerate
  neutrinos from an Abelian family symmetry}},
  \href{https://doi.org/10.1016/S0550-3213(97)00211-3}{\emph{Nucl. Phys. B}
  {\bfseries 496} (1997) 3}
  [\href{https://arxiv.org/abs/hep-ph/9610481}{{\ttfamily hep-ph/9610481}}].

\bibitem{Altmannshofer:2014cfa}
W.~Altmannshofer, S.~Gori, M.~Pospelov and I.~Yavin, \emph{{Quark flavor
  transitions in $L_\mu-L_\tau$ models}},
  \href{https://doi.org/10.1103/PhysRevD.89.095033}{\emph{Phys. Rev. D}
  {\bfseries 89} (2014) 095033}
  [\href{https://arxiv.org/abs/1403.1269}{{\ttfamily 1403.1269}}].

\bibitem{Crivellin:2015mga}
A.~Crivellin, G.~D'Ambrosio and J.~Heeck, \emph{{Explaining
  $h\to\mu^\pm\tau^\mp$, $B\to K^* \mu^+\mu^-$ and $B\to K \mu^+\mu^-/B\to K
  e^+e^-$ in a two-Higgs-doublet model with gauged $L_\mu-L_\tau$}},
  \href{https://doi.org/10.1103/PhysRevLett.114.151801}{\emph{Phys. Rev. Lett.}
  {\bfseries 114} (2015) 151801}
  [\href{https://arxiv.org/abs/1501.00993}{{\ttfamily 1501.00993}}].

\bibitem{Petcov:2018snn}
S.T.~Petcov and A.V.~Titov, \emph{{Assessing the Viability of $A_4$, $S_4$ and
  $A_5$ Flavour Symmetries for Description of Neutrino Mixing}},
  \href{https://doi.org/10.1103/PhysRevD.97.115045}{\emph{Phys. Rev. D}
  {\bfseries 97} (2018) 115045}
  [\href{https://arxiv.org/abs/1804.00182}{{\ttfamily 1804.00182}}].

\bibitem{Frampton:2004ud}
P.~Frampton, S.~Petcov and W.~Rodejohann, \emph{{On deviations from bimaximal
  neutrino mixing}},
  \href{https://doi.org/10.1016/j.nuclphysb.2004.03.014}{\emph{Nucl. Phys. B}
  {\bfseries 687} (2004) 31}
  [\href{https://arxiv.org/abs/hep-ph/0401206}{{\ttfamily hep-ph/0401206}}].

\bibitem{King:2005bj}
S.~King, \emph{{Predicting neutrino parameters from SO(3) family symmetry and
  quark-lepton unification}},
  \href{https://doi.org/10.1088/1126-6708/2005/08/105}{\emph{JHEP} {\bfseries
  08} (2005) 105} [\href{https://arxiv.org/abs/hep-ph/0506297}{{\ttfamily
  hep-ph/0506297}}].

\bibitem{Girardi:2015vha}
I.~Girardi, S.~Petcov and A.~Titov, \emph{{Predictions for the Leptonic Dirac
  CP Violation Phase: a Systematic Phenomenological Analysis}},
  \href{https://doi.org/10.1140/epjc/s10052-015-3559-6}{\emph{Eur. Phys. J. C}
  {\bfseries 75} (2015) 345}
  [\href{https://arxiv.org/abs/1504.00658}{{\ttfamily 1504.00658}}].

\bibitem{Ballett:2013wya}
P.~Ballett, S.F.~King, C.~Luhn, S.~Pascoli and M.A.~Schmidt, \emph{{Testing
  atmospheric mixing sum rules at precision neutrino facilities}},
  \href{https://doi.org/10.1103/PhysRevD.89.016016}{\emph{Phys. Rev. D}
  {\bfseries 89} (2014) 016016}
  [\href{https://arxiv.org/abs/1308.4314}{{\ttfamily 1308.4314}}].

\bibitem{Barry:2010yk}
J.~Barry and W.~Rodejohann, \emph{{Neutrino Mass Sum-rules in Flavor Symmetry
  Models}}, \href{https://doi.org/10.1016/j.nuclphysb.2010.08.015}{\emph{Nucl.
  Phys. B} {\bfseries 842} (2011) 33}
  [\href{https://arxiv.org/abs/1007.5217}{{\ttfamily 1007.5217}}].

\bibitem{Dorame:2011eb}
L.~Dorame, D.~Meloni, S.~Morisi, E.~Peinado and J.~Valle, \emph{{Constraining
  Neutrinoless Double Beta Decay}},
  \href{https://doi.org/10.1016/j.nuclphysb.2012.04.003}{\emph{Nucl. Phys. B}
  {\bfseries 861} (2012) 259}
  [\href{https://arxiv.org/abs/1111.5614}{{\ttfamily 1111.5614}}].

\bibitem{King:2013psa}
S.F.~King, A.~Merle and A.J.~Stuart, \emph{{The Power of Neutrino Mass Sum
  Rules for Neutrinoless Double Beta Decay Experiments}},
  \href{https://doi.org/10.1007/JHEP12(2013)005}{\emph{JHEP} {\bfseries 12}
  (2013) 005} [\href{https://arxiv.org/abs/1307.2901}{{\ttfamily 1307.2901}}].

\bibitem{Grimus:1995zi}
W.~Grimus and M.~Rebelo, \emph{{Automorphisms in gauge theories and the
  definition of CP and P}},
  \href{https://doi.org/10.1016/S0370-1573(96)00030-0}{\emph{Phys. Rept.}
  {\bfseries 281} (1997) 239}
  [\href{https://arxiv.org/abs/hep-ph/9506272}{{\ttfamily hep-ph/9506272}}].

\bibitem{Feruglio:2012cw}
F.~Feruglio, C.~Hagedorn and R.~Ziegler, \emph{{Lepton Mixing Parameters from
  Discrete and CP Symmetries}},
  \href{https://doi.org/10.1007/JHEP07(2013)027}{\emph{JHEP} {\bfseries 07}
  (2013) 027} [\href{https://arxiv.org/abs/1211.5560}{{\ttfamily 1211.5560}}].

\bibitem{Holthausen:2012dk}
M.~Holthausen, M.~Lindner and M.A.~Schmidt, \emph{{CP and Discrete Flavour
  Symmetries}}, \href{https://doi.org/10.1007/JHEP04(2013)122}{\emph{JHEP}
  {\bfseries 04} (2013) 122} [\href{https://arxiv.org/abs/1211.6953}{{\ttfamily
  1211.6953}}].

\bibitem{Chen:2014tpa}
M.-C.~Chen, M.~Fallbacher, K.~Mahanthappa, M.~Ratz and A.~Trautner, \emph{{CP
  Violation from Finite Groups}},
  \href{https://doi.org/10.1016/j.nuclphysb.2014.03.023}{\emph{Nucl. Phys. B}
  {\bfseries 883} (2014) 267}
  [\href{https://arxiv.org/abs/1402.0507}{{\ttfamily 1402.0507}}].

\bibitem{Coloma:2018ioo}
P.~Coloma and S.~Pascoli, \emph{{Theory and Phenomenology of Mass Ordering and
  CP Violation}},  vol.~28, pp.~497--542 (2018),
  \href{https://doi.org/10.1142/9789813226098\_0013}{DOI}.

\bibitem{Feruglio:2017spp}
F.~Feruglio, \emph{{Are neutrino masses modular forms?}},  in \emph{{From My
  Vast Repertoire ...}: {Guido Altarelli's Legacy}}, pp.~227--266 (2019),
  \href{https://doi.org/10.1142/9789813238053\_0012}{DOI}
  [\href{https://arxiv.org/abs/1706.08749}{{\ttfamily 1706.08749}}].

\bibitem{Penedo:2018nmg}
J.~Penedo and S.~Petcov, \emph{{Lepton Masses and Mixing from Modular $S_4$
  Symmetry}},
  \href{https://doi.org/10.1016/j.nuclphysb.2018.12.016}{\emph{Nucl. Phys. B}
  {\bfseries 939} (2019) 292}
  [\href{https://arxiv.org/abs/1806.11040}{{\ttfamily 1806.11040}}].

\bibitem{Criado:2018thu}
J.C.~Criado and F.~Feruglio, \emph{{Modular Invariance Faces Precision Neutrino
  Data}}, \href{https://doi.org/10.21468/SciPostPhys.5.5.042}{\emph{SciPost
  Phys.} {\bfseries 5} (2018) 042}
  [\href{https://arxiv.org/abs/1807.01125}{{\ttfamily 1807.01125}}].

\bibitem{Kobayashi:2018vbk}
T.~Kobayashi, K.~Tanaka and T.H.~Tatsuishi, \emph{{Neutrino mixing from finite
  modular groups}},
  \href{https://doi.org/10.1103/PhysRevD.98.016004}{\emph{Phys. Rev. D}
  {\bfseries 98} (2018) 016004}
  [\href{https://arxiv.org/abs/1803.10391}{{\ttfamily 1803.10391}}].

\bibitem{Kobayashi:2018scp}
T.~Kobayashi, N.~Omoto, Y.~Shimizu, K.~Takagi, M.~Tanimoto and T.H.~Tatsuishi,
  \emph{{Modular A$_{4}$ invariance and neutrino mixing}},
  \href{https://doi.org/10.1007/JHEP11(2018)196}{\emph{JHEP} {\bfseries 11}
  (2018) 196} [\href{https://arxiv.org/abs/1808.03012}{{\ttfamily
  1808.03012}}].

\bibitem{Ding:2019zxk}
G.-J.~Ding, S.F.~King and X.-G.~Liu, \emph{{Modular A$_{4}$ symmetry models of
  neutrinos and charged leptons}},
  \href{https://doi.org/10.1007/JHEP09(2019)074}{\emph{JHEP} {\bfseries 09}
  (2019) 074} [\href{https://arxiv.org/abs/1907.11714}{{\ttfamily
  1907.11714}}].

\bibitem{Petcov:1976ff}
S.~Petcov, \emph{{The Processes $\mu \rightarrow e + \gamma, \mu \rightarrow e
  + \overline{e}, \nu' \rightarrow \nu + \gamma$ in the Weinberg-Salam Model
  with Neutrino Mixing}}, {\emph{Sov. J. Nucl. Phys.} {\bfseries 25} (1977)
  340}.

\bibitem{deGouvea:2013zba}
A.~de~Gouvea and P.~Vogel, \emph{{Lepton Flavor and Number Conservation, and
  Physics Beyond the Standard Model}},
  \href{https://doi.org/10.1016/j.ppnp.2013.03.006}{\emph{Prog. Part. Nucl.
  Phys.} {\bfseries 71} (2013) 75}
  [\href{https://arxiv.org/abs/1303.4097}{{\ttfamily 1303.4097}}].

\bibitem{Calibbi:2017uvl}
L.~Calibbi and G.~Signorelli, \emph{{Charged Lepton Flavour Violation: An
  Experimental and Theoretical Introduction}},
  \href{https://doi.org/10.1393/ncr/i2018-10144-0}{\emph{Riv. Nuovo Cim.}
  {\bfseries 41} (2018) 71} [\href{https://arxiv.org/abs/1709.00294}{{\ttfamily
  1709.00294}}].

\bibitem{Chun:2003ej}
E.J.~Chun, K.Y.~Lee and S.C.~Park, \emph{{Testing Higgs triplet model and
  neutrino mass patterns}},
  \href{https://doi.org/10.1016/S0370-2693(03)00770-6}{\emph{Phys. Lett. B}
  {\bfseries 566} (2003) 142}
  [\href{https://arxiv.org/abs/hep-ph/0304069}{{\ttfamily hep-ph/0304069}}].

\bibitem{Baldini:2018nnn}
{\scshape MEG II} collaboration, \emph{{The design of the MEG II experiment}},
  \href{https://doi.org/10.1140/epjc/s10052-018-5845-6}{\emph{Eur. Phys. J. C}
  {\bfseries 78} (2018) 380}
  [\href{https://arxiv.org/abs/1801.04688}{{\ttfamily 1801.04688}}].

\bibitem{Blondel:2013ia}
A.~Blondel et~al., \emph{{Research Proposal for an Experiment to Search for the
  Decay $\mu \to eee$}},  \href{https://arxiv.org/abs/1301.6113}{{\ttfamily
  1301.6113}}.

\bibitem{Bartoszek:2014mya}
{\scshape Mu2e} collaboration, \emph{{Mu2e Technical Design Report}},
  \href{https://arxiv.org/abs/1501.05241}{{\ttfamily 1501.05241}}.

\bibitem{Adamov:2018vin}
{\scshape COMET} collaboration, \emph{{COMET Phase-I Technical Design Report}},
  \href{https://doi.org/10.1093/ptep/ptz125}{\emph{PTEP} {\bfseries 2020}
  (2020) 033C01} [\href{https://arxiv.org/abs/1812.09018}{{\ttfamily
  1812.09018}}].

\bibitem{Hiller:2014yaa}
G.~Hiller and M.~Schmaltz, \emph{{$R_K$ and future $b \to s \ell \ell$ physics
  beyond the standard model opportunities}},
  \href{https://doi.org/10.1103/PhysRevD.90.054014}{\emph{Phys. Rev. D}
  {\bfseries 90} (2014) 054014}
  [\href{https://arxiv.org/abs/1408.1627}{{\ttfamily 1408.1627}}].

\bibitem{Bauer:2015knc}
M.~Bauer and M.~Neubert, \emph{{Minimal Leptoquark Explanation for the
  R$_{D^{(*)}}$ , R$_K$ , and $(g-2)_g$ Anomalies}},
  \href{https://doi.org/10.1103/PhysRevLett.116.141802}{\emph{Phys. Rev. Lett.}
  {\bfseries 116} (2016) 141802}
  [\href{https://arxiv.org/abs/1511.01900}{{\ttfamily 1511.01900}}].

\bibitem{Cata:2019wbu}
O.~Cat\`a and T.~Mannel, \emph{{Linking lepton number violation with $B$
  anomalies}},  \href{https://arxiv.org/abs/1903.01799}{{\ttfamily
  1903.01799}}.

\bibitem{Adhikari:2016bei}
M.~Drewes et~al., \emph{{A White Paper on keV Sterile Neutrino Dark Matter}},
  \href{https://doi.org/10.1088/1475-7516/2017/01/025}{\emph{JCAP} {\bfseries
  01} (2017) 025} [\href{https://arxiv.org/abs/1602.04816}{{\ttfamily
  1602.04816}}].

\bibitem{Dodelson:1993je}
S.~Dodelson and L.M.~Widrow, \emph{{Sterile-neutrinos as dark matter}},
  \href{https://doi.org/10.1103/PhysRevLett.72.17}{\emph{Phys. Rev. Lett.}
  {\bfseries 72} (1994) 17}
  [\href{https://arxiv.org/abs/hep-ph/9303287}{{\ttfamily hep-ph/9303287}}].

\bibitem{Shi:1998km}
X.-D.~Shi and G.M.~Fuller, \emph{{A New dark matter candidate: Nonthermal
  sterile neutrinos}},
  \href{https://doi.org/10.1103/PhysRevLett.82.2832}{\emph{Phys. Rev. Lett.}
  {\bfseries 82} (1999) 2832}
  [\href{https://arxiv.org/abs/astro-ph/9810076}{{\ttfamily
  astro-ph/9810076}}].

\bibitem{Boyarsky:2009ix}
A.~Boyarsky, O.~Ruchayskiy and M.~Shaposhnikov, \emph{{The Role of sterile
  neutrinos in cosmology and astrophysics}},
  \href{https://doi.org/10.1146/annurev.nucl.010909.083654}{\emph{Ann. Rev.
  Nucl. Part. Sci.} {\bfseries 59} (2009) 191}
  [\href{https://arxiv.org/abs/0901.0011}{{\ttfamily 0901.0011}}].

\bibitem{Laine:2008pg}
M.~Laine and M.~Shaposhnikov, \emph{{Sterile neutrino dark matter as a
  consequence of nuMSM-induced lepton asymmetry}},
  \href{https://doi.org/10.1088/1475-7516/2008/06/031}{\emph{JCAP} {\bfseries
  06} (2008) 031} [\href{https://arxiv.org/abs/0804.4543}{{\ttfamily
  0804.4543}}].

\bibitem{Abazajian:2017tcc}
K.N.~Abazajian, \emph{{Sterile neutrinos in cosmology}},
  \href{https://doi.org/10.1016/j.physrep.2017.10.003}{\emph{Phys. Rept.}
  {\bfseries 711-712} (2017) 1}
  [\href{https://arxiv.org/abs/1705.01837}{{\ttfamily 1705.01837}}].

\bibitem{Boyarsky:2018tvu}
A.~Boyarsky, M.~Drewes, T.~Lasserre, S.~Mertens and O.~Ruchayskiy,
  \emph{{Sterile neutrino Dark Matter}},
  \href{https://doi.org/10.1016/j.ppnp.2018.07.004}{\emph{Prog. Part. Nucl.
  Phys.} {\bfseries 104} (2019) 1}
  [\href{https://arxiv.org/abs/1807.07938}{{\ttfamily 1807.07938}}].

\bibitem{Bulbul:2014sua}
E.~Bulbul, M.~Markevitch, A.~Foster, R.K.~Smith, M.~Loewenstein and
  S.W.~Randall, \emph{{Detection of An Unidentified Emission Line in the
  Stacked X-ray spectrum of Galaxy Clusters}},
  \href{https://doi.org/10.1088/0004-637X/789/1/13}{\emph{Astrophys. J.}
  {\bfseries 789} (2014) 13} [\href{https://arxiv.org/abs/1402.2301}{{\ttfamily
  1402.2301}}].

\bibitem{Boyarsky:2014jta}
A.~Boyarsky, O.~Ruchayskiy, D.~Iakubovskyi and J.~Franse, \emph{{Unidentified
  Line in X-Ray Spectra of the Andromeda Galaxy and Perseus Galaxy Cluster}},
  \href{https://doi.org/10.1103/PhysRevLett.113.251301}{\emph{Phys. Rev. Lett.}
  {\bfseries 113} (2014) 251301}
  [\href{https://arxiv.org/abs/1402.4119}{{\ttfamily 1402.4119}}].

\bibitem{Jeltema:2014qfa}
T.E.~Jeltema and S.~Profumo, \emph{{Discovery of a 3.5 keV line in the Galactic
  Centre and a critical look at the origin of the line across astronomical
  targets}}, \href{https://doi.org/10.1093/mnras/stv768}{\emph{Mon. Not. Roy.
  Astron. Soc.} {\bfseries 450} (2015) 2143}
  [\href{https://arxiv.org/abs/1408.1699}{{\ttfamily 1408.1699}}].

\bibitem{Carlson:2014lla}
E.~Carlson, T.~Jeltema and S.~Profumo, \emph{{Where do the 3.5 keV photons come
  from? A morphological study of the Galactic Center and of Perseus}},
  \href{https://doi.org/10.1088/1475-7516/2015/02/009}{\emph{JCAP} {\bfseries
  02} (2015) 009} [\href{https://arxiv.org/abs/1411.1758}{{\ttfamily
  1411.1758}}].

\bibitem{Baur:2015jsy}
J.~Baur, N.~Palanque-Delabrouille, C.~Y\`eche, C.~Magneville and M.~Viel,
  \emph{{Lyman-alpha Forests cool Warm Dark Matter}},
  \href{https://doi.org/10.1088/1475-7516/2016/08/012}{\emph{JCAP} {\bfseries
  08} (2016) 012} [\href{https://arxiv.org/abs/1512.01981}{{\ttfamily
  1512.01981}}].

\bibitem{Irsic:2017ixq}
V.~Ir\v{s}i\v{c} et~al., \emph{{New Constraints on the free-streaming of warm
  dark matter from intermediate and small scale Lyman-$\alpha$ forest data}},
  \href{https://doi.org/10.1103/PhysRevD.96.023522}{\emph{Phys. Rev. D}
  {\bfseries 96} (2017) 023522}
  [\href{https://arxiv.org/abs/1702.01764}{{\ttfamily 1702.01764}}].

\bibitem{Yeche:2017upn}
C.~Y\`eche, N.~Palanque-Delabrouille, J.~Baur and H.~du~Mas~des Bourboux,
  \emph{{Constraints on neutrino masses from Lyman-alpha forest power spectrum
  with BOSS and XQ-100}},
  \href{https://doi.org/10.1088/1475-7516/2017/06/047}{\emph{JCAP} {\bfseries
  06} (2017) 047} [\href{https://arxiv.org/abs/1702.03314}{{\ttfamily
  1702.03314}}].

\bibitem{Baur:2017stq}
J.~Baur, N.~Palanque-Delabrouille, C.~Yeche, A.~Boyarsky, O.~Ruchayskiy,
  E.~Armengaud et~al., \emph{{Constraints from Ly-$\alpha$ forests on
  non-thermal dark matter including resonantly-produced sterile neutrinos}},
  \href{https://doi.org/10.1088/1475-7516/2017/12/013}{\emph{JCAP} {\bfseries
  12} (2017) 013} [\href{https://arxiv.org/abs/1706.03118}{{\ttfamily
  1706.03118}}].

\bibitem{Chulia:2016ngi}
S.~Centelles~Chuli\'a, E.~Ma, R.~Srivastava and J.W.F.~Valle, \emph{{Dirac
  Neutrinos and Dark Matter Stability from Lepton Quarticity}},
  \href{https://doi.org/10.1016/j.physletb.2017.01.070}{\emph{Phys. Lett. B}
  {\bfseries 767} (2017) 209}
  [\href{https://arxiv.org/abs/1606.04543}{{\ttfamily 1606.04543}}].

\bibitem{losHeros:2020csi}
C.~P\'erez de~los Heros, \emph{{Status of direct and indirect dark matter
  searches}}, \href{https://doi.org/10.22323/1.364.0694}{\emph{PoS} {\bfseries
  EPS-HEP2019} (2020) 694} [\href{https://arxiv.org/abs/2001.06193}{{\ttfamily
  2001.06193}}].

\bibitem{Billard:2013qya}
J.~Billard, L.~Strigari and E.~Figueroa-Feliciano, \emph{{Implication of
  neutrino backgrounds on the reach of next generation dark matter direct
  detection experiments}},
  \href{https://doi.org/10.1103/PhysRevD.89.023524}{\emph{Phys. Rev. D}
  {\bfseries 89} (2014) 023524}
  [\href{https://arxiv.org/abs/1307.5458}{{\ttfamily 1307.5458}}].

\bibitem{Davis:2014ama}
J.H.~Davis, \emph{{Dark Matter vs. Neutrinos: The effect of astrophysical
  uncertainties and timing information on the neutrino floor}},
  \href{https://doi.org/10.1088/1475-7516/2015/03/012}{\emph{JCAP} {\bfseries
  03} (2015) 012} [\href{https://arxiv.org/abs/1412.1475}{{\ttfamily
  1412.1475}}].

\bibitem{Ruppin:2014bra}
F.~Ruppin, J.~Billard, E.~Figueroa-Feliciano and L.~Strigari,
  \emph{{Complementarity of dark matter detectors in light of the neutrino
  background}}, \href{https://doi.org/10.1103/PhysRevD.90.083510}{\emph{Phys.
  Rev. D} {\bfseries 90} (2014) 083510}
  [\href{https://arxiv.org/abs/1408.3581}{{\ttfamily 1408.3581}}].

\bibitem{Grothaus:2014hja}
P.~Grothaus, M.~Fairbairn and J.~Monroe, \emph{{Directional Dark Matter
  Detection Beyond the Neutrino Bound}},
  \href{https://doi.org/10.1103/PhysRevD.90.055018}{\emph{Phys. Rev. D}
  {\bfseries 90} (2014) 055018}
  [\href{https://arxiv.org/abs/1406.5047}{{\ttfamily 1406.5047}}].

\bibitem{OHare:2015utx}
C.A.J.~O'Hare, A.M.~Green, J.~Billard, E.~Figueroa-Feliciano and L.E.~Strigari,
  \emph{{Readout strategies for directional dark matter detection beyond the
  neutrino background}},
  \href{https://doi.org/10.1103/PhysRevD.92.063518}{\emph{Phys. Rev. D}
  {\bfseries 92} (2015) 063518}
  [\href{https://arxiv.org/abs/1505.08061}{{\ttfamily 1505.08061}}].

\bibitem{Dent:2016iht}
J.B.~Dent, B.~Dutta, J.L.~Newstead and L.E.~Strigari, \emph{{Effective field
  theory treatment of the neutrino background in direct dark matter detection
  experiments}}, \href{https://doi.org/10.1103/PhysRevD.93.075018}{\emph{Phys.
  Rev. D} {\bfseries 93} (2016) 075018}
  [\href{https://arxiv.org/abs/1602.05300}{{\ttfamily 1602.05300}}].

\bibitem{Boehm:2018sux}
C.~Boe{}hm, D.~Cerde\~no, P.~Machado, A.~Olivares-Del~Campo, E.~Perdomo and
  E.~Reid, \emph{{How high is the neutrino floor?}},
  \href{https://doi.org/10.1088/1475-7516/2019/01/043}{\emph{JCAP} {\bfseries
  01} (2019) 043} [\href{https://arxiv.org/abs/1809.06385}{{\ttfamily
  1809.06385}}].

\bibitem{Hambye:2012fh}
T.~Hambye, \emph{{Leptogenesis: beyond the minimal type I seesaw scenario}},
  \href{https://doi.org/10.1088/1367-2630/14/12/125014}{\emph{New J. Phys.}
  {\bfseries 14} (2012) 125014}
  [\href{https://arxiv.org/abs/1212.2888}{{\ttfamily 1212.2888}}].

\bibitem{Dick:1999je}
K.~Dick, M.~Lindner, M.~Ratz and D.~Wright, \emph{{Leptogenesis with Dirac
  neutrinos}}, \href{https://doi.org/10.1103/PhysRevLett.84.4039}{\emph{Phys.
  Rev. Lett.} {\bfseries 84} (2000) 4039}
  [\href{https://arxiv.org/abs/hep-ph/9907562}{{\ttfamily hep-ph/9907562}}].

\bibitem{Buchmuller:2004nz}
W.~Buchmuller, P.~Di~Bari and M.~Plumacher, \emph{{Leptogenesis for
  pedestrians}}, \href{https://doi.org/10.1016/j.aop.2004.02.003}{\emph{Annals
  Phys.} {\bfseries 315} (2005) 305}
  [\href{https://arxiv.org/abs/hep-ph/0401240}{{\ttfamily hep-ph/0401240}}].

\bibitem{Dev:2017wwc}
B.~Dev, M.~Garny, J.~Klaric, P.~Millington and D.~Teresi, \emph{{Resonant
  enhancement in leptogenesis}},
  \href{https://doi.org/10.1142/S0217751X18420034}{\emph{Int. J. Mod. Phys. A}
  {\bfseries 33} (2018) 1842003}
  [\href{https://arxiv.org/abs/1711.02863}{{\ttfamily 1711.02863}}].

\bibitem{Drewes:2017zyw}
M.~Drewes, B.~Garbrecht, P.~Hernandez, M.~Kekic, J.~Lopez-Pavon, J.~Racker
  et~al., \emph{{ARS Leptogenesis}},
  \href{https://doi.org/10.1142/S0217751X18420022}{\emph{Int. J. Mod. Phys. A}
  {\bfseries 33} (2018) 1842002}
  [\href{https://arxiv.org/abs/1711.02862}{{\ttfamily 1711.02862}}].

\bibitem{Chun:2017spz}
E.~Chun et~al., \emph{{Probing Leptogenesis}},
  \href{https://doi.org/10.1142/S0217751X18420058}{\emph{Int. J. Mod. Phys. A}
  {\bfseries 33} (2018) 1842005}
  [\href{https://arxiv.org/abs/1711.02865}{{\ttfamily 1711.02865}}].

\bibitem{Bodeker:2020ghk}
D.~Bodeker and W.~Buchmuller, \emph{{Baryogenesis from the weak scale to the
  GUT scale}},  \href{https://arxiv.org/abs/2009.07294}{{\ttfamily
  2009.07294}}.

\bibitem{Pascoli:2006ci}
S.~Pascoli, S.~Petcov and A.~Riotto, \emph{{Leptogenesis and Low Energy CP
  Violation in Neutrino Physics}},
  \href{https://doi.org/10.1016/j.nuclphysb.2007.02.019}{\emph{Nucl.\ Phys.\ B}
  {\bfseries 774} (2007) 1}
  [\href{https://arxiv.org/abs/hep-ph/0611338}{{\ttfamily hep-ph/0611338}}].

\bibitem{Anisimov:2007mw}
A.~Anisimov, S.~Blanchet and P.~Di~Bari, \emph{{Viability of Dirac phase
  leptogenesis}},
  \href{https://doi.org/10.1088/1475-7516/2008/04/033}{\emph{JCAP} {\bfseries
  04} (2008) 033} [\href{https://arxiv.org/abs/0707.3024}{{\ttfamily
  0707.3024}}].

\bibitem{Moffat:2018smo}
K.~Moffat, S.~Pascoli, S.~Petcov and J.~Turner, \emph{{Leptogenesis from Low
  Energy $CP$ Violation}},
  \href{https://doi.org/10.1007/JHEP03(2019)034}{\emph{JHEP} {\bfseries 03}
  (2019) 034} [\href{https://arxiv.org/abs/1809.08251}{{\ttfamily
  1809.08251}}].

\bibitem{Dolan:2018qpy}
M.J.~Dolan, T.P.~Dutka and R.R.~Volkas, \emph{{Dirac-Phase Thermal Leptogenesis
  in the extended Type-I Seesaw Model}},
  \href{https://doi.org/10.1088/1475-7516/2018/06/012}{\emph{JCAP} {\bfseries
  06} (2018) 012} [\href{https://arxiv.org/abs/1802.08373}{{\ttfamily
  1802.08373}}].

\bibitem{Akhmedov:1998qx}
E.K.~Akhmedov, V.~Rubakov and A.~Smirnov, \emph{{Baryogenesis via neutrino
  oscillations}},
  \href{https://doi.org/10.1103/PhysRevLett.81.1359}{\emph{Phys. Rev. Lett.}
  {\bfseries 81} (1998) 1359}
  [\href{https://arxiv.org/abs/hep-ph/9803255}{{\ttfamily hep-ph/9803255}}].

\bibitem{Canetti:2012kh}
L.~Canetti, M.~Drewes, T.~Frossard and M.~Shaposhnikov, \emph{{Dark Matter,
  Baryogenesis and Neutrino Oscillations from Right Handed Neutrinos}},
  \href{https://doi.org/10.1103/PhysRevD.87.093006}{\emph{Phys. Rev. D}
  {\bfseries 87} (2013) 093006}
  [\href{https://arxiv.org/abs/1208.4607}{{\ttfamily 1208.4607}}].

\bibitem{Canetti:2012vf}
L.~Canetti, M.~Drewes and M.~Shaposhnikov, \emph{{Sterile Neutrinos as the
  Origin of Dark and Baryonic Matter}},
  \href{https://doi.org/10.1103/PhysRevLett.110.061801}{\emph{Phys. Rev. Lett.}
  {\bfseries 110} (2013) 061801}
  [\href{https://arxiv.org/abs/1204.3902}{{\ttfamily 1204.3902}}].

\bibitem{Alekhin:2015byh}
S.~Alekhin et~al., \emph{{A facility to Search for Hidden Particles at the CERN
  SPS: the SHiP physics case}},
  \href{https://doi.org/10.1088/0034-4885/79/12/124201}{\emph{Rept. Prog.
  Phys.} {\bfseries 79} (2016) 124201}
  [\href{https://arxiv.org/abs/1504.04855}{{\ttfamily 1504.04855}}].

\bibitem{Agrawal:2021dbo}
P.~Agrawal et~al., \emph{{Feebly-Interacting Particles:FIPs 2020 Workshop
  Report}},  \href{https://arxiv.org/abs/2102.12143}{{\ttfamily 2102.12143}}.

\bibitem{Asaka:2005cn}
T.~Asaka, K.~Ishiwata and T.~Moroi, \emph{{Right-handed sneutrino as cold dark
  matter}}, \href{https://doi.org/10.1103/PhysRevD.73.051301}{\emph{Phys. Rev.
  D} {\bfseries 73} (2006) 051301}
  [\href{https://arxiv.org/abs/hep-ph/0512118}{{\ttfamily hep-ph/0512118}}].

\bibitem{EliasMiro:2011aa}
J.~Elias-Miro, J.R.~Espinosa, G.F.~Giudice, G.~Isidori, A.~Riotto and
  A.~Strumia, \emph{{Higgs mass implications on the stability of the
  electroweak vacuum}},
  \href{https://doi.org/10.1016/j.physletb.2012.02.013}{\emph{Phys. Lett. B}
  {\bfseries 709} (2012) 222}
  [\href{https://arxiv.org/abs/1112.3022}{{\ttfamily 1112.3022}}].

\bibitem{Vissani:1997ys}
F.~Vissani, \emph{{Do experiments suggest a hierarchy problem?}},
  \href{https://doi.org/10.1103/PhysRevD.57.7027}{\emph{Phys. Rev. D}
  {\bfseries 57} (1998) 7027}
  [\href{https://arxiv.org/abs/hep-ph/9709409}{{\ttfamily hep-ph/9709409}}].

\bibitem{Casas:2004gh}
J.~Casas, J.~Espinosa and I.~Hidalgo, \emph{{Implications for new physics from
  fine-tuning arguments. 1. Application to SUSY and seesaw cases}},
  \href{https://doi.org/10.1088/1126-6708/2004/11/057}{\emph{JHEP} {\bfseries
  11} (2004) 057} [\href{https://arxiv.org/abs/hep-ph/0410298}{{\ttfamily
  hep-ph/0410298}}].

\bibitem{Abada:2007ux}
A.~Abada, C.~Biggio, F.~Bonnet, M.~Gavela and T.~Hambye, \emph{{Low energy
  effects of neutrino masses}},
  \href{https://doi.org/10.1088/1126-6708/2007/12/061}{\emph{JHEP} {\bfseries
  12} (2007) 061} [\href{https://arxiv.org/abs/0707.4058}{{\ttfamily
  0707.4058}}].

\bibitem{Clarke:2015gwa}
J.D.~Clarke, R.~Foot and R.R.~Volkas, \emph{{Electroweak naturalness in the
  three-flavor type I seesaw model and implications for leptogenesis}},
  \href{https://doi.org/10.1103/PhysRevD.91.073009}{\emph{Phys. Rev. D}
  {\bfseries 91} (2015) 073009}
  [\href{https://arxiv.org/abs/1502.01352}{{\ttfamily 1502.01352}}].

\bibitem{Mohapatra:1980qe}
R.N.~Mohapatra and R.~Marshak, \emph{{Local B-L Symmetry of Electroweak
  Interactions, Majorana Neutrinos and Neutron Oscillations}},
  \href{https://doi.org/10.1103/PhysRevLett.44.1316}{\emph{Phys. Rev. Lett.}
  {\bfseries 44} (1980) 1316}.

\bibitem{Marshak:1979fm}
R.~Marshak and R.N.~Mohapatra, \emph{{Quark - Lepton Symmetry and B-L as the
  U(1) Generator of the Electroweak Symmetry Group}},
  \href{https://doi.org/10.1016/0370-2693(80)90436-0}{\emph{Phys. Lett. B}
  {\bfseries 91} (1980) 222}.

\bibitem{Masiero:1982fi}
A.~Masiero, J.~Nieves and T.~Yanagida, \emph{{$B^-$l Violating Proton Decay and
  Late Cosmological Baryon Production}},
  \href{https://doi.org/10.1016/0370-2693(82)90024-7}{\emph{Phys. Lett. B}
  {\bfseries 116} (1982) 11}.

\bibitem{Mohapatra:1982xz}
R.N.~Mohapatra and G.~Senjanovic, \emph{{Spontaneous Breaking of Global $B^-$l
  Symmetry and Matter - Antimatter Oscillations in Grand Unified Theories}},
  \href{https://doi.org/10.1103/PhysRevD.27.254}{\emph{Phys. Rev. D} {\bfseries
  27} (1983) 254}.

\bibitem{Buchmuller:1991ce}
W.~Buchmuller, C.~Greub and P.~Minkowski, \emph{{Neutrino masses, neutral
  vector bosons and the scale of B-L breaking}},
  \href{https://doi.org/10.1016/0370-2693(91)90952-M}{\emph{Phys. Lett. B}
  {\bfseries 267} (1991) 395}.

\bibitem{Pati:1974yy}
J.C.~Pati and A.~Salam, \emph{{Lepton Number as the Fourth Color}},
  \href{https://doi.org/10.1103/PhysRevD.10.275}{\emph{Phys. Rev. D} {\bfseries
  10} (1974) 275}.

\bibitem{Mohapatra:1974gc}
R.~Mohapatra and J.C.~Pati, \emph{{A Natural Left-Right Symmetry}},
  \href{https://doi.org/10.1103/PhysRevD.11.2558}{\emph{Phys. Rev. D}
  {\bfseries 11} (1975) 2558}.

\bibitem{Mohapatra:1974hk}
R.N.~Mohapatra and J.C.~Pati, \emph{{Left-Right Gauge Symmetry and an
  Isoconjugate Model of CP Violation}},
  \href{https://doi.org/10.1103/PhysRevD.11.566}{\emph{Phys. Rev. D} {\bfseries
  11} (1975) 566}.

\bibitem{Senjanovic:1975rk}
G.~Senjanovic and R.N.~Mohapatra, \emph{{Exact Left-Right Symmetry and
  Spontaneous Violation of Parity}},
  \href{https://doi.org/10.1103/PhysRevD.12.1502}{\emph{Phys. Rev. D}
  {\bfseries 12} (1975) 1502}.

\bibitem{Senjanovic:1978ev}
G.~Senjanovic, \emph{{Spontaneous Breakdown of Parity in a Class of Gauge
  Theories}}, \href{https://doi.org/10.1016/0550-3213(79)90604-7}{\emph{Nucl.
  Phys. B} {\bfseries 153} (1979) 334}.

\bibitem{Senjanovic:2016bya}
G.~Senjanovic, \emph{{Is left-right symmetry the key?}},
  \href{https://doi.org/10.1142/S021773231730004X}{\emph{Mod. Phys. Lett. A}
  {\bfseries 32} (2017) 1730004}
  [\href{https://arxiv.org/abs/1610.04209}{{\ttfamily 1610.04209}}].

\bibitem{Ross:1985ai}
G.G.~Ross, \emph{{GRAND UNIFIED THEORIES}} (1, 1985).

\bibitem{Dutta:2004zh}
B.~Dutta, Y.~Mimura and R.~Mohapatra, \emph{{Suppressing proton decay in the
  minimal SO(10) model}},
  \href{https://doi.org/10.1103/PhysRevLett.94.091804}{\emph{Phys. Rev. Lett.}
  {\bfseries 94} (2005) 091804}
  [\href{https://arxiv.org/abs/hep-ph/0412105}{{\ttfamily hep-ph/0412105}}].

\bibitem{Dutta:2005ni}
B.~Dutta, Y.~Mimura and R.~Mohapatra, \emph{{Neutrino mixing predictions of a
  minimal SO(10) model with suppressed proton decay}},
  \href{https://doi.org/10.1103/PhysRevD.72.075009}{\emph{Phys. Rev. D}
  {\bfseries 72} (2005) 075009}
  [\href{https://arxiv.org/abs/hep-ph/0507319}{{\ttfamily hep-ph/0507319}}].

\bibitem{Senjanovic:2006nc}
G.~Senjanovic, \emph{{SO(10): A Theory of fermion masses and mixings}},  in
  \emph{{GUSTAVOFEST: Symposium in Honor of Gustavo C. Branco: CP Violation and
  the Flavor Puzzle}}, 12, 2006
  [\href{https://arxiv.org/abs/hep-ph/0612312}{{\ttfamily hep-ph/0612312}}].

\bibitem{Bertolini:2006pe}
S.~Bertolini, T.~Schwetz and M.~Malinsky, \emph{{Fermion masses and mixings in
  SO(10) models and the neutrino challenge to SUSY GUTs}},
  \href{https://doi.org/10.1103/PhysRevD.73.115012}{\emph{Phys. Rev. D}
  {\bfseries 73} (2006) 115012}
  [\href{https://arxiv.org/abs/hep-ph/0605006}{{\ttfamily hep-ph/0605006}}].

\bibitem{Takenaka:2020vqy}
{\scshape Super-Kamiokande} collaboration, \emph{{Search for proton decay via
  $p\to e^+\pi^0$ and $p\to \mu^+\pi^0$ with an enlarged fiducial volume in
  Super-Kamiokande I-IV}},
  \href{https://doi.org/10.1103/PhysRevD.102.112011}{\emph{Phys. Rev. D}
  {\bfseries 102} (2020) 112011}
  [\href{https://arxiv.org/abs/2010.16098}{{\ttfamily 2010.16098}}].

\bibitem{Drewes:2013gca}
M.~Drewes, \emph{{The Phenomenology of Right Handed Neutrinos}},
  \href{https://doi.org/10.1142/S0218301313300191}{\emph{Int. J. Mod. Phys. E}
  {\bfseries 22} (2013) 1330019}
  [\href{https://arxiv.org/abs/1303.6912}{{\ttfamily 1303.6912}}].

\bibitem{Asaka:2005pn}
T.~Asaka and M.~Shaposhnikov, \emph{{The $\nu$MSM, dark matter and baryon
  asymmetry of the universe}},
  \href{https://doi.org/10.1016/j.physletb.2005.06.020}{\emph{Phys. Lett. B}
  {\bfseries 620} (2005) 17}
  [\href{https://arxiv.org/abs/hep-ph/0505013}{{\ttfamily hep-ph/0505013}}].

\bibitem{Ohlsson:2012kf}
T.~Ohlsson, \emph{{Status of non-standard neutrino interactions}},
  \href{https://doi.org/10.1088/0034-4885/76/4/044201}{\emph{Rept. Prog. Phys.}
  {\bfseries 76} (2013) 044201}
  [\href{https://arxiv.org/abs/1209.2710}{{\ttfamily 1209.2710}}].

\bibitem{Farzan:2017xzy}
Y.~Farzan and M.~Tortola, \emph{{Neutrino oscillations and Non-Standard
  Interactions}}, \href{https://doi.org/10.3389/fphy.2018.00010}{\emph{Front.
  in Phys.} {\bfseries 6} (2018) 10}
  [\href{https://arxiv.org/abs/1710.09360}{{\ttfamily 1710.09360}}].

\bibitem{Dev:2019anc}
P.S.~Bhupal~Dev et~al., \emph{{Neutrino Non-Standard Interactions: A Status
  Report}},  \href{https://arxiv.org/abs/1907.00991}{{\ttfamily 1907.00991}}.

\bibitem{Babu:2020nna}
K.S.~Babu, D.~Gon\c{c}alves, S.~Jana and P.A.N.~Machado, \emph{{Neutrino
  Non-Standard Interactions: Complementarity Between LHC and Oscillation
  Experiments}},
  \href{https://doi.org/10.1016/j.physletb.2021.136131}{\emph{Phys. Lett. B}
  {\bfseries 815} (2021) 136131}
  [\href{https://arxiv.org/abs/2003.03383}{{\ttfamily 2003.03383}}].

\bibitem{Esteban:2018ppq}
I.~Esteban, M.~Gonzalez-Garcia, M.~Maltoni, I.~Martinez-Soler and J.~Salvado,
  \emph{{Updated constraints on non-standard interactions from global analysis
  of oscillation data}},
  \href{https://doi.org/10.1007/JHEP08(2018)180}{\emph{JHEP} {\bfseries 08}
  (2018) 180} [\href{https://arxiv.org/abs/1805.04530}{{\ttfamily
  1805.04530}}].

\bibitem{Farzan:2015hkd}
Y.~Farzan and I.M.~Shoemaker, \emph{{Lepton Flavor Violating Non-Standard
  Interactions via Light Mediators}},
  \href{https://doi.org/10.1007/JHEP07(2016)033}{\emph{JHEP} {\bfseries 07}
  (2016) 033} [\href{https://arxiv.org/abs/1512.09147}{{\ttfamily
  1512.09147}}].

\bibitem{Ge:2018uhz}
S.-F.~Ge and S.J.~Parke, \emph{{Scalar Nonstandard Interactions in Neutrino
  Oscillation}},
  \href{https://doi.org/10.1103/PhysRevLett.122.211801}{\emph{Phys. Rev. Lett.}
  {\bfseries 122} (2019) 211801}
  [\href{https://arxiv.org/abs/1812.08376}{{\ttfamily 1812.08376}}].

\bibitem{Babu:2019iml}
K.S.~Babu, G.~Chauhan and P.S.~Bhupal~Dev, \emph{{Neutrino nonstandard
  interactions via light scalars in the Earth, Sun, supernovae, and the early
  Universe}}, \href{https://doi.org/10.1103/PhysRevD.101.095029}{\emph{Phys.
  Rev. D} {\bfseries 101} (2020) 095029}
  [\href{https://arxiv.org/abs/1912.13488}{{\ttfamily 1912.13488}}].

\bibitem{Kovalenko:2013eba}
S.~Kovalenko, M.I.~Krivoruchenko and F.~Simkovic, \emph{{Neutrino propagation
  in nuclear medium and neutrinoless double-beta decay}},
  \href{https://doi.org/10.1103/PhysRevLett.112.142503}{\emph{Phys. Rev. Lett.}
  {\bfseries 112} (2014) 142503}
  [\href{https://arxiv.org/abs/1311.4200}{{\ttfamily 1311.4200}}].

\bibitem{Cirgiliano:2019nyn}
V.~Cirigliano, A.~Garcia, D.~Gazit, O.~Naviliat-Cuncic, G.~Savard and A.~Young,
  \emph{{Precision Beta Decay as a Probe of New Physics}},
  \href{https://arxiv.org/abs/1907.02164}{{\ttfamily 1907.02164}}.

\bibitem{Bischer:2019ttk}
I.~Bischer and W.~Rodejohann, \emph{{General neutrino interactions from an
  effective field theory perspective}},
  \href{https://doi.org/10.1016/j.nuclphysb.2019.114746}{\emph{Nucl. Phys. B}
  {\bfseries 947} (2019) 114746}
  [\href{https://arxiv.org/abs/1905.08699}{{\ttfamily 1905.08699}}].

\bibitem{Bernal:2016gxb}
J.L.~Bernal, L.~Verde and A.G.~Riess, \emph{{The trouble with $H_0$}},
  \href{https://doi.org/10.1088/1475-7516/2016/10/019}{\emph{JCAP} {\bfseries
  10} (2016) 019} [\href{https://arxiv.org/abs/1607.05617}{{\ttfamily
  1607.05617}}].

\bibitem{Aprile:2020tmw}
{\scshape XENON} collaboration, \emph{{Excess electronic recoil events in
  XENON1T}}, \href{https://doi.org/10.1103/PhysRevD.102.072004}{\emph{Phys.
  Rev. D} {\bfseries 102} (2020) 072004}
  [\href{https://arxiv.org/abs/2006.09721}{{\ttfamily 2006.09721}}].

\bibitem{Antusch:2006vwa}
S.~Antusch, C.~Biggio, E.~Fernandez-Martinez, M.~Gavela and J.~Lopez-Pavon,
  \emph{{Unitarity of the Leptonic Mixing Matrix}},
  \href{https://doi.org/10.1088/1126-6708/2006/10/084}{\emph{JHEP} {\bfseries
  10} (2006) 084} [\href{https://arxiv.org/abs/hep-ph/0607020}{{\ttfamily
  hep-ph/0607020}}].

\bibitem{Fernandez-Martinez:2016lgt}
E.~Fernandez-Martinez, J.~Hernandez-Garcia and J.~Lopez-Pavon, \emph{{Global
  constraints on heavy neutrino mixing}},
  \href{https://doi.org/10.1007/JHEP08(2016)033}{\emph{JHEP} {\bfseries 08}
  (2016) 033} [\href{https://arxiv.org/abs/1605.08774}{{\ttfamily
  1605.08774}}].

\bibitem{Parke:2015goa}
S.~Parke and M.~Ross-Lonergan, \emph{{Unitarity and the three flavor neutrino
  mixing matrix}},
  \href{https://doi.org/10.1103/PhysRevD.93.113009}{\emph{Phys. Rev. D}
  {\bfseries 93} (2016) 113009}
  [\href{https://arxiv.org/abs/1508.05095}{{\ttfamily 1508.05095}}].

\bibitem{deGouvea:2019ozk}
A.~De~Gouv\^ea, K.J.~Kelly, G.~Stenico and P.~Pasquini, \emph{{Physics with
  Beam Tau-Neutrino Appearance at DUNE}},
  \href{https://doi.org/10.1103/PhysRevD.100.016004}{\emph{Phys. Rev. D}
  {\bfseries 100} (2019) 016004}
  [\href{https://arxiv.org/abs/1904.07265}{{\ttfamily 1904.07265}}].

\bibitem{Escrihuela:2016ube}
F.~Escrihuela, D.~Forero, O.~Miranda, M.~T\'ortola and J.~Valle, \emph{{Probing
  CP violation with non-unitary mixing in long-baseline neutrino oscillation
  experiments: DUNE as a case study}},
  \href{https://doi.org/10.1088/1367-2630/aa79ec}{\emph{New J. Phys.}
  {\bfseries 19} (2017) 093005}
  [\href{https://arxiv.org/abs/1612.07377}{{\ttfamily 1612.07377}}].

\bibitem{Giunti:2008ve}
C.~Giunti and A.~Studenikin, \emph{{Neutrino electromagnetic properties}},
  \href{https://doi.org/10.1134/S1063778809120126}{\emph{Phys. Atom. Nucl.}
  {\bfseries 72} (2009) 2089}
  [\href{https://arxiv.org/abs/0812.3646}{{\ttfamily 0812.3646}}].

\bibitem{Raffelt:1999gv}
G.G.~Raffelt, \emph{{Limits on neutrino electromagnetic properties: An
  update}}, \href{https://doi.org/10.1016/S0370-1573(99)00074-5}{\emph{Phys.
  Rept.} {\bfseries 320} (1999) 319}.

\bibitem{Joshipura:2003jh}
A.S.~Joshipura and S.~Mohanty, \emph{{Constraints on flavor dependent long
  range forces from atmospheric neutrino observations at super-Kamiokande}},
  \href{https://doi.org/10.1016/j.physletb.2004.01.057}{\emph{Phys. Lett. B}
  {\bfseries 584} (2004) 103}
  [\href{https://arxiv.org/abs/hep-ph/0310210}{{\ttfamily hep-ph/0310210}}].

\bibitem{Grifols:2003gy}
J.A.~Grifols and E.~Masso, \emph{{Neutrino oscillations in the sun probe long
  range leptonic forces}},
  \href{https://doi.org/10.1016/j.physletb.2003.10.078}{\emph{Phys. Lett. B}
  {\bfseries 579} (2004) 123}
  [\href{https://arxiv.org/abs/hep-ph/0311141}{{\ttfamily hep-ph/0311141}}].

\bibitem{Heeck:2010pg}
J.~Heeck and W.~Rodejohann, \emph{{Gauged $L_\mu - L_\tau$ and different Muon
  Neutrino and Anti-Neutrino Oscillations: MINOS and beyond}},
  \href{https://doi.org/10.1088/0954-3899/38/8/085005}{\emph{J. Phys. G}
  {\bfseries 38} (2011) 085005}
  [\href{https://arxiv.org/abs/1007.2655}{{\ttfamily 1007.2655}}].

\bibitem{Bustamante:2018mzu}
M.~Bustamante and S.K.~Agarwalla, \emph{{Universe's Worth of Electrons to Probe
  Long-Range Interactions of High-Energy Astrophysical Neutrinos}},
  \href{https://doi.org/10.1103/PhysRevLett.122.061103}{\emph{Phys. Rev. Lett.}
  {\bfseries 122} (2019) 061103}
  [\href{https://arxiv.org/abs/1808.02042}{{\ttfamily 1808.02042}}].

\bibitem{Smirnov:2019cae}
A.Y.~Smirnov and X.-J.~Xu, \emph{{Wolfenstein potentials for neutrinos induced
  by ultra-light mediators}},
  \href{https://doi.org/10.1007/JHEP12(2019)046}{\emph{JHEP} {\bfseries 12}
  (2019) 046} [\href{https://arxiv.org/abs/1909.07505}{{\ttfamily
  1909.07505}}].

\bibitem{Coloma:2020gfv}
P.~Coloma, M.C.~Gonzalez-Garcia and M.~Maltoni, \emph{{Neutrino oscillation
  constraints on U(1)' models: from non-standard interactions to long-range
  forces}}, \href{https://doi.org/10.1007/JHEP01(2021)114}{\emph{JHEP}
  {\bfseries 01} (2021) 114}
  [\href{https://arxiv.org/abs/2009.14220}{{\ttfamily 2009.14220}}].

\bibitem{Bustamante:2016ciw}
M.~Bustamante, J.F.~Beacom and K.~Murase, \emph{{Testing decay of astrophysical
  neutrinos with incomplete information}},
  \href{https://doi.org/10.1103/PhysRevD.95.063013}{\emph{Phys. Rev. D}
  {\bfseries 95} (2017) 063013}
  [\href{https://arxiv.org/abs/1610.02096}{{\ttfamily 1610.02096}}].

\bibitem{Denton:2018aml}
P.B.~Denton and I.~Tamborra, \emph{{Invisible Neutrino Decay Could Resolve
  IceCube\textquoteright{}s Track and Cascade Tension}},
  \href{https://doi.org/10.1103/PhysRevLett.121.121802}{\emph{Phys. Rev. Lett.}
  {\bfseries 121} (2018) 121802}
  [\href{https://arxiv.org/abs/1805.05950}{{\ttfamily 1805.05950}}].

\bibitem{Fogli:1999qt}
G.L.~Fogli, E.~Lisi, A.~Marrone and G.~Scioscia, \emph{{Super-Kamiokande data
  and atmospheric neutrino decay}},
  \href{https://doi.org/10.1103/PhysRevD.59.117303}{\emph{Phys. Rev. D}
  {\bfseries 59} (1999) 117303}
  [\href{https://arxiv.org/abs/hep-ph/9902267}{{\ttfamily hep-ph/9902267}}].

\bibitem{Gago:2017zzy}
A.M.~Gago, R.A.~Gomes, A.L.G.~Gomes, J.~Jones-Perez and O.L.G.~Peres,
  \emph{{Visible neutrino decay in the light of appearance and disappearance
  long baseline experiments}},
  \href{https://doi.org/10.1007/JHEP11(2017)022}{\emph{JHEP} {\bfseries 11}
  (2017) 022} [\href{https://arxiv.org/abs/1705.03074}{{\ttfamily
  1705.03074}}].

\bibitem{Choubey:2018cfz}
S.~Choubey, D.~Dutta and D.~Pramanik, \emph{{Invisible neutrino decay in the
  light of NOvA and T2K data}},
  \href{https://doi.org/10.1007/JHEP08(2018)141}{\emph{JHEP} {\bfseries 08}
  (2018) 141} [\href{https://arxiv.org/abs/1805.01848}{{\ttfamily
  1805.01848}}].

\bibitem{Porto-Silva:2020gma}
Y.P.~Porto-Silva, S.~Prakash, O.L.G.~Peres, H.~Nunokawa and H.~Minakata,
  \emph{{Constraining visible neutrino decay at KamLAND and JUNO}},
  \href{https://doi.org/10.1140/epjc/s10052-020-08573-9}{\emph{Eur. Phys. J. C}
  {\bfseries 80} (2020) 999}
  [\href{https://arxiv.org/abs/2002.12134}{{\ttfamily 2002.12134}}].

\bibitem{Picoreti:2015ika}
R.~Picoreti, M.M.~Guzzo, P.C.~de~Holanda and O.L.G.~Peres, \emph{{Neutrino
  Decay and Solar Neutrino Seasonal Effect}},
  \href{https://doi.org/10.1016/j.physletb.2016.08.007}{\emph{Phys. Lett. B}
  {\bfseries 761} (2016) 70}
  [\href{https://arxiv.org/abs/1506.08158}{{\ttfamily 1506.08158}}].

\bibitem{Barenboim:2020vrr}
G.~Barenboim, J.Z.~Chen, S.~Hannestad, I.M.~Oldengott, T.~Tram and Y.Y.Y.~Wong,
  \emph{{Invisible neutrino decay in precision cosmology}},
  \href{https://doi.org/10.1088/1475-7516/2021/03/087}{\emph{JCAP} {\bfseries
  03} (2021) 087} [\href{https://arxiv.org/abs/2011.01502}{{\ttfamily
  2011.01502}}].

\bibitem{Frieman:1987as}
J.A.~Frieman, H.E.~Haber and K.~Freese, \emph{{Neutrino Mixing, Decays and
  Supernova Sn1987a}},
  \href{https://doi.org/10.1016/0370-2693(88)91120-3}{\emph{Phys. Lett. B}
  {\bfseries 200} (1988) 115}.

\bibitem{Anamiati:2019maf}
G.~Anamiati, V.~De~Romeri, M.~Hirsch, C.A.~Ternes and M.~T\'ortola,
  \emph{{Quasi-Dirac neutrino oscillations at DUNE and JUNO}},
  \href{https://doi.org/10.1103/PhysRevD.100.035032}{\emph{Phys. Rev. D}
  {\bfseries 100} (2019) 035032}
  [\href{https://arxiv.org/abs/1907.00980}{{\ttfamily 1907.00980}}].

\bibitem{Joshipura:2000ts}
A.S.~Joshipura and S.D.~Rindani, \emph{{Phenomenology of pseudoDirac
  neutrinos}}, \href{https://doi.org/10.1016/S0370-2693(00)01148-5}{\emph{Phys.
  Lett. B} {\bfseries 494} (2000) 114}
  [\href{https://arxiv.org/abs/hep-ph/0007334}{{\ttfamily hep-ph/0007334}}].

\bibitem{Beacom:2003eu}
J.F.~Beacom, N.F.~Bell, D.~Hooper, J.G.~Learned, S.~Pakvasa and T.J.~Weiler,
  \emph{{PseudoDirac neutrinos: A Challenge for neutrino telescopes}},
  \href{https://doi.org/10.1103/PhysRevLett.92.011101}{\emph{Phys. Rev. Lett.}
  {\bfseries 92} (2004) 011101}
  [\href{https://arxiv.org/abs/hep-ph/0307151}{{\ttfamily hep-ph/0307151}}].

\bibitem{deGouvea:2009fp}
A.~de~Gouvea, W.-C.~Huang and J.~Jenkins, \emph{{Pseudo-Dirac Neutrinos in the
  New Standard Model}},
  \href{https://doi.org/10.1103/PhysRevD.80.073007}{\emph{Phys. Rev. D}
  {\bfseries 80} (2009) 073007}
  [\href{https://arxiv.org/abs/0906.1611}{{\ttfamily 0906.1611}}].

\bibitem{Allahverdi:2010us}
R.~Allahverdi, B.~Dutta and R.N.~Mohapatra, \emph{{Schizophrenic Neutrinos and
  $\nu$-less Double Beta Decay}},
  \href{https://doi.org/10.1016/j.physletb.2010.11.006}{\emph{Phys. Lett. B}
  {\bfseries 695} (2011) 181}
  [\href{https://arxiv.org/abs/1008.1232}{{\ttfamily 1008.1232}}].

\bibitem{Fogli:2003th}
G.L.~Fogli, E.~Lisi, A.~Marrone and D.~Montanino, \emph{{Status of atmospheric
  nu(mu) ---\ensuremath{>} nu(tau) oscillations and decoherence after the first
  K2K spectral data}},
  \href{https://doi.org/10.1103/PhysRevD.67.093006}{\emph{Phys. Rev. D}
  {\bfseries 67} (2003) 093006}
  [\href{https://arxiv.org/abs/hep-ph/0303064}{{\ttfamily hep-ph/0303064}}].

\bibitem{Anchordoqui:2005gj}
L.A.~Anchordoqui, H.~Goldberg, M.C.~Gonzalez-Garcia, F.~Halzen, D.~Hooper,
  S.~Sarkar et~al., \emph{{Probing Planck scale physics with IceCube}},
  \href{https://doi.org/10.1103/PhysRevD.72.065019}{\emph{Phys. Rev.}
  {\bfseries D72} (2005) 065019}
  [\href{https://arxiv.org/abs/hep-ph/0506168}{{\ttfamily hep-ph/0506168}}].

\bibitem{Mavromatos:2007hv}
N.E.~Mavromatos, A.~Meregaglia, A.~Rubbia, A.~Sakharov and S.~Sarkar,
  \emph{{Quantum-Gravity Decoherence Effects in Neutrino Oscillations: Expected
  Constraints From CNGS and J-PARC}},
  \href{https://doi.org/10.1103/PhysRevD.77.053014}{\emph{Phys. Rev. D}
  {\bfseries 77} (2008) 053014}
  [\href{https://arxiv.org/abs/0801.0872}{{\ttfamily 0801.0872}}].

\bibitem{Stuttard:2020qfv}
T.~Stuttard and M.~Jensen, \emph{{Neutrino decoherence from quantum
  gravitational stochastic perturbations}},
  \href{https://doi.org/10.1103/PhysRevD.102.115003}{\emph{Phys. Rev. D}
  {\bfseries 102} (2020) 115003}
  [\href{https://arxiv.org/abs/2007.00068}{{\ttfamily 2007.00068}}].

\bibitem{Diaz:2016xpw}
J.S.~Diaz, \emph{{Testing Lorentz and CPT invariance with neutrinos}},
  \href{https://doi.org/10.3390/sym8100105}{\emph{Symmetry} {\bfseries 8}
  (2016) 105} [\href{https://arxiv.org/abs/1609.09474}{{\ttfamily
  1609.09474}}].

\bibitem{Murayama:2000hm}
H.~Murayama and T.~Yanagida, \emph{{LSND, SN1987A, and CPT violation}},
  \href{https://doi.org/10.1016/S0370-2693(01)01136-4}{\emph{Phys. Lett. B}
  {\bfseries 520} (2001) 263}
  [\href{https://arxiv.org/abs/hep-ph/0010178}{{\ttfamily hep-ph/0010178}}].

\bibitem{Barenboim:2001ac}
G.~Barenboim, L.~Borissov, J.D.~Lykken and A.Y.~Smirnov, \emph{{Neutrinos as
  the Messengers of CPT Violation}},
  \href{https://doi.org/10.1088/1126-6708/2002/10/001}{\emph{JHEP} {\bfseries
  10} (2002) 001} [\href{https://arxiv.org/abs/hep-ph/0108199}{{\ttfamily
  hep-ph/0108199}}].

\bibitem{Barenboim:2004wu}
G.~Barenboim and N.E.~Mavromatos, \emph{{CPT violating decoherence and LSND: A
  Possible window to Planck scale physics}},
  \href{https://doi.org/10.1088/1126-6708/2005/01/034}{\emph{JHEP} {\bfseries
  01} (2005) 034} [\href{https://arxiv.org/abs/hep-ph/0404014}{{\ttfamily
  hep-ph/0404014}}].

\bibitem{deGouvea:2017yvn}
A.~de~Gouv\^ea and K.J.~Kelly, \emph{{Neutrino vs. Antineutrino Oscillation
  Parameters at DUNE and Hyper-Kamiokande}},
  \href{https://doi.org/10.1103/PhysRevD.96.095018}{\emph{Phys. Rev. D}
  {\bfseries 96} (2017) 095018}
  [\href{https://arxiv.org/abs/1709.06090}{{\ttfamily 1709.06090}}].

\bibitem{Liao:2017yuy}
J.~Liao and D.~Marfatia, \emph{{IceCube’s astrophysical neutrino energy
  spectrum from CPT violation}},
  \href{https://doi.org/10.1103/PhysRevD.97.041302}{\emph{Phys.\ Rev.\ D}
  {\bfseries 97} (2018) 041302}
  [\href{https://arxiv.org/abs/1711.09266}{{\ttfamily 1711.09266}}].

\bibitem{Cohen:2011hx}
A.G.~Cohen and S.L.~Glashow, \emph{{Pair Creation Constrains Superluminal
  Neutrino Propagation}},
  \href{https://doi.org/10.1103/PhysRevLett.107.181803}{\emph{Phys. Rev. Lett.}
  {\bfseries 107} (2011) 181803}
  [\href{https://arxiv.org/abs/1109.6562}{{\ttfamily 1109.6562}}].

\bibitem{Adam:2011faa}
{\scshape OPERA} collaboration, \emph{{Measurement of the neutrino velocity
  with the OPERA detector in the CNGS beam}},
  \href{https://doi.org/10.1007/JHEP10(2012)093}{\emph{JHEP} {\bfseries 10}
  (2012) 093} [\href{https://arxiv.org/abs/1109.4897}{{\ttfamily 1109.4897}}].

\bibitem{Kostelecky:2003cr}
V.A.~Kostelecky and M.~Mewes, \emph{{Lorentz and CPT violation in neutrinos}},
  \href{https://doi.org/10.1103/PhysRevD.69.016005}{\emph{Phys. Rev. D}
  {\bfseries 69} (2004) 016005}
  [\href{https://arxiv.org/abs/hep-ph/0309025}{{\ttfamily hep-ph/0309025}}].

\bibitem{Kostelecky:2003fs}
V.A.~Kostelecky, \emph{{Gravity, Lorentz violation, and the standard model}},
  \href{https://doi.org/10.1103/PhysRevD.69.105009}{\emph{Phys. Rev. D}
  {\bfseries 69} (2004) 105009}
  [\href{https://arxiv.org/abs/hep-th/0312310}{{\ttfamily hep-th/0312310}}].

\bibitem{Diaz:2009qk}
J.S.~Diaz, V.A.~Kostelecky and M.~Mewes, \emph{{Perturbative Lorentz and CPT
  violation for neutrino and antineutrino oscillations}},
  \href{https://doi.org/10.1103/PhysRevD.80.076007}{\emph{Phys. Rev. D}
  {\bfseries 80} (2009) 076007}
  [\href{https://arxiv.org/abs/0908.1401}{{\ttfamily 0908.1401}}].

\bibitem{Katori:2016eni}
T.~Katori, C.A.~Argüelles and J.~Salvado, \emph{{Test of Lorentz Violation
  with Astrophysical Neutrino Flavor}},  in \emph{{7th Meeting on CPT and
  Lorentz Symmetry (CPT 16) Bloomington, Indiana, United States, June 20-24,
  2016}}, 2016 [\href{https://arxiv.org/abs/hep-ph/1607.08448}{{\ttfamily
  hep-ph/1607.08448}}].

\bibitem{Agarwalla:2019rgv}
S.~Kumar~Agarwalla and M.~Masud, \emph{{Can Lorentz invariance violation affect
  the sensitivity of deep underground neutrino experiment?}},
  \href{https://doi.org/10.1140/epjc/s10052-020-8303-1}{\emph{Eur. Phys. J. C}
  {\bfseries 80} (2020) 716}
  [\href{https://arxiv.org/abs/1912.13306}{{\ttfamily 1912.13306}}].

\end{thebibliography}\endgroup

\end{document}